%% file: version_02_06_19.tex
\documentclass[a4paper,nobind]{ociamthesis} 
\usepackage{pdfpages}
\correctionstrue

\include{./journalsVaps}
\usepackage{braket}

\usepackage{hyperref}
\usepackage{cleveref}
\usepackage[noadjust]{cite}
\usepackage{enumitem}

\begin{document}

\includepdf[pages=1]{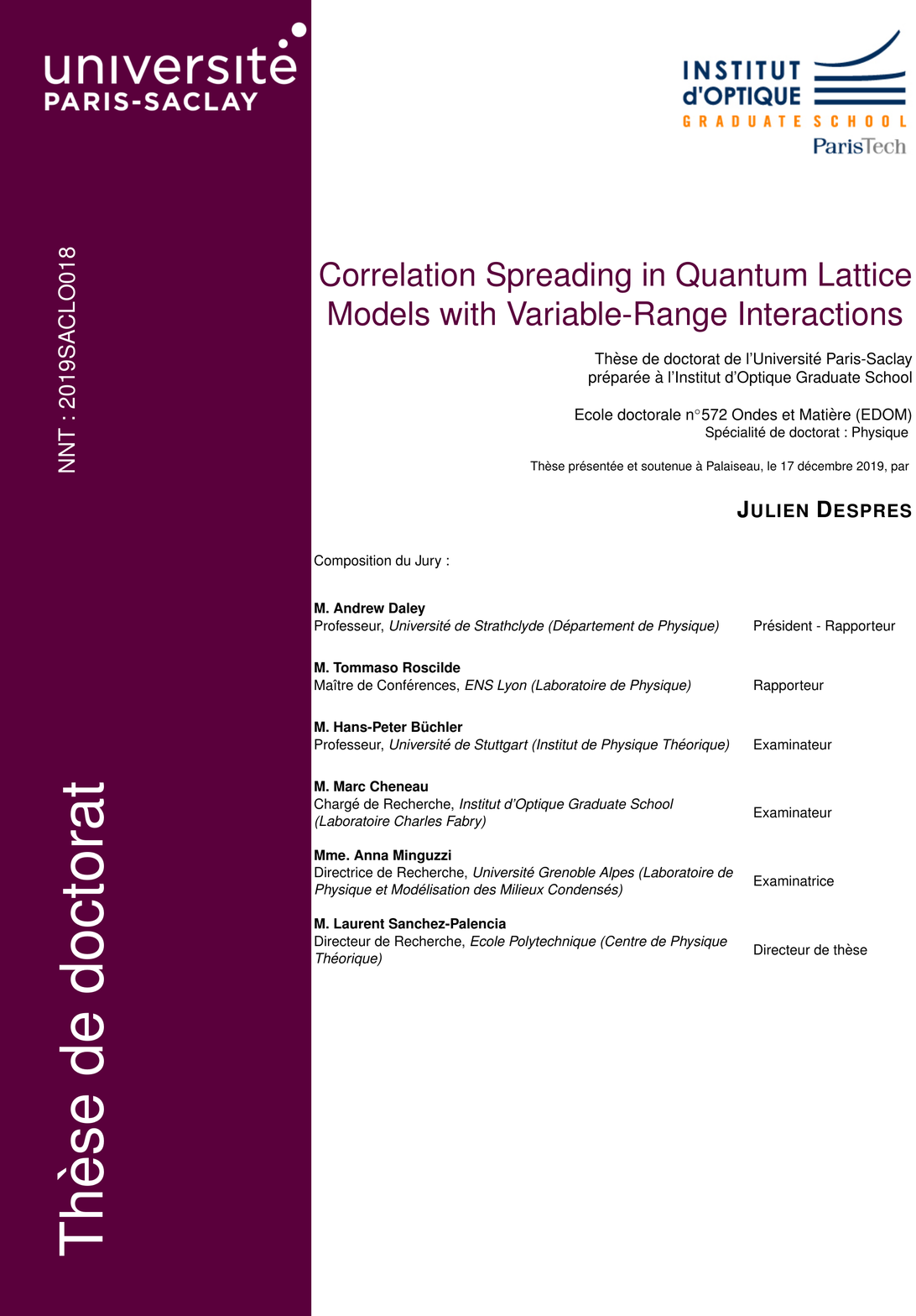}

\setlength{\textbaselineskip}{22pt plus2pt}
\setlength{\frontmatterbaselineskip}{17pt plus1pt minus1pt}
\setlength{\baselineskip}{\textbaselineskip}
\setcounter{secnumdepth}{2}
\setcounter{tocdepth}{2}

\begin{romanpages}


\begin{dedication}
This thesis is dedicated to \\
\textit{C\'ecile}, \textit{Edwige} and \textit{Josiane} \\
for their benevolence and unfailing support \\
\end{dedication}

\begin{remerciements}
 	\input{text/appendix_15}

\end{remerciements}

\begin{publications}
 	\input{text/publications}

\end{publications}

\dominitoc 

\flushbottom

\tableofcontents




\include{text/abbreviations}

\include{text/symbols}

\end{romanpages}

\flushbottom
\include{text/ch1_introduction}
\include{text/ch2_far_from_equilibrium_dynamics}

\include{text/ch3_universal_scaling_laws}

\include{text/ch4_bose_hubbard_chain}

\include{text/ch5_long_range_ising_chain}

\include{text/ch6_conclusion}

\startappendices
\include{text/appendix_1}

\include{text/appendix_2}

\include{text/appendix_5}

\include{text/appendix_6}

\include{text/appendix_9}

\include{text/appendix_8}

\include{text/appendix_7}

\include{text/appendix_3}

\include{text/appendix_4}
\include{text/appendix_10}

\include{text/appendix_12}

\include{text/appendix_13}

\include{text/appendix_14}



\setlength{\baselineskip}{0pt} 
\bibliographystyle{revtexlsp}
\bibliography{biblioLSP}

\includepdf[pages=1]{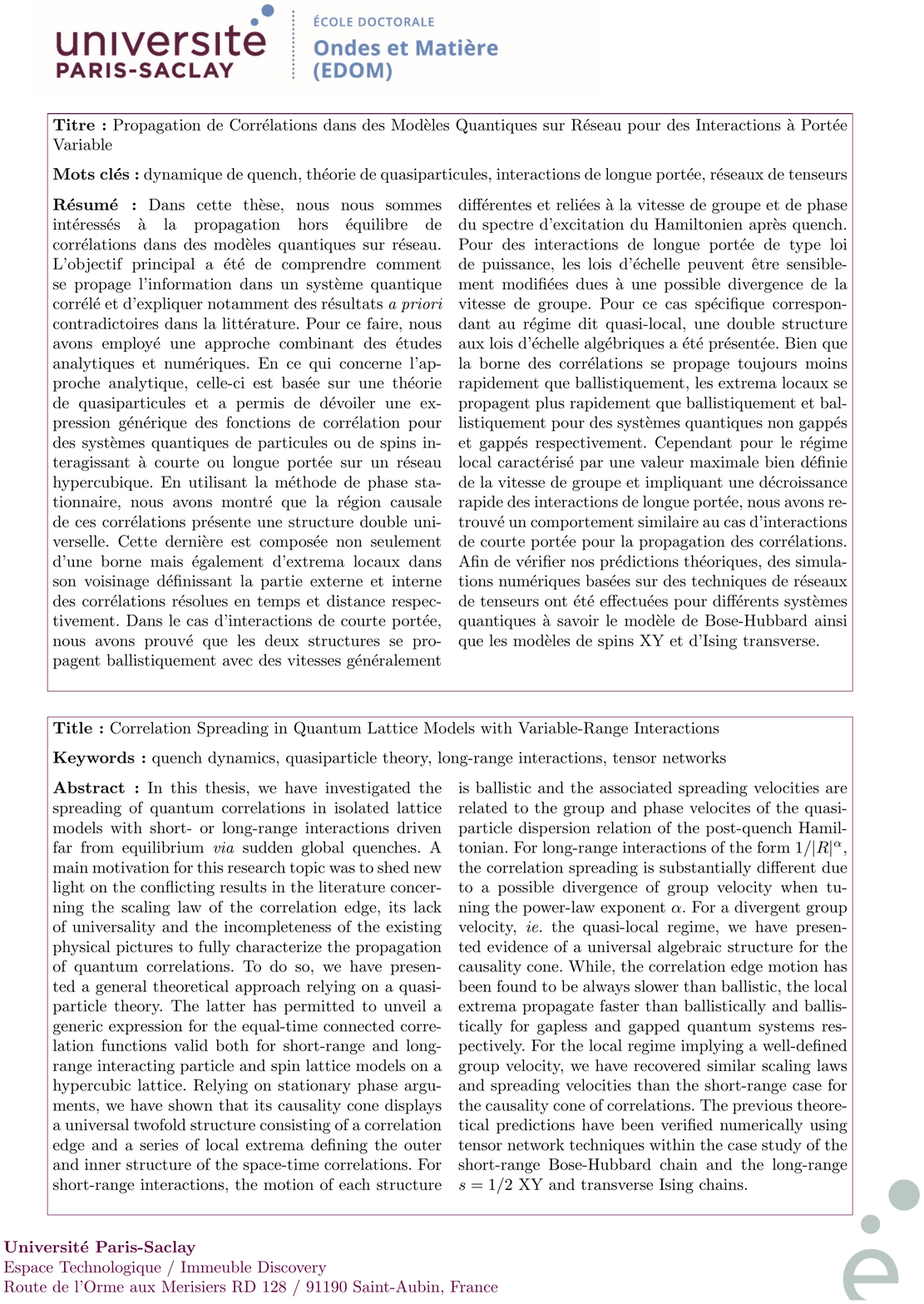}

\end{document}

%% file: text/appendix_15.tex
\setstretch{1.0} 

This manuscript presents my research work centred on the quantum many-body dynamics performed during my PhD thesis from September 2017 to August 2019. 
First of all, I would like to thank the members of my PhD committee for whom I have great respect and admiration, Pr. Daley Andrew, Pr. Roscilde Tommaso,
Pr. Büchler Hans-Peter, Dr. Cheneau Marc and Pr. Minguzzi Anna for agreeing to judge my thesis work.
Special thanks to Pr. Daley Andrew and Pr. Roscilde Tommaso for accepting to act as referees. \\

I wish to acknowledge my PhD advisor Pr. Sanchez-Palencia Laurent and more precisely for his support, his advice, his scientific spirit, his attention to detail
which has greatly influenced me. I would also like to thank him for the trust and the freedom of action he gave me throughout the thesis. \\

During my three years of doctoral studies, I had the privilege of collaborating with Pr. Tagliacozzo Luca, Dr. Carleo Giuseppe. I would like to express my 
gratitude to them for their advice concerning the numerical aspects of my research work. I would also like to thank Villa Louis for the research work 
done in close collaboration. In the continuity, I would like to express my thanks to the different members of the 'Quantum Matter Theory' team
led by Pr. Sanchez-Palencia Laurent, Crosnier-Bellaistre Cécile, Yao Hepeng, Schneider Jan, Gautier Ronan and Thomson Steven. Thank you for the countless discussions concerning not only
quantum physics but also philosophy, politics, religion and many others. \\

During the PhD thesis, I had the great opportunity to work in two different laboratories. Indeed, I spent the first six months in the 'Quantum Gases' 
group at the Charles Fabry laboratory (Institut d'Optique Graduate School) before moving to the Center for Theoretical Physics (\'Ecole Polytechnique). \\
Let me start by thanking all the people at the Institut d'Optique that I have been fortunate enough to rub shoulders with. I would also like to thank all 
the permanent staff of the 'Quantum gases' group and especially Dr. Cheneau Marc and Pr. Boiron Denis for the very nice discussions and their benevolence towards the 
PhD students. I would also like to express my gratitude to the postdoctoral researchers and PhD students working (who worked) for the same group, Dr. Molineri Anaïs, Dr. 
Denechaud Vincent, Dr. Carcy Cécile, Dr. Berthet Guillaume, Dr. Perrier Maxime, Dr. Musawwadah Mukhtar, Dr. Boissé Amaudric, Briosne-Fréjaville Clémence, Lecoutre Baptiste,
Amodjee Ziyad, Hercé Gaétan and Ténart Antoine. During my thesis, my time was also shared with several teaching duties at Institut d'Optique (mathematics and programming).
Hence, I would like to express my thanks to Pr. Goudail François, Pr. Boffety Matthieu, Pointard Benjamin, Prengère Léonard, Pr. Bourassin-Bouchet Charles and
Pr. Lebrun Sylvie. \\
Let me now thank all the people at the Center for Theoretical Physics. I would like to start by the permanent researchers of the 'Condensed Matter' group,
Pr. Biermann Silke, Pr. Ferrero Michel, Pr. Backes Steffen, Pr. Georges Antoine, Pr. Pourovskii Leonid and Pr. Subedi Alaska who provided me precious advice. 
I also wish to acknowledge all the (former) postdoctoral researchers and PhD students of the group, Dr. Pasek Michael, Dr. Lenz Benjamin, Dr. Steinbauer Jakob, Dr. Galler Anna,
Dr. Bhandary Sumanta, Dr. Schäfer Thomas, Bacq-Labreuil Benjamin, Boust James, Moutenet Alice and Turtulici Marcello. Furthermore, I would like to express my gratitude 
to the director of the laboratory, Pr. Chazottes Jean-Ren\'e, the administrative department and especially Mrs. Lang Malika, Mrs. Auger Florence and
Mrs. Debbou Fadila. Finally, I would like to thank the IT department, Mr. Bellon Jean-Luc, Mr. Pham Kim Danh and Mr. Fitamant Yannick for the discussions,
their great help and explanations to use the clusters and their invaluable support during the PhD defense. \\

In conclusion, I would like to thank my wife Cécile, my parents Edwige and Frédéric, my brother Mickaël and Sarah. 

%
%
%

%% file: text/publications.tex
\setstretch{1.0} 

\begin{enumerate}[leftmargin = *]
\item  L. Cevolani, \textbf{J. Despres}, G. Carleo, L. Tagliacozzo and L. Sanchez-Palencia, \textit{Universal Scaling Laws for Correlation Spreading in Quantum 
Systems with Short- and Long-Range Interactions}, Phys. Rev. B \textbf{98}, 024302 (2018) - Editors' suggestions. 
\item \textbf{J. Despres}, L. Villa and L. Sanchez-Palencia, \textit{Twofold Spreading of Correlations in a Strongly-Correlated Lattice Bose Gas}, Sci. Rep. \textbf{9} (4135), (2019). 
\item \textbf{J. Despres}, L. Tagliacozzo and L. Sanchez-Palencia, \textit{Spreading of Correlations and Entanglement in the Long-Range Transverse Ising Chain}, in preparation. 
\item L. Villa, \textbf{J. Despres} and L. Sanchez-Palencia, \textit{Unraveling the Excitation Spectrum of Many-Body Systems from Quantum Quenches}, Phys. Rev. A \textbf{100} (6), 063632 (2019).
\end{enumerate}

%% file: text/abbreviations.tex
\begin{mclistof}{List of Abbreviations}{3.2cm}

\item[ETH] Eigenstate Thermalization Hypothesis, refers to the proposed explanation for the thermalization process in generic non-integrable isolated quantum many-body systems.

\item[GGE] Generalized Gibbs Ensemble, denotes a statistical ensemble constructed from all the local and quasi-local conservation laws fulfilled by an integrable quantum model.  

\item[1D, 2D] One- or two-dimensional, referring in this thesis to the spatial dimension(s) of the lattice.

\item[SRBH] Short-Range Bose-Hubbard, denoting a specific short-range interacting bosonic lattice model.

\item[LRTI] Long-Range Transverse Ising, a long-range interacting spin lattice model.

\item[LRXY] Long-Range XY model, another long-range interacting spin lattice model.

\item[FBZ] First Brillouin Zone, the primitive cell of the reciprocal space (momentum space).

\item[t-MPS] time-dependent Matrix Product State, numerical approach based on tensor networks to perform real time evolution for one-dimensional quantum lattice models.

\item[sp] Stationary Phase, approximation to extract the asymptotic behavior of specific $D$-dimensional integrals. 

\item[CE] Correlation Edge, refers to the border separating the causal and non-causal regions of correlations. 

\item[SF] Superfluid phase, the gapless quantum phase of the one-dimensional short-range Bose-Hubbard model.

\item[MI] Mott-Insulating phase, the gapped phase of the one-dimensional short-range Bose-Hubbard model.

\item[MF] Mean field regime, corresponding to a specific regime in the superfluid phase of the one-dimensional short-range Bose-Hubbard model.

\item[t-VMC] time-dependent Variational Monte Carlo, numerical approach to deduce static and dynamical properties of one-dimensional and two-dimensional
short- and long-range interacting lattice models.

\item[t-DMRG] time-dependent Density Matrix Renormalization Group, another numerical approach based on tensor networks.

\item[LL] Lieb-Liniger model, a quantum model representing a bosonic gas interacting \textit{via} a two-body potential.

\item[HP] Holstein-Primakoff, refers to the Holstein-Primakoff transformation permitting to express spin operators into bosonic ones.

\item[MPS] Matrix Product State, denotes the local representation of a general quantum state in terms of tensor networks.

\item[SVD] Singular Value Decomposition, corresponds in linear algebra to a specific factorization of a real or complex matrix. 

\item[MPO] Matrix Product Operator, denotes the local representation of a general quantum operator in terms of tensor networks.

\item[BA] Bethe ansatz, corresponds to a ansatz method to characterize the low-energy properties of several one-dimensional many-body models at equilibrium.

\item[TLL] Tomonaga-Luttinger liquid, refers to the theory describing the low-energy properties of many one-dimensional gapless many-body quantum models at equilibrium.

\item[SI] Strongly Interacting regime, corresponds to the regime of small non-integer fillings and large two-body repulsive interactions for the superfluid phase
of the Bose-Hubbard model. 

\item[LSWT] Linear Spin Wave Theory, corresponds to the theory permitting to deduce the excitation spectrum of many different spin lattice models at low temperatures. 
The latter assumes small deviations from the ordered polarization axis.

\item[TDVP] Time-Dependent Variational Principle, numerical approach based on tangent spaces with respect to a matrix product state manifold to perform imaginary or 
real time evolution. 

\item[DES] Dynamical Excitation Spectrum, excitation spectrum deduced from an analysis of the space-time correlations.

\end{mclistof}

%% file: text/symbols.tex
\begin{mclistof}{List of Symbols}{3.2cm}

\item[$D$] Dimension of the quantum lattice model.
\item[$a$] Lattice spacing.
\item[$\hbar$] Planck constant.
\item[$k_B$] Boltzmann constant.
\item[$\beta$] Thermodynamic beta, $\beta = 1/k_BT$. 
\item[$Z$] Partition function.
\item[$\tau$] Imaginary time variable.
\item[$t$] Real time variable.
\item[$R$] Distance variable (index for the lattice sites).
\item[$k$] (Quasi-)momentum of a (quasi-)particle.
\item[$\hat{H}_{\mathrm{i}}$] Pre-quench (initial) Hamiltonian.
\item[$\hat{H}_{\mathrm{f}}$] Post-quench (final) Hamiltonian.
\item[$\ket{\Psi(t)}$] Time-evolved many-body quantum state.
\item[$E_{k}$] Excitation spectrum or quasiparticle dispersion relation.
\item[$\hat{a}, \hat{b}$] Annihilation operator acting on the lattice site $R$ for bosonic particles.
\item[$\hat{a}^{\dag},\hat{b}^{\dag}$] Creation operator for bosonic particles.
\item[$\hat{c}$] Annihilation operator for fermionic particles.
\item[$\hat{c}^{\dag}$] Creation operator for fermionic particles.
\item[$\hat{\beta}$] Annihilation operator for (bosonic) Bogolyubov particles.
\item[$\hat{\beta}^{\dag}$] Creation operator for (bosonic) Bogolyubov particles.
\item[$\hat{n}$] Local density operator, $\hat{n}^{(\hat{a})}_{\lambda} = \hat{a}^{\dag}_{\lambda} \hat{a}_{\lambda}$, $\lambda \in \{R,k\}$.
\item[$\hat{\sigma}^{(x,y,z)}$] Pauli matrices.
\item[$\hat{S}^{(x,y,z)}$] Spin operators.
\item[$J$] Hopping amplitude (for particles) or spin exchange coupling (for spins).
\item[$U$] Two-body interaction strength.
\item[$h$] Transverse magnetic field.
\item[$\epsilon$] Long-range antiferromagnetic exchange coupling along the $z$ axis of the Bloch sphere. 
\item[$L$] Length of the lattice chain.
\item[$N$] Number of particles on the lattice.
\item[$N_s$] Number of lattice sites.
\item[$\bar{n}$] Filling of the lattice.
\item[$s$] Spin. 
\item[$\theta$] Heaviside step function.
\item[$\delta_{i,j}$] Kronecker symbol.
\item[$\epsilon^{xyz}$] Levi-Civita symbol.
\item[$\hat{\rho}$] Density matrix.
\item[$\hat{\rho}_A$] Reduced density matrix associated to the subsystem $A$.
\item[$\chi$] Matrix product state bond dimension.
\item[$\tilde{\chi}$] Matrix product operator bond dimension.
\item[$\xi$] Correlation length.
\item[$V_{\mathrm{g}}$] Group velocity.
\item[$V_{\varphi}$] Phase velocity.
\item[$V_s~(c)$] Sound velocity.
\item[$V_{\mathrm{CE}}$] Velocity of the correlation edge.
\item[$V_{\mathrm{m}}$] Velocity of the local maxima. 
\item[$\beta_{\mathrm{CE}}$] Exponent characterizing the scaling law for the correlation edge spreading.
\item[$\beta_{\mathrm{m}}$] Exponent characterizing the scaling law for the maxima spreading.
\item[$G_1(R,t)$] Equal-time connected one-body correlation function. 
\item[$G_2(R,t)$] Equal-time connected density-density correlation function.
\item[$G_x(R,t)$] Equal-time connected spin-spin correlation function along the $x$ axis of the Bloch sphere. 
\item[$G_z(R,t)$] Equal-time connected spin-spin correlation function along the $z$ axis of the Bloch sphere. 
\item[$\mathbb{H}$] Many-body Hilbert space.
\item[$\mathbb{H}_R$] Local Hilbert space associated to the lattice site $R$. 
\item[$\alpha$] Power-law exponent for the long-range (algebraic) interactions. 
\item[$\mathcal{S}$] Entanglement entropy also called von Neumann entropy. 
\item[$\mathcal{S}_n$] R\'enyi entropy of order $n$. 
\item[$\mathcal{M}$] Matrix product state manifold.
\item[$\mathcal{T}_{\ket{\Psi}}$] Tangent space associated to the matrix product state $\ket{\Psi}$ living in a manifold $\mathcal{M}$.

\end{mclistof}

%% file: text/ch1_introduction.tex
\setstretch{1.0} 

\begin{savequote}[8cm]
\textlatin{“There is nothing more difficult to take in hand, more perilous to conduct, or more 
uncertain in its success, than to take the lead in the introduction of a new order of things.”}
  \qauthor{--- Niccolo Machiavelli}
\end{savequote}

\chapter{\label{ch:1-intro}Introduction} 

In 1982 the well-known theoretical physicist Richard Feynman has proposed the idea of a universal quantum simulator consisting of a special device to simulate any many-body 
quantum system, hence, permitting to provide new insight on the many-body problem \cite{feynman1982}. A many-body system, \textit{ie.} a quantum system composed of many
interacting particles, is described mathematically by a Hilbert space whose dimension depends exponentially on the number of particles. Feynman has shown that the simulation
of such system on a classical computer would necessary lead to an exponential time while his hypothetical quantum simulator would not. His approach consists of simulating
a many-body quantum system using a fictional quantum computer based on qubits mimicking the behaviour of the particles, see Refs.~\cite{buluta2009,georgescu2014,NaturePhysicsInsight2012cirac,bloch2012,NaturePhysicsInsight2012blatt,NaturePhysicsInsight2012aspuru-guzik,NaturePhysicsInsight2012houck,ward2013,lsp2018}
for a non-exhaustive list of reviews on quantum simulation. In the last decades, with the dramatic progress realized in the experimental control of quantum matter in
condensed matter, atomic, molecular, and optical physics, such quantum simulators are not fictional anymore. Indeed, they have been realized
experimentally on a large variety of platforms including ultracold atomic gases \cite{gross2017, bloch2012} and trapped ions \cite{deng2005,islam2011,lanyon2011,schneider2012b,NaturePhysicsInsight2012blatt}. 
For instance, a possible experimental realization of the trapped-ion quantum simulator consists of confining by a large homogeneous magnetic field, \textit{ie.} a Penning trap, cold atomic ions. The outermost electron
of each ion stores the information of a qubit (a two-state quantum-mechanical system corresponding here to an effective $s=1/2$ spin) \textit{via} the two quantized values of its spin projection. Another possibility
is to take advantage of a transition between two internal hyperfine states of the ions to encode the two local states of a qubit \cite{richerme2014, jurcevic2014}. \\

Due to the simultaneous progress realized in the field of ultracold atomic gases and trapped ions, the two corresponding experimental platforms have permitted to simulate
a large variety of isolated particle and spin lattice models initially introduced in condensed matter physics with unprecedented control possibilities of the parameters 
in time. In particular, this experimental opportunity in the tuning of interaction parameters has permitted to provide new insights on
one important aspect of the many-body problem, namely the far-from-equilibrium dynamics \cite{bakr2010, esslinger2010, chen2011, trotzky2012, cheneau2012, schneider2012, pertot2014, richerme2014, jurcevic2014}. Indeed, via a sudden modification of one or several interaction parameters of the Hamiltonian
governing an isolated quantum system, the latter can be driven far from equilibrium. Experimentally, these quenches can be performed in different ways. For isolated particle 
lattice models, the amplitude of the laser beams used to create the artificial lattice can be changed abruptly to modify the hopping amplitude \cite{cheneau2012, trotzky2012}. In the case of isolated spin lattice models, 
laser-induced optical dipolar forces can be generated suddenly to create long-range interactions \cite{richerme2014, jurcevic2014}. \\

Understanding the far-from-equilibrium dynamics of isolated quantum lattice models has become a central subject of the many-body theory where one specific issue concerns the 
spreading of quantum correlations \cite{barmettler2012,natu2013,hauke2013,cevolani2015,cevolani2016,buyskikh2016,frerot2018,cevolani2018,despres2019,despres2019bis}. The latter is at
the center of many fundamental phenomena including the propagation of information, relaxation and thermalization. \\
For isolated short-range interacting lattice systems, the existence of the so-called Lieb-Robinson bound implies the emergence of a linear causality
cone beyond which the quantum correlations are exponentially suppressed \cite{lieb1972,bravyi2006,hastings2006}. However, this bound resulting in a ballistic propagation, in particular, of the equal-time correlation
functions \cite{bravyi2006} is not sufficient to fully characterize the causal correlation cone \cite{cevolani2018}. \\
Another class of quantum systems particularly interesting in the context of the far-from-equilibrium dynamics corresponds to long-range interacting lattice models. Indeed, the latter induce rich dynamical behaviour 
due to the breakdown of fundamental concepts such as the equivalence of the thermodynamic ensembles, the Lieb-Robinson propagation bound and the notion of group and phase velocities which can also lead to peculiar behaviours.
For these quantum systems, crucial open questions are still debated, in particular how the correlation and information spreading is modified in the presence of long-range interactions. Naturally, generalized Lieb-Robinson bounds for 
the correlation spreading have been derived in the case of long-range quantum systems with power-law decaying interactions of the form $1/R^\alpha$ \cite{hastings2006,foss-feig2015}. Here, $\alpha$ refers to the power-law
exponent controlling the decay of the long-range interactions. These investigations of the transport of quantum correlations and information in long-range interacting lattice models gained a lot of momentum in the last decades
due to the development of various experimental platforms offering the possibility to simulate a large class of quantum systems where the strength and the decay of the long-range interactions can be accurately controlled. For instance,
long-range interacting quantum systems have been realized experimentally using Rydberg gases \cite{bendkowsky2009,weimer2010,schausz2012,browaeys2016}, polar molecules \cite{micheli2006,yan2013,moses2017}, nonlinear optical 
media \cite{firstenberg2013}, magnetic atoms \cite{griesmaier2005,beaufils2008, lu2011,baier2016,lahaye2009}, solid-state defects \cite{childress2006,balasubramanian2009,dolde2013} and artificial ion crystals.
However, the related experiments and numerical investigations have led to conflicting pictures \cite{hauke2013,eisert2013,jurcevic2014,richerme2014,cevolani2015,cevolani2016,buyskikh2016}.
For example, trapped-ion experiments \cite{richerme2014} and numerical simulations \cite{schachenmayer2015b} for the long-range XY chain point towards bounded, super-ballistic, propagation
for all values of $\alpha$. In contrast, experiments on the long-range transverse Ising chain have reported ballistic propagation of correlation maxima \cite{jurcevic2014}. Moreover, different
powerful numerical approaches to simulate low-dimensional quantum lattice models, such as the time-dependent density matrix renormalization group and variational Monte-Carlo techniques have 
indicated the existence of three distinct regimes, namely instantaneous, sub-ballistic, and ballistic, for increasing values of the power-law exponent $\alpha$ \cite{schachenmayer2013,hauke2013,eisert2013,cevolani2015,cevolani2016,buyskikh2016}. \\

Hence, while the correlation spreading in isolated quantum lattice models has been extensively studied in the literature, a clear general approach and physical picture are missing.
The aim of this thesis is to contribute to establish a generic approach and to draw a comprehensive physical picture for the spreading of correlations in short- and long-range interacting lattice models.  
To do so, we investigate the spreading of quantum correlations in isolated lattice models with short- or long-range interactions, driven far from equilibrium \textit{via} sudden global or local quenches,
both theoretically and numerically. We propose a generic approach relying on a quasiparticle theory that can be applied both to short-range and long-range interacting particle and spin models on a
hypercubic lattice. This quasiparticle approach permits to unveil a generic form of the equal-time correlation functions. Our main result is that the correlation pattern displays a twofold structure in 
the vicinity of the correlation edge. It is characterized by local maxima and a correlation edge that usually propagate differently. For short-range interacting lattice models, both structures propagate
ballistically but with different velocities. For quantum systems with intermediate-range interactions, we show that the local maxima and the correlation edge propagate algebraically. However, the scaling
laws are completely different and essentially unrelated. On the one hand, the correlation edge propagates sub-ballistically with a non-universal scaling law. On the other hand, the correlation maxima may
propagate either ballistically or super-ballistically depending on the properties of the spectrum. This behaviour is universal in the sense that it only depends on the spectrum and not on the observables. These
results shed new light on the propagation of correlations in short- and long-range interacting quantum lattice models. They also have significant consequences for the interpretation of numerical and experimental
observations, in particular in long-range systems. \\

The manuscript is organized as follows : 

\begin{itemize}
\item \textbf{Chapter 2} : \textit{Far-from-equilibrium dynamics in many-body quantum systems} \\ 
We start by giving a general introduction to the far-from-equilibrium dynamics of isolated quantum lattice models. Important phenomena occuring during such dynamics,
such as the relaxation, equilibration and the thermalization, are presented and illustrated by a numerical or experimental result extracted from the literature. In this chapter,
a special care is devoted to the correlation spreading in short-range interacting quantum systems driven far from equilibrium \textit{via} quantum quenches. Important 
insights on this many-body problem including the Lieb-Robinson bound and the Calabrese-Cardy quasiparticle picture are discussed. Then, we turn to a presentation of
the experimental realization of isolated quantum systems, quantum quenches and their subsequent dynamics. Firstly, the quench dynamics of a short-range interacting
bosonic lattice model using ultracold atoms loaded in an optical lattice is discussed. Secondly, we move on to the case of long-range interactions where the experimental
implementation of a spin lattice model with power-law decaying interactions relying on trapped atomic ions and its response to a quantum quench is analyzed. Finally,
we discuss several powerful numerical and theoretical techniques permitting to study the quench dynamics and, more generally, the far-from-equilibrium dynamics of isolated quantum
lattice models. For the numerical part, the time-dependent matrix product state, time-dependent variational Monte Carlo and the exact diagonalization approaches are discussed. For the theoretical part, a quasiparticle approach
based on a mean field approximation and the so-called quench action are briefly introduced. \\

\item \textbf{Chapter 3} : \textit{Universal scaling laws for the correlation spreading in quantum lattice models with variable-range interactions} \\ 
In this chapter, the correlation spreading in isolated quantum lattice models driven far from equilibrium \textit{via} sudden global quenches 
is investigated theoretically. We introduce a quasiparticle approach relying both on a mean field approximation and the bosonic Bogolyubov theory, which is applicable to short-range and long-range interacting bosonic and spin systems.
We show that the latter permits to unveil a generic form for the correlation functions whose corresponding space-time pattern is analyzed using stationary phase arguments.
It turns out that the causal region of the generic space-time correlations is characterized by a twofold linear structure for short-range interacting quantum lattice models and by a twofold algebraic structure for long-range systems
with intermediate-range interactions. The corresponding spreading velocities and scaling laws of the double structure for short-range and long-range interacting lattice models respectively are also determined. 
For the case of short-range interactions, the theoretical predictions are confirmed by a detailed study of the one-dimensional Bose-Hubbard model in both the gapped Mott-insulating and gapless superfluid phases.
For long-range interactions, the scaling laws predicted by our quasiparticle approach are verified for two different one-dimensional long-range interacting $s=1/2$ spin lattice models, namely the XY chain in the
$x$ polarized phase and the transverse Ising chain in the $z$ polarized phase corresponding to a gapless and a gapped quantum system respectively. \\
 
\item \textbf{Chapter 4} : \textit{Twofold correlation cone in a short-range interacting quantum lattice model} \\ 
This chapter is devoted to a numerical investigation of the correlation spreading induced by sudden global quenches in the short-range interacting lattice model
considered previously, the Bose-Hubbard chain. The purpose of this study is twofold. On the one hand, we aim at testing the analytic theory presented in Chapter 3 against 
a numerically exact approach, beyond the mean field approximation. On the other hand, we aim at extending the general picture to quantum regimes that are not amenable to
analytic treatments. After presenting the time-dependent matrix product state approach we use, we discuss the results in detail. Our numerical results fully confirm our analytical predictions
in their respective regimes of validity, namely in the superfluid mean field regime and deep in the Mott-insulating phase. Furthermore, we show that the twofold structure of the 
correlation spreading survives in strongly interacting regimes, in particular in the vicinity of the Mott transitions. We also extend the study to sudden local quenches in the Mott phase, as well as
in the Heisenberg model in order to treat the case of a short-range interacting gapped and gapless quantum system respectively. \\

\item \textbf{Chapter 5} : \textit{Spreading of correlations and entanglement in the $s=1/2$ long-range transverse Ising chain} \\
In the last chapter, we turn to long-range interacting lattice models. We present the numerical approach we use, relying on the matrix product state representation
within the time-dependent variational principle. We then focus on the long-range transverse Ising chain. The numerical results confirm the predictions of the analytic treatment presented in Chapter 3 :
For a sudden global quench and intermediate-range interactions, the correlations show a clear twofold behaviour. While the maxima spread ballistically for this gapped system, the correlation edge displays 
a sub-ballistic behaviour. For the local regime, equivalent to the case of short-range interactions, a twofold ballistic behaviour is found and the associated spreading velocities are fully characterized by
the excitation spectrum. Moreover, we extend our study to the case of sudden local quenches, considering both the local magnetization and several Rényi entropies to characterize the entanglement
spreading. Finally, we turn to another long-range interacting spin lattice model, namely the long-range XY chain, which is expected to behave differently owing to its gapless spectrum. In this case, we 
find a twofold algebraic behaviour where the maxima spread super-ballistically but the correlation edge spreads sub-ballistically.
\end{itemize}

%% file: text/ch2_far_from_equilibrium_dynamics.tex
\setstretch{1.0} 

\begin{savequote}[8cm]
\textlatin{“It doesn't matter how beautiful your theory is, it doesn't matter how smart you are. If it doesn't agree with experiment, it's wrong.”}
  \qauthor{--- Richard P. Feynman}
\end{savequote}

\chapter{\label{ch:2-far_from_equilibrium_dynamics}Far-from-equilibrium dynamics in many-body quantum systems}
\minitoc

\newpage

\section{Dynamics in quantum mechanics}
\label{dynamic_intro}
In the last decades, simultaneous progress realized in the experimental control of quantum matter in condensed matter and atomic,
molecular, and optical physics has given dramatic momentum to the investigation of the far-from-equilibrium dynamics of isolated and correlated quantum 
systems (see Refs.~\cite{polkovnikov2011,eisert2015,pekola2015,langen2015,lewenstein2007,bloch2008, NaturePhysicsInsight2012bloch,
NaturePhysicsInsight2012blatt,NaturePhysicsInsight2012aspuru-guzik,NaturePhysicsInsight2012houck, lsp2018, tarruell2018, aidelsburger2018, lebreuilly2018, lehur2018, bell2018, alet2018} and references therein). In particular,
the major achievement performed in the manipulation of ultracold quantum gases and trapped ions has turned them into ideal experimental
platforms for the quantum simulation. These platforms can reproduce the quantum behavior of a wide class of isolated particle and spin lattice models initially introduced in condensed matter physics with unprecedented 
control possibilities of the corresponding Hamiltonian parameters in time, see Sec.~\ref{ER}. This experimental opportunity in the tuning of interaction
parameters has permitted to drive isolated lattice models far from equilibrium \textit{via} quantum quenches. The latter consist of preparing initially the quantum
model in a many-body quantum state and then modifying abruptly Hamiltonian parameters acting on a specific lattice site or on the full quantum system and 
corresponding to the so-called sudden local or global quenches respectively. \\
The atomic-gas and trapped-ion simulators permitting to drive many-body quantum systems far from equilibrium have triggered a renewed interest for the
research topic and shed new light on long-standing open questions presented at Sec.~\ref{open_q}. 
For instance, a fundamental question about the out-of-equilibrium dynamics concerns the possibility for closed quantum systems driven far from equilibrium
to relax and equilibrate at long
times. An important question raised by these dynamical processes is then to characterize the
corresponding stationary state and to know whether the latter can be described by a thermal equilibrium state in the framework of statistical quantum mechanics.
Moreover, another important aspect of the far-from-equilibrium dynamics concerns the transport properties. The main purpose is to understand how quantum 
information, including quantum correlations and entanglement, can propagate into isolated quantum lattice models and to determine how fast is this information spreading depending
on the dimensionality of the lattice, the interactions and the considered quantum phase. \\
At the moment, a universal picture to answer these fundamental questions is missing. Furthermore, powerful theoretical generic approaches widely used 
to investigate near-equilibrium dynamics are not valid anymore to explain the physical properties of the far-from-equilibrium dynamics. One can mention for example
the well-known linear response theory assuming weak dynamical perturbations of the quantum system around its equilibrium state, see discussion below. Nevertheless, 
important theoretical progress has been accomplished to solve these important questions for low-dimensional lattice models, typically one- and two-dimensional
quantum lattice models, which are directly relevant to quantum simulations. Indeed, the peculiarities of this class of lattice models make them amenable to a variety of analytical and numerical techniques which
are discussed at Sec.~\ref{NTA}. 

\subsection*{Near-equilibrium dynamics - The linear response theory}

As discussed previously, the far-from equilibrium dynamics is a relatively new branch in the research field of quantum dynamics and goes beyond the near-equilibrium one.
The latter consists of investigating the dynamical behavior of a quantum system submitted to a weak external time-dependent 
perturbation around its equilibrium state. In what follows, we briefly outline the widely-used theory to describe such dynamics : the linear response theory within the 
Kubo formula, see Ref.~\cite{kubo1957} for more details. As famous results of this theory, one can mention the spin (charge) susceptibilities of spin models 
submitted to a weak external magnetic field (fermionic models submitted to a small external electric field). \\

From a general point of view, this theory states that the response of the quantum system to a weak external perturbation, which can be expanded in a power 
series of the perturbation, is accurately described by performing a first-order approximation. In other words, the 
response is proportional to the perturbation and hence the problem consists of understanding the proportionality factor. Physically, considering a given weak
external perturbation symbolized by the operator $\hat{V}$, how is modified the expectation value at equilibrium $\langle \hat{A} \rangle_0$ at first order
in $\hat{V}$? This question is fully solved by the Kubo formula, characterizing the linear response of the quantum system to any weak external perturbation, which is 
briefly discussed in what follows. \\

Considering a quantum model governed by the time-dependent Hamiltonian $\hat{H} = \hat{H}_0 + \hat{V}(t) \theta(t-t_0)$. $\hat{H}_0$ denotes the time-independent 
Hamiltonian before applying the weak external perturbation and $\hat{V}(t) \theta(t-t_0)$ to the time-dependent perturbation applied at time $t=t_0$ [$\theta(t-t_0)$
corresponds to the translated Heaviside function defined as $\theta(t-t_0) = 1$, $\forall t\geq t_0$ and $0$ otherwise]. Relying on the 
interacting picture particularly adapted for such non-equilibrium problem, it can be shown that the time-dependent expectation value for a relevant physical
observable $\hat{A}$, \textit{ie.} $\langle \hat{A}(t) \rangle$, at first order in the perturbative term $\hat{V}$, may be written as
(see Refs.~\cite{bruus2004, mahan2000} for a complete derivation) 

\begin{equation}
\langle \hat{A}(t) \rangle \approx \langle \hat{A} \rangle_0 -i \int_{t_0}^{t} \mathrm{d}t' \langle [\hat{A}(t), \hat{V}(t')] \rangle_0.
\label{at}
\end{equation}

\noindent
The brackets $\langle ... \rangle_0$ represents the equilibrium average with respect to the time-independent Hamiltonian $\hat{H}_0$. We stress that 
Eq.~\eqref{at} is valid for any relevant observable $\hat{A}$ and weak external perturbation $\hat{V}$. Note also that the analytical expression 
of the non-equilibrium time-dependent expectation value 
$\langle \hat{A}(t) \rangle$ has been expressed in terms of a retarded correlation function at equilibrium. Indeed, by defining the time-dependent 
connected expectation value $\langle \hat{A}_c(t) \rangle = \langle \hat{A}(t) \rangle - \langle \hat{A} \rangle_0$, allowing us to get rid of the 
equilibrium contribution, one obtains

\begin{equation}
\langle \hat{A}_c(t) \rangle \approx \int_{t_0}^{+\infty}  \mathrm{d}t' C^{R}_{\hat{A},\hat{V}}(t,t'),
\end{equation}

\noindent
where $C^{R}_{\hat{A},\hat{V}}(t,t')$ denotes the retarded ($R$) correlation function depending on the observables $\hat{A}$ and $\hat{V}$ at time $t$ and $t'$ respectively.
The latter reads as 

\begin{equation}
C^{R}_{\hat{A},\hat{V}}(t,t') = -i \theta(t-t') \langle [\hat{A}(t), \hat{V}(t')] \rangle_0.
\label{kubo}
\end{equation}

\noindent
The causal response function at Eq.~\eqref{kubo} corresponds to the well-known Kubo formula representing the linear response of the quantum model
to the perturbation $\hat{V}$ acting on the relevant observable $\hat{A}$. The latter states that the dynamical contribution of the time-dependent
expectation value $\langle \hat{A}(t) \rangle$ can be related to the expectation value at equilibrium of the commutator involving the relevant
observable $\hat{A}$ and the perturbation $\hat{V}$, \textit{ie.} $\langle [\hat{A}(t), \hat{V}(t')] \rangle_0$. \\

However, we stress that the Kubo formula does not hold anymore to investigate the far-from-equilibrium dynamics of quantum models generated for instance 
\textit{via} sudden quantum quenches. The latter consist of preparing a highly excited initial state by applying suddenly a strong global or local
perturbation to the quantum system. Note that the protocol associated to sudden global and local quenches are briefly discussed in the next section and presented 
in more details at Subsec.~\ref{qd}. The strength of the linear response theory lies in the assumption of a weak external perturbation permitting to truncate
the response of the quantum system, displaying a power series of the perturbation, at first order.
Hence, this yields to the powerful and simple Kubo formula at Eq.~\eqref{kubo} only requiring to evaluate a commutator involving 
the relevant observable and the perturbation. As a consequence, due to the assumption that the external perturbation is weak, this theoretical approach does not apply 
to describe the dynamics of quantum models induced by sudden quantum quenches, except very small ones.  
  
\section{Far-from-equilibrium dynamics}
\label{open_q}
In this section, we give a general introduction to the main research activities on the far-from-equilibrium dynamics for isolated quantum lattice models.
We first start by discussing the relaxation and thermalization processes. They consist of characterizing the long-time dynamics of far-from-equilibrium
quantum systems to know whether the quantum model will equilibrate or not, \textit{ie.} if the long-time-evolved many-body quantum state can be seen as a 
stationary state (relaxation), and if the latter can be described by a thermal state in the framework of statistical mechanics (thermalization). Finally, 
we turn to a presentation of the transport properties for such lattice models. More precisely, the information spreading, including the propagation
of entanglement and correlations, for isolated short-range interacting quantum lattice models is discussed. 

\subsection{Relaxation process}
\label{relaxation}

Before discussing the relaxation process, we start by presenting the general context and more precisely a class of widely-used protocols permitting to generate
far-from equilibrium dynamics for isolated quantum systems : the sudden global \cite{greiner2002b, calabrese2006, kollath2007, rigol2007, flech2008, moeckel2008, rigol2008, cevolani2018, cevolani2016, despres2019, despres2019bis, 
despres2019bis2, polkovnikov2011, calabrese2011, trotzky2012, caux2013, barmettler2009} and local quenches \cite{torres-herrera2014, stephan2003, despres2019bis, sirker2015}. \\

Concerning sudden global quenches, an isolated quantum system (quantum model decoupled from its environment) governed by a Hamiltonian $\hat{H}(\gamma)$, 
depending on a single relevant interaction parameter $\gamma$, is first considered. More precisely, $\hat{H}(\gamma)$ represents a translation invariant,
time independent and (in general) a short-range interacting Hamiltonian. At time $t=0$, the initial state, denoted by $\ket{\Psi_0}$, corresponds to an 
eigenstate of $\hat{H}$, in most cases to the ground state of $\hat{H}$.
Hence, $\ket{\Psi_0} = \ket{\Psi_{\mathrm{gs}}(\hat{H}_{\mathrm{i}})}$ where $\hat{H}_\mathrm{i} = \hat{H}(\gamma_\mathrm{i})$ refers to the pre-quench (initial)
Hamiltonian. The latter is built from $\hat{H}$ for a specific initial interaction parameter $\gamma_\mathrm{i}$. Then, at time $t=0^+$, the interaction parameter 
$\gamma_\mathrm{i}$ is instantaneously modified to a value $\gamma_\mathrm{f}$ and remains constant during the observation time. In other words, the real time
evolution is performed by considering a post-quench (final) Hamiltonian $\hat{H}_\mathrm{f} = \hat{H}(\gamma_\mathrm{f})$, see Fig.~\ref{gb}. Thus, the time-evolved many-body quantum state $\ket{\Psi(t)}$ may be written as 

\begin{equation}
\ket{\Psi(t)} = e^{-i \hat{H}_\mathrm{f}t} \ket{\Psi_0}, ~~~\mathrm{with}~~~ \ket{\Psi_0} = \ket{\Psi_{\mathrm{gs}}(\hat{H}_{\mathrm{i}})}.
\label{ute}
\end{equation}

\noindent
Equation \eqref{ute} represents the unitary \footnote{The quantum system has been assumed to be isolated from its environment.} 
real time evolution of the initial state $\ket{\Psi_0}$ with respect to the post-quench Hamiltonian $\hat{H}_\mathrm{f}$.
A direct consequence for such time evolution is the conservation of both the norm and the energy from time $t=0^+$. Here, the
conserved energy $E_\mathrm{f} = \langle \Psi(t) | \hat{H}_{\mathrm{f}} | \Psi(t) \rangle = \langle \Psi_0 | \hat{H}_{\mathrm{f}} | \Psi_0 \rangle$
of the time-evolved quantum state $\ket{\Psi(t)}$ is higher than $E_\mathrm{i} = \langle \Psi_0 | \hat{H}_{\mathrm{i}} | \Psi_0 \rangle$ the ground state energy of the pre-quench Hamiltonian, \textit{ie.} $E_\mathrm{f} > E_{\mathrm{i}}$.
This is due to the initial state $\ket{\Psi_0}$ which can be seen as a highly excited state since it does not consist of an eigenstate of $\hat{H}_\mathrm{f}$. 
As a consequence, the considered isolated quantum model is well driven far from equilibrium from time $t=0^+$. \\

Note that the protocol for sudden global quenches conserves the translational symmetry during the full evolution process if both 
the pre- and post-quench Hamiltonians are translationally invariant. This is in opposition with the protocol for sudden local quenches which 
is slightly different. Indeed, contrary to global quenches where the isolated quantum system is \textit{globally} quenched by considering a different Hamiltonian (more precisely a different interaction parameter) 
to perform the real time evolution; for local quenches, a same Hamiltonian is considered however the initial state is strongly perturbed \textit{locally}. To present such protocol, 
we consider as previously an initial state $\ket{\Psi_0}$ corresponding to the ground state of $\hat{H}$ for a specific interaction parameter $\gamma$. Then, $\ket{\Psi_0}$ is strongly perturbed locally 
\textit{via} for instance a spin-flip for spin lattice models, or by moving, adding, or again removing a particle for bosonic and fermionic lattice models. The direct consequence
for this protocol step is the breaking of the translational invariance. Finally, this highly excited state $\ket{\tilde{\Psi}_0}$ perturbed locally evolves in time with the
Hamiltonian $\hat{H}$ used to compute the initial non-perturbed state $\ket{\Psi_0}$. \\

We stress that it exists other classes of quantum quenches. One can mention the ramps \cite{polkovnikov2011, delcampo2014, braun2015, bakr2010, moeckel2010, dora2011}
(in opposition to sudden) where the Hamiltonian is changed progressively according to a specific function in time, and the geometric quenches
\cite{rigol2007, mossel2010, alba2014} consisting of a sudden modification of the geometry of the lattice. \\

In what follows, to fix the context to discuss both the relaxation and equilibrium processes, one considers the general case of an isolated quantum lattice 
model driven far from equilibrium \textit{via} a sudden global quench. An essential question is to understand how such quantum model can possibly relax towards
an equilibrium steady state. Indeed, the dynamics being time-reversal invariant [see Eq.~\eqref{ute}] and recurrent for lattice models with a finite size, it is 
relatively surprising for these models to reach equilibrium dynamically. A fundamental insight to the understanding of such phenomena is that generic quantum lattice
models driven far from equilibrium display a relaxation and an equilibration processes when investigating the long-time dynamics of time-dependent expectation values of local observables $\langle \hat{O}(t)
\rangle$. It is important to stress that the locality plays a primordial role for these two long-time physical processes. \\

To illustrate the previous statement, let us
consider the example, provided at Ref.~\cite{fagotti2016}, of a non-local (in space) hermitian operator $\hat{P}$ defined as 
$\hat{P} = \ket{n} \bra{m} + \ket{m} \bra{n}$. Here, $\ket{n},\ket{m}$ denotes two different eigenstates for the post-quench Hamiltonian $\hat{H}_\mathrm{f}$ where the 
corresponding eigenenergies are $E^{\mathrm{f}}_n$ and $E^{\mathrm{f}}_m$ respectively. According to Eq.~\eqref{ute}, the corresponding time-dependent expectation
value of such observable can be written as 

\begin{equation}
 \langle \hat{P}(t) \rangle = \langle \Psi(t) | \hat{P} | \Psi(t) \rangle = e^{i(E^{\mathrm{f}}_n - E^{\mathrm{f}}_m )t} \langle \Psi_0 | n \rangle \langle m | \Psi_0 \rangle + \mathrm{h.c.}~.
\end{equation}

\noindent
Due to the presence of the oscillating terms $e^{i(E^{\mathrm{f}}_n - E^{\mathrm{f}}_m)t}$ and $e^{-i(E^{\mathrm{f}}_n - E^{\mathrm{f}}_m)t}$ in the first 
and second contribution respectively, the long-time behavior of such non-local observable can not reach a stationary (also called steady) value $\bar{P}$
defined as $\bar{P} = \lim_{t \rightarrow + \infty} \langle \hat{P}(t) \rangle = \lim_{t \rightarrow + \infty} \langle \Psi(t) | \hat{P} | \Psi(t) \rangle$.
In other words, for very long times the considered quantum model does not appear as if it had equilibrated to a stationary state.  \\

As a consequence, we stress that isolated quantum systems cannot globally relax to a steady state but only locally. The interpretation is that the rest of the quantum 
system acts as an environment. A natural and rigorous description of such physical property is given in terms of time-dependent reduced density matrices.
They consist of performing a bipartition of the lattice model. Then, to probe the behavior in time of one of the two subsystems, all the degrees of freedom associated 
to the complementary subsystem are traced out for the time-dependent density matrix (the latter characterizing globally the lattice model in time).
Mathematically, let us consider $\hat{\rho}(t) = \ket{\Psi(t)} \bra{\Psi(t)}$ denoting the density matrix at time $t$ associated to the full quantum system,
the reduced density matrix for the subsystem $A$ is given by $\hat{\rho}_A(t) = \mathrm{Tr}_{B}[\hat{\rho}(t)]$ where $B$ is the complementary subsystem
(see also Appendix.~\ref{appendix_entropies} for more details). Using a language based on reduced density matrices, one can formulate a rigorous mathematical 
definition of both the local relaxation and local
stationarity \cite{fagotti2016}. Concerning the local relaxation, one can state that an isolated quantum lattice model, having a length $L$, relaxes locally
if the following mathematical limit is verified : 

\begin{equation}
 \lim_{t \rightarrow +\infty} \lim_{L \rightarrow + \infty} \hat{\rho}_A(t) = \hat{\rho}_A(+\infty),~~~\forall~\mathrm{finite~subsystem}~A.
 \label{relax_formula}
\end{equation}

\noindent
The latter states that an isolated quantum lattice model relaxes locally if the reduced density matrix associated to any finite subsystem converges
towards a time-independent matrix in the thermodynamic limit and for very long times. \\

To define the local stationarity, we consider an isolated quantum system relaxing locally in the previous sense. The latter admits
a stationary state if its long-time dynamics can be characterized locally \textit{via} a time-independent density matrix $\hat{\rho}_\mathrm{SS}$ 
for the full quantum system verifying the following condition

\begin{equation}
 \lim_{L \rightarrow + \infty} \mathrm{Tr}_{B}(\hat{\rho}_{\mathrm{SS}}) = \hat{\rho}_A(+\infty),~~~ \forall~\mathrm{finite~subsystem}~A,
 \label{local_stat}
\end{equation}

\noindent
where $B$ is the complementary subsystem. Equation \eqref{local_stat} states that each time-independent reduced density matrix $\hat{\rho}_A$, found  
by considering the limit of infinite times which exists due to the assumption of the local relaxation defined at Eq.~\eqref{relax_formula},
can be deduced from the time-independent density matrix $\hat{\rho}_{\mathrm{SS}}$ by tracing out all the degrees of freedom of the
associated complementary subsystem $B$, in the thermodynamic limit. The fact that the stationary state is described locally by a time-independent 
density matrix $\hat{\rho}_{\mathrm{SS}}$ implies that the long-time expectation values of local observables are identical to the corresponding
expectation values using the density matrix $\hat{\rho}_{\mathrm{SS}}$. Hence, if one considers $\hat{O}$ a local observable and $\bar{O}$ its corresponding 
steady value, one should obtain

\begin{equation}
 \bar{O} = \lim_{t \rightarrow + \infty} \langle \hat{O}(t) \rangle = \mathrm{Tr}(\hat{\rho}_{\mathrm{SS}} \hat{O}). 
\end{equation}

In what follows, we present an experimental observation of such relaxation and equilibration processes. In Ref.~\cite{trotzky2012}, the authors have investigated the 
far-from-equilibrium dynamics of a density wave of ultracold bosonic atoms loaded in an artificial (optical) lattice for an isolated one-dimensional Bose gas.
The latter is characterized by the Hamiltonian $\hat{H}$ of a non-integrable Bose-Hubbard chain with an external harmonic trap which may be written as follows 

\begin{equation}
 \hat{H} = \sum_R \left[ -J ( \hat{a}_R^{\dag} \hat{a}_{R+1} + \mathrm{h.c.} ) + \frac{U}{2} \hat{n}_R (\hat{n}_R - 1) + \frac{K}{2} R^2 \hat{n}_R \right], 
 \label{h_trotzky}
\end{equation}

\noindent
where $\hat{a}_R$ ($\hat{a}^{\dag}_R$) denotes the bosonic annihilation (creation) operator acting on the lattice site $R \in \mathbb{Z}$ and $\hat{n}_R$ the associated 
particle number operator. $J>0$ corresponds to the hopping amplitude, $U>0$ to the on-site repulsive interaction energy and $K = m \omega^2 a^2 > 0$ to
the external harmonic trap ($m$ refers to the particle mass, $\omega$ to the trapping frequency and $a$ to the lattice spacing).  \\

\begin{figure}[!h]
\centering
\includegraphics[scale = 0.138]{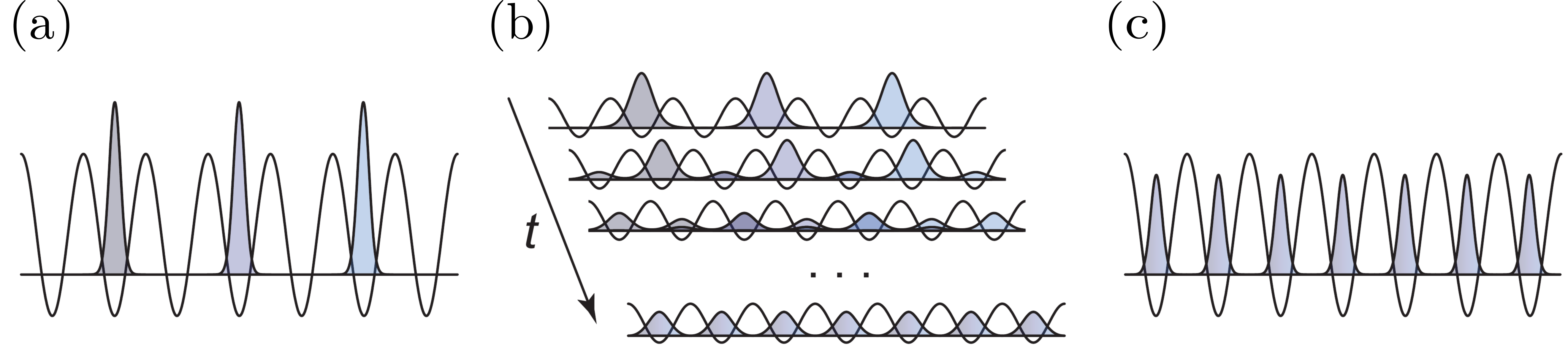}  
\caption{Quench and measurement protocols for the experimental investigation of the far-from-equilibrium dynamics of a density wave for an isolated one-dimensional Bose gas.
(a)~ \textit{Preparation}: the quantum system is initially prepared in a density wave $\ket{\Psi_0}$, see Eq.~\eqref{density_wave}, requiring a large lattice depth to maximize the energy ratio
$U/J$. (b)~ \textit{Evolution}: a global quench is applied to the lattice model at $t=0^+$ by reducing suddenly the lattice depth (or equivalently the ratio $U/J$) permitting a tunneling
process for bosons on nearest lattice sites. (c)~ \textit{Readout}: the properties of the time-evolved quantum state $\ket{\Psi(t)}$ are read out by freezing the time evolution
of the quantum system by suppressing the tunneling process. Figure extracted and adapted from Ref.~\cite{trotzky2012}.}
\label{protocol_trotzky}
\end{figure}

The experimental protocol considered at Ref.~\cite{trotzky2012}, to observe the far-from-equilibrium dynamics of a bosonic density wave in such isolated quantum model, 
is the following, see also Fig.~\ref{protocol_trotzky} :

\begin{enumerate}
 \item At time $t=0$, the quantum system is prepared in a bosonic density wave described by the many-body quantum state 
 
 \begin{equation}
\ket{\Psi_0} = \ket{...,1_{R=-2},0_{R=-1},1_{R=0},0_{R=1},1_{R=2},...},  
\label{density_wave}
 \end{equation}
 
 where only the even lattice sites are occupied by a single bosonic particle. Note that, for a same filling $\bar{n} \gtrsim 1/2$, $\ket{\Psi_0}$ can
 be built from the ground state of the extended version of the previous Hamiltonian $\hat{H}$ by adding the interaction term $V \hat{n}_R \hat{n}_{R+1}$ 
 with $V>0$ inside the sum over the lattice sites at Eq.~\eqref{h_trotzky}, and by considering the following condition on the energy ratios
 $V/J, U/J \gg 1$ and $K/J \ll 1$.  
 
 \item At time $t=0^{+}$, the quantum model is suddenly and globally quenched to different interaction parameters $J,U,K$ from those considered 
 initially to prepare the density wave \footnote{The latter implies in particular $U/J \rightarrow + \infty$ in order to suppress the tunneling process.}.
 The new set of parameters is characterized by $U/J > 1$ and $K/J \ll 1$ while the filling $\bar{n}$ is fixed \footnote{Note that the Hamiltonian $\hat{H}$ at
 Eq.~\eqref{h_trotzky} is fully characterized by three physical quantities : the ratio $U/J$, $K/J$ and the filling $\bar{n}$.}. This defines the
 post-quench Hamiltonian used to perform the unitary time evolution and where a tunneling process is allowed.  
 
 \item Finally, at time $t$, the time evolution of the quantum system is frozen by suppressing the tunnel-coupling ($J$) between nearest lattice sites to read out
 the properties of the corresponding time-evolved quantum state $\ket{\Psi(t)}$. 
\end{enumerate}

\begin{figure}[!h]
\centering
\begin{tabular}{c}
\includegraphics[scale = 0.52]{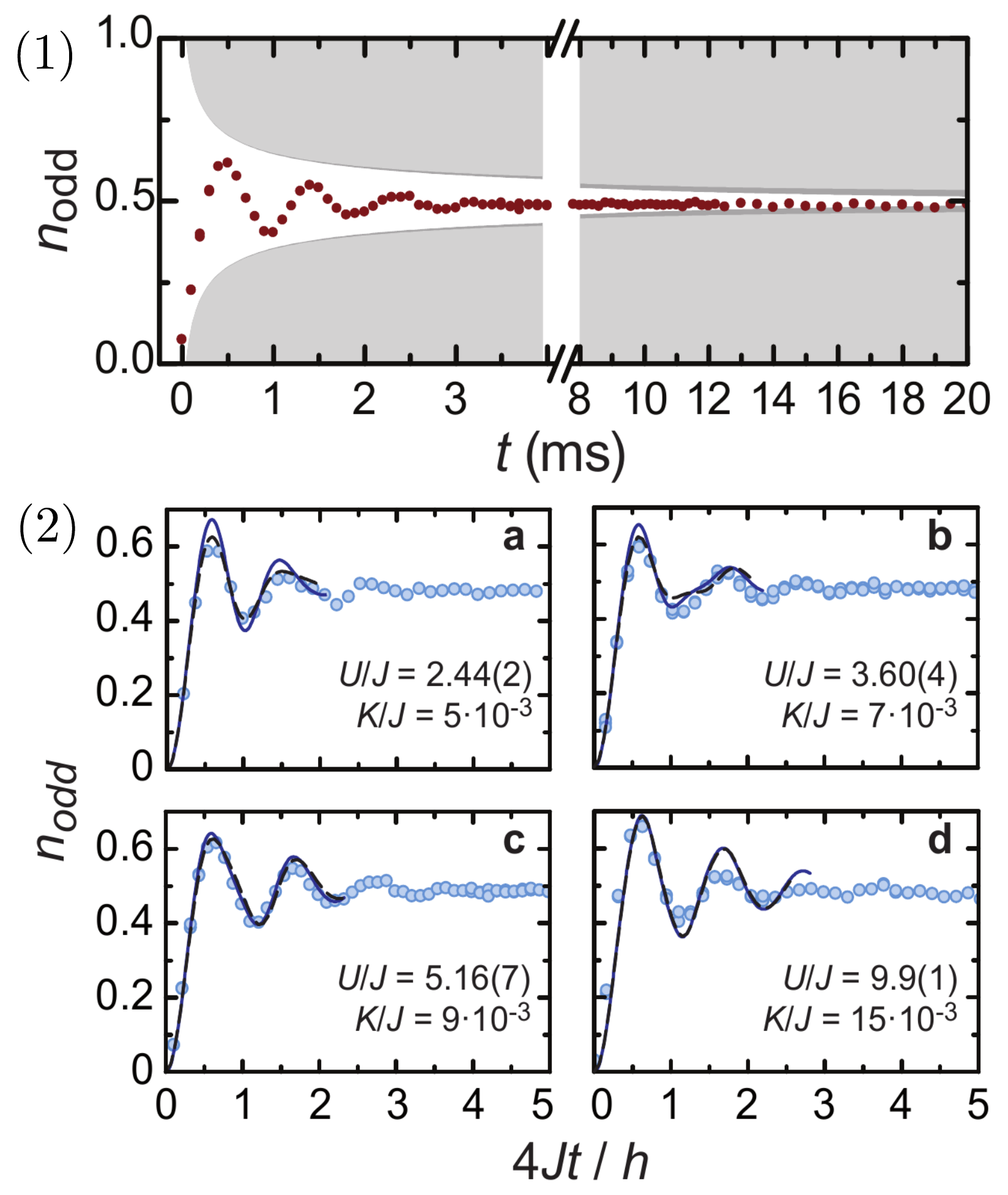} 
\end{tabular}
\caption{Relaxation and equilibration processes in the far-from-equilibrium dynamics of a density wave for an isolated one-dimensional
Bose gas \textit{via} an experimental investigation of the time-dependent odd-density average $n_{\mathrm{odd}}(t)$. (1)~ Odd-density
average $n_{\mathrm{odd}}(t) \propto \sum_R \langle \Psi(t) | \hat{n}_{2R+1} | \Psi(t) \rangle = \sum_R \langle \Psi_0 | \hat{n}_{2R+1}(t) | \Psi_0 \rangle$, 
where $\ket{\Psi_0}$ denotes the state vector of the bosonic density wave defined at Eq.~\eqref{density_wave}, as a function of the time $t$ in milliseconds. 
(2)~ $n_{\mathrm{odd}}(t)$ as a function of the dimensionless time $4Jt/h$, where $h$ is the Planck constant and $J$ the hopping amplitude, for four different
interaction strengths $U/J$ (blue circles). The solid lines represent numerical results obtained from t-DMRG simulations. The dashed lines represents numerical
results using the same tensor-network based technique by adding a next-nearest neighbor hopping term of the form $-J_{\mathrm{NNN}}\sum_R (\hat{a}^{\dag}_R \hat{a}_{R+2} + 
\mathrm{h.c.})$ in the Hamiltonian at Eq.~\eqref{h_trotzky}, where the associated hopping amplitude is given by $J_{\mathrm{NNN}}/J 
\simeq 0.12$ (a), $ 0.08$ (b), $0.05$ (c) and $0.03$ (d). Both figures (1) and (2) are extracted from Ref.~\cite{trotzky2012}.}
\label{odd_density_trotzky}
\end{figure}

On Fig.~\ref{odd_density_trotzky} extracted from Ref.~\cite{trotzky2012}, we display several experimental measurements of the time-dependent odd-density average 
$n_{\mathrm{odd}}(t) \propto \sum_R \langle \Psi(t) | \hat{n}_{2R+1} | \Psi(t) \rangle = \sum_R \langle \Psi_0 | \hat{n}_{2R+1}(t)
| \Psi_0 \rangle$, where $\ket{\Psi_0}$ denotes the initial many-body quantum state defined at Eq.~\eqref{density_wave} and representing the bosonic density wave. \\
The experimental results on Fig.~\ref{odd_density_trotzky}(1) shows the evolution of $n_{\mathrm{odd}}(t)$ as a function of the time $t$ (in milliseconds) for 
a post-quench Hamiltonian defined by the following ratios: $h/4J \simeq 0.9~\mathrm{ms}$, $U/J = 5.16(7) \gg K/J \simeq 9 \times 10^{-3}$. 
As expected, $n_\mathrm{odd} = 0$ at $t=0$. The latter is due to the initial state $\ket{\Psi_0}$ where only the even sites of the lattice chain are 
occupied by a single bosonic particle. Then, just after the global quench where the lattice depth is significantly decreased (and thus allowing a tunneling 
process to occur), the odd-density average rapidly increases to reach a maximal value around $0.6$ at time $t= 0.5~\mathrm{ms}$.
Finally, for $t>0.5~\mathrm{ms}$, clear damped oscillations for $n_\mathrm{odd}$ with a temporal period $T \simeq h/4J$ can be observed. Consequently, such behavior 
is a clear experimental evidence of the local relaxation process for a specific isolated quantum lattice model. Furthermore, note that these damped oscillations 
on several periods (consisting of a time of $3-4~\mathrm{ms}$ approximately) reach the value $\bar{n}_{\mathrm{odd}} \simeq 1/2$ which remains constant for longer
times. Hence, the previous value can be seen as the steady (or stationary) one and characterize the equilibration process. 
This stationary value was expected since the post-quench Hamiltonian is confined in the superfluid phase ($\bar{n} \notin \mathbb{N}$ and $J \neq 0$),
leading to a similar occupation probability of the bosonic particles between the even and odd sites of the lattice chain. \\
Similar experimental investigations have been performed by the authors of Ref.~\cite{trotzky2012} for various values of the interaction parameter $U/J$, as shown
on Fig.~\ref{odd_density_trotzky}(2), where the experimental data are found to be in very good agreement with t-DMRG numerical simulations. For the four different 
interaction strengths $U/J$, the previous statements concerning the time evolution of $n_{\mathrm{odd}}(t)$ deduced from Fig.~\ref{odd_density_trotzky}(1) are 
still valid. Indeed, for each panel of Fig.~\ref{odd_density_trotzky}(2), both a relaxation and a equilibration process occur characterized by damped oscillations
on several periods and a steady odd-density average $\bar{n}_{\mathrm{odd}} \simeq 1/2$ respectively. \\

On Fig.~\ref{odd_density_trotzky}, when investigating the time-dependent odd-density average $n_{\mathrm{odd}}(t)$ for different interaction strengths $U/J$,
both a relaxation and equilibration process have been unveiled. Besides, since the Hamiltonian at Eq.~\eqref{h_trotzky} is non-integrable, the corresponding 
stationary states are expected to be indistinguishable locally from a thermal equilibrium state described by a standard Gibbs ensemble.
This assumption is supported by the eigenstate thermalization hypothesis conjecturing that only isolated non-integrable quantum systems driven far from equilibrium
will evolve in time to a thermal Gibbs state, \textit{ie.} will thermalize. Such thermalization process can not occur for integrable models due to their  
additional local or quasi-local conservation laws \footnote{Hence, the allowed portion of the many-body Hilbert space will be drastically
decreased biasing the quench dynamics.}. Nevertheless, the stationary states for integrable quantum models can still be described in the framework of statistical quantum mechanics by relying on the
so-called generalized Gibbs ensemble. This statistical ensemble consists of building a density matrix to describe the stationary state [see Eq.~\eqref{local_stat}] by 
taking into account all the additional conservation laws. These features concerning the thermalization process are discussed in more details at 
Subsec.~\ref{thermalization}.

\subsection{Thermalization process}
\label{thermalization}

Within the study of the equilibration process for isolated quantum lattice models driven far from equilibrium, one can wonder
how to characterize the corresponding long-time equilibrium state, described locally by a time-dependent density matrix $\hat{\rho}_{\mathrm{SS}}$
verifying Eq.~\eqref{local_stat}. The significant success of statistical mechanics \cite{gibbs1902, toman1938, balescu1975} to study 
physical systems with a large number of degrees of freedom indicates that equilibrium properties can be in general well captured by few global parameters such as
the temperature and possibly the chemical potential, \textit{ie.} by considering the standard Gibbs ensembles. \\

The term \textit{thermalization} refers to the physical process where the stationary state can be accurately described by a thermal equilibrium state. The latter is 
thus expected to be proportional to the operator $e^{-\beta_{\mathrm{eff}} \hat{H}}$ fully characterized by an effective inverse temperature $\beta_{\mathrm{eff}}$
\cite{eisert2015}. Indeed, since isolated quantum systems are considered from the beginning, the total energy is always a conserved physical quantity.
To illustrate the previous statement and for the following discussions, let us consider a sudden global quench at a time $t=0^+$ where $\hat{H}$ denotes the
post-quench Hamiltonian of an isolated many-body quantum system. The total energy $E(t)$ during the unitary time evolution is given by 

\begin{align}
& E(t) = \mathrm{Tr}[\hat{\rho}(t) \hat{H}] = \mathrm{Tr}[\hat{\rho}(0) \hat{H}(t)] = \mathrm{Tr}[\hat{\rho}(0)\hat{H}(0^+)] = \mathrm{Tr}[\hat{\rho}(0^+) \hat{H}] = E(0^+),
 \label{conservedE}
\end{align}

\noindent
and hence is well conserved during the quench dynamics. Let us discuss in more details the definition of the thermal equilibrium state. According 
to the rigorous mathematical definition of the local stationarity provided at Ref.~\cite{fagotti2016} and presented at Eq.~\eqref{local_stat} to describe the 
equilibration process, the time-dependent matrix $\hat{\rho}_{\mathrm{SS}}$ characterizing the stationary state locally can be defined using a standard Gibbs ensemble.
Here, the canonical ensemble is considered since no physical quantity other than the energy is assumed to be conserved during the quench dynamics. 
Consequently, while considering a fixed effective inverse temperature $\beta_{\mathrm{eff}}$, $\hat{\rho}_{\mathrm{SS}}$ may be written as \cite{fagotti2016, vidmar2016}

\begin{equation}
 \hat{\rho}_{\mathrm{SS}} = \hat{\rho}_{\mathrm{Gibbs,c}} = \frac{1}{Z} e^{-\beta_{\mathrm{eff}} \hat{H}},
 \label{rho_gibbs}
\end{equation}

\noindent
with $Z = \mathrm{Tr}(e^{-\beta_{\mathrm{eff}} \hat{H}})$ the canonical partition function. The effective inverse temperature
$\beta_{\mathrm{eff}}$ is deduced by constraining the energy in the Gibbs canonical ensemble to be equal to the one of the time-evolved quantum state $\ket{\Psi(t)}$,
\textit{ie.} $\langle \Psi(t) | \hat{H} | \Psi(t) \rangle = \langle \Psi(0^+) | \hat{H} | \Psi(0^+) \rangle = \mathrm{Tr}(\hat{\rho}_{\mathrm{Gibbs,c}} \hat{H})$. 
A very similar reasoning applies if one has in addition a particle-number conservation. For this specific case, $\hat{\rho}_{\mathrm{SS}}$
is defined \textit{via} the grand canonical ensemble leading to 

\begin{equation}
\hat{\rho}_{\mathrm{SS}} = \hat{\rho}_{\mathrm{Gibbs,gc}} = \frac{1}{Z} e^{-\beta_{\mathrm{eff}} (\hat{H} - \mu_{\mathrm{eff}} \hat{N})}, 
\label{rho_gibbs_2}
\end{equation}

\noindent
with $\hat{N}$ the total particle number operator, $\mu_{\mathrm{eff}}$ the effective chemical potential and $Z =
\mathrm{Tr}[e^{-\beta_{\mathrm{eff}} (\hat{H} - \mu_{\mathrm{eff}} \hat{N})}]$ the grand canonical partition function. $\beta_{\mathrm{eff}}$ is fixed 
in a similar way as before, \textit{ie.} $\langle \Psi(t) | \hat{H} | \Psi(t) \rangle = \langle \Psi(0^+) | \hat{H} | \Psi(0^+) \rangle = 
\mathrm{Tr}(\hat{\rho}_{\mathrm{Gibbs,gc}} \hat{H})$. To set $\mu_{\mathrm{eff}}$, the total number of particles $N$ in the Gibbs 
grand canonical ensemble has to be equivalent to the one of the time-evolved quantum state $\ket{\Psi(t)}$. Hence, the constraint can be formulated as follows, $N = \langle \Psi(t) | \hat{N} | \Psi(t) \rangle = 
\mathrm{Tr}(\hat{\rho}_{\mathrm{Gibbs,gc}} \hat{N})$.\\

We stress again that Eq.~\eqref{rho_gibbs} has to be understood in the following sense : the
stationary state is described locally by the Gibbs canonical ensemble meaning that the long-time expectation values of local observables are identical to the corresponding
thermal expectation values. Mathematically, by considering $\hat{O}$ a local observable and $\bar{O}$ its steady value, we get 

\begin{equation}
 \bar{O} = \lim_{t \rightarrow + \infty} \langle \hat{O}(t) \rangle = \mathrm{Tr}(\hat{\rho}_{\mathrm{SS}} \hat{O}) = \mathrm{Tr}(\hat{\rho}_{\mathrm{Gibbs,c}} \hat{O}).
 \label{th_expec}
\end{equation}

\noindent
Hence, one can be more precise about the underlying physical process behind the thermalization. We recall that for the equilibration process, the complementary subsystem
with respect to the one on which the local observable $\hat{O}$ acts, can be seen as an environment. Therefore, for the thermalization process, this environment is similar to 
a heat bath with the fixed effective inverse temperature $\beta_{\mathrm{eff}}$ \cite{fagotti2016}. Similar statements also hold for Eq.~\eqref{rho_gibbs_2} when
considering the Gibbs grand canonical ensemble.\\

To sum up, the thermalization process, or again the local relaxation toward a thermal equilibrium state, is expected to occur 
for isolated quantum systems in the absence of additional local conserved quantities which are not taken into account by the standard Gibbs ensembles. Such statement 
is conjectured by the eigenstate thermalization hypothesis (ETH) \cite{rigol2008, deutsch1991, srednicki1994, tasaki1998, rigol2012}. Indeed, the latter suggests
that, for generic non-integrable (complex-enough) isolated many-body quantum systems driven far from equilibrium, each eigenstate of the quench Hamiltonian is locally indistinguishable from
a thermal state with the same energy. Hence, this hypothesis implies that the stationary state can be described accurately and locally by the standard Gibbs ensembles, \textit{ie.} is
indistinguishable locally from a thermal equilibrium state, see Ref.~\cite{rigol2008} for more details. \\

\begin{figure}[!h]
\centering
\begin{tabular}{c}
\includegraphics[scale = 0.16]{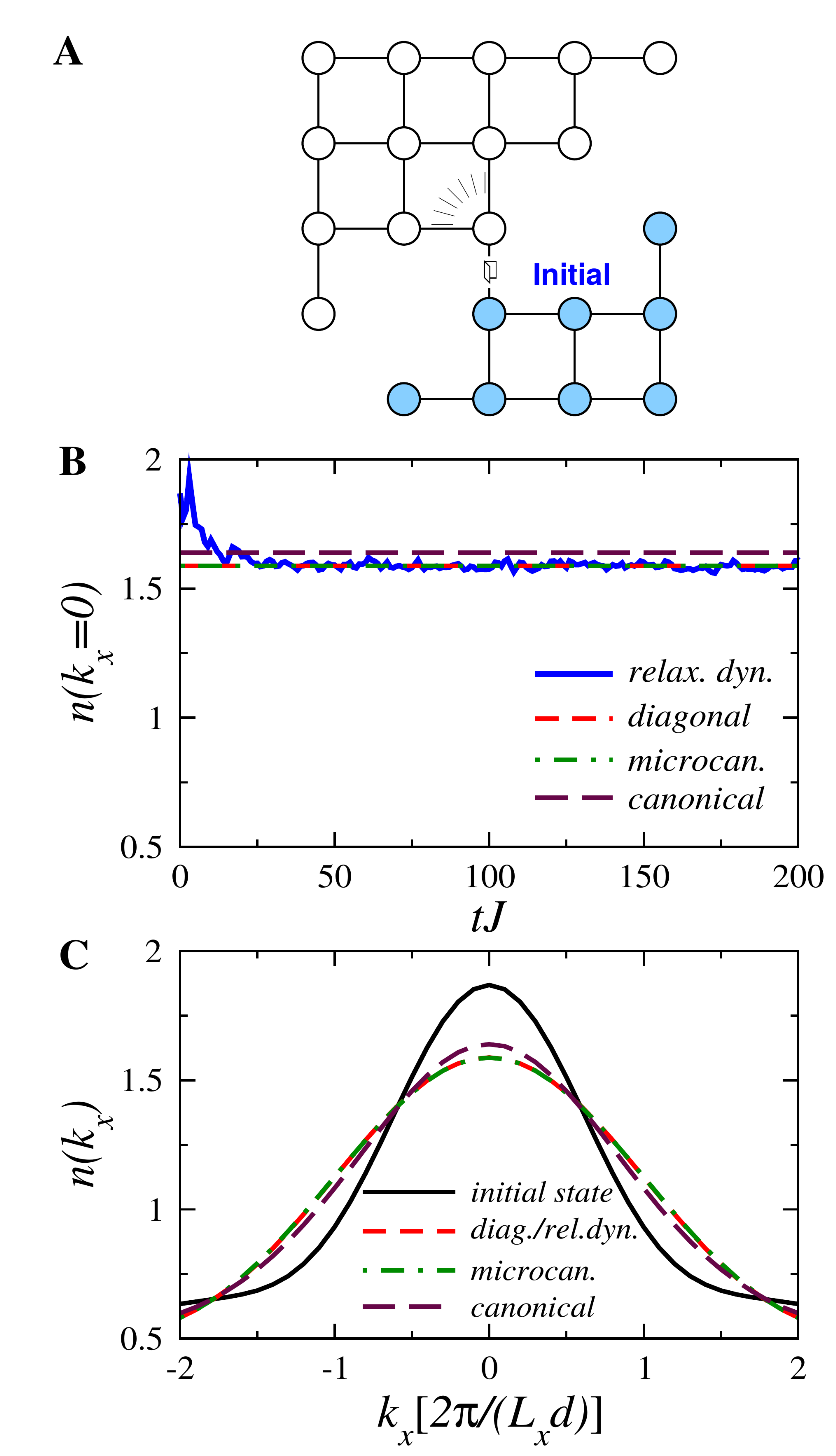}        
\end{tabular}
\caption{Relaxation towards a thermal equilibrium state. (a)~ Two-dimensional lattice on which $5$ hardcore bosons propagate in time. Initially, the hardcore bosons are prepared in the ground state
of the sublattice in the lower-right corner denoted by $\ket{\Psi(0)}$. Then, a sudden geometric quench is performed by releasing them through the indicated link. (b)~ Relaxation dynamics of
$n(k_x = 0)$ the momentum distribution center and compared with the predictions of the diagonal, microcanonical and canonical statistical ensembles. For the microcanonical
ensemble, the average is taken over all eigenstates whose corresponding eigenenergies are lie within a narrow window $[E_0 - \Delta E, E_0 + \Delta E]$ where 
$E_0 = \langle \Psi(0) | \hat{H} | \Psi(0) \rangle = \langle \Psi(t) | \hat{H} | \Psi(t) \rangle = -5.06J$ is the mean energy of the initial state just after the geometric quench ($\hat{H}$ denotes the final 
Hamiltonian), $\Delta E = 0.1J$ and $J$ corresponds to the hopping amplitude. 
The canonical ensemble (effective) temperature is fixed to $k_{B} T = 1.87J$, with $k_{B}$ the Boltzmann constant, so that the ensemble prediction
concerning the energy leads to the value $E_0$. (c)~ Momentum distribution function $n(k_x)$ as a function of the momentum $k_x$ in units of $2\pi/(L_xd)$ in the initial
state, after relaxation, and in the three previous ensembles. $d$ refers to the lattice spacing and $L_x = 5$ to the lattice width along
the horizontal axis. Figures extracted from Ref.~\cite{rigol2008}.}
\label{relax_eth}
\end{figure}

On Fig.~\ref{relax_eth} extracted from Ref.~\cite{rigol2008}, we present numerical results illustrating the thermalization process for 
a non-integrable isolated quantum lattice model driven far from equilibrium. To do so, the authors have investigated, \textit{via} an exact
diagonalization technique, the relaxation dynamics. The considered quantum lattice model consists of $5$ hardcore bosons confined in a two-dimensional lattice where $4$ lattice sites are missing, see Fig.~\ref{relax_eth}(a).  
The hardcore bosons are prepared in the ground state of the sublattice in the lower-right corner denoted by $\ket{\Psi(0)}$. Then, the system is driven far from equilibrium 
\textit{via} a sudden geometric quench performed by releasing them through the indicated link (see the door symbol on Fig.~\ref{relax_eth}). Since the quantum model is 
assumed to be isolated, its time evolution is unitary and the corresponding time-evolved quantum state $\ket{\Psi(t)}$ can be written as $\ket{\Psi(t)} = 
e^{-i\hat{H}t} \ket{\Psi(0)}$ where $\hat{H}$ denotes the final (or the quench) Hamiltonian. The latter reads as ($\hbar = 1$)

\begin{equation}
 \hat{H} = -J \sum_{\langle i,j \rangle} (\hat{b}_i^{\dag} \hat{b}_j + \mathrm{h.c.}) + U \sum_{\langle i,j \rangle} \hat{n}_i \hat{n}_j,
\end{equation}

\noindent
where $\langle i,j \rangle$ indicates a summation over all the nearest-neighbor lattice sites. As usual, $J>0$ refers to the hopping amplitude and $U>0$ to the nearest-neighbor 
repulsion parameter fixed to $U=0.1J$. Since hardcore bosons are considered, they fulfill the usual canonical commutation rules on distinct lattice sites $[\hat{b}_i,
\hat{b}_j] = [\hat{b}_i, \hat{b}_j^{\dag}] = [\hat{b}_i^{\dag}, \hat{b}_j^{\dag}] = 0,~\forall i \neq j$. However, the hardcore constraint imposes 
the canonical anticommutation rules on a same lattice site $\{ \hat{b}_i, \hat{b}_i^{\dag} \} = 1$ and $(\hat{b}_i)^2 = (\hat{b}_i^{\dag})^2 = 0$ for all $i$. $\hat{n}_i
= \hat{b}_i^{\dag} \hat{b}_i$ corresponds to the local density operator acting on the lattice site $i = (i_x,i_y)$. \\

On Fig.~\ref{relax_eth}(b), $n(k_x = 0,t) = \sum_{k_y} n(k_x=0,k_y,t)$ the time-dependent momentum distribution center is represented as a function of the 
dimensionless time $tJ$ ($\hbar=1$). The latter is defined from the time-dependent momentum distribution \footnote{The latter is easily found by performing a 
Fourier transform of the density operator in momentum space $\hat{n}(\mathbf{k}) = \hat{b}^{\dag}_{\mathbf{k}} \hat{b}_{\mathbf{k}}$ and then by considering its expectation value
with respect to the time-evolved quantum state $\ket{\Psi(t)}$.} $n(\mathbf{k},t)$ of the full two-dimensional square lattice,
which may be written as 

\begin{equation}
n(k_x,k_y,t) = (1/L^2) \sum_{i,j} e^{-i2\pi \mathbf{k} (\mathbf{r}_i - \mathbf{r}_j)/L} \langle \Psi(t)|\hat{b}^{\dag}_i \hat{b}_j|\Psi(t)\rangle,
\end{equation}

\noindent
where $L = L_x = L_y = 5$ ($L_y$ : lattice width along the vertical axis) and $\mathbf{r}_i = (i_xd, i_yd)$, the position,
involves the lattice spacing $d$. 
The time-dependent momentum distribution center $n(k_x = 0,t)$ clearly displays a relaxation process and then equilibrates to the steady value $\bar{n}(k_x=0)
\simeq 1.6$, see solid blue line. Hence, it should exist a time-independent density matrix $\hat{\rho}_{\mathrm{SS}}$ describing accurately and locally
this stationary state. The previous results are also compared to the predictions of the diagonal, microcanonical and canonical ensembles (see the dashed red, green, 
purple lines respectively and see caption of Fig.~\ref{relax_eth} and Ref.~\cite{rigol2008} for more details concerning the standard Gibbs ensembles), and show
a very good agreement on a large timescale. Consequently, the density matrix $\hat{\rho}_{\mathrm{SS}}$ previously introduced, can be seen as being equal locally
to the one associated to the standard Gibbs ensembles considered here (microcanonical, canonical), \textit{ie.} $\hat{\rho}_{\mathrm{SS},\mathrm{loc}} 
= \hat{\rho}_{\mathrm{Gibbs},\mathrm{loc}}$ ($\mathrm{loc}$ refers to the locality). As a consequence, the stationary state is indistinguishable locally
from a thermal equilibrium state and this validates the conclusion provided by the ETH. Indeed, one reminds that this hypothesis conjectures that isolated
non-integrable quantum systems, driven far from equilibrium, evolve in time towards a stationary state which appears to be in thermal equilibrium. \\
A second verification is provided at Fig.~\ref{relax_eth}(c) where the momentum distribution $n(k_x)$ is represented as a function of the momentum $k_x$
in units of $2\pi/(L_xd)$, after relaxation, and using the three same ensembles (diagonal, microcanonical, canonical). The numerical results for this observable
after the relaxation process and the corresponding predictions from the two Gibbs ensembles are in very good agreement. This means once again that the properties of the 
stationary state are well captured by the standard Gibbs ensembles. Note that both physical quantities, $n(k_x)$ the full momentum distribution with respect to the 
horizontal axis and the associated time-dependent momentum distribution center $n(k_x = 0, t)$, are accessible from actual experiments using ultracold cold atoms loaded 
in an artificial (optical) lattice, see Ref.~\cite{mandel2003} for instance. \\

However, in the case of integrable isolated quantum systems \cite{caux2011} having additional local conservation laws than
the sole energy (and possibly the particle-number conservation) during the quench dynamics, they do not thermalize and the ETH fails (see Refs.~\cite{rigol2007, rigol2008, rigol2009}
for several examples). More precisely, the ETH breakdown is due to these additional local conserved quantities biasing the dynamics 
and preventing a thermalization in a Gibbs canonical (or grand canonical) ensemble to occur. Hence, the failure of the eigenstate thermalization hypothesis results
in an impossibility to describe the stationary state for integrable isolated systems by a thermal equilibrium state \textit{via} a standard Gibbs ensemble. 
Note that the validity and the breakdown of the ETH have been extensively studied for a large panel of models \cite{rigol2008, rigol2009, polkovnikov2011,
steinigeweg2014, beugeling2014}. \\

Nevertheless, using a so-called generalized Gibbs ensemble (GGE), the stationary state of integrable isolated quantum systems can be accurately described locally,
see Ref.~\cite{vidmar2016} and references therein for more details. The GGE denotes a statistical ensemble characterized by a density matrix 
$\hat{\rho}_{\mathrm{GGE}}$ constructed by taking into account all the local conserved quantities during the quench dynamics.
In the following, we consider a general context to build such density matrix. To do so, a set of local operators $\{\hat{Q}_n \}$ is introduced and denotes
the different local conserved quantities of  
the quench dynamics with an associated set of Lagrange multipliers $\beta_{n}$. Since the set $\{\hat{Q}_n \}$ represents the non-trivial constants of motion, 
they commute mutually, \textit{ie.} $[\hat{Q}_n, \hat{Q}_{n'}]=0,\forall n,n'$, but also each of them commutes with the post-quench Hamiltonian $\hat{H}$,
\textit{ie.} $[\hat{H}, \hat{Q}_n] = 0, \forall n$. Therefore, the associated GGE density matrix is defined as 

\begin{equation}
 \hat{\rho}_{\mathrm{GGE}} = \frac{1}{Z} e^{-\sum_n \beta_n \hat{Q}_n},
\end{equation}

\noindent
with $Z = \mathrm{Tr}(e^{-\sum_n \beta_n \hat{Q}_n})$ the partition function of the generalized Gibbs ensemble. Each Lagrange multiplier $\beta_n$ 
is fixed such that the GGE expectation value of its corresponding operator $\hat{Q}_n$ is equal to the one just after the quench. Once again, we consider
the case of a sudden global quench at time $t=0^+$ to drive the isolated integrable model far from equilibrium.
Mathematically, the effective parameter $\beta_n$ is determined by the following constraint  

\begin{equation}
 \langle \Psi(0^+) | \hat{Q}_n | \Psi(0^+) \rangle = \mathrm{Tr}(\hat{\rho}_{\mathrm{GGE}} \hat{Q}_n).
 \label{cc}
\end{equation}

\noindent
Finally, for an integrable isolated many-body quantum model relaxing and equilibrating locally
to a non-thermal equilibrium state, one expects to find the following relation  

\begin{equation}
 \bar{O} = \lim_{t \rightarrow + \infty} \langle \Psi(t) | \hat{O} | \Psi(t) \rangle = \mathrm{Tr}(\hat{\rho}_{\mathrm{GGE}} \hat{O}),
\end{equation}

\noindent
where $\hat{O}$ is a local observable having a steady value $\bar{O}$. Several examples using the generalized Gibbs ensemble can be found at
Refs.~\cite{vidmar2016, caux2012, fagotti2013}. 

\subsection{Information spreading and the Lieb-Robinson bound}
Another important aspect of the far-from-equilibrium dynamics concerns the transport properties of isolated quantum lattice models and more precisely the spreading of 
information. Indeed, the ability of quantum lattice models to establish long-distance correlations and entanglement, and possibly equilibrate, is determined
by the speed at which information can propagate within the lattice. An important progress for a better understanding of the information spreading is given by the 
Lieb-Robinson bound limiting the speed at which the information can propagate into lattice systems having short-range interactions and a finite local Hilbert space.

\paragraph{Lieb-Robinson bound} In 1972, Lieb and Robinson have unveiled a bound that forms a linear causality cone beyond
which information decays exponentially, see Ref.~\cite{lieb1972}. More precisely, the Lieb-Robinson bound makes a statement about the norm of the commutator of any observables $\hat{O}_A$ and 
$\hat{O}_B$, supported on regions $A$ and $B$ respectively, and taken at different times. The latter may be written as follows 

\begin{equation}
 || [\hat{O}_A(t), \hat{O}_B(0)] || \leq c N_{\mathrm{min}} ||\hat{O}_A|| ~||\hat{O}_B|| e^{- \frac{L-v|t|}{\xi}},
 \label{lrb}
\end{equation}

\noindent
where $L$ denotes the distance between the region $A$ and $B$, and can be seen as the number of edges in the shortest path connecting the two regions.
$N_{\mathrm{min}} = \mathrm{min}(|A|,|B|)$ corresponds to the number of lattice sites in the smallest region, $||\hat{O}_A||$ ($||\hat{O}_B||$) to the operator 
norm of the observable $\hat{O}_A$ ($\hat{O}_B$) and $c,v,\xi$ to positive constants. The positive constant $v$, depending only on the interactions of the Hamiltonian 
governing the quantum model and the lattice structure, plays the role of the group velocity and is generally called the Lieb-Robinson velocity. 
We point out that the Lieb-Robinson bound does not depend on the quantum state of the system but only on the Hamiltonian $\hat{H}$ governing the dynamics.
Indeed, the norm of the commutator $[\hat{O}_A(t), \hat{O}_B(0)] = \hat{O}_A(t)\hat{O}_B(0) - \hat{O}_B(0) \hat{O}_A(t)$ involves the time-evolved operator 
$\hat{O}_A(t)$ which may be written as $\hat{O}_A(t) = e^{i\hat{H}t} \hat{O}_A(0) e^{-i \hat{H}t}$ in the Heisenberg picture and depends only on $\hat{H}$.
Hence, once this operator norm is established, the bound is valid for any quantum state of the system. \\

This Lieb-Robinson bound describes how an observable $\hat{O}_A$ supported on a region $A$ can affect another observable $\hat{O}_B$ living in a different
region $B$ after a time $t$. According to Eq.~\eqref{lrb}, it states that the speed of propagation for the quantum information is bounded ballistically 
and leads for the space-time pattern of the physical quantity $|| [\hat{O}_A(t), \hat{O}_B(0)] ||$ to a linear causality cone (also called \textit{effective light cone})
beyond which the information decays exponentially. Indeed, the quantum information decreases exponentially fast with the distance $L-v|t|$ and becomes negligible 
for any time $|t|$ fulfilling $|t| \ll L/v$. Although the norm of the commutator between two observables $\hat{O}_A$ and $\hat{O}_B$ is not of a general interest in quantum
mechanics, the latter is essential since it implies a similar bound for both the entanglement and the equal-time correlation functions (providing that the initial many-body quantum state
does not contain long-range correlations)~\cite{bravyi2006}. In the following, 
a special care will be devoted to the understanding of the space-time behavior of equal-time correlation functions corresponding to the main research topic of this thesis. \\

In Ref.~\cite{bravyi2006}, the authors have characterized the amount of correlations that can be created by a short-range interacting Hamiltonian evolution. They 
found that the latter vanishes also exponentially outside a linear causality cone. In other words, they unveiled as previously for the quantum information, a ballistic 
bound for the propagation of correlations. To do so, the space-time behavior of equal-time connected correlation functions has been investigated. The authors started by considering 
a many-body quantum state $\ket{\Psi}$ with a finite correlation length $\chi >0$, \textit{ie.} one for which all the connected correlation functions,
$\langle \hat{O}_A \hat{O}_B \rangle_c \equiv \langle \Psi | \hat{O}_A(t) \hat{O}_B(t) | \Psi \rangle - \langle \Psi | \hat{O}_A(t) | \Psi \rangle
\langle \Psi | \hat{O}_B(t) | \Psi \rangle$, decay exponentially. This implies that 

\begin{equation}
| \langle \hat{O}_A \hat{O}_B \rangle_c | \leq \tilde{c} e^{-\frac{L}{\chi}}, 
\end{equation}

\noindent
where $\tilde{c}$ is a positive constant, $\hat{O}_A$ and $\hat{O}_B$ correspond as previously to normalized quantum operators ($|| \hat{O}_A ||, || \hat{O}_B || \leq 1$) 
acting on region $A$ and $B$ respectively where the two regions are separated from each other by a distance $L$. Relying on the Lieb-Robinson bound provided at Eq.~\eqref{lrb},
the authors of Ref.~\cite{bravyi2006} have shed new light on the amount of correlations created after a time $t$ by demonstrating that the corresponding connected
correlation function, \textit{ie.} $| \langle \hat{O}_A(t) \hat{O}_B(t) \rangle_c |$, is bounded by the following quantity 

\begin{equation}
| \langle \hat{O}_A(t) \hat{O}_B(t) \rangle_c | \leq \bar{c} (|A| + |B|) e^{- \frac{L-2vt}{\chi'}},
\label{ccf}
\end{equation}

\noindent
where $\bar{c}$ is a positive constant and $\chi' = \chi + 2 \xi >0$. Hence, from Eq.~\eqref{ccf}, a ballistic bound also exists 
for the propagation of correlations. More precisely, the authors have shown that the correlation spreading is bounded ballistically leading for its associated space-time pattern
to a linear causality cone beyond which the correlations decrease exponentially fast with the distance $L-2vt$. Note that this ballistic bound for the propagation of 
correlations is very similar to the one for the spreading of quantum information given by the Lieb-Robinson bound at Eq.~\eqref{ccf}. However, the major difference 
is related to the bounded velocity at which correlations can be created which characterizes the linear causality cone for the space-time connected correlation functions.
Indeed, the bounded velocity for the correlations is given by $2v$, \textit{ie.} twice the Lieb-Robinson velocity and has been explained by the authors of
Ref.~\cite{calabrese2006} using a quasiparticle picture. \\

\paragraph{Calabrese-Cardy picture} In Ref.~\cite{calabrese2006}, using a conformal field theory (CFT), Calabrese and Cardy have investigated the space-time behavior of correlation functions
following a quantum quench in isolated quantum lattice models. According to their analytical results, they propose a simple picture
based on the spreading of quasiparticle excitations. This quasiparticle picture permits not only to explain the previous factor $2$ for the bounded velocity
characterizing the maximal speed at which the correlations can be created but also, and most importantly, how those correlations are generated providing a more precise 
physical picture and theoretical results than the previous ballistic bound. \\

For the general context of this quasiparticle picture, the authors have considered an isolated quantum lattice model driven far from 
equilibrium \textit{via} a global quantum quench where the subsequent dynamics is analyzed by investigating the generation of correlations. 
For a global quantum quench, the initial state $\ket{\Psi_0}$ having a very high energy compared to the one of the ground state of 
the quench Hamiltonian $\hat{H}$, acts as a source of quasiparticle excitations. Once emitted \textit{via} the quench, these entangled quasiparticles behave semi-classically
and travel at speed $v$. For this initial part of the discussion, a linear and gapless quasiparticle dispersion relation of the form $\omega_k = v |k|$ is considered.
Hence, the quasiparticles of such excitation spectrum are characterized by a single characteristic velocity given by $v$. For those arriving at a same time $t$ 
between points separated by a distance $R$, quantum correlations between local observables are created. Therefore, these quantum correlations in the $R-t$ plane 
display a sharp light-cone effect. Indeed, the connected correlations does not change significantly from their initial values until the so-called activation time
$t^* \sim R/2v$. \\

Furthermore, this quasiparticle picture is not restricted to one-dimensional isolated quantum lattice models but also holds in higher dimensions. 
The latter is also valid for more general excitation spectra taking into account both the properties of the lattice and the presence of a possible finite gap but 
has to be slightly adapted. The semi-classical picture is expected to describe accurately the generation of correlations for such more complex excitation spectrum (denoted 
by $E_k$) as long as each quasiparticle of quasimomentum $k$ is assumed to propagate with its corresponding group velocity $V_{\mathrm{g}}(k) = \partial_k E_k$ 
(with $\hbar$ fixed to unity). For this case, the reasoning concerning the generation of quantum correlations is very similar than the one explained before. The 
space-time correlations still display a linear causality cone. However, the correlation edge built from the different activation times $t^*$ for each distance $R$
is not anymore characterized by twice the sound velocity $2v$ but by twice the maximal group velocity $2V_{\mathrm{g}}^* = 2\mathrm{max}_k[V_{\mathrm{g}}(k)]$. Indeed, the
first correlation for a distance $R$ is expected to occur at the activation time $t^* = R/2V_{\mathrm{g}}^*$ and is generated by the fastest quasiparticles. \\

The ballistic motion of the correlation edge when investigating equal-time connected correlation functions for isolated short-range interacting quantum lattice models 
has been observed not only experimentally~\cite{cheneau2012,fukuhara2013} but also numerically~\cite{manmana2009,barmettler2012,carleo2014}.
In the following, we propose to discuss in details numerical results found using a time-dependent density matrix renormalization group approach (t-DMRG) and extracted from Ref.~\cite{manmana2009}
where a light-cone effect, \textit{ie.} a linear causality cone for the spreading of correlations, has been clearly identified when studying the quench dynamics of a one-dimensional lattice model of spinless fermions. \\

More precisely, the authors of Ref.~\cite{manmana2009} have investigated the quench dynamics of an isolated one-dimensional lattice model of interacting 
spinless fermions at half-filling, implying $\bar{n} = N/N_s = 1/2$ with $N$ the number of spinless fermions and $N_s$ the number of lattice sites, and governed by the following Hamiltonian $\hat{H}$,

\begin{equation}
 \hat{H} = - t_{\mathrm{h}} \sum_j (\hat{c}^{\dag}_{j+1} \hat{c}_j + \mathrm{h.c.}) + V \sum_j \hat{n}_j \hat{n}_{j+1}.
 \label{manHam}
\end{equation}

\begin{figure}[!h]
\centering
\begin{tabular}{c}
\includegraphics[scale = 0.25]{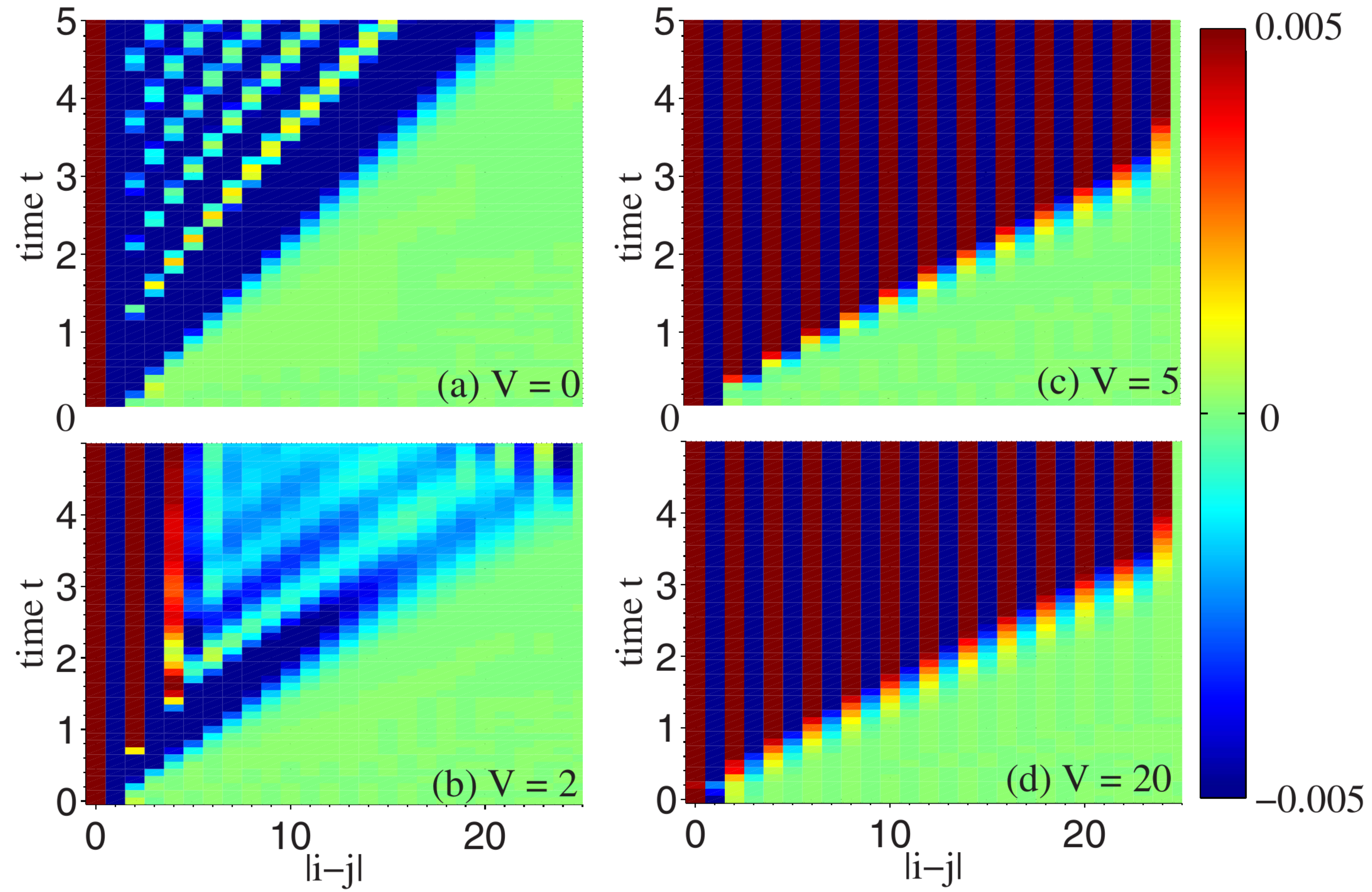}        
\end{tabular}
\caption{Space-time pattern of the equal-time connected density correlation function $C_{i,j}(t) = C_{|i-j|}(t) = \langle \hat{n}_i(t) \hat{n}_j(t) \rangle -
\langle \hat{n}_i(t) \rangle \langle \hat{n}_j(t) \rangle$ of spinless fermions after a sudden global quantum quench. The initial state $\ket{\Psi_0}$ 
is defined by the ground state of $\hat{H}(V_0)$ confined in the charge-density wave regime for a nearest-neighbor repulsion $V_0 = 10$, and then evolves unitarily in time 
with the post-quench Hamiltonian $\hat{H}(V)$ with (a)~$V=0$, (b)~$V=2$, (c)~$V=5$ and (d)~ $V=20$. $C_{i,j}(t) = C_{|i-j|}(t)$ is due to the conservation of the translational 
invariance during the quench dynamics. Figure extracted from Ref.~\cite{manmana2009}.}
\label{manmana_light}
\end{figure}

\noindent
$t_{\mathrm{h}} > 0$ represents the nearest-neighbor hopping amplitude and $V>0$ the nearest-neighbor repulsion. The operators $\hat{c}^{\dag}_j$ ($\hat{c}_j$) denotes
the creation (annihilation) of a spinless fermion on the lattice site of index $j$ and $\hat{n}_j$ the associated local density operator. The convention
adopted by the authors is the following : $\hbar = 1$ (Planck constant) and $a=1$ (lattice spacing) such that the energy is given in units of the hopping amplitude $t_{\mathrm{h}}$
and the time in units of the inverse hopping amplitude $t_{\mathrm{h}}^{-1}$. In what follows, $t_{\mathrm{h}}$ will be also fixed to unity. The Hamiltonian $\hat{H} = \hat{H}(V)$ defined at Eq.~\eqref{manHam} and fully
characterized by the interaction parameter $V$ (the filling is fixed to $\bar{n}=1/2$), displays a quantum phase transition at $V_c = 2$ between a Luttinger-liquid
regime for $V < V_c$ (metallic behavior) and a charge-density-wave (CDW) regime for $V>V_c$ (insulating behavior). The authors have investigated the far-from-equilibrium
dynamics of such quantum lattice model \textit{via} sudden global quenches. The initial state $\ket{\Psi_0}$ is defined as the ground state of $\hat{H}(V_0)$ confined
in the CDW regime where $V_0 = 10$. Then, the sudden global quench is performed by modifying abruptly the interaction parameter from the value $V_0$ to $V$ and remains 
constant in time. The latter defines the quench Hamiltonian $\hat{H}(V)$ governing the unitary time evolution. Since the quantum system is assumed to 
be closed (or isolated), the time-evolved quantum state $\ket{\Psi(t)}$ is governed by the Schrödinger equation and may be written
as $\ket{\Psi(t)} = e^{-i \hat{H}(V)t} \ket{\Psi_0}$. Finally, to characterize the quench dynamics of the lattice model and more precisely the spreading of correlations in 
the space-time plane, the authors have studied the time evolution of the equal-time connected density correlation function $C_{i,j}(t)$ defined as follows

\begin{equation}
C_{i,j}(t) = C_{|i-j|}(t) = \langle \hat{n}_i(t) \hat{n}_j(t) \rangle - \langle \hat{n}_i(t) \rangle \langle \hat{n}_j(t) \rangle.
\label{obs_man}
\end{equation}

\noindent
where $\langle ... \rangle$ represents the expectation value with respect to the initial many-body quantum state $\ket{\Psi_0} = \ket{\Psi_{\mathrm{gs}}[\hat{H}(V_0)]}$. 
On Fig.~\ref{manmana_light}, several numerical results of the space-time pattern of $C_{|i-j|}(t)$ using the t-DMRG approach are displayed as a function of the time $t$
and the distance $|i-j|$. Note that a large panel of post-quench nearest-neighbor repulsion parameters $V$ has been considered allowing to probe the quench dynamics of the 
quantum model for significantly different physical scenarios. Indeed, on Fig.~\ref{manmana_light}(a), $V=0$ hence the post-quench Hamiltonian is confined in a non-interacting 
Luttinger-liquid regime. According to the value $V_0$ given previously, this sudden global quench crosses the quantum critical point at $V_c=2$. 
On Fig.~\ref{manmana_light}(b), $V=V_c=2$ and thus the quantum system is globally quenched from the CDW regime to the quantum critical point. On Figs.~\ref{manmana_light}(c,d),
$V = 5$ and $20$ respectively. For these two last cases, the sudden global quench is confined within the CDW regime. \\

For the four different sudden global quenches presented previously, a clear linear causality cone is visible on each plot as expected from the general quasiparticle
picture provided by Calabrese and Cardy for the spreading of correlations in short-range interacting lattice models. Let us analyze in details Fig.~\ref{manmana_light}(a) 
for instance. For each distance $|i-j|$, one can define an activation time $t^*$ defined as follows : for any time $t<t^*$ the correlations are not activated yet, 
\textit{ie.} $C_{|i-j|} \simeq 0$ (and straightforwardly, for any time $t$ fulfilling $t \geq t^*$, the correlations are present). 
Note that these different times below (above) the activation one for each separation distance $|i-j|$ form the non-causal (causal) region of correlations
\footnote{In general, to get a clear non-causal region of correlations (zero-value correlations for times below the activation one), the equilibrium value
of the equal-time connected correlation function is substracted. Here, this is not necessary since the initial state $\ket{\Psi_0}$ is defined by a simple 
density wave, \textit{ie.} is characterized by an unique ground state of the CDW regime, and thus implying $C_{|i-j|}(0) = 0,~\forall |i-j|$ (see Ref.~\cite{manmana2009}
for the parameters $N$ and $N_s$ considered to obtain such initial quantum state $\ket{\Psi_0}$).}.
Once the activation time $t^*$ for each separation distance $|i-j|$ is located, one can fit them using a linear ansatz. Consequently, it yields for the space-time pattern of
$C_{|i-j|}(t)$ a linear causality cone beyond which the correlations are exponentially suppressed \footnote{The exponential decrease of the density correlations 
as a function of the distance $|i-j|$ during the quench dynamics is due to the initial many-body quantum state $\ket{\Psi_0}$. Indeed, the latter corresponds to the ground state of 
$\hat{H}(V_0)$ where $V_0$ has been carefully chosen such that the quantum model is initially well confined in the charge-density-wave insulating regime. The charge gap in
the CDW ground state $\ket{\Psi_0}$ implies an exponential decay of the density correlations at equilibrium and this property is conserved during the unitary time evolution.}.
While considering the same example given at Fig.~\ref{manmana_light}(a), let us verify that the correlations activated ballistically are characterized by twice the maximal
group velocity with respect to the quench Hamiltonian governing the time evolution, as predicted by the Calabrese-Cardy picture.
On Fig.~\ref{manmana_light}(a), we may remind the reader that a sudden global quench crossing the quantum critical point ($V_c=2$) is considered.
The latter is characterized by a pre-quench (initial) Hamiltonian $\hat{H}(V_0)$ well confined within the CDW regime and a post-quench Hamiltonian
$\hat{H}(V)$ confined in the non-interacting Luttinger-liquid regime implying $V=0$. Hence, the quench Hamiltonian $\hat{H}(V=0)$ can be easily diagonalized by performing a 
Fourier transform of the fermionic operators of creation and annihilation in real space. This yields to the gapless dispersion relation $E_k = -2 t_{\mathrm{h}}
\cos(k)$ ($a$ the lattice spacing is fixed to unity). As a consequence, one can deduce the theoretical group velocity $V_{\mathrm{g}}(k) = \partial_k E_k$ ($\hbar = 1$) 
associated to $E_k$ which may be written as $V_{\mathrm{g}}(k) = 2 t_{\mathrm{h}} \sin(k)$. Finally, the characteristic velocity, \textit{ie.} twice the maximal group 
velocity, is given by $2V_{\mathrm{g}}^* = 2V_{\mathrm{g}}(k^*) = 2 \mathrm{max}[V_{\mathrm{g}}(k)] = 4 t_{\mathrm{h}}$ with $k^* = \pi/2$ corresponding to the edge
of the first Brillouin zone. This analytical velocity is in very good agreement with the slope of the linear correlation edge separating the causal region of the density 
correlations from the non-causal one at Fig.~\ref{manmana_light}(a). This validates the quasiparticle picture of Calabrese and Cardy concerning the spreading 
of correlations for isolated short-range interacting lattice models driven far from equilibrium \textit{via} sudden global quenches, at least for a specific quantum lattice model and for a specific 
sudden global quench. \\

Although the semi-classical picture of Calabrese and Cardy seems to be correct to explain the generation of correlations for short-range interacting lattice models,
it turns out that the latter is not precise enough and incomplete. \\
Indeed, at Chap.~\ref{ch:3-universal_scaling_laws} we present a generic form for the equal-time connected correlation functions for isolated short-range
interacting quantum lattice models \footnote{Note that the generic form presented and analyzed at Chap.~\ref{ch:3-universal_scaling_laws} is also 
valid for isolated quantum lattice models with long-range interactions.} driven far from equilibrium \textit{via} sudden global quenches confined in a same quantum phase (or quantum regime), see Eq.~\eqref{generic_form} and Sec.~\ref{SRC}.
Then, by means of stationary phase arguments, a twofold linear structure in the vicinity of the correlation edge (CE) is unveiled. \\
Consistently with the Calabrese-Cardy picture, one recovers that the CE spreads ballistically with the velocity $2V_{\mathrm{g}}(k^*)$, \textit{ie.} twice the maximal group velocity. 
However, this implies that the quasiparticle picture of Calabrese and Cardy is not precise enough since specific sudden global quenches have been considered to find this important result, \textit{ie.} sudden global quenches such that
both the pre- and post-quench Hamiltonians are confined in a same quantum phase (or regime). If the previous condition is not fulfilled, the velocity of the CE is not necessarily given by $2V_{\mathrm{g}}(k^*)$. This statement has been verified at Ref.~\cite{manmana2009}
where the authors have investigated the correlation edge velocity of the equal-time density correlation function $C_{i,j}(t)$ defined at Eq.~\eqref{obs_man} for sudden global quenches 
characterized by a same pre-quench Hamiltonian confined in the CDW regime and different post-quench Hamiltonians confined in the Luttinger-liquid regime. The numerical spreading velocity associated to the CE has been found to display
significant, but relatively small, deviations from twice the theoretical maximal group velocity in the Luttinger-liquid regime. \\
Moreover, according to our generic form, we also show at Sec.~\ref{SRC} that in the vicinity of the correlation edge, a series of local extrema should also appear. The latter is found to
propagate ballistically with a different velocity given by $2V_{\varphi}(k^*) = 2 E_{k^*}/k^*$ ($\hbar = 1$) corresponding to twice the phase velocity at $k^*$ the quasimomentum for which the group velocity is maximal.
Such linear twofold structure for the spreading of correlations is clearly visible at Figs.~\ref{manmana_light}(c,d) where the density correlations are investigated for sudden global quenches
confined in a similar regime, here the charge-density-wave insulating regime. In both cases, the CE spreads linearly as well as the series of local minima and maxima 
present in its vicinity and characterized by the velocity $V_{\mathrm{m}} = 0$ (vertical extrema). This explains why the Calabrese-Cardy picture is incomplete and hence does not fully characterize the correlation spreading
in short-range lattice models. \\

An interesting extension to the previous investigation of the correlation spreading would be to investigate the case of isolated quantum lattice models with long-range 
interactions. This research topic has already been initiated where generalized Lieb-Robinson bounds have been derived for long-range systems where the interactions 
decay algebraically, $1/R^\alpha$, with the distance $R$~\cite{hastings2006,foss-feig2015}. However, the related experiments and numerical investigations have
lead to conflicting pictures~\cite{hauke2013,eisert2013,jurcevic2014,richerme2014,cevolani2015,cevolani2016,buyskikh2016}. For instance, experiments \cite{richerme2014}
and numerical simulations within truncated Wigner approximation~\cite{schachenmayer2015b} for the one-dimensional long-range XY
model point towards bounded, super-ballistic, propagation for all values of the power-law exponent $\alpha$. In contrast, experiments on the long-range transverse
Ising (LRTI) model have reported a ballistic propagation of the correlation maxima with, however, observable leaks that increase when $\alpha$ decreases~\cite{jurcevic2014}. \\

In the next chapters, we shed new light on these conflicting results. Indeed, a theoretical description of the scaling laws for the correlation spreading 
in long-range interacting quantum lattice models is provided at Chap.~\ref{ch:3-universal_scaling_laws}. The latter relies on stationary-phase arguments applied to our
generic form for the equal-time connected correlation functions. At Chap.~\ref{ch:5-long_range_ising_chain}, these analytical scaling laws are verified numerically using a
tensor network based technique for two different long-range interacting spin lattice chains which can be simulated experimentally using trapped ions, namely the
1D long-range Ising chain and the 1D long-range XY chain.

\section{Experimental realization of quantum lattice models and sudden global quenches}
\label{ER}

The investigation of the correlation spreading in isolated quantum lattice models with short-range or long-range interactions 
driven far from equilibrium is significantly stimulated by experiments. Indeed, the simultaneous progress realized in 
the manipulation of cold atoms or ions have permitted to simulate many different particle and spin Hamiltonians of condensed-matter physics with
an unprecedented control of the interaction parameters in time. Hence, this experimental breakthrough has offered the possibility to investigate the physical properties
of their out-of-equilibrium dynamics \textit{via} quantum quenches and especially to shed new light on the central question of this thesis which concerns the spreading of quantum correlations. \\
For instance, ultracold atoms loaded in artificial (optical) lattices provide a very interesting platform \cite{gross2017, bloch2012} to investigate the out-of-equilibrium (quench)
dynamics of isolated quantum models governed by short-range interacting bosonic and fermionic Hamiltonians such as the Bose- and Fermi-Hubbard Hamiltonian respectively
(see Refs.~\cite{bakr2010, esslinger2010, chen2011, trotzky2012, cheneau2012, schneider2012, pertot2014}) allowing also to emulate effective quantum spin models, such as the 
Ising one \cite{meinert2013}. Another promising platform to investigate the quench dynamics consists of trapped ions \cite{schneider2012b, NaturePhysicsInsight2012blatt} 
where the relevant parameters and interactions can be efficiently controlled. In particular, these experimental setups are well suitable to simulate 
long-range interacting quantum spin models due to the involved Coulomb interaction between the trapped ions \cite{richerme2014, jurcevic2014}. \\

In the following, to discuss how these experimental quantum simulations based on cold atoms and ions are particularly relevant to explore the quench dynamics 
and to probe the spreading of correlations in isolated quantum lattice models with short-range and long-range interactions respectively, two different 
experiments are discussed. Firstly, a one-dimensional short-range interacting bosonic model simulated \textit{via} cold atoms is considered, namely
the Bose-Hubbard chain \cite{cheneau2012}. The experimental results concerning the spreading of density (parity) correlations in its Mott-insulating phase are
analyzed. Finally, we turn to a presentation of the experimental realization using trapped ions of a 1D long-range interacting quantum spin model, the 1D long-range XY model
where the quench dynamics has been investigated by measuring the spin correlations \cite{richerme2014}.  

\subsection{Quantum simulation with ultracold atoms : the Bose-Hubbard chain}
\label{simu_cheneau}
In the following, we discuss the experimental realization of the one-dimensional short-range interacting Bose-Hubbard model using bosonic ultracold atoms loaded 
in an optical lattice, see Refs.~\cite{jaksch1998, greiner2002}. The main physical properties of this bosonic lattice model are discussed in a 
first time. Then, the experimental setup and the quench protocol considered at Ref.~\cite{cheneau2012} to simulate and drive such bosonic lattice model 
far from equilibrium are introduced. Finally, several experimental results concerning its quench dynamics confined in the Mott-insulating phase
are presented where the authors have investigated the parity correlation function to shed new light on the spreading of correlations for short-range interacting
closed quantum lattice models.  

\paragraph{Presentation of the model}
The one-dimensional short-range Bose-Hubbard (1D SRBH) model is described by a Hamiltonian $\hat{H}$ which may be written as follows

\begin{equation}
\hat{H} = -J \sum_{R} \left( \hat{a}^{\dagger}_{R} \hat{a}_{R+1}+\mathrm{h.c.}\right)+\frac{U}{2}\sum_R\hat{n}_{R}(\hat{n}_{R}-1).
\label{first_srbh}
\end{equation}

\noindent
The operators $\hat{a}_R$ and $\hat{a}_{R}^{\dagger}$ denote the annihilation and creation of a bosonic particle on the lattice site $R$ respectively and $\hat{n}_R
= \hat{a}^{\dagger}_R \hat{a}_R$ the local density operator on site $R$. This bosonic lattice model is characterized by three different parameters : the filling $\bar{n}
= N/N_s$ ($N$ refers to the number of bosonic particles on the lattice chain, $N_s$ to the number of lattice sites), the nearest-neighbor hopping amplitude $J>0$ and the repulsive on-site
interaction energy $U>0$. \\

At zero-temperature and at equilibrium, the phase diagram of the 1D SRBH model has been extensively studied and well characterized \cite{sachdev2001,cazalilla2011}.
The latter comprises a superfluid (SF) and a Mott-insulating (MI) phase, 
determined by the competition of the hopping $J$, the on-site interaction $U$, and the average filling $\bar{n}$ (or equivalently the chemical potential $\mu$ if the 
grand canonical statistical ensemble is considered). Furthermore, the 1D SRBH model hosts two different phase transitions : (i) For commensurate filling, $\bar{n} \in 
\mathbb{N}^*$ the SF-MI phase transition (also called Mott-$U$ phase transition) is of the Berezinskii-Kosterlitz-Thouless type, at the critical value $(U/J)_{\mathrm{c}} \simeq 3.3$
for unit filling ($\bar{n}=1$) in 1D~\cite{kuhner2000,kashurnikov1996exact,ejima2011,rombouts2006}. (ii) For incommensurate filling, \textit{ie.} $\bar{n}
\notin \mathbb{N}$, this short-range interacting bosonic lattice model is confined in the superfluid phase for any value of the interaction parameter $U/J$. For strong enough 
interaction ratios $U/J$, the commensurate-incommensurate SF-MI phase transition (also called Mott-$\delta$ phase transition), of the mean field type, 
is driven by doping when $\bar{n}$ approaches a positive integer value. Note that the physical properties of the 1D SRBH model will be discussed in details
at Chap.~\ref{ch:4-bose_hubbard_chain} where the correlation spreading has been investigated numerically in different regimes for each quantum phase and for 
several observables. In the following, we propose to recall the main properties of the Mott-insulating phase since the latter has been considered by the authors of
Ref.~\cite{cheneau2012} to investigate experimentally the quench dynamics of the bosonic lattice model. \\

The Mott-insulating phase is characterized by a gapped excitation spectrum and requires not only a positive integer filling $\bar{n} \in \mathbb{N}^*$ but also a sufficiently
strong interaction parameter $U/J$. The latter tends to pin the bosonic particles in the lattice sites by minimizing the fluctuations of the on-site occupation number.
Furthermore, the Mott-insulating phase displays two different regimes, the weakly interacting and strong-coupling regimes depending on the value of the interaction
parameter $U/J$ at fixed $\bar{n} \in \mathbb{N}^*$. For the strong-coupling regime requiring $U/J \gg 1$, the short-range Bose-Hubbard chain can be diagonalized using
a first-order perturbation theory. The gapped excitation spectrum is then characterized by, see Refs.~\cite{altman2002,huber2007,barmettler2012} for more details,

\begin{equation}
2E_k = U - 2J(2\bar{n}+1)\cos(k).
\label{strongMI}
\end{equation}

\noindent
where $k$ refers to the quasimomentum confined in the first Brillouin zone $\mathcal{B}$.
Deep enough in the MI phase, \textit{ie.} $U/J \gtrsim 2(2\bar{n}+1)$, corresponding to the weakly interacting regime of the Mott-insulating phase,
the gapped excitation spectrum presented at Eq.~\eqref{strongMI} does not hold anymore. In other words, the first-order perturbation theory fails to predict the correct
quasiparticle dispersion relation. To go beyond such theory, one can rely on a fermionization technique. The latter consists of treating the bosonic particles
as hardcore bosons leading finally, using a Jordan-Wigner transformation, to an effective quadratic Hamiltonian of interacting (spinful) fermions for the 1D SRBH model.
This quadratic fermionic Hamiltonian is then diagonalized using a (fermionic) Bogolyubov transformation (see Ref.~\cite{barmettler2012} for a 
complete discussion about the technique) and it yields the following gapped excitation spectrum~\cite{barmettler2012,Ejima2012}

\begin{equation}
2E_k \simeq \sqrt{\left[U-2J(2\bar{n}+1)\cos(k)\right]^2 + 16 J^2 \bar{n}(\bar{n} + 1)\sin^2(k)}.
\label{weakMI}
\end{equation}

\noindent
Note that this quasiparticle dispersion relation is also valid for the strongly interacting regime of the MI phase. Indeed, by developing the latter 
and keeping the first-order terms in $U/J$, one recovers the excitation spectrum at Eq.~\eqref{strongMI} found using a first-order perturbation theory and valid
in the strong-coupling limit $U/J \gg 1$. Most importantly, for both regimes within the MI phase, the elementary excitations correspond to doublon-holon excitation pairs
(which can also be seen as fermionic Bogolyubov quasiparticles when relying on the fermionization technique) where a doublon refers to the occupation of 
$\bar{n}+1$ bosonic particles on a lattice site $R$ ($n_R = \bar{n}+1$) whereas a holon corresponds to $\bar{n}-1$ bosons ($n_R = \bar{n}-1$). The previous discussion
about the Mott-insulating phase, its elementary excitations consisting of doublon-holon excitation pairs and the associated excitation spectrum presented at
Eq.~\eqref{weakMI}, was essential. Indeed, the latter will facilitate the analysis of the experimental results extracted from Ref.~\cite{cheneau2012} characterizing
the far-from-equilibrium dynamics of the 1D SRBH model in the MI phase. 

\paragraph{Experimental setup and quench protocol} In what follows, we present the experimental setup and the quench protocol 
performed by the authors of Ref.~\cite{cheneau2012} in order to simulate the short-range Bose-Hubbard chain and to investigate 
its quench dynamics confined in the Mott-insulating phase respectively. \\

\begin{figure}[!h]
\centering
\begin{tabular}{c}
\includegraphics[scale = 0.18]{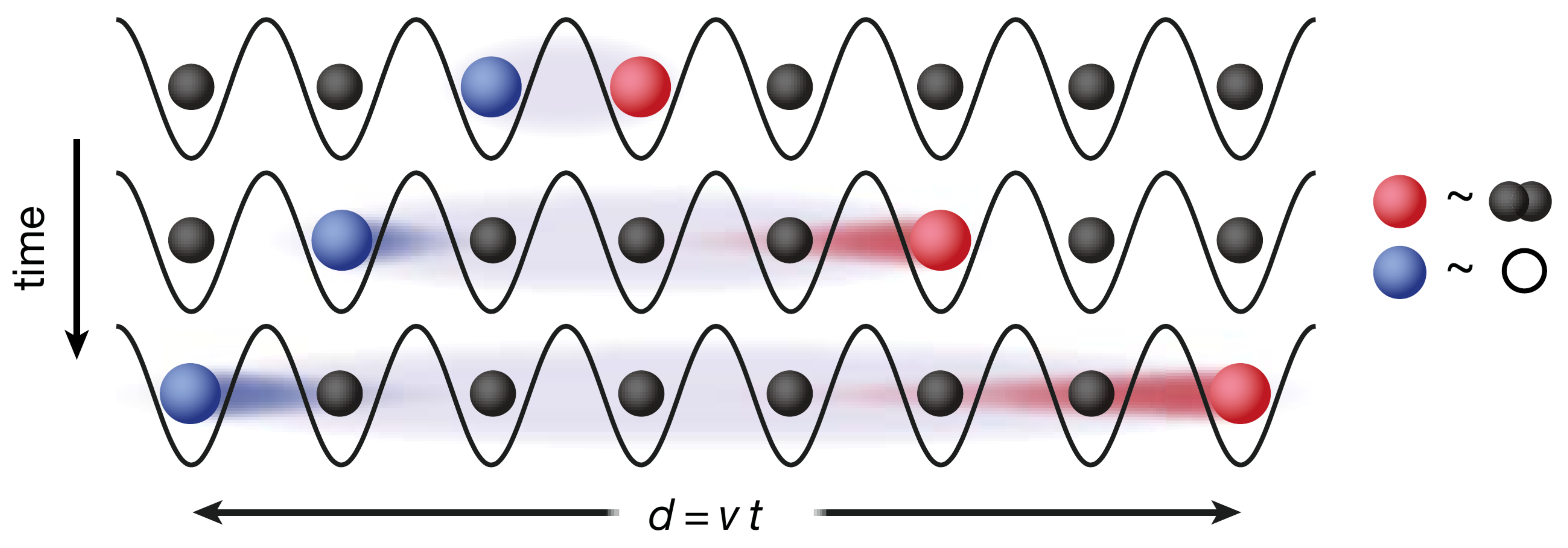}        
\end{tabular}
\caption{Quench dynamics of the 1D short-range Bose-Hubbard model in the Mott-insulating phase. (a)~ Initially, a one-dimensional gas of ultracold bosonic atoms 
(black balls) loaded in an optical lattice is prepared deep in the Mott-insulating phase at unit-filling, $\bar{n}=1$. Then, this bosonic lattice model is driven 
far from equilibrium \textit{via} a sudden global quantum quench performed by lowering the lattice depth. (b)~ Due to the quench, the elementary excitations of
the Mott-insulating phase are emitted at each lattice site. Each elementary excitation of quasimomentum $k$ consists of a quasiparticle pair formed of a
doublon ($n_R=2$ for $\bar{n}=1$, see red ball) and a holon ($n_R=0$ for $\bar{n}=1$, see blue ball) of quasimomentum $k$ and $-k$ respectively (or \textit{vice versa})
and propagating ballistically through the lattice with respect to their associated group velocity. Figure extracted from Ref.~\cite{cheneau2012}.}
\label{cheneau_protocol}
\end{figure}

The experimental sequence consists of loading ultracold bosonic particles in a one-dimensional optical lattice. To do so, the authors have first prepared a 
two-dimensional (2D) degenerate gas of $^{87}\mathrm{Rb}$ confined in an optical lattice along the $z$-axis with a lattice spacing $a_{\mathrm{lat}} = 532~ \mathrm{nm}$ 
\cite{sherson2010, endres2011}. Then, this 2D boson-lattice system is divided into $10$ decoupled chains \textit{via} a second optical lattice along the $y$-axis and by 
fixing both lattice depths to the value $20 E_{\mathrm{r}}$. $E_{\mathrm{r}} = (2\pi \hbar)^2/(8ma_{\mathrm{lat}}^2)$ refers to the recoil energy of the lattice \footnote{The recoil energy $E_{\mathrm{r}}$
represents here the kinetic energy of a $^{87}\mathrm{Rb}$ atom after emitting a lattice photon.} ($m$ to the atomic mass 
of $^{87}\mathrm{Rb}$). The interaction strength $U/J$ of the different chains is tuned \textit{via} a third optical lattice along the $x$-axis. The number of bosonic
atoms per chain has been controlled such that a lattice filling $\bar{n}=1$ is considered (a positive integer filling $\bar{n}$ is mandatory for the bosonic lattice model 
to be confined in the MI phase). \\

Concerning the quench protocol, an initial many-body quantum state $\ket{\Psi_0}$ is prepared deep in the Mott-insulating phase by increasing the $x$-lattice depth 
until an interaction parameter $(U/J)_0 = 40 \gg (U/J)_{\mathrm{c}}^{\bar{n}=1} \simeq 3.3$ is reached. Then, the bosonic lattice model is driven far from equilibrium by suddenly 
lowering the lattice depth. The final lattice depths are chosen such that the quantum system is still confined in the Mott-insulating phase and relatively close to the 
the quantum critical point, \textit{ie.} implying a post-quench interaction parameter $U/J \gtrsim (U/J)_{\mathrm{c}}^{\bar{n}=1}$, see Fig.~\ref{cheneau_protocol}. To sum up, 
the authors of Ref.~\cite{cheneau2012} have investigated the quench (far-from-equilibrium) dynamics of the Bose-Hubbard chain at unit-filling by suddenly and globally 
tuning the effective interaction parameter $U/J$. \\

To read out the properties of the time-evolved quantum state $\ket{\Psi(t)}$, the time evolution is frozen by rapidly increasing the different lattice depths to a value 
$\simeq 80 E_{\mathrm{r}}$. This operation permits to suppress the tunneling process and consequently to freeze the density distribution of the many-body quantum state $\ket{\Psi(t)}$.
Finally, the atoms were detected \textit{via} a fluorescence technique and the occupation number of the bosonic particles on each lattice site deduced using a reconstruction 
algorithm. Due to inelastic (light-assisted) collisions during the imaging process and leading to a loss of atoms pairs, the authors have measured the parity correlation 
function \footnote{Hence, no distinction can be made between a holon and a doublon implying an occupation number $n_R = 0$ and $n_R=2$ at unit-filling $\bar{n}=1$.}.

\paragraph{Quench dynamics}

\begin{figure}[!h]
\centering
\begin{tabular}{c}
\includegraphics[scale = 0.274]{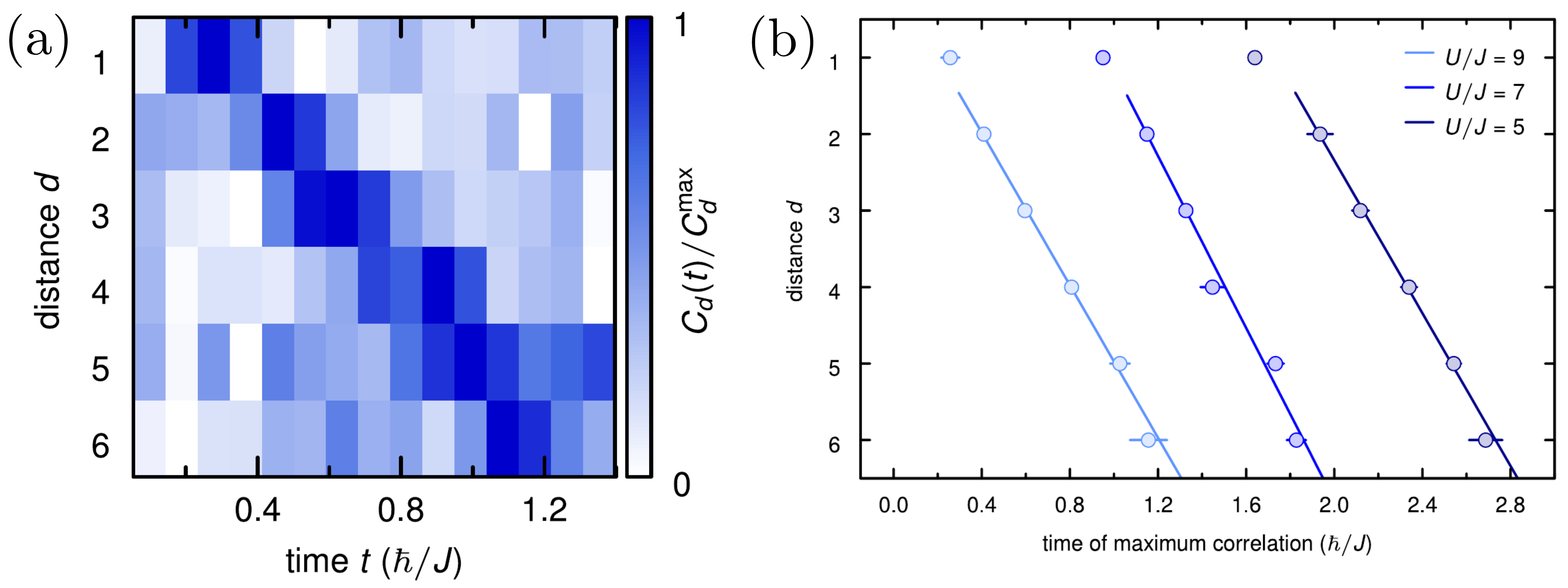}        
\end{tabular}
\caption{Spreading of the two-point parity correlations in the short-range Bose-Hubbard chain for sudden global quenches confined in the Mott-insulating phase
at unit-filling. (a)~ Experimental result for the time evolution of the parity correlations $C_d(t)$ for a quench starting from $(U/J)_0 = 40$ to $U/J = 9$. 
(b)~ Investigation of the propagation velocity for global quenches to $U/J=5$ (dark blue), $U/J=7$ (medium blue) and $U/J=9$ (light blue). For each quench and distance $d$,
the time of the maximum of the correlation signal is extracted from the experimental data of $C_d(t)$. The corresponding propagation velocity is deduced 
\textit{via} a linear fit restricted to $2 \leq d \leq 6$. Note that the data for $U/J=5$ and $U/J=7$ have been shifted horizontally for clarity. Figures
extracted from Ref.~\cite{cheneau2012}.}
\label{cheneau_quench}
\end{figure}

As previously discussed, the authors of Ref.~\cite{cheneau2012} have investigated the two-point parity correlations to characterize the quench dynamics of the 1D SRBH model
in the Mott-insulating phase. This observable is defined as follows

\begin{equation}
C_d(t) = \langle \hat{s}_R(t) \hat{s}_{R+d}(t) \rangle - \langle \hat{s}_R(t) \rangle \langle \hat{s}_{R+d}(t)\rangle, 
\label{parity}
\end{equation}

\noindent
where an average on the lattice sites $R$ has to be understood (permitted by the translational invariance of the considered bosonic model). 
The expectation value $\langle ... \rangle$ is taken with respect to the initial many-body quantum state $\ket{\Psi_0}$ corresponding here to a deep Mott state
at $\bar{n}=1$, \textit{ie.} $\ket{\Psi_0} \simeq \ket{1}^{\otimes N_s}$. The latter is due to the very large initial interaction parameter $(U/J)_0 = 40$ considered 
in the experimental quench protocol. 
The two-point parity correlation function presented at Eq.~\eqref{parity} depends only on the parity operator $\hat{s}_R(t) = e^{i \pi \left[\hat{n}_R(t) - \bar{n} \right]}$ 
characterizing the presence of particles on the lattice site $R$ at time $t$. Indeed, if a doublon ($\langle \hat{n}_R(t) \rangle = \bar{n}+1$) or holon
($\langle \hat{n}_R(t) \rangle = \bar{n}-1$) is present on the lattice site $R$ at time $t$, then $\langle \hat{s}_R(t) \rangle = -1$. Otherwise, $\langle \hat{s}_R(t) \rangle = 1$
corresponding to the case where $\langle \hat{n}_R(t)\rangle = \bar{n}$. Note that since the quench dynamics starts from an almost Mott state at unit-filling, the initial 
two-point parity correlations are equal to zero, \textit{ie.} $C_d(0) = 0, ~\forall d$. \\

On Fig.~\ref{cheneau_quench}, experimental results of the quench dynamics of the 1D SRBH model in the Mott-insulating phase are presented. On Fig.~\ref{cheneau_quench}(a),
a typical experimental result of the two-point parity correlation function $C_d(t)$ rescaled by its maximal value $C_d^{\mathrm{max}}$ is displayed and represented
as a function of the distance $d$ and the time $t$ (in units of $\hbar/J$). Although relatively small distance and time scales are considered, one can clearly distinguish
a signal propagating ballistically coherently with the existence of an effective light cone. \\

On Fig.~\ref{cheneau_quench}(b), the authors have tracked the time of maximum correlation for each distance $d$ for several interaction parameters $U/J$, see blue circles.
As expected, a ballistic motion of the signal has been unveiled and fitted using a linear ansatz. By computing the slope of the latter, the associated spreading velocity has been extracted, see solid blue lines.
Note that this signal is usually associated with the causality cone velocity. However, as discussed in the next chapters, the true light-cone velocity generally differs from the velocity of such maxima.
It turns out that for the specific quenches considered here, the two velocities are very similar because they are performed by considering a post-quench interaction parameter $U/J$ relatively close to the quantum
critical point. In any other case, the two velocities are significantly different, see discussion at Chapter.~\ref{ch:4-bose_hubbard_chain} and Fig.~\ref{fig:numerics_mott_u}. \\

After numerical analysis and treatment, the authors found the following spreading velocity $\hbar v /(Ja_{\mathrm{lat}}) = 5.0(2)$, $5.6(5)$, $5.0(2)$ for the sudden global quench defined by a post-quench
interaction parameter $U/J = 5.02(2)$, $7.0(3)$, $9.0(3)$ respectively. Note that these experimental velocities are in good agreement with $2V_{\mathrm{g}}^*$ corresponding to
the maximal group velocity of the doublon-holon excitations pairs, \textit{ie.} the elementary excitations in the MI phase. To deduce these analytical values, the excitation spectrum
$2E_k$ valid in the weakly interacting regime of the Mott-insulating phase [requiring an interaction parameter $U/J \gtrsim 2(2\bar{n}+1)$ and presented at Eq.~\eqref{weakMI}] is considered.
One finds the following theoretical velocities $2V_{\mathrm{g}}^* \simeq 4.7, 5.8, 5.9$ for the interaction parameters $U/J = 5, 7, 9$ respectively. \\

These results are consistent with the conclusions provided by our quasiparticle picture concerning the spreading of correlations in short-range interacting lattice models driven far from 
equilibrium \textit{via} sudden global quenches. Indeed, the latter predicts that the space-time correlations display a twofold causality cone characterized by a ballistic correlation edge
and a series of local maxima propagating at the velocity $2V_{\mathrm{g}}^*$ and $2V_{\varphi}^*$ respectively. For the previous interaction parameters $U/J$, $2V_{\mathrm{g}}^* \simeq 2V_{\varphi}^*$. 
Hence, one expects to find that the experimental spreading velocity associated to the first local maxima is very close to $2V_{\mathrm{g}}^*$. Note that the quench dynamics of the 1D SRBH model confined in 
the gapped Mott-insulating quantum phase will be discussed in details at Chap.~\ref{ch:3-universal_scaling_laws} and Chap.~\ref{ch:4-bose_hubbard_chain} using theoretical and numerical techniques respectively.

\subsection{Quantum simulation with trapped ions : the long-range XY chain}
In what follows, we turn to a discussion concerning the quantum simulation of isolated lattice models with long-range interactions. In particular, 
the experimental realization of the one-dimensional $s=1/2$ long-range interacting XY (1D LRXY) model using trapped atomic ions performed by Richerme
\textit{et al.} at Ref.~\cite{richerme2014} is presented. To do so, we introduce first the main physical properties of this $s=1/2$ spin lattice model
and the sudden global quench considered by the authors of Ref.~\cite{richerme2014}. Then, the experimental sequence and the quench protocol to simulate and to drive
the 1D LRXY model far from equilibrium are discussed. Finally, a typical experimental result of the space-time spin-spin correlations is presented 
and analyzed. The purpose of this discussion is to point out that, experimentally, new light can be shed on the central research topic of the correlation spreading
in isolated long-range interacting quantum lattice models. 

\paragraph{Presentation of the model and the quench}
In the following, the long-range interacting $s=1/2$ XY spin lattice chain is considered. The latter is defined by a Hamiltonian $\hat{H}$ which may be written as follows ($\hbar$
is fixed to unity)

\begin{equation}
\hat{H} = \frac{1}{2} \sum_{i<j} J_{i,j} (\hat{\sigma}^x_i \hat{\sigma}^x_j + \hat{\sigma}^z_i \hat{\sigma}^z_j) = \sum_{i<j} 2J_{i,j}
(\hat{S}^x_i \hat{S}^x_j + \hat{S}^z_i \hat{S}^z_j). 
\label{h_richerme}
\end{equation}

\noindent
This quantum lattice model is fully characterized by the isotropic long-range exchange coupling term $J_{i,j}$ between the spins on lattice sites $i$ and $j$.
The latter, tunable experimentally, is assumed to display a power-law decaying behavior, \textit{ie.} $J_{i,j} = J_0 /|i-j|^{\alpha}$. $J_0>0$ denoting a constant and 
$\alpha$ the power-law exponent characterize the strength and the decay of the long-range spin-spin interactions in both directions $x$ and $z$ of the Bloch sphere.  
The operator $\hat{\sigma}_i^{a}$ ($\hat{S}_i^{a} = (1/2) \hat{\sigma}_i^{a}$, $\hbar = 1$), with $a \in \{x,y,z\}$, refers to the $s=1/2$ Pauli matrix
($s=1/2$ spin operator) acting on the $i^{\mathrm{th}}$ spin in the $a$ direction. They obey the following commutation rule $\left[\hat{\sigma}^{a}_i, \hat{\sigma}^{b}_{j}
\right] = 2i \epsilon^{abc} \delta_{i,j} \hat{\sigma}^{c}_i$ with $(a,b,c) \in \{x,y,z\}^3$, $\epsilon$ the Levi-Civita symbol defined as follows

\[
  \epsilon^{abc}=\begin{cases}
               = 1 ~~~\mathrm{if}~~abc \in \{xyz, yzx, zxy\} \\
               = -1 ~\mathrm{if}~~ abc \in \{xzy, yxz, zyx\} \\
               = 0 ~~~\mathrm{if}~~ a=b ~~\mathrm{or}~~b = c~~\mathrm{or}~~a=c
            \end{cases}
\]

\noindent
and $\delta_{i,j}$ the Kronecker delta symbol ($\delta_{i,j}=1$ if $i=j$ and $0$ otherwise). The Pauli matrices for spins $s=1/2$ 
also obey the anticommutation rule $\{ \hat{\sigma}_i^a, \hat{\sigma}_i^b \} = 2 \delta_{a,b} I$ with $I$ the $2\times2$ identity matrix. \\

To investigate the far-from-equilibrium dynamics of this spin lattice model, a sudden global quench has been considered by the authors of Ref.~\cite{richerme2014}. 
The quantum system is initially prepared in a many-body quantum state $\ket{\Psi_0}$ where each spin is in its 'down' state along the $z$ direction,
\textit{ie.} $\ket{\Psi_0} = \ket{\downarrow \downarrow ... \downarrow \downarrow}_z$. This initial product state can be realized as the ground state of the Hamiltonian 
$\hat{H}$ of the LRXY chain [see Eq.~\eqref{h_richerme}] with an additional term $+ h \sum_i \hat{\sigma}_i^{z}$ ($h>0$ represents a homogeneous transverse
field in the $z$ direction) and by considering the limit regime where $h \gg J_0$. Then, suddenly at time $t=0$, the quantum model is quenched globally by switching on
the long-range interactions in order to generate the Hamiltonian of the LRXY chain. This quench Hamiltonian fully characterized by a fixed and specific value of $J_0$ 
and $\alpha$ governs the unitary time evolution of the quantum system. Hence, the time-evolved quantum state at time $t$ denoted by $\ket{\Psi(t)}$ may be written as
$\ket{\Psi(t)} = e^{-i \hat{H}t} \ket{\Psi_0}$ according to the time-dependent Schrödinger equation for isolated quantum models. \\

\begin{figure}[!h]
\centering
\begin{tabular}{c}
\includegraphics[scale = 0.11]{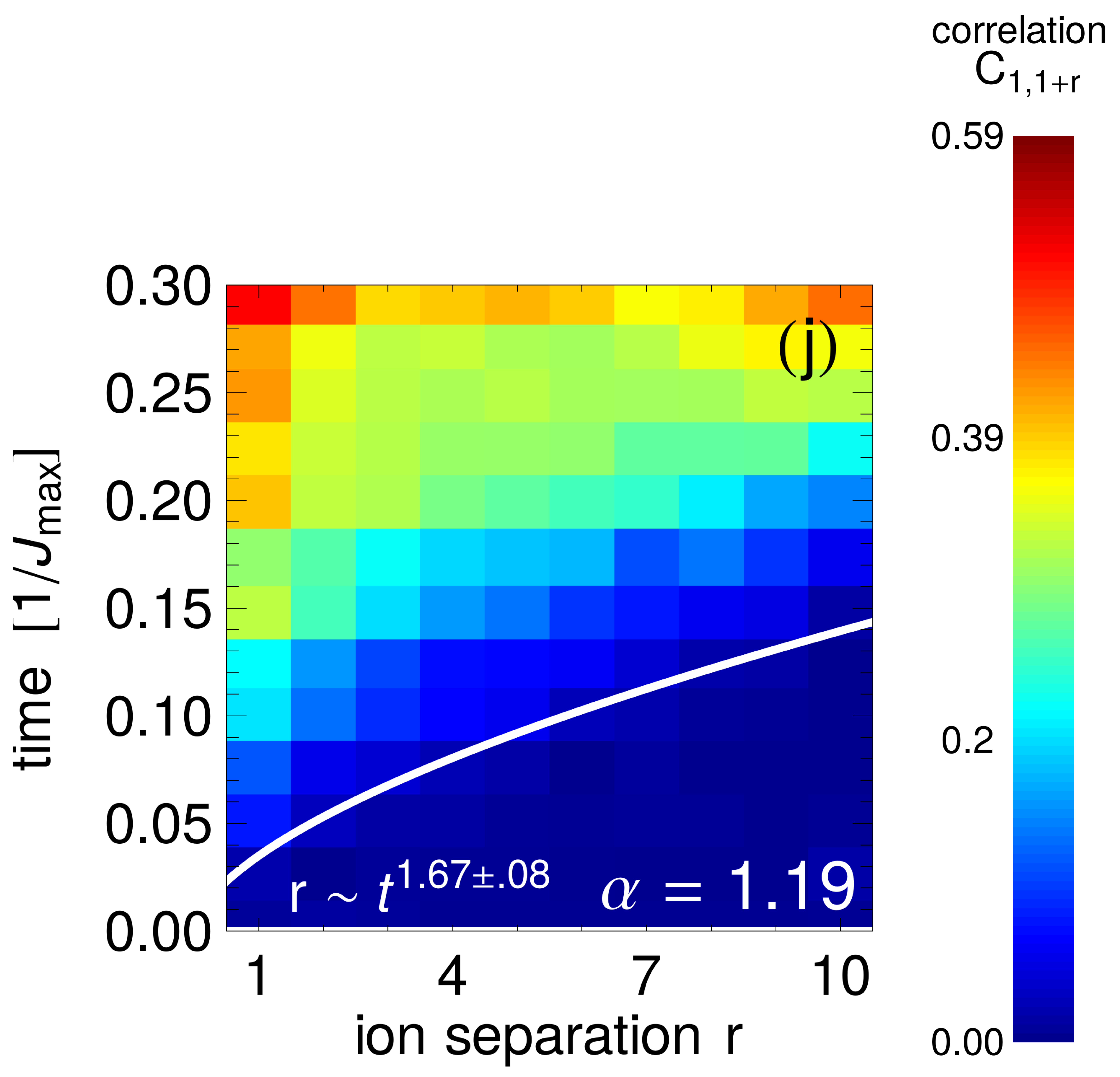}        
\end{tabular}
\caption{Spreading of spin correlations in the long-range XY $s=1/2$ spin chain. The experimental data of $C_{1,1+r}(t)$ the space-time spin-spin correlations
along the $z$ direction of the Bloch sphere are represented for a specific sudden global quench (see Eq.~\eqref{corr_richerme} for the general
expression of the space-time spin correlations along the $z$ axis denoted by $C_{i,j}(t)$). The connected correlation function $C_{1,1+r}(t)$ characterizes 
the spin correlations while considering the $1^{\mathrm{st}}$ spin (or equivalently the $1^{\mathrm{st}}$ lattice site) as a reference point, $r$ refers to the ion separation from the reference spin.
Here, the sudden global quench is defined as follows : The ion chain is initially prepared in the product state $\ket{\Psi_0} = \ket{\downarrow \downarrow ... 
\downarrow \downarrow}_z$. Then, at time $t=0$, the quantum system is suddenly and globally quenched by performing a time evolution of the initial product state
$\ket{\Psi_0}$ with respect to the Hamiltonian $\hat{H}$ of the LRXY chain at Eq.~\eqref{h_richerme} with a power-law exponent $\alpha = 1.19$. Figure extracted from Ref.~\cite{richerme2014}.}
\label{richerme_quench}
\end{figure}

In the aim of characterizing the properties of this time-dependent many-body quantum state $\ket{\Psi(t)}$, the authors of Ref.~\cite{richerme2014} have investigated the
equal-time connected spin-spin correlation function along the $z$ direction, denoted by $C_{i,j}(t)$, which reads as 

\begin{equation}
C_{i,j}(t) = \langle \hat{\sigma}^z_i(t) \hat{\sigma}^z_j(t) \rangle - \langle \hat{\sigma}^z_i(t) \rangle \langle \hat{\sigma}^z_j(t) \rangle,
\label{corr_richerme}
\end{equation}

\noindent
where the expectation value $\langle ... \rangle$ is taken with respect to the initial product state $\ket{\Psi_0}$.
Besides, according to the Heisenberg picture, $\hat{\sigma}^z_i(t) =  e^{i\hat{H}t} \hat{\sigma}^z_i e^{-i \hat{H}t}$ and consequently one can rewrite the previous 
correlation function as $C_{i,j}(t) = \langle \Psi(t) | \hat{\sigma}^z_i \hat{\sigma}^z_j | \Psi(t) \rangle - \langle \Psi(t) | \hat{\sigma}^z_i | \Psi(t) \rangle
\langle \Psi(t) | \hat{\sigma}^z_j | \Psi(t) \rangle$. Note also that due to the specific expression of the long-range spin exchange coupling $J_{i,j} = J_0/|i-j|^{\alpha}$,
the Hamiltonian $\hat{H}$ of the 1D LRXY model is translationally and parity invariant leading to $C_{i,j}(t) = C_{|i-j|}(t), \forall t$. Furthermore, one can also stress that it
is not necessary to subtract the equilibrium value $C_{i,j}(0)$ to the connected correlation function at Eq.~\eqref{corr_richerme} to obtain a clear space-time pattern.
Indeed, since $\ket{\Psi_0}$ corresponds to a product state ($\ket{\Psi_0} = \ket{\downarrow \downarrow ... \downarrow \downarrow}_z$), it implies 
$C_{i,j}(0) = 0,~\forall i,j$. A typical experimental result of the space-time spin-spin correlation function $C_{i,j}(t)$ is displayed at Fig.~\ref{richerme_quench}. The latter allows us to 
characterize the quench dynamics of the considered spin lattice model and to shed new light on the correlation spreading for isolated long-range interacting 
lattice models. 

\paragraph{Experimental setup}

\begin{figure}[!h]
\centering
\begin{tabular}{c}
\includegraphics[scale = 0.15]{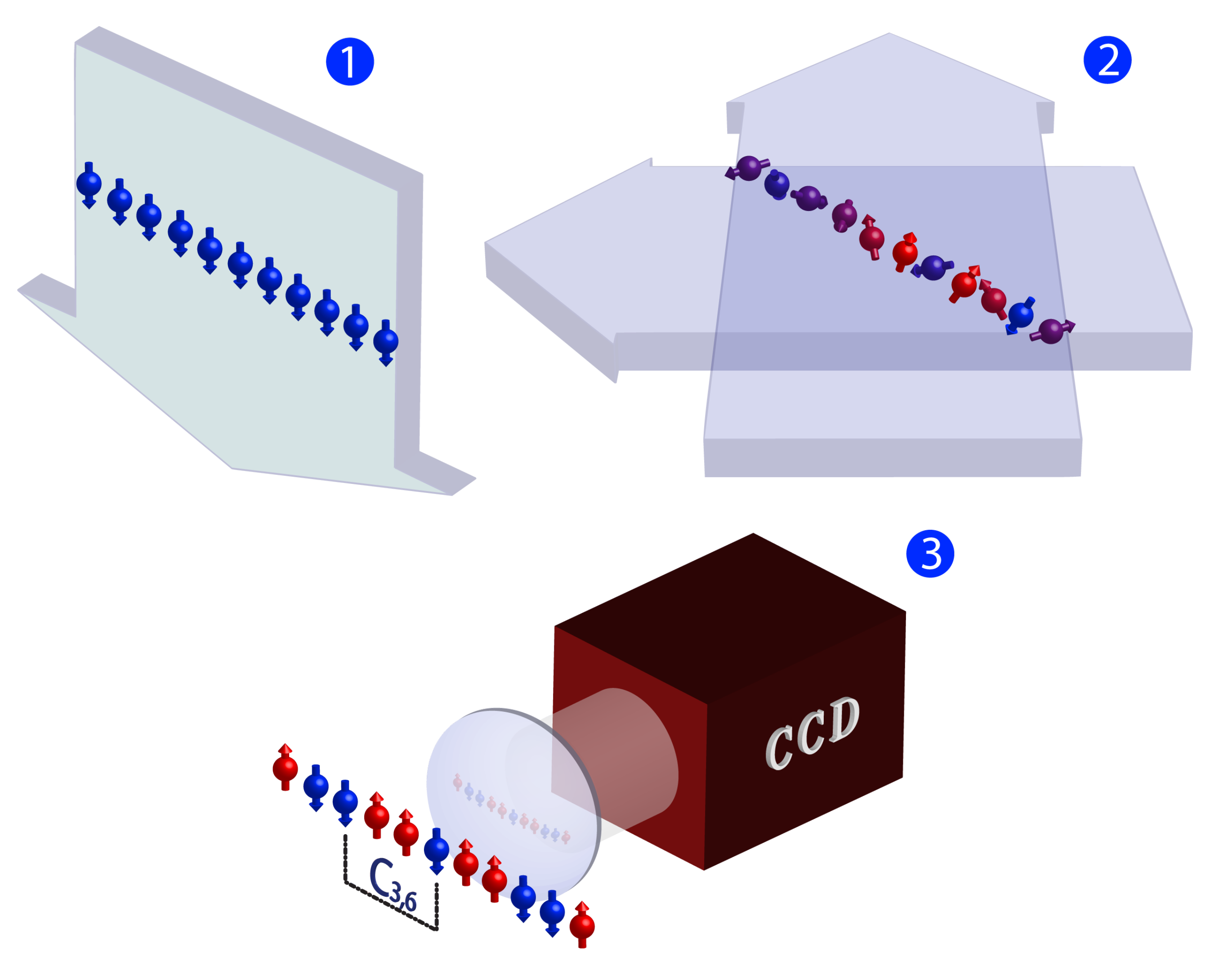}        
\end{tabular}
\caption{Experimental sequence for the investigation of the quench dynamics in the long-range XY $s=1/2$ spin chain. (1) Initially, the quantum spin model 
is prepared by optically pumping all the spins ($11$ spins, or equivalently $11$ lattice sites, have been considered in this experiment) in their 'down' state 
leading to the initial many-body quantum state $\ket{\Psi_0} = \ket{\downarrow}_z^{\otimes 11}$. (2) Once the spin lattice model has been initialized, the latter 
is quenched suddenly and globally by generating the long-range (power-law decaying) interactions \textit{via} laser-induced optical dipolar forces on the 
trapped ions to simulate the Hamiltonian of the LRXY spin chain $\hat{H}$ defined at Eq.~\eqref{h_richerme}. Finally, the initial many-body product state
$\ket{\Psi_0}$ evolves unitarily in time with the Hamiltonian $\hat{H}$ leading to the time-evolved quantum state $\ket{\Psi(t)} = e^{-i\hat{H}t}\ket{\Psi_0}$.
(3) During the unitary time evolution, the projection of each of the $11$ spins along the $z$ direction is imaged onto a CCD camera. These experimental data allows us to 
have access to the space-time spin-spin correlations along the $z$ axis, denoted by $C_{i,j}(t)$. Such correlation function permits to characterize the properties of
the time-evolved quantum state $\ket{\Psi(t)}$ and more precisely the correlation spreading for a long-range interacting spin lattice model.
Figure extracted from Ref.~\cite{richerme2014}.}
\label{richerme_protocol}
\end{figure}

Before presenting and analyzing the experimental data extracted from Ref.~\cite{richerme2014}, we briefly discuss the corresponding experimental setup to simulate
and to drive far from equilibrium the LRXY spin chain. \\

In the experiment of Richerme \textit{et al.} at Ref.~\cite{richerme2014}, trapped atomic $^{171}\mathrm{Yb}^{+}$ ions are manipulated in order to implement 
qubits, \textit{ie.} effective $s=1/2$ spins. More precisely, their hyperfine states $^{2}S_{1/2} \ket{F=0, m_F=0}$ and $\ket{F=1, m_F = 0}$ permit to encode
the local quantum state $\ket{\downarrow}_z$ and $\ket{\uparrow}_z$ respectively, see also Ref.~\cite{monroe2007}. In this experiment, the Ytterbium ions 
are cooled and arranged by using a Paul trap and the long-range interactions between them are mediated \textit{via} phonons coupling the internal hyperfine
states with the collective vibrational modes \cite{porras2004}. \\

Concerning the quench protocol, the ion chain is initially prepared into the product state $\ket{\Psi_0} = \ket{\downarrow ... \downarrow}_z$ \textit{via} an optical pumping, see 
Fig.~\ref{richerme_protocol}(1). Then, a sudden global quench is performed at time $t=0$ by applying laser-induced optical dipolar forces on the ions to yield the Hamiltonian
$\hat{H}$ of the LRXY chain at Eq.~\eqref{h_richerme}. The latter defines the quench Hamiltonian governing the time evolution of the quantum system, see
Fig.~\ref{richerme_protocol}(2).
Finally, to characterize the properties of the time-evolved quantum state $\ket{\Psi(t)} = e^{-i\hat{H}t} \ket{\Psi_0}$, the projection of each spin along the $z$ direction is measured.
To do so, a fluorescence technique is considered by applying a laser beam, addressing the transition $^{2}S_{1/2} \ket{F=1}$
to $^{2}P_{1/2} \ket{F=0}$, to the Ytterbium ions. The latter fluoresce only if they are in the effective local quantum state $\ket{\uparrow}_z$, 
\textit{ie.} the hyperfine state $^{2}S_{1/2} \ket{F=1}$. The (spontaneous) emission of light due to the fluorescence is collected through an objective and imaged onto a 
CDD camera having a single-site resolution, see Fig.~\ref{richerme_protocol}(3). Using this experimental scheme, the space-time spin-spin correlations along the $z$ axis, denoted
by $C_{i,j}(t)$ and presented at Eq.~\eqref{corr_richerme}, can be reconstructed.
 
\paragraph{Quench dynamics}

On Fig.~\ref{richerme_quench}, we present an experimental result extracted from Ref.~\cite{richerme2014} of the space-time spin correlations along the $z$ direction, 
$C_{i,j}(t)$. More precisely, the connected correlation function $C_{1,1+r}(t)$ is considered and characterizes the spin correlations with respect to the first spin.
These space-time spin correlations are represented as a function of $r$, denoting the ion-ion separation
distance according to the previous reference spin, and the time $t$ in units of \footnote{See Ref.~\cite{richerme2014} for the value of $J_{\mathrm{max}}$.}
$1/J_{\mathrm{max}}$ ($\hbar=1$). Besides, the long-range decaying interactions of the LRXY Hamiltonian $\hat{H}$ are defined by a power-law exponent $\alpha=1.19$, see  
caption of Fig.~\ref{richerme_quench} for a detailed description of the sudden global quench considered here. \\

An important observation about the spreading of the spin correlations at Fig.~\ref{richerme_quench} is the presence of a correlation edge (CE). Indeed, the space-time pattern of 
$C_{1,1+r}(t)$ displays both a causal and a non-causal region corresponding to non-zero and zero values for the quantum correlations respectively. The CE associated to these space-time
spin correlations corresponds to the separation between the two regions. Contrary to the quench dynamics of isolated short-range interacting lattice models where a linear 
causality cone for the space-time correlations is expected, the latter is characterized by a non-ballistic motion. The latter has been extracted from the experimental data by tracking the activation time $t^*$, \textit{ie.} 
the first time for which the spin correlation is different from zero, for each ion separation distance $r$ (see Ref.~\cite{richerme2014} for more details about the tracking technique).
Then, these activation times as a function of the ion-ion separation distance $r$ have been fitted by the authors \textit{via} an algebraic ansatz of the form $r \sim t^{1/\beta}$,
where the value $1/\beta = 1.67 \pm 0.08$ is found. Hence, their experimental result suggests the existence of a super-ballistic (faster-than ballistic)
growth of the light-cone boundary. In other words, the measurements seem to point in the direction of a CE propagating algebraically with a faster-than-ballistic motion, \textit{ie.} $1/\beta > 1$ (or $\beta < 1$). \\

However, one needs to take a step back from these experimental results since very small distance and time scales are considered, \textit{ie.} $\mathrm{max}(r) = 10$
and $\mathrm{max}(t) = 0.3/J_{\mathrm{max}}$. The adverse consequence is that the quench dynamics of the LRXY chain, and more precisely the motion of the CE, are biased by strong 
finite-size effects. Furthermore, the considered connected correlation function $C_{1,1+r}(t)$ is responsible for additional errors. Indeed, the
latter considers the first spin, located on one of two edges of the lattice chain, as the reference spin which is the most affected by the finite-size effects. \\

This specific research topic concerning the correlation spreading in long-range interacting lattice models will be extensively studied in this thesis. At Chap.~\ref{ch:3-universal_scaling_laws},
by relying on our generic form for the equal-time connected correlation functions, a sub-ballistic motion of the CE ($r \sim t^{1/\beta}$, with $\beta > 1$) is unveiled for the correlation spreading
in isolated lattice models with strong power-law decaying interactions. Besides, this scaling law for the spreading of the CE is found to not depend on the quantum phase of the quench Hamiltonian, 
\textit{ie.} if the latter is gapped or gapless. Note that for a gapless quench Hamiltonian, a super-ballistic scaling law has to be associated to the spreading of the local maxima and not to the motion of the CE.
The previous statements, deduced from our quasiparticle theory presented at Chap.~\ref{ch:3-universal_scaling_laws}, are verified numerically at Chap.~\ref{ch:5-long_range_ising_chain}. 

\section{Numerical and theoretical approaches}
\label{NTA}

In this section, we present several numerical techniques and theoretical approaches allowing us to investigate the far-from-equilibrium dynamics
of closed quantum lattice models. More precisely, these techniques permit to shed new light on the remaining open questions presented above at Sec.~\ref{open_q}.
In what follows, we first briefly discuss the time-dependent matrix product state (t-MPS) algorithm. More details about this numerical approach are provided at
Chap.~\ref{ch:4-bose_hubbard_chain} and \ref{ch:5-long_range_ising_chain} for one-dimensional quantum systems with short- and long-range interactions respectively. 
The latter will be used to tackle the question of the transport of correlations and entanglement in isolated one-dimensional quantum lattice models numerically. Then,
we present other numerical techniques such as the time-dependent variational Monte-Carlo (t-VMC) and the exact diagonalization (ED) 
whose advantages and disadvantages are discussed. Finally, we turn to a brief overview of the analytic quasiparticle approach, which relies on a mean field approximation.
A second theoretical technique, namely the quench action (QA), is also briefly presented.

\subsection{Time-dependent matrix product state}

In the following, we discuss the time-dependent matrix product state (t-MPS) technique. The latter consists of a powerful numerical approach
to compute the static and dynamical physical properties of closed 1D quantum lattice models based on an analysis of the entanglement entropy (see
Appendix.~\ref{appendix_entropies}, Refs.~\cite{schollwock2005, schollwock2011, paeckel2019} and references therein). More precisely, the t-MPS 
approach gives access not only to the ground state but also to the (real-) time-evolved quantum state. Both algorithms, allowing us to deduce the
static and dynamical properties, rely on the optimization of a many-body ansatz of the form (see Fig.~\ref{MPS})

\begin{equation}
\ket{\Psi} = \sum_{\boldsymbol{\sigma}} \sum_{\boldsymbol{a}} A^{\sigma_1}_{1,a_1} A^{\sigma_2}_{a_1,a_2} ...~ A^{\sigma_{L-1}}_{a_{L-2},a_{L-1}} 
A^{\sigma_L}_{a_{L-1},1} \ket{\boldsymbol{\sigma}}.
\label{ansatz_mps}
\end{equation}

\noindent
with $\boldsymbol{\sigma} = \sigma_1, \sigma_2, ..., \sigma_{L-1}, \sigma_L$ and $\boldsymbol{a} = a_1,a_2, ..., a_{L-2}, a_{L-1}$.
This ansatz \footnote{Here, we have considered the so-called left-canonical representation of a matrix product state where each tensor is left-normalized 
\textit{ie.} each tensor fulfills the condition $\sum_{\sigma_R} (A^{\sigma_R})^{\dag} A^{\sigma_R} = I$ (see Chapter.~\ref{ch:4-bose_hubbard_chain}
and Ref.~\cite{schollwock2011} for more details).} $\ket{\Psi}$ is valid for the description of a general many-body quantum state for 1D quantum models containing $L$ lattice sites. 
$\ket{\Psi}$ lives in the full Hilbert space $\mathbb{H} = \mathbb{H}_R^{\otimes L}$ of dimension $\mathrm{dim}(\mathbb{H}_R)^L = d^L$ where $\mathbb{H}_R$ denotes
the local Hilbert space. This $d$-dimensional local Hilbert space is described by the local basis $\{\ket{\sigma_R}, \sigma_R = 1,...,d \}$. Hence, the set of parameters
$\{\sigma_R, R = 1,...,L, \sigma_R = 1,...,d \}$ (or equivalently $\boldsymbol{\sigma}$) corresponds to the different physical indices of the ansatz. Note that the ansatz
$\ket{\Psi}$ at Eq.~\eqref{ansatz_mps} is also described by a second set of indices. This second set, corresponding to 
$\{ a_{\tilde{R}}, \tilde{R} = 1,...,L-1, a_{\tilde{R}} = 1,..., \bar{a}_{\tilde{R}} \}$ (or equivalently $\boldsymbol{a}$), denotes the so-called virtual indices
and $\chi = \mathrm{max}_{\tilde{R}}(\bar{a}_{\tilde{R}})$ corresponds to the MPS bond dimension. \\

The set of maximal values $\{ \bar{a}_{\tilde{R}}, \tilde{R} = 1,...,L-1 \}$, characterized by the condition $\bar{a}_{\tilde{R}} 
\leq \mathrm{min}(d^{\tilde{R}},d^{L-\tilde{R}})$ (see Appendix.~\ref{appendix3_mps_app}) gives us some information about the amount
of entanglement in the lattice model. Indeed, the previous set contains $L-1$ values where each of them is related to one of the $L-1$ different bipartitions
of the 1D quantum model containing $L$ lattice sites \footnote{For a 1D quantum model containing $L$ lattice sites, there are $L-1$ different possibilities to cut 
the lattice chain.}. More precisely, 
the value $\bar{a}_{\tilde{R}}$ ($\tilde{R} \in [|1,L-1|]$) corresponds to the number of singular values kept in the $\tilde{R}$-th Schmidt matrix \textit{ie.} 
to the dimension of the $\tilde{R}$-th Schmidt matrix.  This $\tilde{R}$-th Schmidt matrix is found when performing the $\tilde{R}$-th singular value decomposition (SVD)
to the many-body quantum state $\ket{\Psi}$ written under a general form to get its corresponding MPS form, see Appendix.~\ref{appendix3_mps_app} and Ref.~\cite{schollwock2011}. 
Most importantly, the latter fully determines the amount of entanglement between the
two subsystems $A$ and $B$ of the finite chain, where $A$ is defined as the first $\tilde{R}$ lattice sites and $B$ as the $L-\tilde{R}$ last lattice sites. Considering
that the $\tilde{R}$-th Schmidt matrix $S(\tilde{R})$ has a dimension $\bar{a}_{\tilde{R}} \times \bar{a}_{\tilde{R}}$ with entries 
$S_{a_{\tilde{R}},a_{\tilde{R}}} \geq 0$, the corresponding entanglement entropy (also called von Neumann entropy) is defined as $\mathcal{S}(\tilde{R}) = -
\sum_{a_{\tilde{R}}=1}^{\bar{a}_{\tilde{R}}} S^2_{a_{\tilde{R}},a_{\tilde{R}}} \mathrm{log}(S^2_{a_{\tilde{R}},a_{\tilde{R}}})$, see Ref.~\cite{schollwock2011} and 
Appendices.~\ref{appendix3_mps_app}-\ref{appendix_entropies} for more details. \\

For low-entangled quantum states \textit{ie.} for $\mathcal{S}(\tilde{R})$ relatively small ($\forall \tilde{R} \in [|1,L-1|]$), only few singular values
with a significant weight in the $\tilde{R}$-th Schmidt matrix will contribute to the entanglement \footnote{The $\tilde{R}$ Schmidt matrices are defined as diagonal
square matrices with positive entries also called singular values. Here, we assume a descending order for the singular values.}.
Hence, the latter can be truncated to reproduce the same amount of entanglement 
implying $\bar{a}_{\tilde{R}} \ll \mathrm{min}(d^{\tilde{R}}, d^{L-\tilde{R}})$, $\forall \tilde{R} \in [|1,L-1|]$. Consequently, the MPS form presented at
Eq.~\eqref{ansatz_mps} consists of a local \footnote{The local representation of any many-body quantum state, without breaking its entanglement, 
is particularly interesting from a numerical point of view. Indeed, one can associate a graphical representation in terms of tensor
networks so that the implementation is drastically simplified.} and very compact \footnote{Indeed, the MPS form at Eq.~\eqref{ansatz_mps} involves $L$ $3$-rd order 
tensors of dimension $\bar{a}_{\tilde{R}} \times \bar{a}_{\tilde{R}+1} \times d$.} representation without breaking the non-locality property (in other words the
entanglement) of the state $\ket{\Psi}$. \\

\begin{figure}[!h]
\centering
\begin{tabular}{c}
\includegraphics[scale = 0.77]{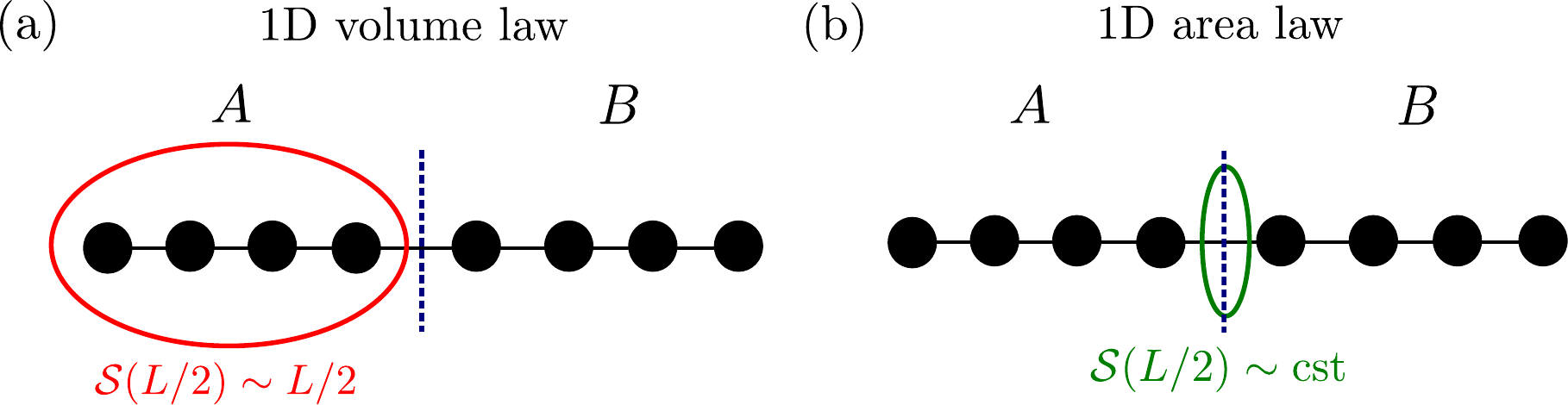}  
\end{tabular}
\caption{Boundary law for the entanglement (von Neumann) entropy in the case of a 1D quantum model containing $L$ lattice sites.
(a)~Volume law for a typical many-body quantum state $\ket{\Psi}$, the entanglement entropy scales with the length of the subsystem 
($\mathcal{S}(\tilde{R}) \sim \tilde{R}$, with $\tilde{R} \in [|1,L-1 |]$). (b)~ Area law for the low-lying excited states of gapped and short-range interacting Hamiltonians, the entanglement
entropy scales with the area of the cuts ($\mathcal{S}(\tilde{R}) \sim \mathrm{cst}$). The dashed blue line represents a cut of the 1D quantum lattice model leading to 
a bipartition where each subsystem $A$ and $B$ contains half of the total number of lattice sites $L$.}
\label{area_volume}
\end{figure}

Note that for a generic state $\ket{\Psi}$ living in the full Hilbert space $\mathbb{H}$, its entanglement entropy follows
a volume law \textit{ie.} the latter scales with the length of the subsystem [$\mathcal{S}(\tilde{R}) \sim \tilde{R}$] until reaching a maximal value 
[$\mathcal{S}(L/2) \sim L/2$] corresponding to the specific bipartition where both subsystems $A$ and $B$ contains a same number of lattice sites (have the 
same length), see Ref.~\cite{page1993} and Fig.~\ref{area_volume}(a). \\
However, for the ground state and the low-lying excited states of gapped Hamiltonians containing relatively short-range interactions 
\footnote{Interactions which couple a finite and small number of lattice sites.}, the entanglement entropy does not follow anymore a volume law but an area law 
\textit{ie.} scales with the area of the cuts \cite{eisert2010}. Therefore, for 1D quantum lattice models, the area law can be reformulated as 
$\mathcal{S}(\tilde{R}) \sim \mathrm{cst}$, $\forall \tilde{R} \in [|1,L-1|]$, see Fig.~\ref{area_volume}(b). \\

To sum up, the MPS ansatz presented at Eq.~\eqref{ansatz_mps} is characterized by two main parameters : $d$ the dimension of the local Hilbert space 
(corresponding to the maximal value of each physical index ($\{\ket{\sigma_R}, \sigma_R = 1,...,d \}$) and $\chi$ the MPS bond dimension related to the amount of entanglement
in the lattice model. As discussed previously, the MPS ansatz is optimal for 1D gapped and short-range interacting Hamiltonians which satisfy the area law ($\chi$ small) 
and more precisely for fermionic or spin lattice models ($d$ small \footnote{Indeed, for spinless (spinful) fermions $d = \mathrm{dim}(\mathbb{H_R}) = 2$ ($=4$). For 
$s=1/2$ ($s=1$) spin models $d=2$ ($d=3$).}). Note that it is still possible to rely on the t-MPS approach to describe accurately the static and dynamical properties of 1D bosonic
lattice models in gapless phases, see Refs.~\cite{despres2019, barmettler2012} for instance. However, the latter, corresponding to the worst case (large $d$ due to the bosonic nature of the particles and large $\chi$ due to the volume law), 
require much more efforts from a numerical point of view to be simulated correctly. \\

This numerical approach has been significantly used to investigate the information spreading in 1D quantum lattice models, see Refs.~\cite{manmana2009, despres2019,
barmettler2012, cheneau2012, buyskikh2016, frerot2018}. In this manuscript, all the numerical results concerning the correlation and entanglement spreading 
in different 1D quantum lattice models are found using this method. In Chap.~\ref{ch:4-bose_hubbard_chain}, we investigate the correlation spreading in the 
1D short-range Bose-Hubbard model. In Chap.~\ref{ch:5-long_range_ising_chain}, we turn to 1D long-range interacting lattice models (namely the 1D long-range XY model
and the 1D long-range Ising model) where the correlation and entanglement spreadings are studied. 

\subsection{Time-dependent variational Monte Carlo}
\label{t-vmc}
The time-dependent Variational Monte Carlo (t-VMC) corresponds to another powerful numerical technique allowing us to deduce the static and dynamical properties
of (closed) quantum lattice models. Note that contrary to the t-MPS algorithm, this approach is not limited to one-dimensional lattice models and can be used
in higher dimensions. The technique gives the optimal real time evolution of a variational quantum state which is generally parametrized as \cite{carleo2014,carleo2012}

\begin{equation}
 \ket{\Psi(t)} = e^{\sum_{R} \lambda_R(t) \hat{O}_R} \ket{\Psi(0)}.
 \label{tvmc_ansatz}
\end{equation}

\noindent
$\ket{\Psi(0)}$ refers to an initial many-body quantum state, the set $\{ \lambda_R(t) \}$ denotes the complex and time-dependent variational parameters. 
The set $\{\hat{O}_R \}$ also called set of excitation operators corresponds to time-independent local operators. The latter is expected to describe 
accurately the relevant excitations involved in the unitary dynamics of the closed quantum lattice model. Therefore, these operators defining the variational
ansatz need to be chosen carefully and accordingly to the Hamiltonian $\hat{H}$ and the quantum phase considered. \\

The variational parameters $\lambda_R(t)$ are deduced \textit{via} the Dirac-Frenkel time-dependent variational principle
\footnote{Note that this principle is also used at Chap.~\ref{ch:5-long_range_ising_chain} to perform real time evolution of
long-range interacting lattice models using a tensor-network language.}. This principle consists of minimizing the distance in the Hilbert 
space between the exact and the variational time dynamics. In other words, the goal is to minimize the distance between the exact generator
\footnote{Since a closed quantum lattice model is considered, its exact dynamics is governed by an unitary evolution given by $\ket{\Psi(t)} = e^{-i\hat{H}t}\ket{\Psi(0)}$.}, $\partial_t
\ket{\Psi_{\mathrm{exact}}(t) } = -i \hat{H} \ket{\Psi(t)}$, and the variational generator, $\partial_t \ket{\Psi_{\mathrm{variational}}(t)} = \left(\sum_R 
\dot{\lambda}_{R}(t) \hat{O}_R \right) \ket{\Psi(t)}$, of the dynamics where $\ket{\Psi(t)}$ is the variational
wavefunction at time $t$ defined at Eq.~\eqref{tvmc_ansatz}. To fulfill the previous minimization problem, it can be shown that the variational parameters
$\lambda_R(t)$ have to satisfy the following motion equation at each iteration in time,

\begin{equation}
 i \sum_{R'} \langle \hat{O}^*_{R} \hat{O}_{R'} \rangle_t^c \dot{\lambda}_{R'}(t) = \langle \hat{O}^*_{R} \hat{H} \rangle_t^c,~~\forall R.  
\end{equation}

\noindent
$\langle ...\rangle_t^c$ denotes the connected average defined as, $\langle \hat{A} \hat{B} \rangle_t^c = \langle (\hat{A} - \langle \hat{A} \rangle_t )(\hat{B} - \langle 
\hat{B} \rangle_t ) \rangle_t = \langle \hat{A} \hat{B} \rangle_t - \langle \hat{A} \rangle_t \langle \hat{B} \rangle_t$, where $\hat{A}$ and $\hat{B}$ are two local 
operators. $\langle ... \rangle_t = \langle \Psi(t) | ... | \Psi(t) \rangle$ corresponds to the expectation value with respect to the variational ansatz at time $t$ denoted by $\ket{\Psi(t)}$
while assuming that the ansatz is well normalized, \textit{ie.} $\langle \Psi(t) | \Psi(t) \rangle = 1$. \\

As discussed previously, t-VMC permits to deduce the static and dynamical properties of closed quantum lattice models. It can simulate both the dynamics of short- and
long-range interacting lattice models at short and long times with an accuracy similar to the one obtained using tensor network based techniques. Consequently, one can 
rely on this numerical approach to investigate for instance the correlation spreading in quantum lattice models generated by sudden global quenches. One reminds the reader that a sudden 
global quench corresponds to a sudden modification of at least one parameter in the Hamiltonian $\hat{H}$ at a specific time. More precisely, at $t=0$, the initial many-body quantum state
$\ket{\Psi(0)}$ corresponds to the ground state of the pre-quench (initial) Hamiltonian $\hat{H}_{\mathrm{i}}$. Then, the initial Hamiltonian $\hat{H}_{\mathrm{i}}$ 
is suddenly quenched at $t=0^+$ to a post-quench (final) Hamiltonian $\hat{H}_{\mathrm{f}}$ governing the real time evolution of the quantum model. This process to drive quantum lattice models far from 
equilibrium is described at Chap.~\ref{ch:3-universal_scaling_laws} in more details. \\

\begin{figure}[h!]
\centering
\includegraphics[scale = 0.235]{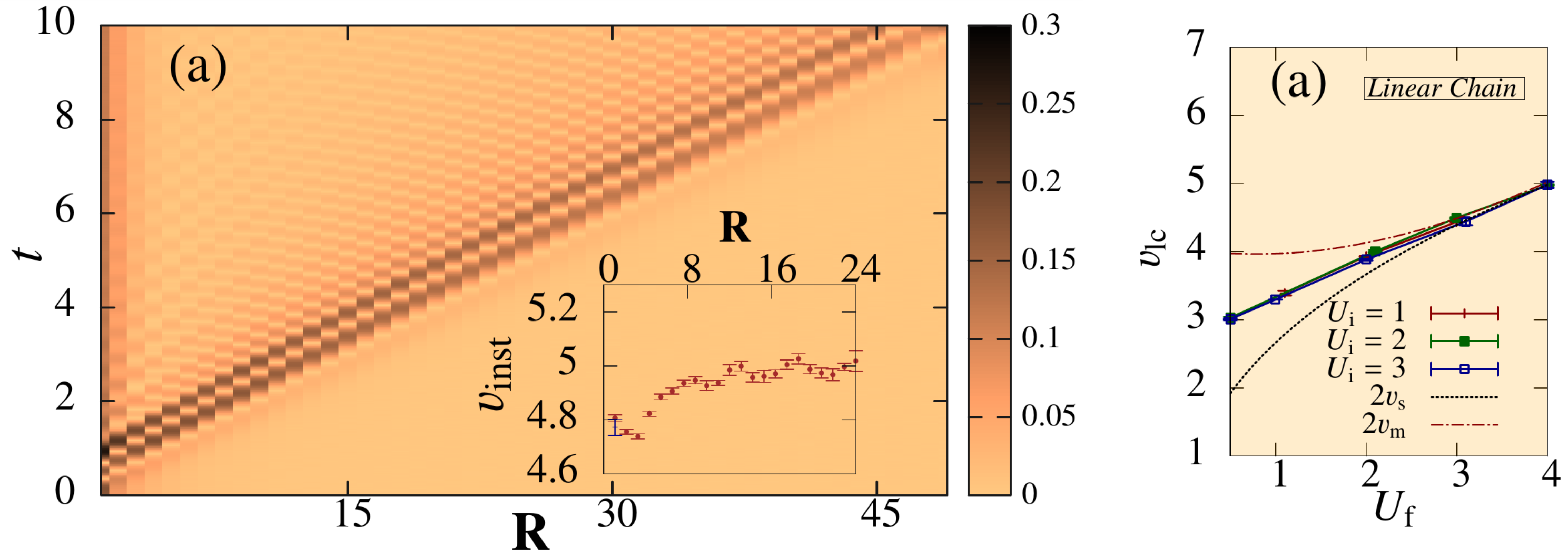}
\caption{\label{carleo_cone} Spreading of correlations in the 1D short-range Bose-Hubbard model. (Left panel) Density-density correlations $N(R,t)$ 
versus dimensionless distance $R$ ($a$ the lattice spacing is fixed to unity) and dimensionless time ($J=\hbar=1$) for a sudden global quench on the interaction strength from $U_{\mathrm{i}}=2$ 
[defining the pre-quench Hamiltonian $\hat{H}_{\mathrm{i}} = \hat{H}(U_{\mathrm{i}})$] to $U_{\mathrm{f}} = 4$ [characterizing the post-quench Hamiltonian
$\hat{H}_{\mathrm{f}} = \hat{H}(U_{\mathrm{f}})$]. (Right panel) Evolution of the dimensionless light-cone velocity $v_{\mathrm{lc}}$ ($\hbar = J = a = 1$)
as a function of the final interaction strength $U_{\mathrm{f}}$ (and of several initial interaction strengths $U_{\mathrm{i}} \in [|1,3|]$).
These numerical results are also compared to twice the sound velocity, $2v_{\mathrm{s}}$ (see dashed black line), and twice the maximum excitation (group) velocity, $2v_{\mathrm{m}}$
(see dashed red line) computed with respect to the post-quench Hamiltonian $\hat{H}_{\mathrm{f}}$. Figures extracted from Ref.~\cite{carleo2014}.}
\end{figure}

On Fig.~\ref{carleo_cone}, we display t-VMC numerical results extracted from Ref.~\cite{carleo2014} and concerning the spreading of correlations in the 1D short-range Bose-Hubbard model (see Eq.~\eqref{H_bhm}
and Subsec.~\ref{sf_chap3} for more details about this bosonic lattice model). The authors have investigated the quench dynamics 
of this 1D generic bosonic lattice model in the gapless superfluid (SF) phase at average density $\langle \hat{n}_R \rangle = 1$ (unit-filling of the lattice chain).
The latter implies a relatively small pre- ($U_{\mathrm{i}}$) and post-quench ($U_{\mathrm{f}}$) two-body interaction strength compared to the hopping amplitude $J$.
In order to investigate the space-time response of the quantum model to a sudden global quench confined in the SF phase, the density-density correlation 
function $N(R,t)$ defined as 

\begin{equation}
 N(R,t) = \langle \hat{n}_R(t) \hat{n}_{0}(t) \rangle - \langle \hat{n}_R(0) \hat{n}_{0}(0) \rangle 
 \label{density_carleo}
\end{equation}

\noindent
has been considered. At Eq.~\eqref{density_carleo}, $\langle ...\rangle$ corresponds to the expectation value with respect to the initial state $\ket{\Psi(0)}$, the ground state of $\hat{H}_{\mathrm{i}}$. \\

On Fig.~\ref{carleo_cone}(left panel), a typical t-VMC result for $N(R,t)$ the density-density correlations is shown on relevant spatial and temporal scales. The latter displays
a clear linear causality structure : a non-causal region of correlations, characterized by $N(R,t) \simeq 0$, is separated from the causal region, \textit{ie.} $N(R,t) >0$, by a
correlation edge (CE) propagating ballistically. The propagation velocity associated to the CE, called $v_{\mathrm{lc}}$ in Ref.~\cite{carleo2014}, is extracted from the slope. \\

For short-range interacting lattice systems, the existence of a correlation edge velocity $V_{\mathrm{CE}}$ implies the emergence of a linear causality cone beyond which the quantum correlations 
are exponentially suppressed, see Refs.~\cite{cevolani2018,lieb1972,bravyi2006,hastings2006}. For equal-time correlation functions and sudden global quenches confined in a same quantum phase, our quasiparticle
picture predicts that $V_{\mathrm{CE}}$ is characterized by twice the maximal group velocity associated to the post-quench Hamiltonian $\hat{H}_{\mathrm{f}}$ (the Hamiltonian driving the real time evolution 
of the quantum lattice model), \textit{ie.} $V_{\mathrm{CE}} = 2V_{\mathrm{g}}^* = 2 \mathrm{max}(\partial_k E_k^{\mathrm{f}})$ with $\hbar=1$, see Chap.~\ref{ch:3-universal_scaling_laws} for more details. \\

On Fig.~\ref{carleo_cone}(right panel), the velocity $v_{\mathrm{lc}}$ extracted from t-VMC numerical calculations is plotted with respect to the post-quench
interaction strength $U_{\mathrm{f}}$ (and for three different pre-quench interaction strengths $U_{\mathrm{i}}$). This so-called light-cone velocity is also
compared to twice the sound velocity $2v_{\mathrm{s}}$ and twice the maximal group velocity $2v_{\mathrm{m}}$ (see Ref.~\cite{carleo2014} and references therein). As expected, 
they found that $v_{\mathrm{lc}}$ is independent of the initial Hamiltonian $\hat{H}_{\mathrm{i}}$. However, the fitted velocities $v_{\mathrm{lc}}$ show significant deviations
from $2v_{\mathrm{m}}$. The results can be explained by the fact that the authors have tracked the motion of the local maxima instead of the one related to the CE. Indeed, we will 
show in the next chapter that the causality cone of equal-time correlation functions displays a twofold linear structure for short-range interacting lattice models driven far from 
equilibrium \textit{via} sudden global quenches. This twofold structure is defined not only by a CE propagating with the velocity $V_{\mathrm{CE}} = 2V_{\mathrm{g}}(k^*) = 2\mathrm{max}(\partial_k E_k^{\mathrm{f}})$ 
but also by a series of local extrema, located in the vicinity of the CE, spreading with a different velocity in general (see Figs.~\ref{fig:BHm} and \ref{fig:BHm2} as examples of such twofold linear
causality structure for the 1D short-range Bose-Hubbard model). This second velocity, defining the motion of the series of local extrema and denoted by $V_{\mathrm{m}}$, is characterized by twice the
phase velocity at $k^*$, the quasimomentum for which the group velocity is maximal \textit{ie.} $V_{\mathrm{m}} = 2V_{\varphi}(k^*) = 2 E_{k^*}^{\mathrm{f}}/k^{*}$ ($\hbar = 1$). Consequently, one can not track the
motion of the local extrema to deduce the one of the CE. Otherwise, one is characterizing the inner structure (the local extrema) of the causality cone and not the propagation front (correlation edge). 
By comparing the behavior of the fitted velocity $v_{\mathrm{lc}}$ [see Fig.~\ref{carleo_cone}(right panel)] with twice the theoretical phase velocity at $k^*$, \textit{ie.} $2V_{\varphi}^*$, [see Fig.~\ref{sf_exc_vel}(b)] 
as a function of $U_{\mathrm{f}}$, one finds a very good agreement between them. This suggests that the authors have extracted from the t-VMC calculations the velocity $V_{\mathrm{m}}$ instead of $V_{\mathrm{CE}}$.

\subsection{Exact diagonalization}
\label{exact_diago}
The quench dynamics of a Hamiltonian $\hat{H}$ can also be investigated numerically using the exact diagonalization. To get information about the static properties (low-lying 
eigenstates and eigenvectors of the Hamiltonian $\hat{H}$) of the quantum model, the latter consists of solving directly the Schrödinger equation given by $\hat{H} \ket{\Psi} = E \ket{\Psi}$. \\

To do so, it first requires to define a basis for the many-body Hilbert space. Then, the corresponding 
Hamiltonian matrix is built and has to be diagonalized. Finally, both the spectrum and the eigenstates of the Hamiltonian $\hat{H}$ can be analyzed. However, a strong limitation
of this method lies in the size of the full Hilbert space $\mathbb{H}$ increasing exponentially with the system size $L$ ($\mathrm{dim}(\mathbb{H}) = 
\mathrm{dim}(\mathbb{H}_R)^{L}$ with $\mathbb{H}_R$ denoting the local Hilbert space). To reduce the complexity (computational cost) of such problem, the
symmetries of the Hamiltonian $\hat{H}$ can be used in order to restrict ourselves to a relevant subset of the many-body Hilbert space. Note that one can also rely
on a Lanczos algorithm (see Ref.~\cite{dagotto1994} for more details) to decrease the complexity of the method. The main idea behind the latter is to construct
the Hamiltonian in a special basis such that it displays a tridiagonal form (to obtain a sparse Hamiltonian matrix) which can be then easily diagonalized using the standard
subroutines. \\

To deduce the dynamical properties, one needs to compute the time-evolved quantum state $\ket{\Psi(t)}$ evolving unitarily \textit{via} the time evolution 
operator $e^{-i \hat{H}t}$, \textit{ie.} $\ket{\Psi(t)} = e^{-i \hat{H} t} \ket{\Psi(0)}$, since closed quantum lattice models are considered here. Let us consider in 
what follows an infinitesimal iteration in time $\mathrm{d}t$. The time-evolved quantum state $\ket{\Psi(t+\mathrm{d}t)}$ can be deduced from $\ket{\Psi(t)}$ using   
for instance a truncated Taylor expansion of the infinitesimal time evolution operator $e^{-i \hat{H} \mathrm{d}t}$ ~\cite{carleo2012}. As a consequence, the 
many-body quantum state $\ket{\Psi(t+\mathrm{d}t)}$ may be written as 

\begin{equation}
 \ket{\Psi(t+\mathrm{d}t)} = e^{-i \hat{H} \mathrm{d}t} \ket{\Psi(t)} \simeq \sum_{n=0}^{n_{\mathrm{max}}} \frac{(-i\mathrm{d}t)^n}{n!} \hat{H}^n \ket{\Psi(t)}.
\end{equation}

\noindent
This technique is particularly simple to implement since each term of the truncated Taylor expansion is constructed by applying the Hamiltonian $\hat{H}$ 
several times to the many-body quantum state at time $t$, $\ket{\Psi(t)}$. In general, the series is expected to converge rapidly with $n$. Furthermore, at each
infinitesimal iteration in time $\mathrm{d}t$, the corresponding cutoff $n_{\mathrm{max}}$ is fixed such that the desirable truncation error is reached. This truncation 
error, determining the numerical accuracy of the unitary real time evolution, can be characterized by implementing a condition on the different conserved quantities during the 
dynamics. For instance, since the real time evolution is unitary here, the total energy is a conserved quantity ($E(t) = \langle \Psi(t) | \hat{H} | \Psi(t) \rangle = 
\langle \Psi(0) | \hat{H} | \Psi(0) \rangle = E_0$, with $E_0$ the energy associated to the initial state $\ket{\Psi(0)}$) and, hence, corresponds to a possible choice to
formulate the condition of convergence in order to fix the cutoff $n_{\mathrm{max}}$. One can state for example that $\ket{\Psi_{\mathrm{truncated}}(t)}$ has converged if it 
fulfills the following condition: $|E(\ket{\Psi_{\mathrm{truncated}}(t)}) - E_0 | \leq \epsilon$, with $\epsilon \ll E_0$.  \\

As discussed previously, the exact diagonalization consists of a numerical method permitting to deduce both the static and dynamical properties of closed quantum lattice models.
The latter has the advantage to give very accurate results but is restricted to relatively small system sizes. Nevertheless, the typical length scale (to give an order 
of magnitude $L_{\mathrm{max}} \lesssim 50$ for $s=1/2$ spin lattice chains \footnote{Note however that this typical length scale can drastically decrease if one
considers lattice models with a dimension of the local Hilbert space higher than two. For instance, one can think to spinful fermions ($\mathrm{dim}(\mathbb{H}_R) = 4$)
or typical bosonic lattice models ($\mathrm{dim}(\mathbb{H}_R) \gg 1$).}) is still relevant to investigate the quench dynamics of a large panel of one- and two-dimensional 
quantum lattice models.  \\

\begin{figure}[h!]
\centering
\includegraphics[scale = 0.22]{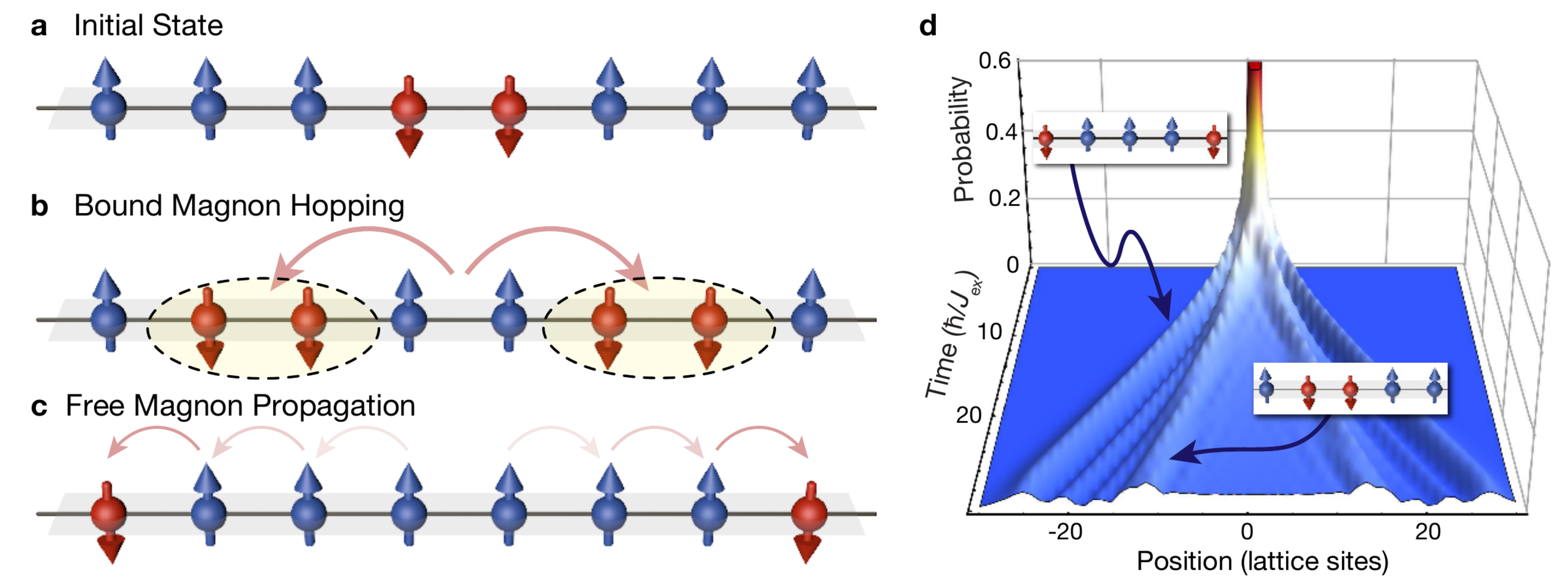}
\caption{\label{fuku_cone} Spreading of magnonic excitations in the 1D short-range Heisenberg model. (a)~Initial many-body quantum state 
where both central 'up' spins are flipped, \textit{ie.} $\ket{\Psi(0)} = \ket{\uparrow ... \uparrow \downarrow \downarrow \uparrow ... \uparrow}$, and its
decomposition into bound (b) and free (c) magnons propagating through the lattice chain. (d)~ Numerical 
result obtained from exact diagonalization showing the probability to find a flipped ('down') spin at a given lattice site following the initial state preparation. 
Note that on panel (d), two different wavefronts corresponding to bound and free magnons can be identified (see insets). The maximum probability ($P_{\mathrm{max}} = 1$ at 
$t=0 \times (\hbar/J_{\mathrm{ex}})$ for both central lattice sites) was clipped in the graph for clarity. Figure extracted from Ref.~\cite{fukuhara2013}.}
\end{figure}

On Fig.~\ref{fuku_cone}, we display a numerical result found using exact diagonalization (extracted from Ref.~\cite{fukuhara2013}) and concerning the spreading of magnonic 
excitations in the $s=1/2$ short-range Heisenberg chain (1D SRH) whose Hamiltonian $\hat{H}$ is defined as 

\begin{align}
& \hat{H} = -J_{\mathrm{ex}} \sum_R \left[ \frac{1}{2} \left( \hat{S}^{+}_R \hat{S}^{-}_{R+1} + \hat{S}^{-}_R \hat{S}^{+}_{R+1} \right) + \hat{S}^{z}_R \hat{S}^z_{R+1} \right], \nonumber \\
& \hat{H} = -J_{\mathrm{ex}} \sum_R \left( \hat{S}^{x}_R \hat{S}^{x}_{R+1} + \hat{S}^{y}_R \hat{S}^{y}_{R+1} + \hat{S}^{z}_R \hat{S}^z_{R+1} \right). \label{H_heis_into}
\end{align}

\noindent
$\hat{S}^{\alpha}_R$ denotes the spin operator acting on the lattice site $R \in \mathbb{Z}$ along the $\alpha \in \{x,y,z\}$ direction,
$J_{\mathrm{ex}}>0$ corresponds to the isotropic spin-exchange coupling (see Subsec.~\ref{1d_heis_local_quench} for more details). 
This model comprises an unique quantum phase
\textit{ie.} the quasi-long range order ferromagnetic phase along the $z$ direction \footnote{Here, the direction has been fixed arbitrarily.}. This gapless 
quantum phase is characterized by the following excitation spectrum $E_k^{\mathrm{free}} = J_{\mathrm{ex}}[1-\cos(k)]$ 
which can be deduced from a Holstein-Primakoff transformation [see Subsec.~\ref{1d_heis_local_quench} for the derivation]. 
The previous excitation spectrum $E_k^{\mathrm{free}}$ characterizes the low-energy excitations of this 1D short-range $s=1/2$ spin lattice model consisting
of free spin-wave excitations, \textit{ie.} free magnonic quasiparticles. \\

On Fig.~\ref{fuku_cone}, the authors have investigated the quench dynamics of the 1D SRH model by studying its response to a 
so-called local quench that we discuss now (see Chap.~\ref{ch:3-universal_scaling_laws} for a more detailed discussion about the protocol associated to sudden 
local quenches in order to drive quantum lattice models far from equilibrium). They first consider an initial quantum state where the two central spins have been flipped \footnote{According to Fig.~\ref{fuku_cone}(d), it seems that the two central lattice sites are $R=0$ and
$R=1$.} leading to $\ket{\Psi(0)} = \ket{\uparrow ... \uparrow \downarrow \downarrow \uparrow ... \uparrow}$. Note that the previous state can be built from 
the ground state of the 1D SRH model, $\ket{\Psi_{\mathrm{gs}}} = \ket{\uparrow ... \uparrow}$, and where the two central spins
located on the lattice site $R=0$ and $R=1$ are flipped \textit{via} $\hat{S}^{-}_0$ and $\hat{S}^{-}_1$ respectively. Hence, it yields the initial state $\ket{\Psi(0)}=
\hat{S}^{-}_0 \hat{S}^{-}_1 \ket{\Psi_{\mathrm{gs}}} = \ket{\uparrow ... \uparrow ~~\downarrow_{R=0}~~ \downarrow_{R=1} ~~ \uparrow ... \uparrow}$ 
[see Fig.~\ref{fuku_cone}(a)]. Then, this initial state evolves unitarily in time with the Hamiltonian $\hat{H}$ of the 1D SRH model presented at Eq.~\eqref{H_heis_into},
$\ket{\Psi(t)} = e^{-i \hat{H}t} \ket{\Psi(0)}$. \\

It is essential to stress that due to the specific initial state $\ket{\Psi(0)}$ considered here, where two neighbouring spins have been flipped, two species of magnonic excitations will be 
generated after the sudden local quench : (i)~free magnons whose associated excitation spectrum $E_k^{\mathrm{free}}$
has already been discussed previously [see also Fig.~\ref{fuku_cone}(c)] (ii)~ bound magnons whose excitation spectrum \footnote{Note that for a same 
quasimomentum $k$, the bound magnonic excitation has a lower energy than the free magnonic excitation. This is due to the term $-J_{\mathrm{ex}} 
\hat{S}^{z}_R \hat{S}^{z}_{R+1}$ in the Hamiltonian of the 1D SRH model, see Eq.~\eqref{H_heis_into}. The latter tells us that two flipped spins can lower their 
energy if and only if they are located on adjacent lattice sites (in a nearest-neighbor configuration).} is characterized by $E_k^{\mathrm{bound}}
= (J_{\mathrm{ex}}/2)[1-\cos(k)]$ (see Ref.~\cite{wortis1963}). \\

To characterize the spreading of both magnonic excitations, the probability $P(R,t)$
to find a flipped ('down') spin at a given lattice site $R$ and for a time $t$ (in units of $\hbar/J_{\mathrm{ex}}$) following the
initial state preparation $\ket{\Psi(0)}$ has been studied numerically using exact diagonalization [see Fig.~\ref{fuku_cone}(d)]. Indeed, investigating the space-time behavior
of local (on-site) observables, for a lattice model in a specific quantum phase and submitted to a local quench, allows us to characterize the spreading of
the corresponding species of quasiparticles. As expected, the authors found two wavefronts related to the two different species of quasiparticle presented previously.
More precisely, at time $t=0^{+}$, all the different free and bound magnons (for each quasimomentum $k$ in the first Brillouin zone $[-\pi,\pi]$) are generated due 
to the highly excited initial state $\ket{\Psi(0)}$ (with respect to the Hamiltonian $\hat{H}$ used to perform the unitary time evolution). Their spreading is governed by 
the shape of their corresponding excitation spectrum [$E_k^{(\mathrm{free},\mathrm{bound})}$] and more precisely by their corresponding group velocity
[$V_{\mathrm{g}}^{(\mathrm{free},\mathrm{bound})}(k) = \hbar^{-1} \partial_k E_k^{(\mathrm{free},\mathrm{bound})}$]. 
The fastest wavefront is associated to the spreading of the free magnonic excitation with the highest group velocity given by $V_{\mathrm{g}}^{\mathrm{free}}(k^*) = 
\mathrm{max}_k[V_{\mathrm{g}}^{\mathrm{free}}(k)] = J_{\mathrm{ex}}/\hbar$ ($k^* = \pi/2$). Straightforwardly, the slowest wavefront is determined by the spreading of the bound magnonic
excitation with the highest group velocity $V_{\mathrm{g}}^{\mathrm{bound}}(k^*) = \mathrm{max}_k[V_{\mathrm{g}}^{\mathrm{bound}}(k)] = J_{\mathrm{ex}}/2\hbar$ ($k^* = 
\pi/2$). \\

Note that for the same lattice model, we found similar results concerning its local quench dynamics both numerically [using a tensor-network technique, see 
Fig.~\ref{fig:heis_local_quench}] and analytically (\textit{via} the linear spin-wave theory, see Subsec.~\ref{1d_heis_local_quench} and 
Appendix.~\ref{appendix_local_quench_spin} for more details). However, in our study, we have investigated the local magnetization along the 
$z$ axis, \textit{ie.} $\langle \Psi(0) | \hat{S}^z_R(t) | \Psi(0) \rangle$ for an initial state $\ket{\Psi(0)}$ defined as
$\ket{\Psi(0)} = \ket{\uparrow ... \uparrow \downarrow \uparrow ... \uparrow}$. Since only one spin has been flipped for the locally perturbed initial many-body 
quantum state $\ket{\Psi(0)}$, only the free magnonic excitations are generated \footnote{Indeed, to generate the bound magnonic excitations, one would have to flip 
two adjacent spins as done previously.}. One recovers that the spreading of the free magnonic excitations in the $R-t$ plane lead to a symmetric \footnote{Indeed, the
wavefront is symmetric since the velocity associated to each of the two edges is equal in absolute value. The latter is due to the symmetry $k \rightarrow -k$ of the excitation
spectrum $E_k^{\mathrm{free}}$ associated to the free magnonic quasiparticles. As a consequence, the slowest quasiparticle will propagate at the same velocity than the fastest
one with a minus sign, $\min_k[V_{\mathrm{g}}^{\mathrm{free}}(k)] = -\max_k[V_{\mathrm{g}}^{\mathrm{free}}(k)] = -V_{\mathrm{g}}^{\mathrm{free}}(k^*) = 
V_{\mathrm{g}}^{\mathrm{free}}(-k^*)$.} wavefront characterized by the velocity $\pm V_{\mathrm{g}}^{\mathrm{free}}(k^*) = \pm J_{\mathrm{ex}}/\hbar$. 

\subsection{Mean field approximation and quasiparticle description}

To investigate theoretically the far-from-equilibrium dynamics of quantum lattice models, one can rely on a quasiparticle picture within a mean field approximation 
~\cite{cevolani2018,cevolani2016}. This theoretical method permits to shed new light on the different open questions presented at Sec.~\ref{open_q} and in 
particular on the correlation spreading in bosonic and spin lattice models submitted to sudden global or local quantum quenches (see for instance Subsec.~\ref{t-vmc} 
and \ref{exact_diago} for an example of a sudden global and local quench respectively). This picture is based on the assumption that immediately after the 
local (global) quench, free counter-propagating quasiparticles (quasiparticles pairs) are emitted (see Fig.~\ref{corr}) and spread into the lattice. The generation of these quasiparticles 
is a consequence of the initial state preparation where a strong local or global perturbation has been applied to the quantum system. Hence, the initial state can be seen as a highly excited state 
for the Hamiltonian considered to perform the real time evolution, and acts a source of quasiparticles (for sudden local quenches) or quasiparticle pairs (for sudden global quenches).
However, this theoretical picture contains several assumptions on the quasiparticles \textit{e.g.} an infinite lifetime, and neglected quasiparticle-quasiparticle interactions. Consequently,
all the analytical predictions provided by this approach and presented in this manuscript are compared to numerical results obtained \textit{via} tensor network based techniques \footnote{Note that other
numerical methods can be considered as discussed previously.} to verify the validity of the different assumptions. \\

This quasiparticle theory, briefly outlined here but discussed in details at Chap.~\ref{ch:3-universal_scaling_laws}, has been found to describe accurately the space-time dynamics of a wide class 
of relevant observables for a large panel of quantum lattice models in different phases. The latter comprises one-dimensional \footnote{We stress that our quasiparticle theory to explain the correlation spreading in 
isolated quantum lattice models is expected to work in higher dimensions and more precisely for a $D$-dimensional hypercubic lattice, see Chap.~\ref{ch:3-universal_scaling_laws}.} bosonic and spin lattice models \footnote{We 
also expect that our quasiparticle theory can be extended straightforwardly to fermionic lattice models.} with short-range [see Chaps.~\ref{ch:3-universal_scaling_laws} and
\ref{ch:4-bose_hubbard_chain}] or long-range interactions [see Chaps.~\ref{ch:3-universal_scaling_laws} and \ref{ch:5-long_range_ising_chain}] in gapped or gapless 
quantum phases. \\

This analytical method is essentially based on a mean field approximation together with the bosonic Bogolyubov theory. Firstly, the mean field 
approximation allows us to write the considered Hamiltonian in a generic quadratic Bose form in the momentum space (up to a possible shift in energy),

\begin{equation}
\hat{H} = \frac{1}{2} \sum_{\mathbf{k}} \mathcal{A}_{\mathbf{k}} \left( \hat{a}^{\dag}_{\mathbf{k}} \hat{a}_{\mathbf{k}} + \hat{a}_{-\mathbf{k}} 
\hat{a}^{\dag}_{-\mathbf{k}} \right) + \mathcal{B}_{\mathbf{k}} \left( \hat{a}^{\dag}_{\mathbf{k}} \hat{a}^{\dag}_{-\mathbf{k}} + \hat{a}_{\mathbf{k}} \hat{a}_{-\mathbf{k}}
\right),
\label{bose_form}
\end{equation}

\noindent
where $\mathcal{A}_{\mathbf{k}}$ and $\mathcal{B}_{\mathbf{k}}$ are momentum-dependent quantities depending on the lattice model, $\hat{a}_\mathbf{k}$
($\hat{a}_{\mathbf{k}}^{\dag}$) denotes the bosonic annihilation (creation) operator. Since the Bose form at Eq.~\eqref{bose_form} is quadratic, it exists an unique 
canonical transformation diagonalizing the bosonic Hamiltonian $\hat{H}$. The diagonalized form of $\hat{H}$ is obtained relying on the bosonic Bogolyubov theory [see Eq.~\eqref{bog_tr}] 
and reads as

\begin{equation}
 \hat{H} = \sum_{\mathbf{k}} E_{\mathbf{k}} \hat{\beta}^{\dag}_{\mathbf{k}} \hat{\beta}_{\mathbf{k}},
\end{equation}

\noindent
with $\hat{\beta}_{\mathbf{k}}$ ($\hat{\beta}^{\dag}_{\mathbf{k}}$) denotes the annihilation (creation) of a Bogolyubov quasiparticle. $E_{\mathbf{k}}$
corresponds to the excitation spectrum associated to the Bogolyubov quasiparticle excitations of the considered lattice model. Finally, the time-dependent
expectation values of the relevant physical observables to describe the quench dynamics are expanded onto the elementary excitations of the quantum system 
previously described. In the case of sudden global quenches, this theoretical method has permitted to unveil a generic expression for the equal-time connected correlation functions $G(\mathbf{R},t)$
of the form 

\begin{equation}
 G(\mathbf{R},t) \equiv \langle \hat{A}_\mathbf{X}(t)\hat{B}_\mathbf{Y}(t)\rangle - \langle \hat{A}_\mathbf{X}(t) \rangle
\langle  \hat{B}_\mathbf{Y}(t) \rangle - \langle \hat{A}_\mathbf{X}(0)\hat{B}_\mathbf{Y}(0)\rangle + \langle \hat{A}_\mathbf{X}(0) \rangle
\langle  \hat{B}_\mathbf{Y}(0) \rangle,
\end{equation}

\noindent
where $\hat{A}$ and $\hat{B}$ are two local operators acting on the local Hilbert space $\mathbb{H}_{\mathbf{X}}$ and $\mathbb{H}_{\mathbf{Y}}$ (on the lattice site 
$\mathbf{X}$ and $\mathbf{Y}$) respectively, where $\mathbf{R} = |\mathbf{X} - \mathbf{Y}|$ is the distance between both lattice sites. The generic form \cite{cevolani2018}
of $G(\mathbf{R},t)$ is given by the following expression  

\begin{equation}
 G(\mathbf{R},t) = g(\mathbf{R}) - \int_{\mathcal{B}} \frac{\mathrm{d}\mathbf{k}}{(2\pi)^D} \mathcal{F}(\mathbf{k})
 \left \{ \frac{e^{i(\mathbf{k}.\mathbf{R}+2E_\mathbf{k}t)} + e^{i(\mathbf{k}.\mathbf{R}-2E_\mathbf{k}t)}}{2} \right \},
\end{equation}

\noindent
where the $D$-dimensional integral spans the first Brillouin zone $\mathcal{B}$. $g(\mathbf{R})$ corresponds to a real-space-dependent 
quantity and $\mathcal{F}(\mathbf{k})$ to a quasimomentum-dependent function giving the amplitude (the contribution weight) of each Bogolyubov quasiparticle. The latter
depends not only on the quench strength but also on both local observables ($\hat{A}$, $\hat{B}$). This generic form, extensively studied theoretically
at Chap.~\ref{ch:3-universal_scaling_laws} and compared to numerical results at Chap.~\ref{ch:4-bose_hubbard_chain} and \ref{ch:5-long_range_ising_chain}, permits to 
understand in details the space-time pattern of correlation functions by reasoning in terms of quasiparticle pairs spreading into the lattice.

\subsection{Quench action}
The quench action \cite{caux2013, caux2016} corresponds to another theoretical technique to investigate the far-from-equilibrium dynamics of closed 
quantum lattice models. It consists of an effective representation for the calculation of time-dependent expectation values (of the form $\langle \Psi(t) | \hat{O} | 
\Psi(t) \rangle$) of a wide class of local observables (here represented by the operator $\hat{O}$) following a quantum quench. More precisely, the latter is based on an 
encoding of the relevant information contained in the overlaps $\langle \phi^{\mathrm{f}}_n | \Psi_0 \rangle$ between the initial state $\ket{\Psi_0}$ and the
eigenstates of the post-quench Hamiltonian $\ket{\phi^{\mathrm{f}}_n}$. Note that these overlaps are essential since they fully characterize the time-dependent 
many-body quantum state $\ket{\Psi(t)}$, and by extension, the quench dynamics of the lattice model. Considering $\ket{\Psi_0}$ the initial state corresponding to the ground state
of the pre-quench (initial) Hamiltonian $\hat{H}_{\mathrm{i}}$ and $\{ \ket{ \phi^{\mathrm{f}}_n } \}$ the eigenbasis for the post-quench (final) 
Hamiltonian $\hat{H}_{\mathrm{f}}$, the time-evolved quantum state $\ket{\Psi(t)}$ (evolving unitarily with $\hat{H}_{\mathrm{f}}$ since closed quantum lattice models are 
considered) may be written as 

\begin{equation}
 \ket{\Psi(t)} = e^{-i \hat{H}_{\mathrm{f}}t} \ket{\Psi_0} = \sum_n \langle \phi^{\mathrm{f}}_n | \Psi_0 \rangle e^{-i E^{\mathrm{f}}_n t} \ket{\phi^{\mathrm{f}}_n}.
\end{equation}

\noindent
$E_n^{\mathrm{f}}$ corresponds to the (energy) eigenvalue associated to the eigenvector $\ket{\phi^{\mathrm{f}}_n}$.
Finally, it can be shown that the expectation values can be reduced to a single sum over the Hamiltonian eigenstates $\ket{\phi^{\mathrm{f}}_n}$, decreasing 
drastically the computational cost, see Ref.~\cite{caux2013} for more details. This method is expected to be generally valid and to describe accurately the quench dynamics of local observables at 
arbitrary times after the quantum quench. In other words, this effective representation can be used to investigate both
the short-time and long-time dynamics. As a consequence, one can rely on the quench action method to characterize the 
stationary (or steady) state and hence investigating both the local relaxation and thermalization processes, see Subsections.~\ref{relaxation} and \ref{thermalization}. \\ \\

In the next chapters, the quench dynamics of one-dimensional quantum lattice models is extensively studied. The latter is characterized by analyzing the 
propagation of information, including the correlations and entanglement, in situations relevant to the modern experiments on ultracold atoms and ions. 
Firstly, we focus on the correlation spreading for short-range interacting lattice models, in gapped and gapless quantum phases for sudden global
and local quenches. Then, we turn to a similar analysis for quantum lattice models with long-range interactions, particularly relevant to experiments on 
molecular condensates and ultracold ions and where well-known theorems break down. 
These fundamental issues about the correlation spreading are tackled using the most modern many-body approaches. Concerning, the analytical point of view, a
quasiparticle picture relying on the Bogolyubov theory within a mean field approximation is developed. For the numerical part, tensor network based techniques
adapted to describe both the static and dynamical properties of one-dimensional quantum lattice models are considered.

%% file: text/ch3_universal_scaling_laws.tex
\setstretch{1.0} 

\begin{savequote}[8cm]
\textlatin{“The important thing about electrons and protons is not what they are but how they behave,
how they move. I can describe the situation by comparing it to the game of chess. In chess, we have various chessmen, kings, knights,
pawns and so on. If you ask what chessman is, the answer would be that it is a piece of wood, or a piece of ivory, or perhaps just a
sign written on paper, or anything whatever. It does not matter. Each chessman has a characteristic way of moving and this is all that
matters about it.”}
  \qauthor{--- Paul A.M. Dirac}
\end{savequote}

\chapter{\label{ch:3-universal_scaling_laws}Universal scaling laws for correlation spreading in quantum lattice models with variable-range interactions} 

\minitoc

\newpage

\section{Position of the problem and generic approach}
\label{PROB}

In the last decades, simultaneous progress realized in the field of ultracold atoms permitted to simulate
many particle and spin Hamiltonians of condensed-matter physics with unprecedented control possibilities 
of the parameters in time allowing to investigate their out-of-equilibrium dynamics. One fundamental issue
of the out-of-equilibrium dynamics 
concerns the spreading of quantum correlations. It is at the center of many fundamental phenomena including the propagation of information and
entanglement, thermalization, and the area laws for entanglement entropy. For short-range interacting lattice systems,
the existence of Lieb-Robinson (LR) bounds implies the emergence of a linear causality cone beyond which the quantum 
correlations are exponentially suppressed ~\cite{lieb1972,bravyi2006,hastings2006}. However, this bound resulting in
a ballistic propagation, in particular, of the equal-time connected correlations \cite{bravyi2006} is not sufficient
to fully characterize the causal correlation cone and does not hold for long-range interacting quantum systems. 
Hence, generalized LR bounds have been derived for long-range systems where the interactions 
decay algebraically, $1/R^\alpha$, with the distance $R$~\cite{hastings2006,foss-feig2015}. The 
related experiments and numerical investigations have, however, lead to conflicting
pictures~\cite{hauke2013,eisert2013,jurcevic2014,richerme2014,cevolani2015,cevolani2016,buyskikh2016}.
For instance, experiments on ion chains~\cite{richerme2014} and numerical simulations within truncated
Wigner approximation~\cite{schachenmayer2015b}
for the one-dimensional (1D) long-range XY (LRXY) model
point towards bounded, super-ballistic, propagation for all values of $\alpha$. In contrast, experiments 
on the long-range transverse Ising (LRTI) model reported ballistic propagation of correlation 
maxima \cite{jurcevic2014}. Moreover, time-dependent Density Matrix Renormalization Group (t-DMRG) and Variational Monte-Carlo (t-VMC) 
numerical simulations indicate the existence of three distinct regimes, namely instantaneous, sub-ballistic, and ballistic, for increasing 
values of the exponent $\alpha$~\cite{schachenmayer2013,hauke2013,eisert2013,cevolani2015,cevolani2016,buyskikh2016}. \\
In this chapter, we shed new light on these apparently conflicting results concerning the scaling laws of the LR bound
and on the lack of universality. In order to give a general and complete description of the correlation spreading, we
propose a generic approach based on a quasiparticle picture that can be applied both to short-range and 
long-range interacting particle and spin models on a hypercubic lattice. Using the latter, we unveil a 
universal twofold structure for the causality cone of correlations, an outer and inner structure determining 
the correlation edge (CE) and the propagation of local extrema respectively. The associated scaling laws  
mainly depend on the interactions between particles and are characterized both for short-range
interactions (see Sec.~\ref{SRC}) and long-range interactions (see Sec.~\ref{LRC}). For particle systems, 
the equal-time connected one-body and density-density correlation functions are studied whereas for spin systems,
we consider equal-time connected spin-spin correlation functions. Moreover, an illustration of the scaling laws for
the correlation spreading within a mean field approximation is provided here for the case study of the one-dimensional
(1D) Short-Range Bose-Hubbard (SRBH) model and two distinct long-range spin
lattice models, namely the 1D Long-Range Transverse Ising (LRTI) and 1D Long-Range XY (LRXY) models. In the next two
chapters, a numerical confirmation of the scaling laws for the same lattice models is presented relying on tensor-network
techniques.

\subsection{Generic Hamiltonian}
Consider a generic quantum system defined on a hypercubic lattice of dimension $D$ and governed by a translation-invariant
Hamiltonian $\hat{H}$ of the following form

\begin{equation}
 \hat{H} = \sum_{\mathbf{R},\mathbf{R}'} J(\mathbf{R},\mathbf{R'}) \hat{O}_1(\mathbf{R}) \hat{O}_2(\mathbf{R}')
 + \sum_{\mathbf{R}} h(\mathbf{R}) \hat{O}_3(\mathbf{R}),
 \label{generic_ham}
\end{equation}

\noindent
where $(\mathbf{R}, \mathbf{R}') \in \mathbb{Z}^{2D}$ are two lattice sites. The first term accounts for a two-site coupling
term built from two operators $\hat{O}_1(\mathbf{R})$ and $\hat{O}_2(\mathbf{R}')$ with an amplitude
given by $J(\mathbf{R},\mathbf{R}')$. The second term denotes a local interaction \textit{via} the operator 
$\hat{O}_3(\mathbf{R})$ and the associated energy $h(\mathbf{R})$. The translational invariance of 
the quantum model implies $J(\mathbf{R},\mathbf{R}') = J(\mathbf{R}-\mathbf{R}')$ and $h(\mathbf{R}) = h$, $\forall ~
(\mathbf{R}, \mathbf{R}') \in \mathbb{Z}^{2D}$. \\

The generic form of Eq.~\eqref{generic_ham} applies to a variety of particle and spin lattice models, including 
the 1D short-range Bose-Hubbard model, the 1D long-range transverse Ising and 1D long-range XY models, which are
particularly relevant for experimental investigations of the correlation spreading based on ultracold atoms or trapped ions,
see for instance Refs.~\cite{cheneau2012,jurcevic2014,richerme2014}, and considered in the following. 
For instance, we show below that the 1D SRBH and the 1D Short-Range Transverse Ising (SRTI) models corresponding to a 
generic short-range interacting lattice and spin model respectively can be written under the general form of 
Eq.~\eqref{generic_ham}. 

\subsubsection{One-dimensional short-range Bose-Hubbard model}

The 1D SRBH model, represented on Fig.~\ref{fig_bh_ti}(a), is built from the bosonic operators $\hat{a}_R$ ($\hat{a}^{\dag}_R$)
corresponding to the bosonic annihilation (creation) operator and $\hat{n}_R = \hat{a}^{\dag}_R \hat{a}_R$ to the 
number operator on the lattice site $R$. They obey the usual commutation rules \textit{ie.}
$\left[\hat{a}_R, \hat{a}^{\dag}_{R'} \right] = \delta_{R,R'}$ and 
$\left[ \hat{a}_R, \hat{a}_{R'} \right] = \left[ \hat{a}^{\dag}_R, \hat{a}^{\dag}_{R'} \right] = 0$.
For the two-site couplings, there are two terms defined by $\hat{O}_1(\mathbf{R}) = \hat{a}^{\dag}_R$,
$\hat{O}_2(\mathbf{R}') = \hat{a}_{R'}$ with $J(\mathbf{R},\mathbf{R}') = -J(\delta_{R+1,R'} + \delta_{R-1,R'})$ corresponding 
to the hopping amplitude. For the local interaction, $h(\mathbf{R}) = U/2$ and $\hat{O}_3(\mathbf{R})
= \hat{n}_R \left( \hat{n}_R - 1 \right)$. Consequently, equation \eqref{generic_ham} leads to the Hamiltonian of the 
1D SRBH model given by 

\begin{equation}
 \hat{H} = -J \sum_{R} \left(\hat{a}^{\dag}_R \hat{a}_{R+1} + \mathrm{h.c} \right) +
 \frac{U}{2} \sum_R \hat{n}_R \left( \hat{n}_R - 1 \right).
 \label{H_bhm}
\end{equation}

\subsubsection{One-dimensional short-range Transverse Ising model}

\begin{figure}[h!]
\centering
\includegraphics[scale = 0.4]{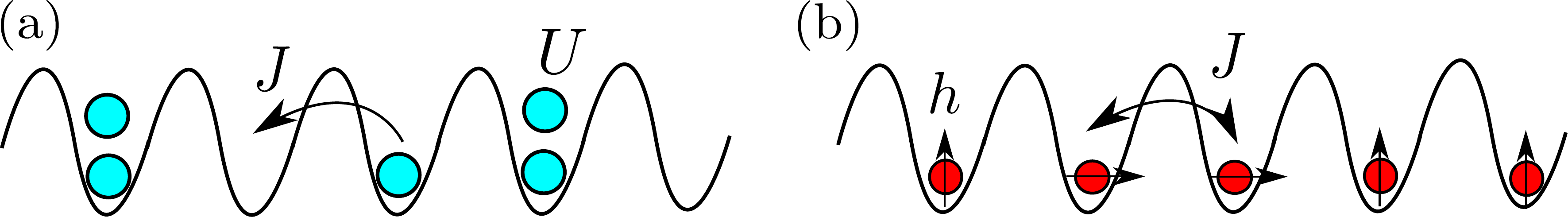}
\caption{\label{fig_bh_ti}
Short-range interacting lattice models. (a)~ Sketch of the 1D short-range Bose-Hubbard model 
representing interacting bosonic particles in an optical lattice. (b)~Sketch of the 1D short-range interacting
transverse Ising model corresponding to spins $s=1/2$ interacting \textit{via} an antiferromagnetic interaction
$J>0$ while an uniform and constant transverse magnetic field $h>0$ is applied along the $z$ axis favorizing the
spins to align along this direction.}
\end{figure}

Concerning the 1D SRTI model represented on Fig.~\ref{fig_bh_ti}(b), it is built from spin operators $\hat{S}^{\alpha}(R) = (\hbar/2) \hat{\sigma}^{\alpha}(R)$
fulfilling the following commutation rule $\left[\hat{S}^{\alpha}_R, \hat{S}^{\beta}_{R'}\right] = i
\epsilon^{\alpha \beta \gamma} \delta_{R,R'} \hat{S}^{\gamma}_R$ with $(\alpha,\beta,\gamma)\in \{x,y,z\}^3$ and
$\epsilon$ the Levi-Civita symbol defined as follows

\[
  \epsilon^{\alpha \beta \gamma}=\begin{cases}
               = 1 ~~~\mathrm{if}~~\alpha \beta \gamma \in \{xyz, yzx, zxy\} \\
               = -1 ~\mathrm{if}~~ \alpha \beta \gamma  \in \{xzy, yxz, zyx\} \\
               = 0 ~~~\mathrm{if}~~ \alpha=\beta ~~\mathrm{or}~~\beta = \gamma~~\mathrm{or}~~\alpha=\gamma
            \end{cases}.
\]

\noindent
The two-site coupling of the 1D SRTI model is characterized by $\hat{O}_1(\mathbf{R}) = \hat{S}^{x}_R$, $\hat{O}_2(\mathbf{R}')
= S^{x}_{R'}$ and $J(\mathbf{R},\mathbf{R}') = 2J \delta_{R+1,R'}$. The local interaction is defined by $\hat{O}_3(\mathbf{R}) = \hat{S}^z_R$
and $h(\mathbf{R}) = -2h$ leading to the well-known Hamiltonian of the 1D SRTI model which reads as follows

\begin{equation}
\hat{H} = 2J\sum_{R} \hat{S}^x_R \hat{S}^x_{R+1} -2h\sum_R \hat{S}^z_R.
\label{short_range_ising}
\end{equation}

\subsection{Quench dynamics}
\label{qd}
To characterize the far-from-equilibrium dynamics of the generic model represented at Eq.~\eqref{generic_ham}, we analyze its response
to a so-called \textit{global or local quench}. In the following, we consider the specific case of \textit{sudden global and local quenches}. 

\subsubsection{Sudden global quenches}
A global quench denotes a modification of, at least, one of the parameters in the Hamiltonian at a given time for a specific duration. 
For a sudden global quench, starting from $\ket{\Psi_0}$ the ground state of $\hat{H}$ for an initial set of parameters, we quench the quantum system 
out-of-equilibrium at time $t=0$ by considering the same Hamiltonian $\hat{H}$ with a different set of parameters or another Hamiltonian, see Fig.~\ref{gb}(a).
We will refer to the initial and final Hamiltonians as the pre-quench Hamiltonian $\hat{H}_\mathrm{i}$
and post-quench Hamiltonian $\hat{H}_{\mathrm{f}}$ respectively. These sudden global quenches have the particularity to conserve the translational 
invariance of the model during the full process, providing that $\hat{H}_\mathrm{i}$, $\hat{H}_{\mathrm{f}}$ are translationally invariant. 
Except whenever mentionned, the pre- and post-quench Hamiltonians are identical, only the interaction parameter is modified. 

\subsubsection{Sudden local quenches}
For sudden local quenches, a single Hamiltonian $\hat{H}$ is considered. In general, one starts from the ground state of $\hat{H}$ whose interaction parameter is
choosen such that the latter corresponds to a product (not entangled) state. Then, a local perturbation is applied \textit{e.g.} a spin-flip for spin lattice models 
or moving a particle for bosonic or fermionic lattice models resulting in an initial state $\ket{\Psi_0}$ which is not translationally invariant anymore. Finally, 
the initial state is suddenly quenched at $t=0$ \textit{via} the Hamiltonian $\hat{H}$ with the same set of parameters as before. \\

\subsubsection{Driving lattice models out of equilibrium}
In the following, we provide more details about the quenching process. According to the time-dependent Schrödinger equation, which reads as 

\begin{equation}
 i \partial_t \ket{\Psi(t)} = \hat{H} \ket{\Psi(t)},
\end{equation}

\noindent
the time-evolved quantum state $\ket{\Psi(t)}$ with respect to the post-quench Hamiltonian $\hat{H}_{\mathrm{f}}$ can
be written as follows ($\hbar$ fixed to unity)

\begin{equation}
 \ket{\Psi(t)} = e^{-i \hat{H}_{\mathrm{f}}t} \ket{\Psi_0}.
 \label{MBWV}
\end{equation}

\begin{figure}[b!]
\centering
\includegraphics[width = \linewidth]{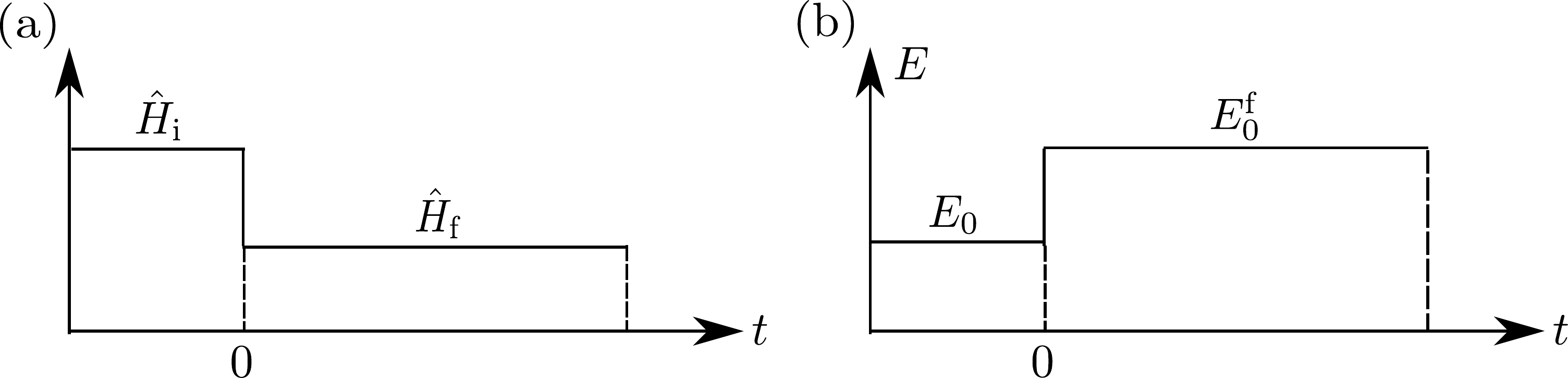}
\caption{\label{gb} Sudden global quench for a generic Hamiltonian $\hat{H}$. (a)~ 
Representation of a global quench where the Hamiltonian $\hat{H}$ is prepared in the pre-quench Hamiltonian 
$\hat{H}_\mathrm{i}$ then a global quench is suddenly applied on $\hat{H}_\mathrm{i}$ at $t=0$ and the
quantum system evolves with the post-quench Hamiltonian $\hat{H}_{\mathrm{f}}$. (b)~Evolution of the total energy 
as a function of time $t$. Sudden global quenches allow to drive quantum systems far from equilibrium by simply
changing a parameter of a Hamiltonian $\hat{H}$. This results in an initial highly excited state (not even an eigenvector)
with respect to the post-quench Hamiltonian $\hat{H}_\mathrm{f}$ whose far-from-equilibrium dynamics can be observed in time.}
\end{figure}

\noindent
This unitary time evolution of the initial quantum state $\ket{\Psi(t)}$ conserves both the norm, 
$ \braket{\Psi(t)|\Psi(t)} = \braket{\Psi_0|\Psi_0} = 1 $, and the total energy, $ \braket{\Psi(t)| 
\hat{H}_{\mathrm{f}} | \Psi(t)} = \braket{\Psi_0 | \hat{H}_{\mathrm{f}} | \Psi_0} = E_0^{\mathrm{f}}$
assuming that $\ket{\Psi_0}$ is well normalized. The energy is conserved from $t=0^+$, corresponding to the time
right after having applied the sudden global quench to a final observation time. Indeed, the sudden global quench allows
to drive the quantum system far from equilibrium by starting from a highly excited state \textit{ie.}
$E_{0}^{\mathrm{f}} > E_0 = \langle \Psi_0 | \hat{H}_{\mathrm{i}} | \Psi_0 \rangle $ where $E_0$ denotes the ground 
state energy with respect to the pre-quench Hamiltonian $\hat{H}_\mathrm{i}$, see Fig.~\ref{gb}(b).
The latter is due to the initial quantum state $\ket{\Psi_0}$ which is not an eigenvector of the post-quench
Hamiltonian $\hat{H}_\mathrm{f}$. 

\subsection{The connected correlation functions}

The real time evolution of quantum lattice systems can be characterized by computing equal-time connected correlation
functions. Those observables read as

\begin{equation}
G_0(\mathbf{R},t) \equiv \langle \hat{A}_X(t)\hat{B}_Y(t)\rangle - \langle \hat{A}_X(t) \rangle
\langle  \hat{B}_Y(t) \rangle
\label{generic_corr}
\end{equation}

\noindent
where $\hat{A}_{X}$, $\hat{B}_{Y}$ denote operators with support in region $X$ and $Y$ separated by a distance $\mathbf{R}$. 
In practice, $\hat{A}_{X}$, $\hat{B}_{Y}$ correpond to local operators acting on a single lattice site $\mathbf{X}$
and $\mathbf{Y}$ respectively. In order to conserve only the dynamical part of $G_0(\mathbf{R},t)$, we substract 
its equilibrium value $G_0(\mathbf{R},0)$ and the resulting connected correlation function denoted by $G(\mathbf{R},t)
= G_0(\mathbf{R},t) - G_0(\mathbf{R},0)$ reads as

\begin{equation}
 G(\mathbf{R},t) \equiv \langle \hat{A}_\mathbf{X}(t)\hat{B}_\mathbf{Y}(t)\rangle - \langle \hat{A}_\mathbf{X}(t) \rangle
\langle  \hat{B}_\mathbf{Y}(t) \rangle - \langle \hat{A}_\mathbf{X}(0)\hat{B}_\mathbf{Y}(0)\rangle + \langle \hat{A}_\mathbf{X}(0) \rangle
\langle  \hat{B}_\mathbf{Y}(0) \rangle
\end{equation}

\noindent
Such correlations for particle (one-body, density-density correlation functions) and spin 
lattice models (spin-spin correlation functions along different directions) can be measured in state-of-the-art
experiments based on ultracold atoms or artifical ion crystals for instance \cite{cheneau2012,fukuhara2013,richerme2014,jurcevic2014}. \\

We describe global quenches where the dynamics is driven by the low-energy sector of $\hat{H}$ that may be assumed to
consist of quasiparticle excitations. Due to the translational invariance, they are characterized by a well-defined
quasimomentum $\mathbf{k}$ and an associated energy $E_\mathbf{k}$. As we shall see later, many relevant observables for particle and spin lattice models in 
various phases and regimes, may be cast into a generic form, see for instance Refs.~\cite{barmettler2012,natu2013, hauke2013,cevolani2015,cevolani2016,buyskikh2016,
frerot2018}. The latter may be written as \cite{cevolani2018}

\begin{equation}
 G(\mathbf{R},t) = g(\mathbf{R}) - \int_{\mathcal{B}} \frac{\mathrm{d}\mathbf{k}}{(2\pi)^D} \mathcal{F}(\mathbf{k})
 \left \{ \frac{e^{i(\mathbf{k}.\mathbf{R}+2E_\mathbf{k}t)} + e^{i(\mathbf{k}.\mathbf{R}-2E_\mathbf{k}t)}}{2} \right \},
 \label{generic_form}
\end{equation}

\noindent
where the $D$-dimensional integral spans the first Brillouin zone $\mathcal{B}$. The real-space-dependent quantity $g(\mathbf{R})$
can be dropped since one wants to describe the far-from-equilibrium dynamics of the lattice model. Equation
\eqref{generic_form} describes
the motion of free counterpropagating quasiparticle pairs, with velocities determined by the excitation spectrum $E_\mathbf{k}$.
The amplitude function $\mathcal{F}(\mathbf{k})$ encodes the overlap between the initial state and the
different quasiparticle wavefunctions with the matrix elements of both the local operators $\hat{A}$, $\hat{B}$. 
In other words, it can be seen as a weight associated to each quasiparticle with a quasimomentum $k$ within the first Brillouin zone. This
weight depends not only on the considered observables but also on the global quench. \\
This explicit form of the correlations can be derived in exactly solvable models and quadratic systems, 
which can be diagonalized by means of canonical transformations \textit{e.g.} bosonic or fermionic Bogolyubov
transformation. The concept of quasiparticles also applies to models that are not exactly solvable, where
they can be determined using tensor-network techniques \cite{haegeman2012,nakatani2014} and
our theory on correlation spreading also holds for such quantum systems. \\

It is convenient to distinguish the case of Hamiltonians having short-range couplings from those having long-range couplings 
where the scaling laws of the universal causality cone are significantly different. 

\section{Lattice models with short-range couplings}
\label{SRC}

We start with the case of lattice models with short-range interactions and more precisely nearest-neighbor interactions. 
It is the simplest case and it applies to a variety of particle and spin models. 

\subsection{Stationary phase approach}
In order to extract the scaling laws of the causality cone, we rely on the stationary phase approximation to investigate 
the long-distance and long-time behavior of the generic form of the equal-time connected correlation
function at Eq.~\eqref{generic_form}. \\

In order to investigate the asymptotic behavior \textrm{ie.} long-time and long-distance behavior along 
the line $R/t = \mathrm{const}$ of the generic connected correlation function $G(\mathbf{R},t)$ at Eq.~\eqref{generic_form}, 
one can rely on the stationary phase approximation. This approximation is based on the cancellation of sinusoids with rapidly
varying phase. Indeed, a sum of many sinusoids having the same phase will add constructively, otherwise they will add incoherently.
Therefore, the $D$-dimensional integral at Eq.~\eqref{generic_form} is dominated by the quasimomentum contributions 
with a stationary phase (sp) corresponding to the following condition,

\begin{equation} 
\partial_{\mathbf{k}} \left( \mathbf{k}.\mathbf{R} \mp 2E_\mathbf{k}t \right) = 0 ~~~\textrm{equivalent~to}~~~ 2V_{\mathrm{g}}(\mathbf{k}_{\mathrm{sp}}) = \pm R/t.
\label{ksp}
\end{equation}

\noindent
For short-range lattice systems, the quasiparticle group velocity $V_\mathrm{g}(\mathbf{k})= \partial_\mathbf{k}E_\mathbf{k}$ is bounded 
\footnote{This statement is not valid for long-range lattice systems. The group velocity can diverge within the first Brillouin zone leading to strong modifications
of the scaling laws for the causality cone, see Sec.~\ref{LRC}.} within
the first Brillouin zone where $E_\mathbf{k}$ corresponds to the excitation spectrum of the short-range interacting lattice model. In other words, this statement implies that one can find a quasimomentum $\mathbf{k}^*$ such that 

\begin{equation}
\mathbf{k}^* = \mathrm{arg~max}\left[V_\mathrm{g}(\mathbf{k})\right]. 
\label{kstar}
\end{equation}

\noindent
Since the group velocity is upper bounded by a value $V_\mathrm{g}^* = V_{\mathrm{g}}(\mathbf{k}^*) = \mathrm{max}\left[
V_{\mathrm{g}}(k) \right]$, Eq.~\eqref{ksp} has a solution only for $R/t \leq 2V_\mathrm{g}^*$.
When applying the stationary phase approximation to Eq.~\eqref{generic_form}, the asymptotic behavior of the connected correlation function $G(\mathbf{R},t)$ is
given by, see Appendix.~\ref{appendix2_sp} for more details about the derivation,

\begin{equation}
G(\mathbf{R},t) \propto \frac{\mathcal{F}(\mathbf{k}_{\mathrm{sp}})}{\left(|2\partial^2_{\mathbf{k}} E_{\mathbf{k}_{\mathrm{sp}}}|t \right)^{\frac{D}{2}}} 
\cos \left(\mathbf{k}_{\mathrm{sp}}.\mathbf{R} - 2E_{\mathbf{k}_{\mathrm{sp}}}t + \sigma D \frac{\pi}{4} \right),
~~~\sigma = \mathrm{sgn}\left[ -2\partial^2_\mathbf{k} E_{\mathbf{k}_{\mathrm{sp}}}t \right].
\label{spa}
\end{equation}

\noindent
The latter contains all the necessary information to predict the space-time behavior of the correlation front, as described below. 

\subsubsection{Twofold causality cone for the space-time correlations}
For $R/t > 2V_{\mathrm{g}^*}$, Eq.~\eqref{ksp} has no solution meaning that the different sinusoids are added incoherently leading to vanishingly small correlations. 
Therefore, the space-time correlations have a so-called \textit{causal} (for $R/t < 2V_{\mathrm{g}^*}$) and \textit{non-causal} (for $R/t > 2V_{\mathrm{g}}^*$) regions
where the values are non-zero and zero respectively. More precisely, the correlations are activated at the time $t^* = R/2V_{\mathrm{g}}^*$, also called activation time,
separating the causal and non-causal regions. The previous equation defines a correlation edge (CE) whose ballistic spreading is characterized by the velocity
$V_{\mathrm{CE}} = 2V_{\mathrm{g}}^* = 2 \partial_{\mathbf{k}} E_{\mathbf{k}^*}$ consistently
with the Calabrese-Cardy picture and the Lieb-Robinson bound, see Refs.~\cite{calabrese2006, lieb1972}. \\

In addition, Eq.~\eqref{spa} leads to a CE but also to a series of local maxima and minima whose space-time propagation
can be determined by the cosine argument. Indeed, in the vicinity of the CE, only the quasiparticles with a quasimomentum
$\mathbf{k}$ close to $\mathbf{k}^*$, moving at the maximal group velocity $V_{\mathrm{g}}^*$, contribute to the correlations. The other quasiparticles
with a quasimomentum $\mathbf{k}$ significantly different from $\mathbf{k}^*$ also propagate through the lattice but with a smaller group velocity. As a consequence,
these quasiparticles affect the correlations of the inner structure of the correlation cone but are irrelevant in the vicinity of the CE.
Then, the maxima (m) are defined by the following motion equation 

\begin{equation}
\mathbf{k}^*.\mathbf{R} - 2E_{\mathbf{k}^*}t + \sigma D \frac{\pi}{4} = 2\pi n~~~\mathrm{with}~~~ n \in \mathbb{Z}.
\end{equation}

\noindent
Concerning the minima, the motion equation will be slightly different. Indeed, the cosine argument has to be equal to $(2n+1)\pi$ with $n \in \mathbb{Z}$.
Hence, both the series of local maxima and minima propagate at the velocity $V_{\mathrm{m}} = 2V_{\varphi}^* = 2 E_{\mathbf{k}^*}/\mathbf{k}^*$
corresponding to twice the phase velocity at $\mathbf{k}^*$ defined by Eq.~\eqref{kstar}. \\

In general, for interacting lattice models, the excitation spectrum $E_\mathbf{k}$ is not linear with the quasimomentum $\mathbf{k}$ over the
first Brillouin zone, thus the group and phase velocities differ, see for instance Fig.~\ref{sf_exc_vel} and inset of Fig.~\ref{fig:BHm}(a). 
Consequently, for short-range interacting lattice models, the \textit{correlation cone} is expected to feature a universal twofold structure 
characterized by a ballistic motion of the CE and the series of local maxima with
an associated velocity $V_{\mathrm{CE}} = 2V_{\mathrm{g}}^*$ and $V_\mathrm{m} = 2 V_\varphi^*$ respectively. \\

\begin{figure}[h!]
\centering
\includegraphics[scale = 0.51]{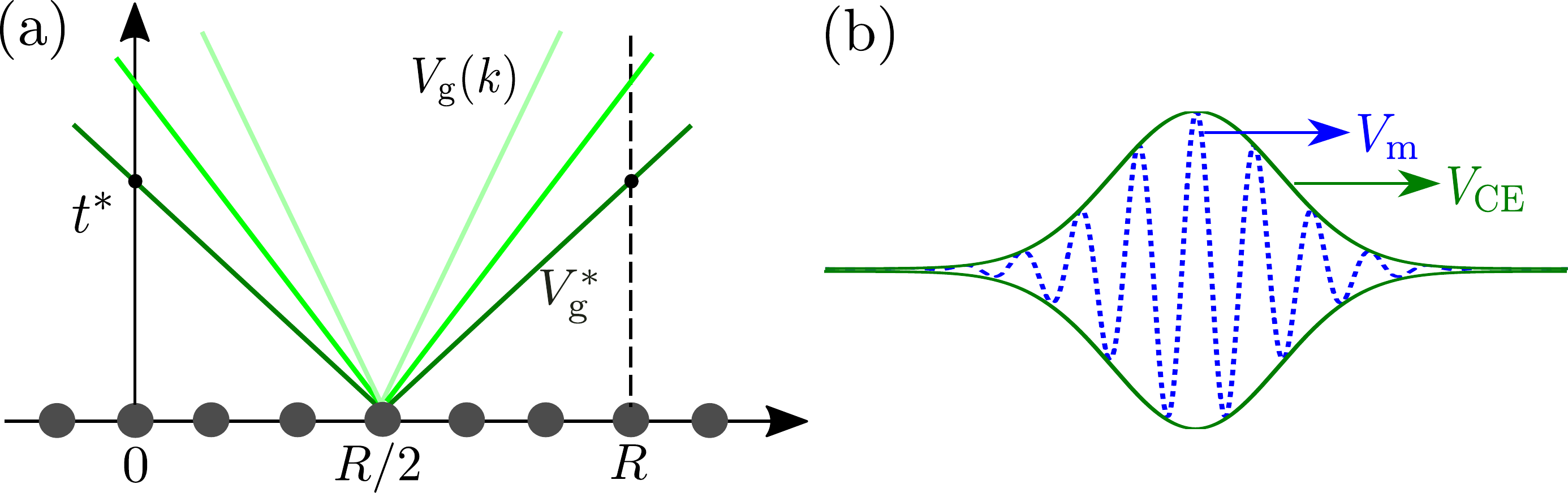}
\caption{\label{corr}
Quantum quench in a 1D short-range lattice model. (a)~Generation of correlations between two points at a distance $R$ by pairs of counter-propagating quasiparticles
emitted at the mid-point $R/2$. The first correlation is generated by the fastest quasiparticles at the activation time $t^*=R/2V_\mathrm{g}^*$. (b)~Correlation spreading
in the vicinity of the correlation edge (CE). The correlation function forms a periodic series of maxima moving at the velocity $V_{\mathrm{m}} = 2 V_\varphi^*$, 
with an envelope moving at the velocity $V_{\mathrm{CE}} = 2V_{\mathrm{g}}^*$. Figure adapted from Ref.~\cite{despres2019}.}
\end{figure}

The twofold structure for the correlation spreading in short-range interacting models found \textit{via} the stationary phase approximation is clarified now. 
For simplicity, let us first consider a 1D lattice model. A correlation between two distinct points separated by a distance $R$ is seeded
when two correlated, counterpropagating quasiparticles emanating from the center reach the two points, see Fig.~\ref{corr}(a). The fastest ones are
those propagating at the maximal group velocity $V_\mathrm{g}^*$. Therefore, it leads the effective velocity $2V_\mathrm{g}^*$ in order to reach the 
corresponding activation time $t^*$. More precisely, a correlation at a distance $R$ and time $t$ is built from a coherent superposition of the 
contributions of all the quasiparticles in the first Brillouin zone. As discussed previously, in the vicinity of the CE, only the fastest ones with a quasimomentum $k^*$ contribute. It creates a sine-like signal at the driving spatial frequency $k^*$ 
whose maxima propagate at the velocity $V_\mathrm{m} = 2V_\varphi^*$. Then, the contribution of quasiparticles with a quasimomentum around $k^*$ modulates the sine-like
signal by an envelope moving at the CE velocity $V_{\mathrm{CE}} = 2V_{\mathrm{g}}^*$, see Fig.~\ref{corr}(b). This behavior is reminiscent of the propagation of a coherent wave-packet in 
a dispersive medium~\cite{brillouin1960,lighthill1965,born1999}. \\

Our theoretical prediction on the linear twofold structure for the correlation spreading in short-range interacting lattice models is explicitly tested 
in the following for the 1D SRBH model in both phases, namely the gapless superfluid and gapped Mott-insulating phases and more precisely in their limit regime
($U \ll J$ for the SF phase and $J \ll U$ for the MI phase) where the model is solvable using a mean field approximation.

\subsection{Case study of the Bose-Hubbard chain: Superfluid phase}
\label{sf_chap3}
\begin{figure}[h!]
\centering
\includegraphics[scale = 0.4]{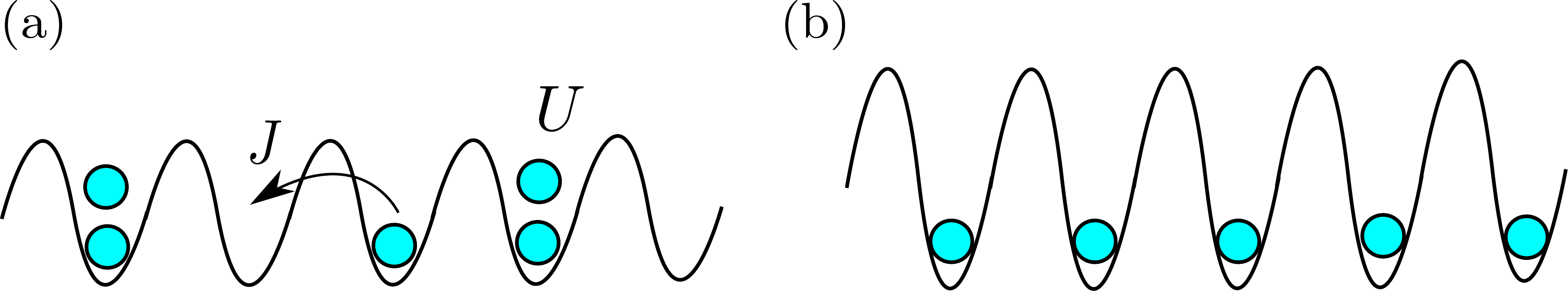}
\caption{\label{sketch_bhm}
Sketch of the 1D short-range Bose-Hubbard model representing interacting bosonic particles at unit-filling $\bar{n} = N/N_s = 1$, in an optical lattice.
(a)~Gapless superfluid phase for $J \gg U$ . In the SF phase, the bosons are delocalized over the lattice sites. (b)~ Gapped Mott-insulating phase for $U \gg J$.
In the MI phase, the bosons are pinned in the lattice sites and the particle-number fluctuations are suppressed.}
\end{figure}

In the following, we illustrate the scaling laws, and more exactly the velocities, characterizing the twofold structure of the correlation cone for short-range interacting
lattice models. To do so, we investigate a one-dimensional bosonic lattice model namely the 1D short-range Bose-Hubbard model whose Hamiltonian
is given at Eq.~\eqref{H_bhm}. For the latter, a competition occurs between a delocalization of the bosons on the lattice (favored by the hopping
amplitude $J>0$) and a localization (\textit{via} the repulsive two-body interaction $U>0$, which tends to localize the bosonic particles by minimizing
the fluctuations of the number of bosons per lattice site). The Bose-Hubbard model possesses two different quantum phases at integer fillings $\bar{n} \in \mathbb{N}^*$ : 
the superfluid (SF) phase at $J\gg U$ and the Mott-insulating (MI) phase at $U \gg J$, see Fig.~\ref{sketch_bhm}. The superfluid phase is characterized by a gapless excitation spectrum and a 
non-zero compressibility whereas the Mott-insulating phase is incompressible and has a gapped excitation spectrum. More details about this lattice model
will be given in the next chapter at Sec.~\ref{1dbhm}.

\subsubsection{Excitation spectrum in the mean field regime}

We first consider the superfluid phase (SF) and the mean field regime corresponding to a high-enough average particle density denoted by $\bar{n} \gg U/2J$, 
see Appendix.~\ref{appendix1_mf} for a derivation of the mean field condition. In this regime, the Hamiltonian of the Bose-Hubbard chain
can be diagonalized using a (bosonic) Bogolyubov approximation, see for instance Refs.~\cite{bogoliubov1947, bogoliubov1958, natu2013,cevolani2015}.
In the following, we provide a brief outline for the derivation of the excitation spectrum. \\

In the limit $J/U \rightarrow 0$ of the SF phase, the bosons are fully delocalized over the lattice. Hence, the many-body wavefunction $\ket{\Psi}$ may be written as

\begin{equation}
 \ket{\Psi} = \frac{1}{\sqrt{N!}} \left( \frac{1}{\sqrt{N_s}}  \sum_{R=1}^{N_s} \hat{a}_R^{\dag} \right)^N \ket{0} =  \frac{1}{\sqrt{N!}} (\hat{a}^{\dag}_{k=0})^N \ket{0}.
\end{equation}

\noindent
where $N$ denotes the total number of bosons and $N_s$ the number of lattice sites ($a$ the lattice spacing is fixed to unity by convention).
All the bosonic particles are in the mode $k=0$ due to their perfect delocalization in real space. Therefore, when considering a finite and non-zero ratio $U/J$ such
that $U \ll J$, one may assume that the mode $k=0$ is occupied by a macroscopic number of bosonic particles. The latter can be formulated as follows

\begin{equation}
\hat{a}_{k=0} \ket{N_0} \simeq \sqrt{N_0} \ket{N_0},~~~ \hat{a}^{\dag}_{k=0} \ket{N_0} \simeq \sqrt{N_0} \ket{N_0}  
\end{equation}

\noindent
where $N_0$ refers to the number of particles in the mode $k=0$. Furthermore, the total number
of bosonic particles $N$ is linked to the one for the mode $k=0$, \textit{ie.} $N_0$, by both following equations

\begin{equation}
N = N_0 + \sum_{k \neq 0} n_k, ~~~ N_0^{2} \simeq N^{2} - 2N \sum_{k \neq 0} n_k,
\label{small}
\end{equation}

\noindent
where the second is found by assuming small density fluctuations. To obtain a quadratic form of the Bose-Hubbard chain at Eq.~\eqref{H_bhm} in the SF mean field regime,
it requires to express bosonic operators in momentum space. The transformation is obtained \textit{via} a Fourier transform and leads for
the bosonic operators of creation (annihilaton) in real space $\hat{a}^{\dag}_{R}$ ($\hat{a}_{R}$) to 

\begin{equation}
\hat{a}^{\dag}_{R} = \frac{1}{\sqrt{N_s}} \sum_{k} e^{ikR} \hat{a}^{\dag}_{k},~~~ \hat{a}_{R} = \frac{1}{\sqrt{N_s}} \sum_{k} e^{-ikR} \hat{a}_{k}.
\label{a_rs}
\end{equation}

\noindent
where $\hat{a}^{\dag}_{k}$, $\hat{a}_{k}$, correspond to the bosonic operator of creation and annihilation in the momentum space respectively.
Then, by injecting Eq.~\eqref{a_rs} in the Bose-Hubbard chain and using Eq.~\eqref{small}, the interaction term can be developed up to quadratic order (in terms of 
$\hat{a}^{\dag}_{k}$ and $\hat{a}_{k}$). Finally, it yields the following quadratic form for the BH Hamiltonian in the Fourier space,

\begin{equation}
\hat{H} = e_0 + \frac{1}{2} \sum_{k \neq 0} \mathcal{A}_{k} \left( \hat{a}^{\dag}_k \hat{a}_k + \hat{a}_{-k} \hat{a}^{\dag}_{-k} \right) +
\mathcal{B}_{k} \left( \hat{a}^{\dag}_k \hat{a}^{\dag}_{-k} + \hat{a}_k \hat{a}_{-k} \right) 
\label{H_quadratic}
\end{equation}

\noindent
with $e_0$ a constant energy and $\mathcal{A}_k$, $\mathcal{B}_k$ two momentum-dependent coefficients. These coefficients are defined as 

\begin{equation}
\mathcal{A}_{k} = \gamma_k + U \bar{n} = 4J\sin^2(k/2) + U\bar{n},~~~ \mathcal{B}_{k} = U\bar{n}. 
\label{a_b_k}
\end{equation}

\noindent
where $\gamma_k = 4J \sin^2(k/2)$ corresponds to the dispersion relation of the free-particle tight-binding model. The resulting quadratic form at 
Eq.~\eqref{H_quadratic} can then be diagonalized using a (bosonic) Bogolyubov canonical transformation. This transformation is based on the introduction of 
hybrid bosonic operators $\hat{\beta}_{k}$ and $\hat{\beta}^{\dag}_{k}$. These new operators, $\hat{\beta}_{k}$ and $\hat{\beta}^{\dag}_{k}$, denoting the
annihilation and creation of a Bogolyubov quasiparticle at quasimomentum $k$ are related to the initial bosonic operators \textit{via} the two following equations

\begin{align}
& \hat{a}_{k} = u_{k} \hat{\beta}_{k} + v_{-k} \hat{\beta}^{\dag}_{-k}, ~~~ \hat{a}^{\dag}_{k} = u_{k}^{*} \hat{\beta}^{\dag}_{k} + v_{-k}^{*} \hat{\beta}_{-k}.
\label{bog_tr}
\end{align}

\noindent
Besides, since the Bogolyubov (quasiparticle) operators are bosonic, they obey the usual commutation rules given by 
$[\hat{\beta}_{k}, \hat{\beta}^{\dag}_{k'}] = \delta_{k,k'}$ and $[\hat{\beta}_{k}, \hat{\beta}_{k'}] = [\hat{\beta}^{\dag}_{k}, \hat{\beta}^{\dag}_{k'}] = 0$. This implies for the coefficients $u_{k}$ and $v_{k}$ to fulfill the condition
$|u_{k}|^{2} - |v_{-k}|^{2} = 1$. Consequently, they can be expressed as a function of a (unknown) parameter $\alpha_{k}$ where $u_{k} = \cosh(\alpha_{k})$ 
and $v_{-k} = \sinh(\alpha_{k})$. The value of $\alpha_k$ is then determined such that the non-diagonal terms \textit{ie.}
$\hat{\beta}_k \hat{\beta}_{-k}$ and $\hat{\beta}^{\dag}_{k} \hat{\beta}^{\dag}_{-k}$ are suppressed. The latter appear when injecting Eq.~\eqref{bog_tr}
in the quadratic form of the BH model given previously at Eq.~\eqref{H_quadratic}. It yields the condition 

\begin{align}
& 2\mathcal{A}_{k} u_{k} v_{-k} + \mathcal{B}_{k} (u_{k}^{2} + v_{-k}^{2}) = 0.
\end{align}

\noindent
The coefficients $u_{k}= \cosh(\alpha_{k})$ and $v_{-k} = \sinh(\alpha_{k})$ fulfilling the previous condition read as

\begin{equation}
u_{k}, v_{-k} = \pm \left[ \frac{1}{2} \left(\frac{\mathcal{A}_{k}}{E_{k}} \pm 1 \right) \right]^{1/2} = 
\pm \left( \frac{\gamma_{k} + U\bar{n}}{2 \sqrt{\gamma_k \left( \gamma_k + 2U\bar{n} \right)}} \pm \frac{1}{2} \right)^{1/2}.
\end{equation}

\noindent
This specific analytical expression of $u_k$ and $v_{-k}$ permits to suppress the off-diagonal terms in the quadratic form. The BH Hamiltonian in the SF mean field regime
is then diagonalized \textit{ie.} depends only on the quasiparticle number operator $\hat{n}_{\beta,k} = \hat{\beta}^{\dag}_{k}\hat{\beta}_k$. It reads as

\begin{equation}
\hat{H} = E_0 + \sum_{k \neq 0} E_{k} \hat{\beta}^{\dag}_{k} \hat{\beta}_{k},
\label{diag_form_e}
\end{equation}

\noindent
with the excitation spectrum $E_k$ defined as

\begin{equation}
E_{k} = \sqrt{\gamma_{k} (\gamma_{k} + 2U\bar{n})}, 
\label{exc_spec_meanfield}
\end{equation}

\noindent
where $\gamma_k = 4J \sin^2(k/2)$ corresponds to the dispersion relation of the free tight-binding model. \\

\begin{figure}
\centering
\includegraphics[scale = 0.36]{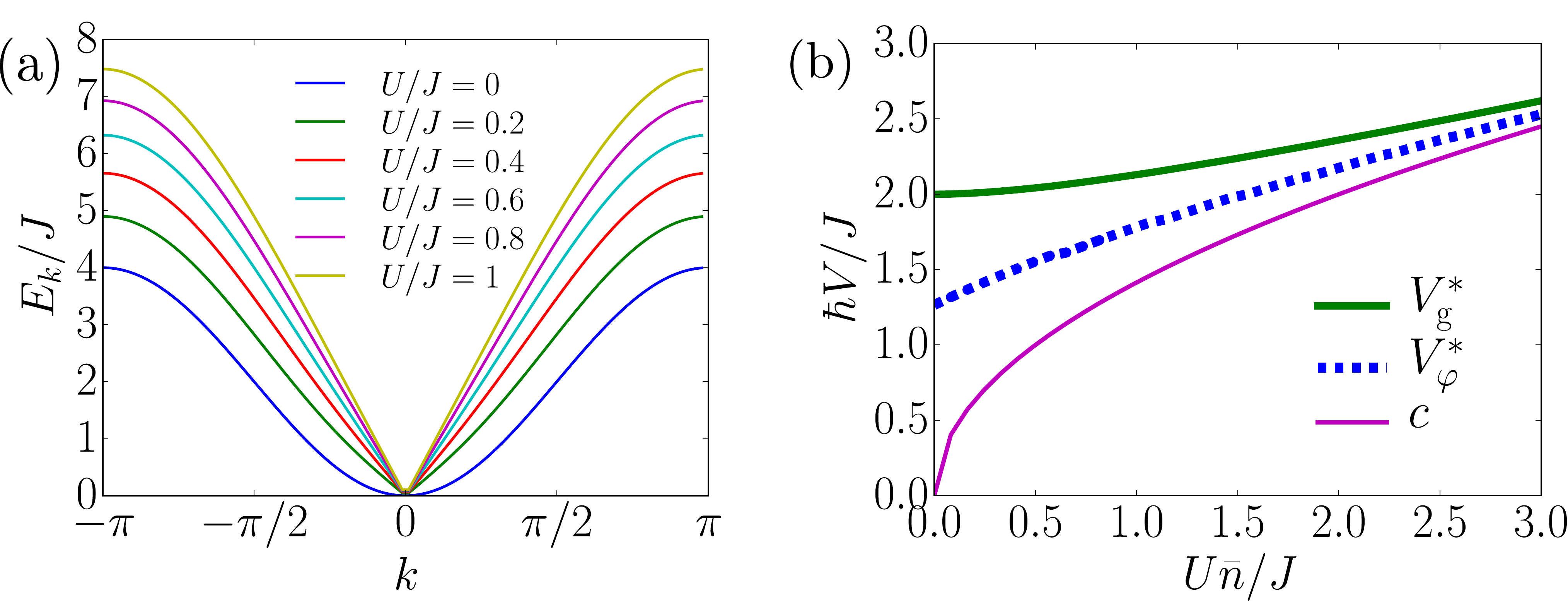}
\caption{\label{sf_exc_vel}
Excitation spectrum properties of the 1D short-range Bose-Hubbard model in the mean field regime of the superfluid phase for $\bar{n}=5$. 
(a)~Energy rescaled by the hopping amplitude $E_k/J$ as a function of the quasimomentum $k$ ($a$ lattice spacing fixed to unity)
in the first Brillouin zone $\mathcal{B} = [-\pi,\pi]$ for several interaction parameters $U/J$. (b)~ Dimensionless physical quantity $\hbar V/J$
as a function of the relevant parameter in the SF mean field regime $U\bar{n}/J$. $V$ denotes three different characteristic velocities : 
the maximal group velocity $V_{\mathrm{g}}^* = V_{\mathrm{g}}(k^*)$, the corresponding phase velocity at $k^*$ denoted by $V_{\varphi}(k^*)$ and 
the sound velocity $c$.}
\end{figure}

According to the diagonal form of the BH model at Eq.~\eqref{diag_form_e}, the ground state at $T=0$ is defined as the vacuum
state $\ket{0}$ for the Bogolyubov quasiparticles. The associated ground state energy is given by $\bra{0} H \ket{0} = E_0$.
The gapless excitation spectrum $E_k$ at Eq.~\eqref{exc_spec_meanfield} is displayed on Fig.~\ref{sf_exc_vel}(a) for several 
interaction parameters $U/J$ and a relatively high filling $\bar{n}=5$. The latter gives the corresponding energy of a Bogolyubov
quasiparticle at quasimomentum $k$ confined in the first Brillouin zone $\mathcal{B} = [-\pi,\pi]$. Two different regimes 
need to be distinguished :

\begin{itemize}
 \item A phononic regime corresponding to the limit of small quasimomenta (or large wavelengths). For $|k| \simeq 0$, $E_k$ is described by an effective linear 
 spectrum consisting of phononic modes, \textit{ie.} $E_k \simeq c|k|$ with $c = \sqrt{2 \bar{n}JU}$ the sound velocity, see Fig.~\ref{sf_exc_vel}(b). 
 
 \item A particle-like regime corresponding to the limit of high quasimomenta (or small wavelengths), provided that $J \gg U\bar{n}$. 
 In this limit, $E_k$ converges towards $\gamma_k = 4J \sin^2(k/2)$ the dispersion relation of the tight-binding model. 
\end{itemize}

\noindent
Considering positive quasimomenta, the two (phononic and particle-like) regimes are separated by an inflection point at $0<k^*<\pi$ depending on the value of $U\bar{n}/J$.
According to the excitation spectrum in the SF mean field regime at Eq.~\eqref{exc_spec_meanfield}, the physical quantity $U\bar{n}/J$ is the single relevant parameter.
Note that the slope of $E_k$ at $k^*$ corresponds to the maximum group velocity, \textit{ie.} $V_{\mathrm{g}}(k^*) = \mathrm{max}[V_{\mathrm{g}}(k)]$. \\

In the following, we briefly analyze the different characteristic velocities of the SF meanfied regime. It will be particularly helpful
in the next paragraphs to analyze the spreading of both the density and phase correlations in this regime. The three investigated characteristic velocities
correspond to (i)~ the group velocity $V_{\mathrm{g}}$ (ii)~ the phase velocity $V_{\varphi}(k)$ and (iii)~ the sound velocity $c$
defined as follows ($\hbar$ is fixed to unity)

\begin{equation}
 V_{\mathrm{g}}(k) = \partial_k E_k,~~~V_{\varphi}(k) = E_k/k,~~~c = \lim_{k\rightarrow 0} V_{\mathrm{g}}(k).
\end{equation}

\noindent
Relying on Eq.~\eqref{exc_spec_meanfield}, the three previous velocities in the SF mean field regime can be determined analytically. It yields respectively
for the group, phase and sound velocities

\begin{align}
& V_{\mathrm{g}}(k) = \frac{4J\cos(k/2)\sin(k/2) \left[4J\sin^2(k/2) + U\bar{n} \right]}{\sqrt{4J\sin^2(k/2) \left[ 4J \sin^2(k/2) + 2\bar{n}U \right]}}, \\ \nonumber \\
& V_{\varphi}(k) = k^{-1} \sqrt{4J\sin^2(k/2) \left[ 4J \sin^2(k/2) + 2\bar{n}U \right]}, \\ \nonumber \\
& c = \sqrt{2\bar{n}JU}.
\end{align}

\noindent
On Fig.~\ref{sf_exc_vel}(b), the maximal value of the group velocity ($V_\mathrm{g}^* = V_\mathrm{g}(k^*)$), the phase velocity at $k^*$ ($V_{\varphi}^*$) 
together with the sound velocity ($c$) are displayed as a function of the dimensionless quantity $U\bar{n}/J$. 
One notices that $V_{\mathrm{g}}^* > V_{\varphi}^* > c,~\forall~ (U\bar{n}/J)\in \mathbb{R}^{+}$. Indeed, due to the convexity of the excitation spectrum $E_k$ and its gapless property, it leads $V_{\mathrm{g}}(k^*) \geq V_{\varphi}(k^*)$ at the
inflexion point $k^*$. However, in the limit of high values of $U\bar{n}/J$, $V_{\mathrm{g}}^* \simeq V_{\varphi}^* \simeq c$. The latter is due to the excitation
spectrum $E_k$ getting phononic modes at higher quasimomentum $k$, see Fig.~\ref{sf_exc_vel}(a). 

\subsubsection{Sudden global quenches in the mean field regime}
In the following, we verify the analytical predictions concerning the spreading velocities of each correlation cone structure for short-range lattice models.
To do so, we investigate the Bose-Hubbard chain in the SF mean field regime driven out of equilibrium \textit{via} sudden global quenches. Consequently, the
pre- and post-quench interaction parameters fulfill the conditions $(U/J)_\mathrm{i}, (U/J)_{\mathrm{f}} \ll  2\bar{n}$ while the filling $\bar{n}$ is fixed during the 
real time evolution. Besides, the spreading of both the phase and density fluctuations are investigated \textit{via} the connected correlation functions $G_1(R,t)$

\begin{align}
& G_1(R,t) = \langle \hat{a}^{\dag}_R(t) \hat{a}_0(t) \rangle - \langle \hat{a}^{\dag}_R(0) \hat{a}_0(0) \rangle
\end{align}

\noindent
and $G_2(R,t)$ respectively,

\begin{align}
& G_2(R,t) = g_2(R,t)-g_2(R,0)~~\mathrm{with}~~g_2(R,t) =  \langle \hat{n}_R(t) \hat{n}_0(t) \rangle - \langle \hat{n}_R(t) \rangle \langle \hat{n}_0(t) \rangle.
\end{align}

\noindent
Both equal-time connected correlation functions are of the form presented at Eq.~\eqref{generic_corr} 
where $\hat{A}_\mathbf{x} = \hat{a}^{\dag}_R$ and $\hat{B}_\mathbf{y} = \hat{a}_0$ for $G_1$. Note that the connected parts for $G_1$, involving the
correlators $\langle \hat{a}^{\dag}_R(t) \rangle$, $\langle \hat{a}^{\dag}_R(0) \rangle$, $\langle \hat{a}_0(t) \rangle $ and $\langle \hat{a}_0(0) \rangle$,
vanish since the canonical statistical ensemble is considered \textit{ie.} implying a fixed total number of bosonic particles on the chain.
For $G_2$, $\hat{A}_\mathbf{x} = \hat{n}_R$ and $\hat{B}_\mathbf{y} = \hat{n}_0$.  \\

To compute analytically both connected correlation functions $G_1$ and $G_2$ for a global quench such that the pre- and
post-quench Hamiltonians are confined in the SF mean field regime, the reasoning is identical :

\begin{itemize}
 \item Firstly, one starts by expressing the correlation functions in terms of the post-quench bosonic operators in the momentum space denoted by
 $\hat{a}_{k,\mathrm{f}}(t), \hat{a}_{k,\mathrm{f}}^{\dag}(t)$.
 \item Secondly, using the Bogolyubov transformation defined at Eq.~\eqref{bog_tr}, one expresses  $\hat{a}_{k,\mathrm{f}}(t)$ and $\hat{a}_{k,\mathrm{f}}^{\dag}(t)$
 as a function of the post-quench
 Bogolyubov operators $\hat{\beta}_{k,\mathrm{f}}(t), \hat{\beta}^{\dag}_{k,\mathrm{f}}(t)$.
 \item Since, the post-quench Bogolyubov operators diagonalize the post-quench Hamiltonian $\hat{H}_{\mathrm{f}} = \hat{H}\left[(U/J)_{\mathrm{f}}\right]$, 
 their time-dependent version take a simple expression. Indeed, relying on the equation of motion $ -i \partial_t \hat{\beta}_{k,\mathrm{f}}^{(\dag)}(t) = \left[\hat{H}, \hat{\beta}_{k,\mathrm{f}}^{(\dag)} \right]_t$,
 one finds
 
 \begin{equation}
  \hat{\beta}_{k,\mathrm{f}}(t) = e^{-i E_{k,\mathrm{f}}t} \hat{\beta}_{k, \mathrm{f}}(0),~~~ \hat{\beta}^{\dag}_{k,\mathrm{f}}(t) =
  e^{i E_{k,\mathrm{f}}t} \hat{\beta}^{\dag}_{k,\mathrm{f}}(0) 
 \end{equation}

 \item Then, the continuity at $t=0$ between the post-quench and pre-quench bosonic operators $\hat{a}^{(\dag)}_{k,\mathrm{f}}(0) = 
 \hat{a}^{(\dag)}_{k,\mathrm{i}}$ permits to find a relation between the post-quench Bogolyubov operators at $t=0$, $\hat{\beta}_{k,\mathrm{f}}(0), 
 \hat{\beta}^{\dag}_{k,\mathrm{f}}(0)$, and the pre-quench Bogolyubov operators $\hat{\beta}_{k,\mathrm{i}}, \hat{\beta}^{\dag}_{k,\mathrm{i}}$. The continuity condition
 is given by
 
 \begin{align}
 & \hat{a}_{k,\mathrm{f}}(0) = u_{k,\mathrm{f}} \hat{\beta}_{k,\mathrm{f}}(0) + v_{-k,\mathrm{f}} \hat{\beta}_{-k, \mathrm{f}}^{\dag}(0) = u_{k,\mathrm{i}} \hat{\beta}_{k,\mathrm{i}}
 + v_{-k,\mathrm{i}}\hat{\beta}_{-k,\mathrm{i}}^{\dag} \\
 & \hat{a}_{k,\mathrm{f}}^{\dag}(0) = u_{k,\mathrm{f}} \hat{\beta}_{k,\mathrm{f}}^{\dag}(0) + v_{-k,\mathrm{f}} \hat{\beta}_{-k,\mathrm{f}}(0)= u_{k,\mathrm{i}}
 \hat{\beta}_{k,\mathrm{i}}^{\dag} + v_{-k,\mathrm{i}}\hat{\beta}_{-k,\mathrm{i}}
 \end{align}
 
 \noindent
 leading for the post-quench Bogolyubov operators at $t=0$ to 
 
 \begin{align}
 & \hat{\beta}_{k,\mathrm{f}}(0) = \left( u_{k, \mathrm{i}} u_{k, \mathrm{f}} - v_{-k,\mathrm{i}} v_{-k,\mathrm{f}} \right) \hat{\beta}_{k, \mathrm{i}} -
 \left( u_{k, \mathrm{i}} v_{-k,\mathrm{f}} - v_{-k,\mathrm{i}} u_{k,\mathrm{f}} \right) \hat{\beta}_{-k, \mathrm{i}}^{\dag} \\
 & \hat{\beta}_{k, \mathrm{f}}^{\dag}(0) = \left( u_{k, \mathrm{i}} u_{k, \mathrm{f}} - v_{-k,\mathrm{i}} v_{-k,\mathrm{f}} \right) \hat{\beta}_{k,\mathrm{i}}^{\dag}
 - \left( u_{k,\mathrm{i}} v_{-k,\mathrm{f}} - v_{-k,\mathrm{i}} u_{k,\mathrm{f}} \right) \hat{\beta}_{-k,\mathrm{i}} 
 \end{align}
 
 \item At this stage, both correlation functions involve only the pre-quench Bogolyubov operators $\hat{\beta}_{k,\mathrm{i}}, \hat{\beta}^{\dag}_{k,\mathrm{i}}$. 
Finally, using the condition $\hat{\beta}_{k, \mathrm{i}} \ket{\mathrm{GS}_\mathrm{i}} = 0$, where $\ket{\mathrm{GS}_\mathrm{i}}$ denotes the ground state of
the pre-quench Hamiltonian $\hat{H}_\mathrm{i} = \hat{H}\left[(U/J)_\mathrm{i}\right]$, one finds the analytical expression of $G_1$ and $G_2$.
\end{itemize}

\subsubsection{The phase fluctuations - $G_1$ correlation function}
We first investigate the phase fluctuations $G_1$ for a sudden global quench confined in the SF mean field regime. 
In the following, the sudden global quench is applied on the two-body repulsive interaction strength $U$ ($U_\mathrm{i}$, $U_\mathrm{f}$ will thus 
denote the pre- and post-quench two-body interaction) and the hopping amplitude $J$ is fixed. Since the Bose-Hubbard chain in this regime can 
be diagonalized using a Bogolyubov canonical transformation, $G_1$ can be calculated
analytically using the previous scheme, see Appendix.~\ref{appendix_g1_sf}. It yields for $G_1$ the following expression 

\begin{align}
& G_1(R,t) \sim -\int_{\mathcal{B}} \frac{\mathrm{d}k}{2\pi} \mathcal{F}_1(k) \left\{ \frac{e^{i(kR+2E_{k,\mathrm{f}}t)} + e^{i(kR-2E_{k,\mathrm{f}}t)}}{2}
\right \},
\label{g1_analytical}
\end{align}

\noindent
with the amplitude function $\mathcal{F}_1$ defined as follows 

\begin{equation}
\mathcal{F}_1(k) = \frac{\bar{n}^2 U_\mathrm{f} \gamma_{k}(U_\mathrm{f} - U_\mathrm{i})}{2 E_{k,\mathrm{i}}
E_{k,\mathrm{f}}^2},
\label{F1}
\end{equation}

\noindent
where $E_{k,\mathrm{i}}$($E_{k,\mathrm{f}}$) denotes the pre-quench (post-quench) excitation spectrum of the 1D short-range Bose-Hubbard model 
in the mean field regime of the SF phase given at Eq.~\eqref{exc_spec_meanfield}. \\

The analytical expression of the phase fluctutations $G_1$ at Eq.~\eqref{g1_analytical} fulfills the generic form presented at Eq.~\eqref{generic_form}.
Moreover, according to Eq.~\eqref{g1_analytical}, the real time evolution of $G_1$ is governed by the 
post-quench Hamiltonian $\hat{H}_{\mathrm{f}}$ through its excitation spectrum $E_{k,\mathrm{f}}$ present in both space-time dependent plane waves.
The previous statement is straightforward since the many-body wavefunction of the quantum system evolves in time with the
post-quench Hamiltonian, see Eq.~\eqref{MBWV}. Finally, based on the analytical expression of $\mathcal{F}_1$ at Eq.~\eqref{F1}, one finds that if $U_\mathrm{i} = U_\mathrm{f}$ \textrm{ie.} the post-quench
Hamiltonian is identical to the pre-quench Hamiltonian then $G_1(R,t) = 0$, $\forall (R,t) \in \mathbb{Z}\times \mathbb{R}^{+}$ meaning that the dynamics 
is suppressed as expected. \\

\begin{figure}[h!]
\centering
\includegraphics[scale = 0.35]{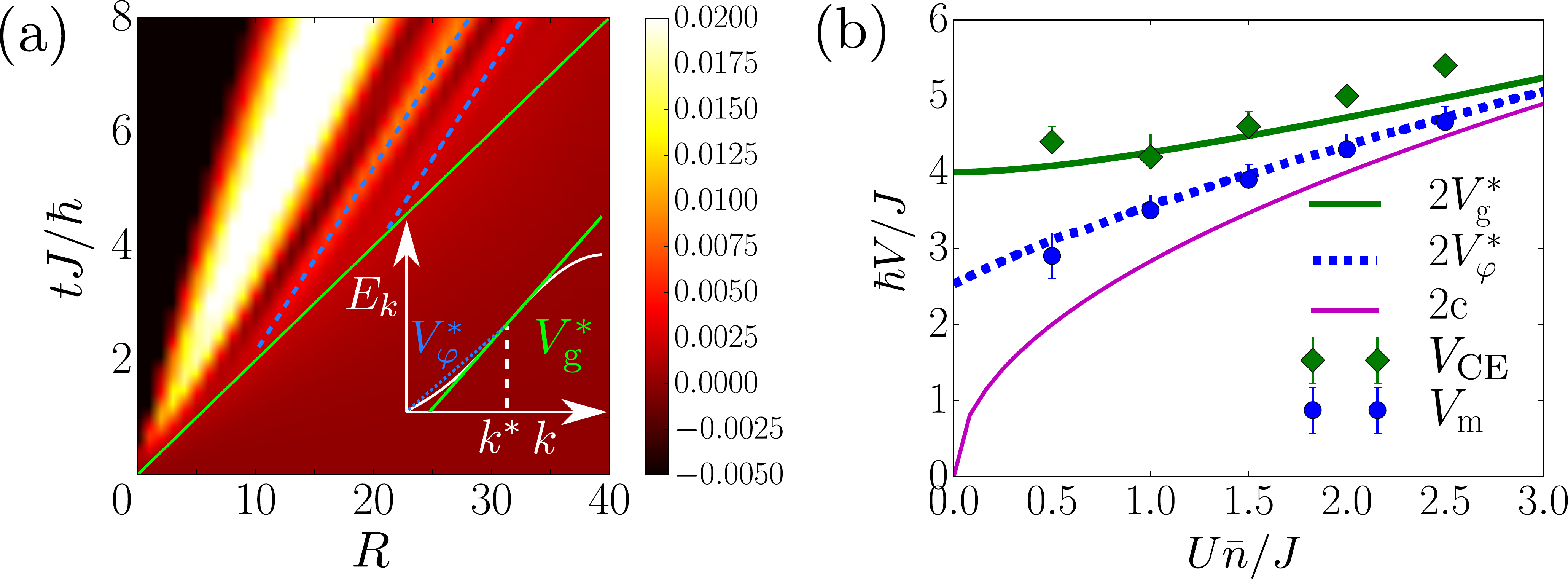}
\caption{\label{fig:BHm}
Spreading of the connected one-body correlation function $G_1(R,t)=\langle \hat{a}^\dag_R (t) \hat{a}_0 (t) \rangle - 
\langle \hat{a}^\dag_R (0) \hat{a}_0 (0) \rangle$ for the 1D Bose-Hubbard model in the mean field regime of the SF phase.
(a)~Analytical result, see Eq.~\eqref{g1_analytical} and \eqref{F1}, for a global quench from the initial value
$U_\mathrm{i}\bar{n}=J$ to the final value $U_\mathrm{f}\bar{n}=0.5J$.
(b)~ Comparison between twice the maximum group velocity ($2V_{\mathrm{g}}^*$, solid green line), twice the corresponding phase velocity ($2V_\varphi^*$, dashed blue line), 
twice the sound velocity ($2c$, solid purple line), and the spreading velocities obtained \textit{via} linear fits to the correlation edge  
($V_{\mathrm{CE}}$, green diamonds) and to several maxima ($V_\mathrm{m}$, blue disks) with the same initial value $U_{\mathrm{i}}\bar{n}/J$ as for (a). For the specific value
$U_{\mathrm{f}}\bar{n}/J = 1$, we considered a different initial ratio $U_{\mathrm{i}}\bar{n}/J$. The inset of (a) represents a sketch of the excitation spectrum
$E_k$ (in the SF mean field regime) as a function of the quasimomentum $k$ showing an inflexion point at $k^*$. The group velocity together with the phase velocity 
at the inflexion point are shown (see solid green and dashed blue line respectively) and denoted by $V_{\mathrm{g}}^*$ and $V_{\varphi}^*$. Figure extracted from Ref.~\cite{cevolani2018}.}
\end{figure}

On Fig.~\ref{fig:BHm}(a), the analytical connected correlation function $G_1$ is displayed as a function of the 
rescaled time $tJ/\hbar$ and distance $R$ for the Bose-Hubbard chain in the SF mean field regime. In order to put the bosonic chain out of equilibrium, a sudden
global quench is considered. Indeed, at $t=0$, while the hopping amplitude $J$ and the filling $\bar{n}$ are fixed, the repulsive two-body interaction is suddenly
modified such that $U_{\mathrm{i}}/2J\bar{n},U_{\mathrm{f}}/2J\bar{n} \ll 1$. For instance, on Fig.~\ref{fig:BHm}(a), the global quench is characterized by
the following pre- and post-quench ratios $U_{\mathrm{i}}\bar{n} = J$ and $U_{\mathrm{f}}\bar{n} = 0.5J$. Note that in this mean field regime, the ratio $U\bar{n}/J$
is the single relevant parameter, see Eq.~\eqref{g1_analytical} and \eqref{exc_spec_meanfield}. Figure~\ref{fig:BHm}(a) clearly shows a twofold structure for the
correlations, characterized by two different velocities.\\

On the one hand, a series of parallel extrema move along straight lines corresponding to a constant propagation velocity $V_{\mathrm{m}}$. The latter is found using
a linear fit, as shown by the dashed blue lines on Fig.~\ref{fig:BHm}(a). The values of $V_{\mathrm{m}}$ associated to different global quenches,
by varying the post-quench interaction parameter $U_{\mathrm{f}}\bar{n}/J = U\bar{n}/J$, confined in the SF mean field regime are reported on Fig.~\ref{fig:BHm}(b)
and symbolized by the blue disks. They are compared to twice the phase velocity at the quasimomentum $k^*$ where the group velocity is maxima, \textit{ie.} $2V_{\varphi}^*$,
see dashed blue line. According to our quasiparticle theory for the correlation spreading, this velocity is expected to characterize the spreading of
the inner structure (local extrema). This statement is confirmed quantitatively on Fig.~\ref{fig:BHm}(b) where $V_{\mathrm{m}} \simeq 2V_{\varphi}^*$.  \\

On the other hand, the various local extrema start at different activation times $t^*(R)$. The latter are aligned along a straight line with a different slope, as shown
by the solid green line on Fig.~\ref{fig:BHm}(a). This line characterizes the outer structure [correlation edge, (CE)] beyond which the correlations are 
exponentially suppressed. In other words, the latter permits to separate the causal and non-causal regions of correlations. However, the CE is defined by a different velocity
$V_{\mathrm{CE}}$ which is not related to the one for the series of local extrema. On Fig.~\ref{fig:BHm}(b), $V_{\mathrm{CE}}$ is reported for the same global quenches 
than those considered previously to study the spreading of the inner structure, see green diamonds. They are compared to twice the maximal group velocity, \textit{ie.}
$2V_{\mathrm{g}}^*$, see solid green line on Fig.~\ref{fig:BHm}(b), which is expected to characterize the motion of the CE according to our theory. Both velocities are
found to be in very good agreement, where the maximal relative error is less than $10\%$. Note that the sound velocity $c$ is also represented on
Fig.~\ref{fig:BHm}(b) by the solid purple line. This permits to conclude that the latter is irrelevant to characterize both the spreading of 
the inner and outer structures. Hence, an effective low-energy theory \textit{e.g.} a Luttinger-liquid approach can not explain the twofold structure for the space-time
correlations. 

\subsubsection{The density fluctuations - $G_2$ correlation function}

We now turn to another physical quantity \textit{ie.} the density fluctuations \textit{via} the study of the $G_2$ connected correlation function. This quantity has been
widely investigated for the 1D SRBH model both theoretically and experimentally but also in other bosonic and fermionic lattice models, see for instance
Refs.~\cite{carleo2014,cheneau2012,barmettler2012,cevolani2018,cevolani2015,despres2019}. The space-time behavior of a second physical quantity is investigated in order to 
confirm that both spreading velocities $V_{\mathrm{CE}}$ and $V_{\mathrm{m}}$ for the CE and the series of local extrema respectively are observable-independent.\\

Similarly to the $G_1$ correlation function to study the phase fluctuations, the $G_2$ density correlations can also be derived analytically in the SF mean field regime
and reads as (see Appendix.~\ref{appendix_g2_sf}),

\begin{align}
& G_2(R,t) \sim -\int_{\mathcal{B}} \frac{\mathrm{d}k}{2\pi} \mathcal{F}_2(k) \left\{ \frac{e^{i(kR+2E_{k,\mathrm{f}}t)} + e^{i(kR - 2E_{k,\mathrm{f}}t)}}{2} \right\},
\label{g2_analytical}
\end{align}

\noindent
with the amplitude function $\mathcal{F}_2$ given by

\begin{align}
& \mathcal{F}_2(k) \simeq \frac{ \bar{n}^2 \gamma_{k}(U_{\mathrm{i}} - U_{\mathrm{f}})}
{E_{k,\mathrm{i}} (\gamma_{k}+ 2 \bar{n} U_{\mathrm{f}})}.
\label{F2}
\end{align}

Similarly to $G_1$, the analytical expression of the density fluctuations $G_2$ at Eq.~\eqref{g2_analytical} fulfills the generic form, see Eq.~\eqref{generic_form}.
Therefore, we expect to find a twofold structure for $G_2$ characterized by a correlation edge and a series of local extrema. Besides, according to our correlation
spreading theory, we predicted that the spreading velocities of the CE and the extrema do not depend on the observable (provided that the generic form is fulfilled).
Both statements are verified below relying on the previous analytical form of $G_2$. \\

\begin{figure}[h!]
\centering
\includegraphics[scale = 0.35]{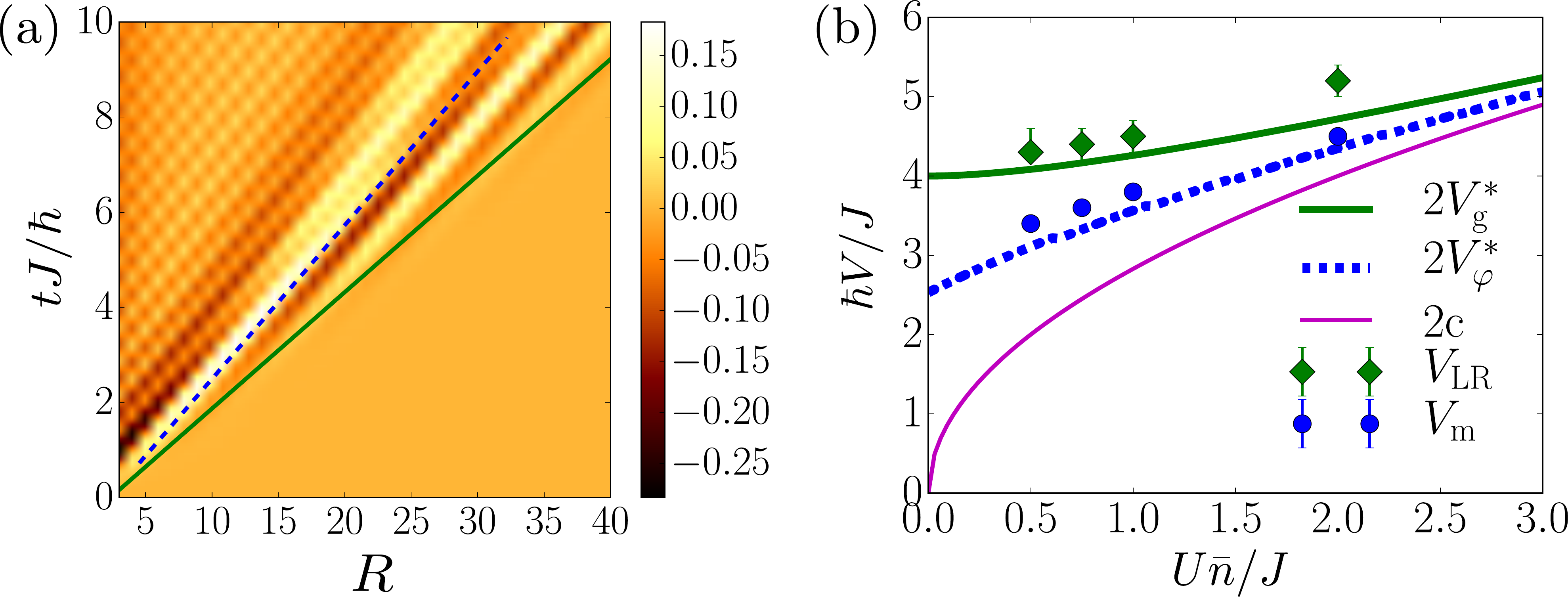}
\caption{\label{fig:BHm2}
Spreading of the connected density-density correlation function $G_2(R,t)=g_2(R,t)-g_2(R,0)$ with $g_2(R,t)=\langle \hat{n}_R(t) \hat{n}_0(t) \rangle
- \langle \hat{n}_R(t) \rangle \langle \hat{n}_0(t) \rangle$ for the 1D Bose-Hubbard model in the superfluid mean field regime.
(a)~Analytical result, see Eq.~\eqref{g2_analytical} and \eqref{F2}, for a global quench from the initial value $U_\mathrm{i}\bar{n}=J$
to the final value $U_\mathrm{f}\bar{n}=0.5J$.
(b)~Comparison between twice the maximum group velocity ($2V_{\mathrm{g}}^*$, solid green line), twice the corresponding phase velocity ($2V_\varphi^*$,
dashed blue line), twice the sound velocity ($2c$, solid purple line) and the spreading velocities obtained \textit{via} linear fits to the correlation edge  
($V_{\mathrm{CE}}$, green diamonds) and to several maxima ($V_\mathrm{m}$, blue disks) with the same initial value $U_{\mathrm{i}}\bar{n}/J$ as for (a).}
\end{figure}

On Fig.~\ref{fig:BHm2}(a), we show the analytical $G_2$ density correlations as a function of the rescaled time $tJ/\hbar$ and distance $R$. The global quench
(confined in the SF mean field regime of the BH chain) is the same as the one considered for $G_1$ at Fig.~\ref{fig:BHm}(a). As expected, $G_2$ displays a double
structure \textit{ie.} a CE and a series of local extrema characterized by two distinct velocities. Note that the double structure of $G_2$ is much clearer compared to the one 
for $G_1$. This effect, discussed in the next chapter, can be attributed to the relatively strong long-range correlations present at equilibrium for $G_1$ compared to
those for $G_2$. \\
As previously for $G_1$, we use linear fits to determine the values of the spreading velocities $V_{\mathrm{CE}}$ and $V_{\mathrm{m}}$ for the CE and the
local extrema respectively. The resulting data are displayed on Fig.~\ref{fig:BHm2}(b) and the values of $V_{\mathrm{CE}}$ ($V_{\mathrm{m}}$) are represented by the
green diamonds (blue disks). The fitted velocities are compared, as previously, to three characteristic velocities determined by the analytical post-quench excitation 
spectrum in the SF mean field regime, \textit{ie.} $2V_{\mathrm{g}}^*$ [twice the (maximal) group velocity at the quasimomentum $k^*$], $2V_\varphi^*$ (twice the phase velocity at $k^*$) and $2c$ 
(twice the sound velocity). As predicted and confirmed at Fig.~\ref{fig:BHm2}, the CE moves at $2V_{\mathrm{g}}^*$ and the local extrema propagate at $2V_{\varphi}^*$
whereas the velocity $2c$ is irrelevant. \\
Finally, the spreading velocities of each structure are similar for $G_1$ and $G_2$ according to Figs.~\ref{fig:BHm}(b) and \ref{fig:BHm2}(b). Indeed, the CE and the local
extrema move at $V_{\mathrm{CE}} \simeq 2V_{\mathrm{g}}^*$ and $V_{\mathrm{m}} \simeq 2V_{\varphi}^*$ respectively. The latter allows us to certify that the spreading
velocities for the twofold structure are observable-independent and fully characterized by the post-quench excitation spectrum. 

\subsubsection{Numerical simulations of the correlation spreading in 1D short-range lattice models} 
The distinction between the CE propagating at the velocity $2V_{\mathrm{g}}^*$ and the extrema emanating from it and propagating at a different velocity 
given by $2V_{\varphi}^*$ allows to understand several unexplained observations. In this paragraph, we shed new light on numerical results in the litterature
concerning the correlation spreading in 1D short-range lattice models. \\

\begin{figure}[h!]
\centering
\includegraphics[scale = 0.4]{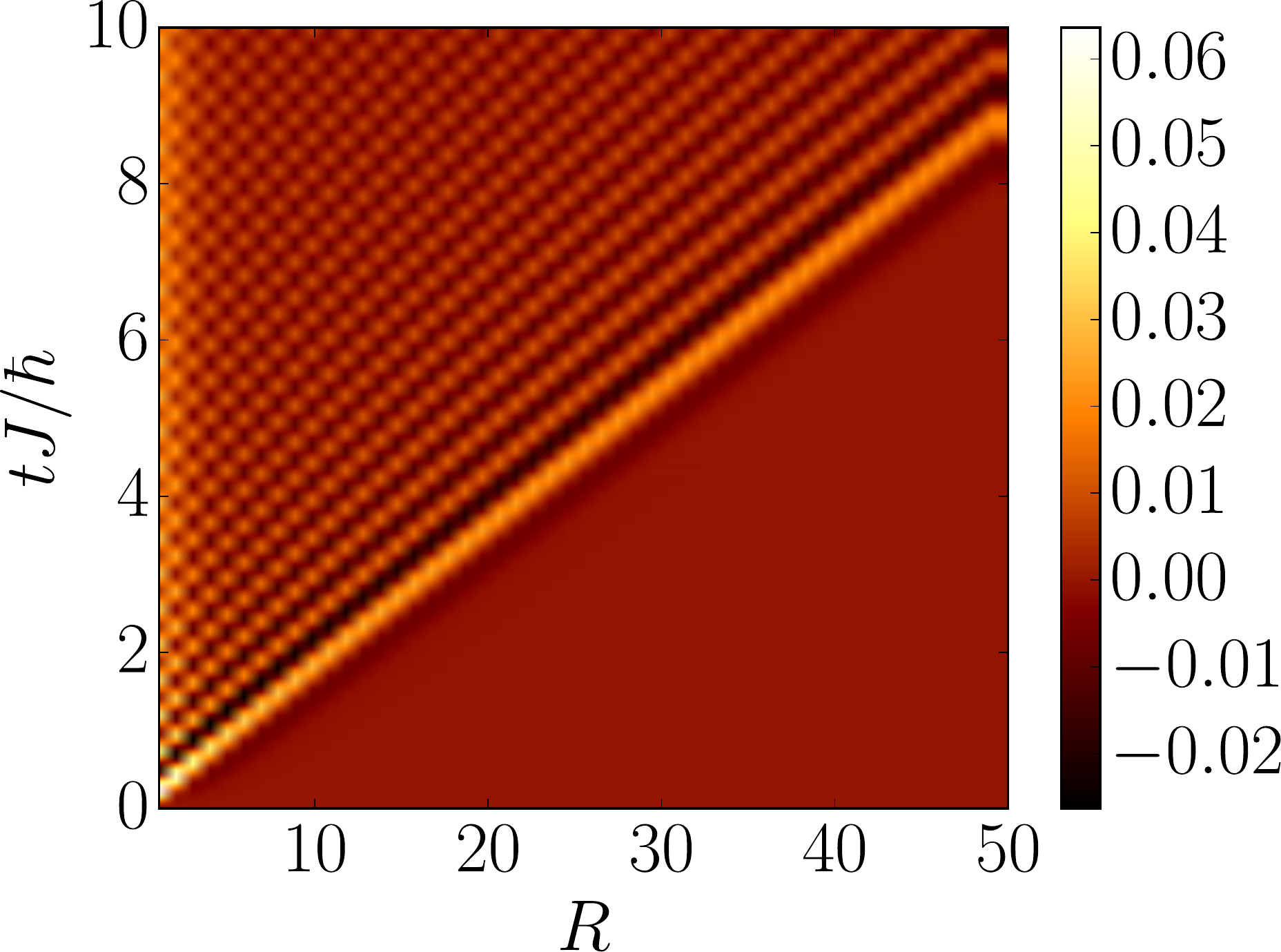}
\caption{\label{fig:check_carleo_by_julien}
Spreading of the connected density-density correlation function $G_2(R,t)=g_2(R,t)-g_2(R,0)$ with $g_2(R,t)=\langle \hat{n}_R(t) \hat{n}_0(t) \rangle
- \langle \hat{n}_R(t) \rangle \langle \hat{n}_0(t) \rangle$ for the 1D Bose-Hubbard model in the superfluid phase, see Eq.~\eqref{g2_analytical} and Eq.\eqref{F2}.
A global quench from the initial value $U_\mathrm{i}\bar{n}=2J$ to the final value $U_\mathrm{f}\bar{n}=4J$ is considered. The latter is in excellent quantitative agreement
with Fig.~1(a) of Ref.~\cite{carleo2014}.}
\end{figure}

In Ref.~\cite{carleo2014}, the 1D short-range Bose-Hubbard model is studied (but also the 2D case). A special care is devoted to the $G_2$ density correlations 
using the time-dependent Variational Monte-Carlo technique (t-VMC) on relevant time and distance scales. On Fig.~1(a) of Ref.~\cite{carleo2014}, 
a single structure \textit{ie.} $V_{\mathrm{m}} \simeq V_{\mathrm{CE}}$ is observed for the correlation cone. The latter is due to the large post-quench
two-body interaction parameter $U_\mathrm{f}=4$. Indeed, for large ratio $U\bar{n}/J$, the analytical spreading velocities of the CE ($2V_{\mathrm{g}}^*$)
and local extrema ($2V_{\varphi}^*$) are almost equal, see Fig.~\ref{fig:BHm}(b). Besides, the correlation edge velocity extracted from the t-VMC calculations
quantitatively agrees with the value $2 V_{\varphi}^*$ calculated previously at Fig.~\ref{fig:BHm2}(b). Our analysis shows that it should thus be assimilated 
to the spreading of the local extrema of the causal region of correlations and not to the one of the correlation edge. In contrast, the CE is determined by
the raise of the envelop of these extrema, and moves at the velocity $2V_{\mathrm{g}}^*$. Finally, we stress that the t-VMC result of $G_2$ at Fig.~1(a) is in very 
good (quantitative) agreement with our analytical expression at Eq.~\eqref{g2_analytical} and represented on Fig.~\ref{fig:check_carleo_by_julien}. Hence, the latter
cross-validates the t-VMC calculations performed on relevant time and distance scales on the one hand and our quasiparticle picture on the other hand. \\

In Ref.~\cite{manmana2009}, the 1D Fermi-Hubbard model is considered where the $G_2$ density correlations are investigated numerically using time-dependent Density
Matrix Renormalization Group (t-DMRG) simulations.
A similar analysis applies, for global quenches in both the SF and CDW (Charge-Density-Wave insulating) regimes of this model, where a twofold spike-like structure 
appears. However, one needs to be prudent since our quasiparticle picture deals only with bosonic lattice models. Nevertheless, we expect that the generic
form of the correlations at Eq.~\eqref{generic_corr} also holds for fermionic lattice models. 

\subsection{Case study of the Bose-Hubbard chain: Mott-in\-su\-la\-ting phase}
\label{mi_chap3}
We now turn to the Mott-Insulating phase (MI) of the 1D short-range Bose-Hubbard model implying that the filling $\bar{n}$ is integer, \textit{ie.} $\bar{n} \in \mathbb{N}^*$.
This quantum phase is characterized by a finite gap in its excitation spectrum contrary to the SF phase. In the following, the correlation spreading is investigated 
in the strong-coupling regime  implying $U \gg J\bar{n}$. To deduce the excitation spectrum of this specific regime, the Hamiltonian can be diagonalized 
using strong-coupling expansions, see Refs.~\cite{altman2002,huber2007,barmettler2012}. In the following, we give more details about the strong-coupling 
perturbation theory working along the lines of Ref.~\cite{barmettler2012}. \\

The low-energy excitations of the MI phase consist of removing one bosonic particle from a lattice site and to put it on another one of the chain.
It yields a so-called doublon-holon excitation pair where the holon (doublon) denotes $\bar{n}-1$ ($\bar{n}+1$) particles on a lattice site.
The main idea of the first-order perturbation theory applied to MI strong-coupling regime  is to consider one doublon-holon excitation pair of energy $U$ and then 
treating the hopping operator $\hat{H}_{\mathrm{hop}}$, defined below, as a perturbative term. Firstly, one starts by rewriting the Hamiltonian of the 
Bose-Hubbard chain at Eq.~\eqref{H_bhm} under the following form

\begin{equation}
 \hat{H} = \hat{H}_\mathrm{hop} + \hat{H}_{\mathrm{int}},~~~ \hat{H}_{\mathrm{int}} = \frac{U}{2} \sum_R \hat{n}_R (\hat{n}_R - 1), 
 ~~~ \hat{H}_{\mathrm{hop}} = -J \sum_R (\hat{a}^{\dag}_R \hat{a}_{R+1} + \mathrm{h.c}).
\end{equation}

\noindent
The ground state of the Bose-Hubbard chain in the limit regime $U/J \rightarrow +\infty$ is given by the pure Mott state defined as follows

\begin{equation}
 \ket{\mathrm{GS}^{(0)}} = \ket{\bar{n}}^{\otimes N_s} \equiv \ket{\bar{n}}.
\end{equation}

\noindent
Then, using the first-order perturbation theory, it leads for the perturbed ground state $\ket{\mathrm{GS}^{(1)}}$ to

\begin{equation}
 \ket{\mathrm{GS}^{(1)}} \simeq \ket{\bar{n}} + \sum_{R,R',R' \neq 0} \frac{\langle \phi_{R,R'}^{(0)}| \hat{H}_{\mathrm{hop}} | \mathrm{GS}^{(0)} \rangle}{E_0 - E_{R,R'}}\ket{\phi_{R,R'}^{(0)}}, 
 \label{gs_1_to_compute}
\end{equation}

\noindent
where $E_0=0$ is the energy associated to the pure Mott state $\ket{\mathrm{GS}^{(0)}}$. $\ket{\phi^{(0)}_{R,R'}} =
\left[\bar{n}(\bar{n}+1)\right]^{-1/2} \hat{a}_{R+R'} \hat{a}^{\dag}_{R} \ket{\bar{n}}$ with $R' \neq 0$ represents the many-body quantum state 
with a doublon-holon pair present on the lattice chain. The condition $R' \neq 0$ enforces that one cannot remove and put the bosonic particle on
the same lattice. Indeed, in the opposite case, the doublon-holon pair is not created. Finally, the prefactor $\left[\bar{n}(\bar{n}+1)\right]^{-1/2}$ comes from the
normalization of the quantum state $\ket{\phi^{(0)}_{R,R'}}$ ($\sqrt{\bar{n}}$, $\sqrt{\bar{n}+1}$ comes from the creation of a holon and doublon respectively).
Performing the calculation starting from Eq.~\eqref{gs_1_to_compute} leads to the following expression for $\ket{\mathrm{GS}^{(1)}}$

\begin{equation}
\ket{\mathrm{GS}^{(1)}} \simeq \ket{\bar{n}}  + \frac{J}{U} \sqrt{\bar{n}(\bar{n}+1)} \sum_{R} \left( \ket{\phi_{R,1}^{(0)} } + \ket{\phi_{R,-1}^{(0)}} \right).
\label{pert_gs}
\end{equation}

\noindent
The equation \eqref{pert_gs} describes the perturbed ground state $\ket{\mathrm{GS}^{(1)}}$ as a superposition of the pure MI state $\ket{\mathrm{GS}^{(0)}} \equiv 
\ket{\bar{n}}$ and the degenerated quantum states $\ket{\phi_{R,\pm 1}^{(0)}}$ which consist of creating a bosonic particle at the lattice site $R$ and to remove one
on the nearest lattice site $R\pm 1$. The previous statement is due to the hopping term $\hat{H}_{\mathrm{hop}}$ where only a tunnelling between the nearest-neighbor
lattice sites is allowed. \\

We now turn to the derivation of the excitation spectrum for the MI strong-coupling regime. To do so, a degenerate pertubation theory needs to be considered. At first order
in $J/U$, the latter consists of diagonalizing the Bose-Hubbard Hamiltonian reduced to the subspace containing all the many-body quantum states having one doublon-holon 
excitation pair.
Due to the lattice translational symmetry, the perturbative term $\hat{H}_{\mathrm{hop}}$ is diagonalized in the Fourier space \textit{via} the eigenbasis
$\{ \ket{\phi^{(0)}_k}\}$. The eigenstate $\ket{\phi^{(0)}_k}$ at quasimomentum $k$ is given by the following form

\begin{equation}
\ket{\phi_{k}^{(0)}} = \frac{\sqrt{2}}{N_s} \sum_{R,R'} \sin\left( kR' \right) \ket{\phi_{R,R'}^{(0)}}, 
\end{equation}

\noindent
and is built from a superposition of the many-body quantum states $ \ket{\phi_{R,R'}^{(0)}}$ denoting the states with one doublon-holon excitation pair (the 
sum over $R'$ is performed over all the lattice sites since the term $\sin(kR')$ implies that the contribution of the non-physical term $R'=0$ vanishes).
Its corresponding eigenvalue is found by 
calculating the energy $\langle \phi_k^{(0)} | \hat{H}_{\mathrm{hop}} | \phi_{k}^{(0)} \rangle$ leading to 

\begin{equation}
\langle \phi_k^{(0)} | \hat{H}_{\mathrm{hop}} | \phi_{k}^{(0)} \rangle =  -2J (2\bar{n}+1) \cos(k).
\end{equation}

\noindent
The last step consists of calculating the contribution in energy coming from the local interaction term $\hat{H}_{\mathrm{int}}$ which gives
$\langle \phi_k^{(0)} | \hat{H}_{\mathrm{int}} | \phi_{k}^{(0)} \rangle = U$. The latter is straightforward since the considered subspace is built
from a superposition of all the quantum states having one doublon-holon excitation pair. Finally, the excitation spectrum $E_k$ reads as

\begin{align}
 & 2E_k = U - 2J(2\bar{n}+1)\cos(k).
 \label{Ek_MI}
\end{align}

\begin{figure}[h!]
\centering
\includegraphics[scale = 0.38]{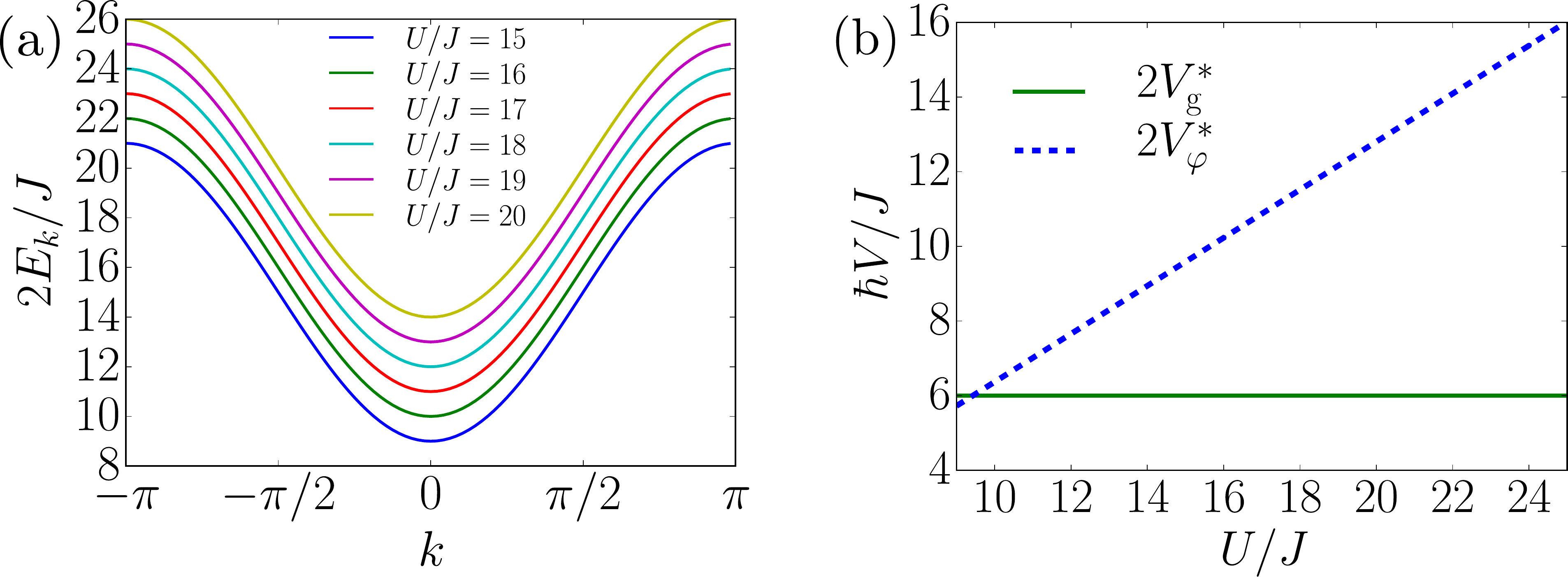}
\caption{\label{mi_exc_vel}
Excitation spectrum properties of the 1D short-range Bose-Hubbard model in the strong-coupling limit of the Mott-insulating phase at unit-filling $\bar{n}=1$. 
(a)~Twice the excitation spectrum rescaled by the hopping amplitude $2E_k/J$ as a function of the quasimomentum $k$ in the first Brillouin zone
$\mathcal{B} = [-\pi,\pi]$ for different interaction parameters $U/J$. (b)~ Dimensionless physical quantity $\hbar V/J$ as a function of $U/J$.
$V$ denotes two distinct velocities namely twice the maximal group velocity $2V_{\mathrm{g}}^*$ and twice the corresponding phase velocity at $k^*$ 
denoted by $2V_{\varphi}^*$.}
\end{figure}

On Fig.~\ref{mi_exc_vel}(a), the excitation spectrum $2E_k$, rescaled by the hopping amplitude $J$, is displayed as a function of the
quasimomentum $k$ for different ratios $U/J$. In the Mott-insulating phase, only the interaction parameter $U/J$ is relevant, see Eq.~\eqref{Ek_MI}.
In the MI strong-coupling regime , $2E_k$ develops a finite gap as expected and determined by $\Delta = U - 2J(2\bar{n}+1)$. \\

In the following, we analyze briefly the group ($V_{\mathrm{g}}$) and phase ($V_{\varphi}$) velocities. Note that contrary to the 
SF phase, the MI phase is incompressible. Therefore, the sound velocity $c$ defined as $c = \lim_{k\rightarrow 0} V_{\mathrm{g}}(k)$ is meaningless.
The analysis of the characteristic velocities will be particularly helpful to characterize the (possible at this stage) twofold structure for the $G_1$
phase fluctuations calculated analytically in the following. According to Eq.~\eqref{Ek_MI}, the analytical expression of the group and phase velocities
are given by

\begin{align}
& V_{\mathrm{g}}(k) = (2\bar{n}+1)J\sin(k) \\ \nonumber \\
& V_{\varphi}(k) = k^{-1} \left[U/2-J(2\bar{n}+1)\cos(k)\right]
\end{align}

\noindent
On Fig.~\ref{mi_exc_vel}(b), the dimensionless quantity $\hbar V/J$ is represented as a function of the interaction parameter $U/J$. $V$ denotes 
two different parameters namely twice the maximal group velocity ($2V_\mathrm{g}^* = 2V_\mathrm{g}(k^*)$) and twice the phase velocity at $k^*$ 
($2V_{\varphi}^* = 2V_{\varphi}(k^*)$). 
In the case where the analytical $G_1$ correlation function displays a twofold structure, the latter are expected to correspond to the spreading velocities of the CE and
the series of local extrema respectively. According to Fig.~\ref{mi_exc_vel}(b) and Eq.~\eqref{Ek_MI}, the maximum of the group velocity $V_\mathrm{g}^*$ 
is at the center of the first Brillouin zone \textit{ie.} $k^* = \pi/2$, where the group and phase velocities are $V_{\mathrm{g}}^*= (2\bar{n}+1)J$ and
$V_\varphi^* = U/\pi$, respectively. Contrary to the SF mean field regime, the phase velocity can exceed the group velocity at $k^*$. The condition is given by 
$U/J>\pi(2\bar{n}+1) > u_c \simeq 3.3$ where $u_c = (U/J)_c^{\bar{n}=1}$ denotes the critical point of the SF-MI phase transition at unit-filling ($\bar{n}=1$),
see Fig.~\ref{mi_exc_vel}(b).

\subsubsection{Sudden global quenches in the strong-coupling limit}
In the following, we verify the theoretical spreading velocities of the twofold causality cone for a sudden global quench confined in
a gapped quantum phase. Previously, a similar analysis has been performed for the case of a gapless phase where the phase $G_1$ and density 
$G_2$ fluctuations in the (gapless) SF mean field regime of the Bose-Hubbard chain were investigated. We have shown that both correlation functions (i) fulfill the generic 
form presented and analyzed at Eq.~\eqref{generic_form} (ii) display a twofold structure whose spreading velocities $V_{\mathrm{CE}}$
(for the correlation edge) and $V_{\mathrm{m}}$ (for the series of local maxima) are determined by $2V_{\mathrm{g}}^*$ and $2V_{\varphi}^*$ respectively. 
Here, while still considering the Bose-Hubbard chain, we turn to the (gapped) Mott-insulating phase in the strong-coupling regime discussed in the 
previous paragraph. Once again, the BH chain is put out-of-equilibrium \textit{via} sudden global quenches confined in a same phase and regime. To investigate the
strong-coupling regime, it implies for the pre- and post-quench interaction parameters to fulfill $(U/J)_\mathrm{i}, (U/J)_{\mathrm{f}} \gg 1$. The filling 
$\bar{n}$ is integer in the MI phase and is fixed to unity during the full real time evolution. Only the phase fluctuations $G_1$ will be investigated.
The density fluctuations $G_2$, presented in details in the next chapter at Sec.~\ref{limit_regimes}, are a pathological case. Indeed, the corresponding analytical
expression is slightly different from the generic form at Eq.~\eqref{generic_form}. This slight difference leads for $G_2$ to a single structure which
can still be explained by our quasiparticle theory. 

\subsubsection{The phase fluctuations - $G_1$ correlation function}
Similarly to the case of the superfluid mean field regime, the spreading of the phase fluctuations is studied \textit{via} the $G_1$ connected
correlation function which reads as

\begin{align}
& G_1(R,t) = \langle \hat{a}^{\dag}_R(t) \hat{a}_0(t) \rangle - \langle \hat{a}^{\dag}_R(0) \hat{a}_0(0) \rangle.
\end{align}

\noindent
Using the time-dependent perturbation theory, see Ref.~\cite{barmettler2012} and Appendix.~\ref{appendix_g1_mi}, the $G_1$ connected correlation
function can be calculated analytically for global quenches confined in the strong-coupling regime. For the specific case of global quenches defined by 
$U_\mathrm{i} \rightarrow +\infty$ and $U_\mathrm{f} \gg J\bar{n}$, its analytical expression can be expressed in the generic form of Eq.~\eqref{generic_form}, 

\begin{align}
& G_1(R,t) \sim -\int_{\mathcal{B}} \frac{\mathrm{d}k}{2\pi} \mathcal{F}_1(k) \left\{ \frac{e^{i(kR-2E_{k,\mathrm{f}}t)} + e^{i(kR+2E_{k,\mathrm{f}}t)}}{2}
\right \},
\label{g1_analytical_mi}
\end{align}

\noindent
where $E_{k,\mathrm{f}}$ denotes the post-quench excitation spectrum valid in the strong-coupling limit, see Eq.~\eqref{Ek_MI} and
$\mathcal{F}_1$ the quasimomentum-dependent amplitude function defined as 

\begin{equation}
\mathcal{F}_1(k) = \frac{4J}{iU_\mathrm{f}}\bar{n}(\bar{n}+1)\sin(k).
\label{F1_mi}
\end{equation}

\begin{figure}[h!]
\centering
\includegraphics[scale = 0.37]{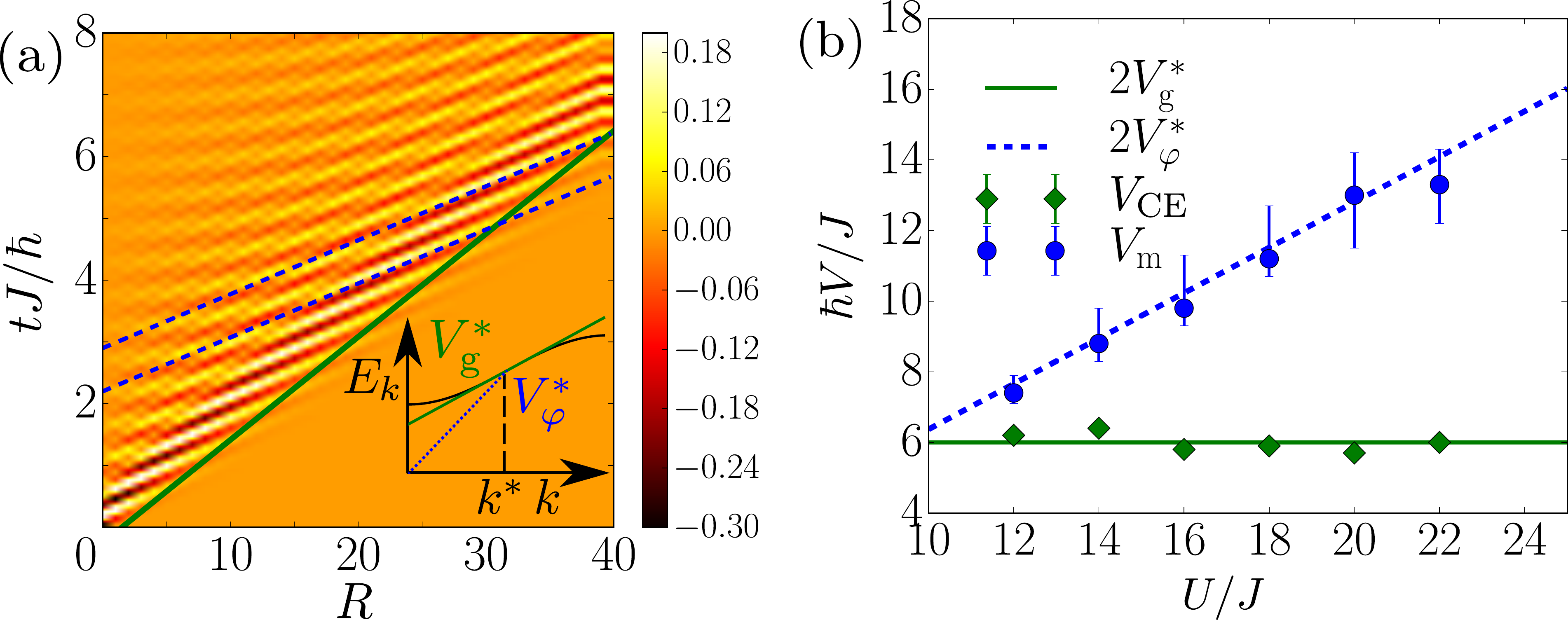}
\caption{\label{fig:BHm3}
Spreading of the connected one-body correlation function $G_1(R,t)=\langle \hat{a}^\dag_R (t) \hat{a}_0 (t) \rangle - 
\langle \hat{a}^\dag_R (0) \hat{a}_0 (0) \rangle$ for the 1D Bose-Hubbard model in the strong-coupling limit of the Mott-insulating phase at 
unit-filling $\bar{n}=1$.
(a)~Analytical result for a global quench from the initial value $U_\mathrm{i} \rightarrow +\infty$, \textit{ie.} from the pure Mott state, 
to the final value $U_\mathrm{f} = 18J$.
(b)~Comparison between twice the maximum group velocity ($2V_{\mathrm{g}}^*$, solid green line), twice the corresponding phase velocity at $k^*$ ($2V_\varphi^*$, 
dashed blue line) and fits to the correlation edge velocity ($V_{\mathrm{CE}}$, green diamonds) and to the velocity of the maxima 
($V_\mathrm{m}$, blue disks) with the same initial values as for (a). The inset of (a) represents a sketch of the excitation spectrum
$E_k$ (in the MI strong-coupling regime) as a function of the quasimomentum $k$ showing an inflexion point at $k^*$. The group velocity
together with the phase velocity at the inflexion point are shown (see solid green and dashed blue line respectively) and denoted by 
$V_{\mathrm{g}}^*$ and $V_{\varphi}^*$. Figures extracted from Ref.~\cite{cevolani2018}.}
\end{figure}

On Fig.~\ref{fig:BHm3}(a), the space-time pattern of the $G_1$ connected correlation function is shown for a sudden global quench confined 
in the strong-coupling limit of the MI phase starting from a pure Mott state $U_\mathrm{i} \rightarrow + \infty$ to $U_\mathrm{f} = 18J$ at 
unit-filling $\bar{n}=1$. As expected, it displays a twofold structure with a CE (solid green line) and a series of local maxima (dashed blue lines).
Besides, contrary to $G_1$ in the SF mean field regime [see Fig.~\ref{fig:BHm}(a,b)], the correlation edge velocity is lower than the one for the extrema, 
\textit{ie.} $V_{\mathrm{CE}} < V_{\mathrm{m}}$. According to Fig.~\ref{fig:BHm3}(b), the latter coincide with twice the group and phase velocities 
at $k^*$ respectively, $V_{\mathrm{CE}} \simeq 2V_{\mathrm{g}}^*$ and $V_{\mathrm{m}} \simeq 2V_\varphi^*$. This is consistent with the form of the
excitation spectrum $2E_k$ in the strong-coupling limit, see Eq.~\eqref{Ek_MI}, displaying a finite gap $\Delta = U - 2J(2\bar{n}+1)$ allowing to reach
a regime where $V_\varphi^* > V_{\mathrm{g}}^*$. \\

\subsubsection{Experimental investigation of the correlation spreading in the Bose-Hubbard chain}
So far, an experimental characterization of the correlation spreading in the Mott-insulating phase 
at unit-filling $\bar{n}=1$ in a 1D optical lattice was only performed close to the critical point\footnote{Note that this
critical point depends on the integer filling $\bar{n}$ but also on the dimensionality of the lattice.} $(U/J)_{\mathrm{c}}^{\bar{n}=1} \simeq 3.3$
~\cite{kuhner2000,kashurnikov1996exact,ejima2011,rombouts2006} and for the two-point parity correlations, see Ref.~\cite{cheneau2012} for more details.
This observable is defined as $C(R,t) = \langle \hat{s}_0(t) \hat{s}_R(t) \rangle - \langle \hat{s}_0(t) \rangle \langle \hat{s}_R(t)
\rangle$. The operator $\hat{s}_R(t) = e^{i\pi\left[\hat{n}_R(t) - \bar{n} \right]}$ gives information on the presence or not of a quasiparticle on the lattice
$R$ at time $t$. Indeed, if a doublon ($\langle \hat{n}_R(t) \rangle = \bar{n}+1$) or holon ($\langle \hat{n}_R(t) \rangle = \bar{n}-1$) is present, then 
$\langle \hat{s}_R(t) \rangle = -1$. In the case where no quasiparticle is detected ($\langle \hat{n}_R(t) \rangle = \bar{n}$), $\langle \hat{s}_R(t) \rangle = 1$.
In the regime of moderate interaction parameters $U/J \gtrsim (U/J)_\mathrm{c}^{\bar{n}=1}$, both characteristic velocities $2V_{\mathrm{g}}^*$ 
and $2V_{\varphi}^*$ are relatively close explaining that no inner structure was observed~\cite{cheneau2012}. Nevertheless, in the next chapter at Sec.~\ref{BKT_ph_tr}
relying on a numerically-exact tensor-network technique, we will show that a twofold structure, with the expected spreading velocities, should be present
in this case even if both characteristic velocities are relatively close. \\

To sum up, the space-time pattern of a generic correlation function for short-range interacting lattice models has been fully characterized
\textit{via} a quasiparticle approach. For the latter, we unveil a twofold spike-like structure defined by two distinct velocities. These spreading velocities
are completely determined by the post-quench excitation spectrum. More precisely, we have shown that the correlation edge moves at $V_{\mathrm{CE}} \simeq 
2V_{\mathrm{g}}^*$ (twice the maximal group velocity) whereas the series of local extrema propagate at $V_{\mathrm{m}} \simeq 2V_\varphi^*$. However, the latter
can be equal in some specific cases, $V_{\mathrm{m}} \simeq V_{\mathrm{CE}}$, leading to a space-time pattern where the two structures are not distinguishable. 
Furthermore, the previous
spreading velocities do not depend on the observable. It only affects the amplitude of the correlations \textit{via} the amplitude 
function $\mathcal{F}$ in our generic form. Indeed, the latter depends on the local observables in the expression of the correlation function.
Finally, considering global quenches confined in a same gapped or gapless phase, it has been found that the spreading velocities $V_{\mathrm{CE}}$
and $V_{\mathrm{m}}$ are always characterized by $2V_{\mathrm{g}}^*$ and $2V_{\varphi}^*$ as expected. The previous statement was
verified analytically for a specific Hamiltonian \textit{ie.} the Bose-Hubbard chain by investigating the phase $G_1$ and density $G_2$ fluctuations in both
the gapless SF and gapped MI phases. \\
However, the previous discussion on the scaling laws, \textit{ie} spreading velocities, of the twofold structure for the causality cone is only valid in the
context of short-range interacting lattice models. In the following, we discuss the counterpart for
long-range interactions \textit{ie.} power-law decaying interactions while still relying on a quasiparticle approach.   

\section{Lattice models with long-range couplings}
\label{LRC}
As previously for short-range interacting lattice models, we start by presenting the scaling laws for the 
correlation spreading in long-range interacting lattice models. These scaling laws defining the twofold structure
of correlations will depend on the presence of a gap in the excitation spectrum. Therefore, in order to give a theoretical
confirmation of the latter, we will investigate two different long-range interacting $s=1/2$ spin
lattice models, namely the 1D Long-Range Transverse Ising (1D LRTI) model in the $z$ polarized phase which displays a
gapped excitation spectrum and the 1D Long-Range XY (1D LRXY) model in the gapless $x$ polarized phase.

\subsection{Scaling laws}
The long-range interactions are represented by a power-law decaying amplitude for the two-site coupling term in the generic Hamiltonian at Eq.~\eqref{generic_ham},
\textit{ie.} $J(\mathbf{R}, \mathbf{R}') \simeq J/|\mathbf{R}-\mathbf{R}'|^{\alpha}$ with $\alpha \in \mathbb{R}^{+*}$. This form preserves the translational
invariance of the lattice model and one can still rely on (i)~ the generic form presented at Eq.~\eqref{generic_corr} and (ii) its asymptotic form at Eq.~\eqref{spa}.
The excitation spectrum $E_k$ is assumed to be regular in the whole first Brillouin zone $\mathcal{B}$, except for a possible cusp at $k=0$. Consequently, it may
be written as $E_k \simeq \Delta + c k^z$ in the infrared limit $k \rightarrow 0$. $z$ denotes the $\alpha$-dependent dynamical exponent and $\Delta$ the possibly vanishing gap.
For a sudden global quench confined in a gapped (gapless) phase, it implies $\Delta \neq 0$ ($\Delta = 0$). Depending on the value of $\alpha$ the power-law exponent
characterizing the decay of the long-range interactions, one can enter in three different regimes \cite{hauke2013, cevolani2015, cevolani2016} presented below.

\begin{itemize}
\item The local regime defined by $z \geq 1$. Both the quasiparticle energy $E_k$ and the group velocity $V_{\mathrm{g}}(k) = \partial_k E_k$ are bounded within 
the first Brillouin zone. 
\item The quasi-local regime defined by $ 0  \leq z < 1$. While the energy is finite within the first Brillouin zone, the group velocity diverges at
the quasimomentum $k=0$. 
\item The instantaneous regime defined by $z < 0$. Both the energy $E_k$ and the group velocity diverge at the quasimomentum $k=0$.
\end{itemize}

\noindent
In the following, the scaling laws describing the quench dynamics is investigated for the first two regimes, namely the local and quasi-local regimes.
The latter are relevant since the excitation spectrum $E_k$ is bounded within the first Brillouin zone $\mathcal{B}$. Besides, the long-range
lattice model is put out of equilibrium \textit{via} sudden global quenches confined in a single phase, without crossing any critical line.

\begin{itemize}
\item We first consider the local regime implying $z \geq 1$. As discussed previously, both the quasiparticle energy $E_k$ and the group velocity $V_{\mathrm{g}}$ are bounded within $\mathcal{B}$.
Similarly to short-range lattice models, one can find a quasimomentum $k^*$ such that $V_{\mathrm{g}}^* = \mathrm{max}[V_{\mathrm{g}}(k)]$, see Eq.~\eqref{kstar}. Hence,
the generic connected correlation function has an asymptotic behavior given at Eq.~\eqref{spa} relying on the stationary phase approximation. Consequently, 
the scaling laws of the twofold structure for the space-time quantum correlations are the same as those for short-range interacting lattice models, \textit{ie.} ballistic implying $t\sim R$, as well as the spreading velocities.
It has been shown previously that both the CE and the series of local extrema spread ballistically with a velocity $V_\mathrm{CE} \simeq 2V_{\mathrm{g}}^*$ 
and $V_{\mathrm{m}} \simeq 2V_{\varphi}^*$ respectively for short-range interacting lattice models. 

\item Concerning the quasi-local regime $ 0  \leq z < 1$, the excitation spectrum $E_k$ can be written as $E_k \simeq \Delta + c |k|^z$ in
the infrared limit $k\rightarrow0$. Assuming $c>0$, the group velocity $V_{\mathrm{g}}(k) = cz|k|^{z-1}(|k|/k)$ diverges\footnote{One can also consider
$c<0$ implying $V_{\mathrm{g}}(k) \rightarrow +\infty$ for $k\rightarrow 0^{-}$. Performing a similar analysis as the one explained in what follows, the
scaling laws have been found to not depend on the sign of this prefactor $c$.}
to $+\infty$ for $k\rightarrow 0^+$. Relying once again on the stationary phase approximation, the equations \eqref{ksp} and \eqref{spa} are still valid. Since $V_{\mathrm{g}}$
is unbounded within $\mathcal{B}$, it exists a quasiparticle with the group velocity $2V_{\mathrm{g}}(k) = R/t$ for each combinaison of $R$ and $t$.
The latter is characterized by the corresponding quasimomentum $k_\mathrm{sp} = (2czt/R)^{1/1-z}$. The correlation edge (CE) is thus dominated by the
infrared divergence $k\rightarrow 0$ implying to evaluate the asymptotic behavior of $G(R,t)$, given at Eq.~\eqref{spa}, around $k_{\mathrm{sp}}$.
To do so, it is also necessary to analyze the amplitude function $\mathcal{F}$ in the infrared limit $k\rightarrow 0$. Assuming the scaling 
$\mathcal{F}(k) \sim |k|^{\nu}$ with $\nu \geq 0$, $G(R,t)$ may be written as follows 

\begin{equation}
 G(R,t) \propto \frac{t^{\gamma}}{R^{\chi}}\cos\left[A_z\left(\frac{t}{R^{z}}\right)^{\frac{1}{1-z}}-2\Delta t + \sigma D \frac{\pi}{4}\right],
 \label{spa_lr}
\end{equation}

\noindent
with the following parameters 

\begin{equation}
\gamma=\frac{\nu+D/2}{1-z}, 
\label{gamma}
\end{equation}

\begin{equation}
\chi=\frac{\nu+D(2-z)/2}{1-z} = \gamma + \frac{D}{2}, 
\label{chi}
\end{equation}

\begin{equation}
A_{z}=2c (1-z) (2cz)^{\frac{z}{1-z}},
\end{equation}

\begin{equation}
\sigma = \mathrm{sgn} \left(-2\partial^2_k E_{k_{\mathrm{sp}}}t \right) = \mathrm{sgn} \left[2cz(1-z) |k_{\mathrm{sp}}|^{z-2} t\right] = 1.  
\end{equation}


\noindent
Finally, the scaling laws of the correlation spreading for long-range interacting lattice models are found by analyzing the previous form of 
$G(R,t)$ at Eq.~\eqref{spa_lr}. 

\paragraph{Motion of the correlation edge} The motion of the CE is determined by the amplitude (prefactor) of the connected correlation function $G(R,t)$ at 
Eq.~\eqref{spa_lr}. 
It is found by imposing that the prefactor should be constant in the vicinity of the CE. This leads to $t^{\gamma} \propto R^{\chi}$ implying
$t \propto R^{\chi/\gamma} = R^{\beta_{\mathrm{CE}}}$. Besides, $\chi = \gamma + D/2$ and $\gamma > 0$ due to $0 \leq z < 1$ and $\nu \geq 0$. 
Finally, it leads for $\beta_{\mathrm{CE}}$ to $\beta_{\mathrm{CE}} = \chi/\gamma > 1$ meaning that the motion of the CE is always sub-ballistic
($\beta_\mathrm{CE} > 1$) and does not depend on whether the excitation spectrum $E_k$ is gapped or gapless. However, it depends on the observable \textit{via} 
the exponent $\nu$, on the dimensionality of the lattice with $D$ and on the long-range interactions with $z$ the $\alpha$-dependent dynamical exponent ($\alpha$ is
the power-law exponent of the long-range interactions). This contrasts with the short-range case where a ballistic spreading for the CE is found, independently
of the dimension and of the observable. 

\paragraph{Motion of the series of local extrema} The motion of the series of local maxima (m) is defined by the maximum of the cosine function. It is now necessary
to distinguish the case where the excitation spectrum $E_k$ is gapless from the gapped case. 

\begin{itemize}
\item For a gapless excitation spectrum ($\Delta = 0$), the maximum of the cosine function is determined by the following equation 

\begin{equation}
A_z \left( \frac{t}{R^z} \right)^{\frac{1}{1-z}} + \sigma D \frac{\pi}{4} = 2\pi n,~~ n\in \mathbb{Z}.
\label{scaling_maxima}
\end{equation}

\noindent
The latter implies $t \propto R^z$. Since $0 \leq z < 1$, the motion of the maxima is always super-ballistic ($\beta_\mathrm{m} = z < 1$). 

\item For a gapped excitation spectrum ($\Delta > 0$), the motion of the maxima is defined by 

\begin{equation}
A_z \left( \frac{t}{R^z} \right)^{\frac{1}{1-z}} -2\Delta t + \sigma D \frac{\pi}{4} = 2\pi n,~~ n\in \mathbb{Z},
\end{equation}

\noindent
which can be brought into the following form

\begin{equation}
A_z \left( \frac{t^z}{R^z} \right)^{\frac{1}{1-z}} - 2\Delta + \frac{\sigma D \pi}{4t} = \frac{2\pi n}{t}.
\end{equation}

\noindent
The latter can be simplified into 

\begin{equation}
A_z \left( \frac{t}{R} \right)^{\frac{z}{1-z}} - 2\Delta \rightarrow 0,
\end{equation}

\noindent
using the limit where the stationary phase approximation is valid \footnote{One reminds that the stationary phase approximation is valid 
for $R,t \rightarrow + \infty$ along the line $R/t \rightarrow \mathrm{ct}$.} ($t\rightarrow +\infty$).
Finally, the motion of the series of local maxima is defined by the condition $t \propto R$. In other words, the series of local maxima always propagate ballistically
($\beta_\mathrm{m} = 1$) for a sudden global quench confined in a gapped quantum phase. Contrary to a gapless phase, the motion of the maxima does not 
depend anymore on $z$ the dynamical exponent, meaning that the quasimomentum dependence of the excitation spectrum $E_k$ becomes irrelevant.
The scaling laws and velocities for the twofold correlation spreading in long-range lattice
models is summarized in Tab.~\ref{tab:SL}.
\end{itemize}

\begin{table}
  \centering
  \begin{tabular}{|c|c|c|c|}
   \hline
   Phase & Regime & Correlation Edge & Maxima \\
   \hline
   Gapped & local & ballistic : $V_{\mathrm{CE}} = 2V_{\mathrm{g}}^*$ & ballistic : $V_{\mathrm{m}} = 2 V_{\varphi}^*$ \\
   \hline 
   Gapless & local & ballistic : $V_{\mathrm{CE}} = 2V_{\mathrm{g}}^*$ & ballistic : $V_{\mathrm{m}} = 2 V_{\varphi}^*$ \\
   \hline 
   Gapped & quasi-local & sub-ballistic : $\beta_{\mathrm{CE}} > 1$ & ballistic : $\beta_{\mathrm{m}} = 1$ \\
   \hline 
   Gapless & quasi-local & sub-ballistic : $\beta_{\mathrm{CE}} > 1$ & super-ballistic : $\beta_{\mathrm{m}} < 1$ \\
   
   \hline
\end{tabular}
  \caption{Scaling laws and velocities of the twofold structure for the correlation spreading in long-range interacting spin and particle lattice 
  models for a sudden global quench confined in a single gapped or gapless phase.} \label{tab:SL}
\end{table}

\end{itemize}

\subsection{Gapless phase : case study of the long-range XY chain in the $x$ polarized phase}
\label{xy_gapless}
In what follows, the scaling laws for the twofold correlation spreading in long-range lattice models, predicted \textit{via} a quasiparticle approach, are
verified for two long-range interacting spin $s=1/2$ models.
Since long-range spin lattice models are considered, the operators $\hat{O}_j(\mathbf{R}),~j \in \{1,2,3 \}$ at Eq.~\eqref{generic_ham} represent spin operators,
the parameter $J(\mathbf{R},\mathbf{R}')$ denotes a long-range spin exchange coupling and $h(\mathbf{R})$ a possible transverse magnetic field.
We start by testing the scaling laws associated to the correlation spreading for sudden global quenches confined in a gapless phase. 
As an example, the long-range XY (LRXY) model is considered. It is defined by $\hat{O}_1(\mathbf{R}) \equiv \hat{S}_{\mathbf{R}}^{x(y)}$ and 
$\hat{O}_2(\mathbf{R}') \equiv \hat{S}_{\mathbf{R}'}^{x(y)}$ in each direction of the $xy$ plane. Besides, the two-site coupling term has an amplitude 
characterized by $J(\mathbf{R},\mathbf{R}') = (-J/2)/ |\mathbf{R}-\mathbf{R}'|^\alpha$ and no local interaction (magnetic field) is present implying 
$\hat{O}_3(\mathbf{R})=0$. The latter can be extended to the long-range XXZ (LRXXZ) model by including $\epsilon$ ($\epsilon > 0$) a long-range antiferromagnetic exchange
coupling in the $z$ direction, which yields the Hamiltonian

\begin{equation}\label{eq:XXZ}
 \hat{H} = \sum_{\mathbf{R} < \mathbf{R}'} \frac{J/2}{|\mathbf{R}-\mathbf{R}'|^\alpha} \left[ 
 -\left( \hat{S}_{\mathbf{R}}^x \hat{S}_{\mathbf{R}'}^x + \hat{S}_{\mathbf{R}}^y \hat{S}_{\mathbf{R}'}^{y} \right) +
 \epsilon \hat{S}_{\mathbf{R}}^{z} \hat{S}_{\mathbf{R}'}^{z} \right].
\end{equation}

\subsubsection{Gapless excitation spectrum in the $x$ polarized phase}
In the following, the 1D LRXY model is considered ($D=1$) with a sudden global quench performed from the ground state of the 1D LRXXZ model ($\epsilon_{\mathrm{i}}
\neq 0$) to the 1D LRXY model ($\epsilon_{\mathrm{f}}=0$). The pre-quench antiferromagnetic exchange coupling and the power-law exponent $\alpha$ are choosen such that
the global quench is confined in the $x$ polarized phase, \textit{ie.} when long-range order ferromagnetic $xy$ phase is established. 
The latter implies small-enough power-law exponent $\alpha$ and $-1<\epsilon_{\mathrm{i}} < \epsilon_c(\alpha)$ where $\epsilon_c(\alpha$) is the critical point separating
the $x$ polarized phase from the $z$ N\'eel phase, see Ref.~\cite{frerot2017} for the phase diagram of the 1D LRXXZ model at equilibrium and zero-temperature. \\
Indeed, when both previous conditions on $\epsilon$ and $\alpha$ are fulfilled, the 1D LRXXZ model displays two different 
quantum phases\footnote{ \vspace{-0.3cm} \begin{itemize} \item In the case of negative values of $\epsilon$, inducing long-range ferromagnetic interactions along the $z$ axis, 
an additional quantum phase appears corresponding to the long-range order ferromagnetic phase along the $z$ axis for $\epsilon < -1,~\forall \alpha \in \mathbb{R}^+$. 
\item For high $\alpha$, fastly decaying long-range interactions, and $|\epsilon|<1$, the 1D LRXXZ can enter a gapless disordered (Luttinger-liquid) phase.\end{itemize}}.
For small-enough $\alpha$ and $-1< \epsilon < \epsilon_c(\alpha)$, the continuous rotational symmetry around the $z$ axis is spontaneously broken and the spins
are polarized along the $x$ axis corresponding to an arbitrary choice. Indeed, the spins can align along any direction of the $x-y$ plane. This defines the twofold-degenerate long-range order $x$
polarized phase characterized by $\langle \hat{S}_R^z \rangle = 0$ and $\langle \hat{S}^x_R \hat{S}^x_{R+R'} \rangle > 0$. 
In the opposite case, \textit{ie.} $\epsilon > \epsilon_c(\alpha)$, the antiferromagnetic interaction along the $z$ axis dominates and the 1D LRXXZ model
enters the twofold-degenerate long-range order $z$ N\'eel phase. The latter is characterized by $\langle \hat{S}^x_R \rangle = 0$ and $(-1)^{R'}
\langle \hat{S}_R^z \hat{S}_{R+R'}^z \rangle > 0$. \\

To find the excitation spectrum in the $x$ polarized phase of the LRXXZ chain, the Hamiltonian can be diagonalized using a standard Holstein-Primakoff (HP)
transformation~\cite{holstein1940,auerbach1994} given by,

\begin{equation}
\hat{S}_{R}^{x} = \frac{1}{2}-\hat{a}_{R}^\dagger \hat{a}_{R},~~~ \hat{S}_{R}^{y} = \simeq -\frac{\hat{a}_{R}^\dagger-\hat{a}_{R}}{2i}, ~~~
\hat{S}_{R}^{z} \simeq -\frac{\hat{a}_{R} + \hat{a}_{R}^\dagger}{2}, 
\end{equation}

\noindent
where interacting terms beyond second order in the boson operators $\hat{a}_{R}$ and $\hat{a}_{R}^\dagger$ are neglected. Inserting the following 
HP transformations into Eq.~\eqref{eq:XXZ} yields a quadratic bosonic form for the 1D LRXXZ Hamiltonian in the momentum space. The latter may be written as 

\begin{equation}
\hat{H} = e_{0} + \frac{1}{2} \sum_{k \neq 0} \mathcal{A}_{k} \left( \hat{a}^{\dag}_k \hat{a}_k + \hat{a}_{-k} \hat{a}^{\dag}_{-k} \right) +
\mathcal{B}_{k} \left( \hat{a}^{\dag}_k \hat{a}^{\dag}_{-k} + \hat{a}_k \hat{a}_{-k} \right),
\label{H_quadra_lrxxz}
\end{equation}

\noindent
with $e_0$ a finite constant energy and the momentum-dependent prefactors $\mathcal{A}_k$ and $\mathcal{B}_k$ defined by 

\begin{equation}
 \mathcal{A}_k =  \frac{J}{2}\left[ P_{\alpha}(0) + P_{\alpha}(k) \frac{\epsilon-1}{2} \right],~~~ \mathcal{B}_k = \frac{J P_{\alpha}(k)}{4}(\epsilon+1),
 \label{1d_lrxxz_a_b}
\end{equation}

\noindent
where $P_\alpha(k) = \int \mathrm{d}R~e^{-i kR}/|R|^\alpha$ is the Fourier transform of the long-range term. Then, the quadratic Hamiltonian at
Eq.~\eqref{H_quadra_lrxxz} can be diagonalized using a bosonic Bogolyubov transformation, see for instance Ref.~\cite{frerot2017} and the previous section devoted to
the 1D BH model in the SF phase. Finally, it leads for the excitation spectrum to
$E_k = \sqrt{\mathcal{A}_k^2 - \mathcal{B}_k^2}$ with the quasimomentum-dependent terms defined previously at Eq.~\eqref{1d_lrxxz_a_b}. By developping the latter,
one obtains the following form for the gapless excitation spectrum $E_k$ of the LRXXZ chain in the $x$ polarized phase

\begin{equation}
 E_k = \frac{JP_{\alpha}(0)}{2} \sqrt{ \left( 1 - \frac{P_{\alpha}(k)}{P_{\alpha}(0)} \right) \left( 1 + \epsilon \frac{P_{\alpha}(k)}{P_{\alpha}(0)} \right)}.
\end{equation}

\begin{figure}[h!]
\centering
\includegraphics[scale = 0.37]{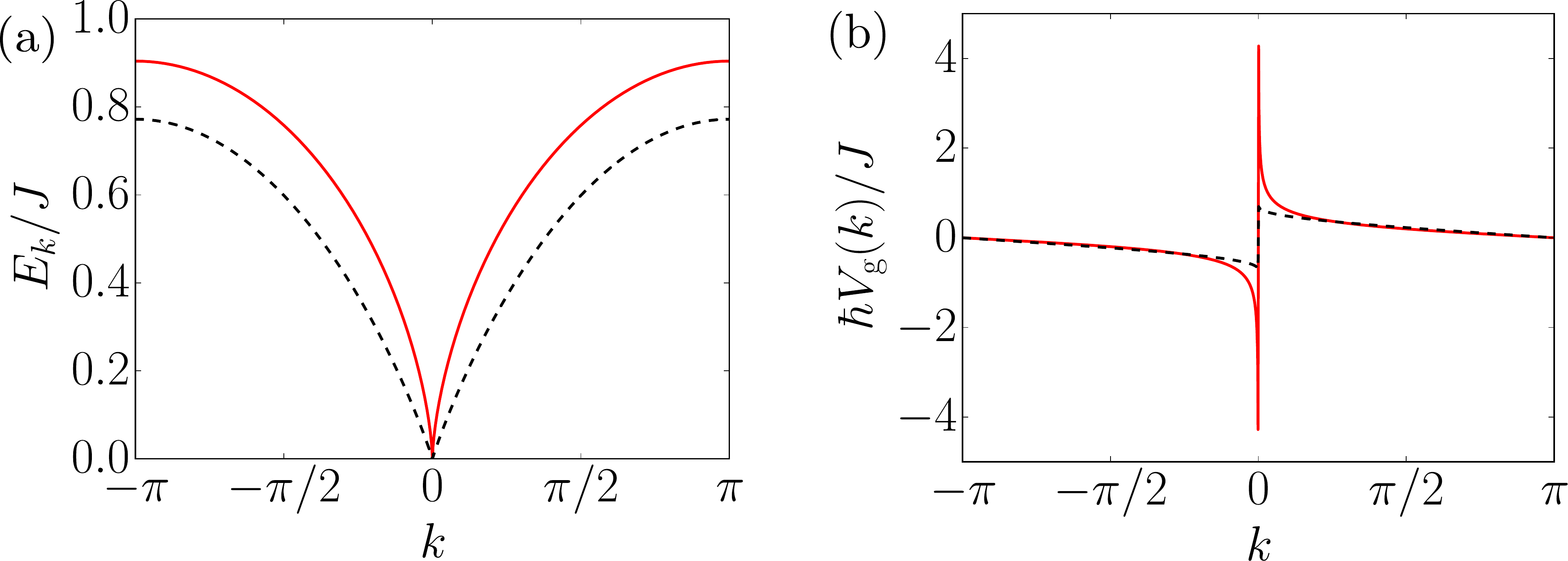}
\caption{\label{ek_vg_lrxy} 
Excitation spectrum properties of the 1D LRXY model in the $x$ polarized phase. 
(a)~Excitation spectrum rescaled by the spin exchange coupling $E_k/J$ as a function of the quasimomentum $k$ in the first Brillouin zone $\mathcal{B}
= [-\pi,\pi]$ in the quasi-local regime for $\alpha = 2.3$ (red solid line) and in the local regime for $\alpha = 3.3$ (black dashed line). (b)~ Dimensionless quantity
$\hbar V_{\mathrm{g}}(k)/J$, where $V_{\mathrm{g}}$ denotes the group velocity, as a function of the quasimomentum $k$ for the same power-law exponents
$\alpha$. For $\alpha = 2.3$ (red solid line), $V_{\mathrm{g}}$ diverges at $k \rightarrow 0$ whereas for $\alpha = 3.3$ (black dashed line), it is finite over
the first Brillouin zone.}
\end{figure}

\noindent
Using the previous excitation spectrum, one immediately finds the gapless one for the LRXY chain ($\epsilon=0$), see Fig.~\ref{ek_vg_lrxy}(a),

\begin{equation}\label{eq:XXZ.Ek}
 E_{k} = \sqrt{\mathcal{A}_k^2 - \mathcal{B}_k^2} = \frac{J P_\alpha(0)}{2} \sqrt{1-\frac{P_\alpha (k)}{P_\alpha (0)}}.
\end{equation}

\noindent
Indeed, for both the LRXY and LRXXZ spin chains in the $x$ polarized phase, the spontaneous rotational symmetry breaking around the $z$ axis induces low-energy excitations,
the so-called Goldstone modes, corresponding to magnons here and implying $\Delta = 0$. In the infrared limit $k \rightarrow 0$, the Fourier transform
of the long-range term $P_\alpha(k)$ can be written as, see supplemental material of \cite{frerot2017},

\begin{equation}\label{eq:Pinfrared}
P_\alpha(k) \approx P_\alpha(0) + P^\prime_\alpha |k|^{\alpha-1},
\end{equation}

\noindent
where $P_\alpha(0)$ and $P_\alpha^\prime$ are finite constants. Hence, in the limit of small quasimomenta,
the excitation spectrum $E_k$ for the LRXY chain behaves as $E_k \propto \vert k \vert^z$ with $z=(\alpha-1)/2$. Moreover, for $1 \leq \alpha < 3$ ($\alpha \geq 3)$, the
1D LRXY model is in the quasi-local (local) regime where the quasiparticle energy $E_k$ is finite and the group velocity $V_{\mathrm{g}}$ diverges (is also finite), see
Fig.~\ref{ek_vg_lrxy}(b).

\subsubsection{Spin-spin correlations along the $z$ axis}
In order to characterize the out-of-equilibrium dynamics of the LRXY chain in the quasi-local regime of the $x$ polarized phase, a sudden global quench 
is considered. It is defined by a pre- and post-quench antiferromagnetic interaction parameter along the $z$ axis fulfilling $-1<\epsilon_\textrm{i} < \epsilon_c(\alpha)$
(1D LRXXZ model) and $\epsilon_\textrm{f}=0$ (1D LRXY model) respectively. The power-law exponent $\alpha$ is fixed during the real time evolution, and confined in the interval $\alpha \in [1,3[$ so that
the quasi-local regime is considered.
Besides, the connected spin-spin correlation function along the $z$ axis, denoted by $G_z$, are investigated. The $G_z$ spin-spin correlations can be 
expressed in the generic form of Eq.~\eqref{generic_corr} where $G_z(R,t) = G_{z,0}(R,t) - G_{z,0}(R,0)$ with 

\begin{equation}
G_{z,0}(R,t) = \langle \hat{S}_R^z(t) \hat{S}_0^z(t) \rangle - \langle \hat{S}_R^z(t)\rangle \langle \hat{S}_0^z(t) \rangle.
\end{equation}

\noindent
The latter can be calculated analytically using the quasiparticle picture relying on the bosonic Bogolyubov transformation. The general scheme is identical to the one
followed to compute the one-body $G_1$ and density-density $G_2$ correlations of the 1D Bose-Hubbard model in the SF mean field regime [see Appendix.~\ref{appendix_g1_sf} and
\ref{appendix_g2_sf}]. Finally, it turns out that
the $G_z$ spin-spin correlations can be expressed in the generic form presented at Eq.~\eqref{generic_form} with a quasimomentum-dependent amplitude function $\mathcal{F}$
given by [see Appendix.~\ref{appendix_gz_lrxy}]

\begin{equation}\label{eq:LRXY.Fk}
 \mathcal{F}(k) = \frac{\epsilon_\textrm{i}}{8}\frac{P_\alpha\left(k\right)}{P_\alpha\left(0\right)}\sqrt{\frac{P_\alpha(0)-P_\alpha\left(k \right)}{P_\alpha(0)+
 \epsilon_\textrm{i}P_\alpha\left(k\right)}}.
\end{equation}

\begin{figure}[t!]
\centering
\includegraphics[scale = 0.35]{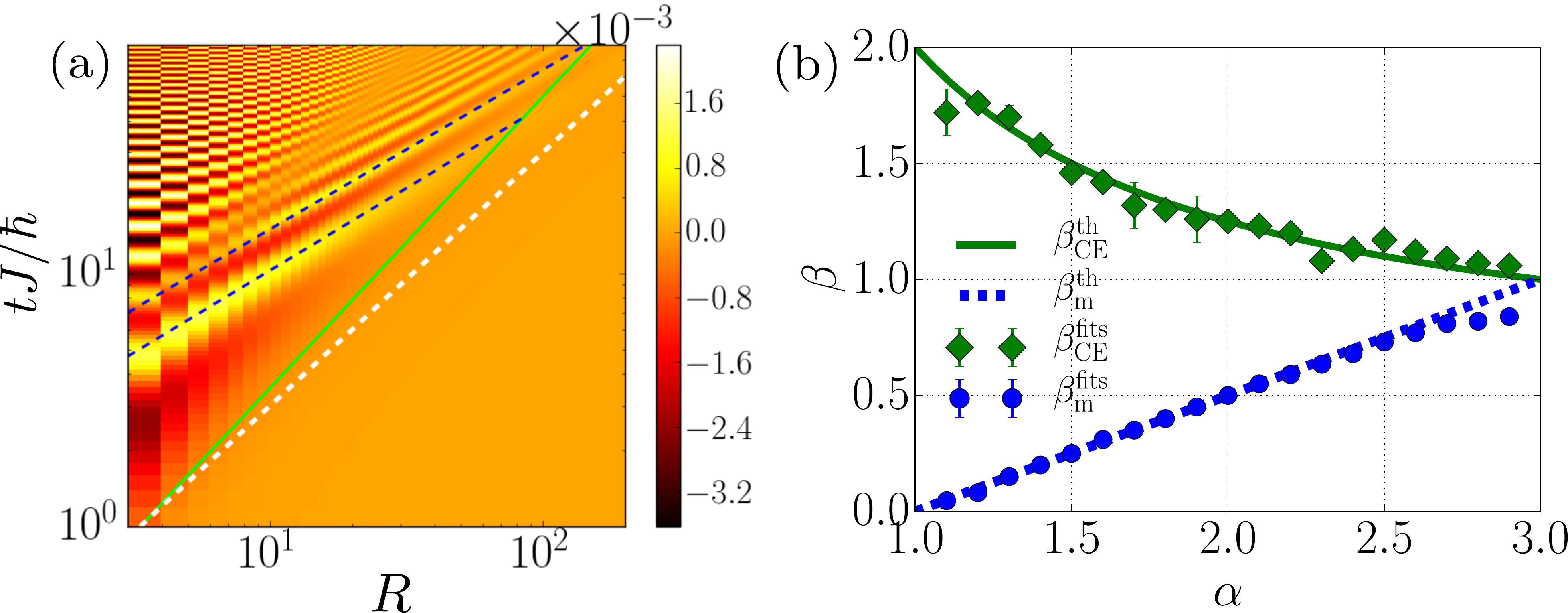}
\caption{
Spreading of the connected spin-spin correlation function $G_z(R,t)=G_{z,0}(R,t)-G_{z,0}(R,0)$ with $G_{z,0}(R,t) =
\langle \hat{S}_R^z(t) \hat{S}_0^z(t) \rangle - \langle \hat{S}_R^z(t)\rangle \langle \hat{S}_0^z(t) \rangle$ for the 1D LRXY model.
(a)~ Analytical result for a global quench confined in the $x$ polarized phase from the ground state of the 1D LRXXZ model at
$\epsilon_\mathrm{i} = 0.2$ to the 1D LRXY model ($\epsilon_{\mathrm{f}} = 0$) and in the quasi-local regime at $\alpha=2.3$.
The space-time spin-spin correlations feature a double algebraic structure (straight lines in log-log scale) with a sub-ballistic correlation edge (solid green line)
and super-ballistic spreading of the series of local maxima (dashed blue lines). The white dotted line indicates ballistic spreading for reference.
(b)~ Evolution of $\beta_{\mathrm{CE}}^{\mathrm{fits}}$ and its corresponding theoretical value $\beta_{\mathrm{CE}}^{\mathrm{th}}$ characterizing the spreading
of the correlation edge with $\beta_{\mathrm{m}}^{\mathrm{fits}}$ and $\beta_{\mathrm{m}}^{\mathrm{th}}$ for the spreading of the series of local maxima as a function
of the power-law exponent $\alpha$ defining the long-range spin exchange couplings. Figure~(a) extracted from Ref.~\cite{cevolani2018}.}
\label{fig:spins_LRXY}
\end{figure}

\noindent
In the infrared limit $k \rightarrow 0^{+}$, it scales as $\mathcal{F}(k)\sim k^\nu$ with $\nu=z=(\alpha-1)/2$. The linearization of the Holstein-Primakoff
transformation holds for $\vert 1/2 - \langle \hat{S}^x_\mathbf{R} \rangle \vert \ll 1$~\cite{auerbach1994}. For the calculations corresponding to the 1D LRXY model
at Fig.~\ref{fig:spins_LRXY}(a),
we find $\max\{|1/2 - \langle \hat{S}^x_R \rangle | \} \simeq 0.12$. It validates the linear spin-wave approximation and this result agrees with the predictions 
for the same model made in Ref.~\cite{frerot2017} where the validity of the spin wave approach for that model is extensively studied. \\

For the 1D LRXY model and the spin-spin correlations along the $z$ axis, we have $\nu=z=(\alpha-1)/2$ which yields, according to Eqs.~\eqref{chi} and \eqref{gamma}, to

\begin{equation}
\beta_{\mathrm{CE}}= \frac{\chi}{\gamma} = 1+ \left(3-\alpha\right)/2\alpha.
\end{equation}

\begin{figure}[t!]
\centering
\includegraphics[scale = 0.57]{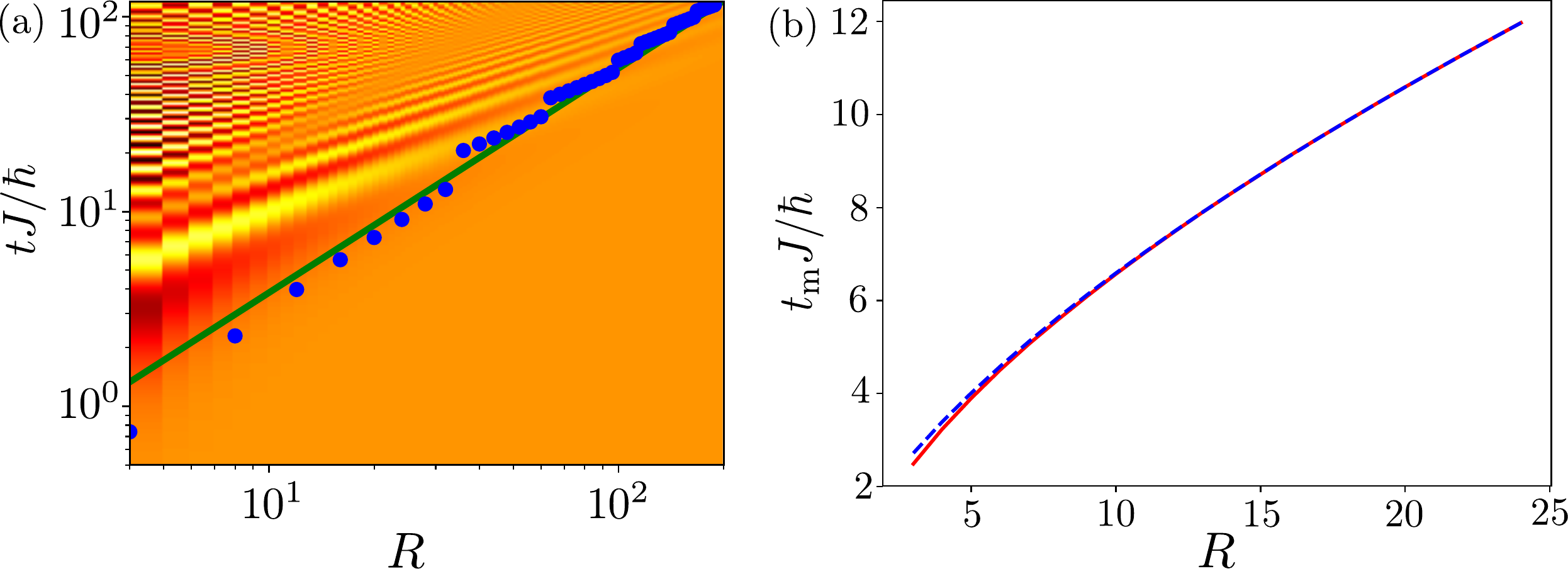}
\caption{Numerical analysis of the correlation edge and the local extrema for the connected spin-spin correlation function $G_z(R,t)$ for the 
1D LRXY model in the $x$ polarized phase at $\alpha = 2.3$. (a)~ Analytical result of $G_z(R,t)$ (same data as in Fig.~\ref{fig:spins_LRXY}(a), log-log scale).
The filled blue points correspond for each distance $R$ to the first time where the correlation reaches $2\%$ of its maximum value. The solid green line shows the
power law predicted theoretically with a fitted multiplicative factor. (b)~Trajectory of the first extremum of $G_z(R,t)$ (lin-lin scale). The figure shows the 
numerical result found from Fig.~(a) (solid red line) together with a fitted power law (dashed blue line). Figures adapted from Ref.~\cite{cevolani2018}.}
\label{fig:fits_lrxy}
\end{figure}

\noindent
In the calculations, the CE is found by tracking the points in the $R-t$ plane where the correlations reach $\epsilon=2\%$ of the maximal value. 
For instance at $\alpha = 2.3$, it yields the filled blue points on Fig.~\ref{fig:fits_lrxy}(a). The latter feature a linear trajectory in the log-log scale, that is
a power law behavior in lin-lin scale. The latter is in excellent agreement with the theoretical prediction $\beta_{\mathrm{CE}} = 1.15$ shown as a solid green line
on Fig.~\ref{fig:fits_lrxy}(a) [see also Fig.~\ref{fig:spins_LRXY}(a)]. The activation time $t^{*}$ as a function of the distance $R$ is then fitted by a power law,
$t^{*} \sim R^{\beta_{\mathrm{CE}}^{\mathrm{fits}}}$ to the blue points. It yields $\beta_{\mathrm{CE}}^{\mathrm{fits}} \simeq 1.08 \pm 0.01$, see supplemental
material of Ref.~\cite{cevolani2018} for more details. A similar tracking technique for the CE is used for different power-law exponents $\alpha$ as shown at
Fig.~\ref{fig:spins_LRXY}(b) by the green diamonds. \\
Concerning the spreading of the series of local maxima, the theoretical value of the exponent $\beta_{\mathrm{m}}$ is given by, see Eq.~\eqref{scaling_maxima},

\begin{equation}
\beta_{\mathrm{m}} = z = (\alpha-1)/2.  
\label{maxima_lr_gapless}
\end{equation}

\noindent
The spreading of the inner structure of the correlation function is analyzed by tracking the position of the first local extremum as a function of time.
We then fit the corresponding function by $t_\mathrm{m} = a R^{\beta_{\mathrm{m}}^{\mathrm{fits}}}+b$. 
Considering the same example as previously for the numerical analysis of the CE at Fig.~\ref{fig:fits_lrxy}(a),
the result for the spreading of the first local extremum is plotted on Fig.~\ref{fig:fits_lrxy}(b) and represented by the solid red line together with
the fitted power law corresponding to the dashed blue line [see also dashed blue lines on Fig.~\ref{fig:spins_LRXY}(a)]. The fit yields 
$\beta_{\mathrm{m}}^{\mathrm{fits}} \simeq 0.63 \pm 0.01$, in excellent agreement with the
theoretical value $\beta_{\mathrm{m}} = 0.65$. A similar tracking technique for the inner structure (series of local extrema) is used for different
power-law exponents $\alpha$ as shown by the blue disks at Fig.~\ref{fig:spins_LRXY}(b). \\

On Fig.~\ref{fig:spins_LRXY}(b), the evolution of the theoretical exponents $\beta_{\mathrm{CE}}$, $\beta_{\mathrm{m}}$ and those fitted 
$\beta_{\mathrm{CE}}^{\mathrm{fits}}$, $\beta_{\mathrm{m}}^{\mathrm{fits}}$ as a function of the power-law exponent $\alpha$ is investigated.
$\alpha$ is contained in the interval $[1,3[$ so that the LRXY chain is in the quasi-local regime of the $x$ polarized phase. 
The results confirm the theoretical scaling laws for the correlation spreading in long-range lattice models confined in the quasi-local regime of a gapless phase.
More precisely, they show clearly a sub-ballistic propagation of the CE ($\beta_{\mathrm{CE}} > 1$) and a super-ballistic propagation of the series
of local maxima ($\beta_{\mathrm{m}} < 1$). \\

These analytical results are also consistent with the experimental observation of a super-ballistic dynamics in the 1D LRXY model realized with trapped ion chains 
for $\alpha>1$~\cite{richerme2014}. In the following, we will interpret this faster-than-ballistic dynamics as the one resulting from the spreading of the series
of local extrema. However, one needs to take a step back from these experimental results since small distance and time scales are considered. Indeed, the dynamics of 
the LRXY chain unveiled experimentally (i) is biased by strong finite-size effects due to small chain lengths (ii) can not be used to certify our theoretical scaling laws 
(implying to investigate the asymptotic behavior of the spin correlations) due to small observation times. To perform a precise comparison, increasing the chain length 
is absolutely necessary. Relevant distance scales (more than $50$ ions) are now accessible, see Ref.~\cite{zhang2017} for instance. 
The previous theoretical results are also in rough agreement with the analysis of numerical calculations performed within the truncated Wigner 
approximation for the 1D LRXY model~\cite{schachenmayer2015b}. Our theoretical result about the maxima spreading at Eq.~\eqref{maxima_lr_gapless}
was found in Ref.~\cite{frerot2018}. Nevertheless, we stress that our analysis shows that this super-ballistic behavior characterizes the spreading of the inner structure
but not the one associated to the outer structure (the correlation edge).

\subsection{Gapped phase : case study of the long-range Ising chain in the $z$-polarized phase}

We now wish to confirm the theoretical scaling laws for the correlation spreading in the second case, \textit{ie.} for sudden global quenches confined 
in the quasi-local regime of a gapped phase. To do so, we rely on the long-range transverse Ising (LRTI) model where the spin operators $\hat{O}_j(\mathbf{R})$, $j \in \{1,2,3\}$,
are defined by $\hat{O}_1(\mathbf{R}) \equiv \hat{S}_{\mathbf{R}}^x$, $\hat{O}_2(\mathbf{R}') \equiv \hat{S}_{\mathbf{R}'}^x$ and
$\hat{O}_3(\mathbf{R}) \equiv \hat{S}_{\mathbf{R}}^z$ with a constant and uniform transverse magnetic field $h(\mathbf{R})=-2h$
and an algebraically decaying spin exchange amplitude $J (\mathbf{R},\mathbf{R}') = 2J/|\mathbf{R}-\mathbf{R}'|^\alpha$. The Hamiltonian of the LRTI model
is thus given by 

\begin{equation}\label{eq:LRTI}
\hat{H} =  \sum_{\mathbf{R} < \mathbf{R}'} \frac{2J}{|\mathbf{R} - \mathbf{R}'|^\alpha} \hat{S}_{\mathbf{R}}^{x}\hat{S}_{\mathbf{R}'}^{x}
- 2h \sum_{\mathbf{R}} \hat{S}_{\mathbf{R}}^{z}.
\end{equation}

\subsubsection{Gapped excitation spectrum in the $z$ polarized phase}
The latter has two quantum gapped phases separated by a second-order transition, see Ref.~\cite{koffel2012} for its $T=0$ phase diagram based on an analysis
of the entanglement entropy. When the transverse magnetic field is dominant ($h \gg J$), the spin model is in the non-degenerate $z$ polarized phase exhibiting
quasi-long-range order and characterized by $\langle \hat{S}^z_R \rangle > 0$ (due to the minus sign in front of the transverse magnetic field $h>0$), $(-1)^{R'} \langle \hat{S}^x_R \hat{S}^x_{R+R'} \rangle = 0$. However, when the 
long-range antiferromagnetic interaction dominates ($h \ll J$), it is confined in the twofold-degenerate $x$ N\'eel phase possessing long-range order where 
$\langle \hat{S}^z_R \rangle = 0$ and $(-1)^{R'} \langle \hat{S}^x_R \hat{S}^x_{R+R'} \rangle > 0$. \\

To find the excitation spectrum $E_k$ in the $z$ polarized phase, one can perform a Holstein-Primakoff (HP) transformation defined as follows

\begin{equation}
\hat{S}_{R}^{x}\simeq \frac{\hat{a}_{R} + \hat{a}_{R}^\dagger}{2},~~~
\hat{S}_{R}^{y} \simeq -\frac{\hat{a}_{R}^\dagger-\hat{a}_{R}}{2i},~~~
\hat{S}_{R}^{z} = \frac{1}{2}-\hat{a}_{R}^\dagger \hat{a}_{R}. 
\end{equation}

\noindent
As previously for the 1D LRXY model, the HP transformation allows to get a quadratic Bose Hamiltonian in the momentum space. Then, using the bosonic Bogolyubov 
transformation, one obtains the same form as that of Eq.~\eqref{H_quadra_lrxxz} with different momentum-dependent prefactors $\mathcal{A}_k$, $\mathcal{B}_k$ given by

\begin{equation}
\mathcal{A}_k = 2h + JP_\alpha(k),~~~\mathcal{B}_k = JP_\alpha(k).
\end{equation}

\begin{figure}[h!]
\centering
\includegraphics[scale = 0.37]{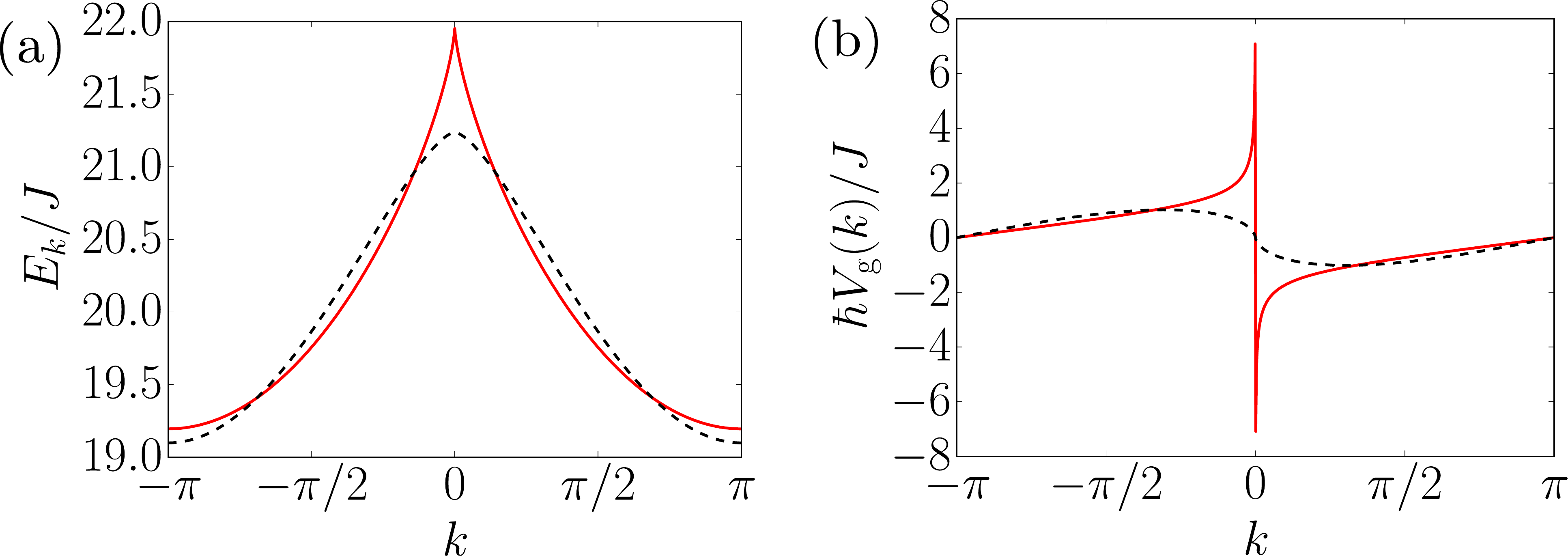}
\caption{\label{ek_vg_lrti} 
Excitation spectrum properties of the 1D LRTI model in the $z$ polarized phase. 
(a)~Excitation spectrum rescaled by the spin exchange coupling $E_k/J$ as a function of the quasimomentum $k$ in the first Brillouin zone
$\mathcal{B} = [-\pi,\pi]$ in the quasi-local regime for $\alpha = 1.7$ (red solid line) and in the local regime for $\alpha = 2.7$ (black dashed line) for $h = 10J$.
(b)~ Dimensionless quantity $\hbar V_{\mathrm{g}}(k)/J$, where $V_{\mathrm{g}}$ is the group velocity, as a function of the quasimomentum $k$ for the same power-law 
exponents $\alpha$. For $\alpha = 1.7$ (red solid line), $V_{\mathrm{g}}$ diverges at $k \rightarrow 0$ whereas for $\alpha = 2.7$ (black dashed line), it is finite
over the first Brillouin zone.}
\end{figure}

\noindent
One then finds the following theoretical expression for $E_k$ the gapped excitation spectrum in the $z$ polarized phase, see
Fig.~\ref{ek_vg_lrti}(a),

\begin{equation}
E_k = \sqrt{\mathcal{A}_k^2 - \mathcal{B}_k^2} = 2 \sqrt{h \left[ h + J P_\alpha (k) \right]}.
\label{ek_lrti_anal}
\end{equation}

\noindent
In the infrared limit $k \rightarrow 0$ where Eq.~(\ref{eq:Pinfrared}) is valid, $E_k$ can be rewritten as

\begin{equation}
E_k \simeq \Delta + c \vert k\vert^z,
\label{ek_lrti_infrared}
\end{equation}

\noindent
where the following parameters $\Delta$ (the gap), $c$ (a prefactor) and $z$ (the dynamical exponent) read as, see Refs.~\cite{cevolani2016, cevolani2018} 

\begin{equation}
\Delta = 2\sqrt{h\left[h+JP_\alpha(0)\right]},~~~c = \sqrt{\frac{h}{h+JP_\alpha(0)}}JP_\alpha^\prime,~~~z=\alpha-1.
\end{equation}

\noindent
According to Eq.~\eqref{ek_lrti_infrared}, the quasiparticle energy $E_k$ is finite and the group velocity diverges
for $0 \leq z < 1$ implying $1 \leq \alpha<2$, see Fig.~\ref{ek_vg_lrti}(a,b).

\subsubsection{Spin-spin correlations along the $x$ axis}
As previously, we study the spin-spin correlations perpendicular to the polarization axis of the quantum phase considered for the sudden global quench. 
For the 1D LRTI model, global quenches confined in the quasi-local regime of the $z$ polarized phase are investigated. As a consequence, 
the spin fluctuations are studied along the $x$ axis \textit{via} $G_x$ the connected spin-spin correlation function defined by 
$G_x(R,t) = G_{x,0}(R,t) - G_{x,0}(R,0)$ where

\begin{equation}
G_{x,0}(R,t) = \langle \hat{S}_R^x(t) \hat{S}_0^x(t) \rangle - \langle \hat{S}_R^x(t)\rangle \langle \hat{S}_0^x(t) \rangle.
\end{equation}

\begin{figure}[t!]
\centering
\includegraphics[scale = 0.34]{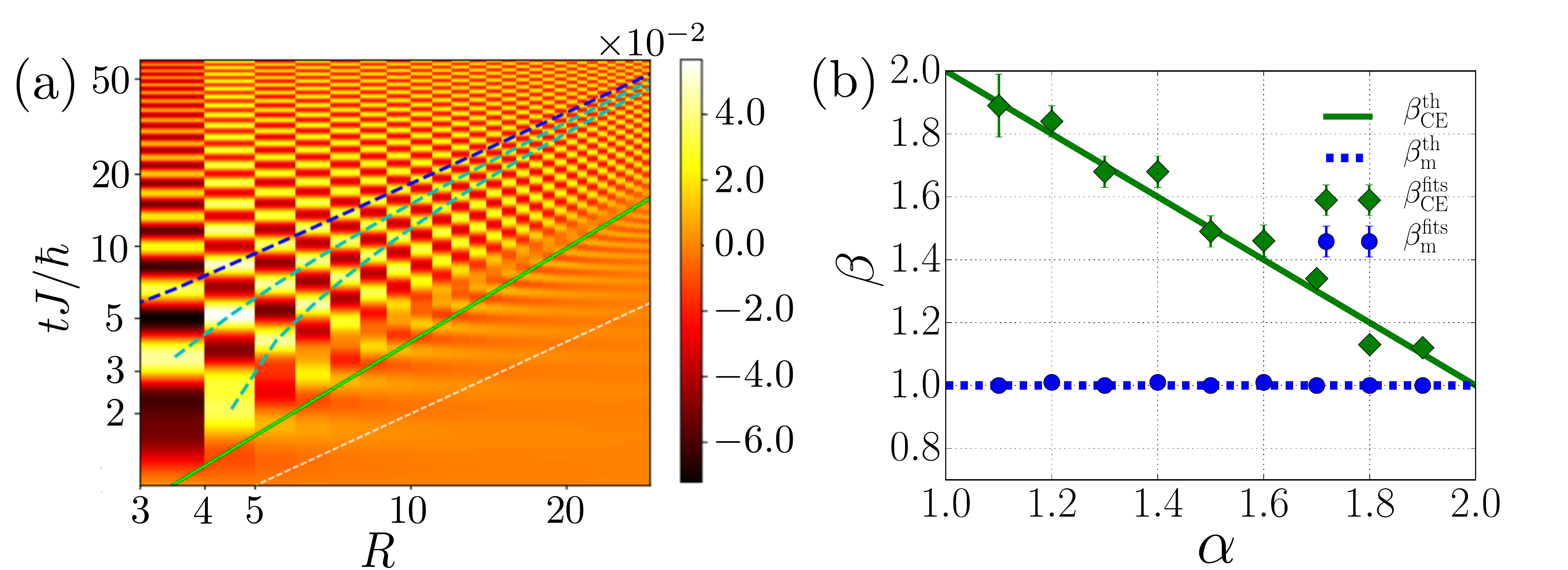}
\caption{
Spreading of the connected spin-spin correlation function $G_x(R,t)=G_{x,0}(R,t)-G_{x,0}(R,0)$ with 
$G_{x,0}(R,t) = \langle \hat{S}_R^x(t) \hat{S}_0^x(t) \rangle - \langle \hat{S}_R^x(t)\rangle \langle \hat{S}_0^x(t) \rangle$ for the 1D LRTI model with 
$\alpha=1.7$. (a)~ Analytical result for a sudden global quench confined in the $z$ polarized phase from $J_{\mathrm{i}} = 0.02h$ to $J_\mathrm{f} = h$. For the 1D LRTI model, at $\alpha = 1.7$, the
critical point separating the $z$ polarized phase from the $x$ N\'eel phase is located at $J/h \simeq 3$ \cite{koffel2012}. 
The space-time spin-spin correlations feature a double algebraic structure (straight lines in log-log scale) with a sub-ballistic correlation edge (solid green line)
and ballistic spreading of the series of local maxima (dashed blue lines). The white dotted line indicates ballistic spreading for reference.
(b)~ Evolution of $\beta_{\mathrm{CE}}^{\mathrm{fits}}$ and its corresponding theoretical value $\beta_{\mathrm{CE}}^{\mathrm{th}}$ characterizing the spreading
of the correlation edge with $\beta_{\mathrm{m}}^{\mathrm{fits}}$ and $\beta_{\mathrm{m}}^{\mathrm{th}}$ for the spreading of the series of local maxima as a function
of the power-law exponent $\alpha$ defining the long-range spin exchange coupling along the $x$ axis. Figure~(a) extracted from Ref.~\cite{cevolani2018}.}
\label{fig:spins_LRTI}
\end{figure}

\noindent
Using again the quasiparticle theory, $G_x$ can be written in the form of Eq.~\eqref{generic_form} with an amplitude function $\mathcal{F}$ defined as follows 
(see Appendix.~\ref{appendix_gx_lrti})

\begin{equation}\label{eq:LRTI.Fk}
 \mathcal{F}(k) = \frac{h\left(J_\textrm{i}-J_\textrm{f}\right)P_\alpha(k)}{8\left[h+J_\textrm{f}P_\alpha(k)\right]\sqrt{h \left[ h + J_\textrm{i} P_\alpha(k) \right]}}.
\end{equation}

\noindent
A sudden global quench on $J$, where $J_\mathrm{i}$ ($J_{\mathrm{f}}$) refers to the pre-quench (post-quench) spin exchange coupling along the $x$ axis, is 
considered while maintaining constant both the transverse magnetic field $h$ and the power-law exponent $\alpha$.
According to the analytical expression of $P_{\alpha}$ in the infrared limit $k \rightarrow 0$ given at Eq.~\eqref{eq:Pinfrared},
the amplitude function $\mathcal{F}$ at Eq.\eqref{eq:LRTI.Fk} converges to a finite value. Consequently, $\mathcal{F}(k)
\sim k^\nu$ with $\nu=0$. For the calculations corresponding to the 1D LRTI model at Fig.~\ref{fig:spins_LRTI}, we find $\max\{\vert 1/2 - \langle \hat{S}^z_R
\rangle \vert\}=0.11$, which also validates the spin-wave approximation for this model. \\

To sum up, concerning the 1D LRTI model, we have $z=\alpha-1$, $\nu=0$ and $D=1$. It yields the theoretical exponent $\beta_{\mathrm{CE}}$ 

\begin{equation}
 \beta_{\mathrm{CE}} = \frac{\chi}{\gamma} = 2-z = 3-\alpha,
 \label{ce_lrti}
\end{equation}

\noindent
defining the spreading of the CE for the spin-spin correlations along the $x$ axis. The latter is completely characterized by the dynamical exponent $z$ (or equivalently
by the power-law exponent $\alpha$). For $\alpha=1.7$, see Fig.~\ref{fig:spins_LRTI}(a), by analyzing the analytical result, one can extract the CE exponent 
which is found to be equal to $\beta_{\mathrm{CE}}^{\mathrm{fits}} = 1.28 \pm 0.02$, in excellent agreement with the theoretical value $\beta_{\mathrm{CE}} = 1.3$.
On Fig.~\ref{fig:spins_LRTI}(b), the fitted correlation edge exponents are compared to the theoretical predictions of Eq.~\eqref{ce_lrti} for different power-law
exponents $\alpha$ confined in the quasi-local regime of the $z$ polarized phase. The numerical and theoretical results are found to be in very good agreement.
The latter permits to certify that the CE propagates sub-ballistically ($\beta_{\mathrm{CE}} > 1$) with an exponent given by the mean field value $3-\alpha$.
Note that the general formula for $\beta_{\mathrm{CE}}$ matches the analytical result of Ref.~\cite{cevolani2015} using the linear spin-wave theory and confirmed
by t-VMC calculations for $\alpha=3/2$ (\textit{ie.} $z=1/2$) where the authors found $\beta_{\mathrm{CE}} \simeq \alpha = 3/2$. It is also in fair agreement
with the analysis of Ref.~\cite{cevolani2016} for the same exponent $\alpha$. \\
Concerning the spreading of the series of local extrema, we found that the dynamical exponent $z$ (or equivalently $\alpha$) in the excitation spectrum 
$E_k$ of a gapped phase is irrelevant. Indeed, the series of local maxima spread always ballistically, 

\begin{equation}
t \propto  R^{\beta_{\mathrm{m}}}~~~\mathrm{with}~~~\beta_\mathrm{m}=1.
\end{equation}

\noindent
This case applies to the 1D LRTI model in the $z$ polarized case where the transverse magnetic field will open a finite gap in the excitation spectrum $E_k$, 
see Fig.~\ref{ek_vg_lrti}(a). It is confirmed in Fig.~\ref{fig:spins_LRTI}(a), where we observe that the local maxima converge to a ballistic
propagation for sufficiently long times. Performing the same analysis as for the LRXY chain, we find $\beta_{\mathrm{m}}^{\mathrm{fits}} \simeq 1.0045 \pm 0.0003$, in 
excellent agreement with the theoretical prediction. The scaling law for the maxima spreading is confirmed at Fig.~\ref{fig:spins_LRTI}(b) by scanning the
power-law exponent $\alpha$ where $\beta_{\mathrm{m}}^{\mathrm{fits}}$ remains the same. \\

In this paragraph, we briefly outline the behavior of both exponents ($\beta_{\mathrm{CE}}$ and $\beta_{\mathrm{m}}$) when approaching the local regime. The latter
corresponds to large values of $\alpha$ \textit{ie.} fastly-decaying long-range interactions. It is reached for $\alpha \geq 2$ for the long-range transverse
Ising chain in the $z$ polarized phase and $\alpha \geq 3$ for the long-range XY chain in the $x$ polarized phase. 
On Figs.~\ref{fig:spins_LRTI}(b) and \ref{fig:spins_LRXY}(b), the scaling laws governing the spreading of the maxima and the CE converge towards $\beta = 1$.
The latter correspond to a ballistic motion of each structure. Indeed, both long-range interacting spin chains will enter the local regime, where the scaling laws
are expected to be the same as those for short-range interacting quantum systems \textit{ie.} ballistic motion for the maxima and the CE characterized by the velocity 
$2V_\varphi^*$ and $2V_\mathrm{g}^*$ respectively. \\ \\

We have shown in this chapter that the correlation spreading has a universal twofold structure whose scaling laws can be related 
to different characteristic spectral properties of the quantum lattice model. \\
For short-range interacting quantum systems, they are readily associated to the group and phase velocities at the quasimomentum where the group
velocity reaches its maximum, which generally differ. \\
For long-range interacting quantum systems with a diverging group velocity (\textit{ie.} quasi-local regime), the spreading of the CE will depend on the observable
\textit{via} the exponent $\nu$ but also on the quasimomentum dependence of the excitation spectrum $E_k$ \textit{via} the exponent $z$ and is always sub-ballistic.
In the vicinity of the CE, the series of local maxima propagate ballistically in gapped systems, contrary to the gapless case where the maxima
spread super-ballistically and are fully characterized by the power-law exponent $\alpha$ and the lattice dimensionality $D$. 
As a consequence, these observations can thus be used as an experimental footprint for the presence of a spectral gap for instance. \\
This double structure can also be observed experimentally by investigating the spin correlations in cold ion chains. Typical relevant distance and time scales are
$N_s = 50$ ($N_s$ corresponds to the number of ions) and $T = 30 \hbar/J$ ($T$ denotes the observation time). Our analysis provides just the first step of an important research problem that
aims at unveiling the physical information encoded in correlation spreading and how this can be extracted in the next generation of experiments (see also \cite{kormos2017} for recent
results in this direction). In practice, the dynamics of the local maxima is easier to observe, and, as discussed above, our predictions are consistent with the 
existing observations. Furthermore, our theory shows that in generic experiments, characterizing the spreading of correlations for both particle and spin
lattice models, the data need to be interpreted carefully. Indeed, the propagation of local extrema does not characterize the correlation edge at all.
Both are independent and rely on different physical properties of the model. Identifying the latter requires an accurate scaling analysis of the leaks.
Existing experimental data have been collected either in a regime of parameters where the two structures coincide~\cite{cheneau2012}, or on small systems
where quantitative analysis is obfuscated by strong finite-size effects. However the next generation of experiments based on Rydberg atoms, tampered wave-guides,
and larger trapped-ion systems provide the natural setup to discern between the CE and the local features as our calculations suggest. \\

In the two last chapters, the correlation spreading is investigated numerically for lattice models with short- and long-range interactions using tensor
network based techniques. For the short-range (long-range) case, the Bose-Hubbard (long-range transverse Ising) chain is considered. 
In a first time, the purpose is to confirm our theoretical predictions for the spreading velocities (scaling laws) of the twofold causality cone presented in this chapter. 
In a second time, the discussion is extended to situations beyond the scope of our quasiparticle theory. For short-range interactions,
the correlation spreading is studied for specific interaction parameters such that the Bose-Hubbard chain is not exactly-solvable, \textit{e.g.} sudden global quenches
confined in the superfluid strongly interacting regime or close to critical points. For long-range interactions, the correlation dynamics for strong sudden global quenches 
(corresponding to a pre- and post-quench Hamiltonians in different quantum phases) is investigated. Finally, the out-of-equilibrium 
dynamics induced \textit{via} sudden local quenches is also widely studied. To give an example, both the correlation and entanglement spreading are investigated 
for the long-range transverse Ising chain. The latter allows us to discuss the differences and similarities between the global and local quench dynamics.

%% file: text/ch4_bose_hubbard_chain.tex
\setstretch{1.0} 

\begin{savequote}[8cm]
\textlatin{“If you come from mathematics, as I do, you realize that there are many problems, even classical problems, which cannot be solved by computation alone.”}
  \qauthor{--- Roger Penrose}
\end{savequote}

\chapter{Twofold correlation cone in a short-range interacting quantum lattice model} 
\label{ch:4-bose_hubbard_chain}
\minitoc

\newpage 
\clearpage

In this chapter, we investigate numerically the correlation spreading in 1D short-range interacting particle and spin 
quantum lattice models for sudden global and local quenches.
A special care will be devoted to the 1D short-range Bose-Hubbard model due to the wealth of its phase diagram.
The latter displays two different quantum phases \textit{ie.} the gapless superfluid (SF) and gapped
Mott-insulating (MI) phases. These two phases are separated by two different phase transitions, the Mott-$\delta$ and Mott-$U$ transitions
of mean field and topological type respectively. Finally, each quantum phase has different regimes \textit{e.g.} the mean field, strongly correlated
and strongly interacting regimes for the SF phase. More details about these quantum phases, both phase transitions and the different regimes are
provided in the following. The purpose of the numerical investigation of the correlation spreading in such quantum system is twofold. 
Firstly, for the limit regimes (mean field regime and strong-coupling regime for the SF and MI phase respectively) where the Bose-Hubbard model
can be diagonalized in terms of canonical transformations, we aim at confirming the existence and the spreading velocities of the twofold linear correlation cone 
predicted in the previous chapter. The numerical results will also suggest to extend our quasiparticle theory.
Secondly, we turn to a study of the correlation spreading in quantum regimes where the model is not exactly-solvable \textit{e.g.} close to the critical points or
in the strongly interacting regime of the SF phase. In such regimes, where the quasiparticles can not be characterized theoretically, the numerical approach is
essential and permits to go beyond the scope of our quasiparticle theory for the correlation spreading. The aim is to determine whether a linear twofold structure
subsists or not and to characterize the corresponding spreading velocities. 

\section{The short-range Bose-Hubbard chain}
\label{1dbhm}

\subsubsection{Mott-$U$ and Mott-$\delta$ phase transitions}

\begin{figure}[b!]
\centering
\includegraphics[scale = 0.44]{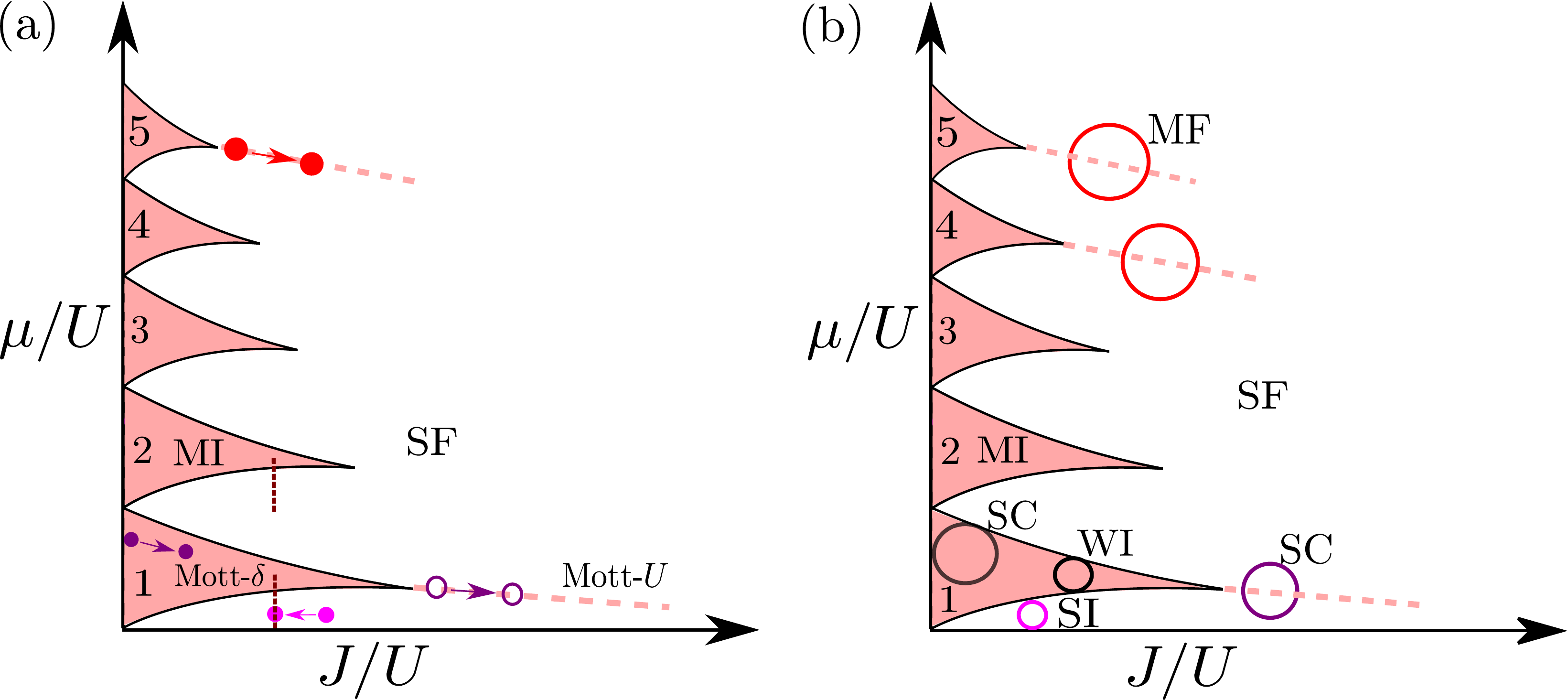}
\caption{\label{fig:phase_diagram_bhm}
Schematic phase diagram of the 1D SRBH model at equilibrium for $T=0$ as a function of the inverse interaction strength $J/U$ and the rescaled chemical
potential $\mu/J$, comprising a MI phase (pink lobes at integer fillings $\bar{n}$) and a SF phase (white region). (a)~ The Mott-$U$ transition at unit filling $\bar{n}=1$ is indicated 
by the dashed pink line and the Mott-$\delta$ transition by the vertical line. The arrows indicate various quenches considered in the following sections
(b)~ Graphical representation of the different regimes in the SF and MI phases. For the SF phase, the mean field regime (MF) is indicated by red circles, the 
strongly correlated regime (SC) at unit-filling by a purple circle and the strongly interacting regime (SI) by a pink one. Concerning the MI phase, the
weakly interacting (WI) and the strong-coupling (SC) regimes at $\bar{n}=1$ are represented by black circles. Figure~(a) extracted from Ref.~\cite{despres2019}.}
\end{figure}

The Hamiltonian of the one-dimensional (1D) Short-Range Bose-Hubbard (SRBH) model reads as

\begin{equation}
\hat{H}=-J \sum_{R} \left( \hat{a}^{\dagger}_{R} \hat{a}_{R+1}+\mathrm{h.c.}\right)+\frac{U}{2}\sum_R\hat{n}_{R}(\hat{n}_{R}-1),
\label{ham_bhm}
\end{equation}

\noindent
where $\hat{a}_R$ and $\hat{a}_{R}^{\dagger}$ are the bosonic annihilation and creation operators on site $R$, $\hat{n}_R = \hat{a}^{\dagger}_R \hat{a}_R$
is the occupation number (filling), $J$ is the hopping amplitude, $U>0$ is the repulsive on-site interaction energy, and the lattice spacing is fixed to unity
($R\in\mathbb{Z}$). At equilibrium and zero-temperature ($T=0$), the phase diagram of the 1D SRBH model is well known ~\cite{sachdev2001,cazalilla2011}, and
sketched on Fig.~\ref{fig:phase_diagram_bhm}. It comprises a superfluid (SF) and a Mott-insulating (MI) phase, determined by the competition of the hopping, the 
interactions, and 
the average filling $\bar{n}$ or, equivalently, the chemical potential $\mu$ depending if the canonical or grand canonical statistical ensemble is considered. \\

This bosonic lattice model hosts two phase transitions of different type. For commensurate filling, $\bar{n}\in\mathbb{N}^*$ the SF-MI, also called 
Mott-$U$, phase transition is of the Berezinskii-Kosterlitz-Thouless (BKT) type, at the critical value $(U/J)_c \simeq 3.3$ for unit filling ($\bar{n}=1$)
in 1D~\cite{kuhner2000,kashurnikov1996exact,ejima2011,rombouts2006}. Indeed, the critical point $(U/J)_c$ depends not only on the integer filling $\bar{n}$ 
but also on the dimensionality of the lattice, $D$. Furthermore, the Bose-Hubbard chain at fixed and commensurate filling $\bar{n}$ and $T=0$ has the same
universality class as the 1D XY model at $T=0$. Therefore, its critical behavior can be mapped onto the one governing the 2D classical XY model at $T\neq 0$.
The latter displays a BKT topological phase transition between a quasi-long-range order phase (algebraic decay for the relevant correlation functions at equilibrium)
at low temperatures (low-$T$) and a short-range order phase (exponential decay for the same correlation functions) at high temperatures (high-$T$). For low-$T$,
the long-range order is prevented due to the Mermin-Wagner theorem stating that a continuous symmetry cannot be spontaneously broken for short-range interacting
lattice models in dimension $D \leq 2$ at $T\neq 0$. \\
For incommensurate filling ($\bar{n} \notin \mathbb{N}$), the Bose gas is a SF for any value of the interaction parameter $U/J$. The
commensurate-incommensurate (Mott-$\delta$) transition, of the mean field type, is then driven by doping when $\bar{n}$ approaches a positive integer value
for sufficiently strong local interactions.

\subsubsection{The superfluid phase}
The superfluid phase, characterized by a gapless excitation spectrum, is found for incommensurate fillings $\bar{n} \notin \mathbb{N}$
for any value of the interaction parameter $U/J$ or at commensurate fillings for sufficiently small two-body repulsive interactions. It corresponds to a 
delocalization of the bosonic particles through the lattice. Indeed, in the limit case of $U/J \rightarrow 0$ and for a commensurate 
or incommensurate filling $\bar{n}$, the bosonic particles are non-interacting and are thus fully delocalized over the lattice. Hence, the ground state of the
lattice model is given by the many-body wavefunction $\ket{\Psi}$ which may be written as

\begin{equation}
 \ket{\Psi} = \frac{1}{\sqrt{N!}} \left( \frac{1}{\sqrt{N_s}}  \sum_{R=1}^{N_s} \hat{a}_R^{\dag} \right)^N \ket{0} =
 \frac{1}{\sqrt{N!}} (\hat{a}^{\dag}_{k=0})^N \ket{0},
 \label{limit_case_sf}
\end{equation}

\noindent
where $\ket{0} \equiv \ket{0}^{\otimes N_s}$ denotes the vacuum state in the Fock space, $N$ the total number of bosons and $N_s$ the number of lattice sites.
Indeed, the fact to be completetly delocalized in real space leads to a full
localization in the reciprocal (momentum) space. Consequently, the mode $k=0$ is macroscopically occupied by the bosons. In the limit of large $N$ and $N_s$, 
Eq.~\eqref{limit_case_sf} is well described by a coherent state which can be expressed as  

\begin{equation}
\ket{\Psi} =  \prod_{R=1}^{N_s} \ket{\alpha_R},~~~ \ket{\alpha_R} = e^{-\frac{|\alpha|^2}{2}}e^{\alpha \hat{a}^{\dag}_R} \ket{0_R} = e^{-\frac{|\alpha|^2}{2}} 
\sum_{n_R=0}^{+\infty} \frac{\alpha^{n_R}}{\sqrt{n_R!}}\ket{n_R},
\label{cgs}
\end{equation}

\noindent
where $n_R$ denotes the bosonic occupation number, $\ket{0_R}$ the vacuum state, on the lattice site $R$. The parameter $\alpha \in \mathbb{C}$ is defined as
$\hat{a}_R \ket{\Psi} = \alpha \ket{\Psi}$ and corresponds to the complex eigenvalues of the non-hermitian annihiliation operator $\hat{a}_R$. An important physical
quantity is $P(n_R)$ the probability that $n_R$ bosons occupy a given lattice site $R$. For the coherent ground state at Eq.~\eqref{cgs} valid in the mean field
regime, the occupation probability $P(n_R)$ is nearly Poissonian and given by 

\begin{equation}
P(n_R) = |\langle n_R| \Psi \rangle |^2 \simeq \bar{n}^{n_R} \frac{e^{-\bar{n}}}{n_R!},
\label{P_mf_sf}
\end{equation}

\noindent
where $\alpha \simeq \sqrt{\bar{n}}$. The latter decreases fastly enough to truncate the local Hilbert space to a specific cutoff denoted by $n_{\mathrm{max}}$.
This permits to perform numerical simulations in the superfluid mean field regime using for instance the time-dependent Matrix Product State approach.
Note that truncating the local Hilbert space in such regime is subtle and requires for instance that $n_{\mathrm{max}} \gg \bar{n}$ (see Sec.~\ref{limit_regimes} for 
more details). \\

The superfluid phase has several regimes depending on the value of the Lieb-Liniger parameter $\gamma = U/2J\bar{n}$ explicitly calculated in Appendix.~\ref{appendix1_mf}.
For $\gamma \ll 1$ and large $\bar{n}$ the Bose gas is confined in the mean field regime. For
$\gamma \lesssim 1$, it is confined in the so-called strongly correlated regime. Finally for $\gamma \gg 1$ and an incommensurate $\bar{n}$, the Bose gas is in the
strongly interacting regime, see Fig.~\ref{fig:phase_diagram_bhm}(b).\\
As discussed in the previous chapter, the short-range Bose-Hubbard chain in the mean field regime can be diagonalized using a (bosonic)
Bogolyubov transformation where the elementary excitations correspond to Bogolyubov quasiparticles. Due to the translational invariance of the model, they are characterized by a well defined quasimomentum $k \in \mathcal{B}
= [-\pi,+\pi]$ where $a$ the lattice spacing is fixed to unity. The associated gapless excitation spectrum reads as, see Sec.~\ref{sf_chap3} for more details, 

\begin{equation}
\label{eq:BogoDisp}
E_k \simeq \sqrt{\gamma_k \left(\gamma_k + 2 \bar{n} U\right)},
\end{equation}

\noindent
where $\gamma_k = 4J \sin^2(k/2)$ is that of the free-particle tight-binding model. In this regime, the Bose gas possesses quasi-long-range order (at $T=0$)
due to relatively high long-range phase fluctuations and corresponding to the low-$T$ phase of the 2D classical XY model. Therefore, at equilibrium, the latter
implies a power-law decay for $G_1$ and $G_2$ the phase and density fluctuations respectively. Indeed, $G_1(R) \sim 1/R^{1/2K}$ and  $G_2(R) \sim -K/R^2$
where $K$ denotes the Luttinger parameter \cite{cazalilla2011}. $K$ decreases from $+\infty$ to $1$ when $U/J$ increases from $0$ to $+\infty$. \\
For the strongly correlated and strongly interacting regimes, the model is not exactly-solvable in terms of canonical transformations. However, the excitation spectrum
$E_k$ valid in the mean field regime is expected to hold for the strongly correlated regime (small integer $\bar{n}$) at sufficiently small $\gamma$. 

\subsubsection{The Mott-insulating phase}
We now turn to the MI phase characterized by a gapped excitation spectrum. It implies a commensurate filling $\bar{n} \in \mathbb{N}^*$ and sufficiently strong local
interactions $U$. This phase tends to pin the bosonic particles in the lattice sites by minimizing the fluctuations of the on-site occupation number, see 
Fig.~\ref{sketch_bhm}(b). Indeed, in the limit case of $U/J \rightarrow \infty$ and a commensurate filling $\bar{n}$, the hopping term is negligible with 
respect to the two-body repulsive interaction strength and the bosons are fully localized and occupy uniformly the lattice sites. Hence, the ground state of the lattice model may be written as

\begin{equation}
 \ket{\Psi} = \prod_{R=1}^{N_s} \frac{\left( \hat{a}_R^{\dag} \right)^{\bar{n}}}{\sqrt{\bar{n}!}}\ket{0} = \ket{\bar{n}}^{\otimes N_s} \equiv \ket{\bar{n}},
 \label{limit_case_mi}
\end{equation}

\noindent
meaning that each lattice site $R$ is occupied by $\bar{n}$ bosons. This gapped phase has short-range order and corresponds to the high-$T$ phase of
the 2D classical XY model. At equilibrium, the latter implies that both correlation functions, $G_1$ and $G_2$, decay exponentially
with respect to a correlation length $\xi_1$, $\xi_2$. Hence, both the phase and density fluctuations may be written as $G_{1,2}(R) \sim e^{-R/\xi_{1,2}}$.\\

Furthermore, the Mott-insulating phase displays two different regimes, the weakly interacting and strong-coupling regimes depending on the value of the
interaction parameter $U/J$ at fixed $\bar{n} \in \mathbb{N}^*$, see Fig.~\ref{fig:phase_diagram_bhm}(b). \\
For the strong-coupling regime at $U/J \gg 1$, the Bose-Hubbard chain can be diagonalized using a first-order perturbation theory already discussed
at Sec.~\ref{mi_chap3}. In this regime, the low-energy excitations correspond to doublon-holon pairs where a doublon denotes a lattice site $R$ 
occupied by $\bar{n}+1$ bosons whereas a holon corresponds to $\bar{n}-1$ bosons on the nearest-neighbor lattice site $R\pm1$. The excitation spectrum 
is then characterized by, see Refs.~\cite{altman2002,huber2007,barmettler2012},

\begin{equation}
 2E_k = U - 2J(2\bar{n}+1)\cos(k)
 \label{ek_chap4_mi}
\end{equation}

\noindent
where $k$ is the quasimomentum confined in the first Brillouin zone $\mathcal{B} = [-\pi, \pi]$. According to Eq.~\eqref{ek_chap4_mi}, the excitation spectrum 
develops a finite gap $\Delta = U-2J(2\bar{n}+1)$ which mainly depends on the local interaction $U$ (since $U \gg J\bar{n}$ in the strong-coupling regime)
and a cosine profile coming from the hopping amplitude $J$ coupling the nearest-neighbor lattice sites. \\
Deep enough in the MI phase, $U/J \gtrsim 2(2\bar{n}+1)$, corresponding to the weakly interacting regime, the bosonic lattice model is exactly-solvable using a fermionization
technique. Indeed, the bosonic particles can be treated as hardcore bosons leading to a effective 
quadratic model of interacting fermions. To do so, one doublon-holon excitation pair is allowed such that the local Hilbert space can be reduced to only three
different quantum states which are $\ket{\bar{n}+1}$, $\ket{\bar{n}}$ and $\ket{\bar{n}-1}$. Then, auxiliary bosons are introduced to switch between the three 
previous quantum states. Finally, the new bosonic operators are fermionized using a Jordan-Wigner transformation where the hardcore constraint is directly encoded.
It yields a effective quadratic Hamiltonian which can be diagonalized using a fermionic Bogolyubov transformation (see Ref.~\cite{barmettler2012} for a 
complete discussion about the technique). As a consequence, the elementary excitations, which still correspond to doublon-holon excitation pairs,
can be seen as fermionic Bogolyubov quasiparticles and are characterized by the following gapped excitation spectrum~\cite{barmettler2012,Ejima2012}

\begin{equation}
2E_k \simeq \sqrt{\left[U-2J(2\bar{n}+1)\cos(k)\right]^2 + 16 J^2 \bar{n}(\bar{n} + 1)\sin^2(k)}.
\label{mott_disp_rel_fermionization}
\end{equation}

\noindent
Note that Eq.~\eqref{mott_disp_rel_fermionization} is also valid in the strong-coupling regime. By developing the latter and keeping the first-order terms in $U/J$, 
one recovers the excitation spectrum at Eq.~\eqref{ek_chap4_mi} valid in the strong-coupling limit.

\section{Time-dependent matrix product state approach}
\label{tmps_approach}
The time-dependent matrix product state (t-MPS) is a powerful numerical approach to compute the static and dynamical physical properties of 1D quantum lattice models
based on an analysis of its entanglement entropy. Moreover, the latter has a graphical representation in terms of tensor networks providing a local
and compact representation of any low-entangled many-body quantum state without breaking its non-locality. We first present the transposition of the 
Dirac formalism of quantum mechanics including kets, bras, overlaps, operators and expectation values into tensor networks. Then, two iterative algorithms are presented.
We first present the one allowing to compute the static properties and in particular the ground state and its associated energy. Then, we turn to the algorithm to perform
real-time evolution for a generic 1D short-range interacting lattice model. The reader not interested in the details of the method can directly jump to 
Sec.~\ref{limit_regimes} where the numerical results for the propagation of correlations in the 1D SRBH model are discussed.

\subsection{Tensor-network based formalism}

\subsubsection{Matrix product states: properties and graphical representation}

We discuss now the Matrix Product State (MPS) form of a general quantum state consisting of a local and compact representation.
The corresponding graphical representation in term of tensor networks and its properties are also presented.\\

Concerning the MPS form, working along the lines of Refs.~\cite{schollwock2005,schollwock2011}, we first start from the most general quantum 
state for a one-dimensional lattice of length $L$. The latter can be expressed as

\begin{equation}
 \ket{\Psi} = \sum_{\sigma_1, \sigma_2,..., \sigma_L} \mathit{\Psi}_{\sigma_1 \sigma_2~ ...~\sigma_L}
 \ket{\sigma_1 \sigma_2~ ...~ \sigma_L},
 \label{general_qs}
\end{equation}

\noindent
with a $d$-dimensional local Hilbert space $\mathbb{H}_R$ described by the local basis $\{\ket{\sigma_R}, \sigma_R = 1,...,d \}$. The state vector denoted by 
$\mathit{\Psi}$ contains $d^L$ components which corresponds to the dimension of the many-body Hilbert space $\mathbb{H} = \otimes_{R=1}^{L} \mathbb{H}_R$.
Relying on the Singular Value Decomposition (SVD), we obtain the following form for the coefficients of the state vector $\mathit{\Psi}_{\sigma_1 \sigma_2~...~\sigma_L}$, 
see Appendix.~\ref{appendix3_mps_app},
\begin{equation}
\Psi_{\sigma_1, (\sigma_2~...~\sigma_L)} = \sum_{a_1=1}^{\bar{a}_1} \sum_{a_2=1}^{\bar{a}_2} ~...~ \sum_{a_{L-1} = 1}^{\bar{a}_{L-1}} A^{\sigma_1}\left[1\right]_{a_1} 
A^{\sigma_2}\left[2\right]_{a_1,a_2}~...~ A^{\sigma_L}\left[L\right]_{a_{L-1}}.
\label{2mps}
\end{equation}

\begin{figure}[!h]
\centering
\begin{tabular}{c}
\includegraphics[scale = 0.33]{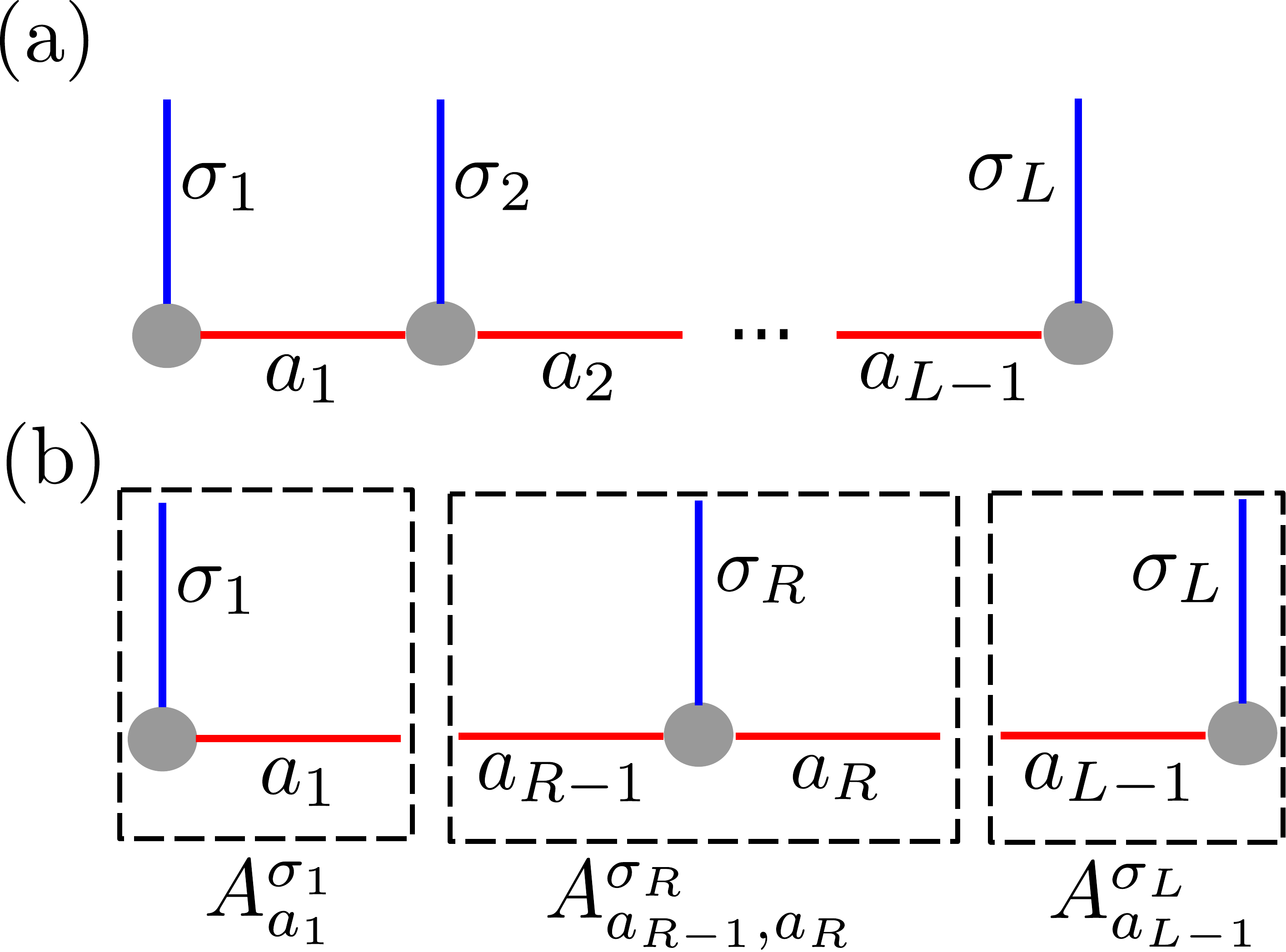}  \\
\end{tabular}
\caption{Graphical representation using tensor networks of a general quantum state $\ket{\Psi}$. (a)~ MPS representation of $\ket{\Psi}$ for a one-dimensional lattice of
$L$ sites. (b)~ Graphical representation of the $A$-matrices at the ends (left and right panels) and in the bulk (central panel) of the chain. The first diagram
represents the row vector $A^{\sigma_1}$ with entries $A^{\sigma_1}_{a_1} = A^{\sigma_1}_{1,a_1}$ and of dimension $1 \times \bar{a}_1$, the last diagram represents 
the column vector $A^{\sigma_L}$ with entries $A^{\sigma_L}_{a_{L-1}} = A^{\sigma_L}_{a_{L-1},1}$ and of dimension $\bar{a}_{L-1} \times 1$. The second 
diagram represents the matrix $A^{\sigma_R}$ with entries $A^{\sigma_R}_{a_{R-1},a_R}$ and of dimension $\bar{a}_{R-1} \times \bar{a}_R$. The red (blue)
line refers to the non-physical (physical) index which is related to the entanglement entropy (local Hilbert space) of the quantum state $\ket{\Psi}$.}
\label{MPS}
\end{figure}

\noindent
Consequently, a general (non translational invariant) quantum state written under its MPS form for a one-dimensional lattice has the
following expression, see Fig.~\ref{MPS}, 
\begin{equation}
\ket{\Psi} = \sum_{\boldsymbol{\sigma}} A^{\sigma_1}\left[1\right]A^{\sigma_2}\left[2\right] ~...~ A^{\sigma_{L-1}}\left[L-1\right] A^{\sigma_L}\left[L\right]
\ket{\boldsymbol{\sigma}},~~~\boldsymbol{\sigma} = \sigma_1 \sigma_2 ~...~\sigma_L.
\label{mps}
\end{equation}

\noindent
In the MPS representation of a quantum state, the first tensor $A^{\sigma_1}\left[1\right]$ consists of a collection of row vectors of dimension $\bar{a}_1$ and
the last tensor $A^{\sigma_L}\left[L\right]$ of a collection of column vectors of dimension $\bar{a}_{L-1}$. $\bar{a}_1$ ($\bar{a}_{L-1}$) corresponds to 
the rank of the first ($L-1$-th) Schmidt matrix, see Appendix.~\ref{appendix3_mps_app}. Besides, a tensor $A^{\sigma_R}[R]$ of dimension $\bar{a}_{R-1}
\times \bar{a}_R$ is associated to each lattice site $R \in [|1,L|]$ of the one-dimensional lattice model. Hence, the MPS form at Eq.~\eqref{mps} leads to a local 
representation of the many-body quantum state $\ket{\Psi}$ without breaking its non-locality, \textit{ie.} its entanglement which is contained in the coefficients
of each tensor. The degree of entanglement of the different tensors of $\ket{\Psi}$ are characterized by the so-called MPS bond dimension
$\chi$ defined by $\chi = \mathrm{max}(\bar{a}_R)$. For the specific case $\chi = 1$, the quantum state $\ket{\Psi}$ corresponds to a (not-entangled) product state. \\

In the following, a translational invariant quantum state will be considered \footnote{Note that this assumption is fulfilled when studying numerically the 
the 1D short-range Bose-Hubbard model.}. This assumption simplifies the notations and leads to 

\begin{equation}
\ket{\Psi} = \sum_{\boldsymbol{\sigma}} A^{\sigma_1}A^{\sigma_2} ~...~ A^{\sigma_{L-1}}
A^{\sigma_L}\ket{\boldsymbol{\sigma}}.
\label{mps2}
\end{equation}

\noindent
The previous MPS form implies that all the different tensors describing the many-body quantum state have the same dimension, \textit{ie.} corresponds
to a collection of $d$ square matrices of dimension $\chi$ (except for the first and last ones which contains a collection of row and column vectors respectively).
Indeed, the translational invariance of the quantum state $\ket{\Psi}$ requires that $A^{\tilde{\sigma}}\left[R\right] = 
A^{\tilde{\sigma}}\left[R+1\right]$, $\forall R\in[[2, L-2|]$ and $\forall \ket{\tilde{\sigma}} \in \mathbb{H}_R$. \\
We turn now to an investigation of the number of complex coefficients to store for the MPS form of the quantum state $\ket{\Psi}$. The latter
contains $d \chi^{2} L$ complex coefficients where $L$ denotes the number of lattice sites, $d$ the dimension of the local Hilbert space $\mathbb{H}_R$ and
$\chi$ refers to the MPS bond dimension. For any low-entangled many-body quantum state, \textit{ie} for $\ket{\Psi}$ corresponding to a superposition
of few quantum states, $\chi$ is relatively small. Besides, one notices that the number of coefficients depends linearly with $L$ and not exponentially 
anymore as for the initial form at Eq.~\eqref{general_qs}. Consequently, the two previous conditions lead for the MPS form to a very compact and local 
representation of any low-entangled quantum state $\ket{\Psi}$. 

\subsubsection{Gauge degree of freedom and normalization}

The local MPS representation is not unique since it is defined up to a gauge degree of freedom. Indeed, if we consider two adjacent tensors at the lattice site $R$ and
$R+1$ denoted by $M^{\sigma_R}$ and $M^{\sigma_{R+1}}$ of dimension $\chi \times \chi \times d$. The MPS form of a quantum state $\ket{\Psi}$ given at 
Eq.~\eqref{mps2} is invariant for any invertible matrix $G$ of dimension $\chi \times \chi$ since $M^{\sigma_R} G G^{-1} M^{\sigma_{R+1}} =
M^{\sigma_{R}} M^{\sigma_{R+1}}$. Fixing this gauge degree of freedom for the tensors allows to simplify significantly the calculations. There are three 
common ways to fix this gauge while enforcing the normalization of the quantum state (i) the left canonical representation by expressing $\ket{\Psi}$ \textit{via}
the left-normalized matrices $U[R]$. The latter appear when performing the different singular value decompositions and are cast into left-normalized tensors
$A^{\sigma_R}$ [see Eq.~\eqref{mps2}] (ii) the right canonical normalization by using the right-normalized matrices $V^{\dag}[R]$ which are then cast into 
right-normalized tensors $B^{\sigma_R}$ leading to the following MPS form 
\begin{equation}
\ket{\Psi} = \sum_{\boldsymbol{\sigma}} B^{\sigma_1}B^{\sigma_2}~...~ B^{\sigma_{L-1}} B^{\sigma_L}\ket{\boldsymbol{\sigma}}~~~\mathrm{where}~~~ B^{\sigma_R}_{a_{R-1},a_R} = 
(V^{\dag})_{a_{R-1},(\sigma_R,a_R)}.
\label{mps3}
\end{equation}

\noindent
(iii) the mixed canonical representation using the left-normalized matrices for the first $R$ lattice sites and the right-normalized ones for the last
$L-R$ lattice sites (see Eq.~\eqref{mps4} where $S$ is the $\mathrm{R}^{\mathrm{th}}$ Schmidt matrix with entries $S_{a_R, a_R}$),

\begin{equation}
\ket{\Psi} = \sum_{\boldsymbol{\sigma}} A^{\sigma_1}~...~ A^{\sigma_R} S B^{\sigma_{R+1}}~ ...~ B^{\sigma_L} 
\ket{\boldsymbol{\sigma}}. 
\label{mps4}
\end{equation}

\begin{figure}[!h]
\centering
\begin{tabular}{c}
\includegraphics[scale = 0.27]{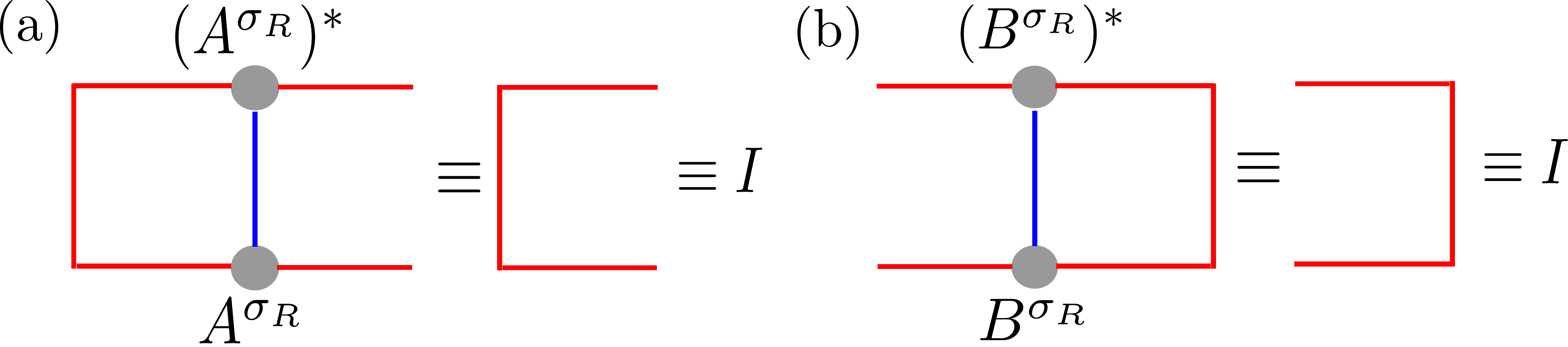}  \\
\end{tabular}
\caption{Graphical representation using tensor networks of the (a)~ left-normalization condition for the tensors $A^{\sigma_R}$ allowing a left-canonical
representation of a MPS $\ket{\Psi}$ [see Eq.~\eqref{left}] (b)~right-normalization condition for the tensors $B^{\sigma_R}$ [see Eq.~\eqref{right}] allowing
a right-canonical representation of a MPS $\ket{\Psi}$.}
\label{left_right}
\end{figure}

\noindent
These three representations of the many-body quantum state $\ket{\Psi}$, corresponding to three different ways to fix the MPS gauge degree of freedom, enforce its
normalization. The left-normalized (right-normalized) matrices $U[R]$ ($V^{\dag}[R]$) lead to a left- (right-) normalization condition for the
tensors $A^{\sigma_R}$ ($B^{\sigma_R}$). For instance, one can write for a left-canonical MPS $\ket{\Psi}$ in terms of $A$ tensors,
\begin{align}
&\delta_{a_R, a'_R} = \sum_{a_{R-1},\sigma_R} (U^{\dag})_{a_R,(a_{R-1}\sigma_R)} U_{(a_{R-1}\sigma_R),a'_R} =
\sum_{a_{R-1},\sigma_R} (A^{\sigma_R})^{\dag}_{a_R,a_{R-1}} A^{\sigma_R}_{a_{R-1},a'_R},
\end{align}

\noindent
leading finally to the so-called left-normalization of the tensors $A^{\sigma_R}$ where the graphical representation is shown on Fig.~\ref{left_right}(a), 

\begin{equation}
 \sum_{\sigma_R} (A^{\sigma_R})^{\dag} A^{\sigma_R} = I.
 \label{left}
\end{equation}

\noindent
For a right-canonical representation of $\ket{\Psi}$, the derivation is similar in order to find the right-normalization condition of the tensors $B^{\sigma_R}$, see 
Fig.~\ref{left_right}(b), given by 

\begin{equation}
 \sum_{\sigma_R} B^{\sigma_R} (B^{\sigma_R})^{\dag} = I.
 \label{right}
\end{equation}

\noindent
The left- and right-normalization condition for the tensors $A^{\sigma_R}$ and $B^{\sigma_R}$ for a left- or right-canonical representation of the MPS $\ket{\Psi}$
enforce its normalization \textit{ie.} $\langle \Psi | \Psi \rangle = 1$. Indeed, let us consider a left-canonical representation of $\ket{\Psi}$ 
given at Eq.~\eqref{mps2}, the overlap $\langle \Psi | \Psi \rangle$ may be written as
\begin{equation}
 \langle \Psi | \Psi \rangle = \sum_{\boldsymbol{\sigma}} (A^{\sigma_1} ~...~A^{\sigma_L})^{\dag} 
(A^{\sigma_1}~...~A^{\sigma_L}) = \sum_{\sigma_L} (A^{\sigma_L})^{\dag} ~...~ \left( \sum_{\sigma_1} 
(A^{\sigma_1})^{\dag} A^{\sigma_1} \right) ...~A^{\sigma_L}.
\end{equation}

\begin{figure}[!h]
\centering
\begin{tabular}{c}
\includegraphics[scale = 0.35]{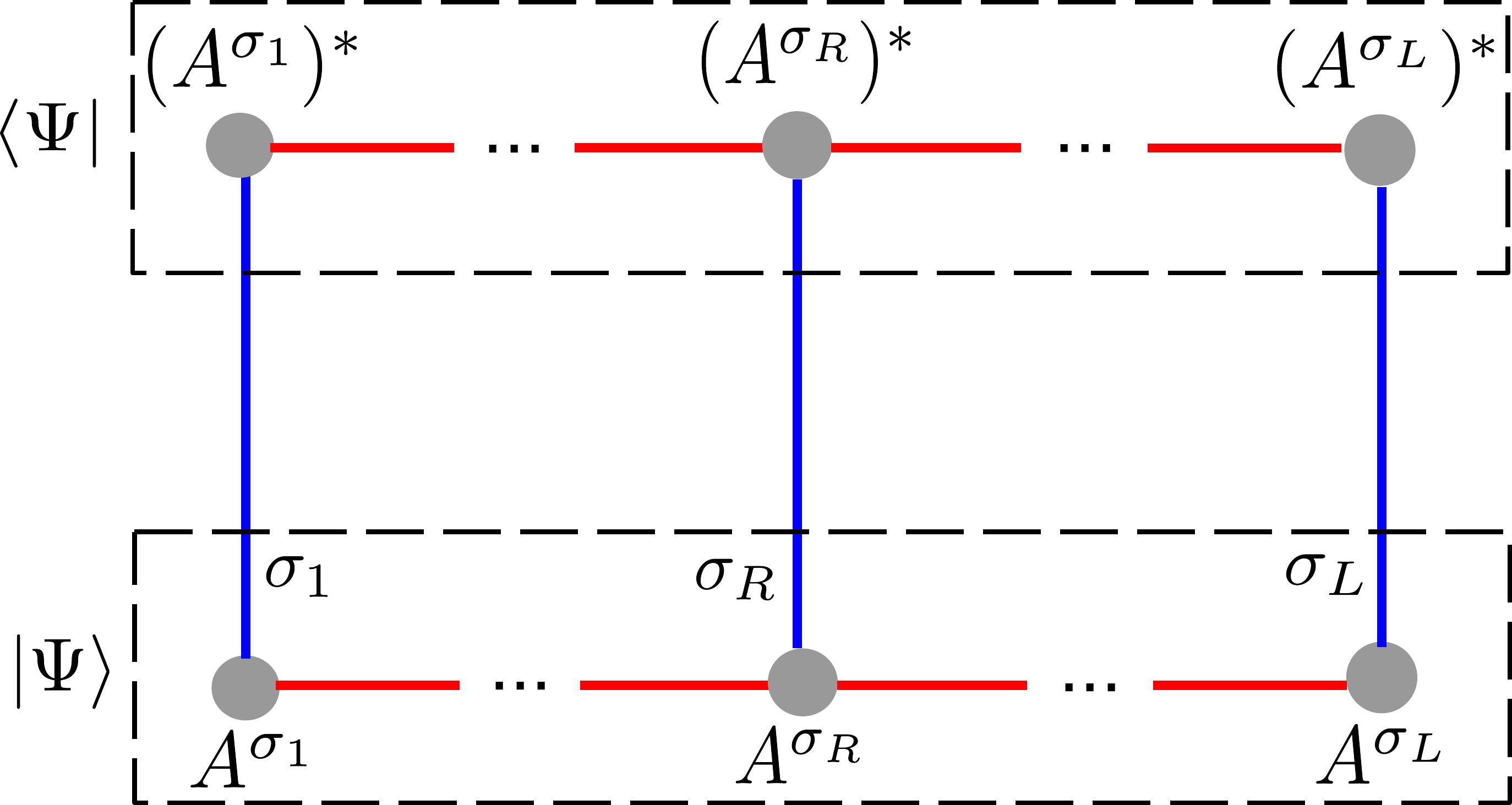}  \\
\end{tabular}
\caption{Graphical representation using tensor networks of the overlap $\langle \Psi | \Psi \rangle$ where $\ket{\Psi}$ is a left-canonical MPS. For simplicity,
the red lines on the left side of $A^{\sigma_1}$, $(A^{\sigma_1})^*$ and on the right side of $A^{\sigma_L}$, $(A^{\sigma_L})^*$ have been omitted since their dimensions
are equal to $1$.}
\label{overlap}
\end{figure}

\noindent
Then, according to Eq.~\eqref{left} for the left-normalization condition of the tensors $A^{\sigma_R}$, it comes directly that
$\langle \Psi | \Psi \rangle = 1$ hence $\ket{\Psi}$ is well normalized. The previous statement is also obvious when representing graphically the
overlap $\langle \Psi | \Psi \rangle$, see Fig.~\ref{overlap}, and the left-normalization condition at Fig.~\ref{left_right}(a).

\subsubsection{Matrix product operators: properties and graphical representation}
In order to get a complete transposition of the Dirac formalism in terms of tensor networks and more precisely to be
able to find the ground state (energy) and to perform real time evolution for 1D short-range interacting lattice models,
it requires to find a matrix product form of operators also called Matrix Product Operators (MPOs), see 
Refs.~\cite{pirvu2010,schollwock2005,schollwock2011}. The most general operator $\hat{O}$ can be expressed under the following form 

\begin{equation}
\hat{O} = \sum_{\boldsymbol{\sigma}',\boldsymbol{\sigma}} \mathit{O}^{\sigma_1' \sigma_2' ~...~ \sigma_{L}'}_{\sigma_1 \sigma_2 ~...~ \sigma_{L}}
\ket{\boldsymbol{\sigma}'}\bra{\boldsymbol{\sigma}},
\end{equation}

\noindent
where $\mathit{O}$ is a $2L$-th order tensor containing $d^{2L}$ coefficients. Considering a single coefficient $\langle \boldsymbol{\sigma} | \Psi \rangle$
with $\ket{\Psi}$ a left-canonical MPS, it yields the following equation

\begin{equation}
\langle \boldsymbol{\sigma} |\Psi \rangle = \bra{\sigma_1~...~\sigma_L} \sum_{\boldsymbol{\sigma'}} A^{\sigma'_1}
~...~A^{\sigma'_L}\ket{\sigma'_1~...~\sigma'_L}  = A^{\sigma_1}~...~A^{\sigma_L}.
\end{equation}

\noindent
Therefore, it is natural to write the expectation $\bra{\boldsymbol{\sigma'}}\hat{O}\ket{\boldsymbol{\sigma}}$ of an operator
$\hat{O}$ as follows 

\begin{equation}
\bra{\boldsymbol{\sigma}}\hat{O}\ket{\boldsymbol{\sigma}'} = W^{\sigma_1\sigma_1'}~...~W^{\sigma_L\sigma_L'}.
\end{equation}

\noindent
Consequently, an operator $\hat{O}$ represented under the MPO form can be written as 

\begin{equation}
\hat{O} = \sum_{\boldsymbol{\sigma}',\boldsymbol{\sigma}} W^{\sigma_1'\sigma_1}...~W^{\sigma_L'\sigma_L} 
\ket{\boldsymbol{\sigma}'}\bra{\boldsymbol{\sigma}},
\label{MPO}
\end{equation}

\noindent
where the $4$-th order $W$ tensors have two physical indices, one associated to the ket and the bra in this specific
order (contrary to the MPS form having only one physical index) and two virtual indices (related to the range of the interactions for a Hamiltonian)
represented by the indices $a_R$ with $R\in[|1,L-1|]$ where a complete description reads as follows

\begin{equation}
\hat{O} = \sum_{\boldsymbol{\sigma}',\boldsymbol{\sigma}} \sum_{\boldsymbol{a}} W^{\sigma_1'\sigma_1}_{1,a_1}~...~
W^{\sigma_L'\sigma_L}_{a_{L-1},1} \ket{\boldsymbol{\sigma}'}\bra{\boldsymbol{\sigma}}, ~~~\boldsymbol{a} = a_1,a_2,...,a_{L-1}.
\end{equation}

\noindent
The MPO form presented at Eq.~\eqref{MPO} can be cast into a more natural form given by

\begin{equation}
\hat{O} = \hat{W}[1] \hat{W}[2]~ ...~ \hat{W}[L],~~~ \hat{W}[R] = \sum_{\sigma_R', \sigma_R} W^{\sigma_R' \sigma_R} \ket{\sigma_R'} \bra{\sigma_R} 
\label{mpo_form_2}
\end{equation}

\noindent
where the index $R$ refers to the lattice site, and can be found easily for 1D short-range interacting particle
or spin Hamiltonians. Indeed, one needs to find for each lattice site $R$ a matrix of operators acting on the local Hilbert
space $\mathbb{H}_R$. For instance, the MPO form associated to the 1D Bose-Hubbard model is given by the following matrices 

\begin{align}
& \hat{W}[1] = 
\begin{pmatrix}
 \frac{U}{2} \hat{n}_1 \left( \hat{n}_1 - 1 \right) & -J \hat{a}^{\dag}_1 & -J \hat{a}_1 & I
\end{pmatrix} \\
& \hat{W}[L] = 
\begin{pmatrix}
 I & \hat{a}^{\dag}_{L} & \hat{a}_{L} & \frac{U}{2} \hat{n}_{L} \left( \hat{n}_{L} - 1 \right)  
\end{pmatrix}^{T} \\
& \hat{W}[R] = 
\begin{pmatrix}
 I & 0 & 0 & 0 \\
 \hat{a}_R & 0 & 0 & 0 \\
 \hat{a}^{\dag}_R & 0 & 0 & 0 \\
 \frac{U}{2} \hat{n}_R \left( \hat{n}_R - 1 \right) & -J \hat{a}_R & -J \hat{a}^{\dag}_R & I \\
 \end{pmatrix}, \forall R \in [|2,L-1|]
\end{align}

\noindent
Indeed, when calculating the product $\hat{W}[1]~...\hat{W}[R]~...~\hat{W}[L]$, one recovers the 1D short-range Bose Hubbard model
whose Hamiltonian is represented at Eq.~\eqref{H_bhm}. \\

\begin{figure}[!h]
\centering
\begin{tabular}{c}
\includegraphics[scale = 0.41]{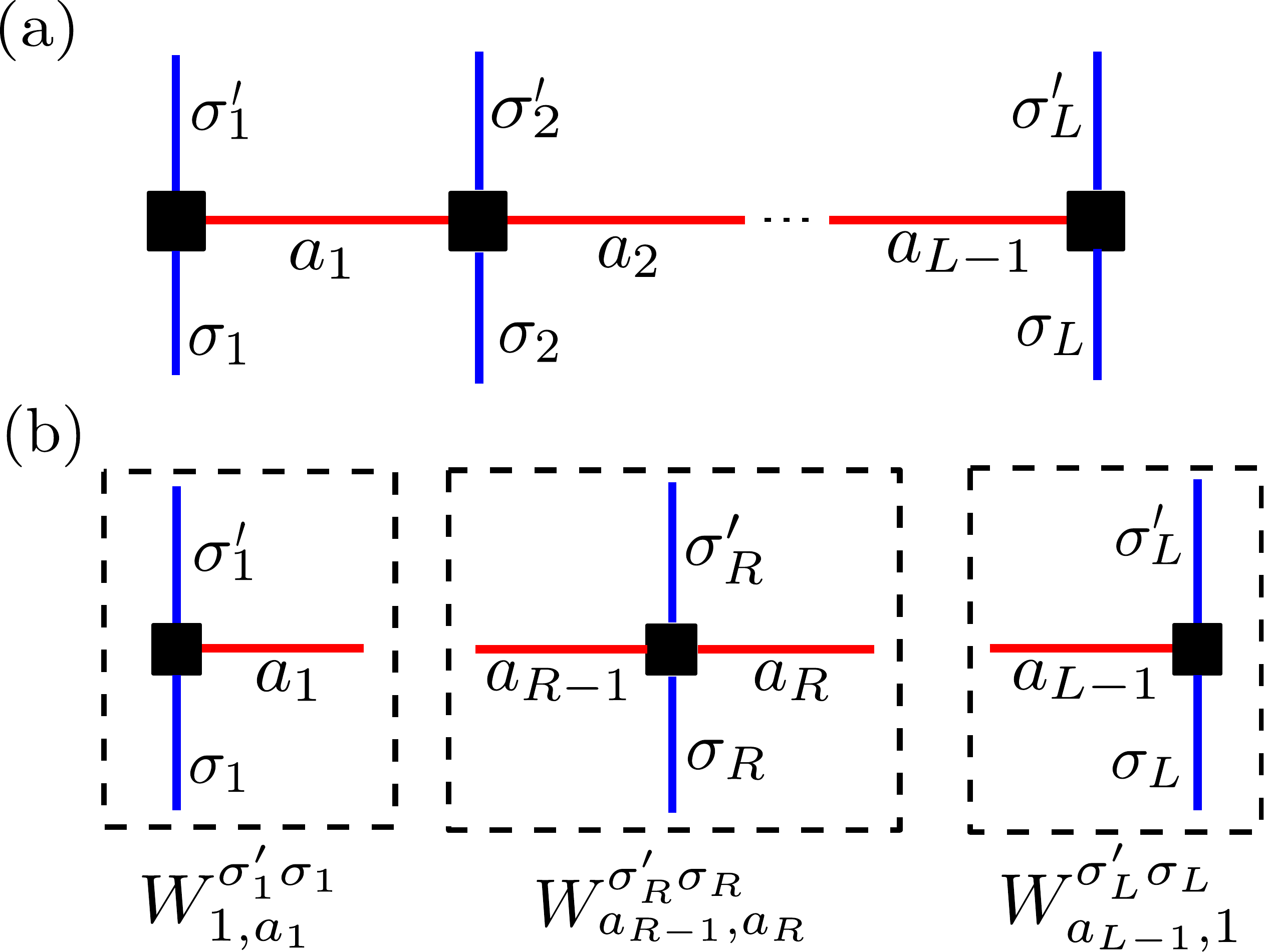}  
\end{tabular}
\caption{Graphical representation of a Matrix Product Operator (MPO) using Tensor Networks. Similarly to the MPS form, the blue (red) line
refers to the physical (non-physical) indices.} 
\label{MPO_form}
\end{figure}

The MPO form, represented at Eq.~\eqref{MPO} and shown on Fig.~\ref{MPO_form} using tensor networks, consists of a local representation (one tensor per lattice site)
of an operator or Hamiltonian similarly to the MPS form for many-body quantum states. It is characterized by a MPO bond dimension $\tilde{\chi}$ corresponding
to the dimension of the square matrices in its representation \footnote{Non-translational operators can also be cast into a MPO form where the tensors
$\hat{W}$ do not have the same dimension. In this case, $\tilde{\chi} = \mathrm{max}\left[\mathrm{dim}(\hat{W}[R])\right]$.}
\textit{ie.} $\tilde{\chi} = \mathrm{dim}(\hat{W})$. It comes directly that for local operators, $\tilde{\chi} = 1$.
For the 1D short-range Bose-Hubbard model, whose one possible MPO form has been characterized previously, $\tilde{\chi} = 4$.

\subsection{Static properties - Ground state}
In the previous paragraphs, we have completely transpose the Dirac formalism including quantum states and operators in terms of tensor networks. 
We now turn to the tensor-network-based algorithm to characterize the static properties of a Hamiltonian and more precisely the ground state 
and ground state energy, while still working along the lines of Refs.~\cite{schollwock2005,schollwock2011}. These static properties correspond to the
first step in order to study the 
far-from-equilibrium dynamics of quantum lattice models submitted to sudden global or local quenches. To do so, several techniques can be used to deduce
these characteristics of a Hamiltonian such that the imaginary time evolution 
(presented in the next chapter) and the Lagrangian minimization that we explicit here. \\

To find the ground state (energy), it requires to get the optimal MPS $\ket{\Psi}$ which minimizes the energy of a given Hamiltonian $\hat{H}$, containing
here only local or short-range (nearest-neighbor) interactions,   

\begin{equation}
 E = \frac{\bra{\Psi}\hat{H} \ket{\Psi}}{\langle \Psi | \Psi \rangle}.
\end{equation}

\noindent
To solve this minimization problem, one can rely on the method of Lagrange multipliers. It consists here in introducing a Lagrangian multiplier $\lambda$ 
and to minimize the previous equation leading to 

\begin{equation}
\underset{|| \ket{\Psi} || = 1}{\mathrm{min}}(\bra{\Psi}\hat{H} \ket{\Psi} - \lambda \langle \Psi | \Psi \rangle).
\end{equation}

\noindent
The latter is done by updating locally the tensors of the MPS $\ket{\Psi}$ in order to decrease the value of the Lagrangian multiplier $\lambda$ hence the energy.
Indeed, we start by keeping all the tensors on all sites constant except one on the lattice site $R$ with entries $M^{\sigma_R}_{a_{R-1},a_{R}}$ as variables.
Then, the Lagrangian minimization condition with respect to the tensor $\left(M^{\sigma_R}\right)^{*}$, corresponding to an eigenvector problem for $M^{\sigma_R}$, has to be solved. We then repeat this operation on another site in the aim of finding once again a state
lower in energy and moving through all lattice sites several times until to reach a convergence for the energy. At the end of this iterative process,
the MPS $\ket{\Psi}$ corresponds to the ground state of $\hat{H}$ and $\lambda$ the associated ground state energy within the ensemble of MPS at a given bond 
dimension (variational space). \\

To find the analytical expression of the Lagrangian minimization condition for the tensor $M^{\sigma_R}$, we need to calculate the Lagrangian 

\begin{equation}
\mathcal{L}( \{ M^{\sigma_R} \}, \{ (M^{\sigma_R})^{\dag} \} ) = \bra{\Psi}\hat{H} \ket{\Psi} - \lambda \langle \Psi | \Psi \rangle,
\label{lagr}
\end{equation}

\noindent
with $\{ M^{\sigma_R} \} = \{M^{\sigma_1},..., M^{\sigma_L} \}$ denotes the set of tensors defining the MPS $\ket{\Psi}$. Hence, the matrix product state $\ket{\Psi}$
has the following expression
 
\begin{equation}
\ket{\Psi} = \sum_{\boldsymbol{\sigma}} M^{\sigma_1} M^{\sigma_2}~...~ M^{\sigma_{L-1}} M^{\sigma_L} \ket{\boldsymbol{\sigma}},~~~\boldsymbol{\sigma} = \sigma_1,\sigma_2,...,
\sigma_{L-1},\sigma_{L}.
\end{equation}

\noindent
Note that no assumption has been made on the normalization of the tensors, \textit{ie.} on how the gauge degree of freedom of $\ket{\Psi}$ has been fixed.
We first calculate the overlap $\langle \Psi | \Psi \rangle$, appearing in the expression of the Lagrangian at Eq.~\eqref{lagr}, while showing explicitly 
the tensor $M^{\sigma_R}$. It yields for the overlap the following expression 
 
\begin{align}
& \langle \Psi | \Psi \rangle = \sum_{\sigma_1,...,\sigma_R,...,\sigma_L} \left(M^{\sigma_1}~...~M^{\sigma_R}~...~M^{\sigma_L} \right)^{\dag} 
\left(M^{\sigma_1} ~...~M^{\sigma_R}~...~ M^{\sigma_L}
\right).
\label{overlap_lagr}
\end{align}

\noindent
Since the overlap is homogeneous to a scalar, one can take the trace of the previous expression. Then, using its cyclicality property, it immediately comes out that
Eq.~\eqref{overlap_lagr} can be rewritten as follows

\begin{align}
& \langle \Psi | \Psi \rangle = \sum_{\sigma_R} \sum_{a_{R-1},a_R} \sum_{a_{R-1}',a_R'} \left( M^{\sigma_R} \right)^{\dag}_{a_R,a_{R-1}} \Psi^{A}_{a_{R-1},a_{R-1}'}
M^{\sigma_R}_{a_{R-1}',a_R'} \Psi^{B}_{a_R',a_R},
\end{align}

\noindent
where the coefficients of the matrices $\Psi^{A}$ and $\Psi^{B}$ read as 

\begin{align}
& \Psi^{A}_{a_{R-1},a_{R-1}'} = \sum_{\sigma_1,...,\sigma_{R-1}} [ \left(M^{\sigma_{R-1}}\right)^{\dag} ...~\left( M^{\sigma_1} \right)^{\dag} M^{\sigma_1}...
~ M^{\sigma_{R-1}} ]_{a_{R-1},a_{R-1}'}, \nonumber \\
& \Psi^{B}_{a_{R}',a_{R}} = \sum_{\sigma_{R+1},...,\sigma_{L}}[M^{\sigma_{R+1}} ...~M^{\sigma_L} \left(M^{\sigma_L}\right)^{\dag}...
~ \left(M^{\sigma_{R+1}} \right)^{\dag}]_{a_{R}',a_{R}} \nonumber.
\end{align}

%

\begin{figure}[!h]
\centering
\begin{tabular}{c}
\includegraphics[scale = 0.36]{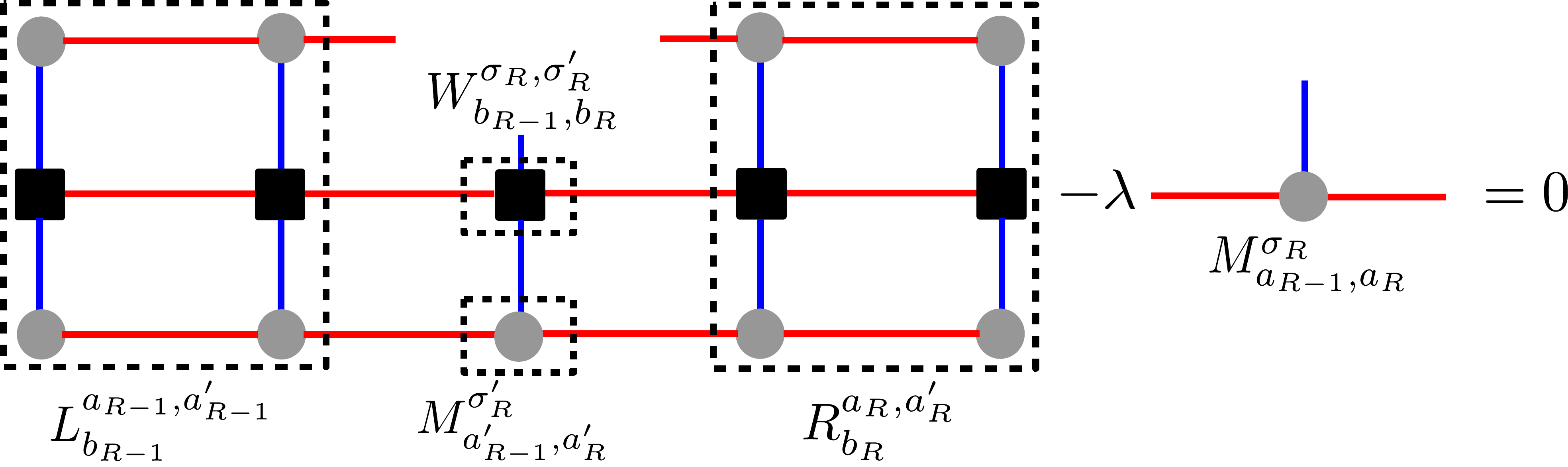}  
\end{tabular}
\caption{Graphical representation using tensor networks of the Lagrangian minimization condition at \eqref{final_eq} to find the tensor $M^{\sigma_R}$ for a MPS 
$\ket{\Psi}$ in the mixed-canonical form.}
\label{lagr_mini_cond}
\end{figure}

\noindent
Relying on Fig.~\ref{overlap}, the matrix $\Psi^A$ ($\Psi^B$) can be seen as the left (right) part of the overlap 
$ \langle \Psi | \Psi \rangle$ with respect to both tensors at the lattice site $R$.
We now discuss the first term of Eq.~\eqref{lagr} with $\hat{H}$ written in the MPO form. As previously for the overlap $\langle \Psi | \Psi \rangle$,
we express the expectation value $\bra{\Psi}\hat{H}\ket{\Psi}$ by showing explicitly the tensors related to the lattice site $R$. 
It yields the following expression for $\bra{\Psi}\hat{H}\ket{\Psi}$ 
   
\begin{equation}
\bra{\Psi}\hat{H}\ket{\Psi} = \sum_{a_{R-1},a_{R-1}'} \sum_{\sigma_R, \sigma_R'} \sum_{a_R,a_R'} 
\sum_{b_{R-1},b_R} L^{a_{R-1}, a_{R-1}'}_{b_{R-1}} W^{\sigma_R \sigma_R'}_{b_{R-1},b_R}
R^{a_R,a_R'}_{b_R} \left(M^{\sigma_R}_{a_{R-1},a_R}\right)^{*} M^{\sigma_R'}_{a_{R-1}',a_R'}.
\label{exp_val_lagr}
\end{equation}  
  
\noindent
where the third-order left ($L$) and right ($R$) tensors are given as follows (see Fig.~\ref{lagr_mini_cond} for their graphical representation)


\begin{align}
& L^{a_{R-1}, a_{R-1}'}_{b_{R-1}} = \sum_{\{ a_{\tilde{R}}, b_{\tilde{R}}, a'_{\tilde{R}}; \tilde{R}<R-1\}} \left( \sum_{\sigma_1, \sigma'_1} W^{\sigma_1 \sigma'_1}_{1,b_1}
(M^{\sigma_1}_{1,a_1})^{*} M^{\sigma'_1}_{1,a'_1} \right) \left( \sum_{\sigma_2, \sigma'_2} W^{\sigma_2 \sigma'_2}_{b_1,b_2} (M^{\sigma_2}_{a_1,a_2})^{*} 
M^{\sigma'_2}_{a'_1,a'_2} \right)  \nonumber \\
& ... \left( \sum_{\sigma_{R-1}, \sigma'_{R-1}} W^{\sigma_{R-1} \sigma'_{R-1}}_{b_{R-2},b_{R-1}} 
(M^{\sigma_{R-1}}_{a_{R-2},a_{R-1}})^{*} M^{\sigma'_{R-1}}_{a_{R-2}',a_{R-1}'} \right)
\end{align}


\begin{align}
& R^{a_R,a_R'}_{b_R} = \sum_{\{ a_{\tilde{R}},b_{\tilde{R}},a'_{\tilde{R}}; \tilde{R}>R\}} \left( \sum_{\sigma_{R+1},\sigma_{R+1}'} W^{\sigma_{R+1} \sigma'_{R+1}}_{b_{R},
b_{R+1}} (M^{\sigma_{R+1}}_{a_R, a_{R+1}})^{*} M^{\sigma'_{R+1}}_{a_R', a_{R+1}'} \right) \nonumber \\
& \left( \sum_{\sigma_{R+2},\sigma_{R+2}'} W^{\sigma_{R+2} \sigma'_{R+2}}_{b_{R+1},
b_{R+2}} (M^{\sigma_{R+2}}_{a_{R+1}, a_{R+2}})^{*} M^{\sigma'_{R+2}}_{a_{R+1}', a_{R+2}'} \right) ... \left( \sum_{\sigma_L, \sigma'_L} W^{\sigma_L \sigma'_L}_{b_{L-1},1} 
(M^{\sigma_L}_{a_{L-1},1})^{*} M^{\sigma'_L}_{a_{L-1}',1} \right) 
\end{align}

\noindent
The expectation value $\bra{\Psi}\hat{H}\ket{\Psi}$ at Eq.~\eqref{exp_val_lagr} is built from $5$ different tensors 
\footnote{In what follows, the tensors are described by showing only the physical indices for simplicity.}
: the third-order left ($L$) and right ($R$) tensors corresponding to the left and right parts of the expectation 
value $\bra{\Psi}\hat{H}\ket{\Psi}$ (with respect the tensors related to the lattice site $R$),
the third-order tensors $M^\sigma_R$ and $(M^{\sigma_R})^{\dag}$ coming from the ket $\ket{\Psi}$ and the bra $\bra{\Psi}$ respectively, and finally the fourth-order
tensors $W^{\sigma_R,\sigma'_R}$ corresponding to the tensor of the Hamiltonian $\hat{H}$ (represented in a MPO form) at the lattice site $R$.
As a consequence, the Lagrangian defined at Eq.~\eqref{lagr} takes the following form

\begin{align}
& \mathcal{L}(\{M^{\sigma_R}\},\{(M^{\sigma_R})^{\dag} \}) = \sum_{a_{R-1},a_{R-1}'} \sum_{\sigma_R, \sigma_R'} \sum_{a_R,a_R'} 
\sum_{b_{R-1},b_R} L^{a_{R-1}, a_{R-1}'}_{b_{R-1}} W^{\sigma_R \sigma_R'}_{b_{R-1},b_R} R^{a_R,a_R'}_{b_R} \nonumber \\
&\left(M^{\sigma_R}_{a_{R-1},a_R}\right)^{*}
M^{\sigma_R'}_{a_{R-1}',a_R'} - \lambda \sum_{\sigma_R} \sum_{a_{R-1},a_R} \sum_{a_{R-1}',a_R'} \left(M^{\sigma_R}_{a_{R-1},a_R}\right)^{*} \Psi^{A}_{a_{R-1},a_{R-1}'}
M^{\sigma_R}_{a_{R-1}',a_R'} \Psi^{B}_{a_R',a_R},
\label{lagr_init}
\end{align}

\noindent
where the tensors on the lattice site $R$ are explicitly shown. One can notice that Eq.~\eqref{lagr_init} depends linearly on the tensors $M^{\sigma_R}$ and 
$(M^{\sigma_R})^{\dag}$. Consequently, the Lagrangian minimization condition for the tensor $M^{\sigma_R}$ is characterized by taking the derivative of the
previous expression with respect to the corresponding hermitian tensor. In other words, the Lagrangian minimization condition for the tensor $M^{\sigma_R}$ is
given by $\partial \mathcal{L}(\{M^{\sigma_R}\},\{ (M^{\sigma_R})^{\dag}\}) / \partial (M^{\sigma_R})^* = 0$, which can be expressed as 

\begin{align}
& \sum_{\sigma_R'}\sum_{a_{R-1}',a_R'} \sum_{b_{R-1},b_R} L^{a_{R-1}, a_{R-1}'}_{b_{R-1}} W^{\sigma_R \sigma_R'}_{b_{R-1},b_R}
R^{a_R,a_R'}_{b_R} M^{\sigma_R'}_{a_{R-1}',a_R'} - \lambda \sum_{a_{R-1}',a_R'} \Psi^{A}_{a_{R-1},a_{R-1}'} M^{\sigma_R}_{a_{R-1}',a_R'} \Psi^{B}_{a_R',a_R} = 0 
\label{igss}
\end{align}

\noindent
According to the previous equation, the minimization condition for the tensor $M^{\sigma_R}$ corresponds to an eigenvalue problem.
Indeed, by introducing $H$, $N$ and $\nu$ defined as follows
\begin{align}
& H_{(\sigma_R,a_{R-1},a_R),(\sigma_R',a_{R-1}',a_R')} = \sum_{b_{R-1},b_R} L^{a_{R-1}, a_{R-1}'}_{b_{R-1}} W^{\sigma_R \sigma_R'}_{b_{R-1},b_R} R^{a_R,a_R'}_{b_R}, \\
& N_{(\sigma_R,a_{R-1},a_R),(\sigma_R',a_{R-1}',a_R')} = \Psi^{A}_{a_{R-1},a_{R-1}'} \Psi^{B}_{a_R',a_R} \delta_{\sigma_R, \sigma_R'}, \\
& \nu_{(\sigma_R,a_{R-1},a_R),1} = M^{\sigma_R}_{a_{R-1},a_R},
\end{align}

\noindent
we obtain a generalized eigenvalue problem $H\nu - \lambda N \nu = 0$ with $\mathrm{dim}(H) = \mathrm{dim}(N) = d \chi^2 \times d\chi^2$ 
and $\mathrm{dim}(\nu) = d\chi^2 \times 1$ where $d$ denotes the dimension of the local Hilbert space and $\chi$ is the MPS bond dimension. 
Note that the normality of the set of tensors $\{M^{\sigma_R}, R=1...L\}$ has not been discussed up to now. Indeed, the previous calculations have been done
without making any assumption on the gauge degree of freedom of the MPS $\ket{\Psi}$. However, the latter can significantly simplify the problem. Indeed, if we consider that the quantum state $\ket{\Psi}$ is left-normalized up to the site $R-1$ and right-normalized
from the site $R+1$ to the last site of the chain $L$, we enforce that $\Psi^{A}_{a_{R-1}, a_{R-1}'} = \delta_{a_{R-1},a_{R-1}'}$ and $\Psi^{B}_{a_R, a_R'} = \delta_{a_R, a_R'}$ and Eq.~\eqref{igss}
can be simplified as, see Fig.~\ref{lagr_mini_cond} for the graphical representation in terms of tensor networks,
\begin{align}
& \sum_{\sigma_R'}\sum_{a_{R-1}',a_R'} \sum_{b_{R-1},b_R} 
L^{a_{R-1}, a_{R-1}'}_{b_{R-1}} W^{\sigma_R \sigma_R'}_{b_{R-1},b_R} R^{a_R,a_R'}_{b_R} M^{\sigma_R'}_{a_{R-1}',a_R'} - \lambda M^{\sigma_R}_{a_{R-1},a_R} = 0
\label{final_eq}
\end{align}

\noindent
This eigenvalue problem is a problem of matrix dimension $d \chi^2 \times d\chi^2$ and can be very demanding in terms of numerical resources if 
the quantum state $\ket{\Psi}$ is not cast into a mixed-canonical form to simplify the expression of the matrix $N$. Moreover, the speed of convergence
\textit{ie.} the number of iterations of the local update governed by Eq.~\ref{final_eq} (also called sweeps where more details are provided in the following)
mainly depends on the initial guess for the quantum state $\ket{\Psi}$. The closer the initial guess is to the ground state, the faster the algorithm is, leading to
a relatively small number of sweeps. The convergence of this algorithm is achieved when there is no modification of the energy \textit{ie.} $\lambda$ the Lagrangian multiplier. Nevertheless, 
the rigorous test is to compute the energy variance $\sigma^2_E(\ket{\Psi})$ associated to the quantum state $\ket{\Psi}$ and defined as 

\begin{equation}
\sigma^2_E(\ket{\Psi}) = \langle \Psi | (\hat{H} - \langle \Psi | \hat{H} | \Psi \rangle )^{2} |\Psi \rangle = 
\langle \Psi | \hat{H}^{2} | \Psi \rangle - (\langle \Psi | \hat{H} | \Psi \rangle)^{2}
\end{equation}

\noindent
to make sure that the final quantum state is an eigenvector for the Hamiltonian $\hat{H}$. Hence, if $\ket{\Psi}$ represents an eigenvector for $\hat{H}$ then
$\sigma^2_E$ should be as close as possible from $0$. \\

In the following, we give more details about the iterative ground state search and more precisely on the left- and right-sweeps
in order to optimize locally the tensors defining $\ket{\Psi}$, 

\begin{itemize}
\item Consider an initial guess for $\ket{\Psi}$ in a right-canonical representation. Consequently, the initial quantum state
consists of a product of $B$-tensors, see Eqs.~\eqref{mps3} which satisfy Eq.~\eqref{right}.
\item Compute the expression of the tensor $R$ iteratively for all sites from $1$ to $L-1$. 
\item \textit{Right-sweep} : Starting from the site $R=1$ until to reach the site $R=L-1$, solve the eigenproblem $H \nu = \lambda N \nu$ previously
discussed for the tensor $M^{\sigma_R}$. Once the solution is obtained, $M^{\sigma_R}$ needs to be cast into a left-normalized tensor $A^{\sigma_R}$
by reshaping the matrix $U[R]$ when applying the SVD. The previous step allows to maintain the desired mixed-canonical representation of the quantum state $\ket{\Psi}$.
Thus, the remaining matrices of the SVD (the Schmidt matrix $S[R]$ and the right-normalized matrix $V^{\dag}[R]$) are multiplied to the tensor $M^{\sigma_{R+1}}$ 
which will be the initial guess for the eigenproblem on the next lattice site $R+1$. The different steps for the previous local update is summarized below where
the lower index for the tensors and matrices denotes the number of local updates performed 

\begin{equation}
M_0^{\sigma_R} B_0^{\sigma_{R+1}} \underset{\mathrm{eig}}{\rightarrow} M_1^{\sigma_R} B_0^{\sigma_{R+1}} \underset{\mathrm{SVD}}{\rightarrow} U_1[R] S_1[R]
V^{\dag}_1[R] B_0^{\sigma_{R+1}} \rightarrow A_1^{\sigma_R} M_0^{\sigma_{R+1}},
\end{equation}

\noindent
with $M_0^{\sigma_{R+1}} = S_1[R] V^{\dag}_1[R] B_0^{\sigma_{R+1}}$.

\item Construct the expression of the tensor $L$ iteratively (by adding the tensors related to one more lattice site). Continue the previous steps in order to reach
the lattice site $L-1$.

\item \textit{Left-sweep} : Similarly to the right-sweep, we first start from the site $R=L$ until to reach the site $R=2$, solve the eigenproblem for
$M^{\sigma_R}$. Once the solution is obtained, $M^{\sigma_R}$ needs to be cast into a right-normalized tensor $B^{\sigma_R}$ by reshaping the matrix 
$V^{\dag}[R]$ when applying the SVD. Similarly to the left-sweep, this step allows to maintain the desired mixed-canonical representation of the quantum 
state $\ket{\Psi}$. Thus, the remaining matrices of the SVD (the left-normalized matrix $U[R]$ and the Schmidt matrix $S[R]$ are multiplied by the tensor
$M^{\sigma_{R-1}}$ on the left which will be the initial guess for the eigenproblem on the next lattice site $R-1$

\begin{equation}
A_1^{\sigma_{R-1}} M_0^{\sigma_R} \underset{\mathrm{eig}}{\rightarrow} A_1^{\sigma_{R-1}} M_1^{\sigma_R}  \underset{\mathrm{SVD}}{\rightarrow}
A_1^{\sigma_{R-1}} U_1[R] S_1[R] V^{\dag}_1[R] \rightarrow M_1^{\sigma_{R-1}} B_1^{\sigma_R},
\end{equation}

\noindent
with $M_1^{\sigma_{R-1}} = A_1^{\sigma_{R-1}} U_1[R] S_1[R]$.

\item Construct iteration after iteration the $R$-expression by adding the tensors related to one more site. Continue the previous steps in order to reach
the lattice site $2$.
\end{itemize}

\subsection{Dynamical properties - Real time evolution}

\subsubsection{Time-discretization problem}
We now move to the numerical technique to perform real time evolution for the same class of quantum lattice models. This algorithm permits
to find the time-dependent quantum state $\ket{\Psi(t)}$ evolving \textit{via} a unitary evolution. Finding the time-evolved quantum state $\ket{\Psi(t)}$
corresponds to the second and last step to perform a sudden global or local quench in order to study the far-from-equilibrium dynamics of quantum lattice models 
numerically. \\

Firstly, let us consider an initial quantum state $\ket{\Psi_0} = \ket{\Psi(t=0)}$.
After a time $t$ and under a unitary evolution, the quantum state $\ket{\Psi(t)}$ has the following expression ($\hbar$ is fixed to unity by convention)

\begin{equation}
 \ket{\Psi(t)} = e^{-i \hat{H} t} \ket{\Psi_0},
\end{equation}

\noindent
solution of the time-dependent Schrödinger equation $ i\partial_t \ket{\Psi(t)} = \hat{H} \ket{\Psi(t)}$. In the following, the Hamiltonian 
$\hat{H}$ represents a 1D short-range interacting lattice model having the form 

\begin{equation}
\hat{H} = \sum_{R=1}^{L-1}\hat{h}(R) ~~\mathrm{where}~~ \hat{h}(R) = \hat{O}_1(R) \hat{O}_2(R+1),
\label{assumed_ham}
\end{equation}

\noindent
with $\hat{h}(R)$ the interacting term acting on two nearest-neighbor lattice sites $R$ and $R+1$ with $\mathrm{dim}[\hat{h}(R)] = d^2 \times d^2$ where $d$ 
denotes the dimension of the local Hilbert space, $\hat{O}_{1,2}$ are two local operators. To deduce $\ket{\Psi(T)}$ where $T$ is the observation time, one has
to discretize the time evolution operator $e^{-i\hat{H}T}$. This can be done using a Trotterization technique which consists of dividing the observation time $T$
into infinitesimal time steps $\mathrm{d}t$ where $T = N \mathrm{d}t$ with $N$ and $\mathrm{d}t$ satisfying $N \rightarrow \infty$ and $\mathrm{d}t \rightarrow 0$ respectively.
Finally, one just needs to apply the infinitesimal time evolution operator $e^{-i \hat{H} \mathrm{d}t}$ many times to the initial state $\ket{\Psi_0}$ and to deduce the final state
$\ket{\Psi(T)}$. Then, using a first-order Trotter decomposition, it leads for the full time evolution operator $e^{-i \hat{H}T}$

\begin{equation}
 e^{-i \hat{H} T} = \prod_{n=1}^{N} e^{-i \displaystyle \sum_{R=1}^{L-1} \hat{h}(R) \mathrm{d}t} \simeq
 \prod_{n=1}^{N} \left[ \prod_{R=1}^{L-1} e^{-i \hat{h}(R) \mathrm{d}t} + \mathcal{O}(\mathrm{d}t^2) \right],
\end{equation}

\noindent
where an error $\mathcal{O}(\mathrm{d}t^2)$ is introduced since the exponential of a sum of operators cannot be factorized if the commutator 
$[\hat{h}(R), \hat{h}(R+1)] \neq 0$ which is the case in general. Indeed, according to the Baker-Campbell-Hausdorff formula, the product of two infinitesimal time evolution operators
may be written as

\begin{equation}
e^{-i \hat{h}(R) \mathrm{d}t -i \hat{h}(R+1) \mathrm{d}t} \simeq e^{-i \hat{h}(R) \mathrm{d}t}e^{-i \hat{h}(R+1) \mathrm{d}t} e^{\frac{\mathrm{d}t^2}{2} [\hat{h}(R), \hat{h}(R+1)]}
\end{equation}

\noindent
Nevertheless, the error coming from the discretization \textit{via} the commutator $[\hat{h}(R), \hat{h}(R+1)]$ is negligible at first order, since it scales as 
$\mathrm{d}t^2$. It yields 

\begin{equation}
e^{-i \hat{h}(R) \mathrm{d}t -i \hat{h}(R+1) \mathrm{d}t} \simeq  e^{-i \hat{h}(R) \mathrm{d}t} e^{-i \hat{h}(R+1) \mathrm{d}t}.
\end{equation}

To apply the infinitesimal time evolution operator $e^{-i\hat{H}\mathrm{d}t}$ to a MPS, it is convenient to distinguish the even bonds from the odd bonds
for $\hat{H}$ in order to decrease the errors coming from the different commutators. This rearrangement allows us to write 

\begin{equation}
 \hat{H} = \hat{H}_{\mathrm{odd}} + \hat{H}_{\mathrm{even}},~~~ \hat{H}_{\mathrm{even}} = \sum_{R=1}^{L/2-1} \hat{h}(2R),~~~\hat{H}_{\mathrm{odd}} = 
 \sum_{R= 1}^{L/2} \hat{h}(2R-1).
\end{equation}

\noindent
Consequently, the infinitesimal time evolution operator $e^{-i\hat{H}\mathrm{d}t}$ can be expressed under the following form

\begin{equation}
e^{-i \hat{H} \mathrm{d}t} \simeq e^{-i \hat{H}_{\mathrm{even}} \mathrm{d}t} e^{-i \hat{H}_{\mathrm{odd}} \mathrm{d}t},~~~ e^{-i \hat{H}_{\mathrm{even}} \mathrm{d}t} = \prod_{R} e^{-i \hat{h}(2R) \mathrm{d}t},
~~~ e^{-i \hat{H}_{\mathrm{odd}} \mathrm{d}t} = \prod_{R} e^{-i \hat{h}(2R-1) \mathrm{d}t}.
\end{equation}

\noindent
Its application on a MPS will increase the bond dimension from $\chi$ up to $d^2 \chi$. $\chi$ denotes the MPS bond dimension and $d$ the dimension of the local Hilbert space.
Therefore, repeating this step a high number of times lead to an exponential growth of the coefficients of the time-evolved quantum state $\ket{\Psi(t)}$. 
As a consequence, the latter has to be truncated to $\chi$ after each time step to avoid such exponential growth and to remain in the same variational MPS space. \\

\subsubsection{Real time evolution algorithm}

\begin{figure}[h!]
\centering
\includegraphics[scale = 0.5]{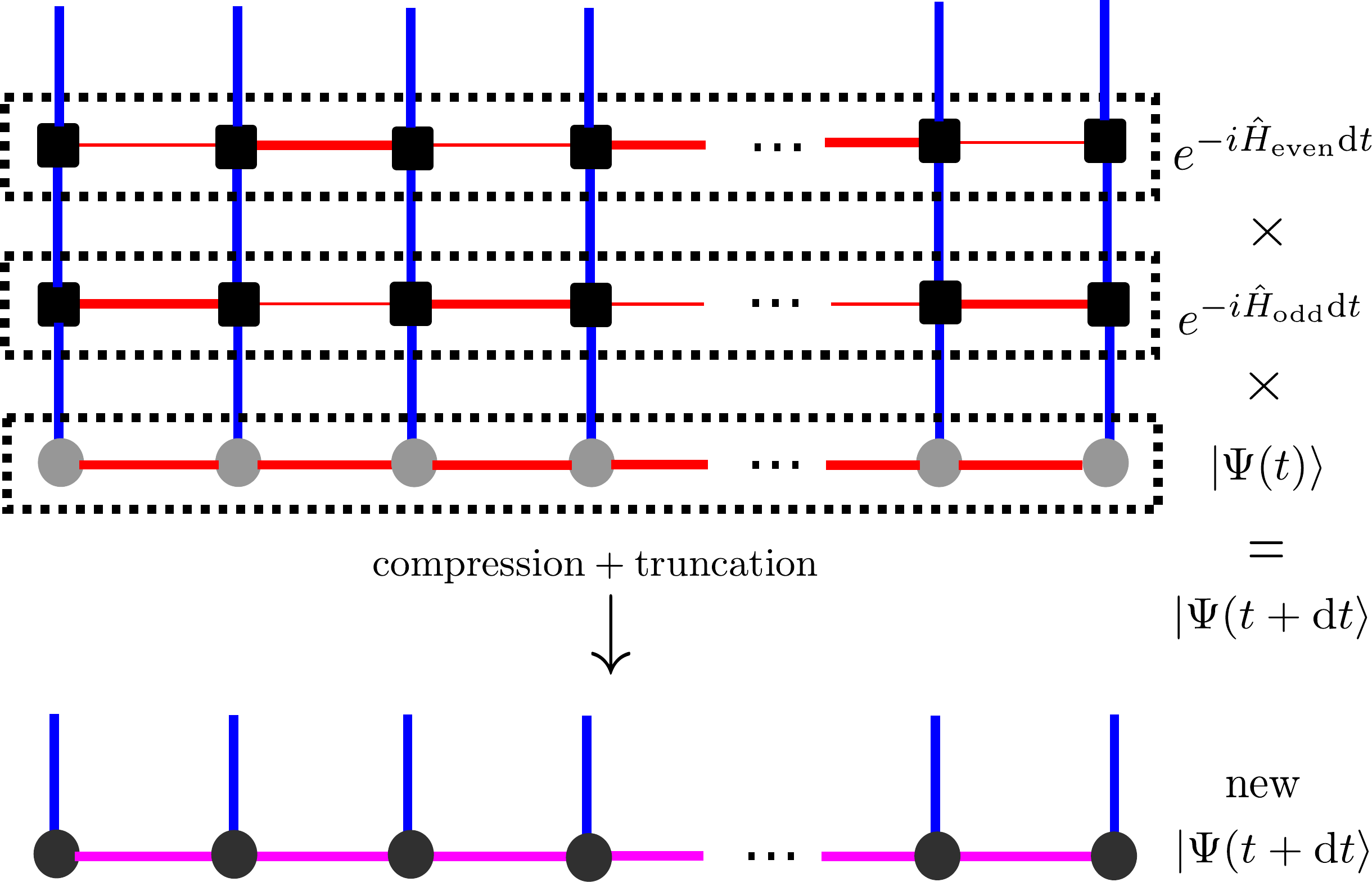}  
\caption{Representation of an infinitesimal Trotter step using tensor networks. The infinitesimal time evolution MPO $e^{-i \hat{H}_{\mathrm{odd}}\mathrm{d}t}$ 
for odd bonds is first applied on the MPS $\ket{\Psi(t)}$, then the one for even bonds $e^{-i\hat{H}_{\mathrm{even}}\mathrm{d}t}$. Finally, the time-evolved quantum 
state $\ket{\Psi(t+\mathrm{d}t)}$ is compressed and truncated to be able to perform real time evolution on longer times.}
\label{time_evolution_MPS}
\end{figure}

We now provide more details about the real time evolution algorithm, see Refs.~\cite{schollwock2005, schollwock2011} and references therein. Starting 
from a MPS $\ket{\Psi(t)}$, the following steps need to be performed in order to get
the time-evolved quantum state after one time step \textit{ie.} $\ket{\Psi(t+\mathrm{d}t)}$, see Fig.~\ref{time_evolution_MPS} for the
graphical representation of the different steps : 

\begin{itemize}
 \item Firstly, apply $e^{-i \hat{H}_{\mathrm{odd}} \mathrm{d}t}$ the infinitesimal time evolution operator for odd bonds on $\ket{\Psi(t)}$.
 \item Secondly, apply $e^{-i \hat{H}_{\mathrm{even}} \mathrm{d}t}$ the infinitesimal time evolution operator for even bonds on the previous quantum state
 $e^{-i \hat{H}_{\mathrm{odd}} \mathrm{d}t} \ket{\Psi(t)}$.
 \item Then, we obtain $\ket{\Psi(t+\mathrm{d}t)} = e^{-i \hat{H} \mathrm{d}t} \ket{\Psi(t)} \simeq e^{-i \hat{H}_{\mathrm{even}} \mathrm{d}t}e^{-i \hat{H}_{\mathrm{odd}} \mathrm{d}t} \ket{\Psi(t)}$. 
 \item Finally, truncate it from dimension $d^2\chi$ to $\chi$ to avoid an exponential growth of the MPS bond dimension.
 \item Repeat the previous steps $N$ times to reach the observation time $T = N\mathrm{d}t$. 
\end{itemize}

\noindent
One possible improvement of the previous algorithm is to consider a second-order Trotter decomposition of the infinitesimal time evolution $e^{-i\hat{H}\mathrm{d}t}$. It is
defined as follows and allows to decrease the error from $\mathcal{O}(\mathrm{d}t^2)$ to $\mathcal{O}(\mathrm{d}t^3)$ \textit{via} a symmetrization :

\begin{equation}
e^{-i \hat{H} \mathrm{d}t} = e^{-i \hat{H}_{\mathrm{odd}} \frac{\mathrm{d}t}{2}} e^{-i \hat{H}_{\mathrm{even}} \mathrm{d}t} e^{-i \hat{H}_{\mathrm{odd}} \frac{\mathrm{d}t}{2}} +
\mathcal{O}(\mathrm{d}t^3).
\end{equation}

\noindent
Indeed, the previous expression of the infinitesimal time evolution is symmetrized, \textit{ie.} fulfills the property
$e^{-i\hat{H}\mathrm{d}t} e^{i\hat{H}\mathrm{d}t} = I$, allowing us to cancel the second-order term in $\mathrm{d}t$.
From this second-order approximant, one can build a symmetrized fourth-order Trotter decomposition, 
see Ref.~\cite{suzuki2005} for more details about the second-order and fourth-order Trotter decompositions, which decomposes the infinitesimal 
time evolution operator $e^{-i \hat{H} \mathrm{d}t}$ under the following form 

\begin{equation}
 e^{-i \hat{H} \mathrm{d}t} = \hat{U}(\mathrm{d}t_1) \hat{U}(\mathrm{d}t_2) \hat{U}(\mathrm{d}t_3) \hat{U}(\mathrm{d}t_2) \hat{U}(\mathrm{d}t_1) + \mathcal{O}(\mathrm{d}t^5),
\end{equation}

\noindent
where the symmetric operator $\hat{U}$ and the three different times $\mathrm{d}t_{1,2,3}$ are defined as

\begin{equation}
\hat{U}(\mathrm{d}t_n) = e^{-i \hat{H}_{\mathrm{odd}} \frac{\mathrm{d}t_n}{2}} e^{-i \hat{H}_{\mathrm{even}} \mathrm{d}t_n} e^{-i \hat{H}_{\mathrm{odd}} \frac{\mathrm{d}t_n}{2}}~~\mathrm{and}~~\mathrm{d}t_1 = \mathrm{d}t_2 = \frac{1}{4-4^{1/3}}\mathrm{d}t,
~\mathrm{d}t_3 = \mathrm{d}t - 2\mathrm{d}t_1 -2 \mathrm{d}t_2.
\end{equation}

\noindent
Relying on the open-source package ALPS \cite{dolfi2014}, the latter will be considered to deduce the time-evolved quantum state. In the following, by using this powerful 
decomposition, we are able to describe accurately the real time evolution of the 1D SRBH model in order to study its out-of-equilibrium dynamics
induced by sudden global and local quenches.

\subsubsection{MPO form of the infinitesimal time evolution operator}
As previously, the Hamiltonian $\hat{H}$ is assumed to contain only nearest-neighbor interactions for a chain with $L$ lattice sites, see Eq.~\eqref{assumed_ham}.
Let us consider an infinitesimal Trotter step $\mathrm{d}t$ for all odd lattice bonds applied to a MPS $\ket{\Psi}$. It is given by the expression

\begin{equation}
e^{-i\hat{H}_\mathrm{odd}\mathrm{d}t} \ket{\Psi} \simeq \prod_{R} e^{-i \hat{h}(2R-1) \mathrm{d}t} \ket{\Psi}.
\end{equation}

\noindent
All the odd bond evolution operators $e^{-i \hat{h}(2R-1) \mathrm{d}t},~\forall R \in [|1,L/2|]$ (or equivalently $e^{-i\hat{h}(R)\mathrm{d}t}$ 
with $R \in \{1,3,..,L-1\}$), coupling two
nearest-neighbor lattice sites, take the form 

\begin{equation}
e^{-i \hat{h}(R)\mathrm{d}t} \equiv \sum_{\sigma_R, \sigma_{R+1}, \sigma'_R, \sigma'_{R+1}} \mathcal{O}^{\sigma_R \sigma_{R+1} \sigma'_R \sigma'_{R+1}}
\ket{\sigma_R \sigma_{R+1}} \bra{\sigma'_R \sigma'_{R+1}}.
\label{obeo}
\end{equation}

\noindent
It is obvious that the previous expression breaks the MPS form since Eq.~\eqref{obeo} is not a local representation of the odd bond evolution operator
\textit{ie.} $e^{-i \hat{h}(R)\mathrm{d}t}$ is not cast into a MPO form. Consequently, it is necessary to express the fourth-order tensor $\mathcal{O}$ with entries
$\mathcal{O}^{\sigma_R \sigma_{R+1} \sigma'_R \sigma'_{R+1}}$ containing $d^4$ coefficients as a product of two
second-order tensors to cast Eq.~\eqref{obeo} in the MPO form in order to maintain a MPS form when it is applied to
a MPS $\ket{\Psi}$, 

\begin{equation}
\mathcal{O}^{\sigma_R \sigma_{R+1} \sigma'_R \sigma'_{R+1}} = \mathcal{O}^{\sigma_R \sigma'_R} \mathcal{O}^{\sigma_{R+1}\sigma'_{R+1}}. 
\end{equation}

\noindent
Let us reshape the fourth-order tensor $\mathcal{O}$ into a matrix $P$ of dimension $d^2 \times d^2$
\begin{equation}
 \mathcal{O}^{\sigma_R \sigma_{R+1} \sigma'_R \sigma'_{R+1}} = P_{(\sigma_R \sigma'_R), (\sigma_{R+1} \sigma'_{R+1})}
\end{equation}

\noindent
Then, by applying the SVD, it yields
\begin{align}
& P_{(\sigma_R \sigma'_R), (\sigma_{R+1} \sigma'_{R+1})} = \sum_{a_R} U_{(\sigma_R \sigma'_R),a_R} S_{a_R, a_R} 
(V^{\dag})_{a_R, (\sigma_{R+1} \sigma'_{R+1})} \\
& P_{(\sigma_R \sigma'_R), (\sigma_{R+1} \sigma'_{R+1})} = \sum_{a_R} U^{\sigma_R \sigma'_R}_{1, a_R}
\bar{U}^{\sigma_{R+1} \sigma'_{R+1}}_{a_R,1}
\end{align}

\noindent
with the fourth-order tensors $U$ and $\bar{U}$,
\begin{align}
& U^{\sigma_R \sigma'_R}_{1, a_R} = U_{(\sigma_R \sigma'_R),a_R} \sqrt{S_{a_R,a_R}} \\
& \bar{U}^{\sigma_{R+1} \sigma'_{R+1}}_{a_R,1} = \sqrt{S_{a_R, a_R}} (V^{\dag})_{a_R,
(\sigma_{R+1} \sigma'_{R+1})} 
\end{align}

\noindent
Thus, the fourth-order tensor $U$ matches a collection of $d^2$ line vectors and $\bar{U}$ a collection of $d^2$
column vectors where the number of coefficients is given by the Schmidt rank of the $S$-matrix. Finally, for the infinitesimal
time evolution operator for all odd bonds $e^{-i\hat{H}_{\mathrm{odd}}\mathrm{d}t} = e^{-i \hat{h}(1) \mathrm{d}t} e^{-i \hat{h}(3) \mathrm{d}t}~...~ e^{-i\hat{h}(L-1)
\mathrm{d}t} $, it yields
\begin{align}
& e^{-i\hat{H}_{\mathrm{odd}}\mathrm{d}t} = \sum_{\boldsymbol{\sigma}, \boldsymbol{\sigma'}} U^{\sigma_1 \sigma'_1} 
\bar{U}^{\sigma_2 \sigma'_2} U^{\sigma_3 \sigma'_3} \bar{U}^{\sigma_4 \sigma'_4}~ ...~ U^{\sigma_{L-1} 
\sigma'_{L-1}} \bar{U}^{\sigma_L \sigma'_L} \ket{\boldsymbol{\sigma}} \bra{\boldsymbol{\sigma'}} \\
& e^{-i\hat{H}_{\mathrm{odd}}\mathrm{d}t} = \prod_{R=0}^{L/2-1} \hat{U}[2R+1] \hat{\bar{U}}[2R+2] 
\end{align}

\noindent
where the operator $\hat{U}[2R+1]$ and $\hat{\bar{U}}[2R+2]$ are given by
\begin{align}
& \hat{U}[2R+1] = \sum_{\sigma_{2R+1}, \sigma'_{2R+1}} U^{\sigma_{2R+1} \sigma'_{2R+1}} \ket{\sigma_{2R+1}}\bra{\sigma'_{2R+1}} \\
& \hat{\bar{U}}[2R+2]  = \sum_{\sigma_{2R+2}, \sigma'_{2R+2}} \bar{U}^{\sigma_{2R+2} \sigma'_{2R+2}} \ket{\sigma_{2R+2}} \bra{\sigma'_{2R+2}} 
\end{align}

\noindent
Following the same steps for the infinitesimal time evolution for all the even bonds $e^{-i \hat{H}_{\mathrm{even}}\mathrm{d}t}
= e^{-i \hat{h}(2) \mathrm{d}t} e^{-i \hat{h}(4) \mathrm{d}t}~ ...~ e^{-i\hat{h}(L-2) \mathrm{d}t} e^{-i \hat{h}(L)\mathrm{d}t}$, one obtains
\begin{align}
& e^{-i \hat{H}_{\mathrm{even}}\mathrm{d}t} = \sum_{\boldsymbol{\sigma}, \boldsymbol{\sigma'}} I^{\sigma_1 \sigma'_1} 
U^{\sigma_2 \sigma'_2} \bar{U}^{\sigma_3 \sigma'_3}U^{\sigma_4 \sigma'_4} \bar{U}^{\sigma_5 \sigma'_5}~ ...~
U^{\sigma_{L-2} \sigma'_{L-2}} \bar{U}^{\sigma_{L-1} \sigma'_{L-1}} I^{\sigma_L \sigma'_L} \ket{\boldsymbol{\sigma}}
\bra{\boldsymbol{\sigma'}} \\
& e^{-i \hat{H}_{\mathrm{even}}\mathrm{d}t} = \hat{I}\hat{U}[2] \left(\prod_{R=1}^{L/2-2} \hat{\bar{U}}[2R+1] \hat{U}[2R+2] \right)
\hat{\bar{U}}[L-1]\hat{I}
\end{align}

\noindent
where $I^{\sigma_1 \sigma'_1}_{1,1} = \delta_{\sigma_1 \sigma'_1}$ and 
$I^{\sigma_L \sigma'_L}_{1,1} = \delta_{\sigma_L, \sigma'_L}$. Indeed, the first and last lattice 
sites do not contribute to the even bonds which is symbolized by the identity matrix. \\

\begin{figure}[h!]
\centering
\includegraphics[scale = 0.43]{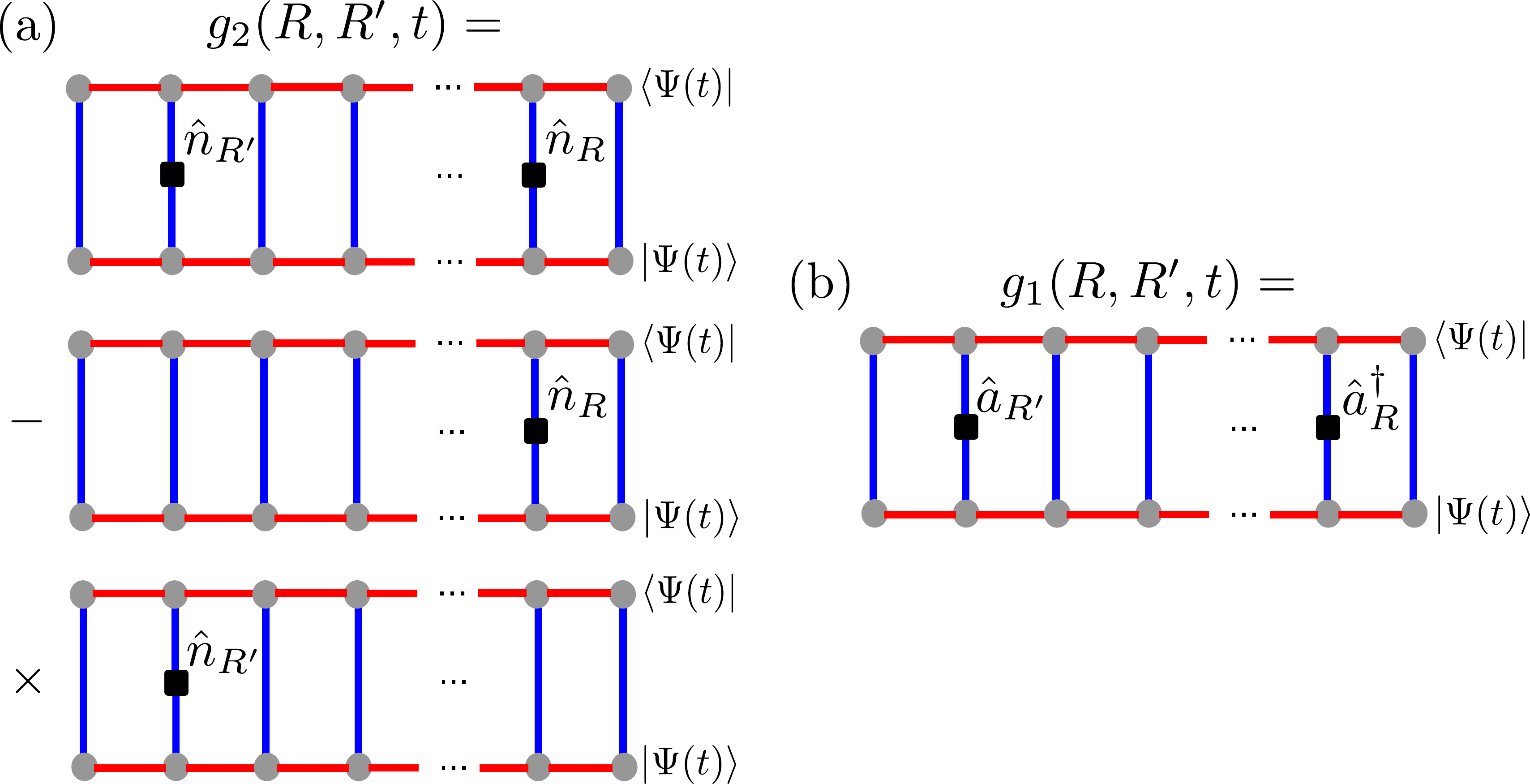}  
\caption{Graphical representation of correlators using tensor networks. (a)~ The equal-time correlation function $g_2(R,R',t) = \langle \hat{n}_R(t) \hat{n}_{R'}(t)
\rangle - \langle \hat{n}_R(t) \rangle \langle \hat{n}_{R'}(t) \rangle$ allowing to characterize the connected density-density correlation
function $G_2(R,t) = g_2(R,t)-g_2(R,0)$ (for $R'$ fixed to $0$) in order to investigate the density fluctuations
of a 1D particle lattice model. (b)~ The equal-time correlation function $g_1(R,R',t) = \langle \hat{a}^{\dag}_R(t) \hat{a}_{R'}(t) \rangle$
to characterize the connected one-body correlation function $G_1(R,t) = g_1(R,t)-g_1(R,0)$ (for $R'$ fixed to $0$) in order to investigate the phase fluctuations
of a 1D particle lattice model. }
\label{correlator}
\end{figure}

Finally, a MPO representation for the infinitesimal time evolution operator for the odd ($e^{-i\hat{H}_{\mathrm{odd}}\mathrm{d}t}$) and 
even ($e^{-i\hat{H}_{\mathrm{even}}\mathrm{d}t}$) bonds has been found and hence for the infinitesimal time evolution operator ($e^{-i \hat{H}\mathrm{d}t}$). Consequently, the 
resulting quantum state $\ket{\Psi(t)}$ is still in the MPS form after each iteration in time. The latter permits to compute spatial- and temporal-dependent
correlation functions, such as the $G_1$ one-body and $G_2$ density-density correlation functions (see Fig.~\ref{correlator}(b) and (a) respectively), relying
on the formalism of tensor networks. The previous space-time observables correspond to the central point of the next sections in order to characterize numerically 
the out-of-equilibrium dynamics of short-range interacting lattice models and in particular the 1D short-range Bose-Hubbard (1D SRBH) model.

\section{Limit regimes in the superfluid and Mott-insulating phases}
\label{limit_regimes}

In the following sections, the out-of-equilibrium dynamics of the one-dimensional short-range Bose-Hubbard model (SRBH) is investigated numerically 
using time-dependent matrix product state approach presented at Sec.~\ref{tmps_approach}. More precisely, we study the response of the 1D SRBH model 
to a variety of sudden global and local quenches ~\cite{calabrese2006,barmettler2012,kollath2007,moeckel2008,manmana2009,roux2010,navez2010,
carleo2014,krutitsky2014}, as can be realized in ultracold-atom experiments~\cite{greiner2002b,cheneau2012,langen2013,geiger2014}. Similarly to the theoretical
results found in the last chapter at Sec.~\ref{SRC}, we start from the ground state of the 1D SRBH Hamiltonian for some initial value of the
interaction parameter $(U/J)_\mathrm{i}$ and let the system evolve with a different value denoted by $(U/J)_\mathrm{f}$. A variety of quenches, spanning the phase diagram, 
see the different arrows on Fig.~\ref{fig:phase_diagram_bhm}(a), is considered. The phase and density fluctuations are studied \textit{via} the computation of the 
equal-time connected correlation functions $G_1$ and $G_2$ respectively defined by $G_1(R,t) = \langle \hat{a}^\dagger_R (t) \hat{a}_0 (t) \rangle -
\langle \hat{a}^\dagger_R (0) \hat{a}_0 (0) \rangle$ and $G_2(R,t) = g(R,t)- g(R,0)$ with $g(R,t) = \langle \hat{n}_R(t) \hat{n}_0(t) \rangle - 
\langle \hat{n}_R(t) \rangle \langle \hat{n}_0(t) \rangle$. Both can be measured in experiments using time-of-flight and fluorescence microscopy imaging,
respectively~\cite{cheneau2012,trotzky2012,langen2013,geiger2014}. \\
In this section, the limit regime of both the superfluid (SF) and Mott-insulating phase (MI) are investigated where the numerical results can be compared to
the analytical predictions presented in the previous chapter at Sec.~\ref{SRC}.

\subsection{Mean field regime of the superfluid phase}
In the following sections, the numerical results are all obtained using the time-dependent density-matrix renormalization group approach (DMRG)
with the matrix-product state representation (t-MPS approach)~\cite{schollwock2005,schollwock2011,dolfi2014}. It yields numerically-exact 
results about the equilibrium and out-of-equilibrium properties of low-dimensional lattice models.

\subsubsection{Time-dependent matrix-product state simulations}
\label{tMPS_params}
We recall that this numerical approach is based on a Schmidt expansion of a many-body quantum state $\ket{\Psi}$ and permits to reduce the full Hilbert space to
a finite, relevant subset, provided that its entanglement
entropy remains sufficiently small \textit{ie.} provided that $\ket{\Psi}$ can be efficiently described by a superposition of a small number of states (only
few Schmidt singular values contributes for a precise description of the latter).
Owing to the area law implying \footnote{The area law implies for the entanglement (von Neumann) entropy a scaling with the area of 
the cuts.} ~\cite{wolf2008,eisert2010}, it is optimal for 1D lattice models with a finite local Hilbert space in gapped phases,
the entanglement of which remains finite and particularly small in the thermodynamic limit. For gapless phases, the area law breaks down and the entanglement entropy
follows a volume law \footnote{The volume law denotes a scaling with the subsystem volume for the entanglement entropy when considering a bipartite quantum system.}
implying more stringent numerical parameters including the high-filling cutoff $n_{\mathrm{max}}$ and the MPS bond dimension $\chi$. 
A careful analysis of the previous numerical cut-offs has been systematically performed to certify the convergence of the results
in all the considered phases and regimes of the 1D SRBH model, see Fig.~\ref{fig:phase_diagram_bhm}(b). This is particularly critical for quenches in the SF 
phase where the numerical requirements are most binding. Indeed, the gapless nature of the excitation spectrum requires a relatively high MPS bond dimension $\chi$ 
and the nearly-Poissonian occupation number probability $P(n_R)$ implies a high-filling cutoff $n_{\mathrm{max}} \gg \bar{n}$ (the filling of the lattice chain).

\paragraph{Truncation of the local Hilbert space}
For the 1D SRBH model, the local Hilbert space spanned by the Fock basis of number states, $\ket{n_R}$, where $n_R \in \mathbb{N}$, is infinite in theory.
However, the probability distribution of the lattice-site occupation $P(n_R)$ decays faster than exponentially in both the SF and MI phases. Accurate results
can thus be obtained by cutting off the local Hilbert space to some maximal value $n_{\mathrm{max}}$. It is important to note that, in some cases, the value
of $n_{\mathrm{max}}$ needs to be significantly much larger than the average filling $\bar{n}$ and its fluctuations. This observation is consistent with
analyses of truncated Bose-Hubbard models \textit{via} quantum Monte-Carlo simulations~\cite{kashurnikov1998zero}. \\

The SF mean field regime, which corresponds to a high filling $\bar{n}$ and a gapless excitation spectrum, has the most binding criteria. It turns out that 
a good estimator for $n_{\mathrm{max}}$ is given by the condition $1 - \sum_{n_R=0}^{n_{\mathrm{max}}} P(n_R) \lesssim 10^{-2}$, where $P(n_R)$
is the probability that $n_R$ bosons occupy the lattice site $R$. 
In the SF mean field regime, the probability distribution $P(n_R)$ is nearly Poissonian, see Eq.~\eqref{P_mf_sf}. For instance, for the filling $\bar{n}=5$ 
used for the data of Fig.~\ref{fig:velocities_superfluid}, it yields $n_{\mathrm{max}} \gtrsim 12$. 
For the strongly correlated SF regime at $\bar{n}=1$ considered in the next section, the density fluctuations are significantly suppressed and using the same 
condition as previously leads to $n_{\mathrm{max}}=5$. 
Concerning the strongly interacting SF regime, owing to the low filling factor $\bar{n}<1$ and the large value of the interaction parameter $U/J$, the
above condition yields $n_{\mathrm{max}}=2$. \\

For the gapped MI phase at $\bar{n}=1$ and moderate values of the interaction parameter $U/J$ ($15 \geq U/J \geq u_\mathrm{c}$), we kept $n_{\mathrm{max}}=5$. 
Finally, for the strong-coupling regime of the MI phase ($U/J \geq 15$) at $\bar{n}=1$ corresponding to the easiest case from a numerical point of view,
truncating the local Hilbert space to $n_{\mathrm{max}}=2$ turns out to be sufficient. \\

In all cases, we have checked that the numerics are converged for these values of $n_{\mathrm{max}}$.

\paragraph{Bond dimension}
Within the MPS formalism, the many-body quantum state for a $L$-site lattice is represented in the tensor network form as follows

\begin{equation}
\ket{\Psi} = \sum_{n_1, n_2,...,n_{L}} A^{n_1}[1] A^{n_2}[2]~...~ A^{n_{L}} [L] \ket{n_1, n_2,..., n_{L}},
\end{equation}

\noindent
where $\ket{n_R}$ spans the local Hilbert space basis. For the 1D SRBH model, it corresponds to a Fock basis truncated at
$n_{\mathrm{max}}$. For each value of $n_R$, the quantity $A^{n_R} [R]$ is a $\chi_{R-1}\times \chi_{R}$ matrix, where $\chi_{R}$ is the rank associated 
to the Schmidt matrix when applying the $R$-th singular value decomposition, see Ref.~\cite{schollwock2011} and Appendix.~\ref{appendix3_mps_app}. The bond
dimension $\chi$ is defined as the maximum rank, $\chi = \mathrm{max}_R \left( \chi_R \right),~R \in [0,...,L]$. Note that for open-boundary conditions, 
the quantities $A^{n_1}[1]$ and $A^{n_{L}}[L]$ denote a collection of several row vectors and column vectors respectively.\\
In the numerics, the maximum value of $\chi$ is chosen sufficiently large so that the truncation does not affect the results.
In practice, the calculations are run for several values of $\chi$ up to convergence of the one-body $G_1$ and density-density $G_2$ correlation functions. 
The required value of $\chi$ significantly depends on the phase, the regime and on the observable. In the following, we give the values used
for the final results presented in the paper. \\

For the SF mean field regime, see Figs.~\ref{fig:velocities_superfluid}(a) and \ref{fig:velocities_g1_superfluid}, we used the values $\chi = 300$ and
$\chi=450$ for the $G_2$ and $G_1$ correlation functions, respectively. The bond dimension used for $G_1$ is higher than the one for $G_2$ due to the
long-range phase correlations already present at equilibrium. For the SF strongly correlated regime at $\bar{n}=1$, we used $\chi = 300$ for both correlation functions.
For the SF strongly interacting regime at $U/J=50$ for $1>\bar{n}>0$, we found that the value $\chi = 100$ is enough to capture the dynamical behavior. \\

Concerning the MI phase at $\bar{n}=1$, $\chi = 300$ was considered for moderate values of $U/J$. However, in the strong-coupling regime, the bond dimension can be
significantly decreased and we consider $\chi = 100$, see Figs.~\ref{fig:g2_single_structure} and \ref{fig:velocities_mott_limit}. 

\subsubsection{The density fluctuations - $G_2$ correlation function}
\begin{figure}[t!]
\centering
\includegraphics[scale=0.37]{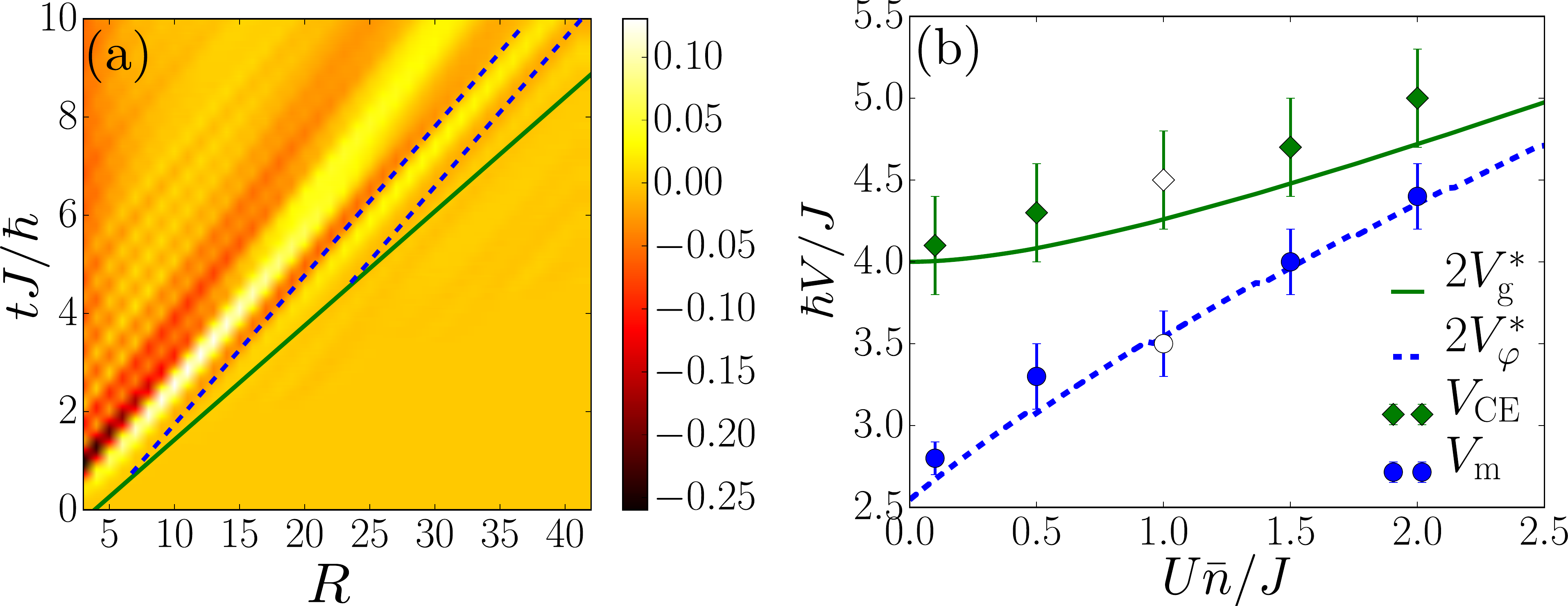}
\caption{\label{fig:velocities_superfluid}
Spreading of correlations in the mean field regime, see red arrow on Fig.~\ref{fig:phase_diagram_bhm}(a). (a)~t-MPS result of $G_2(R,t)$ for a quench 
to $(U/J)_{\mathrm{f}} = 0.1$, together with ballistic fits to the CE (solid, green line) and minima (dashed, blue lines).
(b)~Velocities of the CE ($V_{\mathrm{CE}}$, green diamonds) and minima ($V_{\mathrm{m}}$, blue disks), found from the fits, versus the interaction strength,
and comparison to twice the group velocity $2V_\textrm{g}^*$ (solid green line) and twice the phase velocity $2V_{\varphi}^*$ (dashed blue line). All the quenches
are performed with $\bar{n}=5$ from $(U/J)_\mathrm{i} = 0.2$, except for the points at $U\bar{n}/J=1$ where $(U/J)_\mathrm{f} = 0.2$ and we use a different
initial value, $(U/J)_\mathrm{i} = 0.4$ (open points). Figures extracted from Ref.~\cite{despres2019}.
}
\end{figure}

We first consider the limit regime (mean field regime) of the SF phase characterized by a small Lieb-Liniger parameter, $\gamma = U/2J\bar{n} \ll 1$.
Figure~\ref{fig:velocities_superfluid}(a) displays the t-MPS result for the $G_2$ correlation function versus distance ($R$) and time ($t$) for a quench from $(U/J)_\mathrm{i}
= 0.2$ to $(U/J)_{\mathrm{f}} = 0.1$ and $\bar{n}=5$, \textit{ie.} from $\gamma_\mathrm{i} = 0.02$ to $\gamma_{\mathrm{f}} = 0.01$, see red arrow on
Fig.~\ref{fig:phase_diagram_bhm}(a). It clearly shows a spike-like structure, characterized by two different velocities. On the one hand, a series of parallel maxima
and minima move along straight lines corresponding to a constant propagation velocity $V_{\mathrm{m}}$ (the dashed blue lines show fits to two of these minima).
On the other hand, the various local extrema start at different activation times $t^*(R)$. The latter are aligned along a straight line with a different slope
(solid green line), corresponding to a constant velocity $V_{\mathrm{CE}}$. The latter defines the correlation edge (CE) beyond which the correlations are suppressed. \\

This twofold structure near the CE is readily interpreted using the quasiparticle picture developed at Sec.~\ref{SRC} which we briefly
outline here (for more details, see Ref.~\cite{cevolani2018}) : the $G_1$ and $G_2$ correlation functions are expanded onto the low-energy excitations of the system.
In the mean field regime of the 1D SRBH model, the latter consist of Bogolyubov quasiparticles with a quasimomentum $k \in \mathcal{B} = [-\pi,+\pi]$ and the 
excitation spectrum $E_k \simeq \sqrt{\gamma_k \left(\gamma_k + 2 \bar{n} U\right)}$ (see Subsec.~\ref{sf_chap3} for a derivation of the excitation 
spectrum in the mean field regime). A correlation at a distance $R$ and a time $t$ is built as a coherent superposition of the contributions of the various 
quasiparticles (see generic form of equal-time connected correlation functions at Eq.~\eqref{generic_form}).
In the vicinity of the CE, only the fastest ones implying a quasimomentum $k$ close to $k^*$, contribute, see Fig.~\ref{corr}(a). It creates a sine-like signal
at the spatial frequency $k^\star$, whose extrema move at twice the phase velocity \textit{ie.} $V_{\mathrm{m}} = 2V_{\varphi}^*$~\cite{cevolani2018}. 
Furthermore, this sine-like signal is modulated by an envelope moving at the CE velocity $V_{\mathrm{CE}} = 2V_{\mathrm{g}}^*$ determined by the variations of the
excitation spectrum $E_k$ around the quasimomentum $k^*$, see Fig.~\ref{corr}(b). Finally, the motion of this envelope defines the correlation edge. \\

To test this picture quantitatively, the velocities $V_{\mathrm{m}}$ and $V_{\mathrm{CE}}$ are extracted from the t-MPS results for $G_2(R,t)$ by
tracking, respectively, the local extrema and the activation times. More precisely, $V_{\mathrm{CE}}$ is determined by identifying the value of $t$ (time)
associated to the first correlation for different distances $R$. Then, the velocity $V_{\mathrm{m}}$ is characterized by fitting a local maxima or minima 
in the vicinity of the CE. The results, displayed on Fig.~\ref{fig:velocities_superfluid}(b), show excellent agreement
with the theory, \textit{ie.} $V_{\mathrm{CE}} \simeq 2 V_{\mathrm{g}}^*$ and $V_{\mathrm{m}} \simeq 2 V_{\varphi}^*$ within the fitting errorbars,
see also Fig.~\ref{fig:BHm}. Consequently, the latter support the theoretical predictions for the spreading velocities found using our 
quasiparticle approach presented above. It is important to point out that the t-MPS results are numerically exact and include effects beyond the 
Bogolyubov approximation such as quasiparticle collisions which is not capture by our correlation spreading theory based on the motion of free quasiparticles
with an infinite lifetime. \\

Another important property, predicted by our quasiparticle approach for the correlation spreading in short-range lattice models,
concerns the irrelevance of the observable. Note that the previous statement is valid as long as the considered correlation function can be cast into 
the generic form of Eq.~\eqref{generic_form}. Indeed, we found that the spreading velocities associated to the linear twofold structure of correlations do 
not depend on the local observables $\hat{A}$ and $\hat{B}$, see Eq.~\eqref{generic_corr} and Sec.~\ref{SRC}. As a consequence, the $G_1$ one-body correlation
function is investigated below.

\subsubsection{The phase fluctuations - $G_1$ correlation function}
Previously, we focused on the space-time behavior of the two-body (density-density) correlation function to test the predictions of the quasiparticle approach
concerning the spreading velocities of the linear twofold structure found for Hamiltonians with short-range couplings. To verify their irrelevance with respect to
the local observables defining the equal-time connected correlation function, we turn to the numerical study of the one-body correlations $G_1(R,t)$ using the t-MPS 
technique. 

\begin{figure}[h!]
\centering
\includegraphics[scale=0.37]{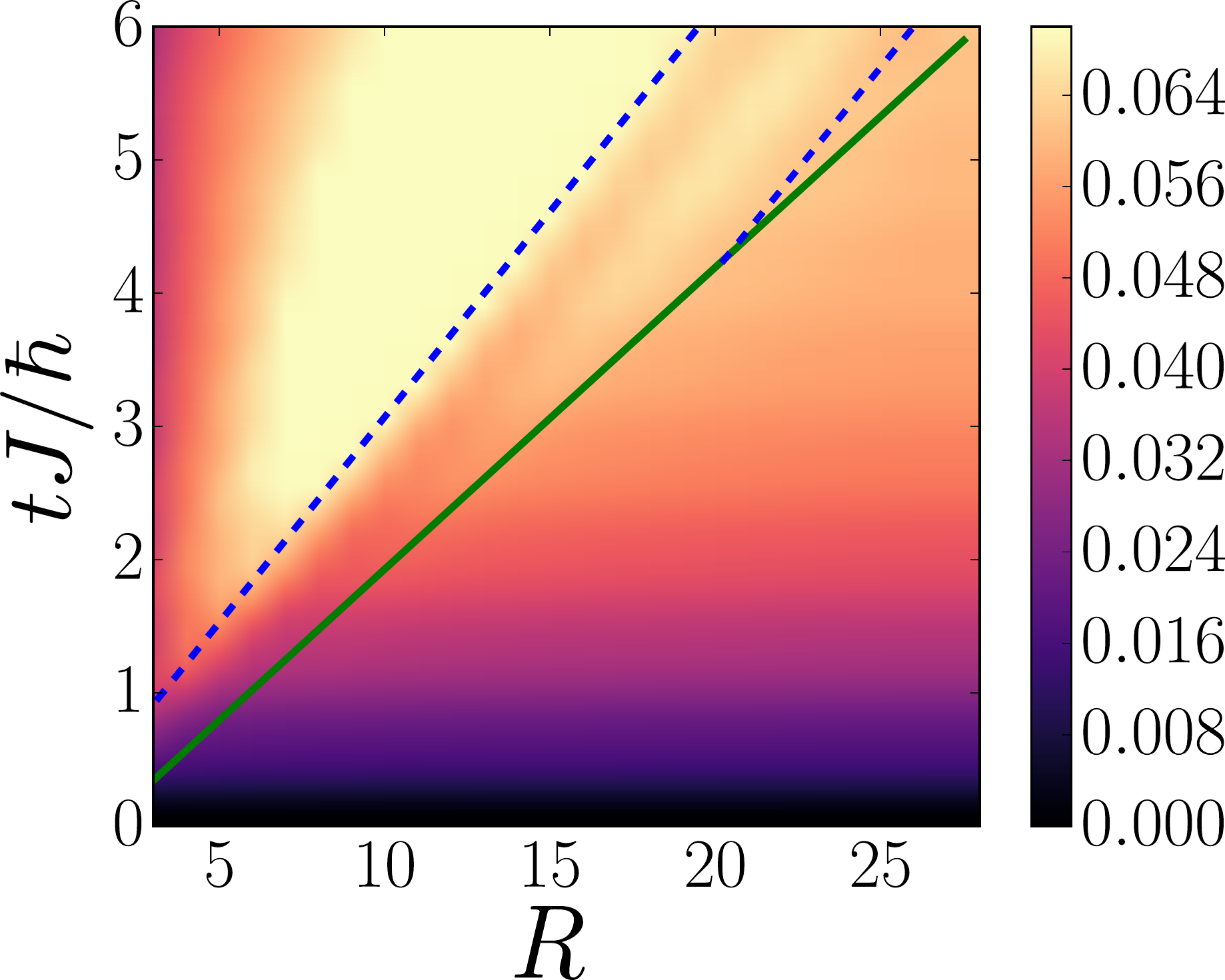}
\caption{\label{fig:velocities_g1_superfluid}
Spreading of the one-body correlation function $G_1(R,t)$ for a global quench in the SF mean field regime from $(U/J)_{\mathrm{i}} = 0.2$ to
$(U/J)_{\mathrm{f}} = 0.1$ and $\bar{n}=5$. The solid-green and dashed-blue lines are fits to the CE and maxima, respectively. For clarity,
the colorbar has been modified from the one used for $G_2$. Figure extracted from the supplemental material of Ref.~\cite{despres2019}.}
\end{figure}

We found that the dynamics of the $G_1$ correlation function shows a spike-like structure, similar to that found for $G_2$. The values of the correlation
edge ($V_{\mathrm{CE}}$) and maxima ($V_{\mathrm{m}}$) velocities agree with those found for the $G_2$ function within less than $10\%$. \\

Figure~\ref{fig:velocities_g1_superfluid} shows an example, for a sudden global quench from $(U\bar{n}/J)_{\mathrm{i}} = 1$ to $(U\bar{n}/J)_{\mathrm{f}} = 0.5$
while fixing the filling to $\bar{n}=5$. The fits to the correlation edge and to the maxima yield the velocities $V_{\mathrm{CE}} = (4.4 \pm 0.3) J/\hbar$
and $V_{\mathrm{m}} = (3.3 \pm 0.2)~ J/\hbar$, in excellent agreement with the corresponding values found from the dynamics of the $G_2$ correlation function,
see Fig.~\ref{fig:velocities_superfluid}(b). It is consistent with the theoretical prediction that the spreading velocities are only characteristic of the 
excitation spectrum and not on the details of the correlation function~\cite{cevolani2018}.\\

Note, however, that the full space-time dependence of the signal
depends on the correlation function \textit{via} the amplitude function $\mathcal{F}$, see Eqs.~\eqref{F1} and \eqref{F2} for its analytical expression
for the $G_1$ and $G_2$ correlation functions respectively. In general, the signal for $G_1$ is less sharp than for $G_2$. This may be attributed to the 
long-range phase correlations present in the initial state, which blur the space-time pattern, see Ref.~\cite{bravyi2006} and discussion at Sec.~\ref{1dbhm}
about the quasi-long-range order of the SF mean field regime and more precisely the spatial behavior at equilibrium of $G_1$ with respect to 
the Luttinger parameter $K$.

\subsection{Strong coupling regime of the Mott-insulating phase}
\subsubsection{The phase fluctuations - $G_1$ correlation function}
The twofold structure for the correlation spreading in short-range lattice models displays several universal features. Indeed, as long as the generic form 
\footnote{This statement is crucial since it exists some pathological cases where the correlation function can not be cast into the
generic form of Eq.~\eqref{generic_form}. However, their space-time patterns can still be explained using our quasiparticle theory but requires further efforts. An example
of such pathological behavior is the $G_2$ density-density correlation function deep in the Mott-insulating phase. The latter is investigated in the present section.} of Eq.~\eqref{generic_form} is fulfilled, the scaling laws and
more precisely the spreading velocities are always characterized by $2V_{\mathrm{g}}^*$ and $2V_{\varphi}^*$ for the CE and the series of local maxima respectively, irrespective of the
correlation function (already discussed in the previous paragraph) and the possible gap in the excitation spectrum. In other words, for sudden global quenches confined in a 
gapless and gapped excitation spectra, the spreading velocities are still given by $2V_{\mathrm{g}}^*$ and $2V_{\varphi}^*$ for the CE and extrema. The latter is verified in the following by
considering the limit regime (strong-coupling) of the gapped MI phase at $\bar{n}=1$ of the 1D SRBH model. \\

\begin{figure}[h!]
\centering
\includegraphics[scale=0.35]{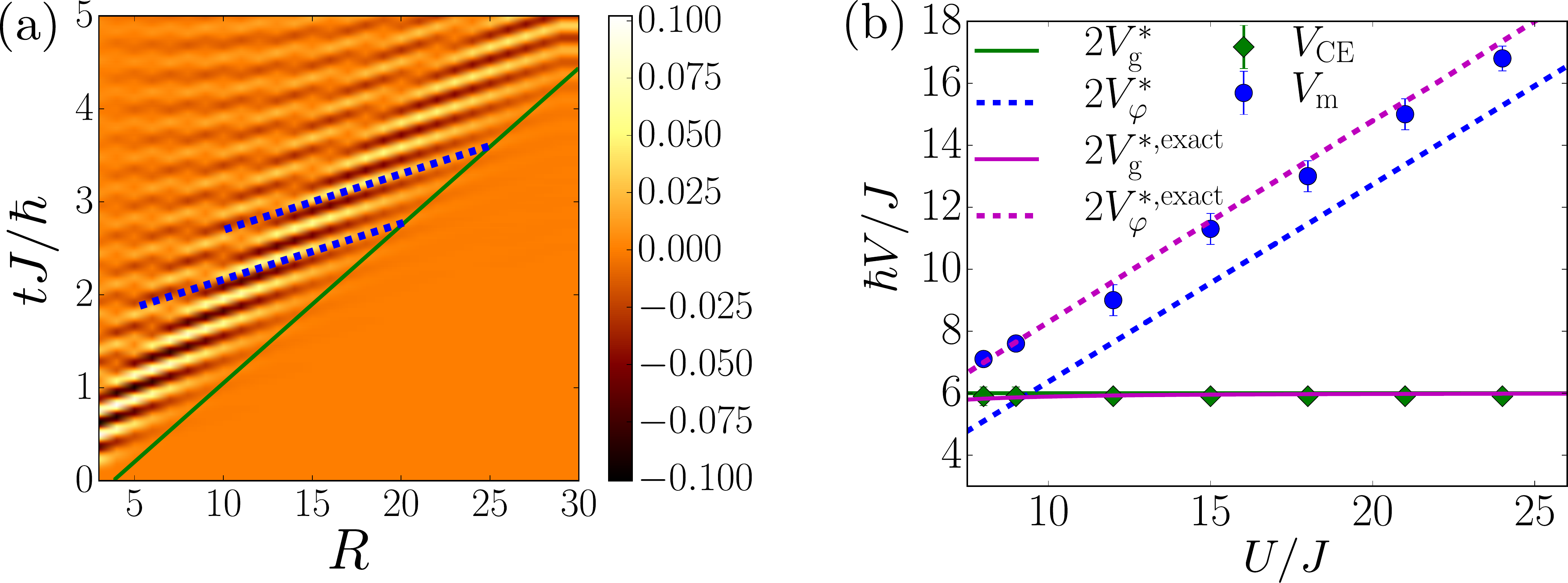}
\caption{\label{fig:velocities_mott_limit}
Spreading of correlations in the MI strong-coupling regime, see purple arrow on Fig.~\ref{fig:phase_diagram_bhm}(a).
(a)~t-MPS result of $G_1(R,t)$ for a quench to $(U/J)_{\mathrm{f}} = 24$, together with ballistic fits to the CE (solid, green line) and minima (dashed, blue lines).
(b)~Velocities of the CE ($V_{\mathrm{CE}}$, green diamonds) and minima ($V_{\mathrm{m}}$, blue disks), found from the fits, versus the interaction strength 
$U/J = (U/J)_{\mathrm{f}}$, and comparison to twice the group velocity $2V_\textrm{g}^*$ (solid lines) and twice the phase velocity $2V_{\varphi}^*$ (dashed lines) obtained
from the excitation spectrum $E_k$ using (blue) the first-order perturbation theory at Eq.~\eqref{ek_chap4_mi} (purple) using the fermionization technique, 
see Eq.~\eqref{mott_disp_rel_fermionization}. All the quenches are performed with $\bar{n}=1$ from $(U/J)_\mathrm{i} \rightarrow +\infty$ (pure Mott state at unit-filling). Figure~(a)
extracted from Ref.~\cite{despres2019}.
}
\end{figure}

Figure~\ref{fig:velocities_mott_limit}(a) displays the t-MPS result of the $G_1$ correlation function for a quench from 
the pure Mott state $\ket{\Psi_{\mathrm{i}}} = \ket{1}^{\otimes N_s}$ (corresponding to the ground state of the 1D SRBH model for a filling $\bar{n}=1$ and 
$(U/J)_\mathrm{i} \rightarrow +\infty$) to $(U/J)_{\mathrm{f}} = 24$ and $\bar{n}=1$ (the filling is fixed during the real-time evolution process), see 
purple arrow on Fig.~\ref{fig:phase_diagram_bhm}(a). As before, the space-time pattern of the $G_1$ correlations shows a linear twofold structure, 
characterized by two different velocities. However, contrary to the SF mean field regime, the series of parallel extrema propagate faster than 
the correlation edge \textit{ie.} $V_{\mathrm{m}} > V_{\mathrm{CE}}$. \\
The spreading velocities $V_{\mathrm{m}}$ and $V_{\mathrm{CE}}$ extracted from the t-MPS results of $G_1(R,t)$, by tracking
the maxima and the activation times as previously, as a function of the post-quench interaction parameter $U/J = (U/J)_{\mathrm{f}}$ are 
reported on Fig.~\ref{fig:velocities_mott_limit}(b). The numerical results (see green diamond and blue disks for $V_{\mathrm{CE}}$ and $V_{\mathrm{m}}$ respectively)
show a very good agreement with the theoretical velocities calculated from the excitation spectrum $E_k$ at Eq.~\eqref{mott_disp_rel_fermionization} obtained from
the fermionization technique (see solid and dashed purple lines for the velocities $2V_{\mathrm{g}}^*$ and $2V_{\varphi}^*$ respectively).
The maximal relative error ($\epsilon_{\mathrm{max}}$) is around $7-8 \%$ concerning the spreading velocity of the local extrema. Besides, the numerical velocities
have also been compared to the analytical velocities deduced from the excitation spectrum $E_k$ at Eq.~\eqref{ek_chap4_mi} valid in the strong-coupling limit (see 
solid and dashed blue lines for $2V_{\mathrm{g}}^*$ and $2V_{\varphi}^*$ respectively). Between these analytical predictions and the numerical results, the maximal
relative error is around $\epsilon_{\mathrm{max}} = 15\%$. However, to get a relatively good agreement with the predictions obtained from the strong-coupling
expansion ($\epsilon_{\mathrm{max}} \leq 10$), sufficiently high values of the post-quench interaction parameter $U/J$ have to be considered ($U/J \geq 18$). 
The latter is in agreement with results of Ref.~\cite{barmettler2012} concerning the accuracy of the strong-coupling expansion to characterize correctly the 
excitation spectrum of the MI phase. \\

According to the agreement between the t-MPS and theoretical results on Fig.~\ref{fig:velocities_mott_limit}(b), the universal feature of the spreading 
velocities concerning the irrelevance of the excitation nature, \textit{ie.} if the excitation spectrum is gapless or gapped, has been verified. In the next paragraph,
we turn to a study of the $G_2$ density fluctuations while still considering a sudden global quench in the MI strong-coupling regime. 

\subsubsection{The density fluctuations - $G_2$ correlation function}

\begin{figure}[h!]
\centering
\includegraphics[scale = 0.36]{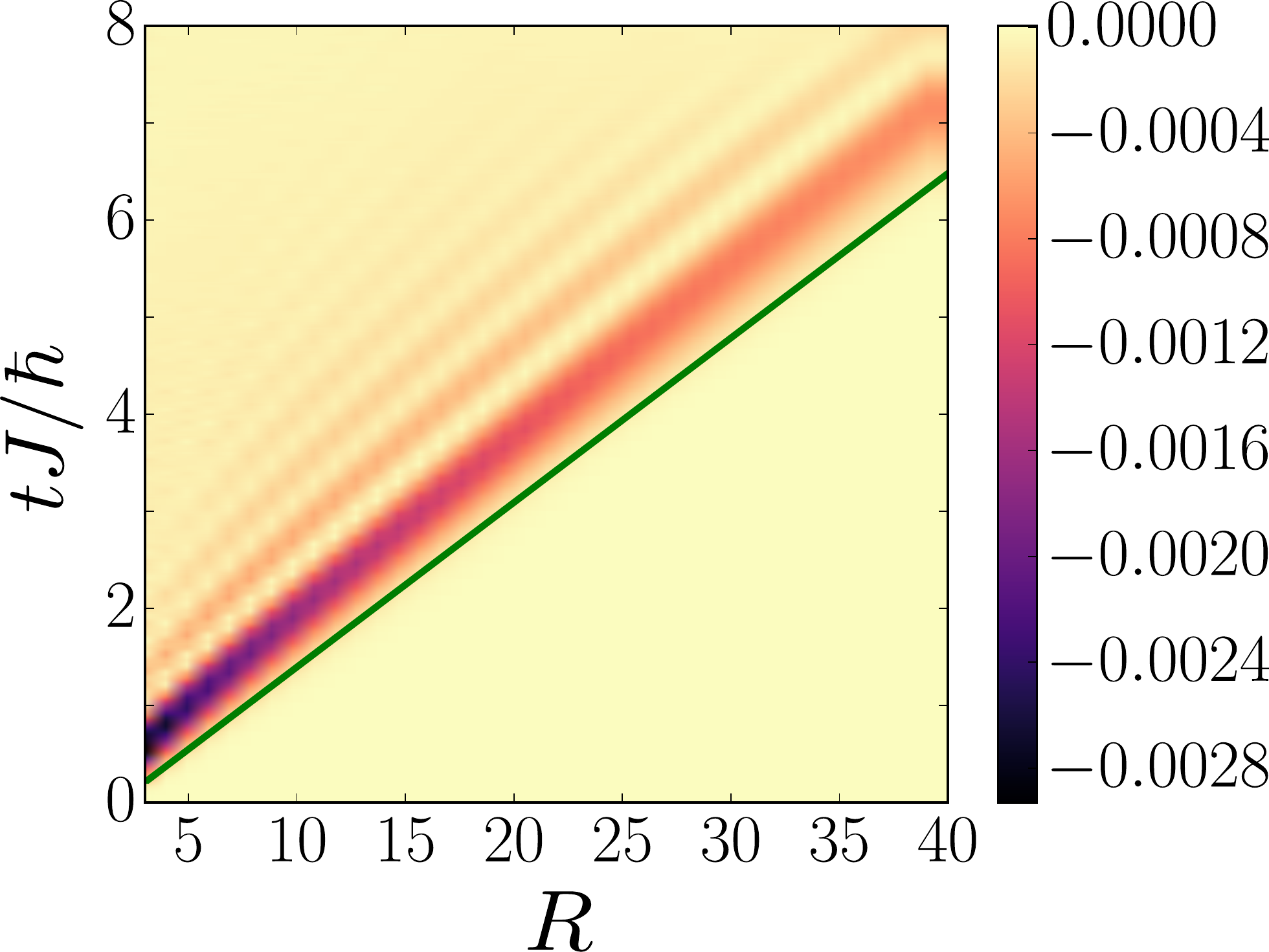}
\caption{\label{fig:g2_single_structure}
t-MPS result of the density-density correlation function $G_2(R,t)$ in the MI strong-coupling regime, see purple arrow on Fig.~\ref{fig:phase_diagram_bhm}(a), for
a sudden global quench from $(U/J)_{\mathrm{i}} \rightarrow +\infty$ to $(U/J)_{\mathrm{f}} = 24$, together with a ballistic fit to the CE (solid, green line).
}
\end{figure}

The $G_2$ density fluctuations in the MI strong-coupling regime are a pathological case of the universal linear twofold structure for the correlation spreading.
On Fig.~\ref{fig:g2_single_structure}, the density-density correlations are displayed for a global quench from the pure Mott state ($(U/J)_{\mathrm{i}} \rightarrow +\infty$)
to $(U/J)_{\mathrm{f}} = 24$ at fixed filling $\bar{n}=1$. The space-time pattern displays a single structure where the CE and the maxima spread with a similar velocity,
\textit{ie.} $V_{\mathrm{CE}} \simeq V_{\mathrm{m}}$. However, one can still rely on our quasiparticle theory to explain the single structure. \\ 

\begin{figure}[h!]
\centering
\includegraphics[scale = 0.36]{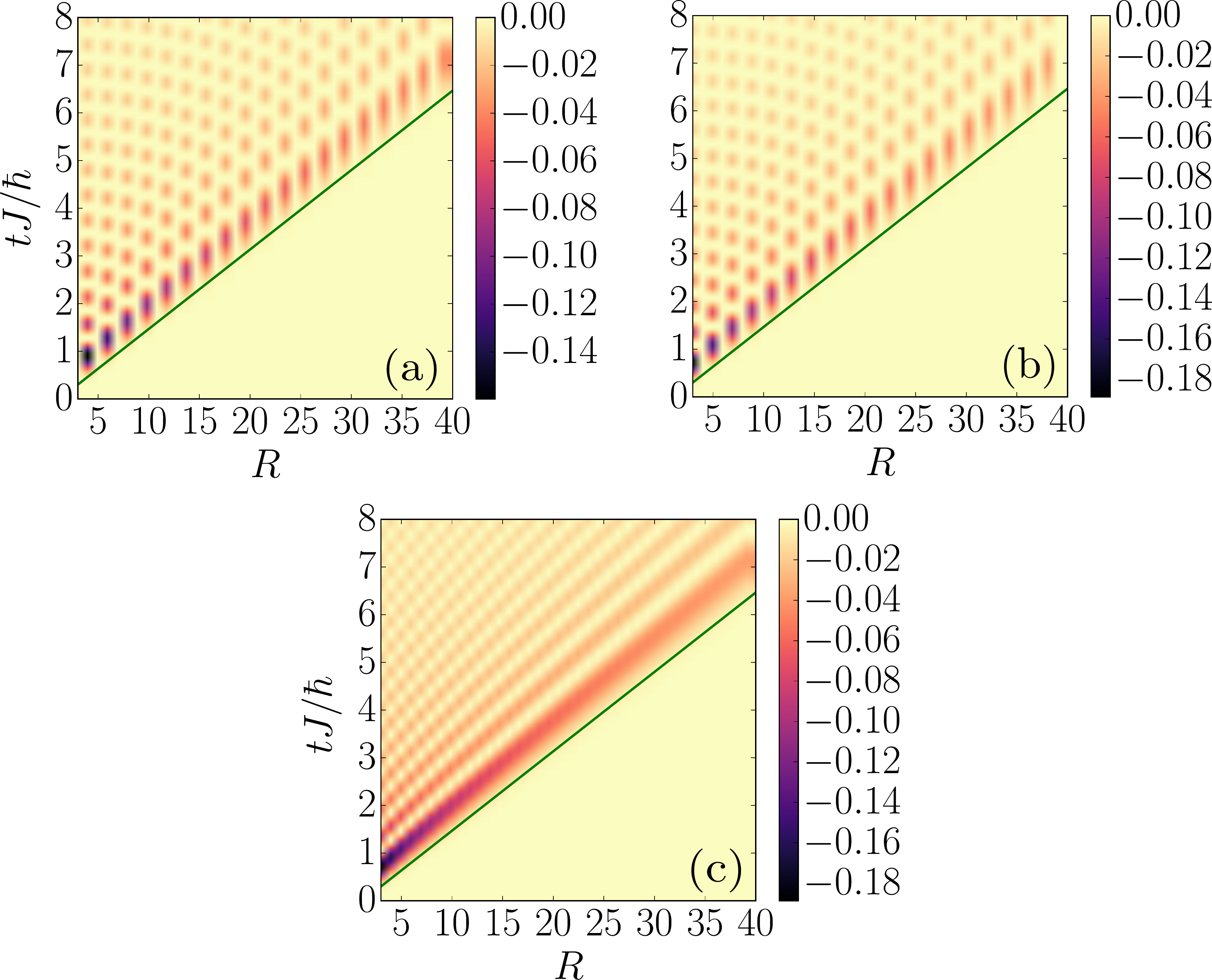}
\caption{\label{fig:g2_analysis}
Analysis of the space-time correlation pattern of $G_2(R,t)$ via $g_2(R,t)$ [see Eq.~\eqref{g2_mott}] at $\bar{n}=1$
for a global quench confined deep into the Mott-insulating phase starting from a pure Mott state $(U/J)_0 \rightarrow \infty$.
Analytical expression, owing to prefactors, of (a)~ $-\Re^2 \left[g_2(R,t) \right]$ (b)~ $-\Im^2 \left[g_2(R,t) \right]$ (c)~ sum
of the two contributions shown at Fig.~(a) and (b). The solid green line corresponds to the theoretical CE velocity characterized by
$2\tilde{V}_{\mathrm{g}}^* = 2J(2\bar{n}+1)$. On Fig.~(c), the first extremum propagates with the same velocity as the one associated to the CE. 
}
\end{figure}

In contrast to $G_1$, see Eq.~\eqref{g1_analytical_mi} and \eqref{F1_mi}, the $G_2$ function can not be cast into the generic
form given at Eq.~\eqref{generic_form}. Instead, combining Jordan-Wigner fermionization and Fermi-Bogolyubov theory to diagonalize the Hamiltonian of the one-dimensional
short-range Bose-Hubbard model in the strong-coupling limit of the Mott-insulating phase~\cite{barmettler2012}, it yields $G_2(R,t) \simeq -2|g_2(R,t)|^2$ with 

\begin{equation}
g_2(R,t) \sim \frac{J}{U}\frac{R}{t} \int_{\mathcal{B}} \frac{\mathrm{d}k}{2 \pi} \left \{e^{i \left(2E_kt + kR\right)} + e^{i \left(2E_kt - kR \right)} \right \}.
\label{g2_mott}
\end{equation}

\noindent
with $\mathcal{B} = [-\pi, \pi]$ the first Brillouin zone. For $U \gg 2(2\bar{n}+1)J$, one can rely on the doublon-holon pair excitation spectrum valid in the
strong-coupling limit given by $2E_k \simeq U - 2(2\bar{n}+1)J\cos(k)$. Owing to the square modulus in the formula $G_2(R,t) \simeq -|g_2(R,t)|^2$, it turns
out that the Mott gap $U$ becomes irrelevant and we are left with the following effective excitation spectrum $2\tilde{E}_k = - 2J(2\bar{n}+1)\cos(k)$.
On the one hand, the group velocity is not affected and we find the maximum value $2V_{\mathrm{g}}^* = 2(2\bar{n}+1)J/\hbar$ at $k^*= \pi/2$. 
The value $2V_{\mathrm{g}}^*=6J/\hbar$ found for $\bar{n}=1$ is in excellent agreement with $V_{\mathrm{CE}}$ fitted from the $G_2$ correlation function
deep in the MI phase, see Fig.~\ref{fig:g2_single_structure}. On the other hand, the corresponding effective phase velocity vanishes, $2\tilde{V}_{\varphi}^* = 0$.
This is consistent with the disappearance of the twofold structure observed in the t-MPS calculations for $G_2$ deep in the MI phase.
More precisely, we find that in the vicinity of the CE both the real and imaginary parts of $g_2$ display a series of static local maxima, 
consistently with $2\tilde{V}_{\varphi}^* = 0$, see Figs.~\ref{fig:g2_analysis}(a,b). These local maxima are shifted by half a period and 
cancel each other when combined for constructing $G_2$, see Fig.~\ref{fig:g2_analysis}(c), see Appendix~\ref{appendix4_single_structure} for more details. \\

In this paragraph, by investigating the $G_2$ density fluctuations in the MI strong-coupling regime of the 1D SRBH, we have shown a pathological case of the expected
twofold structure for the correlation spreading. Indeed, one obtains a single structure where the series of local extrema and the CE propagate at the same velocity
$2V_{\mathrm{g}}^*$  the expected twofold structure. The singularity of $G_2$ deep in the Mott-insulating phase lies in its non-generic form presented at Eq.~\ref{generic_form}.
However, relying on the stationary phase approximation and the quasiparticle approach, the specific space-time pattern of $G_2$ can still be explained,
see Appendix.~\ref{appendix4_single_structure}. \\

In the following, the phase $G_1$ and density $G_2$ fluctuations are investigated while scanning both phase transitions 
(the Mott-$U$ and the Mott-$\delta$ transitions) of the 1D SRBH model. The purpose is to investigate the space-time behavior of the correlations 
relatively close to the critical points to know whether a twofold correlation subsists or not. 
 
\section{Strongly correlated regime at unit-filling : Mott-$U$ transition}
\label{BKT_ph_tr}

\begin{figure}[t!]
\centering
\includegraphics[scale = 0.28]{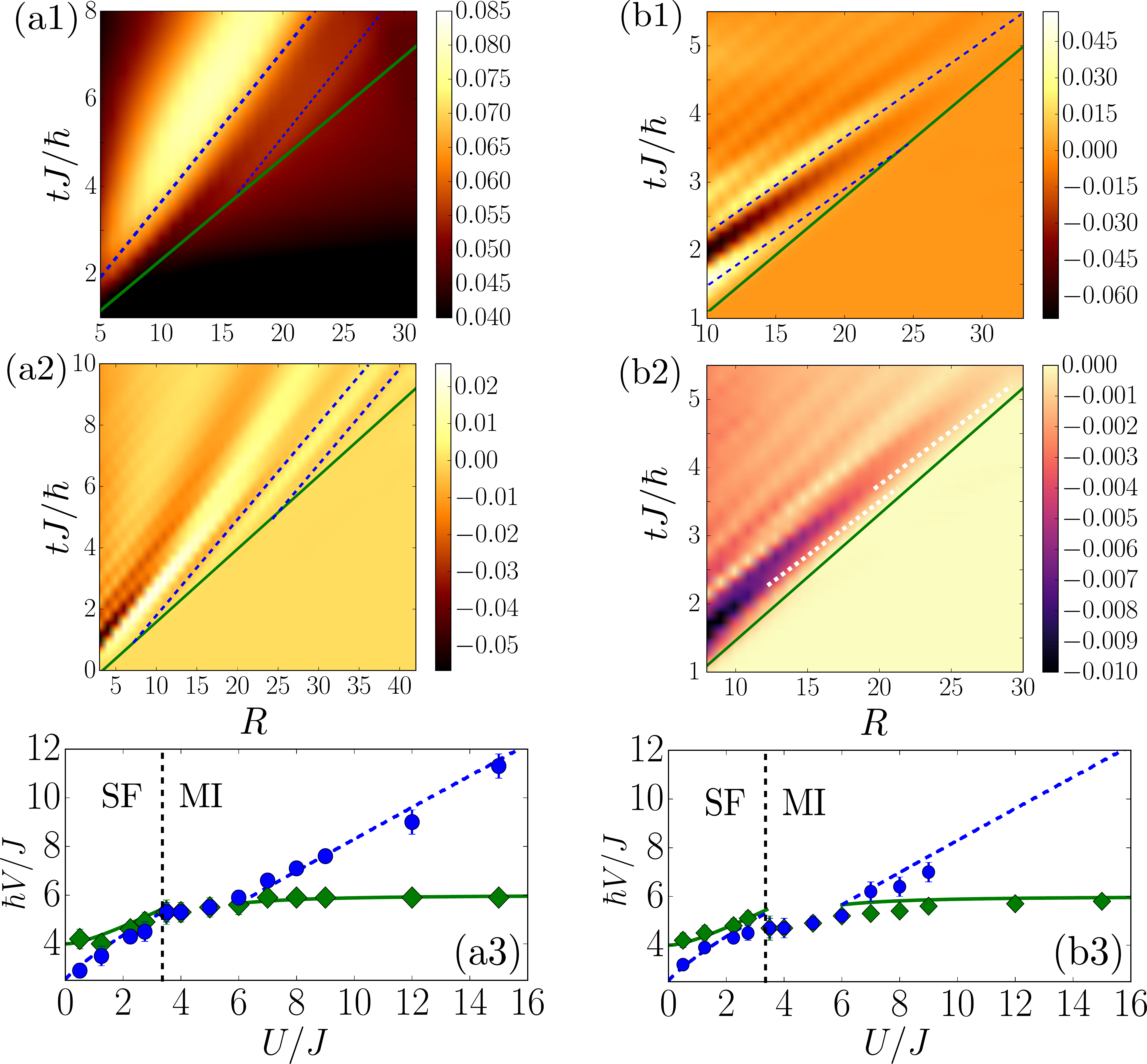}
\caption{\label{fig:numerics_mott_u}
Spreading of the $G_1$ (first row) and $G_2$ (second row) correlations in both the SF and MI phases at $\bar{n}=1$ while scanning the post-quench
interaction $(U/J)_{\mathrm{f}} = U/J$ along the Mott-$U$ transition, see pink dashed line and magenta arrows on Fig.~\ref{fig:phase_diagram_bhm}(a) :
(a)~SF regime with $(U/J)_{\mathrm{f}} = 0.5$. (b)~MI regime near the critical point $u_c \simeq 3.3$ with $(U/J)_{\mathrm{f}} = 8$.
The solid green and dashed blue lines correspond to fits to the CE and extrema, respectively. Note that on panel~(b2), the fits to the maxima are shown 
as dashed white lines for clarity. (d)~Spreading velocities $V_{\mathrm{CE}}$ (green diamonds) and $V_{\mathrm{m}}$ (blue disks), as extracted from fits
to the t-MPS data, and comparison to the characteristic velocities $2V_{\mathrm{g}}^*$ (solid green lines) and $2V_{\varphi}^*$ (dashed blue lines), as
found from the excitation spectrum in the SF~[Eq.~\eqref{eq:BogoDisp}] and MI~[Eq.~\eqref{mott_disp_rel_fermionization}] regimes. All the quenches are performed from
the pre-quench interaction parameter $(U/J)_\mathrm{i} = 1$ for the SF regime and $(U/J)_{\mathrm{i}} = +\infty$ for the MI regime. Figure adapted from
Ref.~\cite{despres2019}.
}
\end{figure}

We now turn to the strongly correlated regime $\gamma = U/2J\bar{n} \sim 1$, where the $G_1$ and $G_2$ correlation functions cannot be systematically calculated.
We first scan the post-quench interaction parameter $(U/J)_{\mathrm{f}}$ from the SF to the MI, along the Mott-$U$ transition at unit filling \textit{ie.} $\bar{n}=1$,
see magenta arrows on Fig.~\ref{fig:phase_diagram_bhm}(a). Note that each global quench is confined in a unique phase : for $(U/J)_{\mathrm{f}} <u_{c}\simeq 3.3$
(SF regime), the initial interaction parameter $(U/J)_{\mathrm{i}}=1$ is considered while for $(U/J)_{\mathrm{f}} > u_{c}$ (MI regime), $(U/J)_{\mathrm{i}} = +\infty$.
Figure~\ref{fig:numerics_mott_u} shows typical numerical results for the spreading of the $G_1$ (first row) and $G_2$ (second row) correlations
for quenches confined in the SF phase [$(U/J)_{\mathrm{f}} = 0.5$, Fig.~\ref{fig:numerics_mott_u}(a1,a2)], and to the MI phase and slightly beyond the critical point 
$u_c \simeq 3.3$ [$(U/J)_{\mathrm{f}} = 8$, Fig.~\ref{fig:numerics_mott_u}(b1,b2)]. In all cases, we find a twofold spike-like structure for both correlation functions.
The corresponding velocities $V_{\mathrm{m}}$ and $V_{\mathrm{CE}}$, extracted as before, are plotted on Fig.~\ref{fig:numerics_mott_u}(a3,b3) and show similar results 
for $G_1$ and $G_2$. This is consistent with one of the universal features of the twofold spike-like structure found for the correlation spreading in short-range lattice models
and verified previously in the limit regime of both phases : the characteristic spreading velocities of the correlation cone are irrespective of the observable, see Ref.~\cite{cevolani2018}. \\

In the SF regime, $(U/J)_{\mathrm{f}} < u_{c}$, the numerical spreading velocities compare very well with the predictions 
$2V_{\varphi}^*$ and $2V_{\mathrm{g}}^*$ as found from the Bogolyubov excitation spectrum at Eq.~\eqref{eq:BogoDisp} 
[see, respectively, the dashed blue and solid green lines on Figs.~\ref{fig:numerics_mott_u}(a3) and (b3)].
Quite surprising, the agreement is fair up to the critical point where $\gamma_c = u_c/2 \simeq 1.6$, far beyond the validity condition of mean field theory implying
$\gamma \ll 1$. \\

Close to the critical point $(U/J)_{\mathrm{f}} \leq u_c$, the quasimomentum $k^*$ decreases down to the phononic regime, \textit{ie.} $k \ll \pi$, according to
the excitation spectrum $E_k$ obtained from the Bogolyubov theory. Consequently, the precise $k$-dependence of $E_k$ beyond this regime becomes irrelevant. 
Moreover, the physics being dominated by long wavelength excitations ($k$ small), the lattice discretization in Eq.~\eqref{ham_bhm} may be disregarded and 
the 1D SRBH model maps onto the continuous 1D Lieb-Liniger model [see Appendix.~\ref{appendix1_mf} for more details]. The latter is integrable by Bethe ansatz
(BA)~\cite{lieb1963a,lieb1963b}. It yields the sound velocity $V_s \simeq 2\bar{n}\sqrt{\gamma}\left(1-\sqrt{\gamma}/4\pi\right)$ at lowest order in the
weak-$\gamma$ expansion. Up to the critical point, the beyond-mean-field correction, $\sqrt{\gamma}/4\pi$, is less than $10\%$, which explains the good agreement
between the numerics and the analytic formula. At the critical point, the numerical results for $V_{\mathrm{m}}$ and $V_{\mathrm{CE}}$ are consistent with
the BA value $2V_s \simeq 4.6$~\cite{despres2019}. \\

The spreading velocities $V_{\mathrm{m}}$ and $V_{\mathrm{CE}}$ are continuous at the Mott-$U$ transition, and do not show any critical behavior.
Right beyond the critical point, they are still nearly equal and we can hardly distinguish two features from the numerics up to $(U/J)_{\mathrm{f}} \simeq 6$.
The latter is in agreement with $t$-VMC calculations where a single structure was found for the $G_2$ density correlations in the MI phase at 
moderate $(U/J)_{\mathrm{f}}$, see Ref.~\cite{carleo2014}. \\

Nevertheless, further entering the MI phase, we recover two distinct features associated to two different velocities. Contrary to the SF regime, 
here we find $V_{\mathrm{m}} > V_{\mathrm{CE}}$, see Fig.~\ref{mi_exc_vel}. These results are readily interpreted from the quasiparticle picture.
Deep enough in the MI phase, $(U/J)_{\mathrm{f}} \gtrsim 6$, the low-energy excitations are doublon-holon pairs, characterized by the excitation spectrum given at 
Eq.~\eqref{mott_disp_rel_fermionization}, see also Refs.~\cite{barmettler2012,Ejima2012}. The comparison between the spreading velocities
$V_{\mathrm{m}}$ and $V_{\mathrm{CE}}$ fitted from the t-MPS results and the theoretical predictions $2V_{\varphi}^*$ and
$2V_{\mathrm{g}}^*$ found from Eq.~(\ref{mott_disp_rel_fermionization}), yields a very good agreement, within less than $5\%$ 
for $G_1$ and $9\%$ for $G_2$, see Figs.~\ref{fig:numerics_mott_u}(a3) and (b3) respectively. The quantitative agreement between the t-MPS
results and the theoretical predictions for the $G_1$ correlations persists up to arbitrary values of $(U/J)_{\mathrm{f}}$. This validates the quasiparticle
analysis also in the strong-coupling regime. \\

Yet, the $G_2$ correlations behave differently. For intermediate interactions, $6 \lesssim (U/J)_{\mathrm{f}} \lesssim 9$, we find a twofold structure consistent
with that found for $G_1$, see Appendix.~\ref{appendix4_single_structure}. However, the latter is relatively hard to distinguish due to a second contribution to the
correlations favouring a single structure. This twofold structure for $G_2$ at moderate $U/J$ has not been identified experimentally in Ref.~\cite{cheneau2012}.
The signal for $G_2$ blurs when entering deeper in the MI regime, and we are not able to identify two distinct features for $U/J \gtrsim 9$, see previous
section and Appendix.~\ref{appendix4_single_structure}. Indeed, the space-time pattern is reduced to a single structure where
$V_{\mathrm{m}} \simeq V_{\mathrm{CE}}$ and characterized by the velocity $2V_{\mathrm{g}}^* = 2J(2\bar{n}+1)$ independently of the post-quench
interaction parameter $U/J = (U/J)_{\mathrm{f}}$.
 
\section{Strongly interacting regime of the superfluid phase}
We finally consider the strongly interacting regime of the SF phase, characterized by $\gamma = U/2J\bar{n} \gg 1$ and $\bar{n} \notin \mathbb{N}$.
More precisely, it implies strong interactions $U/J$ and small non-integer filling $\bar{n}$, see Fig.~\ref{fig:phase_diagram_bhm}(b).
In the following, by investigating the $G_2$ correlations, the purpose is twofold. Firstly, we determine numerically if a twofold dynamics subsists
for a global quench confined in this specific regime. Secondly, we are interested in understanding how the twofold structure (if it exists) evolves when doping the chain (increasing
the filling $\bar{n} \notin \mathbb{N}$) until reaching the critical point of the Mott-$\delta$ \textit{ie.} $\bar{n}=1$. \\

In the strongly interacting regime of the SF phase, the Tomonaga-Luttinger liquid (TLL) theory accurately describes the low-energy physics of the 1D SRBH 
model at equilibrium, including the Mott-$\delta$ transition, see for instance Refs.~\cite{cazalilla2011,haller2010,boeris2016}. The TLL theory
considers an effective harmonic fluid, characterized by a single characteristic velocity, namely the sound velocity $V_s$. Indeed, an effective linear and gapless 
excitation spectrum $E_k$ at low quasimomenta is assumed to describe the low-energy properties of a model, leading to $V_{\mathrm{g}}(k) = V_{\varphi}(k)
= V_s$. As a consequence, the TLL theory can only be used to describe the space-time dynamics of models displaying a single structure for the correlations.     
In contrast, our t-MPS simulations in the strongly interacting SF regime clearly show beyond TLL physics since a twofold dynamics propagating 
at different velocities has been found (at least for relatively small $\bar{n}$). \\

\begin{figure}[t!]
\centering
\includegraphics[scale = 0.35]{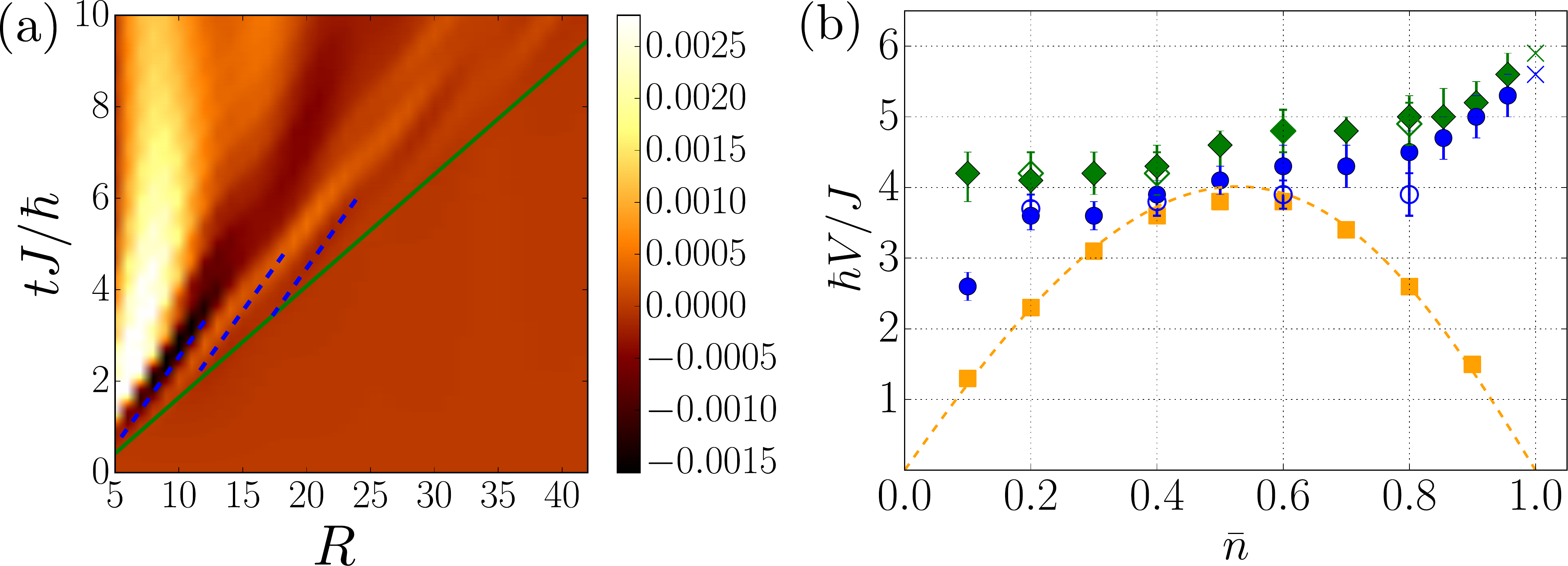}
\caption{ \label{fig:velocities_g2_sir_sf}
Twofold spreading of the $G_2$ density correlations in the strongly interacting SF regime for $(U/J)_{\mathrm{f}} = U/J = 50$ and $0<\bar{n}<1$.
(a) t-MPS result of the connected two-body correlation function $G_2(R,t)$ for a global quench starting from $(U/J)_\mathrm{i} = 1$ to
$U/J = 50$ at filling $\bar{n} \simeq 0.1$. The solid green line and dashed blue lines correspond to ballistic fits to the propagation of the 
correlation edge (CE) and several extrema (m) respectively. The resulting spreading velocities (slope of the linear fits) are displayed on the panel (b). (b) Shown are the
spreading velocities $V_{\mathrm{CE}}$ (green diamonds) and $V_{\mathrm{m}}$ (blue disks) fitted from the t-MPS data,
together with twice the sound velocity $2V_s$ of the Bose-Hubbard model as found from Bose-Fermi mapping (dashed orange line) and from 
MPS simulations (orange squares). Filled symbols correspond to the pre-quench interaction parameter $(U/J)_{\mathrm{i}} = 1$ and open symbols
to $(U/J)_{\mathrm{i}} = 40$. The crosses are linear extrapolations of $V_{\mathrm{CE}}$ and $V_{\mathrm{m}}$ to the Mott-$\delta$ transition at
$\bar{n}=1$.}
\end{figure}

Indeed, we have computed the spreading of the $G_2$ correlations for a fixed and large value of the post-quench interaction parameter, \textit{ie.} $(U/J)_{\mathrm{f}} = 50$,
and varying the filling $\bar{n}$ up to the Mott-$\delta$ transition at $\bar{n}=1$, see pink arrow on Fig.~\ref{fig:phase_diagram_bhm}(a) and Fig.~\ref{fig:velocities_g2_sir_sf}(a)
for a typical t-MPS result of $G_2$. The spreading 
velocities $V_{\mathrm{CE}}$ (green diamonds) and $V_{\mathrm{m}}$ (blue disks), extracted from t-MPS data of $G_2$, are shown on Fig.~\ref{fig:velocities_g2_sir_sf}(b). They show clear deviations from twice the sound velocity of the 1D short-range Bose-Hubbard
model in the strongly interacting limit, $2V_s \simeq (4J/\hbar)\sin(\pi\bar{n}) \big[1-(8J/U)\cos(\pi\bar{n})\big]$
(orange dotted line and squares). The sound velocity $V_s$ has been computed by mapping the Bose-Hubbard model to an equivalent spinless Fermi
model~\cite{cazalilla2003,cazalilla2004} (dotted orange line) and, independently, from the energies of the ground and first excited states using MPS
calculations (see orange squares), showing excellent agreement. \\
Moreover, the emergence of two different characteristic velocities, $V_{\mathrm{CE}} \neq V_{\mathrm{m}}$, indicates that the TLL approach is insufficient 
to describe the spreading of correlations, even upon renormalization of the effective TLL parameters. Note that the two velocities become nearly equal in
the vicinity of the Mott-$\delta$ transition ($\bar{n} \leq 1$) and reach the value $V_{\mathrm{CE}} \simeq V_{\mathrm{m}} \simeq 6 J/\hbar$.
This is consistent with the disappearance of the twofold structure and the value $V_{\mathrm{CE}}$ found for a global quench deep into the MI phase
at $\bar{n}=1$, see Fig.~\ref{fig:numerics_mott_u}(b3) and Appendix.~\ref{appendix4_single_structure}. \\

\section{Introduction to the dynamical excitation spectrum}
In the previous section, the correlation spreading in the strongly interacting (SI) regime of the superfluid phase has been characterized
\textit{via} the study of the $G_2$ density fluctuations for a sudden global quench on the interaction parameter $U/J$. 
We unveiled a twofold linear structure reducing to a single one close to the critical point, \textit{ie.} for $\bar{n} \lesssim 1$, see Fig.~\ref{fig:velocities_g2_sir_sf}.
Nevertheless, the numerical spreading velocities ($V_{\mathrm{m}}$, $V_{\mathrm{CE}}$) have not been compared to theoretical predictions. Indeed, since the
Bose-Hubbard chain is not exactly-solvable in the SI regime, the corresponding excitation spectrum $E_k$ can not be determined analytically. Furthermore, the latter has also 
not been characterized numerically in the literature. \\

However, by considering the problem in the opposite direction, we propose the idea to determine the excitation spectrum $E_k$ \textit{via} an analysis of the 
equal-time connected correlation functions. According to our quasiparticle theory, the latter can be cast (in general) into the generic form 
[see Eq.~\eqref{generic_form}] reading as

\begin{equation}
 G(\mathbf{R},t) = g(\mathbf{R}) - \int_{\mathcal{B}} \frac{\mathrm{d}\mathbf{k}}{(2\pi)^D} \mathcal{F}(\mathbf{k})
 \left \{ \frac{e^{i(\mathbf{k}.\mathbf{R}+2E_\mathbf{k}t)} + e^{i(\mathbf{k}.\mathbf{R}-2E_\mathbf{k}t)}}{2} \right \},
 \label{generic_form_for_des}
\end{equation}

\noindent
where $\mathcal{B}$ denotes the first Brillouin zone and $E_{\mathbf{k}} = E_{\mathbf{k},\mathrm{f}}$ corresponds to the excitation spectrum
of the post-quench Hamiltonian $\hat{H}_{\mathrm{f}}$. The so-called dynamical excitation spectrum (DES) denoted by $S(\mathbf{k},\omega)$ is defined as the $D+1$ (space-time)
Fourier transform of $G(\mathbf{R},t)$ the equal-time connected correlation function and reads as ~\cite{despres2019bis2}

\begin{equation}
S(\mathbf{k},\omega) := \int_{[0,L^D]} \mathrm{d}\mathbf{R} \int_{0}^{T} \mathrm{d}t ~G(\mathbf{R},t) e^{-i(\mathbf{k}.\mathbf{R} + \omega t)},
\label{des}
\end{equation}

\noindent
where $T$ refers to the observation time, $D$ the dimension of the hypercubic lattice and $L$ the length (or equivalently the total number of lattice sites, $a$ the lattice
spacing is fixed to unity) of the quantum lattice model along each spatial dimension. Inserting the generic form of $G(\mathbf{R},t)$ at Eq.~\eqref{generic_form_for_des}
in the definition of the dynamical excitation spectrum (DES) at Eq.~\eqref{des}, it yields in the infinite distance ($L \rightarrow + \infty$)
and observation time limit ($T \rightarrow + \infty$) the following analytical expression for the DES

\begin{equation}
S(\mathbf{k},\omega) \sim \mathcal{F}(\mathbf{k}) \left[\delta(\omega + 2E_{\mathbf{k}}) + \delta(\omega - 2E_{\mathbf{k}}) \right]. 
\label{DES2}
\end{equation}

\noindent
The previous form of the DES allows us to determine the post-quench excitation spectrum $E_{\mathbf{k}}$ renormalized by the amplitude
function $\mathcal{F}(\mathbf{k})$ depending not only on the quench parameters but also on the local observables defining the equal-time correlation function $G(\mathbf{R},t)$.
Indeed, by analyzing Eq.~\eqref{DES2}, it comes immediately that $S(\mathbf{k},\omega)$ represents twice the post-quench excitation spectrum $2E_{\mathbf{k}}$. More 
precisely, the latter is built from the two branches $2E_{\mathbf{k}}$ and $-2E_{\mathbf{k}}$. \\

\begin{figure}[t!]
\centering
\includegraphics[scale = 0.352]{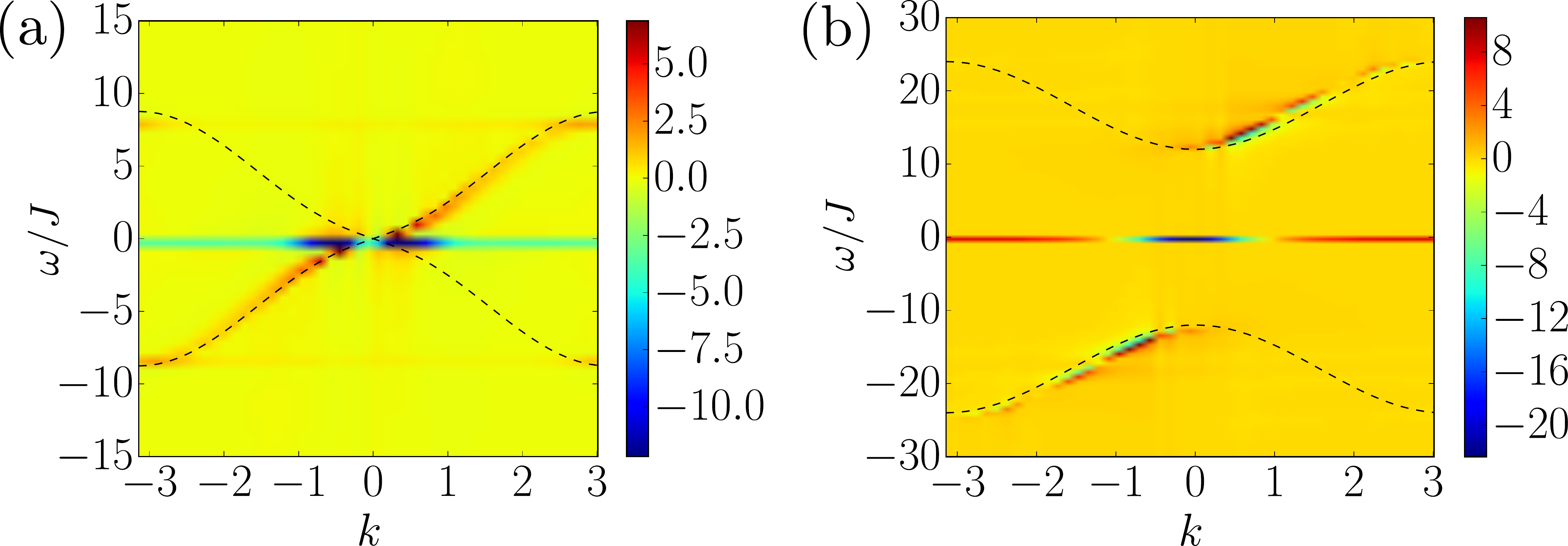}
\caption{ \label{fig:dyn_exc}
Dynamical excitation spectrum $S(k,\omega)$ as a function of the quasimomentum $k$ and the rescaled (dimensionless) energy $\omega/J$ computed from the $t$-MPS data of the correlation spreading
in the short-range Bose-Hubbard chain. (a)~ $S(k,\omega)$ computed from the $G_2(R,t)$ density fluctuations in the superfluid mean field regime at $U_{\mathrm{f}}\bar{n} = 0.5J$,
see Fig.~\ref{fig:velocities_superfluid}(a). (b)~ $S(k,\omega)$ computed from the $G_1(R,t)$ phase fluctuations in the strong-coupling regime of the
Mott-insulating phase at $\bar{n}=1$ and $U_\mathrm{f} = 18J$. The black dashed lines correspond to twice the theoretical post-quench excitation spectrum $E_{k}$ and $-E_{k}$
(a)~ in the SF mean field regime given at Eq.~\eqref{eq:BogoDisp} (b)~ in the strong-coupling limit of the MI phase, see Eq.~\eqref{Ek_MI}. Only the branches with
a positive group velocity are visible since the region $R,t>0$ of the space-time plane is considered for the correlations. Note that the non-zero amplitudes of $S(k,\omega=0)$, containing all the 
time-independent terms of the space-time correlations, are irrelevant for our study.}
\end{figure}

On Fig.~\ref{fig:dyn_exc}, we display several numerical results concerning the dynamical excitation spectrum 
of the short-range Bose-Hubbard chain in both the gapless superfluid and gapped Mott-insulating quantum phases. On Fig.~\ref{fig:dyn_exc}(a), the DES
is deduced by taking the 2D Fourier transform of the $G_2(R,t)$ density fluctuations in the superfluid mean field regime at $U_{\mathrm{f}}\bar{n} = 0.5J$, see
Fig.~\ref{fig:velocities_superfluid}(a). On Fig.~\ref{fig:dyn_exc}(b), $S(k,\omega)$ is computed from the $G_1(R,t)$ phase fluctuations in the strong-coupling regime
of the Mott-insulating phase at $\bar{n}=1$ and $U_\mathrm{f} = 18J$ (see Fig.~\ref{fig:velocities_mott_limit}(a) for $U_{\mathrm{f}} = 24J$).  
Both of them show a very good agreement with the corresponding theoretical excitation spectrum $E_k$ (see black dashed lines) defined at Eq.~\eqref{eq:BogoDisp} for the SF mean field regime
and Eq.~\eqref{Ek_MI} for the strong-coupling limit of the MI phase. Note that only half of the branches $2E_k$ and $-2E_k$ characterized by a positive
group velocity are visible at Fig.~\ref{fig:dyn_exc}. This is due to the $t$-MPS numerical results of the connected correlation functions displaying only positive group velocities since the region $R,t > 0$ of the 
space-time plane is considered. \\

Consequently, the dynamical excitation spectrum corresponds to the first proof that it is possible to investigate the static properties of lattice models
\textit{via} an analysis of their quench dynamics. We point out that this method to probe the excitation spectrum can be extended to far-from-equilibrium 
many-body quantum models in higher dimensions and with long-range interactions (see Ref.~\cite{despres2019bis2} for more details). Furthermore, this technique 
can be applied to the experimental measurements \footnote{From an experimental point of view, one has to keep in mind the consequences of the shot noise and the 
technical noise.} of the correlation functions and can thus be seen as an alternative to the standard experimental
techniques to measure excitation spectra such as the angle-resolved photoemission spectroscopy (ARPES) ~\cite{stewart2008} or Bragg spectroscopy \cite{rey2005}.\\ \\

In summary, working within the case study of the Bose-Hubbard chain and using a numerically-exact many-body approach, we have presented
evidence of a universal twofold dynamics in the spreading of correlations. The latter is characterized by two distinct velocities,
corresponding to the spreading of local maxima on the one hand and to the CE on the other hand. This has been found in all the phases and regimes of the model.
Exceptions appear only in a few cases, for instance (i)~for specific observables in specific regimes \textit{e.g.} the $G_2$ correlations deep in the 
Mott-insulating phase, or (ii)~when the two velocities are (almost) equal, as found at the critical points of the Mott-$U$ and Mott-$\delta$ transitions 
for instance. \\
These predictions are directly relevant to quench experiments using ultracold Bose gases loaded in optical lattices, where the dynamics of the phase and density 
correlations can be observed on space and time scales comparable to our simulations~\cite{lewenstein2007, bloch2008, cheneau2012, trotzky2012,
NaturePhysicsInsight2012bloch}. Most importantly, while in most experiments and numerics the CE is inferred from the behavior of the correlation maxima,
our results show that the two must be distinguished. This is expected to be a general feature of short-range systems and should be relevant to models
other than the sole Bose-Hubbard model. $~~~~~~~~~$

In the following, we investigate the correlation spreading in short-range interacting lattice models for another class of quantum quenches 
\textit{ie.} the sudden local quenches. To do so, space-time local observables are investigated \textit{e.g.} the local density and local magnetization for particle and 
spin lattice models respectively. The purpose of this section is to determine whether a twofold linear structure, similar to the one unveiled for sudden
global quenches, subsists or not and to characterize the corresponding spreading velocities.

\section{Local quenches in short-range interacting models}
\label{local_quench_sec}
\subsection{Local density in the Mott-insulating phase}
In this section, the correlation spreading is investigated for local quenches within the case study of a short-range interacting bosonic and spin lattice model 
\textit{ie.} the Bose-Hubbard and $s=1/2$ Heisenberg chains. As previously discussed, the purpose is to characterize the causality cone and to know whether 
the associated structure is reminiscent of the one unveiled for global quenches. The local quenches considered in the following for the bosonic and spin lattice
models follow a similar scheme presented below.

\begin{enumerate}
 \item One starts from a many-body product state. The latter can be obtained by computing the ground state of a Hamiltonian $\hat{H}$ in a limit regime.
 For the Bose-Hubbard chain, one can rely on the pure Mott state ($\hat{H}$ with $U/J \rightarrow + \infty$ and $\bar{n} \in \mathbb{N}^*$) and defined as $\ket{\Psi}
 = \ket{\bar{n}}^{\otimes N_s}$ with $N_s$ the total number of lattice sites and $\bar{n}$ the filling. On the contrary, for the $s=1/2$ Heisenberg chain, a fully polarized state 
 along an arbitrary direction can be considered. Hence, one can rely on the following many-body product state
 $\ket{\Psi} = \ket{\uparrow}^{\otimes N_s}$. 
 \item The previous product state $\ket{\Psi}$ is then locally perturbed (breaking the translational invariance of the quantum state) which defines the initial
 state $\ket{\Psi_0}$. For the Bose-Hubbard chain, the local perturbation is performed by creating a doublon-holon excitation pair leading for the initial state $\ket{\Psi_0}$ to
 \begin{equation}
 \ket{\Psi_0} = \ket{\bar{n}...
 \bar{n}, \bar{n}-1, \bar{n}+1, \bar{n}, ..., \bar{n}} = \frac{1}{\sqrt{\bar{n}(\bar{n}+1)}} \hat{a}^{\dag}_{N_s/2} \hat{a}_{N_s/2-1} \ket{\Psi}.
 \label{hes_mott}
 \end{equation}
 
 \noindent
 For the Heisenberg chain, it corresponds to a spin-flip applied on a specific lattice site. The initial state $\ket{\Psi_0}$ can thus be written as
 \begin{equation}
 \ket{\Psi_0} = \ket{\uparrow ... \uparrow \downarrow \uparrow ... \uparrow} = \hat{S}^-_{N_s/2} \ket{\Psi}.
 \end{equation}
 \noindent
 
 \item The initial state $\ket{\Psi_0}$ is then evolved unitarily in time with respect to $\hat{H}$, the Hamiltonian considered to deduce the non-perturbed many-body quantum state $\ket{\Psi}$.
 According to the time-dependent Schrödinger equation ($\hbar$ fixed to unity), $\ket{\Psi(t)}$ the time-evolved many-body quantum state has the following expression $\ket{\Psi(t)} = e^{-i \hat{H}t} \ket{\Psi_0}$.
\end{enumerate}

\noindent
Similarly to sudden global quenches, the initial state $\ket{\Psi_0}$ represents a highly-excited state for the Hamiltonian $\hat{H}$ governing the real time evolution.
Indeed, due to the local perturbation, the energy of the initial state $\ket{\Psi_0}$ is much higher than the one associated to the
ground state $\ket{\mathrm{GS}}$ of $\hat{H}$, \textit{ie.} $E_\mathrm{0} = \langle \Psi_\mathrm{0} | \hat{H} | \Psi_\mathrm{0} \rangle \gg E_{\mathrm{GS}} = \langle 
\mathrm{GS} | \hat{H} | \mathrm{GS} \rangle$. Note that the previous scheme to perform a sudden local quench and thus to drive out of equilibrium a lattice model can be generalized
to many others quantum systems including fermionic lattice models. \\

In the following, we first investigate the out-of-equilibrium dynamics of the short-range Bose-Hubbard chain in the strong-coupling regime ($U \gg J\bar{n}$) of
the Mott-insulating phase for $\bar{n} \in [|1,2|]$. The perturbed initial state $\ket{\Psi_0}$ is built from the pure Mott state ($U/J \rightarrow +\infty$)
at an integer filling $\bar{n}$ where a doublon-holon excitation pair is created at the center of the bosonic chain. To characterize the dynamics induced by sudden 
local quenches of the Bose-Hubbard chain, we study the space-time pattern of a local (on-site) observable namely the local density. The corresponding expectation value to compute is 
$\langle \Psi_0 | \hat{n}_R(t) | \Psi_0 \rangle$ where $\hat{n}_R$ denotes the bosonic occupation number operator acting on the local Hilbert space $\mathbb{H}_R$ (on the
lattice site $R$). The latter evolves in time with respect to the Hamiltonian $\hat{H}$ leading to $\hat{n}_R(t) = e^{i\hat{H}t} \hat{n}_R e^{-i\hat{H}t}$ according to the Heisenberg picture. $\hat{H}$ denotes a 
Hamiltonian whose ground state is very close to the pure Mott state $\ket{\bar{n}}^{\otimes N_s}$ implying for the interaction parameter to fulfill $U/J \gg 1$. \\

\begin{figure}[!hb]
\centering
\includegraphics[scale = 0.36]{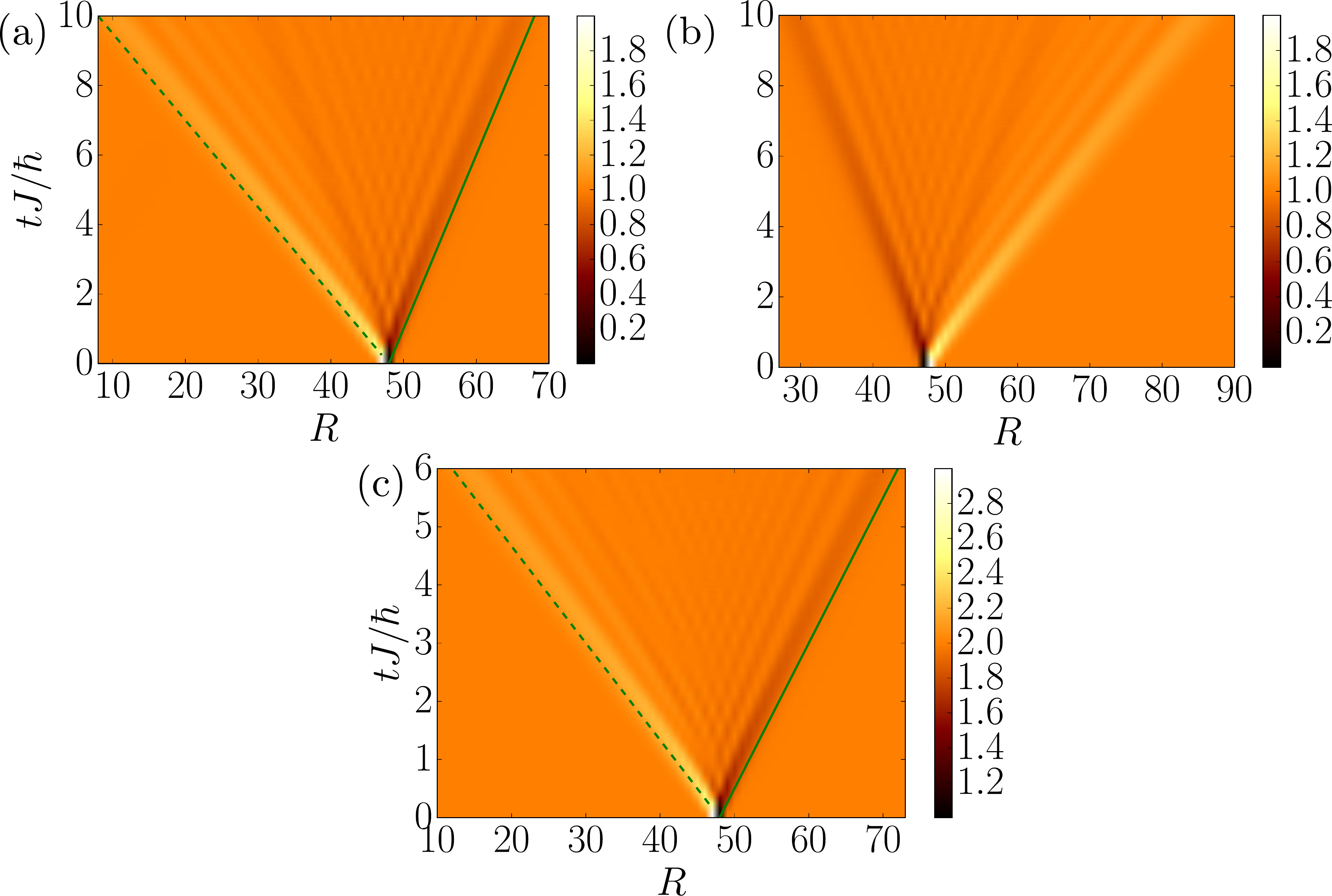}
\caption{\label{fig:mott_local_quench}
t-MPS results of the space-time local density $\langle \Psi_0 | \hat{n}_R(t) | \Psi_0 \rangle$ in the strong-coupling regime ($U \gg J\bar{n}$) of the Mott-insulating phase for different sudden local quenches.
The initial state is defined by (a)~$ \ket{\Psi_0}= \ket{1, ..., 1, 2, 0, 1, ..., 1}$ for $\bar{n}=1$,
(b)~$\ket{\Psi_0} = \ket{1, ..., 1, 0, 2, 1, ..., 1}$, at $\bar{n}=1$ and (c)~$\ket{\Psi_0} = \ket{2, ..., 2, 3, 1, 2, ..., 2}$ for $\bar{n} = 2$.
The dashed and solid green lines represent linear fits to the motion of the CE on the left (governed by the slowest doublon) and on the right (governed by the fastest holon) respectively.}
\end{figure}

On Fig.~\ref{fig:mott_local_quench}, several t-MPS results of the expectation value $\langle \Psi_0 | \hat{n}_R(t) | \Psi_0 \rangle$ are displayed for 
different initial states $\ket{\Psi_0}$ as a function of the distance $R$ and the dimensionless time $tJ/\hbar$. 
On Fig.~\ref{fig:mott_local_quench}(a), the initial state is defined by $\ket{\Psi_0} = \ket{1, ..., 1, 2, 0, 1, ..., 1}$ for $\bar{n}=1$. 
On the panel (b) and (c), $\ket{\Psi_0} = \ket{1, ..., 1, 0, 2, 1, ..., 1}$ and $\ket{\Psi_0} = \ket{2, ..., 2, 3, 1, 2, ..., 2}$ for $\bar{n} = 2$ respectively.
For each case, one immediately notices that the causality cone displays a single linear structure, where the spreading velocities are almost equal ($V_{\mathrm{CE}}
\simeq V_{\mathrm{m}}$). However, the causal region of correlations \footnote{The expectation value $\langle \Psi_0 | \hat{n}_R(t) | \Psi_0 \rangle$ can be seen as a non-equal-time correlation function, see definition of the 
locally-perturbed many-body quantum state $\ket{\Psi_0}$ at Eq.~\eqref{hes_mott}.} is asymmetric contrary to those observed for sudden global quenches. Indeed, the correlation edge
has a different velocity depending on the propagation direction. For instance, on Fig.~\ref{fig:mott_local_quench}(a), the CE velocity is higher for correlations spreading
in the left direction than the one associated to correlations propagating in the right direction. In the following, relying on a quasiparticle approach, we shed new
light about these new and interesting features of the local quench dynamics for the 1D short-range Bose-Hubbard model in the Mott-insulating strong-coupling regime. \\

In the gapped Mott-insulating phase, the elementary excitations are doublon-holon quasiparticle pairs. A doublon (d) corresponds to an excess of particle ($\bar{n}+1$ bosonic
particles on a lattice site) and a holon (h) to $\bar{n}-1$ bosonic particles. In the strong-coupling regime, the excitation spectrum has already been characterized 
and is defined as $2E_{k} = U - 2J(2\bar{n}+1)\cos(k)$, see also the discussion in Appendix~\ref{appendix_g1_mi}. $2E_k$ gives the energy of a doublon-holon
quasiparticle pair for a quasimomentum $k$ confined in the first Brillouin zone. More precisely, the latter is built from the energy of a doublon with 
a quasimomentum $k$ and the one of a holon with an opposite quasimomentum $-k$, \textit{ie.} $2E_{k} = E_{\mathrm{d},k} + E_{\mathrm{h},-k}$ where
$E_{\mathrm{d},k} = U/2 -2J(\bar{n}+1)\cos(k)$ and $E_{\mathrm{h},-k} = U/2 - 2J\bar{n}\cos(k)$ [see Ref.~\cite{barmettler2012} for the analytical expressions of $E_{\mathrm{d},k}$ 
and $E_{\mathrm{h},-k}$ determined using a fermionization technique within a Fermi-Bogolyubov transformation]. \\

Relying on our quasiparticle theory, this asymmetric single structure can be explained. Let us analyze for instance the t-MPS
result at Fig.~\ref{fig:mott_local_quench}(a). The single structure with local density amplitudes close to the value $\bar{n}+1$ gives us information about the spreading
of the doublons into the bosonic chain. The second one with amplitudes close to the value $\bar{n}-1$ is related to the propagation of the holons (the non-causal region
of correlations is characterized by the constant value $\bar{n}$, the filling of the bosonic chain). 
More precisely, the correlations are here created \textit{via} the propagation of free doublons (on the left) with quasimomentum $k \leq 0$ (note that $k$ has also to be confined into the first 
Brillouin zone $\mathcal{B} = [-\pi,\pi]$) at the group velocity $V_{\mathrm{g},\mathrm{d}}(k) = \partial_k E_{\mathrm{d},k} = 2J(\bar{n}+1)\sin(k) \leq 0$ and free holons in the opposite direction (on the right)
with a quasimomentum $k \geq 0$ at the group velocity $V_{\mathrm{g},\mathrm{h}}(k) = \partial_k E_{\mathrm{h},k} = 2J\bar{n}\sin(k) \geq 0$. Hence, the first correlations on the 
left (right) are governed by the doublon (holon) propagating with the lowest (highest) group velocity. The doublon with the lowest group velocity 
$V_{\mathrm{g},\mathrm{d}}(k^*_{\mathrm{d}}) = -2J(\bar{n}+1)$ has a quasimomentum $k^*_{\mathrm{d}} = -\pi/2$ whereas the holon with the highest group velocity 
$V_{\mathrm{g},\mathrm{h}}(k^*_{\mathrm{h}}) = 2J\bar{n}$ is characterized by a quasimomentum $k^*_{\mathrm{h}} = \pi/2$. In other words, the spreading velocity of the CE 
on the left is determined by $V_{\mathrm{g},\mathrm{d}}(k^*_{\mathrm{d}}) = -2J(\bar{n}+1) = -4J$ and the one associated to the CE on the right is given by 
$V_{\mathrm{g},\mathrm{h}}(k^*_{\mathrm{h}}) = 2J\bar{n} = 2J$ [$\bar{n}=1$ for Fig.~\ref{fig:mott_local_quench}(a)]. The correlation edge velocities have been
determined using linear fits and lead to $V_\mathrm{CE}^{\mathrm{l}} = -(4.0 \pm 0.3)J$ and $V_\mathrm{CE}^{\mathrm{r}} = (2.0 \pm 0.2)J$. 
The index $\mathrm{l}$ ($\mathrm{r}$) stands for left (right). The extracted CE velocities are in excellent agreement with the theoretical predictions. \\

A similar analysis allows us to easily interpret the results displayed on Fig.~\ref{fig:mott_local_quench}(b) and (c). For the t-MPS result at Fig.~\ref{fig:mott_local_quench}(b), we
considered the same parameters than those on Fig.~\ref{fig:mott_local_quench}(a). Only the initial state is different between the two simulations where the location
of the doublon and the holon have been exchanged. As expected, the causality cone has the same structure except that now the CE on the left (right) is governed by the 
holon (doublon) with the lowest (highest) group velocity. Hence,  $V_\mathrm{CE}^{\mathrm{l}} \simeq V_{\mathrm{g},\mathrm{h}}(k^*_{\mathrm{h}}) = -2J\bar{n} = -2J$ 
with $k^*_{\mathrm{h}} = -\pi/2$ and $V_\mathrm{CE}^{\mathrm{r}} \simeq V_{\mathrm{g},\mathrm{d}}(k^*_{\mathrm{d}}) = 2J(\bar{n}+1) = 4J$ with $k^*_{\mathrm{d}} = \pi/2$. 
For the numerical result on Fig.~\ref{fig:mott_local_quench}(c), the spreading velocities of the CE are similar to those found for Fig.~\ref{fig:mott_local_quench}(a). 
One just needs to replace the filling by the value $\bar{n}=2$ in the expression of the theoretical CE velocities. The latter are determined by
$V_\mathrm{CE}^{\mathrm{l}} \simeq V_{\mathrm{g},\mathrm{d}}(k^*_{\mathrm{d}}) = -2J(\bar{n}+1) = -6J$ with $k^*_{\mathrm{d}} = -\pi/2$ and 
$V_\mathrm{CE}^{\mathrm{r}} \simeq V_{\mathrm{g},\mathrm{h}}(k^*_{\mathrm{h}}) = 2J\bar{n} = 4J$ with $k^*_{\mathrm{h}} = \pi/2$. \\

Note that the previous analysis can be extended. Indeed, the space-time local density displays interesting physical features which can be used for a better characterization of the local quench dynamics 
of the Bose-Hubbard chain in the Mott-insulating strong-coupling regime. Let us consider once again the t-MPS
result on Fig.~\ref{fig:mott_local_quench}(a). For the other cases, the following discussion still holds but requires however to be slightly adapted. Firstly, since a
same lattice site can not be occupied by a doublon and a holon and knowing that $\mathrm{max}[V_\mathrm{g,\mathrm{d}}(k)] > 
\mathrm{max}[V_\mathrm{g,\mathrm{h}}(k)]$, the doublon has to propagate to the left. Using the same arguments, one can directly conclude that the holon 
can spread in both directions. Both previous conclusions are clearly visible on Fig.~\ref{fig:mott_local_quench}(a). 
Another important aspect concerns the theoretical expression of the space-time local density $\langle \Psi_0 | \hat{n}_R(t) | \Psi_0 \rangle$. 
In theory, the latter can be derived analytically since the Hamiltonian of the Bose-Hubbard chain in the Mott-insulating strong-coupling regime can be diagonalized.
Consequently, one could rely on our general scheme (see for instance Chap.\ref{ch:3-universal_scaling_laws} and Appendices.~\ref{appendix_g1_sf} and \ref{appendix_g2_sf}) based on
the Bogolyubov theory \footnote{The general scheme has to be slightly adapted by considering for this specific case the fermionic Bogolyubov theory.}
to deduce the analytical form. However, in practice, the calculation is relatively complex due to the 
calculation of several correlators involving six fermionic operators. Nevertheless, one can easily guess its analytical expression using the analysis presented
in both previous paragraphs. The main properties of the space-time pattern of the local density $\langle \Psi_0 | \hat{n}_R(t) | \Psi_0 \rangle$ are
recalled here :

\begin{itemize}[noitemsep]
 \item The space-time local density displays an asymmetric single structure \footnote{One can thus think about the analytical expression of the density 
 fluctuations $G_2$ in the Mott-insulating strong-coupling regime where a single linear structure was found (see Figs.~\ref{fig:g2_single_structure}, \ref{fig:g2_analysis} and Appendix.~\ref{appendix4_single_structure}).}. A single linear
 structure is associated to the spreading of the holons for all the quasimomenta $k$ in the first Brillouin zone $\mathcal{B} = [-\pi,\pi]$. The second one is
 associated to the propagation of the doublons.
 \item The single linear structure associated to the doublons can only propagate along one direction contrary to the one for the holons. 
 \item The amplitudes of the non-causal region are characterized by the average density (or filling) of the bosonic chain denoted by $\bar{n}$. 
\end{itemize}

\noindent
Using the analytical expression of $G_2$ valid for a sudden global quench deep in the Mott-insulating phase (see Appendix.~\ref{appendix4_single_structure} and 
Figs.~\ref{fig:g2_single_structure}, \ref{fig:g2_analysis}) displaying a single structure, the generic form for the connected correlation functions at Eq.~\eqref{generic_form} and the previous properties, it yields the
following guess for the theoretical expression of $\langle \Psi_0 | \hat{n}_R(t) | \Psi_0 \rangle$ at Fig.~\ref{fig:mott_local_quench}(a)

\begin{align}
& \langle \Psi_0 | \hat{n}_R(t) | \Psi_0 \rangle = \bar{n} - \left| \int_{\mathcal{B}} \frac{\mathrm{d}k}{2\pi}  \frac{ e^{i\left \{ k(R-N_s/2) + E_{k,\mathrm{h}}t \right \}} + e^{i\left \{ -k(R-N_s/2) + E_{k,\mathrm{h}}t \right \}
}}{2}  \right|^2 \nonumber \\
& + \theta^{1-N_s/2}(-R) \left| \int_{\mathcal{B}} \frac{\mathrm{d}k}{2\pi}  \frac{ e^{i \left \{ k \left[ R-(N_s/2-1) \right] + E_{k,\mathrm{d}}t \right \}} + e^{i \left \{ -k\left[R-(N_s/2-1) \right] + E_{k,\mathrm{d}}
t \right \} } }{2}  \right|^2,
\label{local_density_th}
\end{align}

\noindent
where $\theta^{1-N_s/2}(-R) = \theta(-R -1 + N_s/2)$ with $\theta$ the Heaviside function. $N_s$ is the total number of lattice sites, $E_{k,\mathrm{d}} = 
U/2 -2J(\bar{n}+1)\cos(k)$ and $E_{k,\mathrm{h}} = U/2 - 2J\bar{n}\cos(k)$ denote the quasiparticle dispersion relation associated to the doublons and holons respectively. 
The first term $\bar{n}$ denotes the filling of the chain corresponding to the local density amplitude when no excitation is present. Hence, the latter
characterizes the non-causal region of the space-time pattern. The second (third) term is responsible for the spreading of free holons (free doublons)
into the lattice chain symbolized by plane waves whose spreading properties are governed by $E_{k,\mathrm{h}}$ ($E_{k,\mathrm{d}}$). 
For the third term, thanks to the Heaviside function, the doublons can only propagate to the left part of the chain. Note that this term can be mapped onto 
our generic form presented at Eq.~\eqref{generic_form} up to a modulus square. The latter is responsible for the single structure where the maxima 
and the correlation edge spread into the quantum system with the same velocity (see Appendix.~\ref{appendix4_single_structure} for more details). Note that the third
term has a shift of $N_s/2-1$ in the space-dependent part of the plane waves due to the initial position of the doublon. 
Concerning the second term, it represents the spreading of holons in both directions. Once again, the modulus square is responsible for the single structure 
where its series of local extrema and its correlation edge propagate at the same velocity. Besides, this second term has a shift of $N_s/2$ in the space-dependent 
part of the plane waves which corresponds to the initial position of the holon. \\

We stress that Eq.~\eqref{local_density_th} is in excellent (quantitative) agreement with our t-MPS simulations for any filling $\bar{n} \in \mathbb{N}^*$.  
The latter is also consistent with theoretical predictions found at Ref.~\cite{sirker2015}, where the independent propagation of a holon and a doublon have been investigated 
\textit{via} a sudden local quench at the specific filling $\bar{n} = 1$. \\

We now turn to a theoretical and numerical investigation of the space-time local magnetization for a short-range interacting $s=1/2$ spin lattice model.
As previously, the out-of-equilibrium dynamics is also induced by a local quench but this time the quantum phase considered is gapless.
The purpose of the next section is to know whether a linear single structure also appears for the gapless case and for another lattice model, 
before pointing out the main differences and similarities for the quantum dynamics between global and local quenches in short-range interacting lattice models.   

\subsection{Local magnetization in the short-range Heisenberg chain}
\label{1d_heis_local_quench}
In what follows, the one-dimensional $s=1/2$ short-range Heisenberg model (SRHM), also called $s=1/2$ short-range XXX model, is considered. 
The associated Hamiltonian reads as 

\begin{equation}
\hat{H} = -J \sum_{R} \vec{\hat{S}}_R \vec{\hat{S}}_{R+1} = - J \sum_{R} \hat{S}^x_R \hat{S}^x_{R+1} +  \hat{S}^y_R \hat{S}^y_{R+1} +  \hat{S}^z_R \hat{S}^z_{R+1},
\label{H_heis}
\end{equation}

\noindent
where $\vec{\hat{S}}_R = (\hat{S}^x_R,\hat{S}^y_R,\hat{S}^z_R)$ with $\hat{S}^{\alpha}_R = (\hbar/2) \sigma^{\alpha}_R$ ($\hbar$ is fixed to unity in the following)
is the spin operator acting on the lattice site $R \in [|1,N_s|]$ along the $\alpha \in \{x,y,z\}$ direction. $N_s$ denotes the total number of lattice sites and $J>0$ 
the isotropic spin exchange coupling in the three spatial directions. 
The 1D SRHM displays a gapless ferromagnetic phase $\forall J \in \mathbb{R}^{+*}$. In this quantum phase, the spins align along an arbitrary 
direction which is supposed to be along the $z$ direction in the following. In the ferromagnetic phase along the $z$ direction, the model has a twofold degenerate
ground state which may be written as 

\begin{equation}
 \ket{\Psi} = \frac{1}{\sqrt{2}} \left( \ket{\uparrow}^{\otimes N_s} + \ket{\downarrow}^{\otimes N_s} \right).
\end{equation}

Concerning the excitations of the model, the Hamiltonian $\hat{H}$ at Eq.~\eqref{H_heis} can be diagonalized in the ferromagnetic phase along the $z$ direction.
To do so, the following Holstein-Primakoff transformation needs to be considered

\begin{equation}
\hat{S}_{R}^{x}\simeq \frac{\hat{a}_{R} + \hat{a}_{R}^\dagger}{2},~~~
\hat{S}_{R}^{y} \simeq -\frac{\hat{a}_{R}^\dagger-\hat{a}_{R}}{2i},~~~
\hat{S}_{R}^{z} = \frac{1}{2}-\hat{a}_{R}^\dagger \hat{a}_{R},
\end{equation}

\noindent
yielding to a quadratic Bose form for the 1D SRHM model in momentum space presented below

\begin{equation}
\hat{H} = e_{0} + \frac{1}{2} \sum_{k} \mathcal{A}_{k} \left( \hat{a}^{\dag}_k \hat{a}_k + \hat{a}_{-k} \hat{a}^{\dag}_{-k} \right) +
\mathcal{B}_{k} \left( \hat{a}^{\dag}_k \hat{a}^{\dag}_{-k} + \hat{a}_k \hat{a}_{-k} \right),
\label{H_heis_quadra}
\end{equation}

\noindent
where $\mathcal{A}_k = J[1-\cos(k)]$ and $\mathcal{B}_k = 0$ $\forall k \in \mathcal{B} = [-\pi,\pi]$ the first Brillouin zone. We stress that the Hamiltonian at 
Eq.~\eqref{H_heis_quadra} is already diagonalized in terms of the bosonic operators in momentum space ($\hat{a}_k, \hat{a}^{\dag}_k$). As a consequence, there is no need
to perform a bosonic Bogolyubov transformation which would lead to 

\begin{equation}
 \hat{a}_k^{\dag}= u_k \hat{\beta}_{k}^{\dag} + v_k \hat{\beta}_{-k}, ~~~ \hat{a}_k = u_k \hat{\beta}_k + v_k \hat{\beta}_{-k}^{\dag}, 
\end{equation}

\noindent
with $u_k =1$ and $v_k=0$ implying $\hat{a}_k^{(\dag)} = \hat{\beta}_{k}^{(\dag)}$, $\forall k \in \mathcal{B}$. Finally, up to a constant energy, one
can simplify Eq.~\eqref{H_heis_quadra} into 

\begin{equation}
\hat{H} = \sum_k E_k \hat{a}^{\dag}_k \hat{a}_k,~~~ E_k = J[1-\cos(k)],
\label{exc_spec_ferro}
\end{equation}

\noindent
where $E_k = J[1-\cos(k)]$ corresponds to the gapless excitation spectrum of the short-range Heisenberg chain in the ferromagnetic phase along the $z$ axis. 
The latter gives the energy of a magnon (quantized spin-wave excitation) having a quasimomentum $k$ confined in the first Brillouin zone $\mathcal{B}$. At low energies,
the excitation spectrum at Eq.~\eqref{exc_spec_ferro} is quadratic and characterized by $E_k \simeq (J/2)k^2$. 
Hence, the spin lattice model exhibits free magnons at low quasimomenta ($k\rightarrow 0$). \\

In the following, the out-of-equilibrium dynamics of the short-range Heisenberg chain in the gapless ferromagnetic phase along the
$z$ axis is investigated. The perturbed initial state $\ket{\Psi_0}$ is built from the ground state $\ket{\Psi} = \ket{\uparrow}^{\otimes N_s}$ 
where a spin flip is applied on the central lattice site $N_s/2$ leading to $\ket{\Psi_0} = \hat{S}^{-}_{N_s/2} \ket{\Psi} = 
\ket{\uparrow ... \uparrow \downarrow \uparrow ... \uparrow}$. This initial state then evolves unitarily in time with respect to the Hamiltonian $\hat{H}$
of the Heisenberg chain at Eq.~\eqref{H_heis}. Then, to characterize the dynamics induced \textit{via} the previous local quench, the local 
magnetization along the $z$ axis is studied. The corresponding expectation value to compute is thus $\langle \Psi_0 | \hat{S}^{z}_R(t) | \Psi_0 \rangle$ 
where $\hat{S}^z_R$ denotes the spin operator along the $z$ axis acting on the local Hilbert space $\mathbb{H}_R$ (on the
lattice site $R$). \\

\begin{figure}[!h]
\centering
\includegraphics[scale = 0.33]{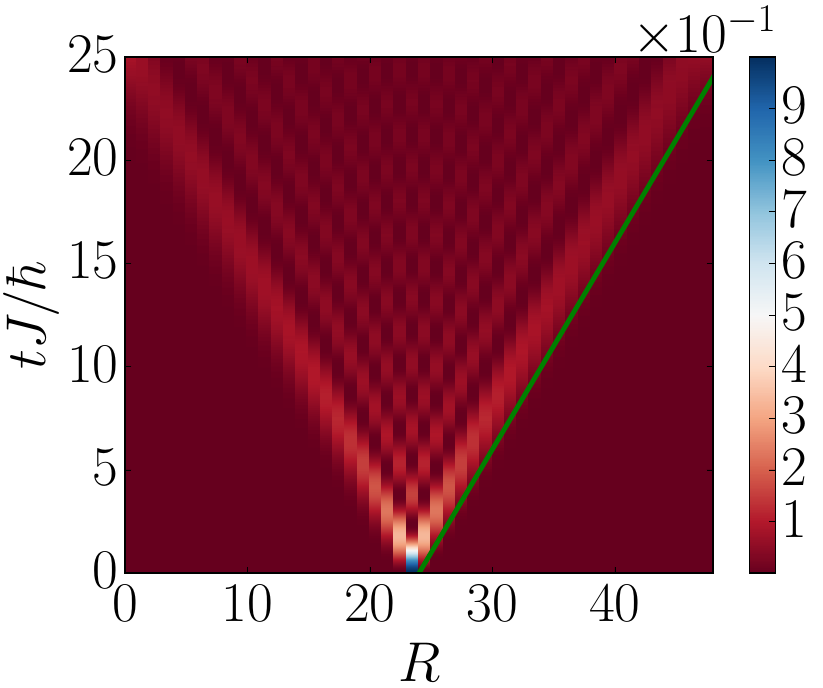} 
\caption{\label{fig:heis_local_quench}
t-MPS result for the spreading of the local magnetization $1/2 - \langle \Psi_0 | \hat{S}^z_R(t) | \Psi_0 \rangle$ in the ferromagnetic phase along
the $z$ axis of the short-range Heisenberg chain. A sudden local quench is considered where the initial state is defined by $\ket{\Psi_0} = 
\ket{\uparrow ... \uparrow \downarrow \uparrow ... \uparrow}$. The solid green line represents a linear fit to the motion of the correlation edge.}
\end{figure}

On Fig.~\ref{fig:heis_local_quench}, a t-MPS result of the local magnetization $1/2 - \langle \Psi_0 | \hat{S}^z_R(t) | \Psi_0 \rangle$ is displayed as a
function of the distance $R$ and the dimensionless time $tJ/\hbar$. The considered initial state is defined as $\ket{\Psi} = \ket{\uparrow ... \uparrow 
\downarrow \uparrow ... \uparrow}$. One immediately notices that the causality cone is symmetric with respect to the central lattice site $N_s/2$ and shows a single linear structure. Indeed, the spreading
velocity associated to the series of local extrema and to the correlation edge is similar, \textit{ie.} $V_{\mathrm{CE}} \simeq V_{\mathrm{m}}$. 
The latter is reminiscent of the behavior found for the local density in the Mott-insulating strong-coupling regime. Indeed, one single linear structure was associated 
to the spreading of doublons and a second one for the spreading of holons. For the Heisenberg chain, the low-energy excitations consist of a single 
species of quasiparticles \textit{ie.} the magnons or spin-wave excitations which is consistent with the fact to find one single linear structure as shown
on Fig.~\ref{fig:heis_local_quench}. The latter also explains that one recovers a symmetric causal structure. Indeed, we show below that the spreading of the
correlation edge on the right (left) is governed by the maximal (minimal) group velocity of the spin-wave excitations. Due to the symmetry $k\rightarrow -k$ of the
excitation spectrum $E_k$ for the 1D Heisenberg model, both velocities are equal in absolute value and characterized by a quasimomentum $k^*$ and $-k^*$ respectively. \\

Relying on our quasiparticle theory, this symmetric single structure can be explained in details. Indeed, using the Bogolyubov theory, the local magnetization
$1/2 - \langle \Psi_0 | \hat{S}^z_R(t) | \Psi_0 \rangle$ can be derived analytically, see Appendix.~\ref{appendix_local_quench_spin} for a full derivation.
The analytical expression of $1/2 - \langle \Psi_0 | \hat{S}^z_R(t) | \Psi_0 \rangle$ reads as 

\begin{equation}
1/2 - \langle \Psi_0 | \hat{S}^z_R(t) | \Psi_0 \rangle = \left| \int_{\mathcal{B}} \frac{\mathrm{d}k}{2\pi} \left \{ \frac{ e^{i[k (R-N_s/2) + E_kt]}
+ e^{i[-k(R-N_s/2) + E_k t]}}{2} \right \} \right|^2,
\label{local_mag_analytics}
\end{equation}

\noindent
where $\mathcal{B} = [-\pi,\pi]$ denotes the first Brillouin zone, $N_s$ the total number of lattice sites and $E_k = J[1-\cos(k)]$ the excitation spectrum 
of the short-range Heisenberg chain in the ferromagnetic phase along the $z$ axis. The latter is in very good (quantitative) agreement with the t-MPS result displayed
on Fig.~\ref{fig:heis_local_quench}. \\

The theoretical expression of the local magnetization at Eq.~\eqref{local_mag_analytics} can be cast into the generic form at Eq.~\eqref{generic_form} up to a modulus square which is responsible for the 
single structure [see Appendix.~\ref{appendix4_single_structure} and also the analytical expression of the space-time local density in the 
Mott-insulating phase at Eq.~\eqref{local_density_th}]. Hence, the correlation edge and the series of local maxima spread into the lattice chain with the
same velocity \textit{ie.} $V_{\mathrm{CE}}^{\mathrm{l}} \simeq V_{\mathrm{m}}^{\mathrm{l}}$ and $V_{\mathrm{CE}}^{\mathrm{r}} \simeq V_{\mathrm{m}}^{\mathrm{r}}$ (the index $\mathrm{l}$ and $\mathrm{r}$ stand for 
'left' and 'right' respectively). The latter are positive (negative) for a spreading on the right (left) side of the chain.
More precisely and according to Eq.~\eqref{local_mag_analytics}, the correlations are created \textit{via} the propagation of free magnons
characterized by the excitation spectrum $E_k = J[1-\cos(k)]$. On the left, the first correlations are created by the spin-wave excitation propagating 
with the minimal group velocity at quasimomentum $-k^* = -\pi/2$ which leads to $V_{\mathrm{g}}(-k^*) = \partial_k E_{-k^*} = -J$. On the right, they are
created by the magnonic excitation spreading with the maximal group velocity $V_{\mathrm{g}}(k^*) = J$ at quasimomentum $k^* = \pi/2$. 
Therefore, the spreading velocities of the correlation edge are determined by $V_{\mathrm{CE}}^{\mathrm{r}(\mathrm{l})} = (-)V_{\mathrm{g}}(k^*) = (-)J$.
Using linear fits to extract the CE velocities on Fig.~\ref{fig:heis_local_quench}, we find that $V_{\mathrm{CE}}^{\mathrm{r}(\mathrm{l})} = (-1)\times (1. \pm 0.05)J$ 
in very good agreement with our theoretical predictions. Note that a very similar space-time pattern has been found for another local observable, namely 
the spin-spin correlator along the $z$ axis defined as $\langle \Psi_0 | \hat{S}^z_R(t) \hat{S}^z_{N_s/2}(t) | \Psi_0 \rangle$.\\ \\

In the two previous studies, we have shed new light on the local quench dynamics for 1D short-range interacting lattice models. The space-time behavior 
of different on-site observables (local density, local magnetization) for sudden local quenches in different quantum phases (gapped phase : Mott-insulating phase for the
1D SRBH model, gapless phase : ferromagnetic phase along the $z$ axis for the 1D SRHM) has been investigated. In all cases, we found that the space-time pattern has
a causal and a non-causal region separated by a correlation edge propagating ballistically. Besides, the causal region is characterized by
a series of local extrema which also spread linearly. At this stage, the previous properties are reminiscent of those for global quenches. 
However, the space-time pattern for sudden local quenches only displays one or several single linear structures depending on the locally-perturbed initial quantum state, \textit{ie.} $\ket{\Psi_0}$, and hence on
the associated local perturbation. Contrary to global quenches, the spreading velocity of the edge is not characterized by twice the maximal group velocity $2V_{\mathrm{g}}^*$.
Indeed, for global quenches, the correlations are governed by the motion of free and counterpropagating quasiparticle pairs. However, for local quenches, they are not governed by the spreading of 
quasiparticle excitation pairs anymore but by the motion of individual quasiparticles. Hence, the latter leads to an edge whose spreading is characterized by the maximal group velocity, \textit{ie.} $V_{\mathrm{g}}^*$. 
We stress that the previous statements concerning the local quench dynamics in isolated short-range interacting lattice models are not restricted to the sole Bose-Hubbard or Heisenberg models. Indeed, they also apply
to other quantum systems, such that the short-range Ising model, and observables, such that spin-spin correlations. \\

Our study concerning the local and global quench dynamics in short-range lattice models may be extended to the case of long-range systems, 
such as spin models as realized in trapped-ion experiments~\cite{jurcevic2014,richerme2014}. In such quantum systems, a possible divergence of the 
group velocity can be tuned. For global quenches and intermediate-range interactions, our correlation spreading theory based on a quasiparticle picture still predicts
a twofold dynamics whose CE and extrema do not propagate ballistically anymore but algebraically, as shown theoretically in Chap.~\ref{ch:3-universal_scaling_laws}.
In this case, the twofold algebraic structure is characterized by the coexistence of a super-ballistic (for a gapless system) or ballistic (gapped case)
signal for the series of local extrema and a sub-ballistic signal for the CE. Furthermore, our numerical results for the Bose-Hubbard chain suggest that the twofold structure of the correlation function may survive
in strongly correlated regimes also for long-range systems. The previous statements are verified numerically in the next chapter relying on the case study of the
long-range XY and long-range transverse Ising chains and shed light on the still debated scaling of the light cone in long-range interacting quantum systems.
For the long-range transverse Ising chain, the study concerning its global quench dynamics is completed by an investigation of the scaling laws
for the correlation and entanglement spreading for a dynamics induced \textit{via} local quenches.

%% file: text/ch5_long_range_ising_chain.tex
\setstretch{1.0} 

\begin{savequote}[8cm]
\textlatin{“When two systems, of which we know the states by their respective representation, enter into a temporary
physical interaction due to known forces between them and when after a time of mutual influence the systems separate 
again, then they can no longer be described as before, viz., by endowing each of them with a representative of its
own. I would not call that one but rather the characteristic trait of quantum mechanics.”}
  \qauthor{--- Erwin Schrödinger}
\end{savequote}

\chapter{Spreading of correlations and entanglement in the $s=1/2$ long-range transverse Ising chain} 
\label{ch:5-long_range_ising_chain}

\vspace{-0.3cm}
\minitoc

\newpage

\section{Position of the problem}
Understanding the out-of-equilibrium dynamics of quantum systems has become a central subject of the many-body theory. It allows 
to characterize many phenomena such as the relaxation, thermalization, transport of matter and of quantum information.
The study of long-range interacting models gained a lot of interest. Indeed, a large class of quantum systems, where the strength 
and the range of the interactions can be controlled, has been realized experimentally, e.g., Rydberg gases~\cite{bendkowsky2009,weimer2010,
schausz2012,browaeys2016}, artificial ion crystals~\cite{deng2005,islam2011,lanyon2011,schneider2012b,NaturePhysicsInsight2012blatt},
polar molecules~\cite{micheli2006,yan2013,moses2017}, nonlinear optical media~\cite{firstenberg2013}, magnetic atoms ~\cite{griesmaier2005,
beaufils2008,lu2011,baier2016,lahaye2009}, and solid-state defects~\cite{childress2006,balasubramanian2009,dolde2013}.
Those long-range quantum systems are especially interesting because they induce rich dynamical behaviors, including dynamical phase transitions.
The breakdown of fundamental concepts such as the equivalence of the thermodynamic ensembles, Lieb-Robinson propagation bound and 
the notions of group and phase velocities can also lead to peculiar behaviors. In the recent years, many long-range interacting lattice models
have been investigated both experimentally and theoretically. However, important open questions are still debated such as how correlations and information
can propagate in a long-range lattice system. It is quite clear in the local regime \textit{ie.} when the group velocity of the quasiparticle dispersion relation is bounded, where a linear cone-structure emerges.
Nevertheless, the question is strongly debated in the quasi-local regime, \textit{ie.} when the group velocity is unbounded. The main question is to determine whether the propagation is super-ballistic, ballistic, or sub-ballistic. 
For a similar context, at Chap.~\ref{ch:3-universal_scaling_laws}, we unveiled theoretically a twofold algebraic structure for the causality cone of equal-time 
connected correlation functions. More precisely, the correlation edge has been found to display always a sub-ballistic motion independently if the considered 
quantum phase is gapped or gapless. \\
In this chapter, we study theoretically and numerically the spreading of information by investigating both the spreading of correlations and entanglement 
for a specific long-range lattice model in a gapped phase. The latter is the paradigmatic, one-dimensional, transverse Ising model with a long-range spin exchange of the 
form $1/R^\alpha$ in its $z$ polarized phase. Using a numerically-exact tensor-network approach, we study a variety of quenches and observables, and determine 
the corresponding dynamical scaling laws. One of the main purposes of this final chapter is to verify numerically whether the CE propagates sub-ballistically or not in the 
quasi-local regime of a gapped phase for a long-range interacting lattice model. This specific discussion is also extended to the case of a gapless quantum phase by considering the quasi-local regime 
of the $x$ polarized phase for the long-range XY chain. 

\section{Time-Dependent Variational Principle within the matrix product state approach}
\label{TDVP_section}

To investigate numerically the spreading of correlations and entanglement in both the 1D long-range $s=1/2$ transverse Ising and in the 1D long-range XY models,
we rely on the time-dependent variational principle (TDVP) within the matrix product state approach \cite{hauke2013,koffel2012,haegeman2011}. The latter corresponds to a new approach to perform imaginary and real time evolutions
and is particularly powerful for long-range interacting lattice models. Hence, this approach permits to deduce their static (imaginary time evolution)
and dynamical (real time evolution) properties. In the following, we first show an efficient representation of the long-range interactions of the form $1/R^{\alpha}$. 
The previous representation allows us to find an optimal matrix product operator (MPO) form of Hamiltonians with power-law decaying interactions.
Then, we define the matrix product state manifold and tangent space at the heart of the TDVP approach.
Finally, we discuss the algorithm based on the TDVP to perform imaginary and real time evolutions. 

\subsection{Matrix product operator representation of long-range $s=1/2$ spin lattice chains}
\begin{figure}[!h]
\centering
\begin{tabular}{c}
\includegraphics[scale = 0.41]{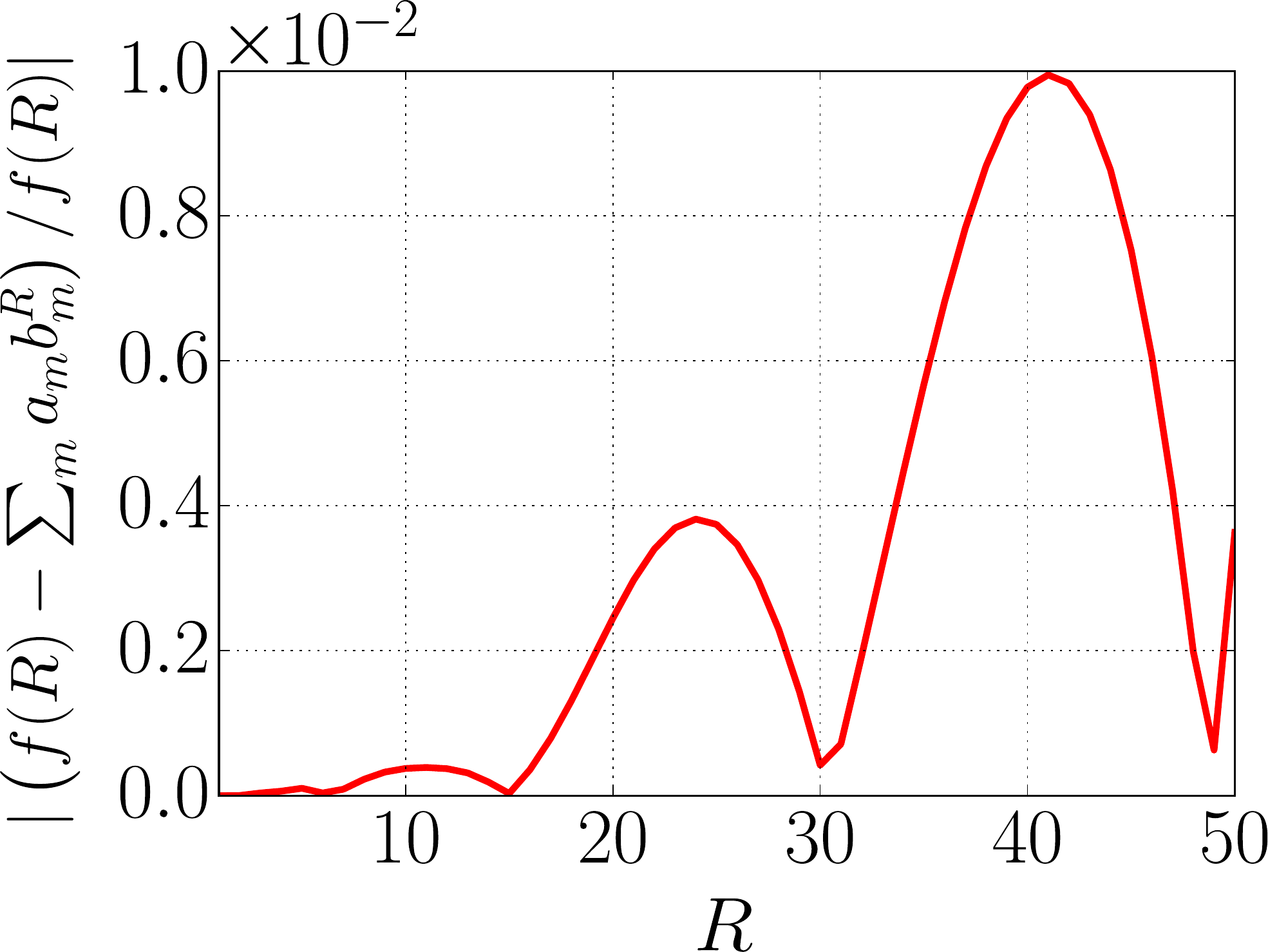}  
\end{tabular}
\caption{Relative error $\epsilon(R) = |(f(R)-\sum_{m=1}^{M} a_m b_m^R)/f(R)|$ between the analytical power-law function $f(R) = 1/R^{\alpha}$ and the approximation in terms
of decaying exponentials $\sum_{m=1}^{M} a_m b_m^R$ as a function of the index (lattice site) $R$, for $\alpha = 1.7$, $L= \mathrm{max}(R) =50$ and $M=6$.} 
\label{decomp_power_law}
\end{figure}

\noindent
To get an optimal MPO representation of Hamiltonians with long-range interactions of the form $1/R^{\alpha}$ \textit{ie.} power-law decaying interactions,
the latter can be expressed as a sum of decaying exponentials \cite{pirvu2010} leading to 

\begin{equation}
f(R-R') = \frac{1}{|R-R^{'}|^{\alpha}} \simeq \sum_{m=1}^{M} a_{m} b_{m}^{|R-R'|},
\end{equation}

\noindent
where $R$ and $R'$ denotes two different lattice sites of the chain, $(R,R') \in \mathbb{Z}^2$ such that $|R-R'| \neq 0$ ($a$ the lattice spacing is fixed to unity).
$\alpha$ refers to the power-law exponent and characterizes the range of the interactions and $M$ to the number of decaying exponentials.
The real coefficients $\{ a_m,m=1...M\}$ and $\{b_m,m=1...M\}$ correspond to the different weights and exponents respectively.
The latter are determined such that they minimize the following quantity

\begin{equation}
 \sum_{R=1}^{L} \left| f(R) - \sum_{m=1}^{M} a_m b_m^R \right|,
 \label{eq2}
\end{equation}

\noindent
with $L$ the chain length (or equivalently the number of lattice sites). More details about the numerical technique to minimize Eq.~\eqref{eq2},
and thus to find the coefficients $\{ a_m\}$ and $\{b_m\}$, can be found at Ref.~\cite{pirvu2010}. The previous decomposition of the power-law decaying interactions 
in terms of decaying exponentials is particularly efficient. Indeed, the latter represents accurately the long-range potential with a relatively small number $M$ of
decaying exponentials. An example of the precision of such decomposition is displayed on Fig.~\ref{decomp_power_law} for $\alpha = 1.7$, $L = 50$ and $M = 6$ 
which are typical values considered in the next sections. The relative error $\epsilon(R)$ between the analytical function $f(R) = 1/R^{\alpha}$ and the approximation
$\sum_{m=1}^{M} a_m b_m^{R}$ is represented as a function of the index (lattice site) $R>0$. One finds that the analytical power-law function $f(R) = 1/R^{\alpha}$ 
is accurately reproduced by the approximation $\sum_{m=1}^{M} a_m b_m^R$ where the maximal relative error is $\mathrm{\epsilon(R)} = 1\%$ for $M=6$. \\

\noindent
As discussed previously, the power-law decaying interactions are accurately reproduced by a sum of decaying exponentials (provided that the number of terms in the sum is
large enough). Most importantly, the approximated form of the long-range interactions allows us to find the optimal MPO representation of $\hat{H}$ the Hamiltonian
describing a long-range interacting model [see Eq.~\eqref{mpo_form_2} for the definition of the MPO representation]. Indeed, the MPO bond dimension $\tilde{\chi}$ can be drastically decreased \textit{via} the 
approximated form, compared to the one where the power-law decaying interactions are naturally implemented. In the following, this statement is verified for two 
1D long-range $s=1/2$ spin lattice models, namely the 1D long-range transverse Ising model and 1D long-range XY models.

\subsubsection{One-dimensional long-range $s=1/2$ transverse Ising model}
The 1D long-range $s=1/2$ transverse Ising model with open boundary conditions is characterized by the following Hamiltonian 

\begin{equation}
\hat{H} = \sum_{R<R'} \frac{2J}{|R-R'|^{\alpha}} \hat{S}_R^x \hat{S}_{R'}^x - 2h \sum_{R=1}^{L} \hat{S}_R^z.
\end{equation}

\noindent
Considering the approximated form of the long-range interactions in terms of decaying exponentials, it yields 

\begin{equation}
\hat{H} \simeq 2J \sum_{R<R'} \sum_{m=1}^{M} a_m b_m^{|R-R'|} \hat{S}_R^x \hat{S}_{R'}^x - 2h \sum_{R=1}^{L} \hat{S}_R^z.
\label{decomp_lrti}
\end{equation}

\noindent
Its associated matrix product operator (MPO) representation is given by  

\begin{equation}
\hat{H} \simeq \prod_{R=1}^{L} \hat{W}[R],
\end{equation}

\noindent
where the matrix $\hat{W}[R]$ containing operators acting on the local Hilbert space $\mathbb{H}_R$ can be defined as

\begin{equation}
\hat{W}[R] = 
\begin{pmatrix}
 I & 2J a_1 b_1 \hat{S}_R^x & 2J a_2 b_2 \hat{S}_R^x & \cdots & 2J a_M b_M \hat{S}_R^x & -2h \hat{S}_R^z \\
 0       & b_1 I  & 0 & \cdots & 0 & \hat{S}_R^x  \\
 0 & 0 & b_2 I & \cdots & 0 & \hat{S}_R^x \\
 \vdots  & \vdots &  \vdots & \ddots & \vdots & \vdots \\ 
 0 & 0 & 0 & \cdots & b_M I & \hat{S}_R^x \\
 0 & 0 & 0 & \cdots & 0 & I
\end{pmatrix}.
\label{bulk_lrti}
\end{equation}

\begin{figure}[!t]
\centering
\begin{tabular}{c}
\includegraphics[scale = 0.53]{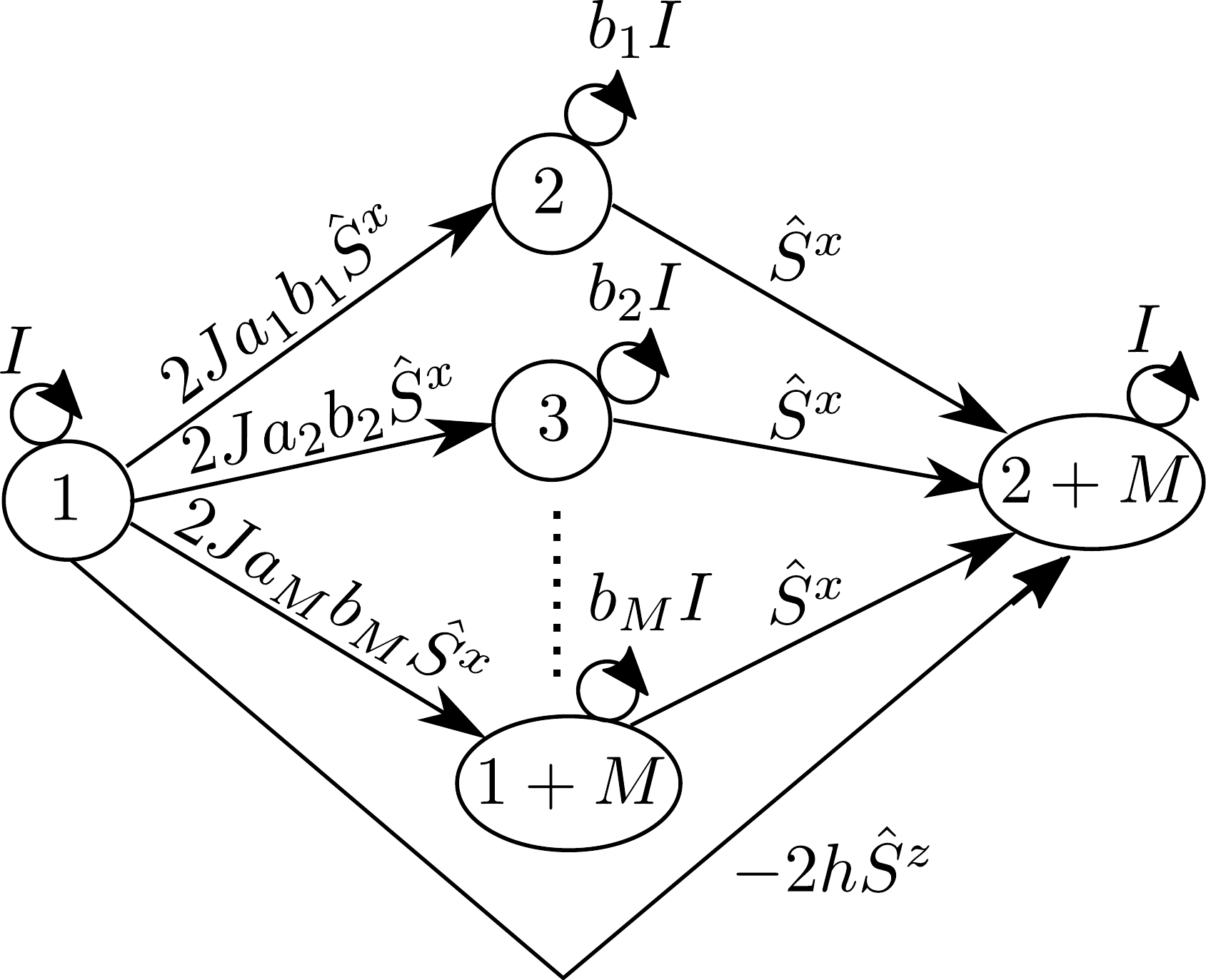}  
\end{tabular}
\caption{Graphical depiction of the bulk matrix $\hat{W}[R]$ [see Eq.~\eqref{bulk_lrti}] of the MPO form for the 1D LRTI model [see Eq.~\eqref{decomp_lrti}]. The latter,
found using a finite automata described in details at Ref.~\cite{crosswhite2008}, consists of summarizing all the different terms in the Hamiltonian by following the arrows.
More precisely, the graph consists of arrows (on the top is associated a term of the Hamiltonian acting on a local Hilbert space) where each of them 
has a starting point (reaches a circle whose number represents the line index of the bulk matrix) and an ending point (reaches the same or a different circle
whose number represents the column index of the bulk matrix). Therefore, to construct the bulk matrix $\hat{W}[R]$, it just requires to investigate each arrow, to deduce both
indices and to write down the corresponding term. Finally, the circle with the highest number corresponds to the dimension of the bulk matrix $\hat{W}[R]$, and consequently
to the MPO bond dimension ($\tilde{\chi} = 2+M$).} 
\label{finite_automata_LRTI}
\end{figure}

\noindent
Note that the bulk matrix $\hat{W}[R]$ is valid $\forall R \in [|2,L-1|]$ due to the translational invariance of the 1D LRTI model. To deduce easily the bulk matrix
$\hat{W}[R]$ of the MPO form for the 1D LRTI model, we refer the reader to Fig.~\ref{finite_automata_LRTI}, see also Ref.~\cite{crosswhite2008}. 
$\hat{W}[1]$ and $\hat{W}[L]$ represent the edges and correspond to a row and column vector. The latter contain operators acting on the local Hilbert space $\mathbb{H}_1$ and 
$\mathbb{H}_L$ respectively. Both vectors are defined as

\begin{align}
& \hat{W}[1] = 
\begin{pmatrix}
  I & 2J a_1 b_1 \hat{S}_1^x & 2J a_2 b_2 \hat{S}_1^x & \cdots & 2J a_M b_M \hat{S}_1^x & - 2h \hat{S}_1^z \\
\end{pmatrix}, \\
& \hat{W}[L] = 
\begin{pmatrix}
 -2h \hat{S}_L^z & \hat{S}_L^x & \hat{S}_L^x  & \cdots & \hat{S}_L^x & I \\
\end{pmatrix}^T.
\end{align}

\noindent
The bond dimension $\tilde{\chi}$ associated to the previous MPO form of $\hat{H}$ is given by $\tilde{\chi} = 2 + M$ where $M$ is the total number of decaying exponentials
to express the power-law decaying interactions.

\subsubsection{One-dimensional long-range $s=1/2$ XY model}
The 1D long-range $s=1/2$ XY model is characterized by the following Hamiltonian 

\begin{equation}
\hat{H} = \sum_{R<R'} \frac{-J/2}{|R-R'|^{\alpha}} \left( \hat{S}_R^x \hat{S}_{R'}^x + \hat{S}_R^y \hat{S}_{R'}^y \right) \simeq  -\frac{J}{2} \sum_{R<R'}
 \sum_{m=1}^{M} a_m b_m^{|R-R'|} \left( \hat{S}_R^x \hat{S}_{R'}^x + \hat{S}_R^y \hat{S}_{R'}^y \right).
 \label{decomp_lrxy}
\end{equation}

\begin{figure}[!t]
\centering
\begin{tabular}{c}
\includegraphics[scale = 0.53]{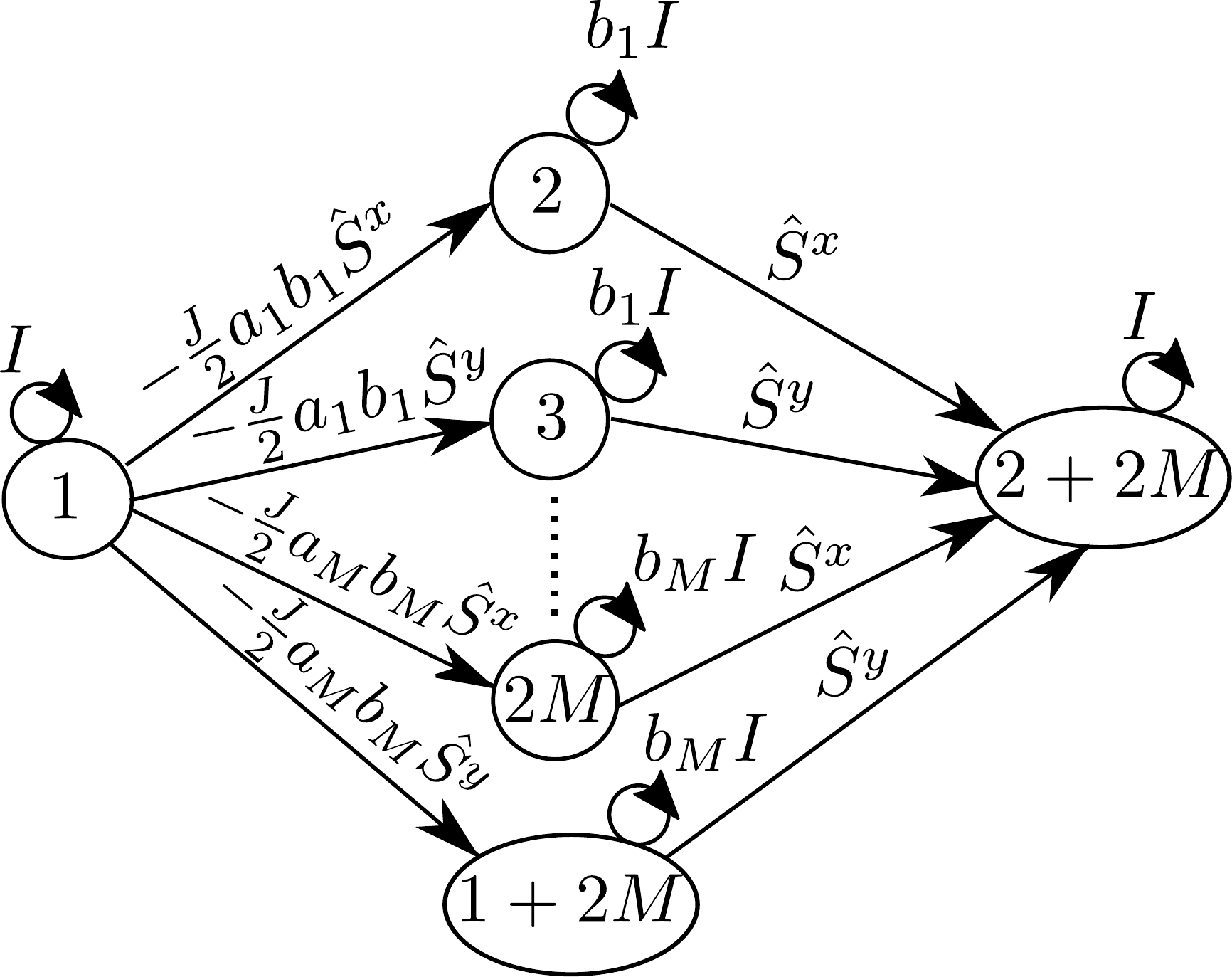}  
\end{tabular}
\caption{Graphical depiction of the bulk matrix $\hat{W}[R]$ [see Eq.~\eqref{bulk_lrxy}] of the MPO form for the 1D LRXY model [see Eq.~\eqref{decomp_lrxy}]. As described
in Ref.~\cite{crosswhite2008}, this finite automata consists of summarizing all the different terms of the Hamiltonian by following the arrows. Similarly to the one
for the 1D LRTI chain presented at Fig.~\ref{finite_automata_LRTI}, the graph consists of arrows (on the top is associated a term of the Hamiltonian acting on a local
Hilbert space) where each of them has a starting point (reaches a circle whose number represents the line index of the bulk matrix) and an ending point 
(reaches the same or a different circle whose number represents the column index of the bulk matrix). According to the finite automata, one immediately finds that the 
MPO bond dimension is $\tilde{\chi} = 2+2M$.} 
\label{finite_automata_LRXY}
\end{figure}

\noindent
Its MPO form is given by

\begin{equation}
\hat{H} \simeq \prod_{R=1}^{L} \hat{W}[R],
\end{equation}

\noindent
where the bulk matrix $\hat{W}[R]$ [see Fig.~\ref{finite_automata_LRXY} for the finite automata to deduce the bulk matrix of the MPO form for the LRXY chain]
and the edge vectors $\hat{W}[1]$ and $\hat{W}[L]$ are defined by  

\begin{equation}
\hat{W}[R] = 
\begin{pmatrix}
 I & -\frac{J}{2} a_1 b_1 \hat{S}_R^x & -\frac{J}{2} a_1 b_1 \hat{S}_R^y & -\frac{J}{2} a_2 b_2 \hat{S}_R^x & -\frac{J}{2} a_2 b_2 \hat{S}_R^y & \cdots &  0 \\
 0       & b_1 I  & 0 & 0 & 0 & \cdots & \hat{S}_R^x \\
 0       & 0 & b_1 I & 0 & 0 & \cdots & \hat{S}_R^y \\
 0 & 0 & 0 & b_2 I &  0 & \cdots & \hat{S}_R^x \\
  0 & 0 & 0 & 0 & b_2 I  & \cdots & \hat{S}_R^y \\
\vdots & \vdots & \vdots & \vdots & \cdots & \ddots & \vdots \\
0 & 0 & 0 & 0 & 0 & \cdots & I 
\end{pmatrix},
\label{bulk_lrxy}
\end{equation}

\begin{align}
& \hat{W}[1] = 
\begin{pmatrix}
 I & -\frac{J}{2} a_1 b_1 \hat{S}_1^x & -\frac{J}{2} a_1 b_1 \hat{S}_1^y & -\frac{J}{2} a_2 b_2 \hat{S}_1^x & -\frac{J}{2} a_2 b_2 \hat{S}_1^y & \cdots &  0 \\
\end{pmatrix}, \\
& \hat{W}[L] = 
\begin{pmatrix}
0 & \hat{S}_L^x & \hat{S}_L^y & \hat{S}_L^x & \hat{S}_L^y & \cdots & I \\
\end{pmatrix}^T.
\end{align}

\noindent
Once again, the matrix $\hat{W}[R]$ is valid $\forall R \in [|2, L-1|]$ due to the translational invariance of the model. One finds that the 
MPO bond dimension $\tilde{\chi}$ is determined by $ \tilde{\chi} = 2 + M + M$. Indeed, each long-range spin exchange coupling (in the plane $x-y$) contributes
to an increase of the MPO bond dimension with a factor $M$. The latter corresponds to the number of decaying exponentials to express the power-law decaying potential. \\

Consequently, from both previous examples, it is straightforward to deduce the general MPO bond dimension for Hamiltonians with power-law decaying interactions represented 
as a sum of decaying exponentials. More precisely, the Hamiltonian $\hat{H}$ is considered to display only long-range two-site coupling terms and possible local interactions
(which have no influence on the MPO bond dimension). In this case, the MPO bond dimension is characterized by the value $\tilde{\chi} = 2 + N_{\mathrm{pl}} M$
where $N_{\mathrm{pl}}$ is the number of long-range two-site couplings and $M$ the number of decaying exponentials to describe accurately the power-law interactions. 
The MPO bond dimension is linear with $M$ and not with $L$, the number of lattice sites, anymore. Indeed, without the approximated form of the long-range interactions,
the MPO bond dimension is characterized by $\tilde{\chi} = 2 + L$. In general, $N_{\mathrm{pl}} M \ll L$ leading to an optimal MPO representation of Hamiltonians 
with power-law interactions. \\
For all the numerical simulations based on the time-dependent variational principle, whose approach is presented below, such 
decomposition of the power-law interactions is considered. Furthermore, the parameter $M$ is choosen such that the maximal relative
error $\epsilon_{\mathrm{max}} = \mathrm{max}[\epsilon(R)] = \mathrm{max}\left[|(f(R)-\sum_{m=1}^{M} a_m b_m^R )/f(R)| \right]$, 
between the analytical power-law function $f(R) = 1/R^{\alpha}$ and the approximation in terms of decaying exponentials $\sum_{m=1}^{M} a_m b_m^R$, is around $1\%$.  

\subsection{Matrix product state manifold and tangent space}

\subsubsection{The matrix product state manifold}
The Matrix Product State (MPS) consists of an efficient local representation of a quantum state living in a many-body Hilbert
space. Although general, it is mostly fruitful for low-dimensional and low-entangled quantum systems. Any quantum state can be represented 
locally without losing its quantum non-locality properties. \\

Considering a lattice chain of $L$ sites within a many-body Hilbert space $\mathbb{H} = \bigotimes_{R=1}^{L} \mathbb{H}_R$, the $d$-dimensional local
Hilbert space $\mathbb{H}_R$ is spaned by the $\ket{\sigma_R}_R$ where $R \in \{1,...,L\}$ and $\sigma_R \in \{1,...,d\}$. In this basis, a general quantum
state $\ket{\Psi}$ living in the full Hilbert space $\mathbb{H}$ may be written as

\begin{equation}
\ket{\Psi} = \sum_{\boldsymbol{\sigma}} \mathit{\Psi}^{\sigma_1 \sigma_2... \sigma_L} \ket{\boldsymbol{\sigma}},~~~\boldsymbol{\sigma} = \sigma_1, ..., \sigma_L,
\end{equation}

\noindent
where $\mathit{\Psi}^{\sigma_1 \sigma_2... \sigma_L} \in \mathbb{C}$. By applying the Singular Value Decomposition (SVD) $L-1$ times to the general vector $\mathit{\Psi}$,
one obtains the so-called Matrix Product State representation of a quantum state, see Appendix.~\ref{appendix3_mps_app}. It consists of a local
form where the state is expressed as a product of tensors where each of them acts on a different local Hilbert space. In the following, the
many-body quantum state $\ket{\Psi}$ can be either translationally invariant or not, leading to the following MPS form

\begin{equation}
 \ket{\Psi} = \sum_{\boldsymbol{\sigma}} A^{\sigma_1}[1] A^{\sigma_2}[2] ...~A^{\sigma_{L-1}}[L-1] A^{\sigma_L}[L] \ket{\boldsymbol{\sigma}},~~~\boldsymbol{\sigma} = \sigma_1, ..., \sigma_L.
\end{equation}

\begin{figure}[!h]
\centering
\begin{tabular}{c}
\includegraphics[scale = 0.40]{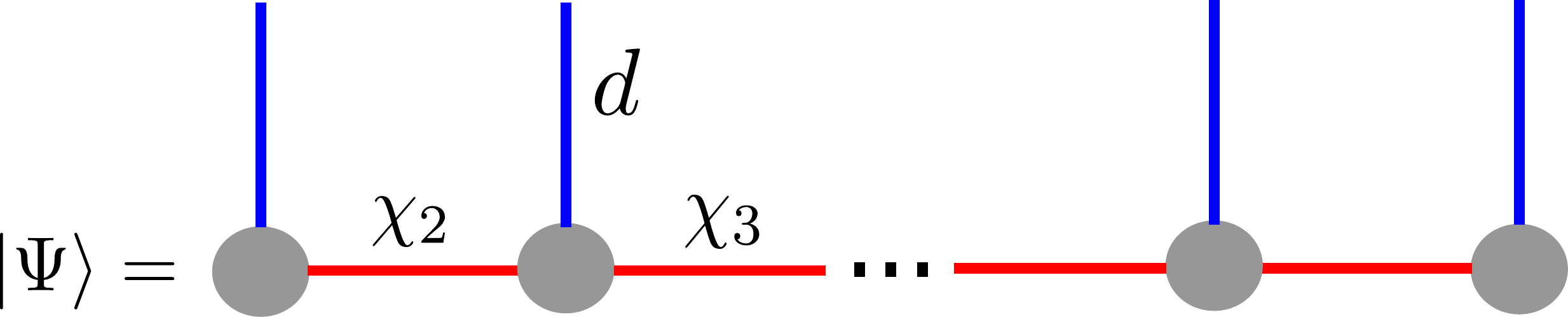}  
\end{tabular}
\caption{Graphical representation using tensor network of a general many-body quantum state
$\ket{\Psi} = \sum_{\boldsymbol{\sigma}} A^{\sigma_1}[1] A^{\sigma_2}[2] ...~ A^{\sigma_{L-1}}[L-1] A^{\sigma_L}[L] \ket{\boldsymbol{\sigma}}$,
$\boldsymbol{\sigma} = \sigma_1,...,\sigma_L$ living in the full Hilbert space $\mathbb{H}$ of a one-dimensional quantum lattice model 
of $L$ sites. Each block is a $3$rd-rank tensor with 2 horizontal virtual legs (solid red lines) and 1 vertical physical leg (solid blue lines). Each tensor 
$A[R]$ has a dimension $\chi_R \times \chi_{R+1} \times d$ except the first and last tensors having a dimension $1 \times \chi_2 \times d$ and $\chi_L \times 1
\times d$ respectively. The MPS bond dimension $\chi$ is given by $\chi = \mathrm{max}_R(\chi_R)$. For homogeneous (translational invariant) one-dimensional lattice models,
the tensors $A[R]$ of the MPS $\ket{\Psi}$ fulfill $A^{\sigma}\left[R \right] = A^{\sigma}\left[R+1\right], \forall R \in [|2,L-2|]$ and $\forall \sigma \in \{1,...,d\}$.}
\label{mps}
\end{figure}

\noindent
Each block $A[R]$, $R \in [|2, L-1|]$, is a third-rank tensor with 2 horizontal legs (virtual legs) and 1 vertical leg (physical leg). It has 
a dimension $\chi_R \times \chi_{R+1} \times d$ where $\chi = \mathrm{max}_R(\chi_R)$ is the bond dimension of the MPS, see Fig.~\ref{mps}. 
The first $A[1]$ and last $A[L]$ tensors have a dimension $1 \times \chi_2 \times d$ and $\chi_{L} \times 1 \times d$ respectively. \\

\noindent
At fixed bond dimension $\chi$, the ensemble of MPS does not define a Hilbert space. Instead, it forms a so-called non-linear (curved) MPS manifold
called $\mathcal{M}_{\chi}$. This non-linearity of the manifold is given by the property that the summation of different MPSs implies to increase
the bond dimension. For instance, considering $\ket{\Psi}$ and $\ket{\Phi}$ two MPSs of bond dimension $\chi$ leaving in the MPS manifold $\mathcal{M}_\chi$,
the sum $\ket{\Psi}+\ket{\Phi}$ can be cast into the MPS form at the prize of a new MPS bond dimension $2\chi$. Indeed, for 
$\ket{\Psi} = \sum_{\boldsymbol{\sigma}} A^{\sigma_1}[1]A^{\sigma_2}[2]...~A^{\sigma_{L-1}}[L-1] A^{\sigma_L}[L] \ket{\boldsymbol{\sigma}}$ 
and $\ket{\Phi} = \sum_{\boldsymbol{\sigma}} \tilde{A}^{\sigma_1}[1] \tilde{A}^{\sigma_2}[2]...~ \tilde{A}^{\sigma_{L-1}}[L-1]\tilde{A}^{\sigma_L}[L]
\ket{\boldsymbol{\sigma}}$, the compression of the sum $\ket{\Psi} + \ket{\Phi}$ to conserve the MPS form is given by
$\ket{\Psi} + \ket{\Phi} = \sum_{\boldsymbol{\sigma}} M^{\sigma_1}[1] M^{\sigma_2}[2]...~M^{\sigma_{L-1}}[L-1]M^{\sigma_L}[L] \ket{\boldsymbol{\sigma}}$ 
where $M^{\sigma_R}[R] = A^{\sigma_R}[R] \oplus \tilde{A}^{\sigma_R}[R]$, $\forall R \in [|1,L|]$. Consequently, $\ket{\Psi}+\ket{\Phi} \in \mathcal{M}_{2\chi}$
and thus does not remain in the MPS manifold $\mathcal{M}_{\chi}$. \\

\subsubsection{The matrix product state tangent space}
\begin{figure}[!h]
\centering
\begin{tabular}{c}
\includegraphics[scale = 0.38]{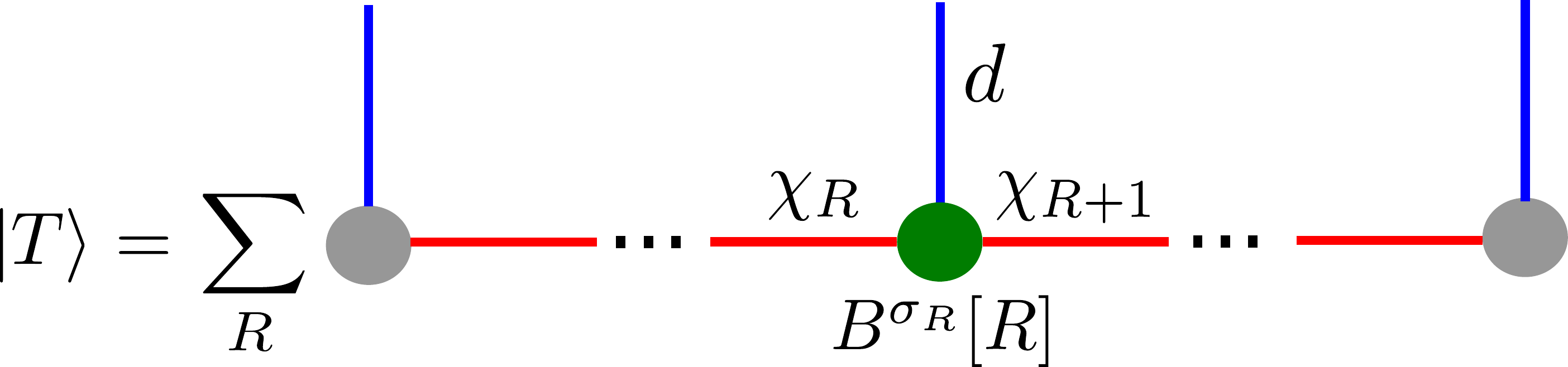}  
\end{tabular}
\caption{Graphical representation using tensor networks of a general tangent vector $\ket{T} = \sum_R B[R] \partial_{A[R]} \ket{\Psi} 
\in \mathcal{T}_{\ket{\Psi}}$ associated to the MPS $\ket{\Psi} = \sum_{\boldsymbol{\sigma}} A^{\sigma_1}[1]...~A^{\sigma_L}[L] \ket{\boldsymbol{\sigma}} 
\in \mathcal{M}_{\chi}$. The local perturbation at the lattice site $R$ is symbolized by the $3$rd-rank tensor $B[R]$ (green disk) of dimension $\chi_R
\times \chi_{R+1} \times d$. The solid blue (red) lines represent the physical (virtual) legs of the different tensors.}
\label{tangent_vector}
\end{figure}

This non-linearity, or curvature of the MPS manifold, allows to define a tangent plane $\mathcal{T}_{\ket{\Psi}}$ associated to a reference
MPS $\ket{\Psi} \in \mathcal{M}_{\chi}$. Indeed, for a MPS $\ket{\Psi}$ of bond dimension $\chi$ and living in the full Hilbert space $\mathbb{H}$,
a tangent vector $\ket{T} \in \mathcal{T}_{\ket{\Psi}}$ is represented as a superposition of $L$ quantum states where each of them is built from 
the reference MPS $\ket{\Psi}$ where a local perturbation (from a MPS point of view) $B[R]$, a third-rank tensor of dimension $\chi_R \times \chi_{R+1} \times d$,
has been applied, see Fig.~\ref{tangent_vector}. For a translational invariant lattice model, the local perturbation fulfills $B^{\sigma}[R] = B^{\sigma}[R+1]$, 
$\forall R \in [|2,L-2|]$ and $\forall \sigma \in \{1,...,d\}$ with $d$ the dimension of the local Hilbert space $\mathbb{H}_R$. \\

Since each tangent vector $\ket{T}$ is defined as a sum of the reference MPS $\ket{\Psi}$ where a local perturbation has been applied, the tangent plane to 
$\ket{\Psi}$ denoted by $\mathcal{T}_{\ket{\Psi}}$ can be seen as the low-energy subspace containing the low-energy dynamics. 
This subspace can be used to integrate the real or imaginary time-dependent Schrödinger equation in order to perform real or imaginary time evolution (depending if 
the purpose is to find the time-evolved quantum state or the ground state of a one-dimensional lattice model) respectively. Note that the physical intuition
of low-energy dynamics given in terms of elementary excitations or quasiparticles (for translational invariant lattice models) is lost when using the standard
technique to perform real or imaginary time evolution. This standard technique, discussed in the previous chapter, consists of writing the time evolution operator in the Matrix Product Operator (MPO)
form and to apply it as many times as necessary on the initial MPS. After one timestep, the bond dimension has been modified from $\chi$ to $d \chi$ where
$d$ is the dimension of the local Hilbert space. As a consequence, the MPS needs to be truncated after each iteration to avoid an exponential growth of the bond dimension.

\subsubsection{Time-Dependent Variational Principle}

The Time-Dependent Variational Principle (TDVP) corresponds to a new strategy using the tangent space to perform real and imaginary time evolutions.
One advantage, which has already been presented previously, is that the low-energy dynamics is given in terms of low-energy excitations. Besides, another advantage 
corresponding to the central point of this method is the fact that the bond dimension of the many-body quantum state does not increase during the process.
Indeed, by using the geometric notion of the tangent plane, the TDVP permits to optimize locally the MPS by solving a differential equation 
(the real or imaginary Schrödinger equation) while remaining confined into a same MPS manifold. Finally, a crucial advantage is that 
the algorithm can deal with Hamiltonians containing long-range interactions. The single requirement is to find the exact MPO form or at least a very
good approximation for the long-range potential. In the following, the previous advantages of the TDVP and the different algorithms to perform real and imaginary
time evolutions are discussed in details.

\subsection{Static properties - Imaginary time evolution}

\subsubsection{General scheme} 
A ground state optimization algorithm is given by solving iteratively the differential Schrödinger equation in
imaginary time ($\tau \rightarrow it$) reading as 

\begin{equation}
 \partial_{\tau}\ket{\Psi(\{ A(\tau) \})} = -\hat{H}\ket{\Psi(\{ A(\tau) \})}.
 \label{ise}
\end{equation}

\noindent
$\hat{H}$ is a Hamiltonian describing a one-dimensional lattice model of length $L$ with short- or long-range interactions. 
$\ket{\Psi(\{ A(\tau) \})}$ corresponds to a many-body quantum state confined in the MPS manifold $\mathcal{M}_{\chi}$ at imaginary time $\tau$.
$\{A(\tau)\}$ denotes the collection of the $3$rd-rank tensors $A[R](\tau)$ with $R \in [|1,...,L|]$.
The latter may be written as 

\begin{equation}
\ket{\Psi(\{ A(\tau) \})} = \sum_{\boldsymbol{\sigma}} A^{\sigma_1}[1](\tau)  A^{\sigma_2}[2](\tau) ...~  A^{\sigma_{L-1}}[L-1](\tau)
A^{\sigma_L}[L](\tau) \ket{\boldsymbol{\sigma}}
\end{equation}

\noindent
where $\boldsymbol{\sigma} = \sigma_1,...,\sigma_L$ and $\sigma_R \in \{1,...,d\}$ with $d$ the dimension of the local Hilbert 
space $\mathbb{H}_R$. By considering a random initial state
$\ket{\Psi_{\mathrm{init}}(\{ A \})} \in \mathcal{M}_{\chi}$, the iterative procedure, in the infinite imaginary time limit, should give rise to a 
projection onto the MPS manifold $\mathcal{M}_{\chi}$ of the best approximation of the ground state of the considered Hamiltonian $\hat{H}$.
Mathematically, it yields 

\begin{equation}
 \ket{\mathrm{GS}} = \lim \limits_{\tau \rightarrow + \infty} \frac{e^{-\tau \hat{H}} \ket{\Psi_{\mathrm{init}}(\{ A \})}}{||
 e^{-\tau \hat{H}} \ket{\Psi_{\mathrm{init}}(\{A\})}||},
 \label{proc_bad}
\end{equation}

\noindent
where $\ket{\Psi_{\mathrm{init}}(\{ A \})} = \ket{\Psi(\{A(0)\})}$ and the factor $|| e^{-\tau \hat{H}} \ket{\Psi_{\mathrm{init}}(\{A\})}||$ enforces the normalization of the ground state. 
As suggested by Eq.~\eqref{proc_bad}, one possible way to perform the imaginary time evolution consists of building the 
infinitesimal imaginary time evolution operator $e^{-\tau \hat{H}}$. Then, this operator has to be applied to the initial state many times
as necessary to converge towards the ground state. However, the latter is non optimal for Hamiltonians containing long-range interactions due to their relatively
large MPO bond dimension $\tilde{\chi}$. Indeed, a truncation step after each imaginary time iteration needs to be performed in order to keep constant the 
MPS bond dimension. Another solution corresponds to solve directly the $1$st order differential equation at Eq.~\eqref{ise} using an Euler integrator 
by taking advantage of the MPS tangent space. \\

In the following, we present the general scheme of an imaginary time evolution relying on the MPS tangent space. By definition, the left-hand side of
\eqref{ise} lives in the tangent space $\mathcal{T}_{\ket{\Psi(\{ A(\tau) \})}}$ so that one can write 

\begin{equation}
 \partial_{\tau}\ket{\Psi(\{ A(\tau) \})} = \ket{\Phi(\{ \partial_\tau{A}(\tau) \}, \{ A(\tau) \})} = \sum_R \partial_{\tau} A[R](\tau) 
 \partial_{A[R](\tau)} \ket{\Psi(\{ A (\tau)\})}
\end{equation}

\noindent
where $\partial_{\tau} A[R](\tau)$ ($\partial_{A[R](\tau)}$) denotes the derivative of the tensor $A[R](\tau)$ (the derivation with respect to $A[R](\tau)$).
However, the right-hand side of Eq.~\eqref{ise} does not live in the MPS tangent space $\mathcal{T}_{\ket{\Psi(\{ A(\tau) \})}}$ due to the Hamiltonian $\hat{H}$.
As a consequence, one has to find the best tangent vector $\ket{T^\star(\tau)}$ associated to the reference state $\ket{\Psi(\{ A(\tau)\})}$
minimizing the distance with respect to $\hat{H}\ket{\Psi(\{ A(\tau) \})}$. In other words, one has to solve the following minimization problem

\begin{equation}
\ket{T^\star(\tau)} = \sum_R B^{\star}[R](\tau) \partial_{A[R](\tau)} \ket{\Psi(\{ A(\tau)\})} := \mathrm{argmin} 
\left(||\ket{T(\tau)} -\hat{H}\ket{\Psi(\{ A(\tau) \})}||^2 \right),
\label{proj}
\end{equation}

\noindent
where $\ket{T^{\star}(\tau)}$ is defined by the collection of the $L$ $3$rd-rank tensors $B^{\star}[R](\tau)$. Then, the ground state can be found by
iteratively integrating the imaginary time Schrödinger equation with a $1$st order Euler scheme. One is doing a steepest-descent optimization
using the gradient method. According to Eq.~\eqref{proj}, the gradient is defined by $-\ket{T^{\star}(\tau)}$ and the direction is
$\ket{D} = -\ket{T^\star(\tau)}/|| \ket{T^\star(\tau)} ||$. At the level of quantum states, the $1$st order Euler scheme reads as 

\begin{align}
 & \ket{\Psi(\{ A(\tau + \mathrm{d}\tau)\})} = \ket{\Psi(\{A(\tau)\})} + \mathrm{d}\tau  \partial_{\tau}\ket{\Psi(\{ A(\tau) \})}\\
 & \ket{\Psi(\{ A(\tau + \mathrm{d}\tau)\})} = \ket{\Psi(\{A(\tau)\})} - \mathrm{d}\tau \ket{T^\star(\tau)} 
 \label{qs_euler}
\end{align}

\noindent
where $\mathrm{d}\tau$ corresponds to the fixed imaginary timestep. The previous Euler scheme can be reformulated under a local version, \textit{ie.}
at the level of the third-rank tensors $A[R](\tau)$ by calculating the following equation

\begin{align}
& \left(\prod_{R', R'\neq R} \partial_{A[R']} \right) \ket{\Psi(\{ A(\tau + \mathrm{d}\tau)\})} = \left(\prod_{R', R' \neq R} \partial_{A[R']}\right) \left(\ket{\Psi(\{A(\tau)\})}
- \mathrm{d}\tau \ket{T^\star(\tau)} \right).
\end{align}

\noindent
Consequently, it locally yields  

\begin{align}
& A[R](\tau + \mathrm{d}\tau) = A[R](\tau) - \mathrm{d}\tau B^{\star}[R](\tau).
\label{loc_proc_ite}
\end{align}

\noindent
Note that the imaginary timestep $\mathrm{d}\tau$ needs to be as small as possible to ensure the convergence of the iterative algorithm.
Using Eq.~\eqref{loc_proc_ite}, each iteration of the imaginary time evolution consists of a local procedure where each tensor $A[R](\tau)$ 
is updated. Most importantly, this method ensures that $\ket{\Psi(\{ A(\tau+\mathrm{d}\tau)\})}$ remains in the same manifold as 
$\ket{\Psi(\{ A(\tau) \})}$, \textit{ie.} $\ket{\Psi(\{ A(\tau+\mathrm{d}\tau)\})} \in \mathcal{M}_{\chi}$. Indeed, the tensor $B^{\star}[R](\tau)$, having the same dimensions as
$A[R](\tau)$, does not modify the MPS bond dimension $\chi$.

\subsubsection{Tensor network formalism}
In the following, we provide more details about the tensor network formalism and the implementation of the imaginary time evolution
for one-dimensional lattice models. \\

\noindent
Let us start by defining a general MPS $\ket{\Psi(\{A\})} = \sum_{\boldsymbol{\sigma}} A^{\sigma_1}[1]...~A^{\sigma_L}[L] 
\ket{\boldsymbol{\sigma}}$ where $\boldsymbol{\sigma} = \sigma_1,...,\sigma_L$ and $L$ the number of lattice sites (or equivalently the length of the chain).
Each $3$rd-rank tensor $A^{\sigma_R}[R]$ has a dimension $\chi_R \times \chi_{R+1} \times d$ and the set of physical states $\ket{\boldsymbol{\sigma}}$
spans the full Hilbert space $\mathbb{H} = \otimes_{R=1}^{L} \mathbb{H}_R$. $\mathbb{H}_R$ denotes the $d$-dimensional local Hilbert space implying that
$\sigma_R \in \{1,...,d\}$, $\forall R \in [|1,L|]$. \\

Any MPS is invariant up to a gauge degree of freedom. Indeed, if one considers the expression of the previous MPS $\ket{\Psi(\{ A \})}$, it is invariant
by the local transformation 

\begin{equation}
\tilde{A}^{\sigma_R}[R] = Y[R] A^{\sigma_R}[R] Y^{-1}[R+1],
\end{equation}

\noindent
where $Y[R]$ refers to a square matrix of dimension $\chi_R \times \chi_R$ and $Y^{-1}[R+1]$ to the inverse square matrix of $Y[R+1]$ with dimension $\chi_{R+1} \times \chi_{R+1}$. 
In the following, the MPS gauge degree of freedom is fixed by considering the isometric gauge. The latter is defined by the left and right gauge-fixing
conditions presented below \cite{koffel2012}. The left gauge-fixing condition is given by 
\begin{equation}
\sum_{\sigma_R} \left(A^{\sigma_R}[R] \right)^{\dag} \Lambda[R] A^{\sigma_R}[R] = \Lambda[R+1].
\end{equation}

\begin{figure}[!h]
\centering
\includegraphics[scale = 0.33]{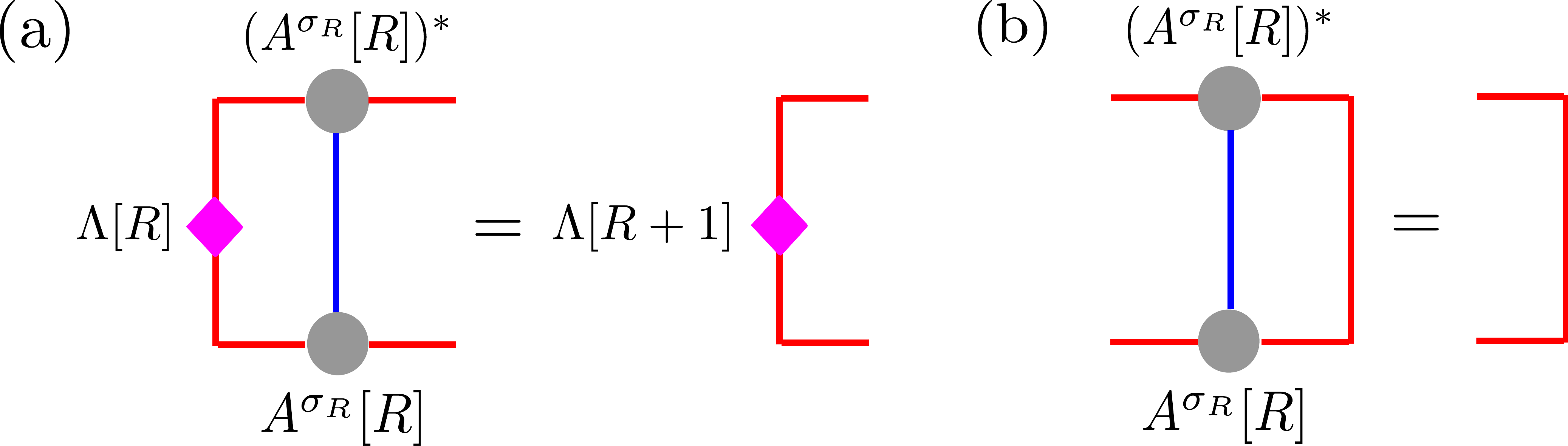}  
\caption{Graphical representation using tensor networks of (a) the left gauge-fixing condition given by $\sum_{\sigma_R} \left(A^{\sigma_R}[R] \right)^{\dag}
\Lambda[R] A^{\sigma_R}[R] = \Lambda[R+1]$ (b) the right gauge-fixing condition $\sum_{\sigma_R} A^{\sigma_R}[R] \left(A^{\sigma_R}[R] \right)^{\dag}
= I$. Both gauge-fixing conditions fix the MPS gauge degree of freedom.}
\label{conditions}
\end{figure}

\begin{figure}[!h]
\centering
\begin{tabular}{c}
\includegraphics[scale = 0.34]{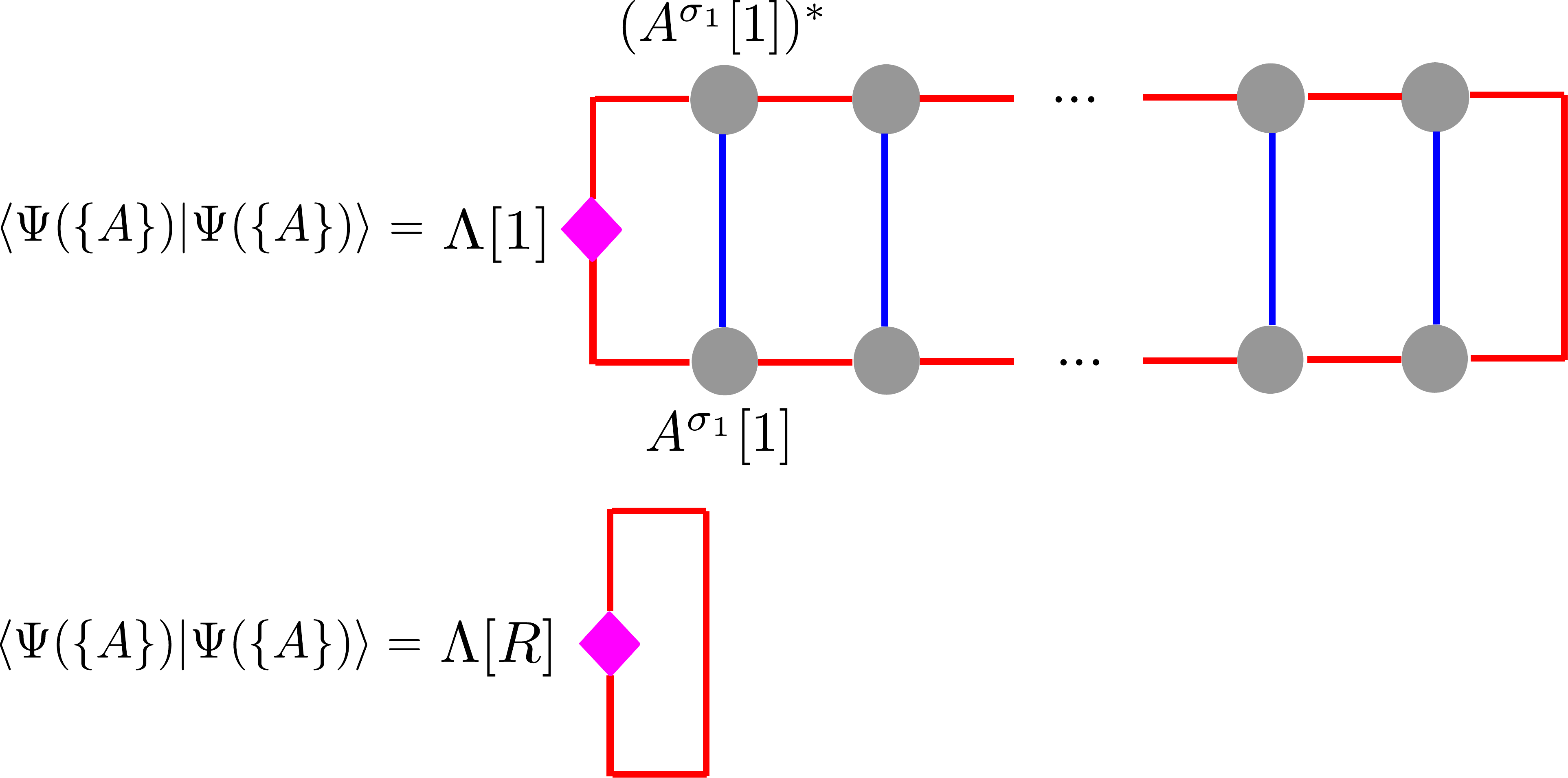}  
\end{tabular}
\caption{Graphical representation using tensor networks of the overlap $\langle \Psi(\{ A \})|\Psi(\{ A \})\rangle$. The MPS $\ket{\Psi (\{A\})}$ is assumed
to be cast into the isometric gauge defined previously at Fig.~\ref{conditions}. It yields for the overlap $\langle \Psi(\{ A \})|\Psi(\{ A \})\rangle =
\mathrm{Tr}(\Lambda[R]) = 1$, $\forall R \in [|1,L|]$. As a consequence, the MPS $\ket{\Psi(\{ A \})}$ is well normalized.} 
\label{normalization}
\end{figure}

\noindent
Figure~\ref{conditions}(a) represents the previous condition in terms of tensor networks. $\Lambda[R]$ denotes the reduced density
matrix for a bipartition of the quantum system given by $\Sigma = L \cup R$ with $L$ the subsystem containing the first $R$ sites
(starting from the left side of the chain) whereas $R$ correspond to the second subsystem containing all the other lattice sites. 
Concerning the right gauge-fixing condition, it is defined by 

\begin{equation}
\sum_{\sigma_R} A^{\sigma_R}[R] \left(A^{\sigma_R}[R] \right)^{\dag} = I,
\end{equation}

\noindent
meaning that the tensors are right-normalized, see Fig.~\ref{conditions}(b). Note that any MPS cast into the isometric gauge
is normalized, see Fig.~\ref{normalization}. Using this specific gauge, the overlap is given by $\langle \Psi( \{A\}) | \Psi(\{A\}) \rangle
= \mathrm{Tr}(\Lambda[R]) = 1$, $\forall R \in [|1,L|]$ leading to the normalization of the MPS $\ket{\Psi(\{A\})}$.
The relation $\mathrm{Tr}(\Lambda[R]) = 1$ is one of the properties of the reduced density matrix which can be easily demonstrated using the Schmidt representation of
any many-body quantum state. We also stress that any entropy measure can be easily computed \textit{via} this specific gauge. Indeed, the
left-gauge fixing condition gives a direct access to the different reduced density matrices. \\

The right gauge-fixing condition of the isometric gauge [see Fig.~\ref{conditions}(b)] implies that the tensor $A^{\sigma_R}[R]$ coarse grain 
the block on its right and projects the tensor product Hilbert space $\mathbb{C}^{\chi_{R+1}} \otimes \mathbb{H}_R$ into the Hilbert space 
$\mathbb{C}^{\chi_R}$ relevant for the description of the state. Consequently, one can define at each lattice site $\chi_{R+1}d - \chi_R$ 
different tangent vectors.
The requirement that those vectors are orthogonal to the original MPS is imposed by defining them through the projection onto
the irrelevant part. Indeed, the isometric gauge corresponds to a unitary part, given by the right gauge-fixing condition for the MPS, 
and an orthogonal part, given by the orthogonality between the tangent vectors and the reference MPS, that we discuss now. \\

\begin{figure}[!h]
\centering
\begin{tabular}{c}
\includegraphics[scale = 0.26]{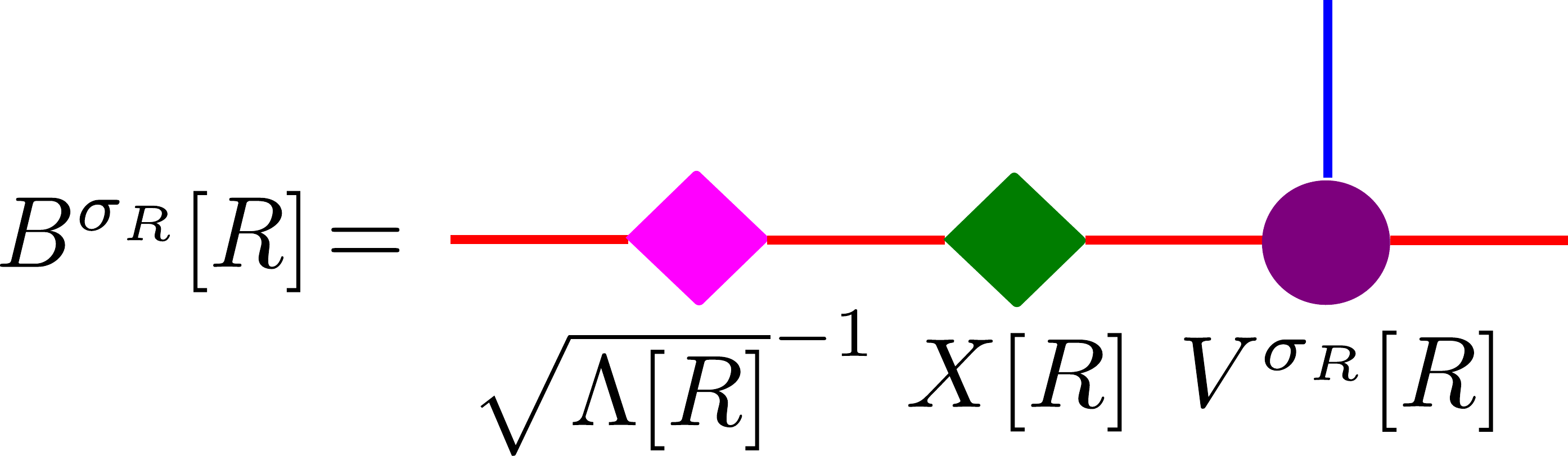}  
\end{tabular}
\caption{Graphical representation using tensor networks of the effective parametrization of the local perturbation $B[R] = \sqrt{\Lambda[R]}^{-1} X[R] V[R]$ 
on the lattice site $R$ and involved in the expression of the tangent vector $\ket{T} = \sum_R B[R] \partial_{A[R]} \ket{\Psi(\{A\})}$.} 
\label{loc_perturb}
\end{figure}

This orthogonal condition between any tangent vector $\ket{T}$ and the reference MPS $\ket{\Psi(\{A\})}$ will fix the gauge degree of freedom for the tangent vectors.
Indeed, any tangent vector has a gauge degree of freedom inherited from the reference MPS. More precisely, the local transformation 

\begin{equation}
\tilde{B}[R]^{\sigma_R} = B[R]^{\sigma_R} - A[R]^{\sigma_R} Y[R+1] + Y[R] A[R]^{\sigma_R},
\end{equation}

\noindent
leaves the tangent vector $\ket{T} = \sum_R B[R] \partial_{A[R]} \ket{\Psi(\{A\})}$ invariant, see Fig.~\ref{tangent_vector} for a representation of $\ket{T}$
in terms of tensor networks. $A[R]$ denotes the $R$-th tensor of the MPS $\ket{\Psi(\{A\})}$ of dimension $\chi_R \times \chi_{R+1} \times d$ and $Y[R]$ a matrix of 
dimension $\chi_R \times \chi_R$. In order to enforce the orthogonality, an effective parametrization of the $B[R]$ tensors as the contraction of auxiliary tensors will be used,
see Fig.~\ref{loc_perturb}. In the following, we adopt the convention

\begin{equation}
\ket{T} = \sum_R  B[R] \partial_{A[R]} \ket{\Psi(\{A\})}~~~\mathrm{with}~~~ B[R]^{\sigma_R} = \sqrt{\Lambda[R]}^{-1} X[R] V[R]^{\sigma_R}.
\end{equation}

\begin{figure}[!h]
\centering
\begin{tabular}{c}
\includegraphics[scale = 0.25]{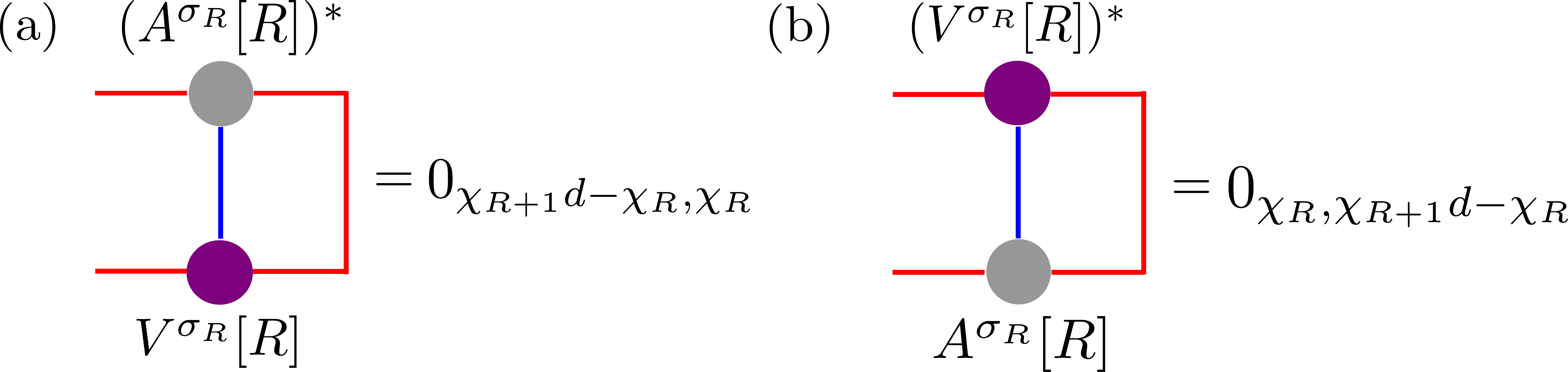}  
\end{tabular}
\caption{Graphical representation using tensor networks of the conditions fulfilled by the projector $V[R]$ so that the tangent vector $\ket{T}$ is 
orthogonal with respect to the reference state $\ket{\Psi(\{ A \})}$. Representation of the condition (a) $\sum_{\sigma_R} V^{\sigma_R}[R] \left( A^{\sigma_R}[R] \right)^{\dag} =
0_{\chi_{R+1}d-\chi_R,\chi_R}$ permitting to get $\langle \Psi(\{ A \})| T \rangle = 0$ (b) $\sum_{\sigma_R} A^{\sigma_R}[R] \left( V^{\sigma_R}[R] \right)^{\dag} =
0_{\chi_R,\chi_{R+1}d-\chi_R}$ implying that $\langle T | \Psi(\{ A \}) \rangle = 0$.}
\label{orthogonality}
\end{figure}

\begin{figure}[!h]
\centering
\begin{tabular}{c}
\includegraphics[scale = 0.26]{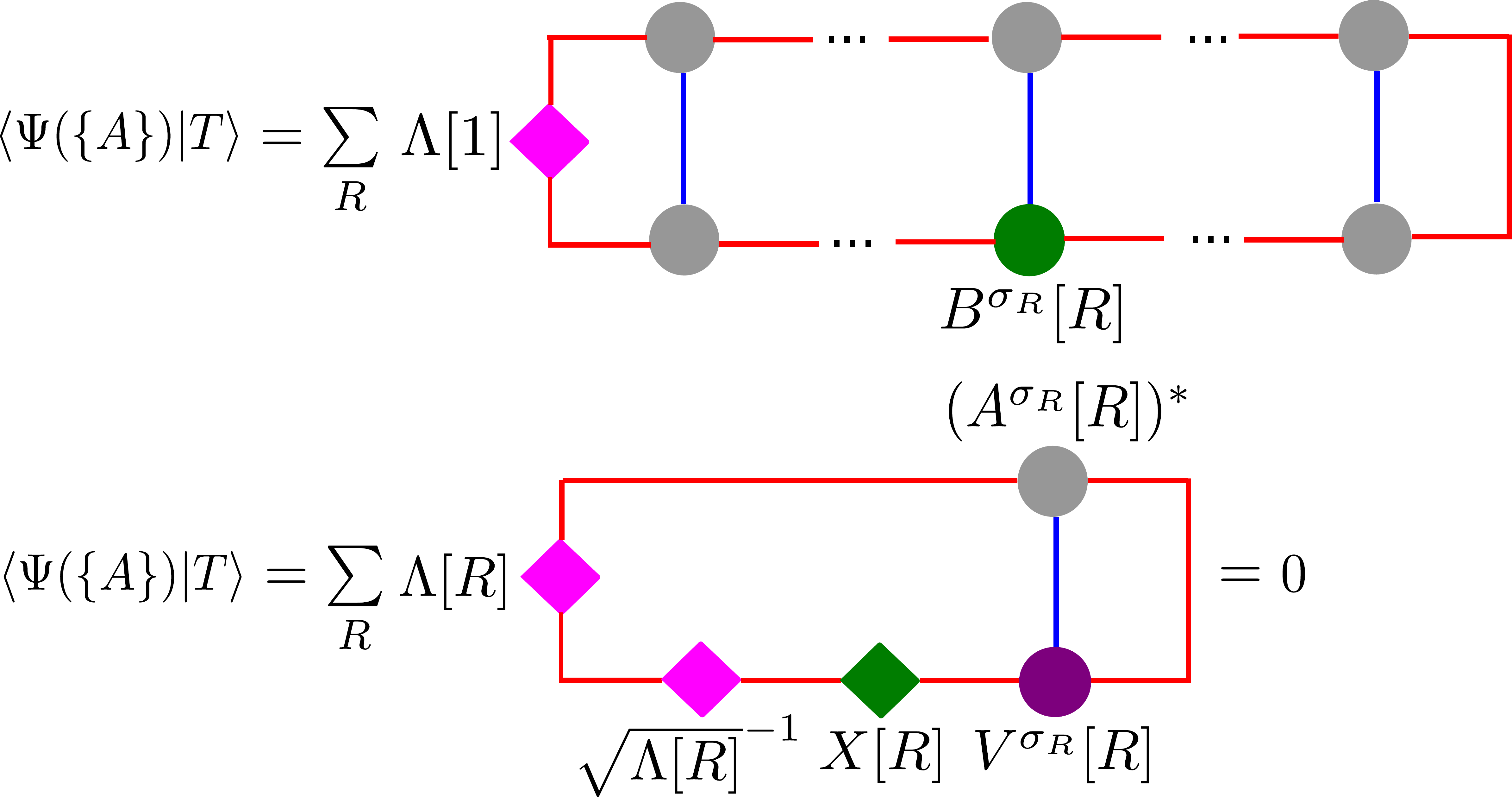}  
\end{tabular}
\caption{Demonstration using tensor network language of the orthogonality between a reference quantum state $\ket{\Psi(\{ A \})}$ and a tangent vector $\ket{T}$
by computing the overlap $\langle \Psi(\{ A \}) | T \rangle$. To demonstrate that $\langle \Psi(\{ A \}) | T \rangle = 0$, one has to used the following condition
$\sum_{\sigma_R} V^{\sigma_R}[R] \left( A^{\sigma_R}[R] \right)^{\dag} = 0_{\chi_{R+1}d-\chi_R,\chi_R}$ fulfilled by the projector $V[R]$.} 
\label{overlap_psi_t}
\end{figure}

\noindent
$\sqrt{\Lambda[R]}^{-1}$ denotes the inverse square root of the reduced density matrix of the subsystem containing the
first $R$ lattice sites (for normalization convenience) and 
corresponds to a matrix of dimension $\chi_R \times \chi_R$. $X[R]$ is the single effective parameter consisting of a matrix of free 
coefficients of dimension $\chi_R \times \chi_{R+1}d - \chi_R$.
$V[R]$ refers to the projector onto the irrelevant space, discarded for the description of the reference MPS $\ket{\Psi(\{A\})}$, of dimension $\chi_{R+1}d-\chi_R \times \chi_{R+1}
\times d$. In other words, the projector $V[R]$ is the one responsible for the orthogonality condition $\langle \Psi(\{A\}) | T \rangle = \langle T | \Psi(\{A\}) \rangle = 0$.
Consequently, it has to fulfill both following conditions, see Fig.~\ref{orthogonality} and \ref{overlap_psi_t},

\begin{equation}
\sum_{\sigma_R} V[R]^{\sigma_R} \left( A[R]^{\sigma_R}\right)^{\dag} =  0_{\chi_{R+1}d-\chi_R,\chi_R},~~~
\sum_{\sigma_R} A[R]^{\sigma_R} \left(V[R]^{\sigma_R} \right)^{\dag} = 0_{\chi_R,\chi_{R+1}d-\chi_R}.
\end{equation}

\begin{figure}[!h]
\centering
\begin{tabular}{c}
\includegraphics[scale = 0.26]{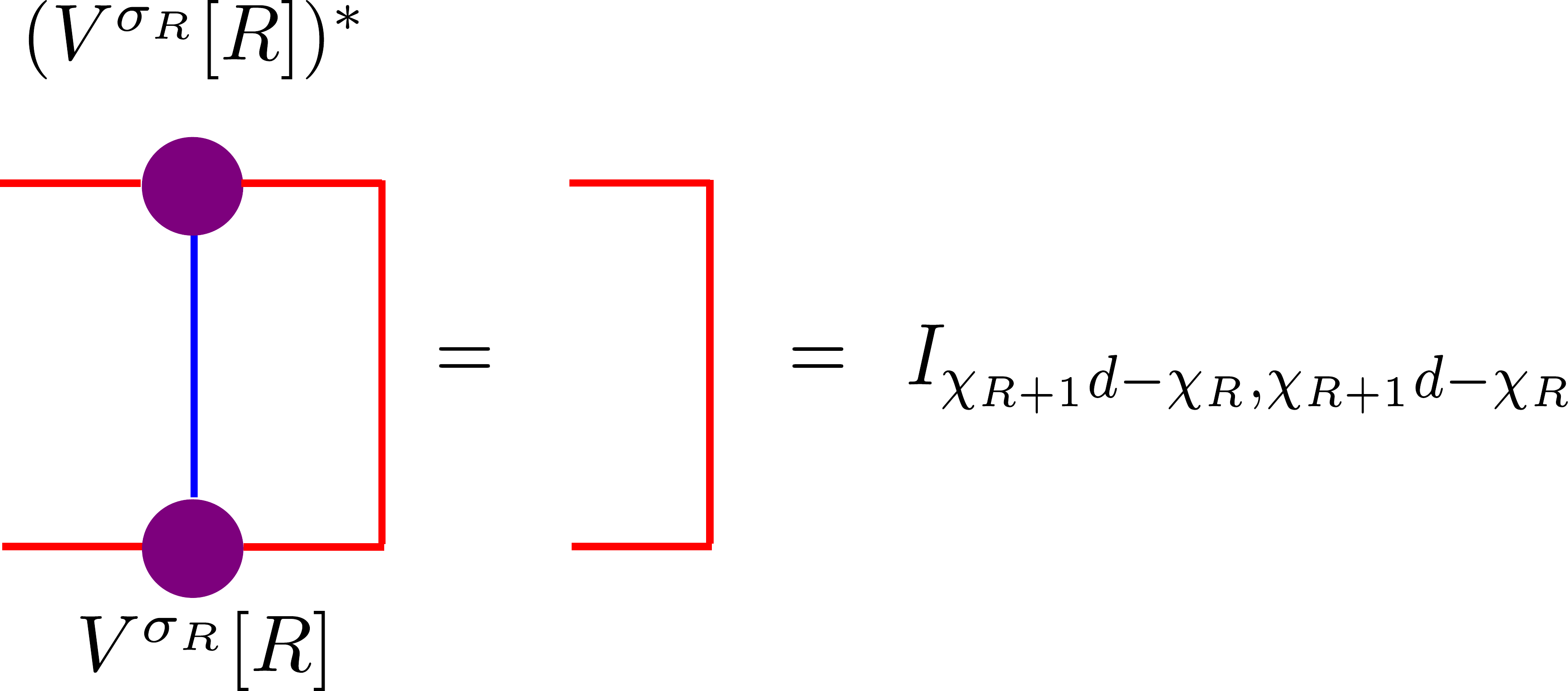}  
\end{tabular}
\caption{Graphical representation using tensor networks of the second condition fulfilled by the projector $V[R]$ in order to find that the  
matrix $X[R]$ corresponds to the single effective parameter of the tensor $B[R]$. The latter allows to explore the MPS tangent plane $\mathcal{T}_{\ket{\Psi(\{ A\})}}$
of the reference state $\ket{\Psi(\{A\})}$. The second condition on $V[R]$ is the right normalization defined as $\sum_{\sigma_R}V^{\sigma_R}[R] 
\left(V^{\sigma_R}[R]\right)^{\dag} = I_{\chi_{R+1}d-\chi_R,\chi_{R+1}d-\chi_R}$.} 
\label{cond_param}
\end{figure} 

\noindent
Note that to find that $X[R]$ is the single effective parameter, the projector $V[R]$ has to be right-normalized, \textit{ie.} has to fulfill the following condition

\begin{equation}
\sum_{\sigma_R}V^{\sigma_R}[R] \left(V^{\sigma_R}[R]\right)^{\dag} = I_{\chi_{R+1}d-\chi_R, \chi_{R+1}d-\chi_R}.
\end{equation}

\noindent
The associated graphical representation in terms of tensor networks is displayed on Fig.~\ref{cond_param}. In the following, we briefly demonstrate the previous statement by 
computing the overlap $\langle T | T \rangle$ where $\ket{T}$ is a tangent vector to the reference MPS $\ket{\Psi(\{A\})}$. The overlap $\langle T | T \rangle$ reads as

 \begin{align}
  & \langle T | T \rangle = \langle \Psi(\{B\},\{A\}) | \Psi(\{B\},\{A\}) \rangle = \sum_{R,R'} \langle \Psi(B[R], A[R]) | \Psi(B[R'], A[R'])\rangle \nonumber \\
  & \langle T | T \rangle = \sum_R \langle \Psi(B[R], A[R]) | \Psi(B[R], A[R]) \rangle + \sum_{R \neq R'} \langle \Psi(B[R], A[R]) | \Psi(B[R'], A[R']) \rangle \nonumber \\
  & \langle T | T \rangle = \sum_R \langle \Psi(B[R], A[R]) | \Psi(B[R], A[R]) \rangle 
  \label{overlap_tt_th}
 \end{align}

\begin{figure}[!h]
\centering
\begin{tabular}{c}
\includegraphics[scale = 0.26]{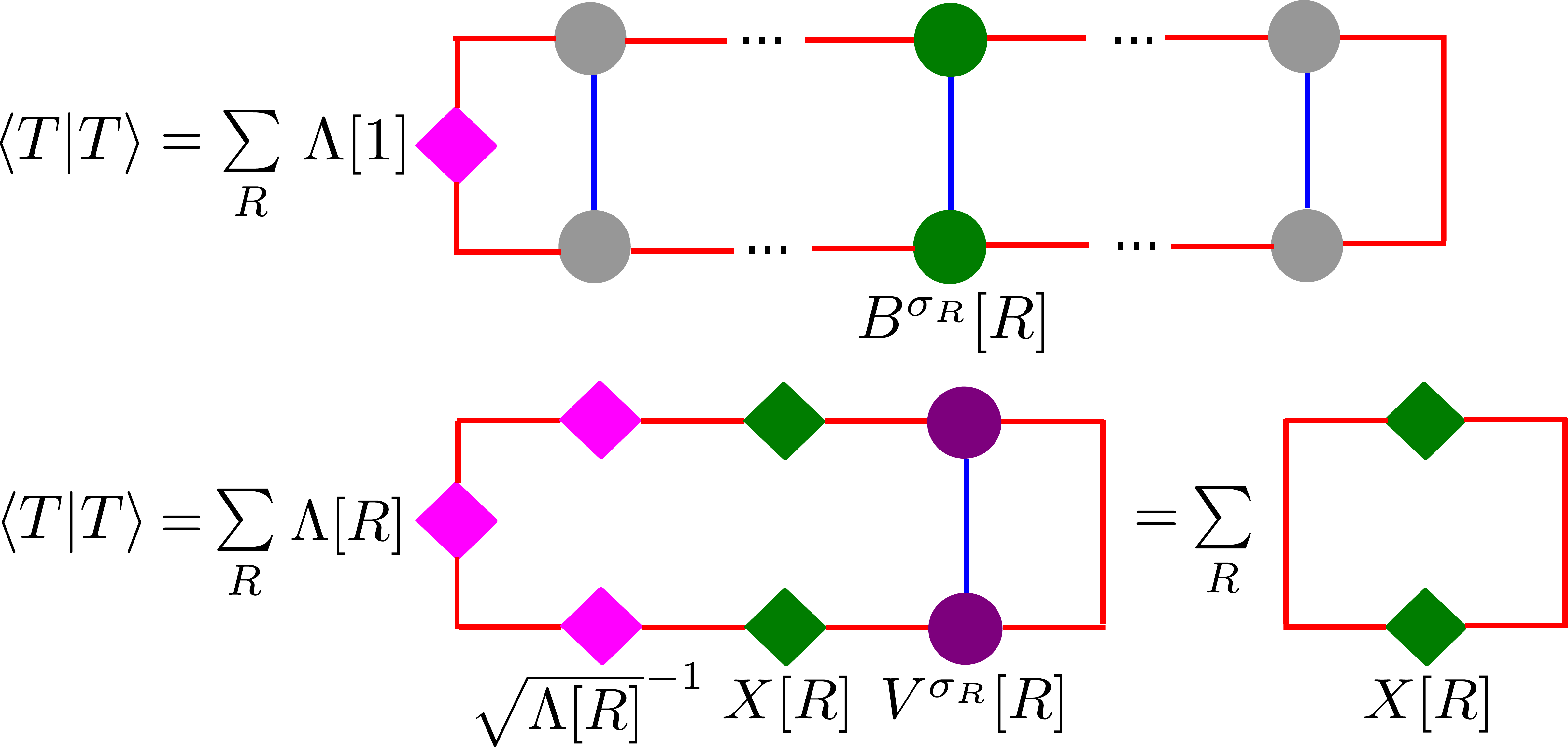}  
\end{tabular}
\caption{Demonstration using tensor networks that the effective parametrization of $B[R]$ the local perturbation on the lattice site $R$ 
for the tangent vector $\ket{T}$ contains a single free parameter corresponding to the matrix $X[R]$ of dimension $\chi_R \times \chi_{R+1}d-\chi_R$.
The overlap $\langle T | T \rangle = \sum_R \langle \Psi(B[R], A[R]) | \Psi(B[R], A[R]) \rangle$ is found to be equal to $\langle T | T \rangle = 
\sum_{R=1}^{L} \mathrm{Tr}\left(X[R] X^{\dag}[R]\right)$ depending only on the matrix $X[R]$. To obtain this result, the right-normalization of the
projector $V[R]$ is assumed, \textit{ie.} $\sum_{\sigma_R} V^{\sigma_R}[R] \left(V^{\sigma_R}[R]\right)^{\dag} = I_{\chi_{R+1}d-\chi_R, \chi_{R+1}d-\chi_R}$. Besides, 
one finds that the norm of a tangent vector $\ket{T}$ is nothing more than the Euclidean inner product of the matrix of free coefficients $X[R]$, \textit{ie.}
$||\ket{T}|| = \sqrt{\langle T | T \rangle} = \sqrt{\sum_{R=1}^{L} \mathrm{Tr}\left(X[R] X^{\dag}[R] \right)}$.} 
\label{overlap_tt}
\end{figure}  
 
\noindent
The term $\sum_{R \neq R'} \langle \Psi(B[R], A[R]) | \Psi(B[R'], A[R']) \rangle$ vanishes due to the condition of orthogonality between
the reference state $\ket{\Psi(\{ A \})}$ and the tangent vector $\ket{T}$. More precisely, for the case $R>R'$, the condition 
$\sum_{\sigma_R} A^{\sigma_R}[R] \left( V^{\sigma_R}[R] \right)^{\dag} = 0_{\chi_R,\chi_{R+1}d-\chi_R}$ permits to cancel the different terms.
For the case $R'>R$, the other condition on the projector is used, \textit{ie.} $\sum_{\sigma_{R'}} V^{\sigma_{R'}}[R'] \left( A^{\sigma_{R'}}[R'] \right)^{\dag} = 
0_{\chi_{R'+1}d-\chi_{R'},\chi_{R'}}$. Equation~\eqref{overlap_tt_th} is represented on Fig.~\ref{overlap_tt} using tensor networks. The latter clearly shows that the
overlap $\langle T | T \rangle$ depends only on the matrix $X[R]$. Hence, for a right-normalized projector $V[R]$, the
single effective parameter of the local perturbation $B[R]$ corresponds to the matrix $X[R]$. 

\subsubsection{Variational principle within the MPS tangent space}
We now turn to the Dirac-Frenkel variational principle within the tangent space. We explicitely derive the
variational dynamical equation for the imaginary time evolution. We also provide details about the Euler-scheme
algorithm to solve the variational equation. \\ 

\begin{figure}[!h]
\centering
\begin{tabular}{c}
\includegraphics[scale = 0.325]{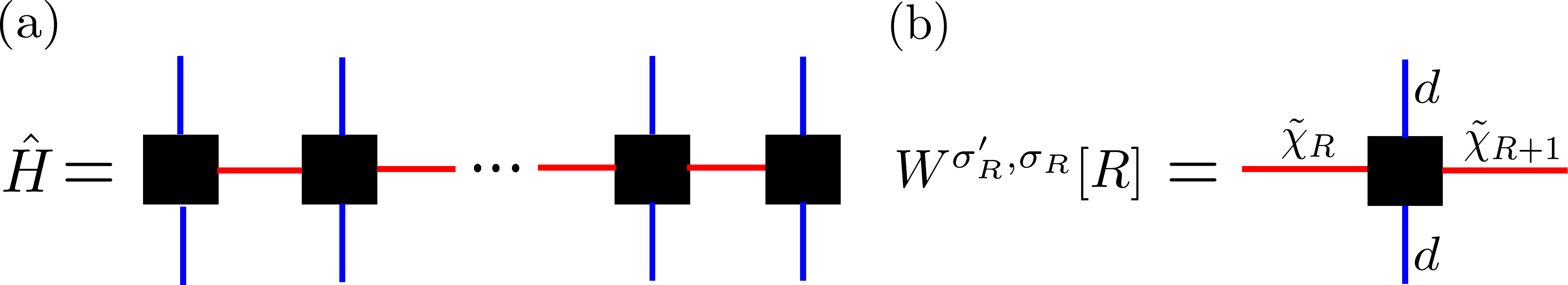}  
\end{tabular}
\caption{Graphical representation using tensor networks of the MPO form for a general Hamiltonian $\hat{H}$ describing
a one-dimensional lattice model of length $L$. (a)~Representation of $\hat{H} = \hat{W}[1]\hat{W}[2] ...~\hat{W}[L-1]
\hat{W}[L]$ under its MPO form where the red (blue) solid lines represent the virtual (physical) indices.
(b)~ Each matrix $\hat{W}[R]$ has a dimension $\tilde{\chi}_Rd \times \tilde{\chi}_{R+1}d$. The latter is built
from $W[R]$ corresponding to a $4$th-rank tensor of dimension $\tilde{\chi}_R \times \tilde{\chi}_{R+1} \times d 
\times d$. $W[R]$ is defined by two virtual (horizontal) legs of dimension $\tilde{\chi}_R$ and 
$\tilde{\chi}_{R+1}$ where $\tilde{\chi} = \mathrm{max}_R (\tilde{\chi}_R)$ corresponds to the MPO
bond dimension associated to the Hamiltonian $\hat{H}$.
Besides, the $4$th-rank tensor $W[R]$ is also characterized by two physical (vertical) legs, where each of them
has a dimension $d$ corresponding to the dimension of the local Hilbert space $\mathbb{H}_R$.}
\label{mpo}
\end{figure}

Firstly, we consider a general Hamiltonian $\hat{H}$ representing a one-dimensional lattice model of length $L$
acting on the many-body Hilbert space $\mathbb{H} = \otimes_{R=1}^{L}\mathbb{H}_R$. $\mathbb{H}_R$ is assumed to denote
a $d$-dimensional local Hilbert space. The Hamiltonian $\hat{H}$ can be represented into a matrix product operator 
(MPO) form, see Fig.~\ref{mpo} for the graphical representation using tensor networks. In other words, $\hat{H}$
is assumed to display the following form, where $\boldsymbol{\sigma}' = \sigma'_1,...,\sigma'_L$ and $\boldsymbol{\sigma} = \sigma_1,...,\sigma_L$
with $\sigma'_R,\sigma_R \in \{1,...,d\}~\forall R \in [|1,L|]$,

\begin{align}
&\hat{H} = \sum_{\boldsymbol{\sigma}',\boldsymbol{\sigma}} W^{\sigma'_1,\sigma_1}[1]  W^{\sigma'_2,\sigma_2}[2]...~
W^{\sigma'_{L-1},\sigma_{L-1}}[L-1]W^{\sigma'_L,\sigma_L}[L] \ket{\boldsymbol{\sigma}'} \bra{\boldsymbol{\sigma}}, \\
& \hat{H} = \hat{W}[1]\hat{W}[2]...~\hat{W}[L-1]\hat{W}[L]~~~\mathrm{with}~~~\hat{W}[R] = \sum_{\sigma'_R,\sigma_R}
W^{\sigma'_R,\sigma_R}[R] \ket{\sigma'_R}\bra{\sigma_R}.
\end{align}

\noindent
$\hat{W}[R]$ is a matrix containing operators acting on the $d$-dimensional local Hilbert space $\mathbb{H}_R$. Each matrix
$\hat{W}[R]$ has a dimension $\tilde{\chi}_R d \times \tilde{\chi}_{R+1}d$. Hence, the bond dimension of $\hat{H}$
symbolized by $\tilde{\chi}$ is defined as $\tilde{\chi} = \mathrm{max}_R(\tilde{\chi}_R)$. In the case 
of translationally invariant Hamiltonians, all the matrices $\hat{W}[R]$ are equal and defined by a dimension
$\tilde{\chi}d \times \tilde{\chi}d$ except the first ($\hat{W}[1]$) and last ($\hat{W}[L]$) matrices having a dimension $1 \times \tilde{\chi}d$
and $\tilde{\chi}d \times 1$ respectively. 

As discussed previously, the ground state of the general Hamiltonian $\hat{H}$ may be found by integrating 
the imaginary time Schrödinger equation ($\hbar = 1$, $\tau = it$) 

\begin{equation}
 \partial_\tau \ket{\Psi(\{ A(\tau)\})} = -\hat{H} \ket{\Psi(\{ A(\tau) \})},
 \label{SE}
\end{equation}

\noindent
with $\ket{\Psi(\{ A(0) \})}$ as the initial condition. The initial state can be chosen arbitrary provided 
it has a non-vanishing overlap with the actual ground state. Within the MPS representation, the integration 
may be performed by successively applying the infinitesimal imaginary time evolution operator
$e^{-\hat{H} \mathrm{d}\tau}$ to the MPS $\ket{\Psi(\{A(\tau)\})}$ for vanishingly small imaginary timesteps
$\mathrm{d}\tau$. In most cases, this operation will lead to an increase of the bond dimension for the
imaginary-time-evolved MPS $\ket{\Psi(\{ A(\tau+\mathrm{d}\tau)\})}$. To solve this issue, the usual solution 
is to truncate the MPS after each iteration to avoid an exponential growth of the MPS bond dimension. \\
An alternative solution, that we consider here, consists of solving iteratively Eq.~\eqref{SE}. At each timestep
$\mathrm{d}\tau$, the right-hand side of the Schr\"odinger equation is projected onto the tangent space. Note that
the left-hand side is automatically in the tangent space by definition of the latter. An immediate consequence is
the conservation of the bond dimension during the full imaginary time evolution. Finally, the imaginary-time-evolved 
quantum state is found using a first-order Euler scheme. In the following, we provide more details about the 
variational dynamical equation for the imaginary time evolution. \\
To find the ground state and its associated energy, the imaginary time Schrödinger equation at Eq.~\eqref{SE}
is integrated using the tangent space. In a first time, it requires to find at each iteration the best projection
$\ket{T^\star(\tau)}$ of $\hat{H}\ket{\Psi(\{ A(\tau)\})}$ onto the tangent space $\mathcal{T}_{\ket{\Psi(\{ A(\tau)\})}}$.
Mathematically, the previous condition may be formulated as follows (be careful to not confound $\star$ denoting the best
approximation and $*$ representing the complex conjugate)
 
\begin{equation}
\ket{T^{\star}(\tau)} := \mathrm{argmin} \left( ||\ket{T(\tau)}-\hat{H}\ket{\Psi(\{A(\tau)\})}||^2 \right),
\label{mini_prob}
\end{equation}

\noindent
where $\ket{T(\tau)}$, corresponding to a tangent vector confined in $\mathcal{T}_{\ket{\Psi(\{ A(\tau)\})}}$ (denoting the tangent space with respect to the reference state $\ket{\Psi(\{A(\tau)\})} \in \mathcal{M}_{\chi}$),
is defined by 

\begin{equation}
\ket{T(\tau)} = \sum_{R} \ket{\Psi(B[R](\tau), A[R](\tau)} =
\sum_R B[R](\tau) \partial_{A[R](\tau)} \ket{\Psi(\{A(\tau)\})}.
\label{tangent_vec}
\end{equation}

\noindent
The minimization problem formulated at Eq.~\eqref{mini_prob} can be reduced to a local problem by finding
the optimal set of tensors $\{B[R](\tau),~R=1,...,L\}$. For instance, the best approximation of the tensor
$B[R](\tau)$, denoted by $B^{\star}[R](\tau)$, is determined by solving the following problem 
 
\begin{equation}
B^{\star}[R](\tau) := \mathrm{argmin}\left(||\ket{\Psi(B[R](\tau), A[R](\tau)}-\hat{H}
\ket{\Psi(\{A(\tau)\})}||^2 \right).
\label{mini_prob2}
\end{equation}
 
\noindent
The latter is determined using the definition of the tangent vectors, see Eq.~\eqref{tangent_vec} 
and Fig.~\ref{tangent_vector}. Furthermore, the isometric gauge is considered [see Figs.~\ref{conditions}, \ref{loc_perturb},
\ref{orthogonality} and \ref{cond_param}] fixing the gauge degree of freedom of both $\ket{T(\tau)}$
and $\ket{\Psi(\{A(\tau)\})}$. As explained previously, a specific parametrization of the
local perturbations symbolized by the set of tensors $\{ B(\tau) \}$ is used to enforce the orthogonality
between the reference state $\ket{\Psi(\{A(\tau)\})}$ and any of its tangent vectors $\{\ket{T(\tau)} \}$,
see Figs.~\ref{loc_perturb},\ref{orthogonality} and \ref{overlap_psi_t}. This parametrization leads to
effective (free) parameters given by the matrices $\{X[R](\tau),~R=1,...,L \}$ of dimension $\chi_R \times
\chi_{R+1}d-\chi_R$. As a consequence, the minimization problem at Eq.~\eqref{mini_prob2} is significantly
simplified into one on the free parameters $X[R](\tau)$ of dimension $\chi_R \times \chi_{R+1} d - \chi_R$
and not on the tensors $B[R](\tau)$ of dimension $\chi_R \times \chi_{R+1} \times d$.
Hence, it yields the following equation for the minimization problem
 
\begin{align}
& X^{\star}[R](\tau) := \mathrm{argmin}\left(||\ket{\Psi ( \sqrt{\Lambda[R](\tau)}^{-1} X[R](\tau) V[R](\tau), A[R](\tau))}
-\hat{H} \ket{\Psi(\{A(\tau)\})}||^2 \right). 
\end{align}

\noindent
The previous equation is linear in terms of ${X}^*[R](\tau)$ and $X[R](\tau)$. Therefore, by taking the
derivative with respect to the tensor $X^*[R](\tau)$, one is able to find a condition on the free 
parameter $X[R](\tau)$. The latter is given by (one still rely on the tensors $B[R](\tau)$ instead of 
using their effective parametrization involving the tensors $X[R](\tau)$ for simplicity) 

\begin{align}
& X^{\star}[R](\tau) : \partial_{X^*[R](\tau)} ||  \ket{\Psi(B[R](\tau), A[R](\tau))}- \hat{H}
\ket{\Psi(\{A(\tau)\})} ||^2 = 0 
\label{mini_cond3}
\end{align}

\noindent
The two terms $\langle \Psi(\{ A(\tau) \}) | \hat{H}^2 \ket{\Psi(\{ A(\tau) \})}$ and  
$-\langle \Psi(\{ A(\tau) \}) |\hat{H}| \Psi(B[R](\tau), A[R](\tau)) \rangle$ vanish since they 
do not involve the tensor $X^*[R](\tau)$. Consequently, the minimization problem at Eq.~\eqref{mini_cond3} reduces
to

\begin{align}
& X^{\star}[R](\tau) : \partial_{X^*[R](\tau)} [ -\langle \Psi(B[R](\tau), A[R](\tau)) | \hat{H} | \Psi(\{ A(\tau) \}) \rangle  \nonumber \\
& ~~~~~~~~~~~~~~ + \langle \Psi(B[R](\tau), A[R](\tau)) | \Psi(B[R](\tau), A[R](\tau)) \rangle ] = 0 
\label{condition_Xm1}
\end{align}

\begin{figure}[!h]
\centering
\begin{tabular}{c}
\includegraphics[scale = 0.3]{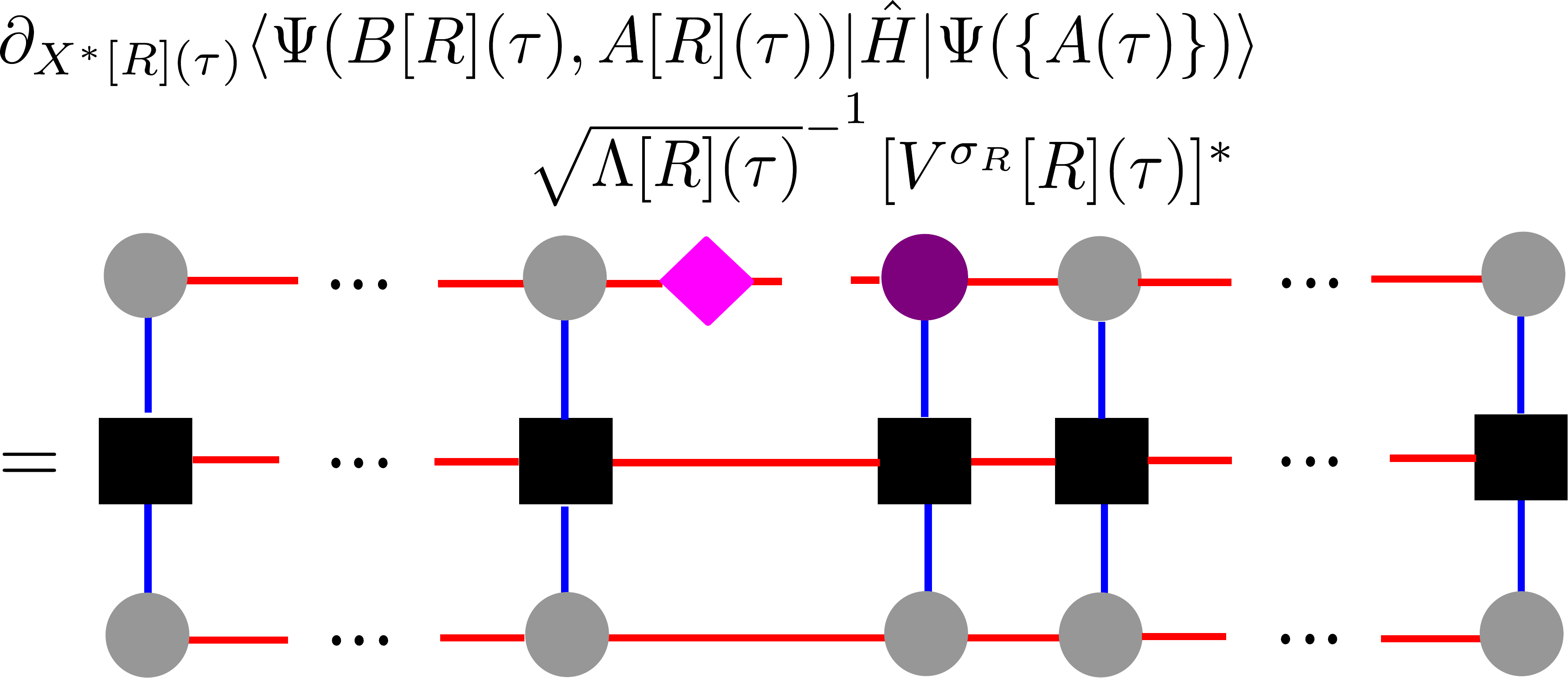}  
\end{tabular}
\caption{Graphical representation using tensor networks of the first term on the left-hand side of Eq.~\eqref{condition_Xm1}. Since the isometric gauge is considered,
one should add to this graph the left and right gauge-fixing conditions of the isometric gauge provided at Fig.~\ref{conditions}.} 
\label{lhs}
\end{figure}

\begin{figure}[!h]
\centering
\begin{tabular}{c}
\includegraphics[scale = 0.25]{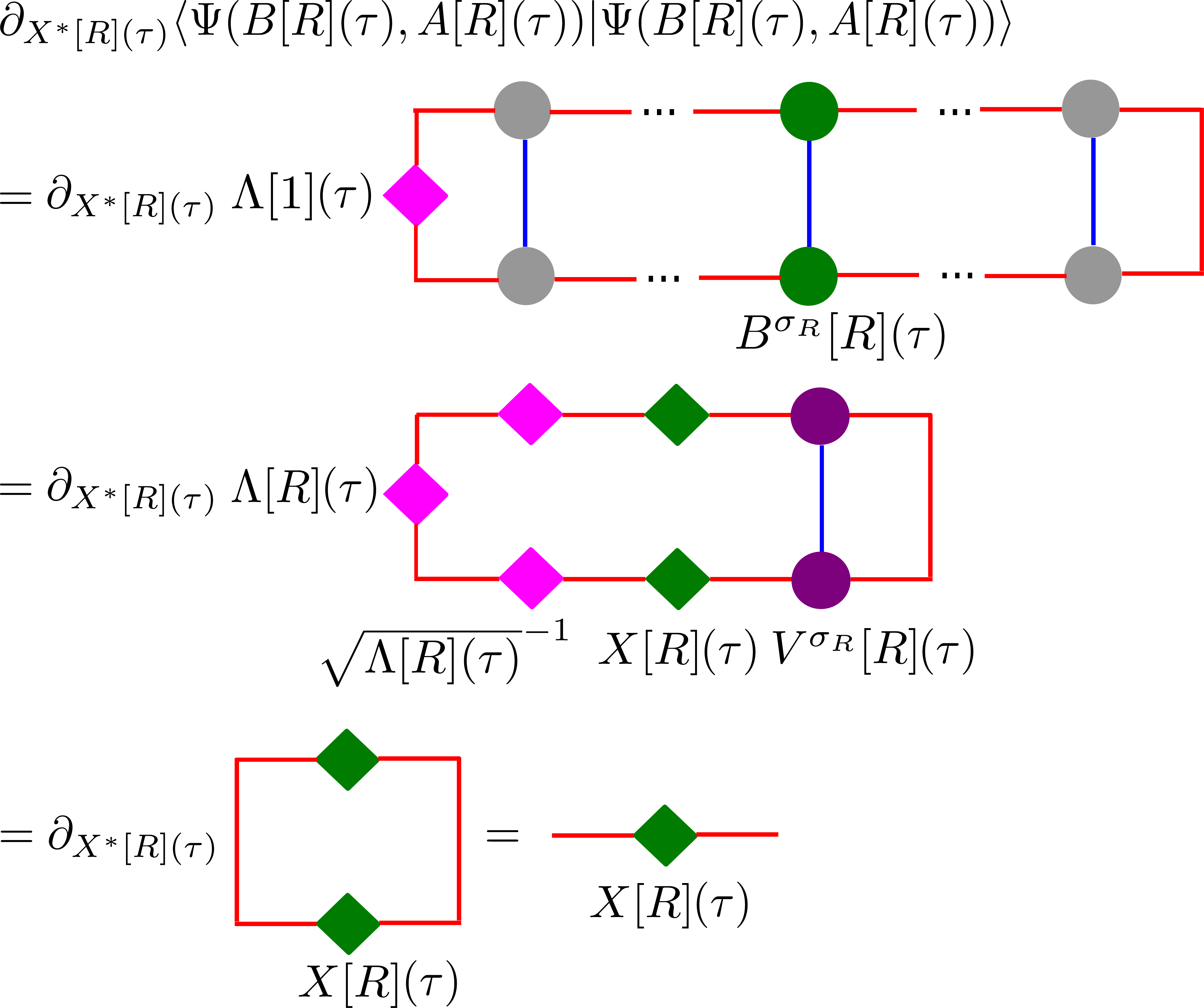}  
\end{tabular}
\caption{Graphical representation using tensor networks of the second term on the left-hand side of Eq.~\eqref{condition_Xm1}.} 
\label{rhs}
\end{figure}

\noindent
Note that the second term on the left-hand side of Eq.~\eqref{condition_Xm1}, representing the derivative with respect 
to $X^*[R](\tau)$ of the square norm

\begin{equation}
||\ket{\Psi(B[R](\tau), A[R](\tau))} ||^2 = \mathrm{Tr}\left[X[R](\tau) X^{\dag}[R](\tau) \right],
\end{equation}

\noindent
is nothing more than the free parameter $X[R](\tau)$. The first and second terms of Eq.~\eqref{condition_Xm1} are represented on Fig.~\ref{lhs} and \ref{rhs}
respectively using tensor networks. Finally, it yields the following equation for the minimization problem

\begin{align}
& X^\star[R](\tau) : \partial_{X^*[R](\tau)} \langle \Psi(B[R](\tau), A[R](\tau)) | \hat{H} 
\ket{\Psi(\{ A(\tau) \})} = X[R](\tau) 
\label{condition_XXX}
\end{align}

\begin{figure}[!h]
\centering
\begin{tabular}{c}
\includegraphics[scale = 0.3]{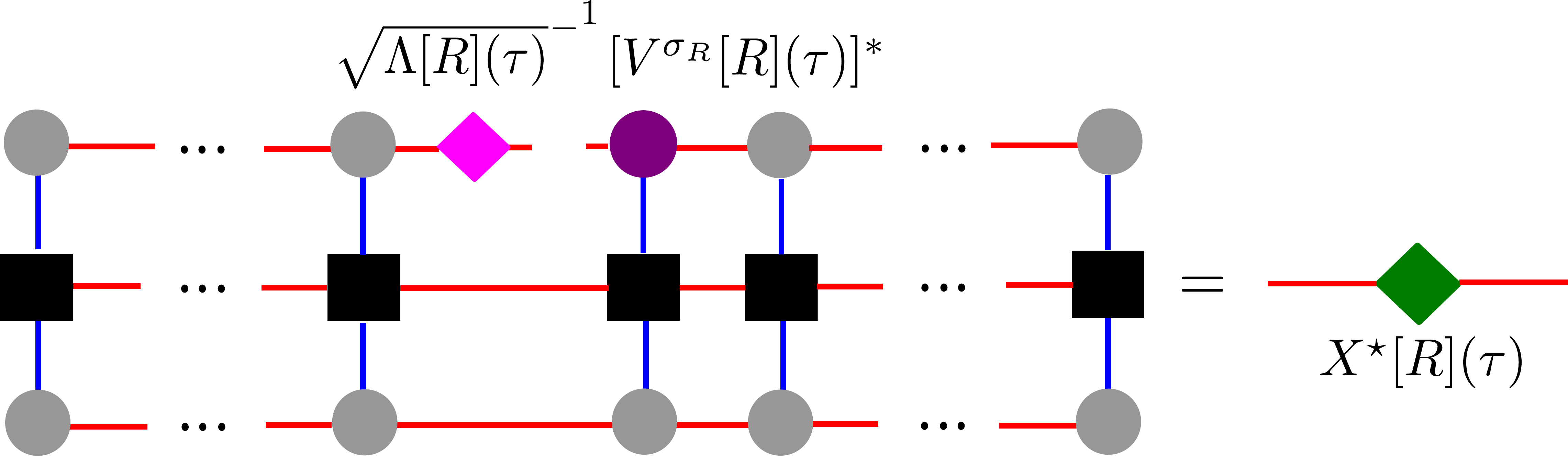}  
\end{tabular}
\caption{Graphical representation using tensor networks of the condition at Eq.~\eqref{condition_XXX}. The latter allows us to find the optimal effective (free) parameter 
$X^{\star}[R](\tau)$ and hence the optimal tangent vector $\ket{T^{\star}(\tau)}$. This tangent vector represents the best approximation of $\hat{H} \ket{\Psi(\{A(\tau)\})}$
living in the tangent space of the reference MPS $\ket{\Psi(\{A(\tau)\})}$ denoted by $\mathcal{T}_{\ket{\Psi(\{ A(\tau) \})}}$. It permits to integrate the imaginary time Schrödinger
equation while confining the imaginary-time-evolved quantum state in a same manifold so that the bond dimension is constant during the evolution process.} 
\label{final_condition}
\end{figure}

\noindent
This minimization condition, allowing to deduce the optimal effective tensor $X^\star[R](\tau)$ is sketched
on Fig.~\ref{final_condition}. \\

As discussed previously, the condition at Eq.~\eqref{condition_XXX} allows us to find the optimal effective (free) parameter 
$X^{\star}[R](\tau)$ and thus the optimal tangent vector $\ket{T^{\star}(\tau)}$. The latter is optimal in the sense that it represents the 
best approximation of $\hat{H} \ket{\Psi(\{A(\tau)\})}$ living in the tangent space of the reference MPS $\ket{\Psi(\{A(\tau)\})}$ denoted by 
$\mathcal{T}_{\ket{\Psi(\{ A(\tau) \})}}$. It permits to integrate the imaginary time Schrödinger equation while confining the imaginary-time-evolved quantum state in a 
same manifold. In other words, during the full evolution process, the bond dimension of the previous many-body quantum state remains constant.
In order to perform the integration of the imaginary time Schrödinger equation defined at Eq.~\eqref{SE}, we rely on the first-order Euler scheme. The latter allows us 
to deduce the many-body quantum state in imaginary time after one timestep $\mathrm{d}\tau$, \textit{ie.} $\ket{\Psi(\{A(\tau + \mathrm{d}\tau)\})}$ when starting from $\ket{\Psi(\{A(\tau)\})}$.
The first-order Euler scheme is defined as follows

\begin{equation}
\ket{\Psi(\{A(\tau + \mathrm{d}\tau)\})} = \ket{\Psi(\{A(\tau)\})} - \mathrm{d}\tau \ket{T^\star(\tau)}.
\end{equation}

\noindent
By differentiating the previous equation with respect to the tensors $A[R'], \forall R' \in \{1,...,R-1,R+1,...,L\}$, it yields a local condition on the 
tensors $\{A[R](\tau + \mathrm{d}\tau),~R=1,...,L\}$. This local update of the tensors permits to deduce the MPS $\ket{\Psi(\{A(\tau + \mathrm{d}\tau)\})}$ and is 
characterized by the following equation 

\begin{equation}
A[R]^{\sigma_R}(\tau + \mathrm{d}\tau) = A[R]^{\sigma_R}(\tau) - \mathrm{d}\tau B[R]^{\sigma_R,\star}(\tau),~~\forall \sigma_R \in \{1,...,d\},~~\forall R \in [|1,L|],
\end{equation}

\noindent
or equivalently, 

\begin{equation}
A[R](\tau + \mathrm{d}\tau) = A[R](\tau) - \mathrm{d}\tau B^{\star}[R](\tau),~~~\forall R \in [|1,L|].
\end{equation}

\noindent
However, it has to be stressed that $\mathrm{d}\tau$ needs to be as small as possible to minimize the error on the norm of the imaginary-time-evolved quantum state
$\ket{\Psi(\{A(\tau + \mathrm{d}\tau)\})}$. Indeed, the norm is not conserved during the imaginary time evolution. The latter is due to the non-normalization
of the tangent vectors where $||\ket{T(\tau)}||^2 = \sum_{R=1}^{L} \mathrm{Tr}\left[X[R](\tau)X^{\dag}[R](\tau)\right] \neq 1$, see Fig.~\ref{overlap_tt}. Considering
a single iteration in imaginary time, one gets

\begin{equation}
 \ket{\Psi(\{ A(\tau + \mathrm{d}\tau)\})} = \ket{\Psi(\{ A(\tau) \})} -\mathrm{d}\tau \ket{T^{\star}(\tau)}.
\end{equation}

\noindent
The error on the norm can be estimated by computing the overlap  $\langle \Psi(\{ A(\tau + \mathrm{d}\tau) \})|\Psi(\{ A(\tau + \mathrm{d}\tau)\}) \rangle$.
Using the orthogonality between the optimal tangent vector $\ket{T^{\star}(\tau)}$ and the imaginary-time-evolved quantum state $\ket{\Psi(\{ A(\tau) \})}$, the error
$\epsilon$ after a single imaginary timestep is given by  

\begin{equation}
\epsilon = \mathrm{d}\tau^2 \sum_{R=1}^L \mathrm{Tr}\left[ X^{\star}[R](\tau) X^{\star,\dag}[R](\tau) \right] = \mathrm{d}\tau^2 ||\ket{T^{\star}(\tau)} ||^2.
\end{equation}

\noindent
This non-conservation of the norm requires to bring back the MPS $\ket{\Psi(\{ A(\tau + \mathrm{d}\tau)\})}$ into the isometric gauge after each infinitesimal iteration. 
Finally, the local procedure for the imaginary-time evolution is iterated up to convergence. The convergence is reached for
$\delta_E = E_i-E_{i+1} \simeq 10^{-13}$. The index $i$ denotes the number of imaginary timesteps $\mathrm{d}\tau$ applied on the (random) initial state
$\ket{\Psi(\{A(0)\})}$ and $E_i = \langle \Psi(\{ A(i \times \mathrm{d}\tau) \}) | \hat{H} | \Psi(\{ A(i\times \mathrm{d}\tau) \})\rangle$ refers to the energy of the quantum
system for an imaginary time $i \times \mathrm{d}\tau$. The previous condition on the energy characterizing the convergence of the algorithm implies that the norm of the gradient
$||-\ket{T^*(i\times \mathrm{d}\tau)}||$ has to be very small. 

\subsection{Dynamical properties - Real time evolution through a local update}

\subsubsection{General scheme}
We now turn to the real time evolution process where the variational dynamical equations are explicitely derived. 
Besides, several details about the considered integrator scheme to solve the variational equations are provided.\\

The real-time evolution of a quantum state is found by integrating the (real-time) Sch\"odinger equation ($\hbar = 1$) 

\begin{equation}
 i \partial_t \ket{\Psi(\{A(t)\})} = \hat{H} \ket{\Psi(\{A(t)\})},
 \label{RSE}
\end{equation}

\noindent
and the time-evolved many-body quantum state $\ket{\Psi(\{A(t)\})}$ reads as 

\begin{equation}
 \ket{\Psi(\{A(t)\})} = e^{-i \hat{H}t} \ket{\Psi(\{A(0)\})}.
 \label{teqs}
\end{equation}

\noindent
$\ket{\Psi(\{A(0)\})}$ denotes the initial state and $\hat{U}(t)=e^{-i\hat{H}t}$ the unitary time evolution operator ($\hat{U}(t)
\hat{U}^{\dag}(t) = I$,$\forall t \in \mathbb{R}$). Similarly to the imaginary-time evolution,
instead of directly constructing the evolution operator $\hat{U}(t)$, the time-dependent variational principle 
is considered to perform the real-time evolution so that the time-evolved quantum state $\ket{\Psi(\{A(t)\})}$ is
confined in a MPS manifold, \textit{ie.} has a fixed bond dimension. The general scheme 
is significantly similar to the one used for the imaginary-time evolution. However, a main difference
appears for the real-time evolution process. Indeed, a special care has to be taken to ensure that the algorithm
does not violate the time-reversal symmetry of the real-time Schrödinger equation at Eq.~\eqref{RSE}
($\hat{U}^{\dag}(t) = \hat{U}(-t)$, $\forall t \in \mathbb{R}$). A first-order Euler scheme does not fulfill such symmetry
and we thus have to consider at least a middle-point integrator scheme that is discussed in the following, see Ref.~\cite{hauke2013}.\\

For the real-time evolution, the minimization problem needs to be adapted. The best projection of 
$-\hat{H} \ket{\Psi(\{A(t)\})}$ onto the tangent space $\mathcal{T}_{\ket{\Psi(\{A(t)\})}}$ is defined by 

\begin{align}
& \ket{T^{\star}(t)} = \sum_{R} \ket{\Psi(B^{\star}[R](t), A[R](t))} := \mathrm{argmin} \left(||\ket{T(t)} +
\hat{H}\ket{\Psi(\{A(t)\})}||^2 \right).
\label{mini_cond_real}
\end{align}

\noindent
Consequently, using the same reasoning as before for the imaginary-time evolution, it yields the following condition
on the tensors $\{A[R](t+\mathrm{d}t),R=1,...,L\}$,

\begin{equation}
A[R](t+\mathrm{d}t) = A[R](t) + i\mathrm{d}t B^{\star}[R](t).
\label{local_update_real}
\end{equation}

\noindent
The tensors $\{B^{\star}[R](t)\}$ are determined through the free matrices $\{X^{\star}[R](t)\}$ when using the isometric 
gauge, see Figs.~\ref{conditions} and \ref{orthogonality}. Note that the convention adopted at Eq.\eqref{mini_cond_real} for the minimization problem
implies to slightly modify Fig.~\ref{final_condition}. The latter represents graphically using tensor networks
the optimal matrix $X^{\star}[R](\tau)$ found from the minimization problem for the imaginary-time evolution.
Indeed, one has to replace the imaginary times by real times and to add a minus sign on the left- or
right-hand side of the equation. Indeed, according to Eq.~\eqref{mini_cond_real}, the minimization problem for real times reduces to the following condition

\begin{align}
& X^\star[R](t) : \partial_{X^*[R](t)} \langle \Psi(B[R](t), A[R](t)) | \hat{H} \ket{\Psi(\{ A(t) \})} = -X[R](t).
\label{condition_X}
\end{align}

\noindent
Contrary to the imaginary time evolution, a first-order Euler scheme (as displayed on Eq.~\eqref{local_update_real})
is not precise enough to perform real-time evolution
due to the time-reversal symmetry of the real-time Schrödinger equation. One needs to consider at least a
middle-point integrator characterized by both following equations, 

\begin{align}
& A[R]\left(t+ \mathrm{d}t/2 \right) = A[R]\left(t \right) + i (\mathrm{d}t/2) B^{\star}[R](t), \label{mpi_1} \\
& A[R]\left(t+ \mathrm{d}t/2 \right) = A[R]\left(t + \mathrm{d}t \right) - i (\mathrm{d}t/2) B^{\star}[R](t+\mathrm{d}t). \label{mpi_2}
\end{align}

\noindent
From Eqs.~\eqref{mpi_1} and \eqref{mpi_2}, it immediately comes out that the time-reversal symmetry is well enforced.
Indeed, starting from the tensors $\{A[R](t+\mathrm{d}t)\}$ or $\{A[R](t)\}$, one obtains the same set of tensors
$\{A[R]\}$ at the middle-point $t+\mathrm{d}t/2$. To complete the infinitesimal step, \textit{ie.} to deduce
$\ket{\Psi(t+\mathrm{d}t)}$, one has to integrate $\ket{\Psi(\{A(t+\mathrm{d}t/2)\})}$ for another $\mathrm{d}t/2$
using the condition below
 
\begin{equation}
A[R]\left(t+\mathrm{d}t \right) = A[R] \left(t+\mathrm{d}t/2 \right) + i(\mathrm{d}t/2) B^{\star}[R](t+\mathrm{d}t/2),
\end{equation}

\noindent
equivalent to the following condition at the level of quantum states

\begin{equation}
\ket{\Psi(\{A(t+\mathrm{d}t)\})} = \ket{\Psi(\{A(t+\mathrm{d}t/2)\})} + i(\mathrm{d}t/2)
\ket{T^{\star}(t+\mathrm{d}t/2)}.
\label{final_eq}
\end{equation}

\subsubsection{Unitary evolution - norm and energy conservation}
As previously discussed, the real-time evolution requires to enforce the time-reversal symmetry. This problem was solved
by relying on the so-called middle-point integrator scheme. Furthermore, it is also necessary to verify that the norm and
the energy of the time-evolved many-body quantum state $\ket{\Psi(\{A(t)\})}$ are conserved quantities during the 
real-time evolution process. Indeed, according to Eq.~\eqref{teqs}, the real-time evolution process corresponds to a 
unitary transformation in time. In the following, we briefly discuss whether 
these conditions are fulfilled or not when using the TDVP approach.\\

Concerning the norm of the time-evolved quantum state $\ket{\Psi(\{A(t)\})}$, the latter is not conserved. Similarly to 
the imaginary-time evolution, an error is introduced at each timestep when updating locally the tensors using the 
middle-point integrator scheme due to the non-normalized tangent vector $\ket{T^{\star}(t)}$. Note that the tangent
vectors $\ket{T(t)}$ can not be normalized since their norm depends on free parameters, the matrices $\{X[R](t)\}$, 
whose values are determined such that they optimize the projection of $-\hat{H}\ket{\Psi(\{A(t)\})}$ on the
tangent plane $\mathcal{T}_{\ket{\Psi(\{A(t)\})}}$. The global error $\epsilon$ on the norm for a timestep
starting from $t+\mathrm{d}t/2$ and ending to $t+\mathrm{d}t$ is characterized by

\begin{equation}
\epsilon = (\mathrm{d}t/2)^2 \sum_{R=1}^L \mathrm{Tr}\left[ X^{\star}[R](t+\mathrm{d}t/2) X^{\star,\dag}[R](t+
\mathrm{d}t/2)\right] = (\mathrm{d}t/2)^2 ||\ket{T^{\star}(t+\mathrm{d}t/2)} ||^2.
\end{equation}

\noindent
The latter is found by calculating the overlap $\langle \Psi(\{A(t+\mathrm{d}t)\}) | \Psi(\{A(t+\mathrm{d}t)\}) \rangle$ using Eq.~\eqref{final_eq}.
This non-conservation of the norm requires to bring back the MPS $\ket{\Psi(\{A(t+\mathrm{d}t)\})}$ into the isometric gauge after each infinitesimal iteration
in real time, see Figs.~\ref{conditions}, \ref{orthogonality} and \ref{cond_param}.\\

To test whether the energy is a conserved quantity or not during the real time evolution, one can calculate the energy at time $t+\mathrm{dt}$ given by 

\begin{equation}
 E(t+\mathrm{d}t) = \frac{\langle \Psi(\{A(t+\mathrm{d}t)\}) |\hat{H}| \Psi(\{A(t+\mathrm{d}t)\}) \rangle}{\langle \Psi(\{A(t+\mathrm{d}t)\}) | 
 \Psi(\{A(t+\mathrm{d}t)\}) \rangle}.
\end{equation}

\noindent
The many-body quantum state $\ket{\Psi(\{A(t+\mathrm{d}t)\})}$ is assumed to be normalized ($\langle \Psi(\{A(t+\mathrm{d}t)\}) | 
 \Psi(\{A(t+\mathrm{d}t)\}) \rangle=1$), \textit{ie.} has been brought back into the isometric gauge.
Relying on Eq.~\eqref{final_eq}, the energy $E(t+\mathrm{d}t)$ may be written as

\begin{align}
& E(t+\mathrm{d}t) = E(t+\mathrm{d}t/2) + (\mathrm{d}t/2)^2 \langle T^{\star}(t+\mathrm{d}t/2) |\hat{H} | T^{\star}(t+\mathrm{d}t/2) \rangle \nonumber \\
& + i(\mathrm{d}t/2)\left( \langle \Psi(\{A(t+\mathrm{d}t/2)\}) |\hat{H} | T^{\star}(t+\mathrm{d}t/2) \rangle 
 -  \langle T^{\star}(t+\mathrm{d}t/2) |\hat{H} |  \Psi(\{A(t+\mathrm{d}t/2)\}) \rangle \right),
 \label{e_tpdt}
\end{align}

\noindent
where $E(t+\mathrm{d}t/2) = \langle \Psi(\{A(t+\mathrm{d}t/2)\}) |\hat{H}| \Psi(\{A(t+\mathrm{d}t/2)\}) \rangle$ ($\ket{\Psi(\{A(t+\mathrm{d}t/2)\})}$ is also assumed to be normalized).
The tangent vector $\ket{T^{\star}(t+\mathrm{d}t/2)}$ denotes the best projection of $-\hat{H} \ket{\Psi(\{A(t+\mathrm{d}t/2)\})}$ on the tangent plane $\mathcal{T}_{\ket{\Psi(\{A(t+\mathrm{d}t/2)\})}}$,
hence their overlap is assumed to be real and relatively large. Considering the previous statement and using the hermiticity of the Hamiltonian $\hat{H}$, it yields for the energy $E(t+\mathrm{d}t)$ the simplified form  

\begin{equation}
 E(t+\mathrm{d}t) = E(t+\mathrm{d}t/2) + (\mathrm{d}t/2)^2 \langle T^{\star}(t+\mathrm{d}t/2) |\hat{H} | T^{\star}(t+\mathrm{d}t/2) \rangle.
 \label{e_tpdt2}
\end{equation}

\noindent
According to Eq.~\eqref{e_tpdt2}, the energy is not strictly conserved due to an additional term 
$(\mathrm{d}t/2)^2 \langle T^{\star}(t+\mathrm{d}t/2) |\hat{H} | T^{\star}(t+\mathrm{d}t/2) \rangle$. However, the real timestep $\mathrm{d}t$ is assumed to be
infinitesimal, \textit{ie.} very small compared to the observation time. Finally, one finds that $E(t+\mathrm{d}t) = E(t+\mathrm{d}t/2) + \mathcal{O}(\mathrm{d}t^2)$.

\subsubsection{Local procedure for the middle-point tensors $\{A(t+\mathrm{d}t/2)\}$}

In the following, we briefly outline the procedure in order to deduce the so-called middle-point tensors $\{A(t+\mathrm{d}t/2)\}$, while working along the lines of
Ref.~\cite{hauke2013} (see also references therein). The latter are computed in such a way that the time-reversal symmetry is conserved during the real-time evolution process. The different steps
consist of

\begin{enumerate}[leftmargin=*]
\item Obtaining a trial $A^{(0)}[R](t+\mathrm{d}t/2)$ from the initial state locally described by $A[R](t)$ using 

\begin{equation}
A^{(0)}[R](t+ \mathrm{d}t/2) = A[R](t) + i (\mathrm{d}t/2) B[R](t).
\end{equation}

\noindent
$\ket{T(t)}$, built from the set of tensors $\{ B[R](t),~R=1,...,L\}$, represents
the best projection of $-\hat{H}\ket{\Psi(\{A(t)\})}$ on the tangent plane $\mathcal{T}_{\ket{\Psi(\{A(t)\})}}$. The $\star$ denoting 'the best approximation'
has been removed for simplicity. 
 
\item Obtaining a trial $B^{(0)}[R](t+\mathrm{d}t/2)$ by finding the best tangent vector for the projection of $-\hat{H} \ket{\Psi(\{A^{(0)}(t+\mathrm{d}t/2)\})}$
on the tangent plane $\mathcal{T}_{\ket{\Psi(\{A^{(0)}(t + \mathrm{d}t/2)\})}}$. 
 
\item Evolving back $A^{(0)}[R](t+\mathrm{d}t/2)$ to $ \tilde{A}^{(0)}[R](t)$ using 

\begin{equation}
\tilde{A}^{(0)}[R](t) = A^{(0)}[R](t+\mathrm{d}t/2) -i(\mathrm{d}t/2)
B^{(0)}[R](t+\mathrm{d}t/2). 
\end{equation}

\noindent
Initially, each tensor $\tilde{A}^{(0)}[R](t)$ will differ from $A[R](t)$.

\item Computing the difference $\mathrm{d} A^{(0)}[R](t) = A[R](t) - \tilde{A}^{(0)}[R](t)$ between the initial tensor $A[R](t)$ and the first-time-evolved-back 
tensor $\tilde{A}^{(0)}[R](t)$. Then project $\mathrm{d} A^{(0)}[R](t)$ onto the tangent space defined at $A^{(0)}[R](t + \mathrm{d}t/2)$.
The latter defines the tensor $\tilde{B}^{(0)}[R](t+\mathrm{d}t/2) = \mathcal{P} \{ \mathcal{T}_{A^{(0)}[R](t + \mathrm{d}t/2)} \} [\mathrm{d} A^{(0)}[R](t)]$. 

\item Computing the inversion error $\epsilon_R^{(0)}$ defined by 

\begin{equation}
\epsilon_R^{(0)} = ||\ket{\Psi(\mathrm{d} A^{(0)}[R](t), A[R](t))}|| = ||\mathrm{d}A^{(0)}[R](t) \partial_{A[R](t)} \ket{\Psi(\{A(t)\})}||.
\end{equation}

\noindent
The lower is $\epsilon_R^{(0)}$, the better is the approximation $\tilde{A}^{(0)}[R](t)$ of $A[R](t)$. 
 
\item The improved estimation of the middle-point tensor $A^{(1)}[R](t+\mathrm{d}t/2)$ is obtained by computing 

\begin{equation}
A^{(1)}[R](t+\mathrm{d}t/2) = A^{(0)}[R](t+\mathrm{d}t/2) + \tilde{B}^{(0)}[R](t+\mathrm{d}t/2). 
\end{equation}

\item Repeating the procedure from (2) to (5) by updating the index for the number of loops in order to converge towards the best approximation
(next loop : $0 \rightarrow 1$ and $1 \rightarrow 2$) and iterate until getting an inversion error $\epsilon_R^{(j)}$ below the required precision 
(typically around $10^{-12}$).
  
\item When the set of tensors $\{A^{(j)}[R](t+\mathrm{d}t/2),R=1,...,L\}$ is obtained, the elementary
evolution step $\mathrm{d}t$ is completed by integrating the many-body quantum state $\ket{\Psi(\{A^{(j)}(t+\mathrm{d}t/2)\})}$ 
by another $\mathrm{d}t/2$ using the following equation

\begin{equation}
A[R](t+\mathrm{d}t) = A^{(j)}[R](t+\mathrm{d}t/2) + i (\mathrm{d}t/2) B^{(j)}[R](t+\mathrm{d}t/2).
\end{equation}

\paragraph{Procedure to compute the tensor $\tilde{B}^{(0)}[R](t+\mathrm{d}t/2)$}
In the following, we briefly discuss how to determine the tensor $\tilde{B}^{(0)}[R](t+\mathrm{d}t/2)$, defined previously (see step 4.)
as $\tilde{B}^{(0)}[R](t+\mathrm{d}t/2) = \mathcal{P} \{ \mathcal{T}_{A^{(0)}[R](t + \mathrm{d}t/2)} \} [\mathrm{d} A^{(0)}[R](t)]$. The latter 
represents the optimal projection on $\mathcal{T}_{A^{(0)}[R](t + \mathrm{d}t/2)}$, the tangent space with respect to $A^{(0)}[R](t + \mathrm{d}t/2)$, of the tensor
$\mathrm{d} A^{(0)}[R](t)$. It permits to obtain a better estimation of the middle-point tensor denoted by $A^{(1)}[R](t+\mathrm{d}t/2)$ and characterized
by the following analytical expression $A^{(1)}[R](t+\mathrm{d}t/2) = A^{(0)}[R](t+\mathrm{d}t/2) + \tilde{B}^{(0)}[R](t+\mathrm{d}t/2)$.\\

First of all, to construct the minimization problem while avoiding any misunderstanding, the previous tensor $\tilde{B}^{(0)}[R](t+\mathrm{d}t/2)$ is denoted by
$\tilde{B}^{(0),\star}[R](t+\mathrm{d}t/2)$ in the following. To deduce the tensor $\tilde{B}^{(0),\star}[R](t+\mathrm{d}t/2)$, we first need to construct
the tangent vector $\ket{\Psi(\tilde{B}^{(0)}[R](t+\mathrm{d}t/2), A^{(0)}[R](t + \mathrm{d}t/2))}$ defined as 

\begin{align}
&\ket{\Psi(\tilde{B}^{(0)}[R](t+\mathrm{d}t/2), A^{(0)}[R](t + \mathrm{d}t/2))} = \tilde{B}^{(0)}[R](t+\mathrm{d}t/2) \partial_{A^{(0)}[R](t + \mathrm{d}t/2)}
\nonumber \\
& \ket{\Psi(\{A^{(0)}(t + \mathrm{d}t/2)\})}.
\end{align}

\noindent
The tensor $\tilde{B}^{(0)}[R](t+\mathrm{d}t/2)$ displays the effective parametrization presented at Fig.~\ref{loc_perturb} whose associated projector
$\tilde{V}^{(0)}[R](t+\mathrm{d}t/2)$ fulfills the relations represented in terms of tensor networks at Figs.~\ref{cond_param} and \ref{orthogonality}.
Here, the orthogonality condition presented at Fig.~\ref{orthogonality} involves the tensor $A^{(0)}[R](t + \mathrm{d}t/2)$ of the MPS 
$\ket{\Psi(\{A^{(0)}(t + \mathrm{d}t/2)\})}$ cast into the isometric gauge, see Fig.~\ref{conditions}. In a second time, the many-body quantum state
$\ket{\Psi(\mathrm{d} A^{(0)}[R](t), A^{(0)}[R](t + \mathrm{d}t/2))}$, depending on the tensor $\mathrm{d} A^{(0)}[R](t)$, has also to be constructed
from $\ket{\Psi(\{A^{(0)}(t + \mathrm{d}t/2)\})}$. The latter may be written as 

\begin{align}
&\ket{\Psi(\mathrm{d} A^{(0)}[R](t), A^{(0)}[R](t + \mathrm{d}t/2))} = \mathrm{d} A^{(0)}[R](t) \partial_{A^{(0)}[R](t + \mathrm{d}t/2)} \nonumber \\
& \ket{\Psi(\{A^{(0)}(t + \mathrm{d}t/2)\})}.
\end{align}

\noindent
Finally, to deduce $\tilde{B}^{(0),\star}[R](t+\mathrm{d}t/2)$, we solve the following minimization problem 

\begin{align}
&\tilde{B}^{(0),\star}[R](t+\mathrm{d}t/2) := \mathrm{argmin}(||\ket{\Psi(\tilde{B}^{(0)}[R](t+\mathrm{d}t/2), A^{(0)}[R](t + \mathrm{d}t/2))} \nonumber \\
& - \ket{\Psi(\mathrm{d} A^{(0)}[R](t), A^{(0)}[R](t + \mathrm{d}t/2))} ||^2 ). 
\label{mini_btilde}
\end{align}

\noindent
Since the matrix $\tilde{X}^{(0)}[R](t+\mathrm{d}t/2)$ corresponds to the free parameter of the tensor $\tilde{B}^{(0)}[R](t+\mathrm{d}t/2)$, this minimization problem 
is reduced to find the matrix $\tilde{X}^{(0),\star}[R](t+\mathrm{d}t/2)$, \textit{ie.} the best approximation for $\tilde{X}^{(0)}[R](t+\mathrm{d}t/2)$. According to 
Eq.~\eqref{mini_btilde}, $\tilde{X}^{(0),\star}[R](t+\mathrm{d}t/2)$ is characterized by the following equation 

\begin{align}
& \partial_{\tilde{X}^{(0),*}[R](t+\mathrm{d}t/2)} \langle \Psi(\tilde{B}^{(0)}[R](t+\mathrm{d}t/2), A^{(0)}[R](t + \mathrm{d}t/2)) | \Psi(\mathrm{d} 
A^{(0)}[R](t), A^{(0)}[R](t + \mathrm{d}t/2)) \rangle \nonumber  \\ 
& = \tilde{X}^{(0)}[R](t+\mathrm{d}t/2) .
\label{tildeX}
\end{align}

\noindent
Equation.~\eqref{tildeX} is represented in terms of tensor networks on Fig.~\ref{tildexfig}. 
\end{enumerate}

\begin{figure}[!h]
\centering
\begin{tabular}{c}
\includegraphics[scale = 0.34]{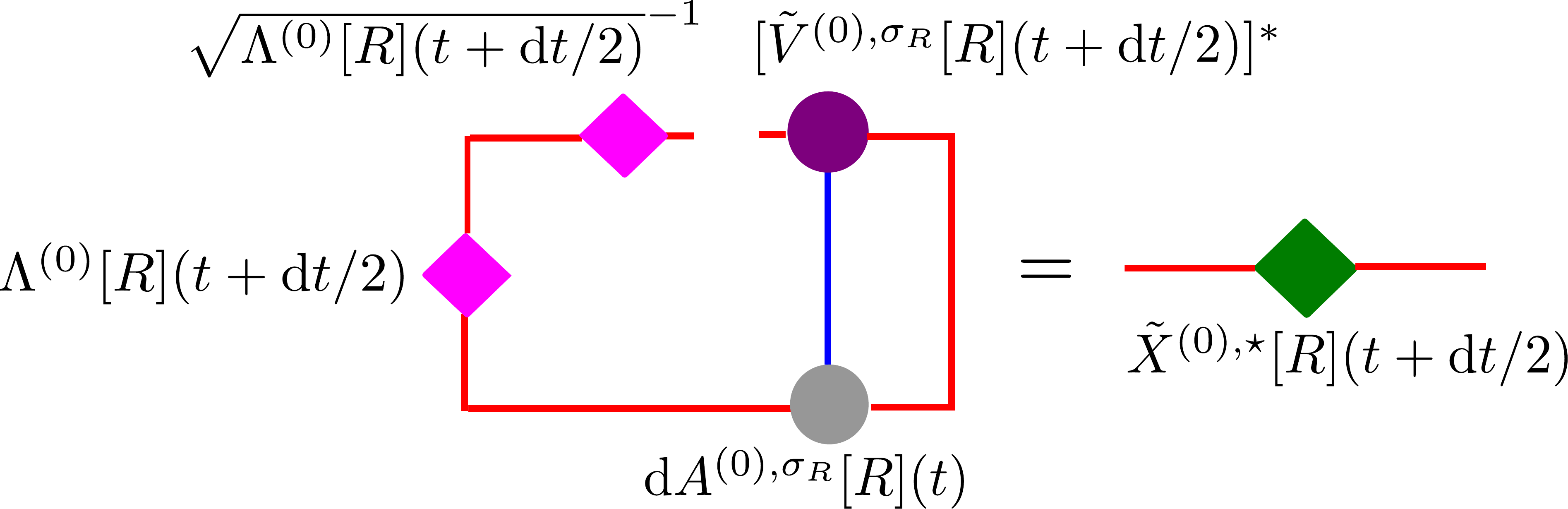}  
\end{tabular}
\caption{Graphical representation using tensor networks of the condition at Eq.~\eqref{tildeX}. The latter allows us to find the optimal matrix 
$\tilde{X}^{(0),\star}[R](t+\mathrm{d}t/2)$ in order to fully characterize the tensor $\tilde{B}^{(0),\star}[R](t+\mathrm{d}t/2)$.} 
\label{tildexfig}
\end{figure}     

\section{Correlation spreading in the $z$ polarized phase}
The purpose of the following sections is to shed new light on the spreading of information in long-range lattice models. 
To do so, we study theoretically and numerically the spreading of correlations and entanglement while focusing on a specific long-range lattice model.
The latter is the paradigmatic, one-dimensional, transverse Ising model with a long-range spin exchange of the 
form $1/R^\alpha$. Its main properties are recalled in the following. The numerical calculations to deduce both the static (ground state, ground state energy) and
dynamical properties (time-evolved quantum state) are performed using the time-dependent matrix product state (t-MPS) approach within
the time-dependent variational principle (TDVP) presented at Sec.~\ref{TDVP_section}. The long-range interactions in the Hamiltonian of the transverse Ising model
are implemented as a sum of decaying exponentials following the method discussed in Ref.~\cite{pirvu2010} and in Sec.~\ref{TDVP_section}.
Concerning the theoretical calculations, they rely on the Holstein-Primakoff transformation and the bosonic Bogolyubov theory defining the linear spin wave theory.
Note that the $s=1/2$ long-range transverse Ising (LRTI) chain has already been
considered in Chap.~\ref{ch:3-universal_scaling_laws}. The latter has been studied theoretically to provide a proof of the scaling laws 
(deduced from the generic form at Eq.~\eqref{generic_form}) for the correlation spreading induced by sudden global quenches confined in the quasi-local regime of gapped quantum phases. \\

In the following, we first investigate the same problem but from a numerical point of view using the t-MPS + TDVP approach. The last section of this chapter is devoted 
to the counterpart, \textit{ie.} to a numerical study of the correlation spreading induced by sudden global quenches confined in the quasi-local regime of gapless
quantum phases. To do so, we consider the exact same example than the one presented at Chap.~\ref{ch:3-universal_scaling_laws} in Subsec.~\ref{xy_gapless} for the theoretical
study. The discussion on the correlation spreading in the 1D LRTI model is extended to the case of sudden global quenches confined in the local regime of the $z$ 
polarized gapped quantum phase. Then, while still considering the local regime of the $z$ polarized phase, we turn to an investigation of the spreading of both the local spin
(\textit{ie.} the local magnetization) and entanglement (\textit{via} an investigation of several R\'enyi entropies corresponding to different 
entropy measures) for sudden local quenches. \\

The $s=1/2$ long-range transverse Ising (LRTI) chain is governed by the following Hamiltonian ($\hbar = 1$)

\begin{equation}\label{H}
 \hat{H} = \sum_{R < R'} \frac{2J}{|R-R'|^{\alpha}} \hat{S}^x_R \hat{S}^x_{R'} - 2h \sum_R \hat{S}^z_R,
\end{equation}

\noindent
where $\hat{S}_R^{j}$ ($j \in \{x,y,z\}$) refers to the $s=1/2$ spin operator acting on the lattice site $R \in\mathbb{Z}$,
$J>0$ corresponds to the exchange energy (antiferromagnetic interaction along the $x$ axis of the Bloch sphere), and $h>0$ denotes the constant 
and homogeneous transverse field along the $z$ axis.
It is the long-range counterpart of the celebrated quantum transverse Ising model, which has been extensively studied in quantum field theory~\cite{sachdev2001}.
Note that the LRTI model can be simulated on a variety of plateforms, including cold Rydberg gases~\cite{browaeys2016,lukin2000,zoller2001} and 
artificial ion crystals, where the power-law exponent $\alpha$ can be controlled via light-mediated interactions, see 
Refs.~\cite{deng2005,lanyon2011, NaturePhysicsInsight2012blatt,schneider2012b,jurcevic2014,richerme2014}.\\

\begin{figure}[t!]
\centering
\includegraphics[scale = 0.55]{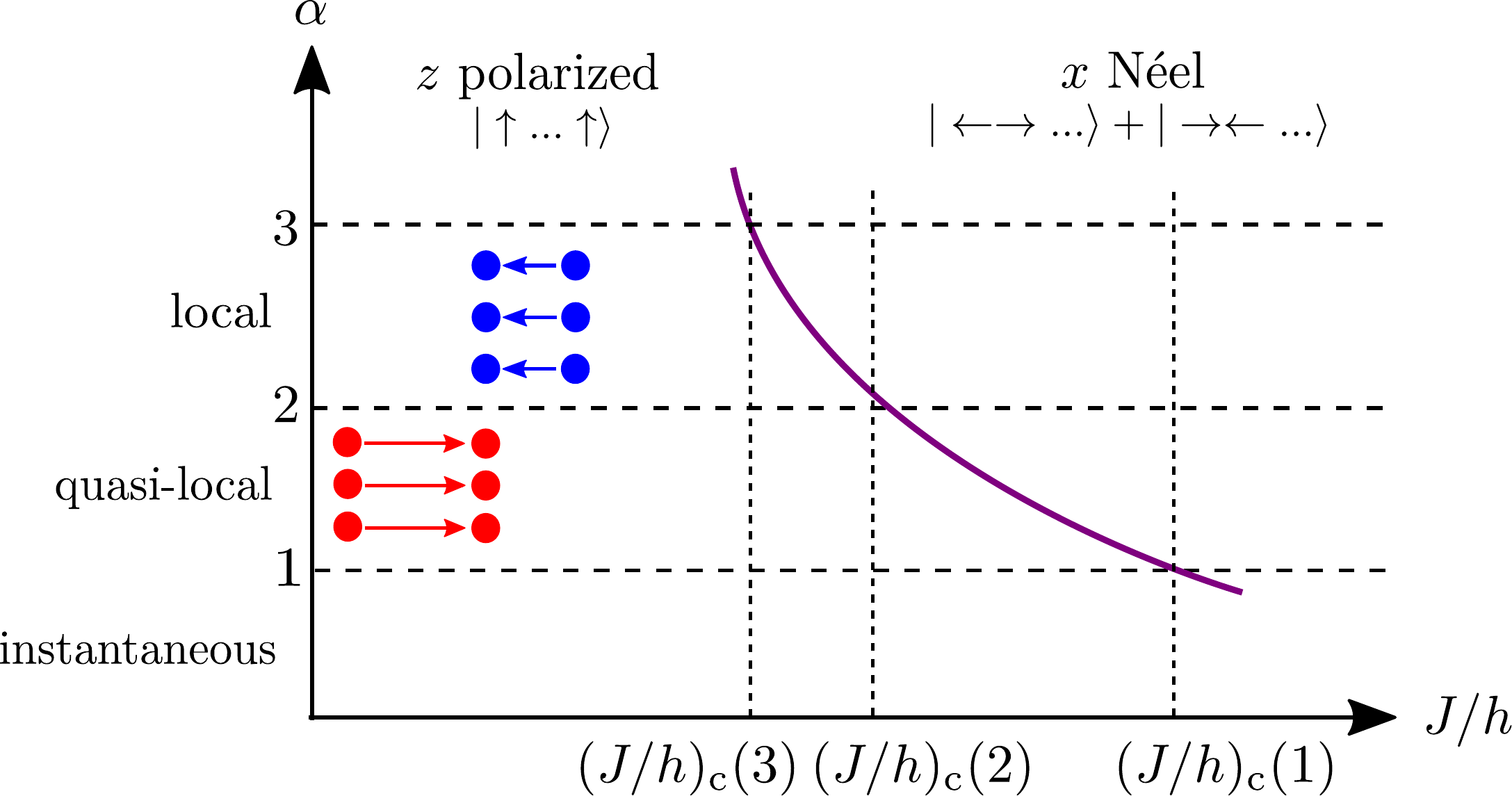}
\caption{\label{fig:phase_diagram}
Schematic phase diagram of the LRTI chain versus the exchange-to-field ratio $J/h$ and the power-law exponent $\alpha$. The critical line separating the two 
gapped quantum phases, namely the $z$ polarized and $x$ N\'eel phases, is represented by the solid purple line. The arrows indicate various sudden global quenches
considered in this chapter. The blue arrows represent sudden global quenches confined in the local regime of the $z$ polarized phase [requiring $\alpha \geq 2$ and 
$(J/h)_{\mathrm{i}},(J/h)_{\mathrm{f}} < (J/h)_\textrm{c}(\alpha)$] whereas the red arrows correspond to sudden global quenches confined in the quasi-local regime of the
$z$ polarized phase [$1 \leq \alpha < 2$ and $(J/h)_{\mathrm{i}},(J/h)_{\mathrm{f}} < (J/h)_\textrm{c}(\alpha)$].}
\end{figure}

Its equilibrium phase diagram is depicted on Fig.~\ref{fig:phase_diagram}, see also Ref.~\cite{koffel2012}.
It comprises two gapped phases separated by a second order quantum phase transition along some critical line $(J/h)_\textrm{c} (\alpha)$.
For large transverse fields $h$, the spin-field interaction dominates and a $z$-polarized phase is formed,
where all the spins predominantly point in the $z$ direction. The ground state of such gapped phase is characterized by $\ket{\Psi_{\mathrm{gs}}} \simeq \ket{\uparrow ... \uparrow}$.
For low values of $h$, the long-range antiferromagnetic couplings dominate and the system forms a staggered N{\'e}el-ordered phase aligned in the $x$ direction. The 
corresponding ground state is thus given by $\ket{\Psi_{\mathrm{gs}}} \simeq \sqrt{2}^{-1}(\ket{\leftarrow \rightarrow ... \leftarrow \rightarrow} + 
\ket{\rightarrow \leftarrow ... \rightarrow \leftarrow})$. The local quantum states $\ket{\leftarrow}_R$ and $\ket{\rightarrow}_R$ are defined by 
$\hat{S}^{x}_R\ket{\rightarrow}_R = (1/2)\ket{\rightarrow}_R$ and $\hat{S}^{x}_R \ket{\leftarrow}_R = (-1/2)\ket{\leftarrow}_R$. Hence, $\ket{\rightarrow}_R = 
\sqrt{2}^{-1}(\ket{\uparrow}_R+\ket{\downarrow}_R)$ and $\ket{\leftarrow}_R = \sqrt{2}^{-1}(\ket{\uparrow}_R - \ket{\downarrow}_R)$. \\

Note that the critical point $(J/h)_\textrm{c}$ for the short-range transverse Ising chain [implying $\alpha \rightarrow +\infty$ for the Hamiltonian $\hat{H}$ at Eq.~\eqref{H}] is always 
smaller than the one for the long-range version. Indeed, the long-range antiferromagnetic interactions along the $x$ axis ($\alpha$ finite)
lead to a frustration of the 1D LRTI model along the $x$ axis. More precisely, the smaller the long-range interactions decay (smaller $\alpha$), the more the spin model
is frustrated along the $x$ axis. Hence, it requires smaller transverse fields $h$ for the LRTI chain to enter its $z$ polarized phase. The previous statement explains
that $(J/h)_\textrm{c}$ increases when $\alpha$ decreases, see Fig.~\ref{fig:phase_diagram}. \\

In the $z$-polarized phase, the Hamiltonian of the LRTI chain at Eq.~\eqref{H} can be diagonalized using a Holstein-Primakoff transformation and the bosonic 
Bogolyubov theory (see Chap.~\ref{ch:3-universal_scaling_laws} and Appendix.~\ref{appendix_gx_lrti}). The low-energy quasiparticles consists of spin-flips 
characterized by the following gapped excitation spectrum $E_k = 2\sqrt{h[h+JP_{\alpha}(k)]}$. $P_{\alpha}(k)$ corresponds to the 
Fourier transform of the long-range potential $1/R^{\alpha}$ and $k$ the quasimomentum confined in the first Brillouin zone $[-\pi,\pi]$ ($a$ the lattice 
spacing is fixed to unity). The linear spin wave  theory (LSWT) predicts three dynamical regimes, see Refs.~\cite{hauke2013,cevolani2015,cevolani2016} and 
Eq.~\eqref{ek_lrti_infrared}. For $\alpha \geq 2$, the quasiparticle dispersion relation is regular in the first Brillouin zone with bounded energies 
$E_k$ and group velocities $V_{\mathrm{g}}(k) = \partial_k E_k$ ($\hbar=1$). The previous properties define the local regime of the $z$ polarized phase. 
This case belongs to the same universal class as the short-range transverse Ising model and the dynamics is characterized by the emergence of a linear causality cone~\cite{lieb1972,hastings2006}.
This expected linear causality cone will be fully characterized in the next section. For $\alpha<1$, both the excitation spectrum $E_k$ and the group velocity
$V_{\mathrm{g}}(k)$ feature an algebraic, infrared divergence defining the instantaneous regime. Indeed, for this specific quantum regime, there is no characteristic time and the spreading is
instantaneous. Finally, for $1 \leq \alpha<2$, the quasiparticle energy $E_k$ is bounded, however, the group velocity $V_{\mathrm{g}}(k)$ diverges as $V_{\mathrm{g}}(k) \sim |k|^{z-1}$ with $z=\alpha-1$, see 
Eq.~\eqref{ek_lrti_infrared}. The latter case corresponds to the quasi-local regime. Whether some form of causality emerges in this specific regime remains strongly
debated~\cite{hauke2013,jurcevic2014,richerme2014,cevolani2015, maghrebi2016,cevolani2016,buyskikh2016}. In Chap.~\ref{ch:3-universal_scaling_laws},
relying on the generic form for the equal-time connected correlation functions at Eq.~\eqref{generic_form}, we have shown that an algebraic twofold
causality cone should emerge. More precisely, for sudden global quenches confined in a gapped quantum phase, the correlation edge (CE) should spread sub-ballistically
and the series of local extrema ballistically. In the following, we start by focusing numerically on this case before investigating the local regime. \\

All the numerical results presented below are obtained using matrix-product state (t-MPS) simulations within the time-dependent variational 
principle (TDVP), see Refs.~\cite{schollwock2005,schollwock2011,haegeman2012,koffel2012}. A careful analysis of the bond dimension has been systematically performed to 
certify the convergence of the t-MPS results in all cases. Note that contrary to the short-range Bose-Hubbard chain, fixing the dimension of the local Hilbert
space is straightforward for the LRTI chain. Indeed, since $s=1/2$ spins are considered $\mathrm{dim}(\mathbb{H}_R) = 2$ $\forall R \in [|1,L|]$ where $\mathbb{H}_R$
refers to the local Hilbert space for the lattice site $R$ ($R$-th spin). The associated local basis is given by $\{\ket{\uparrow},\ket{\downarrow} \}$.
Besides, in the following sections, we study the dynamical behaviour of several observables that are all experimentally accessible.

\subsection{The quasi-local regime}
\label{qlr_ising}

In this section, we consider the spreading of spin correlations in the gapped $z$ polarized phase. We first investigate the quasi-local
regime implying $1 \leq \alpha < 2$, see the red solid lines at Fig.~\ref{ek_vg_lrti} for the shape of the excitation spectrum $E_k$ and the group
velocity $V_{\mathrm{g}}(k)$ as a function of the quasimomentum $k$ in the first Brillouin zone in the quasi-local regime of the $z$ polarized phase at $\alpha = 1.7$
and $J/h = 0.1$. Similarly to the theoretical investigation presented in Chap.~\ref{ch:3-universal_scaling_laws}, we study the spin-spin
correlations perpendicular to the polarization axis of the gapped quantum phase considered for the sudden global quench. In other words, the spin 
fluctuations are studied along the $x$ axis \textit{via} 
$G_x$ the connected spin-spin correlation function defined by $G_x(R,t) = G_{x,0}(R,t) - G_{x,0}(R,0)$ where

\begin{equation}
G_{x,0}(R,t) = \langle \hat{S}_R^x(t) \hat{S}_0^x(t) \rangle - \langle \hat{S}_R^x(t)\rangle \langle \hat{S}_0^x(t) \rangle.
\end{equation}

\noindent
In the following, sudden global quenches on the exchange-to-field ratio $J/h$ such that one-dimensional long-range transverse Ising model is confined 
in the $z$ polarized phase are considered. The exchange-to-field ratios $(J/h)_{\mathrm{i}}$ and $(J/h)_{\mathrm{f}}$ will refer to the pre- and 
post-quench interaction parameters respectively. \\

\begin{figure}[t!]
\centering
\includegraphics[scale = 0.36]{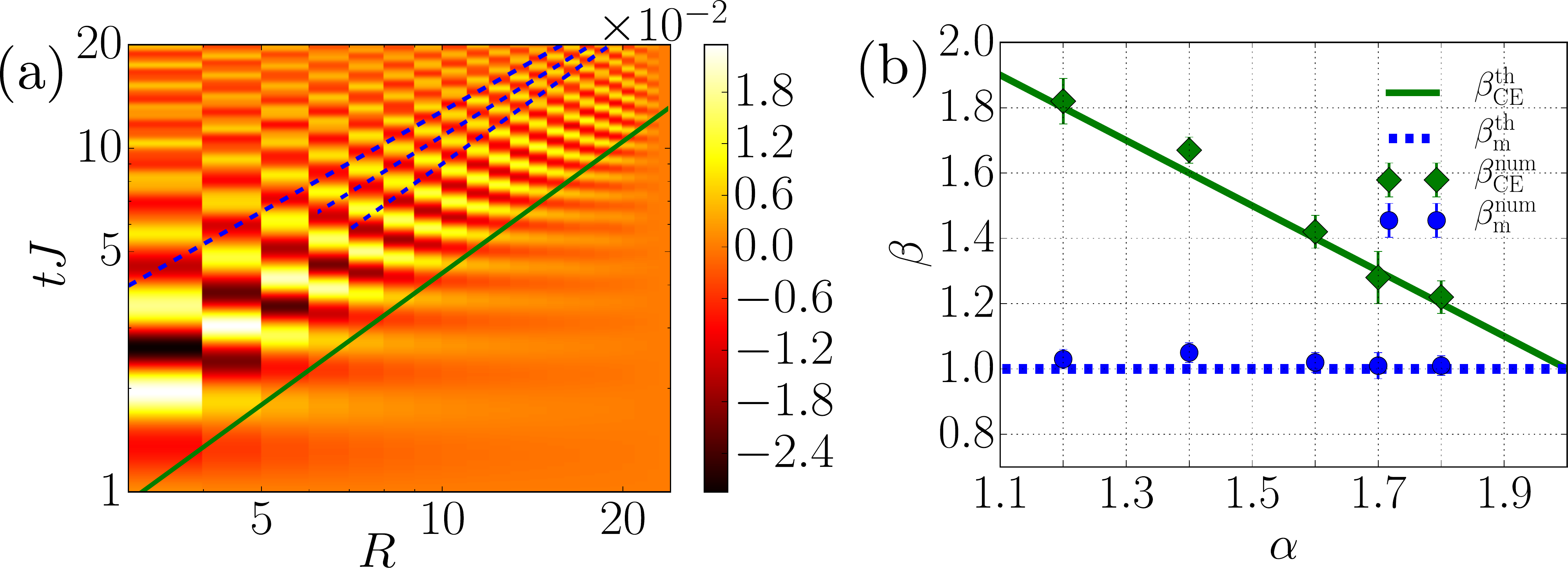}
\caption{
Spreading of the equal-time connected spin-spin correlation function $G_x(R,t)=G_{x,0}(R,t)-G_{x,0}(R,0)$ with 
$G_{x,0}(R,t) = \langle \hat{S}_R^x(t) \hat{S}_0^x(t) \rangle - \langle \hat{S}_R^x(t)\rangle \langle \hat{S}_0^x(t) \rangle$ for the LRTI chain
in the quasi-local regime of the $z$ polarized phase. (a)~ t-MPS calculation of $G_x$ for a sudden global quench from $(J/h)_{\mathrm{i}} = 2 \times 10^{-2}$ 
to $(J/h)_\mathrm{f} = 1$ at $\alpha = 1.7$ (log-log scale for both axis). For the 1D LRTI model, at $\alpha = 1.7$, the critical point separating the $z$ polarized phase from the
$x$ N\'eel phase is located at $J/h \simeq 3$ \cite{koffel2012}. The space-time spin-spin correlations feature a twofold algebraic structure 
(twofold linear structure in log-log scale) with a sub-ballistic correlation edge ($\beta_{\mathrm{CE}}^{\mathrm{num}} > 1$, see solid green line) and 
ballistic local extrema ($\beta_{\mathrm{m}}^{\mathrm{num}} \simeq 1$, see dashed blue lines).
(b)~ Evolution of $\beta_{\mathrm{CE}}^{\mathrm{num}}$ and its corresponding theoretical value $\beta_{\mathrm{CE}}^{\mathrm{th}} = 3-\alpha$ characterizing the spreading
of the correlation edge with $\beta_{\mathrm{m}}^{\mathrm{num}}$ and $\beta_{\mathrm{m}}^{\mathrm{th}} = 1$ for the spreading of the series of local extrema as a function
of the power-law exponent $\alpha$ defining the decay of the long-range spin exchange couplings along the $x$ axis. Figures extracted from Ref.~\cite{despres2019bis}.}
\label{fig:SpinCorr}
\end{figure}

Figure~\ref{fig:SpinCorr}(a) shows a typical result of a t-MPS calculation for the connected correlation function $G_x$ for a quench in the quasi-local 
regime of the $z$ polarized phase. The sudden global quench is defined by $\alpha=1.7$ from $(J/h)_\textrm{i}=2 \times 10^{-2}$ to $(J/h)_\textrm{f}=1$. Hence, the
pre- and post-quench Hamiltonians are well confined in the quasi-local regime of the $z$ polarized phase, see red arrows on Fig.~\ref{fig:phase_diagram}. As expected from the theoretical results
discussed in Chap.~\ref{ch:3-universal_scaling_laws}, the correlation pattern is characterized by an algebraic causality cone. More precisely, this causality cone 
displays two different algebraic structures. \\
(i) A correlation edge (CE) separating the causal region from the non-causal one for the spin correlations and spreading sub-ballistically with the scaling law $t \sim R^{\beta_{\mathrm{CE}}}$, 
$\beta_{\mathrm{CE}}>1$. On Fig.~\ref{fig:SpinCorr}(a), the motion of the CE has been determined by tracking the activation time $t^*$ for different spin separation distances $R$ 
where the correlations reach a same fraction $\epsilon$ of the maximal amplitude of the spin fluctuations (see solid green line). Then, by computing the slope of the solid
green line, one can extract the numerical CE exponent $\beta_{\mathrm{CE}}^{\mathrm{num}}$. For $\alpha = 1.7$, we find $\beta_{\mathrm{CE}}^{\mathrm{num}} = 1.28 \pm 0.08$.
This numerical exponent is in very good agreement with the theoretical value $\beta_{\mathrm{CE}}^{\mathrm{th}} = 3-\alpha = 1.3 > 1$ found using the LSWT, see 
Chap.~\ref{ch:3-universal_scaling_laws} and Ref.~\cite{cevolani2018} for more details. \\
(ii) A series of local extrema in the vicinity of the CE spreading ballistically, and thus implying the scaling law $t\sim R^{\beta_{\mathrm{m}}}$ with 
$\beta_{\mathrm{m}} \simeq 1$ (see Ref.~\cite{cevolani2018} and Chap.~\ref{ch:3-universal_scaling_laws}).
In the long-time and long-distance limit, several local extrema have been tracked [see dashed blue lines on Fig.~\ref{fig:SpinCorr}(a); notice the log-log scale].
We find $\beta_{\mathrm{m}}^{\mathrm{num}} = 1.01 \pm 0.04$ in very good agreement with the theoretical exponent $\beta_{\mathrm{m}}^{\mathrm{th}} = 1$. 
As pointed out in Refs.~\cite{cevolani2018,despres2019}, the motion of the local extrema is significantly different from the one of the CE. In other words, the scaling law
for the spreading of the extrema does not characterize at all the one associated to the correlation edge (CE).\\

\begin{figure}[t!]
\centering
\includegraphics[scale = 0.47]{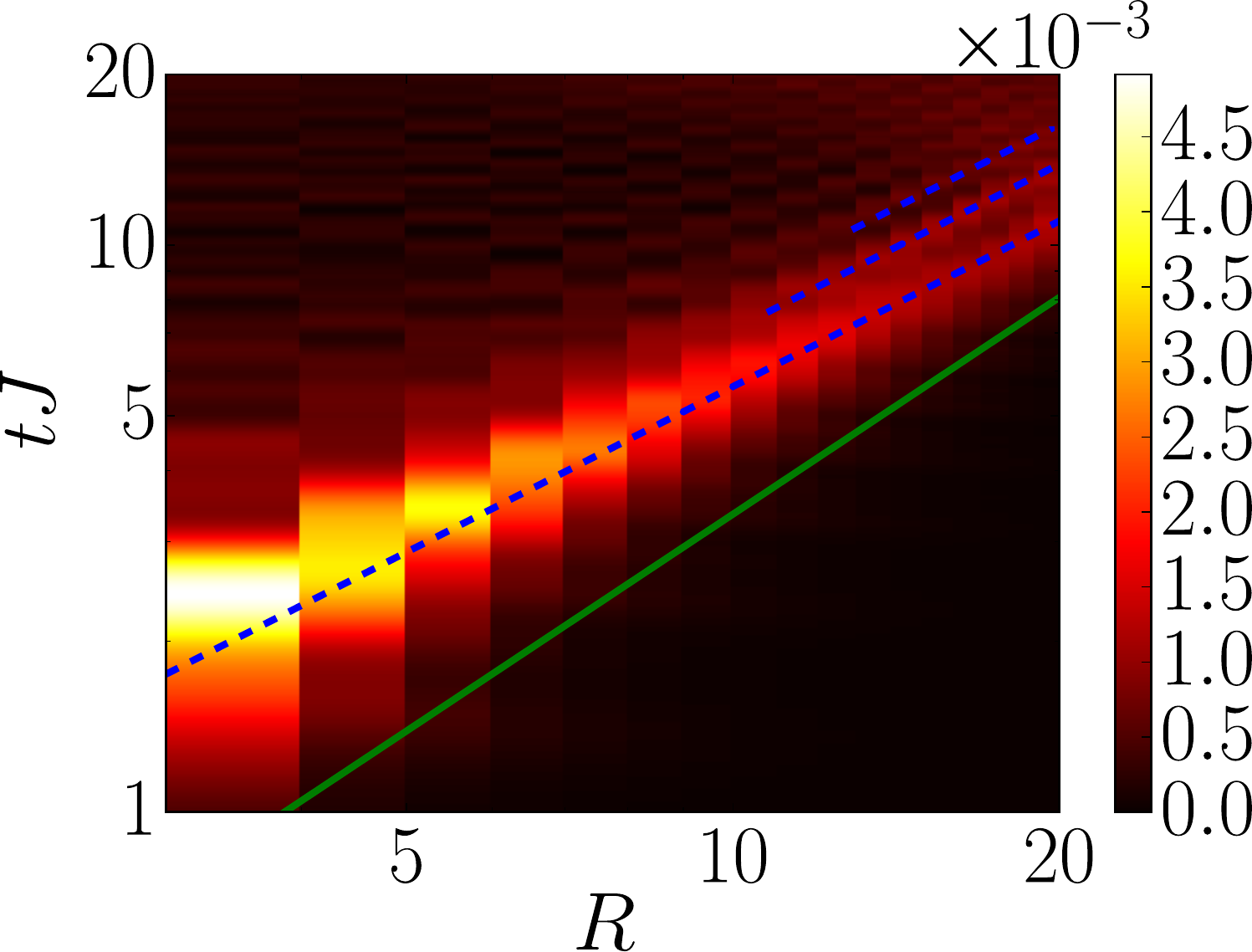}
\caption{Spreading of the equal-time connected spin-spin correlation function $G_z(R,t)=G_{z,0}(R,t)-G_{z,0}(R,0)$ with 
$G_{z,0}(R,t) = \langle \hat{S}_R^z(t) \hat{S}_0^z(t) \rangle - \langle \hat{S}_R^z(t)\rangle \langle \hat{S}_0^z(t) \rangle$ for the LRTI chain
in the quasi-local regime of the $z$ polarized phase. t-MPS calculation of $G_z$ for a sudden global quench from $(J/h)_{\mathrm{i}} = 2 \times 10^{-2}$ 
to $(J/h)_\mathrm{f} = 1$ at $\alpha = 1.7$ (log-log scale for both axis). Similar to $G_x$, the space-time spin-spin correlations along the $z$ axis feature 
a twofold algebraic structure (twofold linear structure in log-log scale) with a sub-ballistic correlation edge ($\beta_{\mathrm{CE}}^{\mathrm{num}} = 1.23 \pm 0.04 \simeq 3-\alpha$, see 
solid green line) and ballistic local extrema ($\beta_{\mathrm{m}}^{\mathrm{num}} = 0.98 \pm 0.02 \simeq 1$, see dashed blue lines). Figure extracted from Ref.~\cite{despres2019bis}.}
\label{fig:gz_qlr_z_pol}
\end{figure}

On Fig.~\ref{fig:SpinCorr}(b), different sudden global quenches confined in the quasi-local regime of the $z$ polarized phase have been considered. They
are defined by a same pre- [$(J/h)_{\mathrm{i}} = 2 \times 10^{-2}$] and post-quench [$(J/h)_{\mathrm{f}} = 1$] interaction parameter while the power-law exponent 
$\alpha$ scans the interval $[1,2[$. The latter corresponds to the condition to remain in the quasi-local regime of the $z$ polarized phase. The exponents $\beta_{\mathrm{CE}}^{\mathrm{num}}$ and $\beta_{\mathrm{m}}^{\mathrm{num}}$
are extracted using the same techniques as previously. We find that while the maxima spread ballistically ($\beta_{\mathrm{m}}^{\mathrm{num}} 
\simeq \beta_{\mathrm{m}}^{\mathrm{th}} = 1$), the CE is sub-ballistic ($\beta_{\mathrm{CE}}^{\mathrm{num}} \simeq \beta_{\mathrm{CE}}^{\mathrm{th}} = 
3 - \alpha >1$ for $1\leq \alpha < 2$). Note that for each algebraic structure defining the causality cone, the maximal relative error between the theoretical
and numerical exponents $\beta$ is around $5\%$. The previous results confirm the emergence of a slower-than-ballistic form of causality and an inner structure
characterized by local extrema in the vicinity of the CE spreading ballistically. A similar behavior of the causality structure \footnote{By similar behavior, we mean that the causality structure is defined by a
sub-ballistic motion of the CE and a ballistic spreading of the local extrema characterized by the spreading exponent $\beta_{\mathrm{CE}} = 3-\alpha$ and $\beta_{\mathrm{m}} = 1$ respectively.} has been found for another observable,
namely the $G_z$ spin-spin correlation function along the $z$ axis, \textit{ie.} along the polarization axis of the gapped quantum phase considered here, see Fig.~\ref{fig:gz_qlr_z_pol} for
a numerical example of the space-time pattern of $G_z$. However, this observable is not discussed in details here, see Ref.~\cite{despres2019bis} for more information. Nevertheless, 
the derivation of the analytical expression of $G_z$ for a sudden global quench confined in the $z$ polarized phase (valid both in the 
local and quasi-local regimes) is provided at Appendix.~\ref{appendix_gz_lrti}. 

\subsection{The local regime}
\label{local_regime_ising}

We now turn to the local regime of the $z$ polarized phase. The latter is characterized by a finite excitation spectrum $E_k$ and group velocity
$V_{\mathrm{g}}(k)$ in the first Brillouin zone ($k \in [-\pi,\pi]$ with $a$ the lattice spacing fixed to unity). Hence, the power-law exponent $\alpha$
has to fulfill the following condition $\alpha \geq 2$ (see the black dashed lines at Fig.~\ref{ek_vg_lrti} for the shape of the excitation spectrum $E_k$ and the group
velocity $V_{\mathrm{g}}(k)$ as a function of the quasimomentum $k$ in the first Brillouin zone in the local regime of the $z$ polarized phase at $\alpha = 2.7$
and $J/h = 0.1$). As previously, we study the space-time spin fluctuations along the $x$ axis \textit{via} the equal-time connected correlation function 
$G_x(R,t)$ and the sudden global quenches, confined in the local regime of the $z$ polarized phase, are performed on the interaction parameter $J/h$ 
while the power-law exponent $\alpha \geq 2$ is fixed. \\

In the following, we recall the main theoretical results concerning the correlation spreading for sudden global quenches confined in the local regime of a gapless 
or gapped quantum phase of long-range interacting lattice models. In Chap.~\ref{ch:3-universal_scaling_laws}, by relying on the generic form for the equal-time connected correlation functions 
presented at Eq.~\eqref{generic_form}, we unveiled a twofold linear causality cone. Once again, the latter is characterized by a CE and a series of local extrema 
whose spreading velocities ($V_{\mathrm{CE}}$ and $V_{\mathrm{m}}$) are reminiscent of those for the case of short-range interacting lattice models, see 
Ref.~\cite{despres2019bis}. Indeed, similarly to short-range lattice models, one can find a quasimomentum $k^*$ such that $V_{\mathrm{g}}^* = V_{\mathrm{g}}(k^*) 
= \mathrm{max}[V_{\mathrm{g}}(k)]$, see Eq.~\eqref{kstar}. Hence, the generic equal-time connected correlation function at Eq.~\eqref{generic_form} has an
asymptotic behavior defined at Eq.~\eqref{spa}. Consequently, the scaling laws and the spreading velocities of the twofold causality cone of correlations
are the same as those for short-range interacting lattice models. Both the CE and the series of local extrema spread ballistically ($\beta_{\mathrm{CE}} = 
\beta_{\mathrm{m}} = 1$) with a velocity $V_\mathrm{CE} = 2V_{\mathrm{g}}^*$ and $V_{\mathrm{m}} = 2V_{\varphi}^*$ respectively. In general $V_\mathrm{g}^*
\neq V_{\varphi}^*$, leading to $V_{\mathrm{CE}} \neq V_{\mathrm{m}}$. The latter is expected for quantum lattice models having a non-phononic quasiparticle dispersion relation~\cite{cevolani2018,despres2019}. \\

\begin{figure}[t!]
\centering
\includegraphics[scale = 0.37]{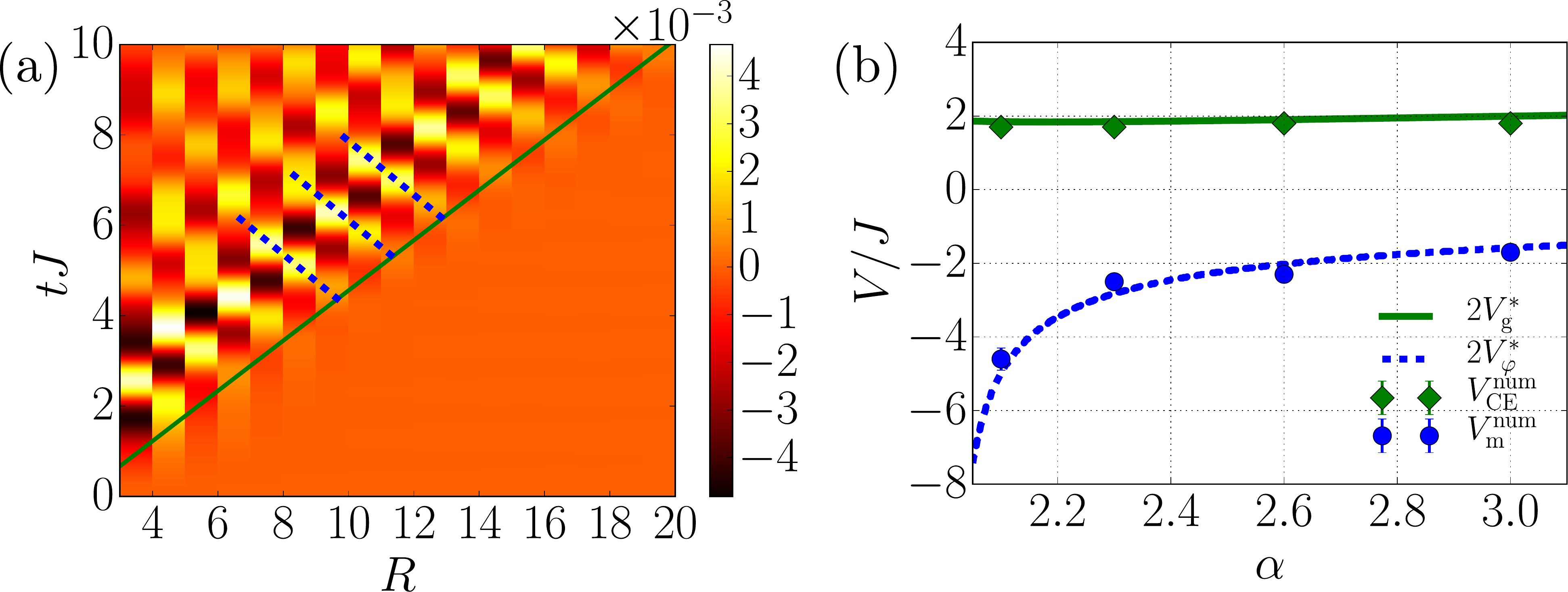}
\caption{
Spreading of the equal-time connected spin-spin correlation function $G_x(R,t)=G_{x,0}(R,t)-G_{x,0}(R,0)$ with 
$G_{x,0}(R,t) = \langle \hat{S}_R^x(t) \hat{S}_0^x(t) \rangle - \langle \hat{S}_R^x(t)\rangle \langle \hat{S}_0^x(t) \rangle$ for the LRTI chain
in the local regime of the $z$ polarized phase. (a)~ t-MPS calculation of $G_x$ for a sudden global quench 
from $(J/h)_{\mathrm{i}} = 1.1$ to $(J/h)_\mathrm{f} = 1$ at $\alpha = 3$ (linear scale for both axis). The space-time spin fluctuations along the $x$ axis 
feature a twofold causality cone. Both the correlation edge (see solid green line) and the series of local extrema (see dashed blue lines) propagate ballistically
($\beta_{\mathrm{CE}}^{\mathrm{num}} = \beta_{\mathrm{m}}^{\mathrm{num}} = 1$) with different velocities. The correlation edge is characterized by the spreading
velocity $V_{\mathrm{CE}}^{\mathrm{num}}>0$ and the local extrema by $V_{\mathrm{m}}^{\mathrm{num}} < 0$. (b)~ Evolution of $V_{\mathrm{CE}}^{\mathrm{num}}$
and the corresponding theoretical value $V_{\mathrm{CE}}^{\mathrm{th}} = 2V_{\mathrm{g}}^*$ (twice the maximal group velocity) characterizing the spreading
of the correlation edge together with $V_{\mathrm{m}}^{\mathrm{num}}$ and $V_{\mathrm{m}}^{\mathrm{th}} = 2 V_{\varphi}^*$ (twice the phase velocity at the quasimomentum 
$k^*$ where the group velocity is maximal) for the spreading of the series of local extrema as a function of the power-law exponent $\alpha \geq 2$ (since the local 
regime of the $z$ polarized phase is considered) defining the decay of the long-range spin exchange couplings along the $x$ axis. Figures extracted 
from Ref.~\cite{despres2019bis}.}
\label{fig:SpinCorrBis}
\end{figure}

Figure~\ref{fig:SpinCorrBis}(a) shows a t-MPS result for the equal-time connected correlation function $G_x$ for a sudden global quench in the local 
regime of the $z$ polarized phase. The spin-spin correlations are plotted as a function of the dimensionless time $tJ$ ($\hbar = 1$) and the distance $R$.
The quench is defined by $\alpha = 3 > 2$ from $(J/h)_\textrm{i} = 1.1$ to $(J/h)_\textrm{f} = 1$. Hence, the pre- and post-quench Hamiltonians are well confined in the $z$ polarized phase, see blue arrows on Fig.~\ref{fig:phase_diagram}.
As expected from the theoretical results presented at Chap.~\ref{ch:3-universal_scaling_laws} and recalled above, the space-time pattern for the $G_x$ spin 
fluctuations is characterized by a twofold linear causality cone, \textit{ie.} a CE and a series of local maxima spreading ballistically [note the lin-lin scale
on Fig.~\ref{fig:SpinCorrBis}(a)]. The latter implies the scaling law $t\sim R^{\beta}$ with $\beta = \beta_{\mathrm{CE}} = \beta_{\mathrm{m}} = 1$. 
On Fig.~\ref{fig:SpinCorrBis}(a), the CE separating the causal region from the non-causal one for the spin correlations propagates at the velocity $V_{\mathrm{CE}}^{\mathrm{num}}
= (1.8 \pm 0.2)J$ in good agreement with the theoretical prediction $V_{\mathrm{CE}} = 2V_{\mathrm{g}}^* \simeq 2J$ (see solid green line). Besides, in the vicinity
of the CE, a series of local extrema having also a ballistic scaling law spread at the velocity $V_{\mathrm{m}}^{\mathrm{num}} = (-1.7 \pm 0.2)J$ (see dashed blue lines).
This numerical spreading velocity for the local extrema is in good agreement with the theoretical velocity $2V_{\varphi}^* \simeq -1.57J$. \\

On Fig.~\ref{fig:SpinCorrBis}(b), different sudden global quenches confined in the local regime of the $z$ polarized phase have been considered. They
are defined by a same pre- [$(J/h)_{\mathrm{i}} = 1.1$] and post-quench [$(J/h)_{\mathrm{f}} = 1$] interaction parameter while the power-law exponent 
$\alpha$ scans the interval $[2,3]$. The spreading velocities $V_{\mathrm{CE}}^{\mathrm{num}}$ and $V_{\mathrm{m}}^{\mathrm{num}}$ are extracted using the same techniques as previously. For the CE,
one tracks the different activation times $t^*$, and for the inner structure, one tracks several local extrema in the vicinity of the CE. By comparing the numerical 
and theoretical velocities on Fig.~\ref{fig:SpinCorrBis}(b), one can certify that both the CE and the series of local extrema propagate ballistically at the velocity 
$V_{\mathrm{CE}} \simeq 2V_{\mathrm{g}}^*$ and $V_{\mathrm{m}} \simeq 2V_{\varphi}^*$ respectively. Indeed, the maximal relative error between the numerical and theoretical velocities
is around $10\%$. \\

We stress that this twofold linear causality cone for the correlation spreading in the local regime of a quantum phase for a long-range interacting lattice model
is reminiscent of the one found for short-range interacting lattice models. Indeed, the local regime corresponds to relatively high power-law exponents $\alpha$ 
so that the long-range interactions have a sufficiently fast decay to be seen as relatively short-range interactions. Note that the discussion above can be 
extended to other long-range lattice models and observables \footnote{More precisely, the discussion can be extended to other equal-time connected correlation functions.
Due to the translational invariance of the pre- and post-quench Hamiltonians, one can not consider on-site observables for which the spatial dependence will vanish.} 
with the assumption that they can be cast into the generic form of Eq.~\eqref{generic_form}. Their space-time pattern will display a twofold linear causality 
cone where each structure will propagate into the lattice with the velocity $2V_{\mathrm{g}}^*$ and $2V_{\varphi}^*$ for the CE and the local extrema respectively.
Besides, the twofold linear causality cone characterized previously is expected to hold in the case of sudden global quenches confined in the local regime of a 
gapless quantum phase for long-range lattice models.  

\section{Local magnetization in the $z$ polarized phase}
\label{local_magnetization}

\subsection*{The local regime}

In the following, while the local regime of the $z$ polarized phase ($\alpha \geq 2$) is still considered, we move to a numerical investigation of the 
far-from-equilibrium dynamics of the long-range $s=1/2$ transverse Ising chain induced by sudden local quenches. Firstly, the local spin 
(or local magnetization) is studied before turning to an analysis of the entanglement spreading \textit{via} the computation of several entropy measures (R\'enyi entropies)
in the next subsection. \\

For the different sudden local quenches considered here, the initial state preparation is reminiscent of the one used in Chap.~\ref{ch:4-bose_hubbard_chain} at
Subsec.~\ref{1d_heis_local_quench}. In this paragraph, we recall the main steps of both the initial state preparation and the local quench protocol. We first consider
the Hamiltonian of the 1D LRTI model (deep in the $z$ polarized phase) with a small exchange-to-field ratio, $J/h \ll 1$. The initial 
state $\ket{\Psi_0}$ for the sudden local quench is built from the ground state $\ket{\Psi_{\mathrm{gs}}}$ of $\hat{H} = \hat{H}(J/h \ll 1)$, $\ket{\Psi_{\mathrm{gs}}} \simeq \ket{\uparrow ... \uparrow}$.
Then, the previous many-body quantum state is locally perturbed by flipping the central spin of the lattice chain, $\ket{\Psi_0} = \hat{S}^{-}_{N_s/2} \ket{\Psi_{\mathrm{gs}}}
\simeq \ket{\uparrow ... \uparrow \downarrow \uparrow ... \uparrow}$ with $N_s$ the number of lattice sites. This locally perturbed initial state $\ket{\Psi_0}$ 
evolves unitarily in time with respect to the same Hamiltonian, \textit{ie.} $\ket{\Psi(t)} = e^{-i \hat{H}t} \ket{\Psi_0}$. Finally, the local quench 
dynamics of the lattice model is characterized by computing the time-dependent expectation values of some relevant local observables, for instance the local magnetization
along the $z$ axis (polarization axis of the quantum phase considered here) represented mathematically by $\langle \Psi(t) | \hat{S}^z_R | \Psi(t) \rangle = \langle \Psi_0 | \hat{S}^z_R(t) | 
\Psi_0 \rangle$. \\

\begin{figure}[t!]
\centering
\includegraphics[scale = 0.373]{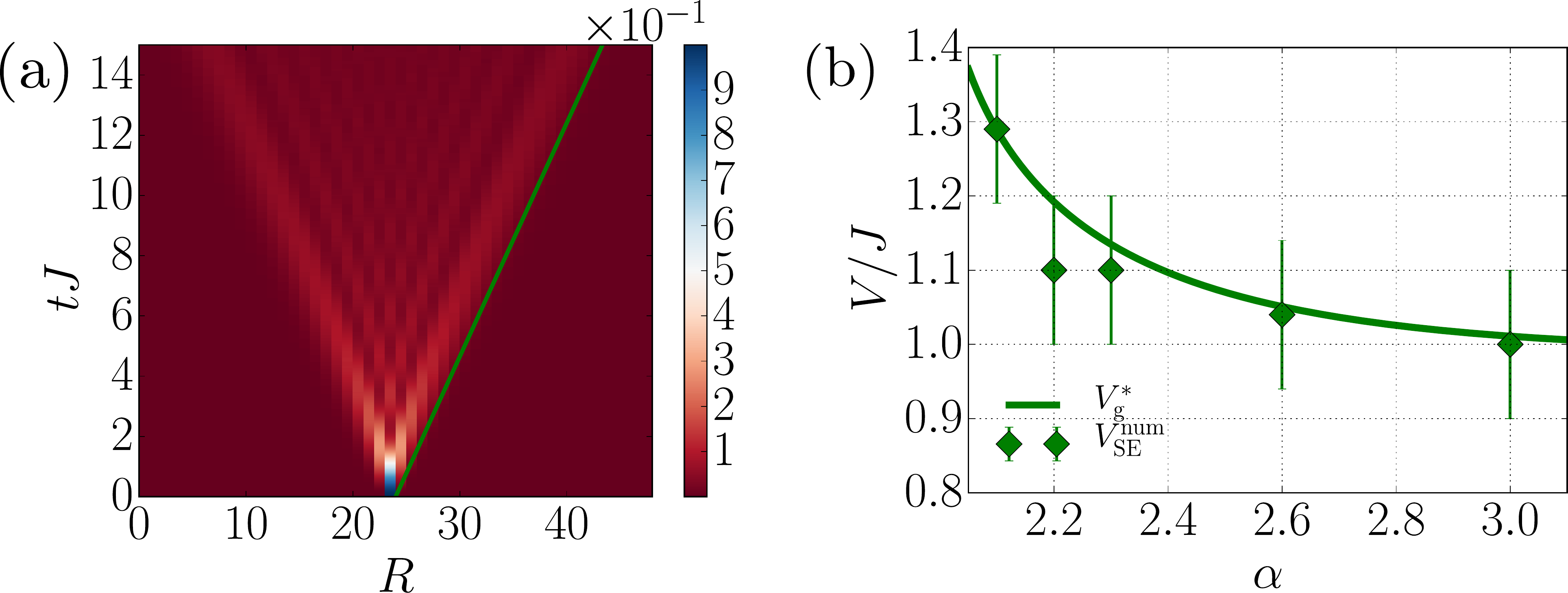}
\caption{\label{local_mag_lr} Spreading of the local magnetization for sudden local quenches in the local regime of the $z$ polarized phase.
(a)~t-MPS result of $1/2 - \langle \hat{S}^z_R(t) \rangle $ (lin-lin scale) for $J/h = 2\times 10^{-2}$ and $\alpha = 2.1 > 2$
with a ballistic fit of the spin edge (SE) motion (solid green line). The spin-flip has been applied on the central lattice site resulting
in a locally perturbed initial state $\ket{\Psi_0} \simeq \ket{\uparrow ... \uparrow \downarrow \uparrow ... \uparrow}$. Note that the time-dependent 
expectation value $1/2 - \langle \hat{S}^z_R(t) \rangle $ has been represented instead of $\langle \hat{S}^z_{R}(t) \rangle$. The latter allows us to
get not only positive values confined in the interval $[0,1]$ but also zero values in the non-causal region. (b)~Spreading velocity 
$V_{\mathrm{SE}} = V_{\mathrm{SE}}^{\mathrm{num}}$ (green diamonds) extracted from the numerical data by computing the slope associated to the ballistic fit to the SE motion, 
and comparison to the theoretical prediction $V_{\mathrm{SE}}^{\mathrm{th}} = V_{\mathrm{g}}^*$ (solid green line) as a function of the power-law exponent $\alpha > 2$
at fixed exchange-to-field ratio $J/h = 2\times 10^{-2}$. Both velocities are given in units of the spin exchange coupling $J>0$ ($\hbar = 1$).
Figure (b) extracted from Ref.~\cite{despres2019bis}.}
\end{figure}

Figure~\ref{local_mag_lr}(a) shows a typical t-MPS result for the space-time local magnetization $1/2 - \langle \hat{S}^{z}_R(t) \rangle$ for a sudden 
local quench in the local regime of the $z$ polarized phase [$\alpha = 2.1 > 2$ and $J/h = 2 \times 10^{-2}$]. The expectation values are represented
as a function of the dimensionless time $tJ$ ($\hbar = 1$) and the distance $R$. As expected from its theoretical expression, the local 
magnetization along the $z$ axis displays a single linear structure fully characterized by a spin edge (SE) spreading at the velocity $V_{\mathrm{SE}} \simeq 
V_{\mathrm{g}}^*$ (see discussion below \footnote{We point out that the theoretical expression of $1/2 - \langle \hat{S}^{z}_R(t) \rangle$ deduced from the linear 
spin-wave theory (see Appendix.~\ref{appendix_local_quench_spin} for more details) is in very good qualitative agreement with the numerical result displayed on 
Fig.~\ref{local_mag_lr}(a).}). Indeed, by fitting the motion of the SE using a ballistic (linear)
ansatz [see solid green line on Fig.~\ref{local_mag_lr}(a)] in order to extract the corresponding velocity, we find $V_{\mathrm{SE}}^{\mathrm{num}} = (1.3 \pm 0.1)J$ in very good 
agreement with the theoretical velocity $V_{\mathrm{g}}^* \simeq 1.29J$ [see Fig.~\ref{local_mag_lr}(b)]. We recall that $V_{\mathrm{g}}^*$ corresponds to 
the maximal group velocity reached for a quasimomentum $k^*$ [hence defined as $V_{\mathrm{g}}^* = V_{\mathrm{g}}(k^*) = 
\mathrm{max}_k[V_{\mathrm{g}}(k)] = \mathrm{max}_k(\partial_k E_k)$] and calculated from the excitation spectrum $E_k=2\sqrt{h[h+JP_{\alpha}(k)]}$ valid in 
the $z$ polarized phase of the $s=1/2$ LRTI chain. \\

On Fig.~\ref{local_mag_lr}(b), different sudden local quenches confined in the local regime of the $z$ polarized phase are considered. They
are defined by a same interaction parameter $J/h = 2 \times 10^{-2}$ while the power-law exponent $\alpha$ scans the interval $[2,3]$.
The spreading velocity $V_{\mathrm{SE}}^{\mathrm{num}}$, extracted using the same technique as previously,
is compared to the theoretical maximal group velocity $V_{\mathrm{g}}^*$. Both velocities, given in units of $J$, are in good agreement within the errorbars coming from 
the ballistic fits. As a consequence, one can certify that the SE propagates ballistically at the specific velocity $V_{\mathrm{SE}} \simeq V_{\mathrm{g}}^*$. Similar 
results were found at Chap.~\ref{ch:4-bose_hubbard_chain} for the same local observable in the case of the short-range $s=1/2$ Heisenberg chain 
in the ferromagnetic phase along the $z$ axis. \\
Note that for large power-law exponents $\alpha$ in the local regime of the $z$ polarized phase ($\alpha \geq 3$), the SE velocity (as well as $V_\mathrm{g}^*$) 
converges toward the value $J$, see Fig.~\ref{local_mag_lr}(b). The latter can be understood from the excitation spectrum $E_k$ of the short-range
$s=1/2$ transverse Ising chain, \textit{ie.} $E_k = 2\sqrt{h[h+J\cos(k)]}$. Indeed, by performing a Taylor expansion of $E_k$ since $J/h \ll 1$, 
one finds $E_k \simeq 2h[1+(J/2h)\cos(k)]$. Hence, the group velocity $V_{\mathrm{g}}(k) = \partial_k E_k$ 
can be written as $V_{\mathrm{g}}(k) \simeq -J\sin(k)$. Consequently, the maximal group velocity $V_{\mathrm{g}}^*$ is reached for $k^* = - \pi/2$ and it yields the 
value $V_{\mathrm{g}}^* \simeq J$. \\

We now turn to an analysis of the analytical expression of the local magnetization along the $z$ axis. The aim is to explain its space-time pattern 
displaying a single structure [see Fig.~\ref{local_mag_lr}(a)] fully characterized by a SE (spin edge) propagating into the lattice with the
velocity $V_{\mathrm{g}}^*$ [see Fig.~\ref{local_mag_lr}(b)]. According to the LSWT, the theoretical expression of $1/2 - \langle \hat{S}^{z}_R(t) \rangle$ may be written as 
(see Appendix.~\ref{appendix_local_quench_spin} for a complete derivation)

\begin{align}
 & 1/2 - \langle \hat{S}^z_R(t) \rangle \simeq \left| \int_{\mathcal{B}} \frac{\mathrm{d}k}{2\pi} \mathcal{F}_1(k)  \left \{ \frac{
 e^{i[k(R- N_s/2) + E_kt]}+ e^{-i[k (R - N_s/2) - E_kt]} }{2} \right \} \right|^2 \nonumber  \\
 & ~~~~~~~~~~~~~~~~~~ + \left| \int_{\mathcal{B}} \frac{\mathrm{d}k}{2\pi} \mathcal{F}_2(k) \left \{ \frac{ e^{i[k(R-N_s/2) + E_kt]} +
 e^{-i[k(R-N_s/2) - E_kt]} }{2} \right \} \right|^2.
 \label{local_mag_first}
\end{align}

\noindent
$N_s$ denotes the total number of lattice sites for the spin chain, $\mathcal{F}_1(k)$ and $\mathcal{F}_2(k)$ correspond to two different quasimomentum-dependent 
amplitude functions, 

\begin{align}
& \mathcal{F}_1(k) = \frac{1}{2}\left( \frac{\mathcal{A}_k}{E_k} +1 \right) = \frac{1}{2} \left( \frac{2h + JP_{\alpha}(k)}{2\sqrt{h[h+JP_{\alpha}(k)]}} + 1\right), \\
& \mathcal{F}_2(k) = -\frac{\mathcal{B}_k}{2E_k} = -\frac{J P_{\alpha}(k)}{4\sqrt{h[h+JP_{\alpha}(k)]}},
\end{align}

\noindent
where $E_k = 2\sqrt{h[h+JP_{\alpha}(k)]} = \sqrt{\mathcal{A}_k^2 - \mathcal{B}_k^2}$ with $\mathcal{A}_k = 2h + JP_{\alpha}(k)$ and $\mathcal{B}_k = JP_{\alpha}(k)$. The 
quasimomentum $k$ is confined in the first Brillouin zone $\mathcal{B} = [-\pi,\pi]$. Equation~\eqref{local_mag_first} is valid for a sudden local quench in the $z$ polarized phase (both for the quasi-local
and local regimes) starting from the initial state $\ket{\Psi_0} \simeq \ket{\uparrow ... \uparrow \downarrow \uparrow ... \uparrow}$.
To characterize the SE and its corresponding velocity, one has to investigate the asymptotic behavior of Eq.~\eqref{local_mag_first} by considering the 
stationary phase approximation. In the following, we recall the main steps of such approximation (see Appendix.~\ref{appendix2_sp} for more details). Considering the phase function $\Phi(k) = -k\tilde{R} + E_k t$ 
($\tilde{R} = R - N_s/2$) of the complex exponential at Eq.~\eqref{local_mag_first} with a positive group velocity, the stationary-phase quasimomentum $k_{\mathrm{sp}}$ 
has to fulfill the condition 

\begin{equation}
V_{\mathrm{g}}(k_{\mathrm{sp}}) = \tilde{R}/t > 0.
\label{condition_spa}
\end{equation}

\noindent
Then, by performing an evaluation of Eq.~\eqref{local_mag_first} around this stationary point, it yields the
following asymptotic form for the local magnetization

\begin{align}
 & 1/2 - \langle \hat{S}^z_{\tilde{R}}(t) \rangle \sim \left|\frac{\mathcal{F}_1(k_{\mathrm{sp}})}{(|\partial^2_k E_{k_{\mathrm{sp}}}|t)^{1/2}} \left[\cos(k_{\mathrm{sp}} \tilde{R}
 -E_{k_{\mathrm{sp}}}t + \phi) - i \sin(k_{\mathrm{sp}} \tilde{R} - E_{k_{\mathrm{sp}}}t + \phi) \right]\right|^2 \nonumber \\
 & ~~~~~~~~~~~~~~~~~~ + \left|\frac{\mathcal{F}_2(k_{\mathrm{sp}})}{(|\partial^2_k E_{k_{\mathrm{sp}}}|t)^{1/2}} \left[\cos(k_{\mathrm{sp}} \tilde{R}
 -E_{k_{\mathrm{sp}}}t + \phi) - i \sin(k_{\mathrm{sp}} \tilde{R} - E_{k_{\mathrm{sp}}}t + \phi) \right]\right|^2,
 \label{local_mag_spa}
\end{align}

\noindent
where $\tilde{R} = R - N_s/2$ and $\phi = - (\pi/4)~\mathrm{sgn}(\partial_k^2 E_{k_\mathrm{sp}}t)$ is a constant phase term irrelevant for our study. The local regime of the $z$ polarized phase
($\alpha \geq 2$) implies both a finite quasiparticle energy $E_k$ and group velocity $V_{\mathrm{g}}(k)$ over the whole Brillouin zone $\mathcal{B}$. Therefore, it exists a quasimomentum 
$k^*$ such that the group velocity is maximal, \textit{ie.} $V_{\mathrm{g}}(k^*) = \mathrm{max}_k [V_{\mathrm{g}}(k)]$. As a consequence, one can deduce that the SE is governed by the velocity
$V_{\mathrm{g}}^* = V_{\mathrm{g}}(k^*)$ according to Eq.~\eqref{condition_spa} for $k_{\mathrm{sp}} = k^*$. Besides, the inner structure of the causality cone in the vicinity of the SE is
determined by Eq.~\eqref{local_mag_spa} for $k_{\mathrm{sp}} = k^*$. More precisely, considering the first term of Eq.~\eqref{local_mag_spa} with $k_{\mathrm{sp}} = k^*$, both the square of
the real and imaginary parts have a SE propagating ballistically with a velocity $V_{\mathrm{SE}} = V_{\mathrm{g}}^*$. In the vicinity of the SE, the series of local maxima spread at the velocity 
$V_{\mathrm{m}} = V_{\varphi}^* = E_{k^*}/k^* \ll 0$, \textit{ie.} at the phase velocity evaluated at the quasimomentum $k^*$.
Note that $V_{\varphi}^* \ll 0$ is due to the negative quasimomentum $k^*$ and the large gap in the excitation spectrum $E_k$ of the $z$ polarized phase for a small exchange-to-field ratio $J/h$. 
However, the space-time pattern of $\langle \hat{S}^z_R(t) \rangle$ near the SE does not display the expected twofold linear structure. Indeed, this is because the first term of Eq.~\eqref{local_mag_spa}
is the sum of two contributions (square of the real and imaginary parts) which are shifted by half a period and cancel each other \footnote{This statement is also valid for the second term of Eq.~\eqref{local_mag_spa}.}.
Consequently, we end up with a SE propagating at the same velocity as the one associated to the local maxima when combining both contributions. In other words, it yields for the space-time local magnetization a linear single
structure characterized by the sole SE velocity $V_{\mathrm{g}}^*$. \\

\begin{figure}[t!]
\centering
\includegraphics[scale = 0.38]{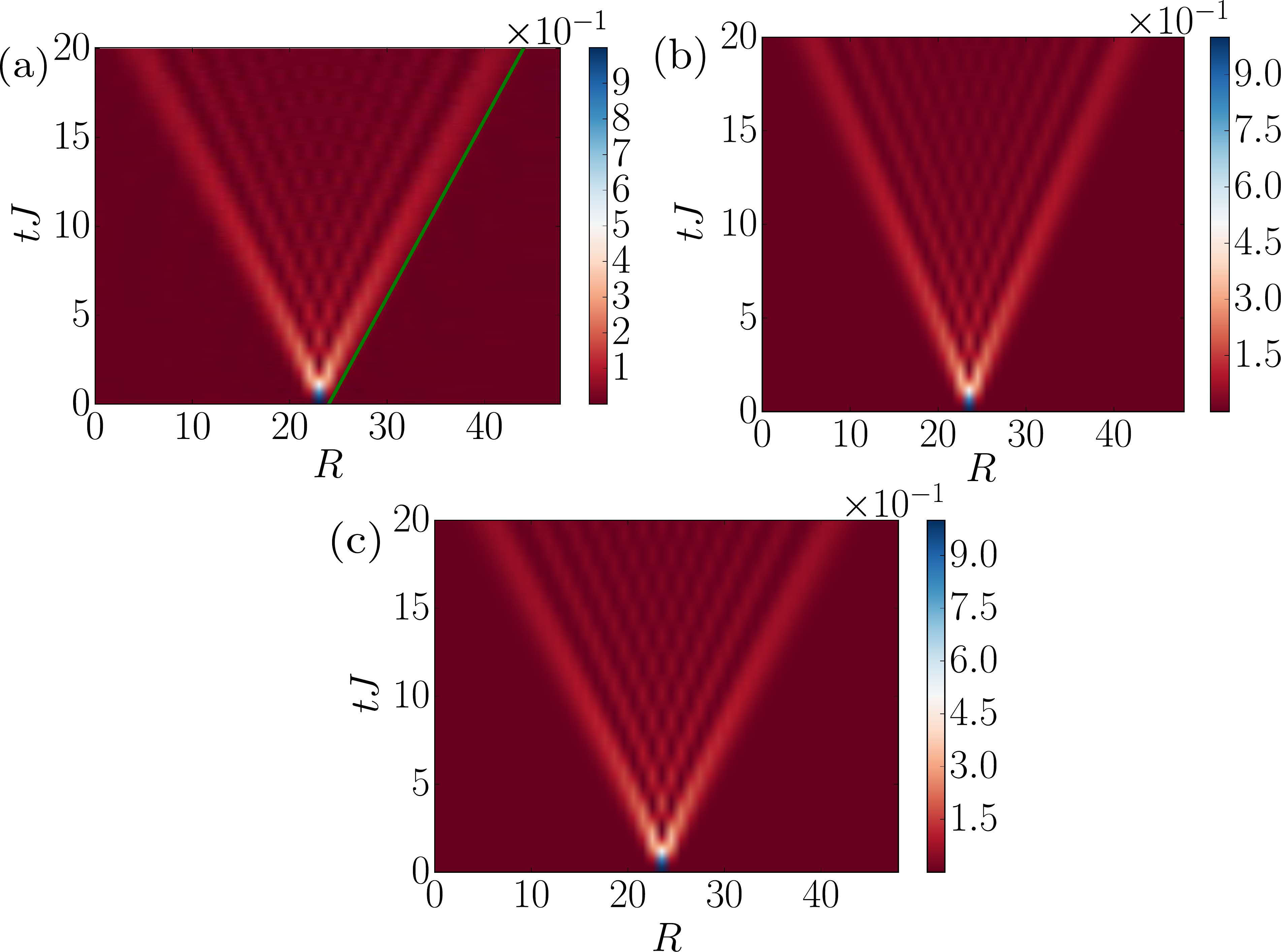}
\caption{\label{fig:full_plot_local_mag} 
Spreading of the local magnetization $1/2 - \langle \hat{S}^z_R(t) \rangle $ for a sudden local quench defined by the locally-perturbed initial state $\ket{\Psi_0} \simeq \ket{\uparrow ... \uparrow \downarrow \uparrow ... \uparrow}$
evolving unitarily in time with the Hamiltonian $\hat{H}$ of the LRTI or SRTI chain (lin-lin scale). Both spin lattice models are confined in the $z$ polarized phase and characterized by a same exchange-to-field ratio $J/h = 2\times 10^{-2}$.
(a)~t-MPS + TDVP result for the long-range transverse Ising chain deep in the local regime of the $z$ polarized phase for $\alpha = 3$ with a ballistic fit of the spin edge motion (see solid green line and Fig.~\ref{local_mag_lr}(b) for the 
corresponding velocity).
(b)~Analytical result for the previous long-range interacting lattice model with $\alpha = 3$ deduced from the linear spin-wave theory, see Eq.~\eqref{local_mag_first}.
(c)~Analytical result for the short-range transverse Ising chain in the $z$ polarized phase, see Eq.~\eqref{local_mag_first} with $E_k = 2\sqrt{h[h+J\cos(k)]} = \sqrt{\mathcal{A}_k^2 - \mathcal{B}_k^2}$ 
where $\mathcal{A}_k = 2h + J\cos(k)$ and $\mathcal{B}_k = J\cos(k)$. A perfect agreement between the three different space-time patterns is found. Figures extracted from Ref.~\cite{despres2019bis}.}
\end{figure}

To sum up, relying both on the LSWT and numerical calculations using the t-MPS + TDVP technique, a linear single structure has been unveiled for the spreading of the local magnetization
in the local regime of the $z$ polarized phase. This single structure is characterized by a SE (spin edge) and local maxima propagating at the maximal group velocity, \textit{ie.} $V_{\mathrm{g}}^*$.
For large power-law exponents $\alpha$ in the local regime, this velocity converges towards the maximal group velocity calculated from the excitation spectrum $E_k$ of the 
short-range transverse Ising (SRTI) model in the $z$ polarized phase, \textit{ie.} 
$E_k = 2\sqrt{h[h+J\cos(k)]}$. Since the space-time local magnetization depends only on the maximal group velocity $V_{\mathrm{g}}^*$, one should recover a similar pattern
between the LRTI chain with a large $\alpha$ in the local regime and the SRTI chain where both quantum lattice models are characterized by a same and small exchange-to-field ratio $J/h$.
To verify such hypothesis, we have compared for an interaction parameter $J/h = 2 \times 10^{-2}$ the space-time pattern of the local magnetization using (i) Eq.~\eqref{local_mag_first} and 
the excitation spectrum $E_k = 2 \sqrt{h[h+JP_{\alpha}(k)]}$ of the 1D LRTI model in the $z$ polarized phase at $\alpha = 3$ (ii) the t-MPS + TDVP technique to simulate the previous long-range interacting 
quantum model for $\alpha = 3$ and (iii)~Eq.~\eqref{local_mag_first} and the excitation spectrum of the 1D SRTI model in the $z$ polarized phase. Knowing that the SRTI chain is considered for the case (iii),
both amplitude functions $\mathcal{F}_1(k)$ and $\mathcal{F}_2(k)$ have to be adapted since the quasimomentum-dependent functions $\mathcal{A}_k$ and $\mathcal{B}_k$ are different from those valid for the LRTI chain. 
For the 1D SRTI model in the $z$ polarized phase, $\mathcal{A}_k = 2h + J\cos(k)$ and $\mathcal{B}_k = J\cos(k)$ $\forall k \in \mathcal{B}$, see Appendix.~\ref{appendix_local_quench_spin} for additional details.
As shown on Fig.~\ref{fig:full_plot_local_mag}, a qualitative agreement has been found between the three different space-time patterns. \\

\begin{figure}[t!]
\centering
\includegraphics[scale = 0.38]{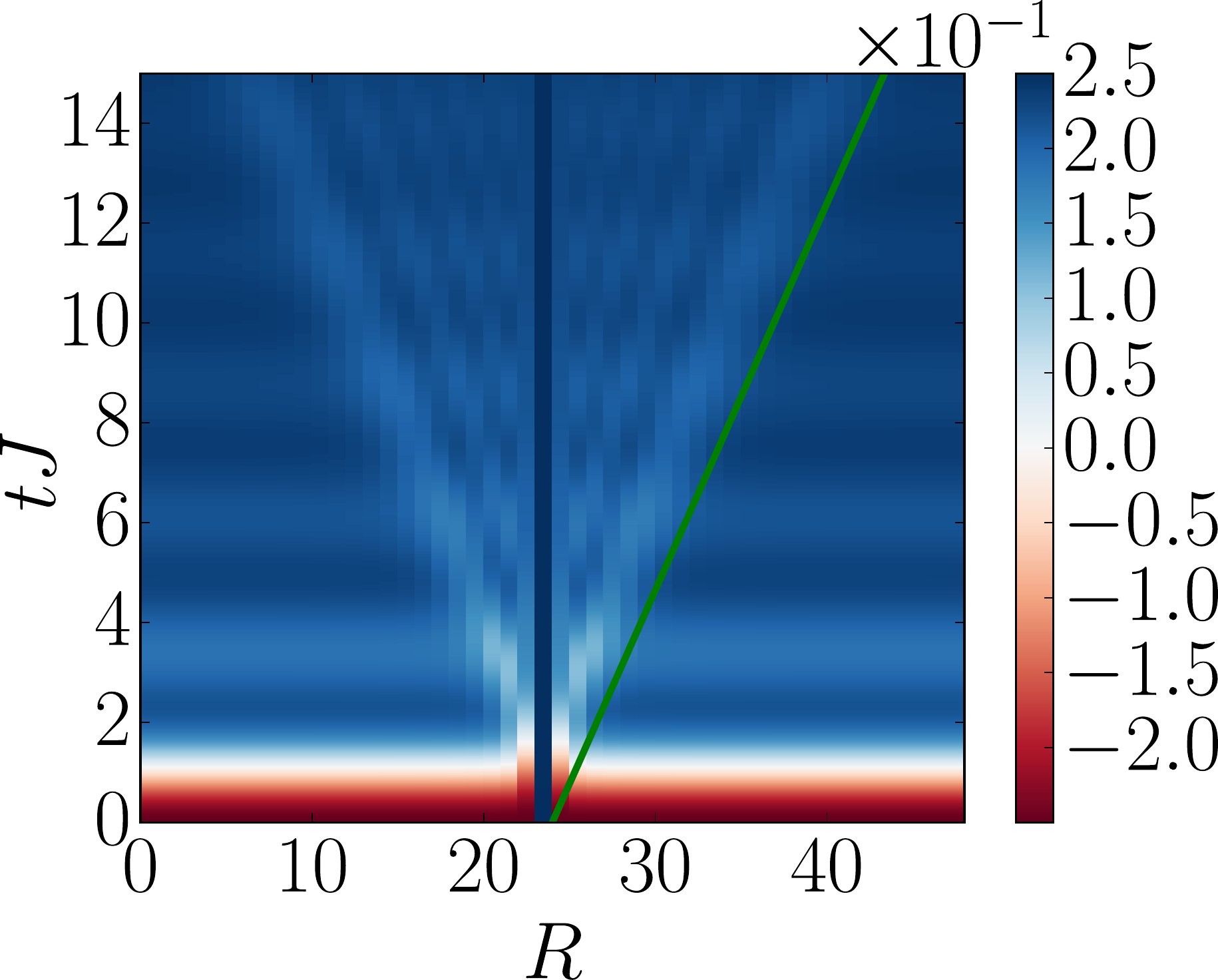}
\caption{\label{corr_local_quench_local_regime} Spreading of the spin correlations along the $z$ axis $\langle \hat{S}^z_R(t)\hat{S}^{z}_{N_s/2}(t) \rangle$
for a sudden local quench in the local regime of the $z$ polarized phase (lin-lin scale). The space-time spin correlations have been computed using the
t-MPS + TDVP technique for $J/h = 2\times 10^{-2}$ and $\alpha = 2.1 > 2$ for a locally perturbed initial state $\ket{\Psi_0} \simeq \ket{\uparrow ...
\uparrow \downarrow \uparrow ... \uparrow}$. The solid green line represents a ballistic fit to the motion of the CE (correlation edge). The associated slope
$V_{\mathrm{CE}}^{\mathrm{num}}$ has been found to be in excellent agreement both with the numerical spin edge velocity $V_{\mathrm{SE}}^{\mathrm{num}}$ and the theoretical
maximal group velocity $V_{\mathrm{g}}^*$ (see Fig.~\ref{local_mag_lr}(b) for the values of $V_{\mathrm{SE}}^{\mathrm{num}}$ and $V_{\mathrm{g}}^*$).
Note that contrary to the local magnetization the spin correlations display another time-dependent term irrelevant for our study, see Fig.~\ref{local_mag_lr}(a).}
\end{figure}

Note that the previous statements are not restricted to the sole local magnetization $\langle \hat{S}^z_R(t) \rangle$ and still apply for other space-time local observables. 
Indeed, by performing the same analysis as the one presented on Fig.~\ref{local_mag_lr}(b) for the spin correlations along the $z$ axis (defined by the space-time expectation 
value $\langle \hat{S}^z_R(t)\hat{S}^{z}_{N_s/2}(t) \rangle$), we have reached similar conclusions. More precisely, the latter has been found to display a single linear structure fully 
characterized by a CE (correlation edge) propagating at $V_{\mathrm{g}}^*$ for any power-law exponent $\alpha$ such that the LRTI chain is confined in the local regime of the $z$ 
polarized phase (see Fig.~\ref{corr_local_quench_local_regime} for an example of the space-time pattern for the spin-spin correlation function $\langle \hat{S}^z_R(t)\hat{S}^{z}_{N_s/2}(t) \rangle$). \\

By considering the local regime of the $z$ polarized phase and according to the previous discussion, the space-time 
pattern of local observables for sudden local quenches [see Fig.~\ref{local_mag_lr}] displays different features from the case of sudden global quenches [see 
Fig.~\ref{fig:SpinCorrBis}]. Indeed, for the sudden global quenches, a linear twofold structure has been unveiled where a CE (correlation edge) and a series of local maxima 
spread linearly with the velocity $2V_{\mathrm{g}}^*$ and $2V_{\varphi}^*$ respectively. Note that another difference between both behaviors corresponds to the factor 
two in the characteristic spreading velocities \footnote{Concerning the single structure of the local magnetization (already explained previously), we have shown that it 
corresponds to the sum of two contributions having a twofold linear structure where the SE and the local maxima propagate at the velocity $V_{\mathrm{g}}^*$ and $V_{\varphi}^*$
respectively.}. The latter can be explained easily in terms of quasiparticles propagating into the lattice. For sudden global quenches, the space-time correlations are governed by
free counter-propagating quasiparticle pairs [see Fig.~\ref{corr}]. However, in the case of sudden local quenches, the causality cone associated to the space-time pattern of local observables
is governed by individual quasiparticles emitted from the reference lattice site \footnote{The reference lattice site refers to the one where the local perturbation has been applied.}. Note that this
factor $2$ for the characteristic spreading velocities between global and local quantum quenches was expected. Indeed, the generic form of the correlation functions for sudden global quenches
[see Eq.~\eqref{generic_form}] depends only on the quasiparticle pair energy $2E_k$, contrary to the one of the expectation value of space-time dependent on-site observables for sudden local quenches
where only the individual-quasiparticle energy $E_k$ is relevant [see for instance Eq.~\eqref{local_mag_first} and Eq.~\eqref{local_mag_analytics}]. \\

\subsection*{The quasi-local regime}

To conclude the analysis of the space-time local magnetization for sudden local quenches in the $z$ polarized phase of the 1D LRTI model, we point out that the 
quasi-local regime \footnote{The latter requires for the power-law exponent $\alpha$ to fulfill the condition $1 \leq \alpha < 2$ for the 1D LRTI model confined in the gapped $z$ polarized phase.}
implying a finite quasiparticle energy $E_k$ and a divergence of the group velocity $V_{\mathrm{g}}(k)$ in the first Brillouin zone has also been investigated. To do so, a same initial state 
preparation as the one considered previously for the characterization of the local magnetization in the local regime has been considered. One recalls that the latter leads to the locally perturbed many-body quantum state $\ket{\Psi_0}
\simeq \ket{\uparrow ... \uparrow \downarrow \uparrow ... \uparrow}$. The main result of this study is that similarly to the correlation spreading for sudden global quenches
confined in the quasi-local regime of the $z$ polarized phase, the SE associated to the space-time pattern of the local magnetization propagates
sub-ballistically \textit{ie.} $t \sim R^{\beta_{\mathrm{SE}}}$ with $\beta_{\mathrm{SE}} > 1$. More precisely, we found that $\beta_{\mathrm{SE}} = \beta_{\mathrm{CE}} = 3 - \alpha > 1$ for $1 \leq \alpha < 2$,
see Ref.~\cite{despres2019bis} for more details. While still considering sudden local quenches confined in the quasi-local regime of the $z$ polarized phase for the 1D LRTI model, the previous statements have
been found to be valid for another local observable, namely the equal-time spin correlator along the $z$ axis defined as $\langle \hat{S}^z_R(t)\hat{S}^{z}_{N_s/2}(t) \rangle$. The latter confirms the existence of a sub-ballistic motion of the edge for the spreading of 
space-time dependent local observables for both sudden global and local quantum quenches confined in the gapped phase of long-range interacting lattice models. \\

\section{Entanglement spreading in the $z$ polarized phase}

\subsection*{The local regime}

In the following, the local regime of the $z$ polarized phase for the long-range transverse Ising chain is still considered. Besides, we rely on a same 
initial state preparation as the one presented at Sec.~\ref{local_magnetization} leading to the many-body quantum state $\ket{\Psi_0} \simeq \ket{\uparrow ... 
\uparrow \downarrow \uparrow ... \uparrow}$. However, we turn to an analysis of the propagation of entanglement \textit{via} a study of the $n$-order 
($n \in \mathbb{R}^+ \backslash \{ 1 \}$) R\'enyi entropy defined as 

\begin{equation}
 \mathcal{S}_n(R,t) = \frac{1}{1-n} \mathrm{log}\left \{ \mathrm{Tr} \left[ \hat{\rho}^n(R,t) \right] \right \}.
\end{equation}

\noindent
$\hat{\rho}(R,t) = \mathrm{Tr}_{\bar{R}}\left[ \hat{\rho}(t) \right] = \mathrm{Tr}_{\bar{R}}\left( \ket{\Psi(t)} \bra{\Psi(t)} \right)$ denotes the reduced
density matrix at time $t$ of the subsystem built from the first $R$ lattice sites. The latter is found by tracing out all the degrees of freedom of the
complementary subsystem $\bar{R} = \{ R+1,...,N_s\}$.
In the following, the space-time behavior of three R\'enyi entropies are characterized \textit{ie.} $\mathcal{S}_{n=1/2}(R,t)$, $\mathcal{S}_{n=2}(R,t)$ and
$\mathcal{S}_{n\rightarrow 1}(R,t)$. The latter corresponds to the von Neumann entropy which may be written as 

\begin{equation}
 \mathcal{S}_{n \rightarrow 1}(R,t) = -\mathrm{Tr}\left \{ \hat{\rho}(R,t) \mathrm{log}\left[ \hat{\rho}(R,t) \right] \right \}.
\end{equation}

\noindent
These three different entropy measures correspond to three different possibilities to characterize the amount of entanglement between two subsystems of a bipartite quantum model. We refer the reader to the 
Appendix.~\ref{appendix_entropies} for more details about the definition and properties of the von Neumann and R\'enyi entropies. The main purpose of the following discussion
is to compare the entanglement spreading with the correlation spreading for sudden local quenches. \\

\begin{figure}[t!]
\includegraphics[width = \columnwidth]{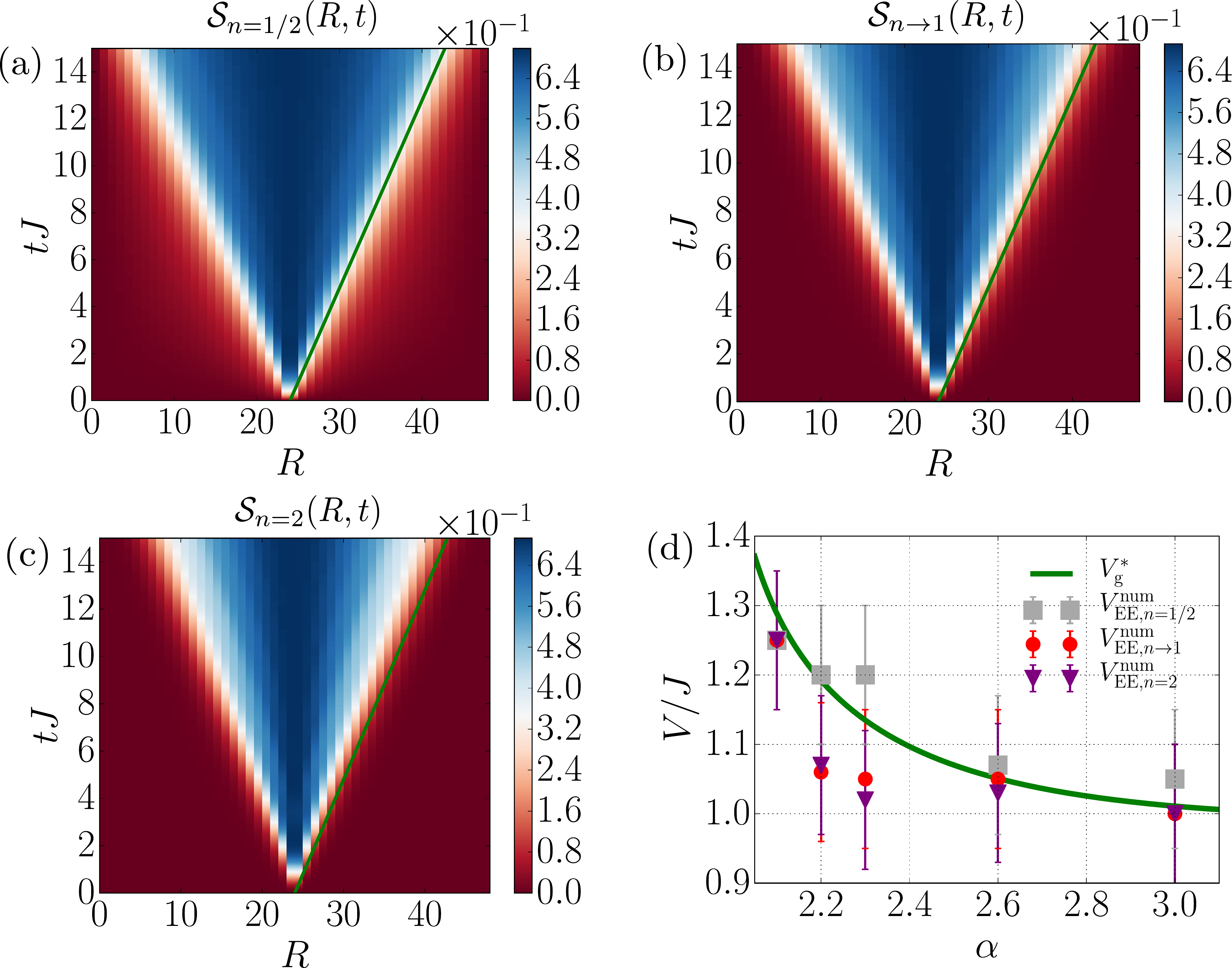}
\caption{\label{ent_entropy_lr}
Spreading of several R\'enyi entropies for sudden local quenches in the local regime of the $z$ polarized phase ($\alpha \geq 2$). (a)~ t-MPS result of the
$n=1/2$ order R\'enyi entropy $\mathcal{S}_{n=1/2}(R,t)$ (b)~ von Neumann entropy $\mathcal{S}_{n \rightarrow 1}(R,t)$ (c)~ the $n=2$ order R\'enyi entropy 
$\mathcal{S}_{n=2}(R,t)$ (lin-lin scale) for $J/h = 2 \times 10^{-2}$ and $\alpha = 2.1 > 2$. The sudden local quench is characterized by a local perturbation 
(a spin-flip) applied on the central lattice site resulting in an initial state $\ket{\Psi_0} \simeq \ket{\uparrow ... \uparrow \downarrow \uparrow ... \uparrow}$.
On Figs.~(a,b,c), the solid green lines correspond to a ballistic fit of the motion of the EE (entropy edge) whose corresponding velocities are reported on 
Fig.~(d). (d)~ Spreading velocities $V_{\mathrm{EE}}$ of the entropy edge for the R\'enyi entropies
of order $n=1/2$ (grey rectangles), $n=2$ (purple triangles) and for the von Neumann entropy ($n \rightarrow 1$, red dots) extracted 
from the t-MPS data, and comparison to the theoretical maximal group velocity $V_{\mathrm{g}}^* = \mathrm{max}_k(\partial_k E_k)$ (solid green line). $E_k$ refers to the excitation spectrum 
of the LRTI chain in the gapped $z$ polarized phase. Figures extracted from Ref.~\cite{despres2019bis}.}
\end{figure}

Figures~\ref{ent_entropy_lr}(a,b,c) show typical t-MPS results \footnote{Note that according to the isometric gauge considered in the t-MPS + TDVP simulations, the reduced density matrix is directly given by 
the left gauge-fixing condition, see Fig.~\ref{conditions}(a).} for the three considered R\'enyi entropies for a sudden 
local quench in the local regime of the $z$ polarized phase ($\alpha = 2.1 > 2$ and $J/h = 2 \times 10^{-2}$). The space-time patterns are represented 
as a function of the dimensionless time $tJ$ ($\hbar = 1$) and the distance $R$ (in a lin-lin scale). Each of them displays a linear causality cone, 
starting from the reference site $N_s/2$ where the local perturbation has been applied, as expected from the Lieb-Robinson bound. Note that the latter
do not show a series of local maxima for their inner structure contrary to the local magnetization, see Fig.~\ref{local_mag_lr}(a). Hence, the causality cones 
are fully characterized by an entropy edge (EE) whose ballistic motion has been fitted [see solid green line on Fig.~\ref{ent_entropy_lr}(a,b,c)]. For the three considered
R\'enyi entropies, the corresponding velocity $V^{\mathrm{num}}_{\mathrm{EE},n}$ is found to be in good agreement with the theoretical maximal group velocity $V_{\mathrm{g}}^*$. Indeed, for 
$\alpha = 2.1$, we find $V^{\mathrm{num}}_{\mathrm{EE},n} = (1.25 \pm 0.1)J$ for $n=1/2,2$ and $n \rightarrow 1$ close to the velocity $V_{\mathrm{g}}^* \simeq 1.29J$. \\

On Fig.~\ref{ent_entropy_lr}(d), several sudden local quenches confined in the local regime of the $z$ polarized phase are considered. They are defined by a same 
small exchange-to-field ratio $J/h = 2 \times 10^{-2}$ while the power-law exponent $\alpha$ scans the interval $[2,3]$. The spreading velocities
$V_{\mathrm{EE,n}}^{\mathrm{num}}/J$ are represented as a function of the power-law exponent $\alpha$ and compared to the theoretical maximal group velocity 
$V_{\mathrm{g}}^*$ calculated from the excitation spectrum $E_k$ of the LRTI chain in the $z$ polarized phase, \textit{ie.} $E_k = 2 \sqrt{h[h+JP_{\alpha}(k)]}$.
The numerical velocities $V_{\mathrm{EE,n}}^{\mathrm{num}}$ are extracted from the t-MPS data using the same technique as previously (by considering a ballistic fit 
to track the motion of the EE). According to the results reported on Fig.~\ref{ent_entropy_lr}(d), these velocities are in good agreement with the theoretical maximal group velocity $V_{\mathrm{g}}^*$. As
a consequence, one can claim that the EE propagates linearly with a velocity $V_{\mathrm{EE,n}} \simeq V_{\mathrm{g}}^*$, at least for $ n \in \{1/2,2\}$ and
$n \rightarrow 1$. Hence, one can stress that the entropy edge (EE) behavior is similar to the one for the spin (SE) [see Fig.~\ref{local_mag_lr}(b)] and correlation
(CE) edges [see Fig.~\ref{corr_local_quench_local_regime} and discussion at Sec.~\ref{local_magnetization}]. This important result can be interpreted in the following 
manner : just after the sudden local quench, all the quasiparticles (for any quasimomentum $k$ in the first Brillouin zone) are emitted and spread into the lattice. 
For the local regime, only the quasiparticle with the highest group velocity ($V_{\mathrm{g}}^*$) is relevant to describe the edge separating the causal and non-causal
regions. This allows us to explain the motion of both the SE and CE. Besides, these quasiparticles spreading into the lattice are also expected to carry the entanglement.
Consequently, the entanglement spreading properties should be similar to those for the correlation spreading, at least concerning the edge.

\section{Correlation spreading in the long-range XY chain}

\subsection*{The quasi-local regime of the $x$ polarized phase}

In the following, the purpose is to verify numerically the theoretical scaling laws of the correlation spreading
for a long-range interacting lattice model in the quasi-local regime of a \textit{gapless} quantum phase. For this context and considering any space-time correlation function fulfilling
the generic form presented at Eq.~\eqref{generic_form}, we unveiled a twofold algebraic structure displaying (i)~a correlation edge (CE), separating both the causal and non-causal region of 
correlations and spreading sub-ballistically ($t \sim R^{\beta_{\mathrm{CE}}}$, $\beta_{\mathrm{CE}} > 1$) (ii)~but also a series of local extrema in the vicinity of 
the CE propagating super-ballistically ($t~\sim R^{\beta_{\mathrm{m}}}$, $\beta_{\mathrm{m}} < 1$), see Sec.~\ref{LRC} for a detailed discussion and Tab.~\ref{tab:SL}
for a summary of the scaling laws. To verify the previous scaling laws, using the MPS + TDVP technique presented at Sec.~\ref{TDVP_section}, we turn to a 
numerical investigation of the correlation spreading in the quasi-local regime of the gapless $x$ polarized phase for the long-range $s=1/2$ XY chain, see also the finite automata
at Fig.~\ref{finite_automata_LRXY} in order to construct the MPO form of the corresponding Hamiltonian. \\

In order to characterize the far-from-equilibrium dynamics of the LRXY chain in the quasi-local regime of the $x$ polarized phase, a sudden global quench 
is considered. It is defined by a pre- and post-quench antiferromagnetic interaction parameter along the $z$ axis fulfilling $-1<\epsilon_\textrm{i} < \epsilon_c(\alpha)$
(1D LRXXZ model) and $\epsilon_\textrm{f}=0$ (1D LRXY model) respectively, see Eq.~\eqref{eq:XXZ} for the Hamiltonian of the 1D LRXXZ model.
The power-law exponent $\alpha$ is fixed and confined in the interval $\alpha \in [1,3[$ so that the quasi-local regime of the $x$ polarized phase (for the 
LRTI chain) is considered to perform the real time evolution. Besides, the connected spin-spin correlation function along the $z$ axis, denoted by $G_z$, are investigated.
The latter reads as $G_z(R,t) = G_{z,0}(R,t) - G_{z,0}(R,0)$ with 

\begin{equation}
G_{z,0}(R,t) = \langle \hat{S}_R^z(t) \hat{S}_0^z(t) \rangle - \langle \hat{S}_R^z(t)\rangle \langle \hat{S}_0^z(t) \rangle.
\end{equation}

\noindent
We already have shown that the $G_z$ spin-spin correlations can be expressed in the generic form of Eq.~\eqref{generic_form}, with a
quasimomentum-dependent amplitude function $\mathcal{F}$ defined at Eq.~\eqref{eq:LRXY.Fk} [see also Appendix.~\ref{appendix_gz_lrxy}].
Then, to extract both exponents $\beta_{\mathrm{CE}}$ and $\beta_{\mathrm{m}}$ associated to the scaling law for the CE and the series of local extrema respectively,
it requires to evaluate both the amplitude function $\mathcal{F}$ and the excitation spectrum $E_k$ [see Eq.~\eqref{eq:XXZ.Ek}] in the infrared limit 
$k \rightarrow 0^{+}$ to deduce the coefficients $\nu$ and $z$ respectively (see Sec.~\ref{LRC} for more details). Indeed, for the quasi-local regime of gapped or gapless
quantum phases, the space-time pattern of the equal-time connected correlation functions is governed by all the quasiparticles spreading with a positive
\footnote{Indeed, only the quasiparticles with a positive group velocity are relevant since the region $R,t>0$ of the space-time plane is considered in the following. Straightforwardly, the space-time pattern
in the region $R<0$ and $t>0$ can be deduced from the symmetry $R \rightarrow -R$ (note the symmetry $k \rightarrow -k$ of the excitation spectrum $E_k$ meaning that a 
quasiparticle at quasimomentum $k$ and $-k$ have a similar energy and an opposite group velocity). For the latter, only the quasiparticles 
with a negative group velocity are relevant.} and divergent group velocity in the vicinity of the CE. In other words, only the quasiparticles with a quasimomentum
in the infrared limit $k \rightarrow 0^{+}$ (according to the shape of the excitation spectrum $E_k$ of the gapless $x$ polarized phase, see for
instance Fig.~\ref{ek_vg_lrxy}) are relevant to describe the space-time behavior of the correlations close to the CE. 
We found that the amplitude function scales as $\mathcal{F}(k)\sim k^\nu$ with $\nu=z=(\alpha-1)/2$. Finally, concerning the spin-spin
correlations along the $z$ axis for the LRXY chain in the quasi-local regime of the $x$ polarized phase, we obtain $\nu=z=(\alpha-1)/2$ which yields according to Eqs.~\eqref{chi} and \eqref{gamma}

\begin{equation}
\beta_{\mathrm{CE}}= \frac{\chi}{\gamma} = 1+ \left(3-\alpha\right)/2\alpha > 1,~~~ \beta_{\mathrm{m}} = z = (\alpha-1)/2 < 1,~~~\forall \alpha \in [1,3[.  
\label{summary_beta}
\end{equation}

\begin{figure}[h!]
\centering
\includegraphics[scale = 0.35]{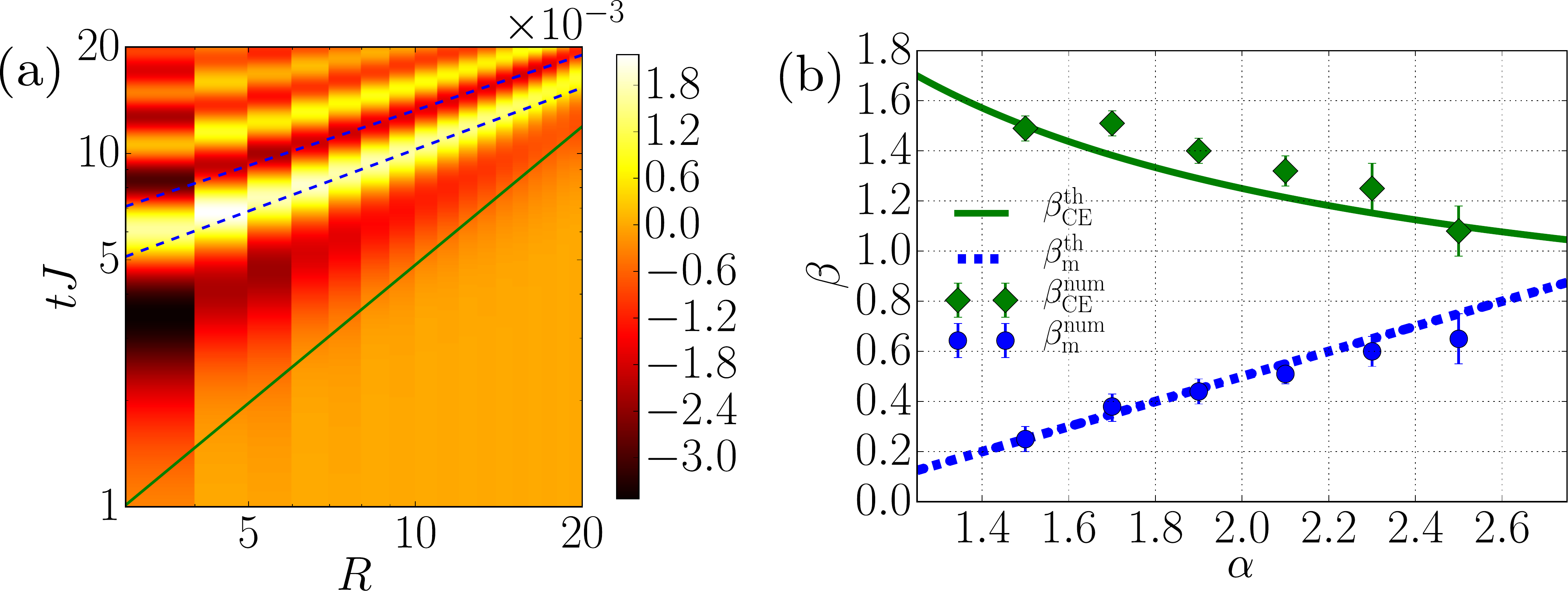}
\caption{
Spreading of the connected spin-spin correlation function $G_z(R,t)=G_{z,0}(R,t)-G_{z,0}(R,0)$ with $G_{z,0}(R,t) =
\langle \hat{S}_R^z(t) \hat{S}_0^z(t) \rangle - \langle \hat{S}_R^z(t)\rangle \langle \hat{S}_0^z(t) \rangle$ for the 1D LRXY model.
(a)~ t-MPS result for a sudden global quench confined in the $x$ polarized phase from the ground state of the 1D LRXXZ model at
$\epsilon_\mathrm{i} = 0.2$ to the 1D LRXY model ($\epsilon_{\mathrm{f}} = 0$) in the quasi-local regime at $\alpha=2.3$.
The space-time spin-spin correlations feature a double algebraic structure (straight lines in log-log scale) with a sub-ballistic correlation edge (see solid green line
for a linear fit to the CE motion) and a super-ballistic spreading of the series of local extrema (see dashed blue lines for linear fits to the motion of one local maxima
and minima). The exponents $\beta_{\mathrm{CE}}^{\mathrm{num}}$ and $\beta_{\mathrm{m}}^{\mathrm{num}}$ are deduced by computing the slope of the linear fit to the CE and 
the minima and/or maxima motions respectively.
(b)~ Evolution of $\beta_{\mathrm{CE}}^{\mathrm{num}}$ and its corresponding theoretical value $\beta_{\mathrm{CE}}^{\mathrm{th}}$ [see Eq.~\eqref{summary_beta}] 
characterizing the spreading of the correlation edge with $\beta_{\mathrm{m}}^{\mathrm{num}}$ and $\beta_{\mathrm{m}}^{\mathrm{th}}$ [see also Eq.~\eqref{summary_beta}]
determining the spreading of the series of local extrema as a function of the power-law exponent $\alpha \in [1,3[$ (quasi-local regime of the $x$ polarized phase for the LRXY chain).}
\label{num_LRXY}
\end{figure}

\noindent
On Fig.~\ref{num_LRXY}(a), we display a typical t-MPS result for $G_z(R,t)$, the space-time spin-spin correlations along the $z$ axis, as a function of the 
distance $R$ and the dimensionless time $tJ$ ($\hbar = 1$). The space-time spin-spin correlations feature a double algebraic structure (straight lines in log-log scale)
with a sub-ballistic correlation edge (see solid green line for a linear fit to the CE motion) and a super-ballistic spreading of the series of local extrema (see dashed blue lines for linear fits to the motion of one local maxima
and minima). The exponents $\beta_{\mathrm{CE}}^{\mathrm{num}}$ and $\beta_{\mathrm{m}}^{\mathrm{num}}$ are deduced by computing the slope of the linear fit to the CE and 
the minima and/or maxima motions respectively. In the following, we discuss in more details the twofold algebraic structure for the $G_z$ space-time spin fluctuations along the $z$ axis. \\

In practice, the motion of the CE is deduced by tracking the points in the $R-t$ plane where the correlations reach several percents of the maximal value. Then, these
points are fitted \textit{via} a ballistic (linear) fit whose slope gives access to $\beta_{\mathrm{CE}}^{\mathrm{num}}$. For instance at $\alpha = 2.3$, it yields the 
solid green line on Fig.~\ref{num_LRXY}(a). The latter features a linear trajectory in the log-log scale corresponding to a power law behavior in the lin-lin scale.
By computing the slope of the solid green line, we find $\beta_{\mathrm{CE}}^{\mathrm{num}} = 1.25 \pm 0.1$ in very good agreement with the theoretical prediction 
$\beta_{\mathrm{CE}}^{\mathrm{th}} \simeq 1.15$.\\

Concerning the inner structure of $G_z$ in the vicinity of the CE, the spreading of the series of local extrema is analyzed by fitting them using an ansatz of the form $t_\mathrm{m} = a R^{\beta_{\mathrm{m}}^{\mathrm{num}}}+b$. It yields the dashed blue lines on Fig.~\ref{num_LRXY}(a). 
Once again, the latter feature a linear trajectory in the log-log scale corresponding to a power-law behavior in the lin-lin scale. As previously for the CE, by
computing their slope, one has access to the spreading exponent $\beta_{\mathrm{m}}^{\mathrm{num}}$ characterizing numerically the scaling law of the inner structure (the series of local extrema). 
For the example displayed on Fig.~\ref{num_LRXY}(a), we find $\beta_{\mathrm{m}}^{\mathrm{num}} = 0.60 \pm 0.06$ which is also in very good agreement with the 
theoretical exponent $\beta_{\mathrm{m}}^{\mathrm{th}} = 0.65$.\\

On Fig.~\ref{num_LRXY}(b), the evolution of the theoretical exponents ($\beta_{\mathrm{CE}}^{\mathrm{th}}$, $\beta_{\mathrm{m}}^{\mathrm{th}}$) and those extracted 
from the t-MPS data ($\beta_{\mathrm{CE}}^{\mathrm{num}}$, $\beta_{\mathrm{m}}^{\mathrm{num}}$) as a function of the power-law exponent $\alpha$ is investigated.
The power-law exponent $\alpha$ is contained in the interval $[1,3[$ so that the LRXY chain is in the quasi-local regime of the $x$ polarized phase. Besides, the 
pre- and post-quench antiferromagnetic interaction parameters are fixed and characterized by the value $\epsilon_\mathrm{i} = 0.2$ and $\epsilon_{\mathrm{f}} = 0$ respectively.
As shown on Fig.~\ref{num_LRXY}(b), the numerical exponents confirm the theoretical scaling laws for the correlation spreading in long-range lattice models confined in the quasi-local regime of a gapless phase.
More precisely, they show clearly a sub-ballistic propagation of the CE ($\beta_{\mathrm{CE}} > 1$) and a super-ballistic propagation of the series
of local extrema ($\beta_{\mathrm{m}} < 1$). Note that for large $\alpha$ in the quasi-local regime of the $x$ polarized phase ($\alpha \lesssim 3$), both 
the numerical and theoretical spreading exponents converge towards $\beta = 1$. For these typical values of $\alpha$, the LRXY chain is close to the local regime 
of the $x$ polarized phase where a twofold linear structure is expected for the space-time spin-spin correlation function. The latter is expected to be defined
by a CE propagating at the velocity $2V_{\mathrm{g}}^*$ (twice the maximal group velocity) and a series of local extrema, in the vicinity of this CE, spreading
at $2V_\varphi^*$ (twice the phase velocity at $k^*$, quasimomentum for which the group velocity is maximal). According to the shape of $E_k$ the gapless excitation spectrum 
of the LRXY chain in the local regime of the $x$ polarized phase [see Fig.~\ref{ek_vg_lrxy}(a)], the quasiparticles having a small quasimomentum $k$ ($k \rightarrow 0^+$)
are those spreading with the highest group velocity [see Fig.~\ref{ek_vg_lrxy}(b)]. Since $E_k$ is gapless and displays a phononic-like (quasi-linear)
behavior in the limit $k \rightarrow 0^+$, both the group and phase velocities are almost equal ($V_{\mathrm{g}}^* \simeq V_{\varphi}^*$). As a consequence,
the space-time pattern of the spin-spin correlations along the $z$ axis should display a single linear structure where both the CE and the series of local extrema propagate 
with a similar velocity. These statements have been verified analytically by investigating the causality cone of the spin fluctuations
$G_z$ for a sudden global quench where the post-quench Hamiltonian (Hamiltonian of the LRXY chain) is confined in the local regime of the $x$ polarized phase (see Appendix.~\ref{appendix_gz_lrxy} 
for the theoretical expression of $G_z$). \\ \\

To sum up, relying on both the t-MPS + TDVP numerical technique and our quasiparticle approach, we have shed new light on the quench dynamics for long-range
interacting lattice models. \\
Working within the case study of the long-range transverse Ising chain, we first verified
numerically, for sudden global quenches, the theoretical predictions concerning the scaling laws of the twofold algebraic structure displayed by the space-time correlations
in the quasi-local regime of a gapped phase. We confirmed the existence of a sub-ballistic motion for the CE (correlation edge), $t\sim R^{\beta_{\mathrm{CE}}}$ with
$\beta_{\mathrm{CE}} > 1$, and a ballistic spreading of the series of local maxima, $t \sim R^{\beta_{\mathrm{m}}}$, $\beta_{\mathrm{m}} = 1$. \\
The previous discussion on the correlation spreading has also been extended to the local regime where a twofold linear structure was expected. This theoretical prediction has
been verified as well as the associated spreading velocities. Indeed, we found that the CE spreads ballistically with the velocity $V_{\mathrm{CE}} \simeq 2V_{\mathrm{g}}^*$ and the series of local maxima with
$V_{\mathrm{m}} \simeq 2V_\varphi^*$. This behavior of the causality cone is reminiscent of the one found (theoretically at Chap.~\ref{ch:3-universal_scaling_laws} and numerically at Chap.~\ref{ch:4-bose_hubbard_chain}) 
for short-range interacting lattice models in a gapped or gapless quantum phase. The latter is due to the suppression of the divergence for the group velocity allowing us to define a maximal group velocity. \\
Then, we turned to a numerical analysis of several relevant physical quantities for sudden local quenches, 
while still considering the LRTI chain in the local regime of the gapped $z$ polarized phase. We first investigated the space-time local magnetization where a single ballistic
structure has been unveiled. It is characterized by a SE (spin edge) and a series of local extrema propagating at the maximal group velocity, \textit{ie.} $V_{\mathrm{SE}} \simeq V_{\mathrm{m}} \simeq V_{\mathrm{g}}^*$. 
Then, we analyzed the entanglement spreading \textit{via} the study of several space-time R\'enyi entropies in the local regime of the $z$ polarized phase. For each of them, we found a similar causality cone
fully characterized by an EE (entropy edge) spreading ballistically with the same velocity as previously, \textit{ie.} $V_{\mathrm{EE}} \simeq V_{\mathrm{g}}^*$.\\
Finally, we moved on another long-range interacting lattice model, namely the long-range XY chain where the quasi-local regime of the gapless 
$x$ polarized phase has been considered. The purpose was to verify the theoretical scaling laws of the twofold algebraic structure for the correlation spreading 
for the quasi-local regime of gapless quantum phases. Relying on the t-MPS + TDVP numerical approach, we have confirmed both the existence of a sub-ballistic motion of the CE 
and a super-ballistic spreading of the local maxima. \\

Note that there are several interesting extensions to this research work regarding the correlation spreading in isolated lattice models. For instance, a first possibility is to investigate numerically the 
correlation spreading in short- and long-range interacting lattice models in higher dimensions. The main purpose would be to verify our theoretical predictions 
concerning the scaling laws (and more precisely the spreading velocities for the short-range case) of the twofold structure for the causality cone. A second extension 
would be to consider open lattice models (interacting with their environment) and to know whether our theoretical scaling laws for the correlation spreading are 
modified or not.

%% file: text/ch6_conclusion.tex
\setstretch{1.0} 

\begin{savequote}[8cm]
\textlatin{“Ends are not bad things, they just mean that something else is about to begin. And there are many things that don't really end, anyway,
they just begin again in a new way. Ends are not bad and many ends aren't really an ending; some things are never-ending.”}
  \qauthor{--- C. JoyBell C.}
\end{savequote}

\chapter{\label{ch:6-conclusion} Conclusion} 

In this thesis, we have investigated theoretically and numerically the spreading of quantum correlations in isolated lattice models with short- or long-range interactions.
The latter are driven far from equilibrium \textit{via} sudden global or local quenches. Such research topic is at the center of many fundamental phenomena occuring in the
framework of the far-from-equilibrium dynamics in quantum mechanics, including the propagation 
of information and entanglement, the relaxation and thermalization processes. Furthermore, a main motivation for this specific study is due to the conflicting
results in the literature concerning the scaling laws of the correlation edge (CE) also called light-cone edge, its lack of universality and the incompleteness 
of the existing physical pictures to fully characterize the propagation of quantum correlations. \\

In order to give a general and complete description of the correlation spreading induced by sudden global quenches, we have introduced a quasiparticle approach 
relying both on a mean field approximation and the bosonic Bogolyubov theory, which is applicable to short-range and long-range interacting bosonic and spin
lattice models on a hypercubic lattice. The latter has permitted to unveil a generic form of the equal-time connected correlation functions whose space-time 
pattern has been analyzed within stationary phase arguments. We have revealed a universal twofold structure for the causality cone of the quantum correlations,
\textit{ie.} an outer and inner structure determining the CE and the propagation of local extrema respectively. \\
For short-range interacting quantum lattice models, they are readily associated to the group and phase velocities at $k^*$, the quasimomentum where the group velocity
reaches its maximum. More precisely, while the CE propagates ballistically at twice the maximal group velocity ($2V_{\mathrm{g}}^*$), the series of local extrema located 
in its vicinity will spread linearly with twice the phase velocity at the quasimomentum where the group velocity is maximal ($2V_{\varphi}^*$).
Since these characteristic velocities generally differ, the correlation spreading in short-range lattice models is fully characterized by a twofold linear structure. \\ 
For long-range quantum systems with fast decaying long-range interactions, defining the so-called the local regime, the correlation spreading
is identical to the case of short-range interactions. The causality cone is characterized by a twofold linear structure, a CE and a series of local extrema
in its vicinity spreading ballistically with twice the maximal group velocity and twice the phase velocity at $k^*$ respectively. However, for long-range quantum systems 
with intermediate-range interactions, defining the so-called quasi-local regime, the correlation spreading and more precisely the scaling laws of its associated
twofold structure are drastically different. By relying on our quasiparticle approach together with the stationary phase approximation, the spreading of the CE is
found to depend not only on the observable but also on the decay of the long-range interactions and on the dimensionality of the lattice. Nevertheless, a general
feature concerning the propagation of the CE has been unveiled. Indeed, independently of whether the excitation spectrum is gapped or gapless, the CE spreads
always sub-ballistically, \textit{ie.} slower than ballistically. However, in the vicinity of the CE, the series of local extrema propagate ballistically in gapped
systems contrary to the gapless case where the extrema spread super-ballistically, \textit{ie.} faster than ballistically. For the latter, the motion of the extrema is fully
determined by the quasimomentum-dependence of the excitation spectrum depending on the decay of the power-law interactions and on the lattice dimensionality. \\
For the case of short-range interactions, the theoretical predictions are confirmed by a detailed study of the one-dimensional (1D) Bose-Hubbard model in both the 
gapped Mott-insulating and gapless superfluid phases. To do so, the equal-time connected one-body ($G_1$) and density-density ($G_2$) correlation functions have been derived
analytically. For long-range interactions, the scaling laws predicted by our quasiparticle approach are verified for two different one-dimensional long-range
interacting $s=1/2$ spin lattice models, namely the XY chain in the $x$ polarized phase and the transverse Ising chain in the $z$ polarized phase
corresponding to a gapless and a gapped quantum system respectively. For these long-range quantum systems, the analytical expression of equal-time 
connected spin-spin correlation functions have been calculated. \\

Another important research work presented in this manuscript has been devoted to a numerical investigation of the correlation spreading induced by sudden
global quenches in the short-range interacting lattice model considered previously, the Bose-Hubbard chain. The purpose of this study is twofold. On the one hand, 
we aim at testing the theoretical predictions of our quasiparticle approach against a numerically exact approach, beyond the mean field approximation.
On the other hand, we aim at extending the general picture to quantum regimes that are not amenable to analytic treatments. \\
Relying on the time-dependent matrix product state (t-MPS) approach, our numerical results fully confirm the analytical predictions provided by our quasiparticle approach 
in their respective regimes of validity, namely in the superfluid mean field regime and deep in the Mott-insulating phase. More generally, we have presented evidence 
of a universal twofold linear dynamics in the spreading of correlations for this bosonic lattice model. Indeed, this twofold linear structure for the causality cone of 
the quantum correlations has been found in all the phases and regimes of the bosonic model and for different relevant observables, namely the $G_1$ and $G_2$
correlation functions characterizing the phase and density fluctuations respectively. Exceptions appear only in a few cases, for instance (i) for specific observables
in specific regimes, \textit{e.g.} the $G_2$ correlations in the strongly interacting regime of the Mott-insulating phase, or (ii) when the two velocities are
(almost) equal, as found at the critical points of the Mott-$U$ and Mott-$\delta$ transitions for instance. \\

We also have extended the study to sudden local quenches. This investigation has been performed both theoretically and numerically by relying on the same methods
as above (quasiparticle theory and t-MPS numerical approach) for two distinct one-dimensional short-range interacting lattice models.
The latter are the Bose-Hubbard chain confined deep in the Mott-insulating phase and the $s = 1/2$ Heisenberg chain in the 
ferromagnetic phase along the $z$ axis to treat the case of a gapped and a gapless quantum system respectively. Furthermore, to characterize the local quench dynamics,
on-site observables have been investigated, \textit{e.g.} the local density for the bosonic lattice chain and the local magnetization for the $s=1/2$
spin chain. \\
For both cases, the space-time pattern has been found to display not only a causal and a non-causal region separated by an edge propagating 
ballistically but also a series of local extrema in the vicinity of this edge which also spread linearly. At this stage, the previous properties are 
reminiscent of those for sudden global quenches. However, for sudden local quenches, the space-time pattern displays a single linear structure implying that the 
edge and the series of local extrema propagate with the same velocity. Besides, the spreading velocity of the edge, and thus the one for the extrema, is not
characterized anymore by twice the maximal group velocity $2V_{\mathrm{g}}^*$ but by $V_{\mathrm{g}}^*$. Indeed, while for sudden global quenches, the causality
cone is governed by free and counterpropagating quasiparticle excitation pairs, for sudden local quenches, it is governed by the spreading of individual quasiparticles.
Consequently, the motion associated to the edge is governed by the fastest individual quasiparticle leading to the spreading velocity $V_{\mathrm{g}}^*$.
The previous statements also apply to other quantum lattice models such as the short-range Ising model and for different local observables, \textit{e.g.} the spin-spin correlations. \\

Our investigations related to the local and global quench dynamics in short-range lattice models have been extended to the case of long-range quantum systems,
where the long-range interactions are characterized by a decaying power-law function of the form $1/|R|^{\alpha}$, such as $s=1/2$ spin lattice models 
realized in trapped-ion experiments. In such quantum systems, a possible divergence of the group velocity can be generated by tuning the power-law exponent
$\alpha$ defining the decay of the long-range interactions. Such divergence defining the so-called quasi-local regime \footnote{In the present discussion 
and in the following ones, the quasiparticle energy is assumed to be always bounded within the first Brillouin zone.} requires intermediate-range interactions,
\textit{ie.} a relatively small value of $\alpha$. \\
For this regime and for sudden global quenches, our quasiparticle approach still predicts a twofold dynamics whose
CE and local maxima do not propagate ballistically anymore but algebraically. In this case, it is characterized by the coexistence of a super-ballistic
(for gapless systems) or ballistic (gapped case) signal for the series of local extrema and a sub-ballistic one for the CE.
These statements have been verified numerically by considering two distinct long-range interacting $s=1/2$ spin lattice models, namely the long-range XY chain in
the $x$ polarized phase and the long-range transverse Ising chain in the $z$ polarized phase to treat the case of a gapless and gapped model respectively. \\
Using the t-MPS approach within the time-dependent variational principle (TDVP), we have not only shed new light on the still debated scaling of the light-cone boundary
but also on the presence of a second structure consisting of a series of local extrema. Besides, our numerical results have confirmed the analytical predictions provided
by our quasiparticle approach concerning the scaling laws of the twofold algebraic structure. For both long-range spin lattice models, we also have shown that the motion of the CE
and the local extrema converge towards a ballistic propagation when increasing $\alpha$. \\

Furthermore, this numerical investigation has been extended to the local regime. The latter requires fast decaying long-range interactions, \textit{ie.}
a relatively large value of $\alpha$, and is characterized by a well-defined maximal group velocity. Within the case study of the 1D LRTI model in the gapped
$z$ polarized phase, the numerical results have confirmed the presence of a twofold linear structure for the correlation spreading predicted by our generic quasiparticle
approach and suggested by the previous numerical study in the quasi-local regime. The latter consists of a CE and a series of local extrema propagating ballistically
with the velocity $2V_{\mathrm{g}}^*$ and $2V_{\varphi}^*$ respectively. This specific behavior of the causality cone for the correlation spreading is reminiscent
of the one for short-range interacting lattice models and can be explained by the divergence suppression of the group velocity in the local regime. \\

While still considering the one-dimensional long-range transverse Ising model in the local regime of the $z$ polarized phase, we also have performed a
numerical investigation of the correlation and entanglement spreading for a dynamics induced \textit{via} sudden local quenches. \\
We first investigated the space-time local magnetization and spin-spin correlations where a single ballistic structure has been found. Both the SE (spin edge)
for the local magnetization or the CE for the spin correlations and the associated series of local extrema propagate at the maximal group velocity $V_{\mathrm{g}}^*$. \\
Then, we analyzed the information spreading \textit{via} the study of several space-time R\'enyi entropies. For each of them, we found a similar causality cone
fully characterized by an EE (entropy edge) spreading ballistically with the same velocity as previously, \textit{ie.} $V_{\mathrm{g}}^*$. Such causality cone is also 
reminiscent of the one for the spreading of entanglement in short-range interacting lattice models and is thus consistent with the Lieb-Robinson bound. \\ 

After completion of this manuscript, we have extended this numerical study to the quasi-local regime. The main message is that, for the spreading of correlations
and entanglement, we find a sub-ballistic motion of the CE and EE respectively. Note that such sub-ballistic motion is also valid for the SE associated to the space-time 
pattern of the local magnetization. To conclude, this slower-than-ballistic propagation of the CE and EE for sudden local quenches is similar to the one for
the correlation spreading induced \textit{via} sudden global quenches and thus can be considered as a generic property of the quench dynamics in isolated 
long-range interacting lattice models.      

\paragraph{Outlook and perspectives} $~~$ \\

This research work devoted to a better understanding of the quench dynamics in isolated quantum lattice models can be extended towards several directions. \\

An important extension would be to verify experimentally the theoretical predictions of our quasiparticle approach for the correlation spreading
in one-dimensional short- and long-range interacting quantum lattice models. \\
For short-range interacting models, the purpose would be to confirm the presence of a twofold linear structure and the associated spreading velocities.
Indeed, while in most experiments and numerics the CE is inferred from the behavior of the correlation maxima, our results stress that the two must be distinguished.
This is expected to be a general feature of short-range systems. Such experimental study can be performed on the Bose-Hubbard chain using ultracold Bose gases 
loaded in optical lattices, where the dynamics of the phase and density correlations can be observed on relevant space and time scales comparable to those
considered in our simulations. \\
For long-range interacting models with intermediate-range interactions, the twofold algebraic structure for the propagation of quantum correlations can also be 
observed experimentally by investigating for instance the spin correlations in cold ion chains. Our analysis provides the first step of an important research problem
that aims at unveiling the physical information encoded in correlation spreading and how this can be extracted in the next generation of experiments. Besides, our theory 
shows that in generic experiments, characterizing the spreading of correlations for both particle and spin lattice models, the data need to be interpreted carefully. As discussed
previously, the propagation of local extrema does not characterize the correlation edge at all. Both are independent and rely on different physical properties of 
the isolated lattice model. For instance, identifying the latter requires an accurate scaling analysis of the leaks and to consider relatively large length chains,
\textit{ie.} having several decades of sites. \\

Furthermore, since our quasiparticle approach is very general and is expected to hold to describe the correlation spreading in a large class of
isolated quantum lattice models, there are different research works to perform in order to test and to possibly extend its validity. \\
For instance, a possible extension would be to investigate the correlation spreading in isolated interacting quantum models for a dimensionality of the lattice
higher than one. The main purpose would consist of verifying our theoretical predictions concerning the scaling laws and spreading velocities for the 
propagation of quantum correlations in long-range and short-range interacting lattice models respectively on a square or cubic lattice. This study can be 
performed not only numerically using tensor network techniques or the time-dependent variational Monte-Carlo approach, at least for the two-dimensional case, but
also experimentally using for example ultracold Bose gases loaded in a two-dimensional or three-dimensional optical lattice. \\
A second extension would concern the generic form of the equal-time correlation functions and more precisely to know if the latter holds to describe accurately 
the correlation spreading in both short- and long-range fermionic lattice models driven far from equilibrium \textit{via} sudden global quenches.
The generic approach to deal with this problem will be substantially similar to the one used for bosonic lattice models. The main difference is to
replace the quadratic Bose form for the generic Hamiltonian by a quadratic Fermi form and to rely on the fermionic Bogolyubov theory. \\ 
And last but not least, another extension would be to consider open lattice models which are coupled to their environment and to know whether 
the theoretical scaling laws provided by our quasiparticle approach for the correlation spreading are modified or not. \\

As pointed out previously, another research direction consists of unveiling the physical information encoded in correlation spreading and how 
this can be extracted in the next generation of experiments. More precisely, the latter aims at extracting equilibrium properties of a quantum system \textit{via}
the study of its quench dynamics. \\
For instance, we have shown in this manuscript that the structure of the quantum correlations in the vicinity of the
causal edge can be related to basic properties of the elementary excitations of the quench Hamiltonian. This includes characteristic velocities ($2V_{\mathrm{g}}^*$, 
$2V_{\varphi}^*$) for short-range lattice models and long-range lattice models with fastly-decaying interactions or again the quasimomentum dependence of the excitation spectrum \footnote{The quasimomentum dependence of the excitation 
spectrum is provided by the scaling law of the local maxima for gapless quench Hamiltonians.} and the presence of a gap \footnote{The presence of a gap in the
excitation spectrum of the quench Hamiltonian is determined by the scaling law of the local maxima. If the latter is ballistic then a gap is present.
However, if the spreading of the local maxima is characterized by a super-ballistic motion then the post-quench Hamiltonian is gapless.} for long-range lattice models 
with intermediate-range interactions. \\
Finally, we also have shown analytically and numerically that the excitation spectrum of the quench Hamiltonian can be fully determined by a two-dimensional
Fourier transform of the equal-time correlation functions, also called quench spectral function (QSF) in Ref.~\cite{despres2019bis2}. From an experimental point of view,
such approach may considerably simplify the determination of excitation spectra in correlated systems compared to standard pump-probe spectroscopy techniques,
such as ARPES (angle-resolved photoemission spectroscopy) or Bragg spectroscopy. The latter consist of exciting the system at a well-defined frequency and
wavevector, and observing the response of the system after some interaction time. In practice, it requires to control the probe and systematically scan both
the frequency and the wavevector. In QSF spectroscopy, the sudden global quench \textit{ie.} the sudden modification of a parameter of the Hamiltonian 
replaces the pump. Indeed, it generates a complete set of excitations that propagate throughout the lattice. At a given time $t$ after the quantum quench,
the spatial dependence of the correlation function is measured by a direct imaging of the full system, as now commonly done in atomic, molecular and optical physics. For
instance, in bosonic lattice models, the one-body correlations can be deduced from standard time-of-flight techniques. For the density-density correlations, the bosonic 
atoms can be detected \textit{via} a fluorescence technique and it also requires a series of images to measure the density fluctuations. The space-time
pattern of the correlations is then reconstructed by scanning the time $t$ from $0$, time where the sudden global quench is applied, to $T$ the observation time. 
An extension of the QSF approach would be for instance to determine if the latter can be generalized to sudden local quenches.

%% file: text/appendix_1.tex
\setstretch{1.0} 

\chapter{\label{appendix2_sp} Asymptotic behavior of the generic connected correlation function}

In this appendix, one presents the derivation of the asymptotic behavior, \textit{ie.} long-time and long-distance behavior along
a constant line $R/t$, of $G(\mathbf{R},t)$ the equal-time connected correlation function having the following generic form 

\begin{equation}
 G(\mathbf{R},t) \sim \int_{\mathcal{B}} \frac{\mathrm{d}\mathbf{k}}{(2\pi)^D} \mathcal{F}(\mathbf{k})
 \left \{ \frac{e^{i(\mathbf{k}.\mathbf{R}+2E_\mathbf{k}t)} + e^{i(\mathbf{k}.\mathbf{R}-2E_\mathbf{k}t)}}{2} \right \}.
 \label{new_generic_form}
\end{equation}

\noindent
According to the stationary phase approximation, the previous $D$-dimensional integral is dominated by the quasimomentum contributions 
with a stationary phase (sp) corresponding to the condition

\begin{equation} 
\mathbf{k}_{\mathrm{sp}} : \partial_{\mathbf{k}} \left( \mathbf{k}.\mathbf{R} \mp 2E_\mathbf{k}t \right) = 0 ~~~\textrm{equivalent~to}~~~ 2V_{\mathrm{g}}(\mathbf{k}_{\mathrm{sp}}) = \pm R/t.
\label{ksp_app}
\end{equation}

\noindent
Using a second order Taylor expansion around $\mathbf{k}_{\mathrm{sp}}$, the stationary-phase quasimomentum fulfilling the previous condition, of the phase function
$\Phi(\mathbf{k}) = \mathbf{k}.\mathbf{R} - 2E_\mathbf{k}t$, it yields

\begin{equation}
 \Phi(\mathbf{k}) = \Phi(\mathbf{k}_{\mathrm{sp}}) + \frac{1}{2} \partial^2_{\mathbf{k}} \Phi(\mathbf{k}_{\mathrm{sp}})
 (\mathbf{k} - \mathbf{k}_{\mathrm{sp}})^2 + \mathcal{O}\left[(\mathbf{k} - \mathbf{k}_{\mathrm{sp}})^3\right],
 \label{phase_fct}
\end{equation}

\noindent
since the first derivative of the phase function $\Phi(\mathbf{k})$ with respect to the quasimomentum $\mathbf{k}$ and evaluated 
at $\mathbf{k}_{\mathrm{sp}}$ is equal to zero, $\partial_{\mathbf{k}} \Phi(\mathbf{k}_{\mathrm{sp}}) = 0$. Hence,
by inserting Eq.~\eqref{phase_fct} into Eq.~\eqref{new_generic_form}, one obtains the following expression for $G(\mathbf{R},t)$,

\begin{align}
& G(\mathbf{R},t) \sim \mathcal{F}(\mathbf{k}_{\mathrm{sp}})  \Re  \left\{ e^{i \Phi(\mathbf{k}_{\mathrm{sp}})} \int_{\mathcal{B}}
\mathrm{d}\mathbf{k}~e^{i\frac{1}{2} \partial^2_{\mathbf{k}} \Phi(\mathbf{k}_{\mathrm{sp}})
 (\mathbf{k} - \mathbf{k}_{\mathrm{sp}})^2} \right\}.
\end{align}

\noindent
Using the following formula for imaginary gaussian integrals,

\begin{equation}
\int_{\mathcal{B}} \mathrm{d} \mathbf{k}~e^{i\frac{1}{2} \partial^2_{\mathbf{k}} \Phi(\mathbf{k}_{\mathrm{sp}}) (\mathbf{k} - \mathbf{k}_{\mathrm{sp}})^2} = 
\left(\frac{2\pi}{|\partial^2_{\mathbf{k}} \Phi(\mathbf{k}_{\mathrm{sp}})|} \right)^{\frac{D}{2}} e^{i\sigma D \frac{\pi}{4}},~~~\sigma = \mathrm{sgn}\left[ 
\partial^2_{\mathbf{k}} \Phi(\mathbf{k}_{\mathrm{sp}}) \right],
\end{equation}

\noindent
the asymptotic behavior of the equal-time connected correlation function $G(\mathbf{R},t)$ is given by

\begin{equation}
 G(\mathbf{R},t) \sim \mathcal{F}(\mathbf{k}_{\mathrm{sp}}) 
 \left(\frac{2\pi}{|\partial^2_{\mathbf{k}} \Phi(\mathbf{k}_{\mathrm{sp}})|} \right)^{\frac{D}{2}} \Re \left\{ e^{i[ \Phi(\mathbf{k}_{\mathrm{sp}}) + \sigma D \frac{\pi}{4}]} \right\}.
\end{equation}

\noindent
Then, using that $\partial^2_{\mathbf{k}} \Phi(\mathbf{k}_{\mathrm{sp}}) = -2\partial^2_{\mathbf{k}} E_{\mathbf{k}_{\mathrm{sp}}}t$, the final form of $G(\mathbf{R},t)$ in the asymptotic limit 
\footnote{For simplicity, we assumed a single solution $\mathbf{k}_{\mathrm{sp}}$. If Eq.~\eqref{ksp_app}
has several solutions, one has to sum over the different contributions where each of them corresponds to a specific solution $\mathbf{k}_{\mathrm{i},\mathrm{sp}}$ \cite{cevolani2015}.} may be written as

\begin{equation}
 G(\mathbf{R},t) \sim \frac{\mathcal{F}(\mathbf{k}_{\mathrm{sp}})}{\left(|2\partial^2_{\mathbf{k}} E_{\mathbf{k}_{\mathrm{sp}}}|t \right)^{\frac{D}{2}}} \cos \left( \mathbf{k}_{\mathrm{sp}}.\mathbf{R} - 2 E_{\mathbf{k}_{\mathrm{sp}}}t
 + \sigma D \frac{\pi}{4} \right),~~~\sigma = \mathrm{sgn}\left(-2 \partial^2_{\mathbf{k}} E_{\mathbf{k}_{\mathrm{sp}}}t \right).
 \label{final_eq_gen}
\end{equation}

\noindent
Finally, by analyzing Eq.~\eqref{final_eq_gen}, the scaling laws for the correlation spreading in short- or long-range interacting quantum lattice systems can be deduced.
The outer structure, characterized by a correlation edge, is determined by the prefactor whereas the inner structure, composed of a series of local extrema, is determined by the 
argument of the cosine function.

%% file: text/appendix_2.tex
\setstretch{1.0} 

\chapter{\label{appendix1_mf} Mean field regime of the superfluid phase in the 1D short-range Bose-Hubbard model}

In this appendix, one presents a derivation of the mean field condition for the one-dimensional short-range Bose-Hubbard
(1D SRBH) model. To do so, the starting point is to find the one associated to the 1D Lieb-Liniger (LL) model and then to
rely on the correspondence between the parameters of the 1D SRBH and 1D LL models by discretizing the 1D LL model
on the length scale defined by the lattice spacing. \\

The 1D continuous-space LL model can be expressed as follows

\begin{equation}
\label{LLmodel}
\hat{H} = \frac{\hbar^2}{2m} \left[ - \sum_{i=1}^{N} \frac{\partial^2}{\partial x_i^2} + c\sum_{i\neq j}
\delta(x_i - x_j) \right].
\end{equation}

\noindent
It represents a one-dimensional gas of $N$ bosons of mass $m$ interacting via a two-body potential and more
precisely with a contact interaction \footnote{\textit{via} the regularized Dirac distribution where $\delta(x_i-x_j) = 1$ if $x_i=x_j$ and $0$ otherwise.}, characterized by the
repulsive interaction strength $c>0$. \\

\noindent
We start by providing a derivation of the dimensionless interaction parameter $\gamma = E_{\mathrm{int}}/E_{\mathrm{kin}}$
with $E_\mathrm{int}$ denoting the typical interaction energy and $E_{\mathrm{kin}}$ the typical kinetic energy. 
The previous energies $E_{\mathrm{int}}$ and $E_{\mathrm{kin}}$ can be written as follows 

\begin{equation}
 E_{\mathrm{int}} = \frac{\hbar^2 c \rho N}{2m},~~~E_{\mathrm{kin}} = \frac{\hbar^2N}{2ma^2},
\end{equation}

\noindent
with $a$ the typical length scale and $\rho = N/L$ the density of the one-dimensional gas. Then, the dimensionless interaction parameter $\gamma$ can be rewritten as

\begin{equation}
 \gamma = \frac{E_{\mathrm{int}}}{E_{\mathrm{kin}}} = \frac{\hbar^2 c \rho N}{2m} \frac{2ma^2}{\hbar^2 N} = c\rho a^2,
\end{equation}

\noindent
and considering that $a \rho = 1$, thus $a^2 = \rho^{-2}$, it yields the following expression for $\gamma$

\begin{equation}
 \gamma = \frac{c}{\rho}.
 \label{gamma_th}
\end{equation}

\noindent
According to Eq.~\eqref{gamma_th}, the 1D LL model possesses a crossover between a mean field regime at large density and
small interactions ($\gamma \ll 1$) and a strongly interacting regime at small density and large interactions
($\gamma \gg 1$). \\

\noindent
Finally, the correspondence between the parameters of the BH and LL models is found by discretizing the LL model, see Eq.~(\ref{LLmodel}),
on the length scale defined by the lattice spacing $a$. It yields both equations

\begin{equation}
J = \frac{\hbar^2}{2ma^2},~~~ U = \frac{\hbar^2 c}{ma},
\label{params_LL_BH}
\end{equation}

\noindent
where $J$ denotes the hopping amplitude and $U$ the repulsive two-body interaction of the 1D SRBH model. Replacing 
both expressions at Eq.~\eqref{params_LL_BH} in the theoretical expression of $\gamma$ leads to

\begin{equation}
\gamma = \frac{U}{2J\rho a} = \frac{U}{2J \bar{n}},~~~\bar{n} = \rho a = \frac{N}{L}a = \frac{N}{N_s},
\label{gamma_final}
\end{equation}

\noindent
with $N_s$ the number of lattice sites. The dimensionless interaction parameter $\gamma$ at Eq.~\eqref{gamma_final} for the 
1D SRBH model allows us to characterize the different quantum regimes of the superfluid (SF) phase. Indeed, similarly to the
1D LL model, the 1D SRBH model in the SF phase is confined in the strongly interacting regime for $\gamma \gg 1$ and in
the mean field regime for $\gamma \ll 1$.

%% file: text/appendix_5.tex
\setstretch{1.0} 

\chapter{\label{appendix_g1_sf} $G_1$ phase fluctuations in the superfluid mean field regime of the Bose-Hubbard chain}
In this section, we give a brief outline of the derivation of the $G_1$ phase fluctuations for the Bose-Hubbard chain
in the SF mean field regime. To drive the bosonic lattice model far from equilibrium, a sudden global quench confined in the SF mean field regime is considered.
While the hopping amplitude $J$ and the filling $\bar{n}$ are fixed, a sudden modification of the repulsive two-body interaction $U$ is performed such that
$U_{\mathrm{i}}/2J\bar{n}, U_{\mathrm{f}}/2J\bar{n} \ll 1$. The calculation is based on a quasiparticle picture using the bosonic Bogoliubov theory. In
the following, the Bose-Hubbard is assumed to contain $N_s$ lattice sites. Besides, $\hbar$ and $a$ the lattice spacing are fixed to unity by convention. \\

One recalls that the $G_1$ connected one-body correlation function may be written as follows

\begin{align}
& G_{1}(R,t) = \langle \hat{a}_{R}^{\dag}(t) \hat{a}_{0}(t) \rangle - \langle \hat{a}_{R}^{\dag}(0) \hat{a}_{0}(0) \rangle.
\end{align}

\noindent
Expressing the latter into the Fourier (momentum) space and using the momentum conservation coming from the translational invariance of the model leads to 

\begin{equation}
G_{1}(R,t) = \frac{1}{N_s}\sum_{k} e^{-i k R} \left( \langle \hat{n}_{k,\mathrm{f}}(t) \rangle - \langle \hat{n}_{k,\mathrm{f}}(0) \rangle \right),
\end{equation}

\noindent
where $\hat{n}_{k,\mathrm{f}}(t) = \hat{a}^{\dag}_{k,\mathrm{f}}(t) \hat{a}_{k,\mathrm{f}}(t)$ denotes the post-quench bosonic occupation number for a momentum $k$ at time $t$. At this stage, $G_1$ depends only on the post-quench
bosonic operators in the momentum space \textit{ie.} $\hat{a}_{k,\mathrm{f}}$ and $\hat{a}_{k,\mathrm{f}}^{\dag}$. Then, relying on the Bogolyubov
transformation for the post-quench operators given as follows,

\begin{equation}
\hat{a}_{k,\mathrm{f}}(t) = u_{k,\mathrm{f}} \hat{\beta}_{k,\mathrm{f}}(t) + v_{-k,\mathrm{f}} \hat{\beta}^{\dag}_{-k,\mathrm{f}}(t), ~~~ 
\hat{a}^{\dag}_{k,\mathrm{f}}(t) = u_{k,\mathrm{f}} \hat{\beta}^{\dag}_{k,\mathrm{f}}(t) + v_{-k,\mathrm{f}} \hat{\beta}_{-k,\mathrm{f}}(t). 
\end{equation}

\noindent
One can express $\hat{a}_{k,\mathrm{f}}(t)$ and $\hat{a}_{k,\mathrm{f}}^{\dag}(t)$ as a function of the post-quench Bogolyubov quasiparticle operators
$\hat{\beta}_{k,\mathrm{f}}(t)$ and $\hat{\beta}^{\dag}_{k,\mathrm{f}}(t)$. Note that the post-quench Bogolyubov operators diagonalize the post-quench
Hamiltonian $\hat{H}_{\mathrm{f}} = \hat{H}\left(U_{\mathrm{f}}/J\right)$. Hence, their time-dependent version takes a simple expression.
Indeed, relying on the equation of motion 

\begin{equation}
-i \partial_t \hat{\beta}_{k,\mathrm{f}}^{(\dag)}(t) = \left[\hat{H}, \hat{\beta}_{k,\mathrm{f}}^{(\dag)} \right]_t, 
\end{equation}

\noindent
it can be shown that 
 
\begin{equation}
\hat{\beta}_{k,\mathrm{f}}(t) = e^{-i E_{k,\mathrm{f}}t} \hat{\beta}_{k, \mathrm{f}}(0),~~~ \hat{\beta}^{\dag}_{k,\mathrm{f}}(t) =
e^{i E_{k,\mathrm{f}}t} \hat{\beta}^{\dag}_{k,\mathrm{f}}(0).
\end{equation}

\noindent
Then, the continuity at $t=0$ between the post-quench and pre-quench bosonic operators, \textrm{ie.} $\hat{a}^{(\dag)}_{k,\mathrm{f}}(0) = \hat{a}^{(\dag)}_{k,\mathrm{i}}$,
permits to find a relation between the post-quench Bogolyubov operators at $t=0$ ($\hat{\beta}_{k,\mathrm{f}}(0), \hat{\beta}^{\dag}_{k,\mathrm{f}}(0)$) and the pre-quench
Bogolyubov operators ($\hat{\beta}_{k,\mathrm{i}}, \hat{\beta}^{\dag}_{k,\mathrm{i}}$). The continuity condition is given by
 
 \begin{align}
 & \hat{a}_{k,\mathrm{f}}(0) = u_{k,\mathrm{f}} \hat{\beta}_{k,\mathrm{f}}(0) + v_{-k,\mathrm{f}} \hat{\beta}_{-k, \mathrm{f}}^{\dag}(0) = u_{k,\mathrm{i}} \hat{\beta}_{k,\mathrm{i}}
 + v_{-k,\mathrm{i}}\hat{\beta}_{-k,\mathrm{i}}^{\dag} \\
 & \hat{a}_{k,\mathrm{f}}^{\dag}(0) = u_{k,\mathrm{f}} \hat{\beta}_{k,\mathrm{f}}^{\dag}(0) + v_{-k,\mathrm{f}} \hat{\beta}_{-k,\mathrm{f}}(0)= u_{k,\mathrm{i}}
 \hat{\beta}_{k,\mathrm{i}}^{\dag} + v_{-k,\mathrm{i}}\hat{\beta}_{-k,\mathrm{i}}
 \end{align}
 
 \noindent
 leading for the post-quench Bogolyubov operators at $t=0$ to 
 
 \begin{align}
 & \hat{\beta}_{k,\mathrm{f}}(0) = \left( u_{k, \mathrm{i}} u_{k, \mathrm{f}} - v_{-k,\mathrm{i}} v_{-k,\mathrm{f}} \right) \hat{\beta}_{k, \mathrm{i}} -
 \left( u_{k, \mathrm{i}} v_{-k,\mathrm{f}} - v_{-k,\mathrm{i}} u_{k,\mathrm{f}} \right) \hat{\beta}_{-k, \mathrm{i}}^{\dag} \\
 & \hat{\beta}_{k, \mathrm{f}}^{\dag}(0) = \left( u_{k, \mathrm{i}} u_{k, \mathrm{f}} - v_{-k,\mathrm{i}} v_{-k,\mathrm{f}} \right) \hat{\beta}_{k,\mathrm{i}}^{\dag}
 - \left( u_{k,\mathrm{i}} v_{-k,\mathrm{f}} - v_{-k,\mathrm{i}} u_{k,\mathrm{f}} \right) \hat{\beta}_{-k,\mathrm{i}} 
 \end{align}
 
\noindent
By following the previous steps, the $G_1$ phase fluctuations involve only the pre-quench Bogolyubov operators $\hat{\beta}_{k,\mathrm{i}}, \hat{\beta}^{\dag}_{k,\mathrm{i}}$. 
Finally, using the condition $\hat{\beta}_{k, \mathrm{i}} \ket{\mathrm{GS}_\mathrm{i}} = 0$, where $\ket{\mathrm{GS}_\mathrm{i}}$ denotes the ground state of
the pre-quench Hamiltonian $\hat{H}_\mathrm{i} = \hat{H}\left(U_{\mathrm{i}}/J\right)$, one finds the following analytical expression for $G_1$

\begin{equation}
G_1(R,t) = \frac{4}{N_s} \sum_{k} e^{-i k R} \xi^{(1)}_{\mathrm{i}, \mathrm{f}}(k) \sin^{2}(E_{k,\mathrm{f}}t),
\end{equation}

\noindent
where $\xi^{(1)}_{\mathrm{i}, \mathrm{f}}(k)$ is a quantity depending on the quasimomentum $k$ and on the pre- (post-) quench parameters \textit{via} the coefficients
$u_{k,\mathrm{i}(\mathrm{f})}$ and $v_{k,\mathrm{i}(\mathrm{f})}$, 

\begin{equation}
\xi^{(1)}_{\mathrm{i}, \mathrm{f}}(k) = u_{k, \mathrm{f}}v_{k, \mathrm{f}} \left( u_{k,\mathrm{i}} u_{k, \mathrm{f}} -
v_{k, \mathrm{i}} v_{k, \mathrm{f}} \right) \left( u_{k,\mathrm{i}} v_{k, \mathrm{f}} - v_{k,\mathrm{i}}u_{k,\mathrm{f}} \right).
\end{equation}

\noindent
We then have to replace $u_{k,\mathrm{i}(\mathrm{f})}$ and $v_{k,\mathrm{i}(\mathrm{f})}$ by their expression 

\begin{equation}
u_{k,\mathrm{i}(\mathrm{f})}, v_{k,\mathrm{i}(\mathrm{f})} = \pm \left[ \frac{1}{2} \left(\frac{\mathcal{A}_{k,\mathrm{i}(\mathrm{f})}}{E_{k,\mathrm{i}(\mathrm{f})}}
\pm 1 \right) \right]^{1/2},
\end{equation}

\noindent
where $E_{k,\mathrm{i}(\mathrm{f})} = \sqrt{\mathcal{A}^2_{k,\mathrm{i}(\mathrm{f})}
-\mathcal{B}^2_{k,\mathrm{i}(\mathrm{f})}}$ with $\mathcal{B}_{k,\mathrm{i}(\mathrm{f})} = \gamma_k + U_{\mathrm{i}(\mathrm{f})}\bar{n}$
and $\mathcal{A}_{k,\mathrm{i}(\mathrm{f})} = U_{\mathrm{i}(\mathrm{f})} \bar{n}$. Besides, $\gamma_k = 4J \sin^2(k/2)$ represents the dispersion
relation of the free tight-binding model. One obtains 

\begin{align}
& \xi^{(1)}_{\mathrm{i}, \mathrm{f}}(k) = \frac{\left(\mathcal{A}_{k, \mathrm{i}} \mathcal{B}_{k,\mathrm{f}} - \mathcal{A}_{k, \mathrm{f}} \mathcal{B}_{k, \mathrm{i}}
\right)\mathcal{B}_{k, \mathrm{f}}}{4 E_{k,\mathrm{i}} E_{k,\mathrm{f}}^{2}} = \frac{\bar{n}^2 U_\mathrm{f} \gamma_{k}(U_\mathrm{f} - U_\mathrm{i})}{4 E_{k,\mathrm{i}}
E_{k,\mathrm{f}}^2}.
\end{align}

\noindent
Considering the thermodynamic limit for the one-dimensional Bose-Hubbard model ($N_s \rightarrow +\infty$), the discrete sum over the quasimomentum can be expressed
in a continuous form leading to 

\begin{equation}
 \frac{1}{N_s} \sum_{k} \underset{N_s \rightarrow +\infty}{\rightarrow} \int_{\mathcal{B}}\frac{\mathrm{d}k}{2 \pi},
\end{equation}

\noindent
where $\mathcal{B} = [-\pi, \pi]$ denotes the first Brillouin zone. Considering also that   

\begin{equation}
e^{-i k R} \sin^{2}(E_{k,\mathrm{f}}t) = \frac{1}{2} \left( e^{-i k R } - \frac{e^{-i (k R - 2E_{k,\mathrm{f}}t)} + e^{-i( k R + 2E_{k,\mathrm{f}}t)} }{2} \right),
\end{equation}

\noindent
the $G_1$ connected correlation function can be written as follows 

\begin{align}
& G_1(R,t) = 2\int_{\mathcal{B}} \frac{\mathrm{d} k}{2 \pi} \xi^{(1)}_{\mathrm{i}, \mathrm{f}}(k) \left( e^{-i k R } - \frac{e^{-i (k R - 2E_{k,\mathrm{f}}t)} + 
e^{-i( k R + 2E_{k,\mathrm{f}}t)} }{2} \right).
\end{align}

\noindent
Finally, one considers only the relevant (time-dependent) part of the latter to describe the far-from-equilibrium dynamics of the phase fluctuations and by fixing
$\mathcal{F}_1(k) = 2\xi^{(1)}_{\mathrm{i}, \mathrm{f}}(k)$, $G_1$ is expressed under the generic form presented at Eq.~\eqref{generic_form} and reads as

\begin{align}
& G_1(R,t) \sim -\int_{\mathcal{B}} \frac{\mathrm{d}k}{2\pi} \mathcal{F}_1(k) \left\{ \frac{e^{i(kR+2E_{k,\mathrm{f}}t)} + e^{i(kR-2E_{k,\mathrm{f}}t)}}{2}
\right \},
\end{align}

\noindent
with the quasimomentum-dependent amplitude function $\mathcal{F}_1$ defined as 

\begin{equation}
 \mathcal{F}_1(k) = \frac{\left(\mathcal{A}_{k,\mathrm{i}} \mathcal{B}_{k,\mathrm{f}} - \mathcal{A}_{k,\mathrm{f}} \mathcal{B}_{k,\mathrm{i}}\right)
\mathcal{B}_{k,\mathrm{f}}}{2 E_{k,\mathrm{i}}E_{k,\mathrm{f}}^{2}} = \frac{\bar{n}^2 U_\mathrm{f} \gamma_{k}(U_\mathrm{f} - U_\mathrm{i})}{2 E_{k,\mathrm{i}}
E_{k,\mathrm{f}}^2}.
\end{equation}

\noindent
The previous form of the $G_1$ phase fluctuations valid in the SF mean field regime is represented on Fig.~\ref{fig:BHm}(a) for a specific global quench.
As expected, it displays a linear twofold spike-like structure in the vicinity of the correlation edge where the spreading velocities are characterized on 
Fig.~\ref{fig:BHm}(b).

%% file: text/appendix_6.tex
\setstretch{1.0} 

\chapter{\label{appendix_g2_sf} $G_2$ density fluctuations in the superfluid mean field regime of the Bose-Hubbard chain}
In this section, we give a brief outline of the derivation of the $G_2$ density fluctuations for the Bose-Hubbard chain
in the SF mean field regime. Similarly to the study of the $G_1$ phase fluctuation, a sudden global quench confined in the SF mean field regime is considered to drive
the 1D Bose-Hubbard model far from equilibrium. During the real time evolution, the hopping amplitude $J$ and the filling $\bar{n}$ are fixed and the quench is
performed on the repulsive two-body interaction $U>0$. The calculation, based on a quasiparticle theory, follows the same steps used in Appendix.~\ref{appendix_g1_sf} to
deduce the analytical form of the $G_1$ connected one-body correlation function. Note that in the following, the Bose-Hubbard chain is assumed to contain $N_s$ lattice sites
and $\hbar$ and $a$ the lattice spacing are fixed to unity by convention. \\

The $G_2$ connected density-density correlation function may be written as follows

\begin{equation}
G_{2}(R,t) = \langle \hat{n}_{R}(t) \hat{n}_{0}(t) \rangle - \langle \hat{n}_{R}(t) \rangle \langle \hat{n}_{0}(t) \rangle - \langle \hat{n}_{R}(0) 
\hat{n}_{0}(0) \rangle + \langle \hat{n}_{R}(0) \rangle \langle \hat{n}_{0}(0) \rangle.
\label{g2_app}
\end{equation}

\noindent
We first investigate the space-time behavior of the correlator $\langle \hat{n}_{R}(t) \hat{n}_{0}(t) \rangle$ before treating the second correlator 
$\langle \hat{n}_{R}(t) \rangle \langle \hat{n}_{0}(t) \rangle$. For the two last terms, one just needs to evaluate both previous expressions at the time $t=0$. 
Concerning the first correlator in the expression of $G_2$, $\langle \hat{n}_{R}(t) \hat{n}_{0}(t) \rangle$ can be expressed in the Fourier space and then simplified 
relying on the bosonic Wick theorem. It yields the following form 

\begin{align}
& \langle \hat{n}_{R}(t) \hat{n}_{0}(t) \rangle = \frac{1}{N_s^2} \sum_{k_1,k_2,k_3,k_4} e^{i(k_2 - k_1)R} \{ \langle \hat{a}_{k_1,\mathrm{f}}^{\dag}(t) \hat{a}_{k_3,\mathrm{f}}
^{\dag}(t) \rangle  \langle \hat{a}_{k_2,\mathrm{f}}(t) \hat{a}_{k_4,\mathrm{f}}(t) \rangle \nonumber \\
& ~~~~~~~~~~~~~~~~~~~~ + \langle \hat{a}_{k_1,\mathrm{f}}^{\dag}(t) \hat{a}_{k_4,\mathrm{f}}(t) \rangle \langle \hat{a}_{k_2,\mathrm{f}}(t) \hat{a}_{k_3,\mathrm{f}}^{\dag}(t) 
\rangle + \langle \hat{a}_{k_1,\mathrm{f}}^{\dag}(t) \hat{a}_{k_2,\mathrm{f}}(t)\rangle \langle \hat{a}_{k_3,\mathrm{f}}^{\dag}(t) \hat{a}_{k_4,\mathrm{f}}(t) \rangle \}.
\end{align}

\noindent
Then, using the second-order approximation (by keeping only the terms proportional to $N_0^2$ and $N_0$ with $N_0$ the number of bosons in the mode $k=0$)
and the momentum conservation (due to the translational invariance of the BH model), the correlator $\langle \hat{n}_{R}(t) \hat{n}_{0}(t) \rangle$ 
can be approximated by the following form

\begin{align}
& \langle \hat{n}_{R}(t) \hat{n}_{0}(t) \rangle \simeq \frac{3N_0^2}{N_s^2} + \frac{N_0}{N_s^2} \sum_{k} e^{-ik R}  \langle \hat{a}_{k,\mathrm{f}}^{\dag}(t) \hat{a}^{\dag}_{-k,\mathrm{f}}(t)
+ \hat{a}_{-k,\mathrm{f}}(t)\hat{a}_{k,\mathrm{f}}(t) + \hat{a}_{k,\mathrm{f}}^{\dag}(t) \hat{a}_{k,\mathrm{f}}(t) \nonumber \\
& ~~~~~~~~~~~~~~~~~~~~~  + \hat{a}_{-k,\mathrm{f}}(t) \hat{a}_{-k,\mathrm{f}}^{\dag}(t)\rangle + 2\langle \hat{a}^{\dag}_{k,\mathrm{f}}(t) \hat{a}_{k,\mathrm{f}}(t) \rangle.
\label{imp2}
\end{align}

\noindent
For the second correlator $\langle \hat{n}_{R}(t) \rangle \langle \hat{n}_{0}(t) \rangle$, the same steps are considered. It yields the analytical expression 

\begin{equation}
\langle \hat{n}_{R}(t) \rangle \langle \hat{n}_{0}(t) \rangle = \frac{1}{N_s^2} \sum_{k_1,k_2,k_3,k_4} e^{i(k_2-k_1)R} \langle \hat{a}_{k_1,\mathrm{f}}^{\dag}(t) \hat{a}_{k_2,\mathrm{f}}(t) \rangle
\langle \hat{a}^{\dag}_{k_3,\mathrm{f}}(t) \hat{a}_{k_4,\mathrm{f}}(t) \rangle,
\end{equation}

\noindent
which simplifies into the following form when considering once again a second-order approximation and the momentum conservation

\begin{equation}
\langle \hat{n}_{R}(t) \rangle \langle \hat{n}_{0}(t) \rangle \simeq \frac{N_0^2}{N_s^2} + \frac{2N_0}{N_s^2} \sum_{k} \langle \hat{a}^{\dag}_{k,\mathrm{f}}(t) \hat{a}_{k,\mathrm{f}}(t) \rangle.
\label{imp1}
\end{equation}

\noindent
For the two last terms of Eq.~\eqref{g2_app}, one just needs to evaluate the previous expressions at $t=0$ (see Eqs.~\eqref{imp1} and \eqref{imp2}). The third term leads to 

\begin{align}
& \langle \hat{n}_{R}(0) \hat{n}_{0}(0) \rangle \simeq \frac{3N_0^2}{N_s^2} + \frac{N_0}{N_s^2} \sum_{k} e^{-ik R}  \langle \hat{a}_{k,\mathrm{f}}^{\dag}(0) \hat{a}^{\dag}_{-k,\mathrm{f}}(0)
+ \hat{a}_{-k,\mathrm{f}}(0)\hat{a}_{k,\mathrm{f}}(0) + \hat{a}_{k,\mathrm{f}}^{\dag}(0) \hat{a}_{k,\mathrm{f}}(0) \nonumber \\
& ~~~~~~~~~~~~~~~~~~~~~  + \hat{a}_{-k,\mathrm{f}}(0) \hat{a}_{-k,\mathrm{f}}^{\dag}(0)\rangle + 2\langle \hat{a}^{\dag}_{k,\mathrm{f}}(0) \hat{a}_{k,\mathrm{f}}(0) \rangle,
\end{align}

\noindent
and the last one to

\begin{align}
& \langle \hat{n}_{R}(0) \rangle \langle \hat{n}_{0}(0) \rangle \simeq \frac{N_0^2}{N_s^2} + \frac{2N_0}{N_s^2} \sum_{k} \langle \hat{a}^{\dag}_{k,\mathrm{f}}(0)
\hat{a}_{k,\mathrm{f}}(0) \rangle.
\end{align}

\noindent
Consequently, the $G_2$ density-density correlation function can be approximated in momentum space by

\begin{align}
& G_2(R,t) \simeq \frac{N_0}{N_s^2} \sum_{k} e^{-ikR} \{ \langle ( \hat{a}_{k,\mathrm{f}}^{\dag}(t) \hat{a}^{\dag}_{-k,\mathrm{f}}(t) + \hat{a}_{-k,\mathrm{f}}(t) \hat{a}_{k,\mathrm{f}}(t) + 
 \hat{a}_{k,\mathrm{f}}^{\dag}(t) \hat{a}_{k,\mathrm{f}}(t) \nonumber \\
& + \hat{a}_{-k,\mathrm{f}}(t) \hat{a}_{-k,\mathrm{f}}^{\dag}(t) ) \rangle - \langle ( \hat{a}_{k,\mathrm{f}}^{\dag}(0) \hat{a}^{\dag}_{-k,\mathrm{f}}(0)
+ \hat{a}_{-k,\mathrm{f}}(0) \hat{a}_{k,\mathrm{f}}(0) + \hat{a}_{k,\mathrm{f}}^{\dag}(0) \hat{a}_{k,\mathrm{f}}(0) \nonumber \\
& + \hat{a}_{-k,\mathrm{f}}(0) \hat{a}_{-k,\mathrm{f}}^{\dag}(0) ) \rangle \}.
\label{fourier_space_g2}
\end{align}

\noindent
Then, by following the same scheme used to calculate the $G_1$ phase fluctuations and based on a quasiparticle theory, see Appendix.~\ref{appendix_g1_sf}, one obtains for
$G_2$ the density fluctuations 

\begin{align}
&G_2(R,t) \simeq \frac{4N_0}{N_s^2} \sum_{k} \xi^{(2)}_{\mathrm{i}, \mathrm{f}}(k) e^{-ikR} \sin^{2}(E_{k,\mathrm{f}}t),
\label{needed_1}
\end{align}

\noindent
where $\xi^{(2)}_{\mathrm{i}, \mathrm{f}}(k)$ denotes a quantity depending on the quasimomentum $k$ 
and on the pre- (post-) quench parameters \textit{via} the coefficients $u_{k,\mathrm{i}(\mathrm{f})}$ and $v_{k,\mathrm{i}(\mathrm{f})}$, 

\begin{equation}
\xi^{(2)}_{\mathrm{i}, \mathrm{f}}(k) = \left( u_{k, \mathrm{f}} + v_{k, \mathrm{f}}\right)^{2} 
\left( u_{k,\mathrm{i}} u_{k, \mathrm{f}} - v_{k, \mathrm{i}} v_{k, \mathrm{f}} \right) \left( u_{k,\mathrm{i}} v_{k,\mathrm{f}} - v_{k,\mathrm{i}}u_{k,\mathrm{f}} \right).
\label{needed_2}
\end{equation}

\noindent
Then, we replace $u_{k,\mathrm{i}(\mathrm{f})}$ and $v_{k,\mathrm{i}(\mathrm{f})}$ by their respective expression given by  

\begin{equation}
u_{k,\mathrm{i}(\mathrm{f})}, v_{k,\mathrm{i}(\mathrm{f})} = \pm \left[ \frac{1}{2} \left(\frac{\mathcal{A}_{k,\mathrm{i}(\mathrm{f})}}{E_{k,\mathrm{i}(\mathrm{f})}}
\pm 1 \right) \right]^{1/2},
\end{equation}

\noindent
coming from the bosonic Bogolyubov transformation in order to diagonalize the BH chain in the SF mean field regime.
$E_{k,\mathrm{i}(\mathrm{f})} = \sqrt{\mathcal{A}^2_{k,\mathrm{i}(\mathrm{f})} - \mathcal{B}^2_{k,\mathrm{i}(\mathrm{f})}}$
denotes the excitation spectrum in this regime for the pre- (post-) quench Hamiltonian 
with $\mathcal{B}_{k,\mathrm{i}(\mathrm{f})} = \gamma_k + U_{\mathrm{i}(\mathrm{f})}\bar{n}$ and
$\mathcal{A}_{k,\mathrm{i}(\mathrm{f})} = U_{\mathrm{i}(\mathrm{f})} \bar{n}$. Besides, $\gamma_k = 4J \sin^2(k/2)$ represents the dispersion
relation of the free tight-binding model. It yields the following expression for the quantity $\xi^{(2)}_{\mathrm{i}, \mathrm{f}}(k)$

\begin{equation}
\xi^{(2)}_{\mathrm{i}, \mathrm{f}}(k) = \frac{\mathcal{A}_{k,\mathrm{f}}\mathcal{B}_{k,\mathrm{i}} - \mathcal{A}_{k,\mathrm{i}}\mathcal{B}_{k,\mathrm{f}}}{2\left(
\mathcal{A}_{k,\mathrm{f}} +  \mathcal{B}_{k,\mathrm{f}}\right)E_{k,\mathrm{i}}} = \frac{ \bar{n} \gamma_{k}(U_{\mathrm{i}} - U_{\mathrm{f}})}
{2E_{k,\mathrm{i}} (\gamma_{k}+ 2 \bar{n} U_{\mathrm{f}})}.
\end{equation}

\noindent
Then, by considering the thermodynamic limit for the one-dimensional Bose-Hubbard model ($N_s \rightarrow +\infty$), the discrete sum over the quasimomentum can be expressed
in a continuous form leading to 

\begin{equation}
 \frac{1}{N_s} \sum_{k} \underset{N_s \rightarrow + \infty}{\rightarrow} \int_{\mathcal{B}}\frac{\mathrm{d}k}{2 \pi},
\end{equation}

\noindent
where $\mathcal{B} = [-\pi, \pi]$ denotes the first Brillouin zone. Considering also that   

\begin{equation}
e^{-i k R} \sin^{2}(E_{k,\mathrm{f}}t) = \frac{1}{2} \left( e^{-i k R } - \frac{e^{-i (k R - 2E_{k,\mathrm{f}}t)} + e^{-i( k R + 2E_{k,\mathrm{f}}t)} }{2} \right),
\end{equation}

\noindent
the $G_2$ connected correlation function can be written as follows 

\begin{align}
& G_2(R,t) \simeq \frac{2 N_0}{N_s} \int_{\mathcal{B}} \frac{\mathrm{d} k}{2 \pi} \xi^{(2)}_{\mathrm{i}, \mathrm{f}}(k) \left( e^{-ikR}-\frac{e^{-i (kR - 2E_{k,\mathrm{f}}t)} + 
e^{-i(kR + 2E_{k,\mathrm{f}}t)}}{2} \right) 
\end{align}

\noindent
Finally, one considers only the relevant (time-dependent) part of the latter to describe the far-from-equilibrium dynamics of the density fluctuations and by fixing
$\mathcal{F}_2(k) = 2(N_0/N_s) \xi^{(2)}_{\mathrm{i}, \mathrm{f}}(k) \simeq 2\bar{n}\xi^{(2)}_{\mathrm{i}, \mathrm{f}}(k)$,
$G_2$ is expressed under the generic form presented at Eq.~\eqref{generic_form} and reads as

\begin{align}
& G_2(R,t) \sim -\int_{\mathcal{B}} \frac{\mathrm{d}k}{2\pi} \mathcal{F}_2(k) \left\{ \frac{e^{i(kR+2E_{k,\mathrm{f}}t)} + e^{i(kR-2E_{k,\mathrm{f}}t)}}{2}
\right \},
\end{align}

\noindent
with the quasimomentum-dependent amplitude function $\mathcal{F}_2$ defined as 

\begin{equation}
 \mathcal{F}_2(k) \simeq \frac{ \bar{n}^2 \gamma_{k}(U_{\mathrm{i}} - U_{\mathrm{f}})}
{E_{k,\mathrm{i}} (\gamma_{k}+ 2 \bar{n} U_{\mathrm{f}})}.
\end{equation}

\noindent
The previous form of the $G_2$ density fluctuations valid in the SF mean field regime is represented on Fig.~\ref{fig:BHm2}(a) for a specific global quench.
As expected and similarly to $G_1$, it displays a linear twofold spike-like structure in the vicinity of the correlation edge where the spreading velocities
are characterized on Fig.~\ref{fig:BHm2}(b).

%% file: text/appendix_9.tex
\setstretch{1.0} 

\chapter{\label{appendix_g1_mi} $G_1$ phase fluctuations in the Mott-insulating strong-coupling regime of the Bose-Hubbard chain}
In this section, we provide the derivation of the $G_1$ phase fluctuations for the Bose-Hubbard (BH) chain in the Mott-insulating phase.
More precisely, the strong-coupling regime of the MI phase is considered implying $U \gg J\bar{n}$ where $\bar{n} \in \mathbb{N}^*$ denotes the integer 
filling of the bosonic chain. To investigate the far-from-equilibrium dynamics of the BH model in this regime, we analyze the $G_1$ phase fluctuations for a sudden global quench defined as follows.
At $t=0$, the initial state $\ket{\Psi_0}$ corresponds to a pure Mott state ($U_{\mathrm{i}} \rightarrow + \infty$). Then, the latter evolves in time with a post-quench Hamiltonian $\hat{H}_{\mathrm{f}}$
characterized by a large on-site interaction $U_{\mathrm{f}} \gg J \bar{n}$. During the real time evolution process, both the hopping amplitude $J$ and the filling $\bar{n}$ are fixed. Besides, the 1D BH model is assumed to contain $N_s$ lattice and both $\hbar$ ($\hbar = h/2\pi$ and $h$ denotes the Planck constant) and $a$ (lattice spacing) are fixed to unity. \\

The $G_1$ correlation function, which may be written as

\begin{align}
& G_1(R,t) = \langle \hat{a}^{\dag}_R(t) \hat{a}_0(t) \rangle - \langle \hat{a}^{\dag}_R(0) \hat{a}_0(0) \rangle,
\label{g1_appendix_intro}
\end{align}

\noindent
is calculated analytically using the time-dependent perturbation theory while working along the lines of Ref.~\cite{barmettler2012}.
To do so, we first need to determine the expression of the time-dependent many-body quantum state $\ket{\Psi(t)}$. The latter can be expressed as follows

\begin{equation}
 \ket{\Psi(t)} = e^{-i\hat{H}_{\mathrm{f}}t} \ket{\Psi_0} = \alpha_{\mathrm{GS}_{\mathrm{f}}} e^{-i E_{\mathrm{GS}_{\mathrm{f}}}t} \ket{\mathrm{GS}_{\mathrm{f}}^{(1)}}
 + \sum_k \alpha_{k,\mathrm{f}} e^{-iE_{k,\mathrm{f}}t} \ket{\phi_{k,\mathrm{f}}^{(1)}},
 \label{psit}
\end{equation}

\noindent
where $\ket{\Psi_0} \equiv \ket{\bar{n}}$ corresponds to the ground state of the pre-quench Bose-Hubbard Hamiltonian $\hat{H}_{\mathrm{i}} = \hat{H}(U_\mathrm{i}
\rightarrow + \infty)$. The overlaps are defined by $\alpha_{\mathrm{GS}_{\mathrm{f}}} = \langle \mathrm{GS}_{\mathrm{f}}^{(1)} | \bar{n} \rangle$ and $\alpha_{k,\mathrm{f}} 
= \langle \phi_{k,\mathrm{f}}^{(1)} | \bar{n} \rangle$. $\ket{\mathrm{GS}_{\mathrm{f}}^{(1)}}$ is the first-order (correction in $J/U_\mathrm{f}$) perturbed ground state of $\hat{H}_{\mathrm{f}} = \hat{H}(U_\mathrm{f}
\gg J\bar{n})$ given by [calculated in the main text, see Eq.~\eqref{pert_gs}] 

\begin{equation}
\ket{\mathrm{GS}_{\mathrm{f}}^{(1)}} \simeq \ket{\bar{n}}  + \frac{J}{U_{\mathrm{f}}} \sqrt{\bar{n}(\bar{n}+1)} \sum_{R} \left( \ket{\phi_{R,1,\mathrm{f}}^{(0)}}
+ \ket{\phi_{R,-1,\mathrm{f}}^{(0)}} \right).
\end{equation}

\noindent
Besides, $\{ \ket{\phi_{k,\mathrm{f}}^{(1)}} \}$ denotes the collection of the first-order (correction in $J/U_\mathrm{f}$) perturbed eigenstates of
$\hat{H}_{\mathrm{hop},\mathrm{f}} = \hat{H}_{\mathrm{hop}}$ due to a coupling with the non-perturbed ground state $\ket{\mathrm{GS}_{\mathrm{f}}^{(0)}}
\equiv \ket{\bar{n}}$. The perturbed state at first order in $J/U_{\mathrm{f}}$ is given as follows 

\begin{equation}
\ket{\phi_{k,\mathrm{f}}^{(1)}} \simeq \ket{\phi_{k,\mathrm{f}}^{(0)}} + \frac{\langle \bar{n} | \hat{H}_{\mathrm{hop},\mathrm{f}} |
\phi_{k,\mathrm{f}}^{(0)} \rangle }{E_{k,\mathrm{f}} - E_0} \ket{\bar{n}},
\end{equation}

\noindent
with $E_0 = 0$ and $E_{k,\mathrm{f}} \simeq U_{\mathrm{f}}$. The post-quench hopping Hamiltonian $\hat{H}_{\mathrm{hop},\mathrm{f}}$ is diagonal in the basis 
$\left\{ \ket{\phi_{k,\mathrm{f}}^{(0)}} \right\}$ where the eigenstate $\ket{\phi_{k,\mathrm{f}}^{(0)}}$ is defined as 

\begin{equation}
\ket{\phi_{k,\mathrm{f}}^{(0)}} = \frac{\sqrt{2}}{N_s} \sum_{R,R'} \sin\left( kR' \right) \ket{\phi_{R,R',\mathrm{f}}^{(0)}}.
\end{equation}

\noindent
Finally, the perturbed state $\ket{\phi_{k,\mathrm{f}}^{(1)}}$ can be written as 

\begin{equation}
\ket{\phi_{k,\mathrm{f}}^{(1)}} \simeq \ket{\phi_{k,\mathrm{f}}^{(0)}} - \frac{J}{U_{\mathrm{f}}} \sqrt{2\bar{n}(\bar{n}+1)} \eta_k \sin(k) \ket{\bar{n}}, ~~~
\eta_k = 1 - \cos(k N_s).
\end{equation}

\noindent
with $k = n\pi/N_s$ and $n \in [0,N_s-1]$. From now, $\ket{\Psi(t)}$ can be explicitly calculated. The characteristic energies present at Eq.~\eqref{psit} are 
$E_{\mathrm{GS}_{\mathrm{f}}} = 0$ and $E_{k,\mathrm{f}} = U_{\mathrm{f}} - 2J(2\bar{n}+1)\cos(k)$. We now turn to the calculation of the two different overlaps 
 $\alpha_{\mathrm{GS}_{\mathrm{f}}} =  \langle \mathrm{GS}_{\mathrm{f}}^{(1)}| \mathrm{GS}^{(0)}_{\mathrm{i}} \rangle = \langle \mathrm{GS}_{\mathrm{f}}^{(1)} | \bar{n} \rangle$ and $\alpha_{k,\mathrm{f}} 
= \langle \phi_{k,\mathrm{f}}^{(1)} | \mathrm{GS}^{(0)}_{\mathrm{i}} \rangle  = \langle \phi_{k,\mathrm{f}}^{(1)} | \bar{n} \rangle$. The first overlap $\alpha_{\mathrm{GS}_{\mathrm{f}}}$ can be expressed as

\begin{align}
& \alpha_{\mathrm{GS}_{\mathrm{f}}} = \left[ \bra{\bar{n}}  + \frac{J}{U_{\mathrm{f}}} \sqrt{\bar{n}(\bar{n}+1)} \sum_{R} \left( \bra{\phi_{R,1,\mathrm{f}}^{(0)}} +
\bra{\phi_{R,-1,\mathrm{f}}^{(0)}} \right) \right] \ket{\bar{n}} = 1,
\end{align}

\noindent
and the second one denoted by $\alpha_{k,\mathrm{f}}$ as 

\begin{align}
& \alpha_{k,\mathrm{f}} = \langle \phi_{k,\mathrm{f}}^{(1)}| \bar{n} \rangle 
 \simeq \left[ \bra{\phi_{k,\mathrm{f}}^{(0)}} - \frac{J}{U_{\mathrm{f}}} \sqrt{2\bar{n}(\bar{n}+1)} \eta_k 
\sin(k) \bra{\bar{n}} \right] \ket{\bar{n}} \\
& \alpha_{k,\mathrm{f}} \simeq - \frac{J}{U_{\mathrm{f}}} \sqrt{2\bar{n}(\bar{n}+1)} \eta_k \sin(k).
\end{align}

\noindent
Finally, the time-dependent many-body quantum state $\ket{\Psi(t)}$ for a global quench confined in the MI strong-coupling regime from $U_{\mathrm{i}} \rightarrow
+ \infty$ to $U_{\mathrm{f}} \gg J\bar{n}$ can be written as 

\begin{align}
& \ket{\Psi(t)} \simeq \ket{\bar{n}}  + \frac{J}{U_{\mathrm{f}}} \sqrt{\bar{n}(\bar{n}+1)} \sum_{R} \left( \ket{\phi_{R,1,\mathrm{f}}^{(0)}} + \ket{\phi_{R,-1,\mathrm{f}}^{(0)}} 
\right) \nonumber \\
& ~~~~~~~~~~~~ -\frac{J}{U_{\mathrm{f}}} \sqrt{2\bar{n}(\bar{n}+1)} \sum_k  \eta_k \sin(k) e^{-iE_{k,\mathrm{f}} t} \ket{\phi_{k,\mathrm{f}}^{(0)}}.
\label{psit_g1}
\end{align}

\noindent
Note that the previous calculations can be easily extended to the case of a sudden global quench in the MI strong-coupling regime with a finite pre-quench two-body interaction $U_{\mathrm{i}} \gg J\bar{n}$. 
It just requires to consider the first-order perturbed ground state $\ket{\mathrm{GS}_{\mathrm{i}}^{(1)}}$ instead of $\ket{\mathrm{GS}_{\mathrm{i}}^{(0)}} = \ket{\bar{n}}$ for the many-body initial state $\ket{\Psi_0}$. \\

\noindent
We now turn to the calculation of the $G_1$ phase fluctuations defined at Eq.~\eqref{g1_appendix_intro} where the factor $\beta_{\mathrm{f}} = J\sqrt{\bar{n}(\bar{n}+1)}/U_\mathrm{f}$ is introduced. The index '$\mathrm{f}$'
for the different many-body quantum states at Eq.~\eqref{psit_g1} is removed for simplicity. \\

\noindent
The second correlator is straighforward and given by 

\begin{equation}
\langle \hat{a}^{\dag}_R(0) \hat{a}_0(0) \rangle = 
\langle \bar{n} | \hat{a}^{\dag}_R \hat{a}_0 | \bar{n} \rangle = \delta_{R,0}.
\end{equation}

\noindent
For the first correlator $\langle \hat{a}^{\dag}_R(t) \hat{a}_0(t) \rangle = \langle \Psi(t) | \hat{a}^{\dag}_R \hat{a}_0 | \Psi(t) \rangle$, it yields

\begin{align}
& \langle \hat{a}^{\dag}_R(t) \hat{a}_0(t) \rangle \simeq \langle \bar{n} | \hat{a}_R^{\dag} \hat{a}_0 | \bar{n} \rangle +
\beta_{\mathrm{f}} \sum_{R'} \{ \langle \bar{n} | \hat{a}^{\dag}_R \hat{a}_0 | \phi_{R',1}^{(0)} \rangle + \langle \phi_{R,1}^{(0)} | \hat{a}^{\dag}_R \hat{a}_0
| \bar{n} \rangle + \langle \bar{n} | \hat{a}^{\dag}_R \hat{a}_0 | \phi_{R',-1}^{(0)} \rangle \nonumber \\
& ~~~~~~~~~~~~~~~~ + \langle \phi_{R',-1}^{(0)} | \hat{a}^{\dag}_R \hat{a}_0 | \bar{n} \rangle \} - \sqrt{2} \beta_{\mathrm{f}} \sum_k e^{-iE_{k,\mathrm{f}} t}
\eta_k \sin(k) \langle \bar{n} | \hat{a}^{\dag}_R \hat{a}_0 | \phi_k^{(0)} \rangle \nonumber \\
& ~~~~~~~~~~~~~~~~ - \sqrt{2} \beta_{\mathrm{f}} \sum_k e^{iE_{k,\mathrm{f}} t} \eta_k \sin(k) \langle \phi_k^{(0)} | \hat{a}^{\dag}_R \hat{a}_0 | \bar{n} \rangle \\
& \langle \hat{a}^{\dag}_R(t) \hat{a}_0(t) \rangle \simeq - \sqrt{2} \beta_{\mathrm{f}} \sum_k \eta_k \sin(k) \left( e^{-iE_{k,\mathrm{f}}t} \langle \bar{n} | \hat{a}^{\dag}_R \hat{a}_0 | \phi_k^{(0)} \rangle + e^{iE_{k,\mathrm{f}}t} \langle \phi_k^{(0)} | \hat{a}^{\dag}_R \hat{a}_0 | \bar{n} \rangle \right) \nonumber \\
& ~~~~~~~~~~~~~~~~ + \delta_{R,0} + \frac{2J}{U_{\mathrm{f}}} \bar{n}(\bar{n}+1) \delta_{R,1}.
\end{align}

\noindent
Finally, the first correlator may be written as 

\begin{align}
& \langle \hat{a}^{\dag}_R(t) \hat{a}_0(t) \rangle \simeq - \frac{2J \bar{n}(\bar{n}+1)}{U_{\mathrm{f}}N_s}\sum_k \eta_k \sin(k) e^{-iE_{k,\mathrm{f}}t} \sin(kR)
+ \delta_{R,0} + \frac{2J \bar{n}(\bar{n}+1)}{U_{\mathrm{f}}}\delta_{R,1}.
\end{align}

\noindent
Hence, the expression of the $G_1$ one-body correlation function has the following form

\begin{align}
& G_1 (R,t) \simeq \frac{2J \bar{n}(\bar{n}+1)}{U_{\mathrm{f}}} \left( \delta_{R,1} - \frac{1}{N_s}\sum_k \eta_k \sin(k) \sin(kR) e^{-iE_{k,\mathrm{f}}t} \right).
\end{align}

\noindent
Considering that the term $\delta_{R,1}$ is irrelevant for the dynamics (constant term for $R=1$) and by doing some calculations,
it can be shown that the latter form of the $G_1$ phase fluctuations is equivalent to 

\begin{equation}
G_1(R,t) \simeq - \frac{4J \bar{n}(\bar{n}+1)}{iU_{\mathrm{f}}N_s} \sum_k \sin(k) e^{ikR} \cos(E_{k,\mathrm{f}} t),
\label{ps_g1}
\end{equation}

\noindent
with $k \in \mathcal{B} = [-\pi,\pi[$. \\

We now stipulate that $2E_{k,\mathrm{f}} = U_{\mathrm{f}} - 2J(2\bar{n}+1)\cos(k)$ corresponding to the convention adopted in the main
text. The factor $2$ is added to clarify how the correlations are created in the Mott-insulating strong-coupling regime of the BH chain. In fact, the excitation
spectrum $2E_{k,\mathrm{f}}$ correspond to the energy of a quasiparticle pair at the quasimomentum $k$ with respect to the interaction parameters of the post-quench Hamiltonian $\hat{H}_{\mathrm{f}}$.
Indeed, one already knows that the low-energy excitations of the MI phase are made of doublon-holon quasiparticle pairs. Hence, the previous excitation spectrum contains the energy associated to 
a doublon (d) of quasimomentum $k$ and a holon (h) of quasimomentum $-k$, \textit{ie.} $2E_{k,\mathrm{f}} = E_{\mathrm{d},k,\mathrm{f}} + E_{\mathrm{h},-k, \mathrm{f}}$, where
$E_{\mathrm{d},k,\mathrm{f}} = U_{\mathrm{f}}/2 -2J(\bar{n}+1)\cos(k)$ and $E_{\mathrm{h},-k, \mathrm{f}} = U_{\mathrm{f}}/2 - 2J\bar{n}\cos(k)$ [see Ref.~\cite{barmettler2012}
for the analytical expressions determined using a fermionization technique with a Fermi-Bogolyubov transformation]. Finally, the correlations in the MI phase are thus created 
\textit{via} the propagation of a doublon with quasimomentum $k$ at the group velocity $V_{\mathrm{g},\mathrm{d},\mathrm{f}}(k) 
= \partial_k E_{\mathrm{d},k,\mathrm{f}} = 2J(\bar{n}+1)\sin(k)$ and a holon in the opposite direction with a quasimomentum $-k$ at the group velocity
$V_{\mathrm{g},\mathrm{h},\mathrm{f}}(-k) = (\partial_k E_{\mathrm{h},k,\mathrm{f}})|_{-k} = -2J\bar{n}\sin(k)$. The latter lead finally to the effective group velocity

\begin{equation}
|V_{\mathrm{g},\mathrm{d},\mathrm{f}}(k)-V_{\mathrm{g},\mathrm{h},\mathrm{f}}(-k)| = 2J(2\bar{n}+1)\sin(k) = 2V_{\mathrm{g},\mathrm{f}}(k) = \partial_k 2E_{k,\mathrm{f}}.
\end{equation}

\noindent
The previous statements concerning the individual spreading of doublons and holons will be supported by numerical calculations presented at Chap.~\ref{ch:4-bose_hubbard_chain} where
the space-time local density $\langle \hat{n}_R(t) \rangle$ is investigated for sudden local quenches confined in the Mott-insulating strong-coupling regime. \\

Finally, according to Eq.~\eqref{ps_g1} and the new convention for the post-quench excitation spectrum, $G_1$ can be easily cast into the generic form presented at Eq.~\eqref{generic_form} and reads as

\begin{equation}
G_1(R,t) \simeq - \int_{\mathcal{B}} \frac{\mathrm{d}k}{2\pi} \mathcal{F}_1(k) \left \{ \frac{e^{i(kR + 2E_{k,\mathrm{f}} t)} + e^{i(kR - 2E_{k,\mathrm{f}}t)}}{2} \right \}
\end{equation}

\noindent
with the quasimomentum-dependent amplitude function $\mathcal{F}_1$ given by 

\begin{equation}
\mathcal{F}_1(k) = \frac{4J}{iU_\mathrm{f}} \bar{n}(\bar{n}+1) \sin(k).
\end{equation}

\noindent
The $G_1$ phase fluctuations in the MI strong-coupling regime are represented on Fig.~\ref{fig:BHm3}(a) 
for a specific global quench on the two-body interaction parameter. As expected, the space-time pattern displays a linear twofold 
structure where the velocity of each structure (the one associated to the correlation edge and the series of local extrema) are characterized 
on Fig.~\ref{fig:BHm3}(b) as a function of the post-quench interaction parameter $U_{\mathrm{f}}/J = U/J$.

%% file: text/appendix_8.tex
\setstretch{1.0} 

\chapter{\label{appendix_gz_lrxy} $G_z$ spin-spin correlations in the $x$ polarized phase of the long-range XY chain}
In this section, we provide more details about the derivation of the $G_z$ spin-spin correlations along the $z$ axis in the $x$ polarized phase of 
the long-range XY (LRXY) chain. A sudden global quench on $\epsilon$ the antiferromagnetic interaction parameter along the $z$ axis is considered.
More precisely, one starts from the ground state of the long-range XXZ chain with $-1<\epsilon_\textrm{i} < \epsilon_c(\alpha)$ ($x$ polarized phase).
The latter evolves in time with the 1D LRXY Hamiltonian where $\epsilon_\textrm{f}=0$. The power-law exponent $\alpha$ is fixed during the real time evolution. 
Besides, it is confined in the interval $\alpha \in [1,3[$ in order to consider the quasi-local regime of the $x$ polarized phase for the LRXY chain.
In the following, the LRXY and LRXXZ spin chains are assumed to contain $N_s$ lattice sites and $\hbar$, $a$ the lattice spacing, are fixed to unity by convention. \\

The connected spin-spin correlation function along the $z$ axis, denoted by $G_z$, are investigated where $G_z(R,t) = G_{z,0}(R,t) - G_{z,0}(R,0)$ with

\begin{equation}
G_{z,0}(R,t) = \langle \hat{S}_R^z(t) \hat{S}_0^z(t) \rangle - \langle \hat{S}_R^z(t)\rangle \langle \hat{S}_0^z(t) \rangle.
\end{equation}

\noindent
In order to use the general scheme (presented in the previous appendices) relying on a bosonic Bogolyubov transformation, we need to express $G_z$ in terms of the
post-quench bosonic operators in momentum space $\hat{a}_{k,\mathrm{f}}$ and $\hat{a}^{\dag}_{k,\mathrm{f}}$. The first step consists of injecting the
Holstein-Primakoff transformation, defined below and valid for the long-range XY and long-range XXZ chains in the $x$ polarized phase, in the expression of $G_z$.  

\begin{equation}
\hat{S}_{R}^{x} = \frac{1}{2}-\hat{a}_{R}^\dagger \hat{a}_{R},~~~ \hat{S}_{R}^{y} = \simeq -\frac{\hat{a}_{R}^\dagger-\hat{a}_{R}}{2i}, ~~~
\hat{S}_{R}^{z} \simeq -\frac{\hat{a}_{R} + \hat{a}_{R}^\dagger}{2}.
\end{equation}

\noindent
This transformation leads for $G_z$ in terms of the post-quench bosonic operators in real space ($\hat{a}_R$, $\hat{a}_R^{\dag}$) to 

\begin{align}
& G_z(R,t) \simeq \frac{1}{4} \big \{\langle \hat{a}_R(t)\hat{a}_0(t) \rangle +  \langle \hat{a}_R(t) \hat{a}_0^{\dag}(t) \rangle +
\langle \hat{a}_R^{\dag}(t) \hat{a}_0(t) \rangle + \langle \hat{a}_R^{\dag}(t) \hat{a}_0^{\dag}(t) \rangle \nonumber \\
&~~~~~~~~~~~~~~ - \langle \hat{a}_R(0) \hat{a}_0(0)
\rangle_0 - \langle \hat{a}_R(0) \hat{a}_0^{\dag}(0) \rangle - \langle \hat{a}_R^{\dag}(0) \hat{a}_0(0) \rangle -\langle \hat{a}_R^{\dag}(0) \hat{a}_0^{\dag}(0)
\rangle \big \}.
\label{post_quench_gz}
\end{align}
 
\noindent
Finally, Eq.~\eqref{post_quench_gz} is expressed into the momentum space \textit{via} a Fourier transform of the bosonic operators. The latter may be written as follows  

\begin{align}
& G_z(R,t) \simeq \frac{1}{4 N_s} \sum_k e^{-ikR} \Big \{ \langle \hat{a}_{k,\mathrm{f}}(t) \hat{a}_{-k,\mathrm{f}}(t) \rangle  +  \langle \hat{a}_{k,\mathrm{f}}(t) 
\hat{a}_{k,\mathrm{f}}^{\dag}(t) \rangle + \langle \hat{a}_{-k,\mathrm{f}}^{\dag}(t) \hat{a}_{-k,\mathrm{f}}(t) \rangle \nonumber \\
& + \langle \hat{a}_{-k,\mathrm{f}}^{\dag}(t)
\hat{a}_{k,\mathrm{f}}^{\dag}(t) \rangle - \langle \hat{a}_{k,\mathrm{f}}(0) \hat{a}_{-k,\mathrm{f}}(0) \rangle - \langle \hat{a}_{k,\mathrm{f}}(0) \hat{a}_{k,\mathrm{f}}^{\dag}(0) \rangle
-\langle \hat{a}_{-k,\mathrm{f}}^{\dag}(0) \hat{a}_{-k,\mathrm{f}}(0) \rangle \nonumber \\
& -\langle \hat{a}_{-k,\mathrm{f}}^{\dag}(0) \hat{a}_{k,\mathrm{f}}^{\dag}(0) \rangle \Big \}.
\label{fourier_space_gz}
\end{align}

\noindent
From now, $G_z$ is expressed in terms of the post-quench bosonic operators in momentum space. Hence, the general scheme considered previously [see for instance 
Appendices.~\ref{appendix_g1_sf} and \ref{appendix_g2_sf}] can be used to deduce the final form of the spin correlations. 
However, one can notice that Eq.~\eqref{fourier_space_gz} has the same expression than the one at Eq.~\eqref{fourier_space_g2} up to a prefactor. Therefore, one can
directly consider Eqs.~\eqref{needed_1} and \eqref{needed_2} resulting from the general scheme based on the properties of the bosonic Bogolyubov transformation. By updating
the prefactor, $G_z$ can be written as 

\begin{align}
& G_z(R,t) \simeq \frac{1}{N_s} \sum_{k} \xi^{(z)}_{\mathrm{i}, \mathrm{f}}(k) e^{-ikR} \sin^{2}(E_{k,\mathrm{f}}t).
\end{align}

\noindent
The quantity $\xi^{(z)}_{\mathrm{i}, \mathrm{f}}(k)$ depends on the quasimomentum $k$ and on the pre- (post-) quench parameters 
\textit{via} the coefficients $u_{k,\mathrm{i}(\mathrm{f})}$ and $v_{k,\mathrm{i}(\mathrm{f})}$ as follows 

\begin{equation}
\xi^{(z)}_{\mathrm{i}, \mathrm{f}}(k) = \left( u_{k, \mathrm{f}} + v_{k, \mathrm{f}}\right)^{2} 
\left( u_{k,\mathrm{i}} u_{k, \mathrm{f}} - v_{k, \mathrm{i}} v_{k, \mathrm{f}} \right) \left( u_{k,\mathrm{i}} v_{k,\mathrm{f}} - v_{k,\mathrm{i}}u_{k,\mathrm{f}} \right).
\end{equation}

\noindent
Then, by replacing $u_{k,\mathrm{i}(\mathrm{f})}$ and $v_{k,\mathrm{i}(\mathrm{f})}$ by their respective expression given by  

\begin{equation}
u_{k,\mathrm{i}(\mathrm{f})}, v_{k,\mathrm{i}(\mathrm{f})} = \pm \left[ \frac{1}{2} \left(\frac{\mathcal{A}_{k,\mathrm{i}(\mathrm{f})}}{E_{k,\mathrm{i}(\mathrm{f})}}
\pm 1 \right) \right]^{1/2},
\end{equation}

\noindent
coming from the bosonic Bogolyubov transformation in order to diagonalize the long-range XXZ (long-range XY) chain in the $x$ polarized phase.
$E_{k,\mathrm{i}(\mathrm{f})} = \sqrt{\mathcal{A}^2_{k,\mathrm{i}(\mathrm{f})} - \mathcal{B}^2_{k,\mathrm{i}(\mathrm{f})}}$ denotes
the excitation spectrum in the $x$ polarized phase of the LRXXZ (LRXY) chain. For a global quench on $\epsilon$ the antiferromagnetic 
interaction along $z$ and maintaining constant both the power-law exponent $\alpha$ and the spin-exchange coupling $J$ in the $xy$ plane,
the pre- (post-) quench coefficients $\mathcal{B}_{k,\mathrm{i}(\mathrm{f})}$ and $\mathcal{A}_{k,\mathrm{i}(\mathrm{f})}$ are given by 

\begin{equation}
 \mathcal{A}_{k,\mathrm{i}(\mathrm{f})} =  \frac{J}{2}\left[ P_{\alpha}(0) + P_{\alpha}(k) \frac{\epsilon_{\mathrm{i}(\mathrm{f})}-1}{2} \right],
 ~~~ \mathcal{B}_{k,\mathrm{i}(\mathrm{f})} = \frac{J P_{\alpha}(k)}{4}(\epsilon_{\mathrm{i}(\mathrm{f})}+1),
\end{equation}

\noindent
Note that for the post-quench parameters (referring to the LRXY chain), one just needs to replace $\epsilon_{\mathrm{f}}$ by zero. Finally, it yields the following
expression for the quantity $\xi^{(z)}_{\mathrm{i}, \mathrm{f}}(k)$

\begin{equation}
 \xi^{(z)}_{\mathrm{i}, \mathrm{f}}(k) =  \frac{\mathcal{A}_{k,\mathrm{f}} \mathcal{B}_{k,\mathrm{i}} - \mathcal{A}_{k,\mathrm{i}} 
 \mathcal{B}_{k,\mathrm{f}}}{2(\mathcal{A}_{k,\mathrm{f}} + \mathcal{B}_{k,\mathrm{f}}) E_{k,\mathrm{i}}} = 
 \frac{\epsilon_\textrm{i}}{4}\frac{P_\alpha\left(k\right)}{P_\alpha\left(0\right)}\sqrt{\frac{P_\alpha(0)-P_\alpha\left(k \right)}{P_\alpha(0)+
 \epsilon_\textrm{i}P_\alpha\left(k\right)}}.
\end{equation}

\noindent
Then, by considering the thermodynamic limit ($N_s \rightarrow +\infty$) and the useful relation, 

\begin{equation}
e^{-i k R} \sin^{2}(E_{k,\mathrm{f}}t) = \frac{1}{2} \left( e^{-i k R } - \frac{e^{-i (k R - 2E_{k,\mathrm{f}}t)} + e^{-i( k R + 2E_{k,\mathrm{f}}t)} }{2} \right),
\end{equation}

\noindent
the $G_z$ connected spin-spin correlation function fulfills the generic form presented at Eq.~\eqref{generic_form} and may be written as 

\begin{align}
& G_z(R,t) \sim -\int_{\mathcal{B}} \frac{\mathrm{d}k}{2\pi} \mathcal{F}(k) \left\{ \frac{e^{i(kR+2E_{k,\mathrm{f}}t)} + e^{i(kR-2E_{k,\mathrm{f}}t)}}{2}
\right \},
\end{align}

\noindent
with an amplitude function $\mathcal{F}(k)$ given by 

\begin{equation}
\mathcal{F}(k) = \frac{\xi^{(z)}_{\mathrm{i}, \mathrm{f}}(k)}{2} = \frac{\epsilon_\textrm{i}}{8}\frac{P_\alpha\left(k\right)}{P_\alpha\left(0\right)}\sqrt{\frac{P_\alpha(0)-P_\alpha\left(k \right)}{P_\alpha(0)+
 \epsilon_\textrm{i}P_\alpha\left(k\right)}}.
\end{equation}

\noindent
The $G_z$ spin-spin correlations along the $z$ axis are represented on Fig.~\ref{fig:spins_LRXY}(a) for a global quench on the antiferromagnetic interaction
$\epsilon$ and in the quasi-local regime of the $x$ polarized phase for the LRXY chain ($\alpha \in [1,3[$). As expected, the associated space-time pattern displays an algebraic (linear in log-log scale)
twofold spike-like structure ($t \sim R^{\beta}$) whose exponent $\beta$ for the CE and the series of local extrema are characterized on Fig.~\ref{fig:spins_LRXY}(b).

%% file: text/appendix_7.tex
\setstretch{1.0} 

\chapter{\label{appendix_gx_lrti} $G_x$ spin-spin correlations in the $z$ polarized phase of the long-range transverse Ising chain}
In this section, the derivation of the $G_x$ spin-spin correlations along the $x$ axis in the $z$ polarized phase of the
long-range transverse Ising (LRTI) chain is provided. A sudden global quench on the spin exchange coupling $J$ in the $x$ direction is considered while remaining
deep in the $z$ polarized phase. The latter implies that $J_\mathrm{i},J_{\mathrm{f}} \ll h$. Besides, the transverse magnetic field $h$ and the power-law exponent $\alpha$ are fixed 
during the real time evolution process. The calculation, based on a quasiparticle theory, follows the same steps used in Appendices.~\ref{appendix_g1_sf}, \ref{appendix_g2_sf}.
In the following, the long-range Ising chain is assumed to contain $N_s$ lattice sites and $\hbar$ and $a$ the lattice spacing are fixed to unity by convention. \\

The $G_x$ connected spin-spin correlation function along the $x$ axis may be written as $G_x(R,t) = G_{x,0}(R,t) - G_{x,0}(R,0)$ where

\begin{equation}
G_{x,0}(R,t) = \langle \hat{S}_R^x(t) \hat{S}_0^x(t) \rangle - \langle \hat{S}_R^x(t)\rangle \langle \hat{S}_0^x(t) \rangle.
\end{equation}

\noindent
In order to use the general scheme (presented in the previous appendices) relying on a bosonic Bogolyubov transformation, we need to express $G_x$ in terms of the
post-quench bosonic operators in momentum space $\hat{a}_{k,\mathrm{f}}$ and $\hat{a}^{\dag}_{k,\mathrm{f}}$. The first step consists of injecting the
Holstein-Primakoff transformation defined below, and valid for the Ising chain in the $z$ polarized phase, in the expression of $G_x$.  

\begin{equation}
\hat{S}_{R}^{x}\simeq \frac{\hat{a}_{R} + \hat{a}_{R}^\dagger}{2},~~~
\hat{S}_{R}^{y} \simeq -\frac{\hat{a}_{R}^\dagger-\hat{a}_{R}}{2i},~~~
\hat{S}_{R}^{z} = \frac{1}{2}-\hat{a}_{R}^\dagger \hat{a}_{R}. 
\end{equation}

\noindent
This transformation leads for $G_x$, in terms of the post-quench bosonic operators in real space, to 

\begin{align}
& G_x(R,t) \simeq \frac{1}{4} \big \{\langle \hat{a}_R(t)\hat{a}_0(t) \rangle +  \langle \hat{a}_R(t) \hat{a}_0^{\dag}(t) \rangle +
\langle \hat{a}_R^{\dag}(t) \hat{a}_0(t) \rangle + \langle \hat{a}_R^{\dag}(t) \hat{a}_0^{\dag}(t) \rangle \nonumber \\
&~~~~~~~~~~~~~~ - \langle \hat{a}_R(0) \hat{a}_0(0) \rangle - \langle \hat{a}_R(0) \hat{a}_0^{\dag}(0) \rangle - \langle \hat{a}_R^{\dag}(0) \hat{a}_0(0) \rangle -\langle \hat{a}_R^{\dag}(0) \hat{a}_0^{\dag}(0)
\rangle \big \}.
\label{post_quench_gx}
\end{align}
 
\noindent
Finally, Eq.~\eqref{post_quench_gx} is expressed into the momentum space \textit{via} a Fourier transform of the bosonic operators in real space and can read as 

\begin{align}
& G_x(R,t) \simeq \frac{1}{4 N_s} \sum_k e^{-ikR} \Big \{ \langle \hat{a}_{k,\mathrm{f}}(t) \hat{a}_{-k,\mathrm{f}}(t) \rangle  +  \langle \hat{a}_{k,\mathrm{f}}(t) 
\hat{a}_{k,\mathrm{f}}^{\dag}(t) \rangle + \langle \hat{a}_{-k,\mathrm{f}}^{\dag}(t) \hat{a}_{-k,\mathrm{f}}(t) \rangle \nonumber \\
& + \langle \hat{a}_{-k,\mathrm{f}}^{\dag}(t)
\hat{a}_{k,\mathrm{f}}^{\dag}(t) \rangle - \langle \hat{a}_{k,\mathrm{f}}(0) \hat{a}_{-k,\mathrm{f}}(0) \rangle - \langle \hat{a}_{k,\mathrm{f}}(0) \hat{a}_{k,\mathrm{f}}^{\dag}(0) \rangle
-\langle \hat{a}_{-k,\mathrm{f}}^{\dag}(0) \hat{a}_{-k,\mathrm{f}}(0) \rangle \nonumber \\
& -\langle \hat{a}_{-k,\mathrm{f}}^{\dag}(0) \hat{a}_{k,\mathrm{f}}^{\dag}(0) \rangle \Big \}.
\label{fourier_space_gx}
\end{align}

\noindent
From now, $G_x$ is expressed in terms of the post-quench bosonic operators in momentum space. Hence, one can rely on the general scheme used in the previous appendices
to deduce the final form of the correlations. 
However, one can notice that Eq.~\eqref{fourier_space_gx} has the same expression than the one at Eq.~\eqref{fourier_space_g2} up to a prefactor. Therefore, one can
directly consider Eqs.~\eqref{needed_1} and \eqref{needed_2} resulting from the general scheme based on the properties of the bosonic Bogolyubov transformation. By updating
the prefactor, $G_x$ can be written as 

\begin{align}
& G_x(R,t) \simeq \frac{1}{N_s} \sum_{k} \xi^{(x)}_{\mathrm{i}, \mathrm{f}}(k) e^{-ikR} \sin^{2}(E_{k,\mathrm{f}}t).
\end{align}

\noindent
The quantity $\xi^{(x)}_{\mathrm{i}, \mathrm{f}}(k)$ depends on the quasimomentum $k$ and on the pre- (post-) quench parameters 
\textit{via} the coefficients $u_{k,\mathrm{i}(\mathrm{f})}$ and $v_{k,\mathrm{i}(\mathrm{f})}$ as follows 

\begin{equation}
\xi^{(x)}_{\mathrm{i}, \mathrm{f}}(k) = \left( u_{k, \mathrm{f}} + v_{k, \mathrm{f}}\right)^{2} 
\left( u_{k,\mathrm{i}} u_{k, \mathrm{f}} - v_{k, \mathrm{i}} v_{k, \mathrm{f}} \right) \left( u_{k,\mathrm{i}} v_{k,\mathrm{f}} - v_{k,\mathrm{i}}u_{k,\mathrm{f}} \right).
\end{equation}

\noindent
Then, by replacing $u_{k,\mathrm{i}(\mathrm{f})}$ and $v_{k,\mathrm{i}(\mathrm{f})}$ by their respective expression given by  

\begin{equation}
u_{k,\mathrm{i}(\mathrm{f})}, v_{k,\mathrm{i}(\mathrm{f})} = \pm \left[ \frac{1}{2} \left(\frac{\mathcal{A}_{k,\mathrm{i}(\mathrm{f})}}{E_{k,\mathrm{i}(\mathrm{f})}}
\pm 1 \right) \right]^{1/2},
\end{equation}

\noindent
coming from the bosonic Bogolyubov transformation in order to diagonalize the long-range Ising chain in the $z$ polarized phase.
$E_{k,\mathrm{i}(\mathrm{f})} = \sqrt{\mathcal{A}^2_{k,\mathrm{i}(\mathrm{f})} - \mathcal{B}^2_{k,\mathrm{i}(\mathrm{f})}}$ denotes
the excitation spectrum in the polarized phase for the pre- (post-) quench Hamiltonian. For a global quench on the spin exchange coupling $J$
in the direction $x$ and maintaining constant both the magnetic field $h$ and the power-law exponent $\alpha$, the pre- (post-) quench coefficients 
$\mathcal{B}_{k,\mathrm{i}(\mathrm{f})}$ and $\mathcal{A}_{k,\mathrm{i}(\mathrm{f})}$ are given by 
$\mathcal{B}_{k,\mathrm{i}(\mathrm{f})} = J_{\mathrm{i}(\mathrm{f})} P_\alpha(k)$, $\mathcal{A}_{k,\mathrm{i}(\mathrm{f})}
= 2h + J_{\mathrm{i}(\mathrm{f})} P_\alpha(k)$. Finally, it yields the following expression for the quantity $\xi^{(x)}_{\mathrm{i}, \mathrm{f}}(k)$

\begin{equation}
\xi^{(x)}_{\mathrm{i}, \mathrm{f}}(k) = \frac{\mathcal{A}_{k,\mathrm{f}}\mathcal{B}_{k,\mathrm{i}} - \mathcal{A}_{k,\mathrm{i}}\mathcal{B}_{k,\mathrm{f}}}{2\left(
\mathcal{A}_{k,\mathrm{f}} +  \mathcal{B}_{k,\mathrm{f}}\right)E_{k,\mathrm{i}}}
=  \frac{h\left(J_\textrm{i}-J_\textrm{f}\right)P_\alpha(k)}{4\left[h+J_\textrm{f}P_\alpha(k)\right]\sqrt{h \left[ h + J_\textrm{i} P_\alpha(k) \right]}}.
\end{equation}

\noindent
Then, by considering the thermodynamic limit ($N_s \rightarrow +\infty$) and the useful relation, 

\begin{equation}
e^{-i k R} \sin^{2}(E_{k,\mathrm{f}}t) = \frac{1}{2} \left( e^{-i k R } - \frac{e^{-i (k R - 2E_{k,\mathrm{f}}t)} + e^{-i( k R + 2E_{k,\mathrm{f}}t)} }{2} \right),
\end{equation}

\noindent
the $G_x$ connected spin-spin correlation function fulfills the generic form presented at Eq.~\eqref{generic_form} and may be written as 

\begin{align}
& G_x(R,t) \sim -\int_{\mathcal{B}} \frac{\mathrm{d}k}{2\pi} \mathcal{F}(k) \left\{ \frac{e^{i(kR+2E_{k,\mathrm{f}}t)} + e^{i(kR-2E_{k,\mathrm{f}}t)}}{2}
\right \},
\end{align}

\noindent
with an amplitude function $\mathcal{F}(k)$ given by 

\begin{equation}
\mathcal{F}(k) = \frac{\xi^{(x)}_{\mathrm{i}, \mathrm{f}}(k)}{2} = 
\frac{h\left(J_\textrm{i}-J_\textrm{f}\right)P_\alpha(k)}{8\left[h+J_\textrm{f}P_\alpha(k)\right]\sqrt{h \left[ h + J_\textrm{i} P_\alpha(k) \right]}}.
\end{equation}

\noindent
The $G_x$ spin-spin correlations along the $x$ axis in the $z$ polarized phase are represented on Fig.~\ref{fig:spins_LRTI}(a) 
for a specific global quench on the spin exchange coupling $J$ and in the quasi-local regime ($\alpha \in [1,2[$). As expected, the associated space-time pattern 
displays an algebraic twofold spike-like structure ($t \sim R^{\beta}$) whose exponent $\beta$ for the CE and the series of local extrema 
are characterized on Fig.~\ref{fig:spins_LRTI}(b).

%% file: text/appendix_3.tex
\setstretch{1.0} 

\chapter{\label{appendix3_mps_app}  Matrix product state form of a general quantum state}

We provide here more details on the derivation of a general quantum state under its matrix product state (MPS) form by working along the lines of 
Refs.~\cite{schollwock2005, schollwock2011}. First, we consider the most general quantum state for a one-dimensional lattice of length $L = N_s a$, 
with $N_s$ the number of lattice sites and $a$ the lattice spacing fixed to unity by convention, which can be expressed as

\begin{equation}
 \ket{\Psi} = \sum_{\sigma_1,\sigma_2,..,\sigma_L} \mathit{\Psi}_{\sigma_1 \sigma_2 ...\sigma_L} \ket{\sigma_1 \sigma_2 ... \sigma_L},
 \label{general_qs_app}
\end{equation}

\noindent
with a $d$-dimensional local Hilbert space described by the local basis $\{\ket{\sigma_R}, \sigma_R = 1,...,d \}$. The first step consists of reshaping the vector
$\mathit{\Psi}$ containing $d^L$ components, \textit{ie.} living in the many-body Hilbert space, into a matrix of dimension $d \times d^{L-1}$ where the
coefficients are related as 
 
\begin{equation}
\mathit{\Psi}_{\sigma_1 \sigma_2 ...\sigma_L} = \Psi_{\sigma_1, (\sigma_2 ... \sigma_L)}.
\end{equation}

\noindent
As a second step, the singular value decomposition (SVD) applied to the rectangular matrix $\Psi$ leads to 

\begin{equation}
\Psi_{\sigma_1, (\sigma_2...\sigma_L)} = \sum_{a_1=1}^{\bar{a}_1} U\left[1\right]_{\sigma_1, a_1} S\left[1\right]_{a_1,a_1}(V^{\dag}\left[1\right])_{a_1,(\sigma_2...\sigma_L)},
\end{equation}

\noindent 
where $U\left[1\right]$ is a left-normalized matrix ($U^{\dag}[1]U[1] = I$) of dimension $d \times \bar{a}_1$ with $\bar{a}_1 \leq d$ the rank associated to 
the diagonal and square Schmidt matrix $S\left[1\right]$ of dimension $\bar{a}_1 \times \bar{a}_1$ where $\bar{a}_1$ corresponds to the number of non-zero positive singular
values and $V^{\dag}\left[1\right]$ is a right-normalized rectangular matrix ($V[1]V^{\dag}[1] = I$) of dimension $\bar{a}_1 \times d^{L-1}$. Then, 
we express the matrix $U$ as a collection of $d$ row vectors of dimension $1\times \bar{a}_1$ denoted by $A^{\sigma_1}\left[1\right]$ where the coefficients
are $A^{\sigma_1}\left[1\right]_{a_1}$ with $a_1 \in [|1,\bar{a}_1|]$. Finally, by reshaping the matrix $S \left[1\right]V^{\dag}\left[1\right]$ of dimension
$\bar{a}_1 \times d^{L-1}$ in a matrix $\tilde{\Psi}$ of dimension $\bar{a}_1 d \times d^{L-2}$ with entries $\tilde{\Psi}_{(a_1 \sigma_2),(\sigma_3...\sigma_L)} =
\left(S\left[1\right] V^{\dag}\left[1\right] \right)_{a_1, (\sigma_2 ...\sigma_L)} $, the previous equation can be rewritten as 

\begin{equation}
\Psi_{\sigma_1, (\sigma_2...\sigma_L)} = \sum_{a_1=1}^{\bar{a}_1} A^{\sigma_1}\left[1\right]_{a_1} \tilde{\Psi}_{(a_1 \sigma_2),(\sigma_3...\sigma_L)}.
\label{1mps_app}
\end{equation}

\noindent
As a third step, we factorize $\tilde{\Psi}$ the matrix of dimension $\bar{a}_1 d \times d^{L-2}$ using once again the SVD which leads to 

\begin{equation}
\tilde{\Psi}_{(a_1 \sigma_2), (\sigma_3...\sigma_L)} = \sum_{a_2=1}^{\bar{a}_2} U\left[2\right]_{(a_1 \sigma_2), a_2} S\left[2\right]_{a_2,a_2}(V^{\dag}\left[2\right])_{a_2,(\sigma_3...\sigma_L)},
\end{equation}

\noindent 
where $U\left[2\right]$ is a matrix of dimension $\bar{a}_1 d \times \bar{a}_2$ with $\bar{a}_2 \leq d^2$ the rank associated to the Schmidt matrix $S\left[2\right]$ 
of dimension $\bar{a}_2 \times \bar{a}_2$ and $V^{\dag}\left[2\right]$ is a matrix of dimension $\bar{a}_2 \times d^{L-2}$. Then, the matrix $U\left[2\right]$ is
decomposed as a collection of $d$ rectangular matrices $A[2]$ of dimension $\bar{a}_1 \times \bar{a}_2$. Finally, $S\left[2\right] V^{\dag}\left[2\right]$ is reshaped
in a matrix $\tilde{\tilde{\Psi}}$ of dimension $\bar{a}_2 d \times d^{L-3}$ with entries $\tilde{\tilde{\Psi}}_{(a_2 \sigma_3),(\sigma_4...\sigma_L)} = 
\left(S\left[2\right] V^{\dag}\left[2\right] \right)_{a_2, (\sigma_3 ...\sigma_L)}$. Consequently, $\tilde{\Psi}$ reads as

\begin{equation}
\tilde{\Psi}_{(a_1 \sigma_2), (\sigma_3...\sigma_L)} = \sum_{a_2=1}^{\bar{a}_2} A^{\sigma_2}\left[2\right]_{a_1,a_2} 
\tilde{\tilde{\Psi}}_{(a_2 \sigma_3),(\sigma_4...\sigma_L)}.
\end{equation}

\noindent
Few points must be underlined. $\tilde{\tilde{\Psi}}$ is a matrix of dimension $\bar{a}_2 d \times d^{L-3}$ with $\bar{a}_2$ the rank of the Schmidt matrix
$S\left[2\right]$ of dimension $\bar{a}_2 \times \bar{a}_2$. The number of singular values on the diagonal of the Schmidt matrix $S\left[2\right]$ is defined
as $\bar{a}_2 \leq \mathrm{min}\left( \bar{a}_1 d, d^{L-2} \right) = \bar{a}_1 d \leq d^{2}$. Therefore, the maximal dimension of the $\tilde{\tilde{\Psi}}$ matrix
is obtained for $\bar{a}_2 = d^2$ implying $\bar{a}_1 = d$ and is equal to $ d^2 d \times d^{L-3}$. Hence, $\tilde{\tilde{\Psi}}$ contains $d^L$ coefficients as expected. \\

\noindent
Finally, in order to get the MPS form of Eq.~\eqref{general_qs_app}, we replace the expression of $\tilde{\Psi}$ in Eq.~\eqref{1mps_app} which yields to

\begin{equation}
\Psi_{\sigma_1, (\sigma_2...\sigma_L)} = \sum_{a_1=1}^{\bar{a}_1} \sum_{a_2 = 1}^{\bar{a}_2} A^{\sigma_1}\left[1\right]_{a_1} A^{\sigma_2}\left[2\right]_{a_1,a_2}
\tilde{\tilde{\Psi}}_{(a_2 \sigma_3),(\sigma_4...\sigma_L)}.
\end{equation}

\noindent
We have to reproduce the previous steps for each lattice site $R$ to build the corresponding tensor $A^{\sigma_R}\left[R\right]$.
Thus, upon further SVDs, we obtain the following form for the coefficients of the initial matrix $\Psi_{\sigma_1, (\sigma_2...\sigma_L)}$ 

\begin{equation}
\Psi_{\sigma_1, (\sigma_2...\sigma_L)} = \sum_{a_1=1}^{\bar{a}_1} \sum_{a_2=1}^{\bar{a}_2} ... \sum_{a_{L-1} = 1}^{\bar{a}_{L-1}} A^{\sigma_1}\left[1\right]_{a_1} A^{\sigma_2}\left[2\right]_{a_1,a_2}
...~ A^{\sigma_L}\left[L\right]_{a_{L-1}}.
\label{2mps_app}
\end{equation}

\noindent
Consequently, a general (non translational invariant) quantum state written under its MPS form for a one-dimensional lattice has the
following expression, see also Fig.~\ref{MPS}, 

\begin{equation}
\ket{\Psi} = \sum_{\boldsymbol{\sigma}} A^{\sigma_1}\left[1\right]A^{\sigma_2}\left[2\right] ...~ A^{\sigma_{L-1}}\left[L-1\right] A^{\sigma_L}\left[L\right]
\ket{\boldsymbol{\sigma}},~~~\boldsymbol{\sigma} = \sigma_1 \sigma_2 ...~\sigma_L. 
\label{mps_app}
\end{equation}

\noindent
The latter consists of a local and compact representation of a general many-body quantum state. The tensor $A^{\sigma_1}\left[1\right]$ consists of a collection
of $d$ row vectors of dimension $\bar{a}_1$ and the last one $A^{\sigma_L}\left[L\right]$ in a collection of $d$ column vectors of dimension $\bar{a}_{L-1}$.
Besides, a tensor $A^{\sigma_R}[R]$ of dimension $\bar{a}_{R-1} \times \bar{a}_R \times d$ is associated to each lattice site $R$ of the chain leading to a local
representation of the quantum state without breaking its non-locality, \textit{ie.} its entanglement, which is contained in the coefficients of each tensor and
characterized by the so-called MPS bond dimension $\chi$ defined as $\chi = \mathrm{max}(\bar{a}_R)$. The MPS form is also a compact representation since the latter 
depends linearly on the length of the chain, \textit{ie.} the number of lattice sites, and not exponentially as for the general form presented at Eq.~\eqref{general_qs_app}.

%% file: text/appendix_4.tex
\setstretch{1.0} 

\chapter{\label{appendix4_single_structure}  Single structure of the $G_2$ density fluctuations deep in the Mott-insulating phase}
In order to explain the suppression of the twofold structure for the $G_2$ density fluctuations for a sudden global quench deep in the Mott-insulating phase 
(requiring $U\gg J$ and $\bar{n} \in \mathbb{N}^{*}$, here $\bar{n}=1$), we compute the function $G_2(R,t)$ while working along the lines of Ref.~\cite{barmettler2012}. Considering the manifold of 
doublon-holon pairs and mapping the resulting Hamiltonian of the 1D SRBH model into a quadratic fermionic one, the latter can be diagonalized \textit{via}
a Fermi-Bogolyubov transformation. Finally, it yields for the density-density correlation function 

\begin{equation}\label{eq:MIG2}
G_2(R,t) \simeq -2 \left[ |g_2(R,t)|^2 +|\bar{g}_2(R,t)|^2 \right],
\end{equation}

\noindent
with $g_2(R,t)$ and $\bar{g}_2(R,t)$ which can be expressed as follows

\begin{align}
& g_2(R,t) \sim \frac{J}{U}\frac{R}{t} \int_{\mathcal{B}} \frac{\mathrm{d}k}{2 \pi} \left \{ e^{i \left(2E_kt + kR\right)} + e^{i \left(2E_kt - kR \right)}
\right \},
\label{g2} \\
& \bar{g}_2(R,t) \sim \left(\frac{J}{U} \right)^2 \int_{\mathcal{B}} \frac{\mathrm{d}k}{2 \pi} \sin^2(k) \left \{ e^{i \left(2E_kt - kR \right)} +
e^{-i \left(2E_kt + kR \right)} \right \},
\label{g2_bar}
\end{align}

\noindent
where $\mathcal{B} = [-\pi,\pi]$ denotes the first Brillouin zone and the excitation spectrum is 
$2E_k \simeq \sqrt{ \left[ U - 2J(2\bar{n}+1)\cos(k)\right]^2 + 16 J^2 \bar{n}(\bar{n}+ 1)\sin^2(k)}$, see Eq.~\eqref{mott_disp_rel_fermionization}.

\paragraph{Quench deep into the Mott-insulating phase}
For a global quench deep in the MI phase ($U \gg J$), the second right-hand-side term (which scales as $(J/U)^2$) in Eq.~(\ref{eq:MIG2}) is much smaller
than the first one (scales as $J/U$) and the former can be neglected. Using Eq.~(\ref{g2}), it yields explicitly for $G_2(R,t) \simeq -2|g_2(R,t)|^2$,

\begin{equation}\label{eq:MIG2bis}
G_2(R,t) \sim - 2 \left(\frac{J}{U}\right)^2 \left(\frac{R}{t}\right)^2 \left| \int_{\mathcal{B}} \frac{\mathrm{d}k}{2 \pi}
\left \{ e^{i \left(2E_kt + kR \right)} + e^{i \left(2E_kt - kR \right)} \right \} \right|^2.
\end{equation}

\noindent
Moreover, the excitation spectrum may be expanded in powers of $J/U$. Up to first-order, it yields $2E_k \simeq U-2J(2\bar{n}+1)\cos(k)$ the excitation
spectrum valid in the strong-coupling regime ($U \gg J$). The gap term $e^{iUt}$ can then be factorized in the two terms under the integral in 
Eq.~(\ref{eq:MIG2bis}) and disappears due to the square modulus. Introducing the effective excitation spectrum $ 2\tilde{E}_k = -2J(2\bar{n}+1)\cos(k)$, we then find
$G_2 \simeq -2 \vert g_2(R,t)\vert^2$ with  

\begin{equation}\label{eq:MIG2bis2}
g_2(R,t) \sim \frac{J}{U} \frac{R}{t} \int_{\mathcal{B}} \frac{\mathrm{d}k}{2\pi}
\left \{ e^{i \left(2\tilde{E}_kt + kR\right)} + e^{i \left(2\tilde{E}_kt - kR \right)} \right \}.
\end{equation}

\noindent
To determine the asymptotic behavior, the previous integral may be evaluated using the stationary phase approximation (see also Appendix.~\ref{appendix2_sp}).
In the infinite time and distance limit along the line $R/t=\mathrm{cst}$, the integral in Eq.~\eqref{eq:MIG2bis2} is dominated by
the momentum contributions with a stationary phase (sp), \textit{ie.} $\partial_k (2\tilde{E}_k t \pm kR) = 0$ or, equivalently,
$2\tilde{V}_{\mathrm{g}}(k_{\mathrm{sp}}) = \pm R/t$ where $\tilde{V}_{\mathrm{g}}(k) =\partial_k \tilde{E}_k$ is the group velocity of the effective excitation spectrum.
Since the latter is upper bounded by the value $\tilde{V}_{\mathrm{g}}^* = \mathrm{max}(\tilde{V}_{\mathrm{g}}) = J(2\bar{n}+1)$, it has a solution only for
$R/t < 2\tilde{V}_{\mathrm{g}}^*$. We then find

\begin{equation}
\label{g2_ksp}
g_2(R,t) \sim \frac{J}{U} \frac{\tilde{V}_{\mathrm{g}}(k_{\mathrm{sp}})}{\left(|\partial_k^{2}\tilde{E}_{k_{\mathrm{sp}}}| t\right)^{1/2}} \left[
\cos\left(2\tilde{E}_{k_{\mathrm{sp}}}t - k_{\mathrm{sp}} R + \sigma \frac{\pi}{4} \right) + i\sin \left(2\tilde{E}_{k_{\mathrm{sp}}}t - k_{\mathrm{sp}} R + \sigma\frac{\pi}{4} \right) \right],
\end{equation}

\noindent
with $\sigma = \mathrm{sgn} \left(\partial_k^2 \tilde{E}_{k_{\mathrm{sp}}} \right)$. For both the real and imaginary parts of $g_2(R,t)$, the
correlations are activated ballistically at the time $t=R/2\tilde{V}_{\mathrm{g}}^*$. It defines a linear correlation edge (CE) with velocity $V_{\mathrm{CE}} = 
2\tilde{V}_{\mathrm{g}}^*$. In addition, Eq.~\eqref{g2_ksp} also yields a series of local maxima, defined by the equation $2\tilde{E}_{k_{\mathrm{sp}}}t - k_{\mathrm{sp}} R = \mathrm{cst}$. In the vicinity of the CE cone,
these maxima (m) propagate at the velocity $V_{\mathrm{m}} = 2\tilde{V}_{\varphi}^* = 2\tilde{E}_{k^*}/k^*$, \textit{ie.} twice the phase velocity at the
maximum of the group velocity, $k^*$. Hence, the real and imaginary parts of $g_2(R,t)$ both display a twofold structure with a CE velocity $2\tilde{V}_{\mathrm{g}}^* = 2J(2\bar{n}+1)$ and  a velocity of the maxima $2\tilde{V}_{\varphi}^* = 0$, as shown on Figs.~\ref{fig:g2_analysis}(a) and (b).
In contrast, $G_2(R,t)$, does not display the twofold structure. This is because it is the sum of the squares of the two latter contributions [see Eq.~\eqref{g2_ksp}], which are shifted by half a period and cancel each other.
It thus gives a single cone structure, characterized by the sole CE velocity $2\tilde{V}_{\mathrm{g}}^*$, as shown on Fig.~\ref{fig:g2_analysis}(c).

\paragraph{Quench into the Mott-insulating phase for moderate $U/J$}
For moderate values of $U/J$, the second term in the right-hand-side of Eq.~(\ref{eq:MIG2}), $|\bar{g}_2(R,t)|^2$, becomes relevant. 
Relying again on the stationary phase approximation to characterize the asymptotic behavior of $\bar{g}_2(R,t)$, it yields

\begin{equation}
\bar{g}_2(R,t) \sim \left(\frac{J}{U}\right)^2 \frac{\sin^2(k_\mathrm{sp})}{\left(|\partial_k^{2}E_{k_\mathrm{sp}}| t\right)^{1/2}} 
\cos\left(2E_{k_\mathrm{sp}}t - k_\mathrm{sp} R + \sigma' \frac{\pi}{4} \right),
\end{equation}

\noindent
with $\sigma' = \mathrm{sgn}\left(\partial_k^2 E_{k_\mathrm{sp}} \right)$ and $E_k$ the excitation spectrum given 
at Eq.~\eqref{mott_disp_rel_fermionization}. Using the same argument as above, we find that $\bar{g}_2(R,t)$ shows a 
twofold structure defined by, now, the CE velocity $2V_{\mathrm{g}}^* = 2 \mathrm{max}\left(\partial_k E_k \right)$ and the maxima
velocity $2V_\varphi^* = 2E_{k^*}/k^* \neq 0$. Since there is a single contribution here, the quantity $\vert\bar{g}_2(R,t)\vert^2$ displays 
a twofold structure with the same characteristic velocities. More precisely, both the length and time scales of the oscillations are divided
by two but the velocities are not affected. \\
For a global quench into the MI phase at a moderate value of $U/J$, both $|g_2(R,t)|^2$ and $|\bar{g}_2(R,t)|^2$ contribute to the density-density
correlation function $G_2(R,t)$. While the $|g_2(R,t)|^2$ contribution is characterized by the sole CE velocity $2V_{\mathrm{g}}^*$,
the $|\bar{g}_2(R,t)|^2$ contribution provides the double structure observed on $G_2$ for $6<U/J<10$ in the $t$-MPS calculations, see Fig.~\ref{fig:numerics_mott_u}(b3). 

%% file: text/appendix_10.tex
\setstretch{1.0} 

\chapter{\label{appendix_local_quench_spin} Local quench dynamics of the local magnetization for 1D $s=1/2$ spin lattice models}
In this section, we derive the general expression of the space-time local magnetization $\langle \Psi_0 | \hat{S}^z_R(t) | \Psi_0 \rangle$ for a sudden
local quench in a polarized phase along the $z$ axis. In the following, the Hamiltonian $\hat{H}$ is assumed to describe a one-dimensional $s=1/2$ 
spin lattice model with short- or long-range interactions in a polarized phase along the $z$ axis. The perturbed initial state $\ket{\Psi_0}$ is built from the ground state (one
of the ground states) of $\hat{H}$ corresponding to a fully polarized state along the $z$ with a positive total magnetization. The latter describes the
following many-body quantum state $\ket{\Psi} = \ket{\uparrow}^{\otimes N_s}$ where $N_s$ refers to the total number of lattice sites. To drive the quantum system far from equilibrium 
\textit{via} a sudden local quench, a local perturbation (a spin-flip) is applied on the central spin for the state $\ket{\Psi}$ using the spin operator $\hat{S}^{-}_{N_s/2}$. This leads to the perturbed initial state $\ket{\Psi_0}
= \ket{\uparrow ... \uparrow \downarrow \uparrow ... \uparrow}$ which corresponds to a highly-excited state with respect to the ground state energy of the Hamiltonian
$\hat{H}$, \textit{ie.} to the energy of the many-body quantum state $\ket{\Psi}$. \\ 

The Hamiltonian $\hat{H}$ is assumed to display a quadratic Bose form in momentum space given by 

\begin{equation}
\hat{H} = e_{0} + \frac{1}{2} \sum_{k} \mathcal{A}_{k} \left( \hat{a}^{\dag}_k \hat{a}_k + \hat{a}_{-k} \hat{a}^{\dag}_{-k} \right) +
\mathcal{B}_{k} \left( \hat{a}^{\dag}_k \hat{a}^{\dag}_{-k} + \hat{a}_k \hat{a}_{-k} \right),
\label{bose_quadra}
\end{equation}

\noindent
where $\hat{a}_k$ ($\hat{a}_k^{\dag}$) refers to the annihilation (creation) of a bosonic particle in the mode $k$ and $e_0$ to a constant energy. 
Since the Hamiltonian $\hat{H}$ is supposed to be in a polarized phase along the $z$ axis, the previous quadratic Bose form at Eq.~\eqref{bose_quadra} is found
using the following Holstein-Primakoff transformation

\begin{equation}
\hat{S}^x_R = \frac{1}{2} \left(\hat{a}^{\dag}_R + \hat{a}_R \right),~~~\hat{S}^y_R = \frac{-1}{2i} \left( \hat{a}^{\dag}_R - \hat{a}_R \right),~~~ \hat{S}^z_R =
\frac{1}{2} - \hat{a}^{\dag}_R \hat{a}_R.
\end{equation}

\noindent
For this Holstein-Primakoff transformation and using the relations $\hat{S}^{\pm}_R = \hat{S}^x_R \pm i \hat{S}^y_R$, one obtains 

\begin{equation}
 \hat{S}^{-}_R = \hat{a}^{\dag}_R, ~~~ \hat{S}^{+}_R = \hat{a}_R,~~~\ket{\uparrow}_R = \ket{0}_R,~~~ \ket{\downarrow}_R = \ket{1}_R,
\end{equation}

\noindent
meaning that the vacuum state of bosonic particles on the lattice site $R$ denoted by $\ket{0}_R$ corresponds to a spin up state $\ket{\uparrow}_R$
and a spin down state $\ket{\downarrow}_R$ to the presence of one bosonic particle, \textit{ie.} $\ket{1}_R$. Indeed, $\ket{\downarrow}_R = \hat{S}^{-}_R\ket{\uparrow}_R = \hat{a}^{\dag}_R \ket{0}_R = \ket{1}_R$ and 
similarly $\ket{\uparrow}_R = \hat{S}^{+}_R \ket{\downarrow}_R = \hat{a}_R \ket{1}_R = \ket{0}_R$. Besides, to get the same dimension of the local Hilbert space 
for the $s=1/2$ spin lattice model ($\mathrm{dim}(\mathbb{H}_R)=2$), the occupation number of bosonic particles on lattice site $R$ is restricted to two possibilities \textit{ie.} $n_R \in \{0,1\}$ where $\hat{n}_R 
\ket{n_R} = n_R \ket{n_R}$. Consequently, the expectation value  $\langle \Psi_0 | \hat{S}^z_R(t) | \Psi_0 \rangle$ may be written as 

\begin{equation}
\langle \Psi_0 | \hat{S}^z_R(t)| \Psi_0 \rangle = \langle \Psi | \hat{S}^+_{N_s/2} \hat{S}^z_R(t) \hat{S}^-_{N_s/2} | \Psi \rangle = \langle \Psi | \hat{a}_{N_s/2}
\left[\frac{1}{2} - \hat{a}_R^{\dag}(t) \hat{a}_R(t) \right] \hat{a}^{\dag}_{N_s/2} | \Psi \rangle,
\label{tmp1}
\end{equation}

\noindent
where $\ket{\Psi} = \ket{\uparrow}^{\otimes N_s}$. Then, since $\hat{H}$ is quadratic in terms of the bosonic operators, the Wick theorem
can be used to simplify Eq.~\eqref{tmp1}. It yields for the local magnetization

\begin{align}
& \langle \Psi_0 | \hat{S}^z_R(t) | \Psi_0 \rangle = \frac{1}{2} - \langle \hat{n}_R(t) \rangle - \langle \hat{a}_{N_s/2} \hat{a}^{\dag}_R(t)
\rangle \langle \hat{a}_R(t) \hat{a}^{\dag}_{N_s/2} \rangle - \langle \hat{a}_{N_s/2} \hat{a}_R(t) \rangle \langle \hat{a}^{\dag}_R(t) \hat{a}^{\dag}_{N_s/2}\rangle,
\nonumber \\
& \langle \Psi_0 | \hat{S}^z_R(t) | \Psi_0 \rangle = \frac{1}{2} - \langle \hat{n}_R(t) \rangle - |\langle \hat{a}_{N_s/2} \hat{a}^{\dag}_R(t)\rangle|^2
 - |\langle \hat{a}_{N_s/2} \hat{a}_R(t) \rangle|^2, \label{tmp2}
\end{align}

\noindent
where $\braket{..}$ represents the expectation value with respect to the quantum state $\ket{\Psi} = \ket{\uparrow}^{\otimes N_s}$. In order to compute Eq.~\eqref{tmp2},
the latter is expressed in terms of the bosonic operators in momentum space ($\hat{a}_k$, $\hat{a}^{\dag}_k$). Then, one can rely on the following properties of
the bosonic Bogolyubov transformation

\begin{align}
& \hat{a}_k^{\dag}(t) = u_k \hat{\beta}_{k}^{\dag}(t) + v_k \hat{\beta}_{-k}(t), ~~~ \hat{a}_k(t) = u_k \hat{\beta}_k(t) + v_k \hat{\beta}_{-k}^{\dag}(t), \label{1} \\ \nonumber \\
& \hat{\beta}_k(t) = e^{-iE_kt}\hat{\beta}_k(0),~~~ \hat{\beta}_{k}^{\dag}(t) = e^{iE_kt}\hat{\beta}^{\dag}_k(0), \label{2} \\ \nonumber \\
& \hat{\beta}_k \ket{\Psi} = 0, ~~~ \bra{\Psi} \hat{\beta}^{\dag}_k = 0, \label{3}
\end{align}

\noindent
where $\hat{\beta}_k$ ($\hat{\beta}^{\dag}_k$) denotes the annihilation (creation) of a Bogolyubov quasiparticle at quasimomentum $k$, 
$E_k$ denotes the excitation spectrum of $\hat{H}$ in the polarized phase along the $z$ axis and has the symmetry $k\rightarrow -k$. 
Note that $u_k$ and $v_k$ are real coefficients satisfying $u_k = u_{-k}$ and $v_k = v_{-k}$, $\forall k \in \mathcal{B} = [-\pi, \pi]$. 
Equation \eqref{1} represents the canonical transformation of the bosonic operators into (bosonic) Bogolyubov operators. The time evolution of the Bogolyubov operators are given at Eq.~\eqref{2}. Finally,
equation \eqref{3} represents mathematically the fact that the non-perturbed many-body quantum state $\ket{\Psi}$ does not support any quasiparticle excitation (such as 
spin-wave excitations for instance since 1D $s=1/2$ spin lattice models are considered here). This Bogolyubov transformation allows us to diagonalize the quadratic Bose form of
$\hat{H}$ at Eq.~\eqref{bose_quadra}. Hence, the Hamiltonian $\hat{H}$ can be expressed as $\hat{H} = \sum_k E_k \hat{\beta}^{\dag}_k \hat{\beta}_k$ up to a constant term,
\textit{ie.} an energy shift, with an excitation spectrum $E_k$ defined as $E_k = \sqrt{\mathcal{A}_k^2 - \mathcal{B}_k^2}$. \\

\noindent
Expressing the bosonic operators in momentum space and relying on the properties of the bosonic Bogolyubov transformation, it yields for 
$\langle \hat{n}_R(t) \rangle$ to 
 
\begin{align}
& \langle \hat{n}_R(t) \rangle = \frac{1}{N_s} \sum_{k_1,k_2} e^{i(k_1 - k_2)R} \langle \hat{a}_{k_1}^{\dag}(t) \hat{a}_{k_2}(t) \rangle = \frac{1}{N_s} \sum_k v_k^2.
\end{align}

\noindent
For the correlator $|\langle \hat{a}_{N_s/2} \hat{a}^{\dag}_R(t) \rangle|^2$, it yields the following expression
\begin{align}
& |\langle \hat{a}_{N_s/2} \hat{a}^{\dag}_R(t) \rangle|^2 = \left| \frac{1}{N_s} \sum_{k_1, k_2} e^{-ik_1(N_s/2)} e^{ik_2R} \langle \hat{a}_{k_1} \hat{a}_{k_2}^{\dag}(t)
\rangle \right|^2 = \left| \frac{1}{N_s} \sum_k u_k^2 e^{i[k(R-N_s/2) + E_kt]} \right|^2,
\end{align}

\noindent
and for the second term $|\langle \hat{a}_{N_s/2} \hat{a}_R(t) \rangle|^2$ in the theoretical expression of $\langle \Psi_0 | \hat{S}^z_R(t) | \Psi_0 \rangle$, 
we find

\begin{align}
& |\langle \hat{a}_{N_s/2} \hat{a}_R(t) \rangle |^2 = \left| \frac{1}{N_s} \sum_{k_1,k_2} e^{-ik_1(N_s/2)} e^{-ik_2R} \langle 
\hat{a}_{k_1} \hat{a}_{k_2}(t) \rangle \right|^2 = \left| \frac{1}{N_s} \sum_k  u_k v_k e^{i[k(R-N_s/2) + E_kt]} \right|^2.
\end{align}

\noindent
Finally, the analytical expression of $\langle \Psi_0 | \hat{S}^z_R(t) | \Psi_0 \rangle$ may be written as  


\begin{align}
& \langle \Psi_0 | \hat{S}^z_R(t) | \Psi_0 \rangle = \left(\frac{1}{2} - \frac{1}{N_s}\sum_k v_k^2 \right) 
 - \left| \frac{1}{N_s} \sum_k u_k^2 e^{i[k(R - N_s/2) + E_kt]} \right|^2 - \left|\frac{1}{N_s} \sum_k u_k v_k e^{i[k(R-N_s/2) + E_kt]} \right|^2.
\end{align}

\noindent
Using the formula $\sum_k = (1/2) \left( \sum_k + \sum_{-k} \right)$ and considering the thermodynamic limit for one-dimensional 
lattice models given by $(1/N_s)\sum_k \equiv \int_{\mathcal{B}} \mathrm{d}k/2\pi$, it yields for the local magnetization the following form 

\begin{align}
& \langle \Psi_0 | \hat{S}^z_R(t) | \Psi_0 \rangle = \left(\frac{1}{2} - \int_{\mathcal{B}} \frac{\mathrm{d}k}{2\pi} v_k^2 \right) 
- \left| \int_{\mathcal{B}} \frac{\mathrm{d}k}{2\pi} u_k^2 \left\{ \frac{e^{i[k(R-N_s/2) + E_kt]}+ e^{i[-k(R-N_s/2) + E_kt]}}{2} \right \} \right|^2 \nonumber \\
& ~~~~~~~~~~~~~~~~~~ - \left| \int_{\mathcal{B}} \frac{\mathrm{d}k}{2\pi} u_k v_k \left \{ \frac{e^{i[k(R-N_s/2) + E_kt]}+ e^{i[-k(R-N_s/2) + E_kt]}}{2} \right \}
\right|^2.
\label{eq_expec_val}
\end{align} 
 
\noindent
In the main text, we have plotted the following observable $1/2 - \langle \Psi_0 | \hat{S}^z_R(t) | \Psi_0 \rangle$ leading to 

\begin{align}
& \frac{1}{2} - \langle \Psi_0 | \hat{S}^z_R(t) | \Psi_0 \rangle = \int_{\mathcal{B}} \frac{\mathrm{d}k}{2\pi} v_k^2 
+ \left| \int_{\mathcal{B}} \frac{\mathrm{d}k}{2\pi} u_k^2 \left \{ \frac{e^{i[k(R-N_s/2) + E_kt]}+ e^{i[-k(R-N_s/2) + E_kt]}}{2} \right \} \right|^2 \nonumber \\
& ~~~~~~~~~~~~~~~~~~~~~~~~ + \left| \int_{\mathcal{B}} \frac{\mathrm{d}k}{2\pi} u_k v_k \left \{ \frac{e^{i[k(R-N_s/2) + E_kt]}+ e^{i[-k(R-N_s/2) + E_kt]}}{2} \right \}
\right|^2,
\label{eq_expec_val_2}
\end{align} 

\noindent
where the coefficients $u_k$, $v_k$ are defined by 

\begin{align}
 u_k,v_k = \pm \sqrt{\frac{1}{2}\left( \frac{\mathcal{A}_k}{E_k} \pm 1 \right)}.
\end{align}

\noindent
Both forms at Eq.~\eqref{eq_expec_val} and \eqref{eq_expec_val_2} are valid for one-dimensional $s=1/2$ spin lattice models displaying a polarized phase along the
$z$ axis. In the following, several examples of such lattice model are provided : 

\begin{itemize}
\item The 1D short-range $s=1/2$ Heisenberg model [see Eq.~\eqref{H_heis}] in the quasi-long-range gapless ferromagnetic phase along the $z$ axis, characterized by $\mathcal{A}_k =
E_k = J[1-\cos(k)]$ and $\mathcal{B}_k = 0$, $\forall k \in \mathcal{B} = [-\pi, \pi]$. Hence, $u_k = 1$ and $v_k = 0$, $\forall k \in \mathcal{B} = [-\pi, \pi]$.
Therefore, equation \eqref{eq_expec_val_2} can be simplified and it yields for the local magnetization the following form 

\begin{equation}
 \frac{1}{2} - \langle \Psi_0 | \hat{S}^z_R(t) | \Psi_0 \rangle = \left| \int_{\mathcal{B}} \frac{\mathrm{d}k}{2\pi} \left \{ \frac{ e^{i[k (R-N_s/2) + E_kt]}
+ e^{i[-k(R-N_s/2) + E_k t]}}{2} \right \} \right|^2.
\end{equation}

\item The 1D short-range $s=1/2$ transverse Ising model in the $z$-polarized phase ($h \gg J$) whose Hamiltonian is represented at Eq.~\eqref{short_range_ising}. 
The latter is characterized by $\mathcal{A}_k = J\cos(k) + 2h$ and $\mathcal{B}_k = J\cos(k)$, $\forall k \in \mathcal{B} = [-\pi,\pi]$ the first Brillouin zone.
The gapped excitation spectrum $E_k$ is thus determined by $E_k = \sqrt{\mathcal{A}_k^2 - \mathcal{B}_k^2} = 2\sqrt{h[h+J\cos(k)]}$. 

\item The 1D long-range $s=1/2$ transverse Ising model (see Eq.~\ref{eq:LRTI}) in the $z$-polarized phase where $\mathcal{A}_k = J P_{\alpha}(k) + 2h$,
$\mathcal{B}_k = J P_{\alpha}(k)$, $\forall k \in \mathcal{B}$ ($P_{\alpha}$ represents the Fourier transform of the power-law function
$|R|^{-\alpha}$). Hence, the gapped excitation spectrum $E_k$ reads as $E_k = \sqrt{\mathcal{A}_k^2 - \mathcal{B}_k^2} = 2\sqrt{h[h+JP_{\alpha}(k)]}$.

\item The 1D long-range $s=1/2$ XXZ (LRXXZ) model (see Eq.~\ref{eq:XXZ}) in the long-range-order gapped ferromagnetic phase along the $z$ axis 
requiring $\epsilon < -1$, $\forall \alpha \in \mathbb{R}^{+*}$. The anisotropy along the $z$ axis opens a finite gap in the excitation spectrum. 
The latter is characterized by $E_k = \mathcal{A}_k = (J/2)\left[-P_{\alpha}(k)- \epsilon P_{\alpha}(0) \right]$, $\mathcal{B}_k = 0$ meaning that $u_k = 1$ and
$v_k = 0$, $\forall k \in \mathcal{B}$. 

\item The 1D long-range $s=1/2$ Heisenberg model (see Eq.~\ref{eq:XXZ} for $\epsilon = -1$) in the long-range-order gapless ferromagnetic phase along the 
$z$ axis. Its excitation spectrum $E_k$ is directly given by the one of the 1D LRXXZ model in the long-range-order gapped ferromagnetic phase along the $z$ axis by replacing $\epsilon$ by $-1$.
Thus, the gapless excitation spectrum of the 1D long-range $s=1/2$ Heisenberg model reads as $E_k = \mathcal{A}_k = (J/2)\left[P_{\alpha}(0) -P_{\alpha}(k)\right]$ knowing that $\mathcal{B}_k = 0$, 
and implies that $u_k = 1$ and $v_k = 0$ $\forall k \in \mathcal{B}$. 
\end{itemize}

%% file: text/appendix_12.tex
\setstretch{1.0} 

\chapter{\label{appendix_entropies} The von Neumann and R\'enyi entropies}
In this appendix, we provide additional details about the definition and properties of the von Neumann and R\'enyi entropies. 
The latter correspond to entropy measures, and not physical observables \footnote{Indeed, the von Neumann and R\'enyi entropies are not physical observables since they 
do not depend linearly on the reduced density matrix, see definitions later.}, allowing us to characterize the amount of entanglement
for any many-body quantum state $\ket{\Psi}$. \\

Let us first fix the notation. In the following, a bipartite quantum system $\Omega = A+B$ is considered. $\Omega$ refers to the full quantum system
whereas $A$ corresponds to a quantum subsystem of $\Omega$ and $B = \bar{A}$ the complementary subsystem. $\hat{\rho} = \hat{\rho}_{AB} = \ket{\Psi} \bra{\Psi}$
denotes the density matrix associated to the many-body quantum state $\ket{\Psi}$ living in the full Hilbert space $\mathbb{H} = \mathbb{H}_{\Omega} = \mathbb{H}_A
\otimes \mathbb{H}_{B}$. From the density matrix $\hat{\rho}$, one can define the so-called reduced density matrix $\hat{\rho}_A$ ($\hat{\rho}_B$) of the subsystem 
$A$ ($B$) by tracing out all the degrees of freedom of the complementary subsystem $B$ ($A$). As a consequence, it yields, 

\begin{equation}
\hat{\rho}_{A} = \mathrm{Tr}_B\left( \hat{\rho} \right)~~~ \mathrm{and}~~~ \hat{\rho}_{B} = \mathrm{Tr}_A\left( \hat{\rho} \right).
\end{equation}

\noindent
In order to characterize the amount of entanglement for the many-body quantum state $\ket{\Psi}$ living in the bipartite system $\Omega = A + B$, it is particularly 
helpful to rely on its Schmidt decomposition given by 

\begin{equation}
 \ket{\Psi} = \sum_{j=1}^{\mathrm{min}[\mathrm{dim}(\mathbb{H}_A), \mathrm{dim}(\mathbb{H}_B)]} \mathrm{\Psi}_j \ket{\phi_j}_A \otimes \ket{\phi_j}_B,
 \label{schmidt}
\end{equation}

\noindent
where the collection of states $\{\ket{\phi}_A \}$ and $\{ \ket{\phi}_B \}$ forms an orthonormal basis of the subsystem $A$ and $B$ respectively.
The latter can be easily found by considering the general many-body quantum state $\ket{\Psi} = \sum_{l,m} \mathit{\Psi}_{l,m}\ket{\varphi_l}_A \otimes \ket{\varphi_m}_B$
and by applying the singular value decomposition on the state matrix $\mathit{\Psi}$ of dimension $\mathrm{dim}(\mathbb{H}_A) \times \mathrm{dim}(\mathbb{H}_B)$.   
The normalization of the quantum state $\ket{\Psi}$ leads to the following condition $\sum_{j} |\mathrm{\Psi}_j |^2  = \sum_{j} \Pi_j = 1$. Straightforwardly, it implies 
$\mathrm{Tr}\left(\hat{\rho} \right) = \mathrm{Tr}\left(\hat{\rho}_A \right) = \mathrm{Tr} \left(\hat{\rho}_B \right) = 1$. \\

The R\'enyi entropy of order $n$ denoted by $\mathcal{S}_n$, and the von Neumann entropy $\mathcal{S}$, associated to the subsystem 
$A$ are defined respectively by

\begin{align}
 & \mathcal{S}_n(\hat{\rho}_A) = \frac{1}{1-n} \mathrm{log} \left[ \mathrm{Tr} \left( \hat{\rho}_A^n \right) \right], 
 ~\forall n \in \mathbb{R}^+\backslash \{1\}, \label{reon} \\ 
 & \mathcal{S}(\hat{\rho}_A) =  \lim_{n \rightarrow 1} \mathcal{S}_n(\hat{\rho}_A) = - \mathrm{Tr}\left[ \hat{\rho}_A \mathrm{log}(\hat{\rho}_A) \right].
\end{align}

\noindent
Note that the previous expressions are exactly the same for the subsystem $B$ (modify the reduced density matrix $\hat{\rho}_A$ by $\hat{\rho}_B$).
It is important to point out that $\mathcal{S}_n(\hat{\rho}_A) = \mathcal{S}_n(\hat{\rho}_B) 
~\forall n \in \mathbb{R}^+ \backslash \{1 \}$ 
and $\mathcal{S}(\hat{\rho}_A) = \mathcal{S}(\hat{\rho}_B)$ according to the Schmidt decomposition of $\ket{\Psi}$ presented at Eq.~\eqref{schmidt}. Indeed, the
entropy measures characterize the amount of entanglement between the two subsystems $A$ and $B$ which corresponds to a boundary property of the quantum model.
Hence, for any bipartition, the value of the entropies does not depend on the choice of the subsystem to perform the calculation. 

\paragraph{von Neumann entropy : special case of the R\'enyi entropy}
In what follows, we provide the proof that the von Neumann entropy (also called entanglement entropy) corresponds to the R\'enyi entropy in the limit $n \rightarrow 1$,
\textit{ie.} $\mathcal{S}(\hat{\rho}_A) =  \lim_{n \rightarrow 1} \mathcal{S}_n(\hat{\rho}_A), \forall \hat{\rho}_A$. The many-body quantum state $\ket{\Psi}$ fulfills the
form presented at Eq.~\eqref{schmidt} and is assumed to be well normalized. We first investigate the term $\mathrm{Tr}\left( \hat{\rho}_A^n \right)$ in the expression
of the R\'enyi entropy of order $n$, see Eq.~\eqref{reon}. It yields 

\begin{align}
& \mathrm{Tr}\left( \hat{\rho}_A^n \right) = \sum_j \left(|\mathrm{\Psi_j}|^2\right)^n = \sum_j \Pi_j^n = 1 - \sum_j \left( \Pi_j - \Pi_j^n \right), \\
&  \mathrm{Tr}\left( \hat{\rho}_A^n \right) = 1 - \sum_j \Pi_j \left(1-\Pi_j^{n-1} \right) = 1 - \sum_j \Pi_j \left[1-e^{(n-1)\mathrm{log}(\Pi_j)} \right].
\end{align}
 
\noindent
Then, by considering the limit $n\rightarrow 1$ and using a Taylor expansion of the exponential, we obtain

\begin{align}  
& \lim_{n\rightarrow1} \mathrm{Tr}\left( \hat{\rho}_A^n \right) = 1 - (1-n) \sum_j \Pi_j \mathrm{log}(\Pi_j).
\end{align}

\noindent
Then, it yields for the R\'enyi entropy in the limit $n\rightarrow 1$ the following expression 

\begin{align}
& \lim_{n\rightarrow1} \mathcal{S}_n(\hat{\rho}_A) = \frac{1}{1-n} \mathrm{log} [ 1 - (1-n)\sum_j \Pi_j \mathrm{log}(\Pi_j)], \\
& \lim_{n\rightarrow1} \mathcal{S}_n(\hat{\rho}_A) = -\frac{1}{1-n} (1-n)\sum_j \Pi_j \mathrm{log}(\Pi_j) = - \sum_j \Pi_j \mathrm{log}(\Pi_j).
\end{align}

\noindent
Finally, we have shown that 

\begin{align}
& \lim_{n\rightarrow1} \mathcal{S}_n(\hat{\rho}_A) = -\mathrm{Tr}\left[ \hat{\rho}_A \mathrm{log}(\hat{\rho}_A) \right] = \mathcal{S}(\hat{\rho}_A),~\forall \hat{\rho}_A.
\end{align}

\paragraph{Extrema of the R\'enyi entropies $\mathcal{S}_n$}
In the following, we investigate the behavior of the R\'enyi entropies and more precisely their extrema. To do so, one considers
a many-body quantum state $\ket{\Psi}$ in a superposition of two different states having the form 
\begin{equation}
 \ket{\Psi} = \alpha e^{i\Phi_1} \ket{\phi_1}_A \otimes \ket{\phi_1}_B  + \beta e^{i\Phi_2} \ket{\phi_2}_A \otimes \ket{\phi_2}_B,
 \label{state2}
\end{equation}

\noindent
where $(\Phi_1,\Phi_2, \alpha, \beta) \in \mathbb{R}^{4}$. The normalization of the quantum state implies the condition $\alpha^2 + \beta^2 = 1$. 
The probability $p$ ($p'$) that $\ket{\Psi}$ is projected on the first (second) quantum state is characterized by
$p = |(\bra{\phi_1}_A \otimes \bra{\phi_1}_B) \ket{\Psi}|^2 = \alpha^2 \in [0,1]$  ($p' = \beta^2 = 1 -\alpha^2 = 1-p$).
For the specific form of $\ket{\Psi}$ at Eq.~\eqref{state2}, the R\'enyi entropy $\mathcal{S}_n(\hat{\rho}_A) = \mathcal{S}_n(\hat{\rho}_B)$ 
can be written as 
\begin{align}
& \mathcal{S}_n(\hat{\rho}_A) =  \frac{1}{1-n} \mathrm{log} \left[ \mathrm{Tr} \left( \hat{\rho}_A^n \right) \right] =  \frac{1}{1-n} \mathrm{log} \left[ \mathrm{Tr} 
\begin{pmatrix}
p & 0 \\
0 & 1-p
\end{pmatrix}^n  \right]  = \frac{1}{1-n} \mathrm{log} \left[ p^n + (1-p)^n \right].
\label{final_re_2}
\end{align}

\noindent
Finally, the derivative with respect to the probability $p$ may be written as follows

\begin{equation}
\frac{\partial \mathcal{S}_n(\hat{\rho}_A)}{\partial p} = \frac{\partial \mathcal{S}_n(p)}{\partial p} =
\frac{n}{1-n} \frac{p^{n-1} - (1-p)^{n-1}}{p^n + (1-p)^n}.
\label{final_dre_2}
\end{equation}

\begin{figure}[!h]
\centering
\begin{tabular}{cc}
\includegraphics[scale = 0.38]{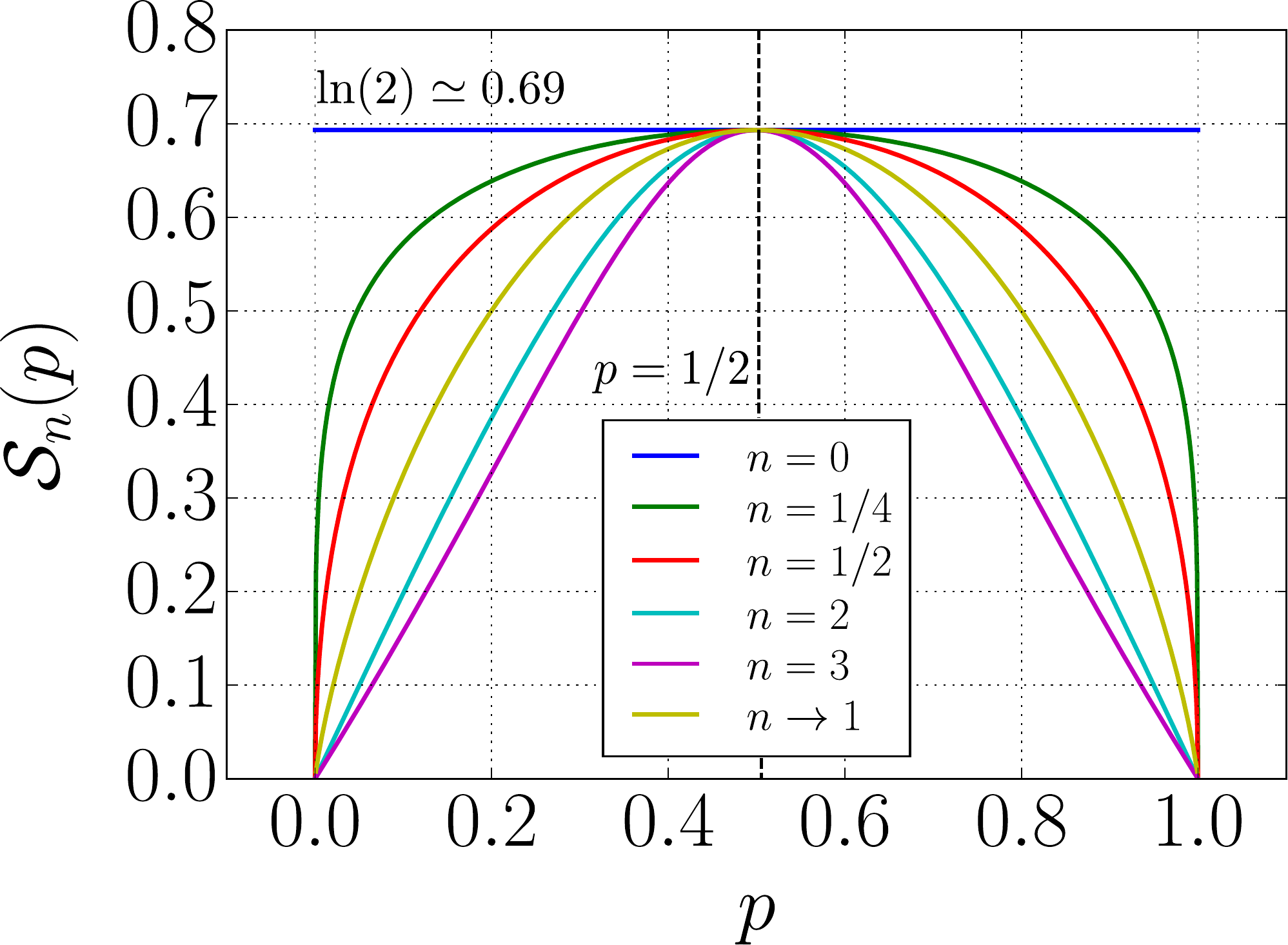}  
\end{tabular}
\caption{Evolution of the R\'enyi entropy $\mathcal{S}_n(p)$ as a function of the probability $p$ and the index $n$ (denoting the order of the R\'enyi entropy)
for a many-body quantum state $\ket{\Psi}$ fulfilling the form presented at Eq.~\eqref{state2}.}
\label{sre}
\end{figure}

\noindent
Using Eq.~\eqref{final_re_2} and \eqref{final_dre_2}, one immediately finds that the R\'enyi entropies \footnote{One can add that the R\'enyi entropy at $n=0$ is constant
and thus irrelevant to characterize the entanglement.}
are minimal (equal to zero) for
$p \in \{0,1\} ~\forall n \in \mathbb{R}^+ \backslash \{1\}$ and maximal for $p=1/2,~\forall n \in \mathbb{R}^+ \backslash \{1\}$, see Fig.~\ref{sre} (note that the 
previous discussion is also valid for the von Neumann entropy). The last statement corresponds to the equiprobability condition where $\ket{\Psi}$ consists of a superposition of two quantum states with 
the same probability and for which the R\'enyi entropies are maximal and characterized by the maximal value $\mathrm{max}_p(\mathcal{S}_n(p)) = \mathcal{S}_n(p=1/2) = \mathrm{log}(2)$. \\

The generalization to $N$ quantum states fulfilling the equiprobability condition for $\ket{\Psi}$ is straightforward. Indeed, it yields for 
the maximal value of the R\'enyi entropies $\mathrm{max}(\mathcal{S}_n) = \mathrm{log}(N)$. In the following, we give the proof that the R\'enyi entropies reach
their maximum for a same set of coefficients $\{\Pi_j\}$, \textit{ie.} the diagonal coefficients of the reduced density matrix $\hat{\rho}_A$, which can be determined using the equiprobability condition.
To do so, one relies on the Lagrangian multiplier method where $\lambda$ is the Lagrangian multiplier, the constraint
is $\sum_j \Pi_j - 1 = 0$ (one recall that the normalization of the quantum state $\ket{\Psi}$ implies the normalization of the reduced density matrix $\hat{\rho}_A$)
and the physical quantity to maximize is $\mathcal{S}_n(\hat{\rho}_A)$. The Lagrangian is defined by 
\begin{equation}
 \mathcal{L}(\Pi_j, \lambda) = \mathcal{S}_n(\hat{\rho}_A) - \lambda \left( 1-\sum_l \Pi_l \right).
\end{equation}

\noindent
To solve this problem, one needs to investigate the different derivatives of the Lagrangian with respect to each variable : 

\begin{align}
& \frac{\partial \mathcal{L}(\Pi_j, \lambda)}{\partial \lambda} = 0 \Rightarrow \sum_l \Pi_l = 1.
\end{align}

\noindent
The previous condition corresponds to the constraint of the problem, \textit{ie.} to the normalization of the many-body quantum state $\ket{\Psi}$. We then investigate 
the derivative with respect to each coefficient $\Pi_j$,

\begin{align}
& \frac{\partial \mathcal{L}(\Pi_j, \lambda)}{\partial \Pi_j} = \frac{\partial}{\partial \Pi_j} \left \{ 
\frac{1}{1-n} \mathrm{log}\left[\mathrm{Tr}(\hat{\rho}_A^{n}) \right] - \lambda \left( 1-\sum_l \Pi_l \right) \right \} = 0, \\
& \frac{\partial \mathcal{L}(\Pi_j, \lambda)}{\partial \Pi_j} = \frac{\partial}{\partial \Pi_j} \left \{ 
\frac{1}{1-n} \mathrm{log}\left(\sum_l \Pi_l^{n} \right) - \lambda \left( 1-\sum_l \Pi_l \right) \right \} = 0, \\
& \displaystyle \frac{\partial \mathcal{L}(\Pi_j, \lambda)}{\partial \Pi_j} = \displaystyle \frac{n}{1-n} \frac{\Pi_j^{n-1}}{\sum_l \Pi_l^n} + \lambda = 0,~\forall \Pi_j.
\end{align}

\noindent
Therefore, one has the two following conditions

\begin{equation}
\Pi_j = \left( \frac{n-1}{n}\lambda \sum_l \Pi_l^n \right)^{\frac{1}{n-1}} \forall \Pi_j,~~~\sum_l \Pi_l = 1. 
\label{conditions_lagr}
\end{equation}

\noindent
Then, by multiplying the first one by $\sum_j$, one obtains

\begin{equation}
 \sum_j \left( \frac{n-1}{n} \lambda \sum_l \Pi_l^n \right)^{\frac{1}{n-1}} = 1 ~~\Rightarrow ~~ \lambda = \mathrm{dim}(\hat{\rho}_A)^{1-n} 
 \left(\frac{n-1}{n}\sum_l \Pi_l^n \right)^{-1}.
\end{equation}

\noindent
Finally, by injecting the latter in the first condition of Eq.~\eqref{conditions_lagr}, it yields for the coefficients $\{\Pi_j\}$ of the reduced density 
matrix $\hat{\rho}_A$,

\begin{align}
& \Pi_j = \left( \frac{n-1}{n} \sum_l \Pi_l^n \right)^\frac{1}{n-1} \frac{1}{\mathrm{dim}(\hat{\rho}_A)}  \left( \frac{n-1}{n} \sum_l \Pi_l^n \right)^\frac{-1}{n-1}, \\
& \Pi_j = \frac{1}{\mathrm{dim}(\hat{\rho}_A)} = \frac{1}{N}.
\end{align}

\noindent
Finally, using a Lagrangian method, we have shown that the R\'enyi entropies reach their maximum for a same set of coefficients $\{\Pi_j \}$
determined by the equiprobability condition. For a many-body quantum state $\ket{\Psi}$ in a superposition of $N$ states, it yields $\Pi_j = 1/N,~\forall j \in [|1,N|]$.  

\paragraph{Properties of the R\'enyi entropies}
We now turn to several properties of the R\'enyi entropies. In what follows, $(\hat{\rho}_A,\hat{\rho}'_A) \in
\mathcal{M}_{N}(\mathbb{R})^2$ denotes two normalized reduced density matrices.

\begin{itemize}
 \item \textit{Continuity} : $\mathcal{S}_n(\hat{\rho}_A)$ continuous in $\hat{\rho}_A$. 
 \item \textit{Normalization} : $\mathcal{S}_n(\hat{\rho}_A) = \mathrm{log}(N)$ for $\hat{\rho}_A = \mathrm{diag}(1/N)$.
 \item \textit{Unitary invariance} : $U \in \mathcal{M}_{N}(\mathbb{C})$ corresponds to a unitary matrix. 
 \begin{align}
 & \mathcal{S}_n(U \hat{\rho}_A U^{\dag}) = \frac{1}{1-n} \mathrm{log} \left \{ \mathrm{Tr}[(U \hat{\rho}_A U^{\dag})^n ] \right \}.
 \end{align}
 
 \noindent
 Then, using the following equation 
 \begin{equation}
 (U \hat{\rho}_A U^{\dag})^n = (U \hat{\rho}_A U^{\dag})(U \hat{\rho}_A U^{\dag})...(U \hat{\rho}_A U^{\dag}) = U \hat{\rho}_A^n U^{\dag},
 \end{equation}

 \noindent
 one obtains for $\mathcal{S}_n(U \hat{\rho}_A U^{\dag})$,
 \begin{align}
 & \mathcal{S}_n(U \hat{\rho}_A U^{\dag}) = \frac{1}{1-n} \mathrm{log}\left[ \mathrm{Tr} \left(U \hat{\rho}_A^n U^{\dag} \right) \right], \\
 & \mathcal{S}_n(U \hat{\rho}_A U^{\dag}) = \frac{1}{1-n} \mathrm{log}\left[ \mathrm{Tr} \left(U^{\dag} U \hat{\rho}_A^n \right) \right], \\
 & \mathcal{S}_n(U \hat{\rho}_A U^{\dag}) = \frac{1}{1-n} \mathrm{log}\left[ \mathrm{Tr} \left( \hat{\rho}_A^n \right) \right] = \mathcal{S}_n(\hat{\rho}_A).
 \end{align}
 
\item \textit{Additivity} : 
\begin{equation}
 \mathcal{S}_n(\hat{\rho}_A \otimes \hat{\rho}'_A) = \frac{1}{1-n} \mathrm{log}\{ \mathrm{Tr} \left[(\hat{\rho}_A \otimes \hat{\rho}'_A)^n \right] \}.
\end{equation}

\noindent
Then, using the following equation,

\begin{equation}
(\hat{\rho}_A \otimes \hat{\rho}'_A)^n = (\hat{\rho}_A \otimes \hat{\rho}'_A)(\hat{\rho}_A \otimes \hat{\rho}'_A)...(\hat{\rho}_A \otimes \hat{\rho}'_A)
 = \hat{\rho}_A^n \otimes (\hat{\rho}'_A)^n,
\end{equation}

\noindent
one finds for $\mathcal{S}_n(\hat{\rho}_A \otimes \hat{\rho}'_A)$

\begin{align}
 & \mathcal{S}_n(\hat{\rho}_A \otimes \hat{\rho}'_A) = \frac{1}{1-n} \mathrm{log}\left[ \mathrm{Tr} \left(\hat{\rho}_A^n \otimes (\hat{\rho}'_A)^n \right) \right], \\
 & \mathcal{S}_n(\hat{\rho}_A \otimes \hat{\rho}'_A) = \frac{1}{1-n} \mathrm{log}\left \{ \mathrm{Tr}(\hat{\rho}_A^n) \mathrm{Tr}[(\hat{\rho}'_A)^n] \right \}, \\
 & \mathcal{S}_n(\hat{\rho}_A \otimes \hat{\rho}'_A) = \frac{1}{1-n} \{ \mathrm{log}\left[ \mathrm{Tr}(\hat{\rho}_A^n)\right] +
 \mathrm{log}\left[ \mathrm{Tr}((\hat{\rho}'_A)^n)\right] \}, \\
 & \mathcal{S}_n(\hat{\rho}_A \otimes \hat{\rho}'_A) = \mathcal{S}_n(\hat{\rho}_A) + \mathcal{S}_n(\hat{\rho}'_A).
\end{align}

 \item \textit{No arithmetic mean} : 
 \begin{align}
  & \mathcal{S}_n(\hat{\rho}_A \oplus \hat{\rho}'_A) = \frac{1}{1-n} \mathrm{log}\{ \mathrm{Tr}\left[(\hat{\rho}_A \oplus \hat{\rho}'_A)^n \right] \}, \\
  & \mathcal{S}_n(\hat{\rho}_A \oplus \hat{\rho}'_A) = \frac{1}{1-n} \mathrm{log}\left \{ \mathrm{Tr}(\hat{\rho}_A^n) + \mathrm{Tr}[(\hat{\rho}'_A)^n]\right \}, \\
  & \mathcal{S}_n(\hat{\rho}_A \oplus \hat{\rho}'_A) \neq \frac{\mathrm{Tr}(\hat{\rho}_A)}{\mathrm{Tr}(\hat{\rho}_A + \hat{\rho}'_A)} \mathcal{S}_n(\hat{\rho}_A)
  + \frac{\mathrm{Tr}(\hat{\rho}'_A)}{\mathrm{Tr}(\hat{\rho}_A + \hat{\rho}'_A)} \mathcal{S}_n(\hat{\rho}'_A), \\
  & \mathcal{S}_n(\hat{\rho}_A \oplus \hat{\rho}'_A) \neq \frac{1}{2} \left[ 
  \mathcal{S}_n(\hat{\rho}_A) + \mathcal{S}_n(\hat{\rho}'_A) \right].
  \label{am}
 \end{align}
\end{itemize}

%% file: text/appendix_13.tex
\setstretch{1.0} 

\chapter{\label{appendix_gz_lrti} $G_z$ spin-spin correlations in the $z$ polarized phase of the long-range transverse Ising chain}

In this section, we derive the expression of the $G_z$ equal-time connected spin-spin correlation function along the $z$ axis in the $z$ polarized phase of the
long-range transverse Ising (LRTI) chain. To drive the spin chain far from equilibrium, a sudden global quench on the spin exchange energy $J>0$ along the $x$ direction of the Bloch sphere 
is considered while both the power-law exponent $\alpha$ and the transverse field $h$ are fixed. The latter implies that $J_\mathrm{i},J_{\mathrm{f}} \ll h$ in order to remain in the $z$ polarized phase,
see Eq.~\eqref{H} and Fig.~\ref{fig:phase_diagram}.
The calculation, based on the linear spin wave theory, follows the same steps used at Appendix.~\ref{appendix_gx_lrti}. In the following, the long-range Ising
chain is assumed to contain $N_s$ lattice sites and $\hbar$ and $a$ the lattice spacing are fixed to unity by convention. \\

The $G_z$ spin fluctuations along the $z$ axis may be written as $G_z(R,t) = G_{z,0}(R,t) - G_{z,0}(R,0)$ where

\begin{equation}
G_{z,0}(R,t) = \langle \hat{S}_R^z(t) \hat{S}_0^z(t) \rangle - \langle \hat{S}_R^z(t)\rangle \langle \hat{S}_0^z(t) \rangle.
\end{equation}

\noindent
The calculation is mainly based on the properties of the bosonic Bogolyubov theory. As a consequence, we first need to express $G_z$ in terms of the
post-quench bosonic operators in momentum space $\hat{a}_{k,\mathrm{f}} = \hat{a}_k$ and $\hat{a}^{\dag}_{k,\mathrm{f}} = \hat{a}_k^{\dag}$. To do so,
the following Holstein-Primakoff transformation, valid in the $z$ polarized phase, is considered  

\begin{equation}
 \hat{S}^z_R = \frac{1}{2} - \hat{a}^{\dag}_R \hat{a}_R,~~~ \hat{S}^x_R = \frac{1}{2}(\hat{a}_R^{\dag}+ \hat{a}_R),~~~
 \hat{S}^y_R = \frac{-1}{2i}(\hat{a}_R^{\dag} - \hat{a}_R).
\end{equation}

\noindent
By injecting this transformation in the previous expression of $G_z$ and relying on the bosonic version of the Wick theorem, we obtain 

\begin{align}
& G_z(R,t) = \langle \hat{a}^{\dag}_R \hat{a}^{\dag}_0 \rangle_t  \langle \hat{a}_R \hat{a}_0 \rangle_t + \langle \hat{a}^{\dag}_R \hat{a}_0 \rangle_t
\langle \hat{a}_R \hat{a}^{\dag}_0 \rangle_t - \langle \hat{a}^{\dag}_R \hat{a}^{\dag}_0 \rangle_0 \langle \hat{a}_R \hat{a}_0 \rangle_0 - \langle \hat{a}^{\dag}_R 
\hat{a}_0 \rangle_0 \langle \hat{a}_R \hat{a}^{\dag}_0 \rangle_0, \nonumber \\
& G_z(R,t) = |\langle \hat{a}^{\dag}_R \hat{a}^{\dag}_0 \rangle_t|^2 + |\langle \hat{a}_R^{\dag} \hat{a}_0 \rangle_t |^2 - |\langle \hat{a}_R^{\dag} \hat{a}_0^{\dag}
\rangle_0 |^2 - |\langle \hat{a}_R^{\dag} \hat{a}_0\rangle_0|^2.
\end{align}

\noindent
To find the previous expression, we assume $R \neq 0$ and use the notation $\langle ... \rangle_t \equiv \langle \Psi(t) | ... | \Psi(t) \rangle$
with $\ket{\Psi(t)} = e^{-i\hat{H}_\mathrm{f}t}\ket{\Psi_0}$ ($\ket{\Psi_0} = \ket{\Psi_{\mathrm{gs},\mathrm{i}}}$ denotes the ground state of the 
pre-quench Hamiltonian $\hat{H}_{\mathrm{i}}$ defined by the pre-quench interaction parameter $J_{\mathrm{i}}/h$).
We now express the latter in the reciprocal space \textit{via} a Fourier transform of the post-quench bosonic
operators in real space. It yields for $G_z$ the following formula 

\begin{align}
& G_z(R,t) = \frac{1}{N_s^2} \left \{ \left|\sum_k e^{ikR} \langle \hat{a}_k^{\dag} \hat{a}_{-k}^{\dag} \rangle_t \right|^2 - \left|\sum_k e^{ikR} \langle \hat{a}_k^{\dag}
\hat{a}_{-k}^{\dag} \rangle_0 \right|^2 + \left|\sum_k e^{ikR} \langle \hat{a}_k^{\dag} \hat{a}_{k} \rangle_t \right|^2 - \left|\sum_k e^{ikR} \langle \hat{a}_k^{\dag} \hat{a}_{k} \rangle_0 \right|^2 \right \}.
\label{gz_inter}
\end{align}

\noindent
In what follows, we present the different properties of the bosonic Bogolyubov theory which are used to deduce the final theoretical expression of the $G_z$
spin fluctuations. 

\begin{enumerate}[leftmargin=*]
\item The Bogolyubov transformation, allowing us to express the post-quench bosonic operators in terms of the post-quench (bosonic) Bogolyubov operators, is given by 
 
 \begin{equation}
\hat{a}_{k,\mathrm{f}}^{\dag}(t) = \hat{a}_k^{\dag}(t) = u_k^\mathrm{f} \hat{\beta}_{k,\mathrm{f}}^{\dag}(t) + v_k^\mathrm{f} \hat{\beta}_{-k,\mathrm{f}}(t),~~~
\hat{a}_{k,\mathrm{f}}(t) = \hat{a}_k(t) = u_k^\mathrm{f} \hat{\beta}_{k,\mathrm{f}}(t) + v_k^\mathrm{f} \hat{\beta}_{-k,\mathrm{f}}^{\dag}(t),
 \end{equation}

\noindent
where the coefficients $u_k^\mathrm{f}$ and $v_k^\mathrm{f}$ are defined at Appendix.~\ref{appendix_gx_lrti} for the LRTI chain in the $z$ polarized phase. 
  
\item Since the post-quench Bogolyubov operators diagonalize the post-quench Hamiltonian $\hat{H}_{\mathrm{f}}$ of the LRTI chain, their real time evolution in the
Heisenberg picture takes a simple form given as 

\begin{equation}
\hat{\beta}_{k,\mathrm{f}}^{\dag}(t) = e^{i E_k^\mathrm{f} t} \hat{\beta}_{k,\mathrm{f}}^{\dag}(0), ~~~ \hat{\beta}_{k,\mathrm{f}}(t) = e^{-i E_k^\mathrm{f} t} 
\hat{\beta}_{k,\mathrm{f}}(0),
\end{equation}

\noindent
where $E_k^{\mathrm{f}}$ refers to the post-quench excitation spectrum of the LRTI chain in the $z$ polarized phase (see Appendix.~\ref{appendix_gx_lrti}).

\item The pre- and post-quench Bogolyubov operators are linked using the continuity of the bosonic operators in momentum space. In other words, using both following 
relations $\hat{a}_{k,\mathrm{f}}(0) = \hat{a}_{k,\mathrm{i}}$ and $\hat{a}_{k,\mathrm{f}}^{\dag}(0) = \hat{a}_{k,\mathrm{i}}^{\dag}$, one obtains 
\footnote{Note that the indices $\mathrm{i}$ and $\mathrm{f}$ for the scalars are now written as exponents for clarity. Indeed, this notation avoids to have,
with the quasimomentum $k$, three aligned indices.} 

\begin{equation}
\hat{\beta}_{k,\mathrm{f}}(0) = M_k^{\mathrm{i},\mathrm{f}} \hat{\beta}_{k,\mathrm{i}} - N_k^{\mathrm{i},\mathrm{f}} \hat{\beta}_{-k,\mathrm{i}}^{\dag},~~~
\hat{\beta}_{k,\mathrm{f}}^{\dag}(0) = M_k^{\mathrm{i},\mathrm{f}} \hat{\beta}_{k,\mathrm{i}}^{\dag} - N_k^{\mathrm{i},\mathrm{f}} \hat{\beta}_{-k,\mathrm{i}},
\end{equation}

\noindent
with $M_k^{\mathrm{i},\mathrm{f}} = u_k^\mathrm{i} u_k^\mathrm{f} - v_k^\mathrm{i} v_k^\mathrm{f}$ and 
$N_k^{\mathrm{i},\mathrm{f}} = u_k^\mathrm{i} v_k^\mathrm{f} - u_k^\mathrm{f} v_k^\mathrm{i}$. 

\item By definition, the initial state $\ket{\Psi_0}$ corresponding to the ground state of $\hat{H}_{\mathrm{i}}$ does not support any quasiparticle leading 
to both relations

\begin{equation}
\hat{\beta}_{k,\mathrm{i}} \ket{\Psi_0} = 0, ~~~ \bra{\Psi_0} \hat{\beta}_{k,\mathrm{i}}^{\dag} = 0. 
\end{equation}
\end{enumerate}

\noindent
From now, one has to calculate each correlator of Eq.~\eqref{gz_inter}. We first start to investigate the correlators 
$\langle \hat{a}_k^{\dag} \hat{a}_{-k}^{\dag} \rangle_t = \langle \hat{a}_k^{\dag}(t) \hat{a}_{-k}^{\dag}(t) \rangle$ and 
$\langle \hat{a}_k^{\dag} \hat{a}_{-k}^{\dag} \rangle_0 = \langle \hat{a}_k^{\dag}(0) \hat{a}_{-k}^{\dag}(0) \rangle$. One obtains the following expressions using 
the previous properties of the bosonic Bogolyubov theory

\begin{align}
 & \langle \hat{a}_k^{\dag} \hat{a}_{-k}^{\dag} \rangle_t = -N_k^{\mathrm{i},\mathrm{f}} M_k^{\mathrm{i},\mathrm{f}} \left[\left(u_k^\mathrm{f}\right)^2
 e^{i2E_k^\mathrm{f} t} + \left(v_k^\mathrm{f}\right)^2 e^{-i2E_k^\mathrm{f}t} \right] + u_k^\mathrm{f} v_k^\mathrm{f} \left[ \left(N_k^{\mathrm{i},\mathrm{f}}
 \right)^2 + \left(M_k^{\mathrm{i},\mathrm{f}} \right)^2 \right], \\
 & \langle \hat{a}_k^{\dag} \hat{a}_{-k}^{\dag} \rangle_0 = -N_k^{\mathrm{i},\mathrm{f}} M_k^{\mathrm{i},\mathrm{f}} \left[ \left(u_k^\mathrm{f}\right)^2 + \left(v_k^\mathrm{f}\right)^2 \right] + 
 u_k^\mathrm{f} v_k^\mathrm{f} \left[ \left(N_k^{\mathrm{i},\mathrm{f}}  \right)^2 + \left(M_k^{\mathrm{i},\mathrm{f}} \right)^2 \right].
\end{align}

\noindent
We now turn to the two last correlators of Eq.~\eqref{gz_inter} corresponding to $\langle \hat{a}_k^{\dag} \hat{a}_{k} \rangle_t = \langle \hat{a}_k^{\dag}(t) 
\hat{a}_{k}(t) \rangle$ and $\langle \hat{a}_k^{\dag} \hat{a}_{k} \rangle_0 = \langle \hat{a}_k^{\dag}(0) \hat{a}_{k}(0)\rangle$. 

\begin{align}
 & \langle \hat{a}_k^{\dag} \hat{a}_{k} \rangle_t = \left( u_k^\mathrm{f} N_k^{\mathrm{i},\mathrm{f}} \right)^2 + \left( v_k^\mathrm{f} M_k^{\mathrm{i},\mathrm{f}}
 \right)^2 -2u_k^\mathrm{f} v_k^\mathrm{f} M_k^{\mathrm{i},\mathrm{f}} N_k^{\mathrm{i},\mathrm{f}} \cos(2E_k^\mathrm{f} t), \\
 & \langle \hat{a}_k^{\dag} \hat{a}_{k} \rangle_0 = \left( u_k^\mathrm{f} N_k^{\mathrm{i},\mathrm{f}} \right)^2 + 
 \left( v_k^\mathrm{f} M_k^{\mathrm{i},\mathrm{f}} \right)^2 -2u_k^\mathrm{f} v_k^\mathrm{f} M_k^{\mathrm{i},\mathrm{f}} N_k^{\mathrm{i},\mathrm{f}}. 
\end{align}

\noindent
We now express the previous correlators in terms of the pre- and post-quench quasimomentum-dependent functions $\mathcal{A}_k^{(\mathrm{i},\mathrm{f})}$ and
$\mathcal{B}_k^{(\mathrm{i},\mathrm{f})}$ defined at Appendix.~\ref{appendix_gx_lrti}. Both functions appear when expressing the pre- and post-quench Hamiltonians
of the LRTI chain in the $z$ polarized phase under the general quadratic Bose form [see Eq.~\eqref{H_quadratic} for instance]. 
In what follows, we provide preliminary results simplifying the calculation of the different correlators  

\begin{align}
& u_k^\mathrm{f} v_k^\mathrm{f}  = - \frac{\mathcal{B}_k^\mathrm{f}}{2 E_k^\mathrm{f}}, \\
& \left(M_k^{\mathrm{i},\mathrm{f}}\right)^2 = \frac{1}{2E_k^{\mathrm{i}}E_k^\mathrm{f}}\left(\mathcal{A}_k^\mathrm{i} \mathcal{A}_k^\mathrm{f} + E_k^\mathrm{i}E_k^\mathrm{f}-
\mathcal{B}_k^\mathrm{i}\mathcal{B}_k^\mathrm{f} \right), \\
& \left(N_k^{\mathrm{i},\mathrm{f}}\right)^2 = \frac{1}{2E_k^{\mathrm{i}}E_k^\mathrm{f}}\left(\mathcal{A}_k^\mathrm{i} \mathcal{A}_k^\mathrm{f} - E_k^\mathrm{i}E_k^\mathrm{f}-
\mathcal{B}_k^\mathrm{i}\mathcal{B}_k^\mathrm{f} \right), \\
& M_k^{\mathrm{i},\mathrm{f}}N_k^{\mathrm{i},\mathrm{f}} = \frac{1}{2E_k^{\mathrm{i}}E_k^\mathrm{f}}\left(\mathcal{B}_k^\mathrm{i}\mathcal{A}_k^\mathrm{f}- \mathcal{B}_k^\mathrm{f}
\mathcal{A}_k^\mathrm{i} \right).
\end{align}

\noindent
Finally, we find that the correlators $\langle \hat{a}_k^{\dag} \hat{a}_k \rangle_t$ and $\langle \hat{a}_k^{\dag} \hat{a}_k \rangle_0$ have the following form

\begin{align}
 & \langle \hat{a}_k^{\dag} \hat{a}_k \rangle_t = \frac{1}{2E_k^{\mathrm{i}} \left(E_k^\mathrm{f}\right)^2} \left \{ \mathcal{A}_k^\mathrm{i} \left( \mathcal{A}_k^\mathrm{f} \right)^2 - 
 \mathcal{B}_k^\mathrm{i} \mathcal{B}_k^\mathrm{f} \mathcal{A}_k^\mathrm{f} - E_k^\mathrm{i} \left(E_k^\mathrm{f}\right)^2  + \left[\mathcal{B}_k^\mathrm{i}\mathcal{A}_k^\mathrm{f} \mathcal{B}_k^\mathrm{f} -\mathcal{A}_k^\mathrm{i} 
 \left(\mathcal{B}_k^\mathrm{f}\right)^2 \right] \cos(2E_k^\mathrm{f}t) \right \} \\ 
 & \langle \hat{a}_k^{\dag} \hat{a}_k \rangle_0 = \frac{1}{2E_k^{\mathrm{i}} \left(E_k^\mathrm{f}\right)^2} \left[ \mathcal{A}_k^\mathrm{i} \left( \mathcal{A}_k^\mathrm{f} \right)^2 - 
 \mathcal{B}_k^\mathrm{i} \mathcal{B}_k^\mathrm{f} \mathcal{A}_k^\mathrm{f} - E_k^\mathrm{i} \left(E_k^\mathrm{f}\right)^2  + \mathcal{B}_k^\mathrm{i}\mathcal{A}_k^\mathrm{f}
 \mathcal{B}_k^\mathrm{f} -\mathcal{A}_k^\mathrm{i} \left(\mathcal{B}_k^\mathrm{f}\right)^2  \right] \nonumber \\
 &  \langle \hat{a}_k^{\dag} \hat{a}_k \rangle_0 = \frac{1}{2E_k^\mathrm{i}}\left( \mathcal{A}_k^\mathrm{i} - E_k^\mathrm{i} \right) = \left(v_k^{\mathrm{i}}\right)^2 
\end{align}

\noindent
For the two last correlators $\langle \hat{a}_k^{\dag} \hat{a}_{-k}^{\dag} \rangle_t$ and $\langle \hat{a}_k^{\dag} \hat{a}_{-k}^{\dag} \rangle_0$, one obtains 

\begin{align}
& \langle \hat{a}_k^{\dag} \hat{a}_{-k}^{\dag} \rangle_t =  \frac{1}{2E_k^{\mathrm{i}} \left(E_k^\mathrm{f}\right)^2} \left \{ \left( \mathcal{B}_k^\mathrm{f}\mathcal{A}_k^\mathrm{i}
-\mathcal{B}_k^\mathrm{i}\mathcal{A}_k^\mathrm{f} \right) \left[ \mathcal{A}_k^\mathrm{f} \cos(2 E_k^\mathrm{f}t) + iE_k^\mathrm{f}\sin(2E_k^\mathrm{f}t) \right] + 
\left[ \mathcal{B}_k^\mathrm{i} \left(\mathcal{B}_k^\mathrm{f}\right)^2 - \mathcal{B}_k^\mathrm{f}\mathcal{A}_k^\mathrm{i}\mathcal{A}_k^\mathrm{f} \right] \right \} \\
& \langle \hat{a}_k^{\dag} \hat{a}_{-k}^{\dag} \rangle_0 =  \frac{1}{2E_k^{\mathrm{i}} \left(E_k^\mathrm{f}\right)^2} \left[ \left( \mathcal{B}_k^\mathrm{f}\mathcal{A}_k^\mathrm{i}
-\mathcal{B}_k^\mathrm{i}\mathcal{A}_k^\mathrm{f} \right) \mathcal{A}_k^\mathrm{f} + 
\mathcal{B}_k^\mathrm{i} \left(\mathcal{B}_k^\mathrm{f}\right)^2 - \mathcal{B}_k^\mathrm{f}\mathcal{A}_k^\mathrm{i}\mathcal{A}_k^\mathrm{f} \right]  =
-\frac{\mathcal{B}_k^\mathrm{i}}{2E_k^\mathrm{i}} = u_k^\mathrm{i} v_k^\mathrm{i} 
\end{align}

\noindent
Note that it is consistent to find $\langle \hat{a}_k^{\dag} \hat{a}_k \rangle_0 = \left(v_k^{\mathrm{i}}\right)^2$ and 
$\langle \hat{a}_k^{\dag} \hat{a}_{-k}^{\dag} \rangle_0 = u_k^\mathrm{i} v_k^\mathrm{i}$. Indeed, both previous correlators can be directly
computed \textit{via} the properties of the bosonic Bogolyubov transformation related to the pre-quench Hamiltonian $\hat{H}_{\mathrm{i}}$.
More precisely, using the continuity of the bosonic operators at time $t=0$ and the Bogolyubov transformation for the pre-quench Hamiltonian 
$\hat{H}_{\mathrm{i}}$ defined by $\hat{a}_{k,\mathrm{i}}^{\dag} = u_k^\mathrm{i} \hat{\beta}_{k,\mathrm{i}}^{\dag}
+ v_k^\mathrm{i} \hat{\beta}_{-k,\mathrm{i}}$ and $\hat{a}_{k,\mathrm{i}} = u_k^\mathrm{i} \hat{\beta}_{k,\mathrm{i}} + v_k^\mathrm{i} \hat{\beta}_{-k,\mathrm{i}}^{\dag}$, 
one can easily verify the previous expressions. \\

\noindent
Finally, for a sudden global quench confined in the $z$ polarized phase of the 1D long-range transverse Ising model, the space-time spin fluctuations $G_z(R,t)$ have the 
following analytical form in the thermodynamic limit

\begin{align}
 &G_z(R,t) = \Big | \int_{\mathcal{B}} \frac{\mathrm{d}k}{2\pi}e^{ikR}  \frac{1}{2E_k^{\mathrm{i}} \left(E_k^\mathrm{f}\right)^2} \Big
 \{ \left( \mathcal{B}_k^\mathrm{f}\mathcal{A}_k^\mathrm{i} -\mathcal{B}_k^\mathrm{i}\mathcal{A}_k^\mathrm{f} \right) \left[
 \mathcal{A}_k^\mathrm{f} \cos(2 E_k^\mathrm{f}t) + iE_k^\mathrm{f}\sin(2E_k^\mathrm{f}t) \right] \nonumber \\
 & ~~~~~~~~~~~~~ + \left[ \mathcal{B}_k^\mathrm{i} \left(\mathcal{B}_k^\mathrm{f}\right)^2 - \mathcal{B}_k^\mathrm{f}\mathcal{A}_k^\mathrm{i}\mathcal{A}_k^\mathrm{f} \right] \Big \} \Big|^2 \nonumber \\
 & ~~~~~~~~~~~~~ + \Big| \int_{\mathcal{B}} \frac{\mathrm{d}k}{2\pi}e^{ikR} \frac{1}{2E_k^{\mathrm{i}} \left(E_k^\mathrm{f}\right)^2} \Big\{ \mathcal{A}_k^\mathrm{i} 
 \left( \mathcal{A}_k^\mathrm{f} \right)^2 - \mathcal{B}_k^\mathrm{i} \mathcal{B}_k^\mathrm{f} \mathcal{A}_k^\mathrm{f} - E_k^\mathrm{i} \left(E_k^\mathrm{f}\right)^2 
 \nonumber \\  
 &  ~~~~~~~~~~~~~   + \left[ \mathcal{B}_k^\mathrm{i}\mathcal{A}_k^\mathrm{f} \mathcal{B}_k^\mathrm{f} -\mathcal{A}_k^\mathrm{i} \left(\mathcal{B}_k^\mathrm{f}\right)^2 \right] \cos(2E_k^\mathrm{f}t)\Big \}
 \Big |^2 
 \nonumber \\
 & ~~~~~~~~~~~~~ - \left| \int_{\mathcal{B}} \frac{\mathrm{d}k}{2\pi}e^{ikR} \left( v_k^\mathrm{i} \right)^2 \right|^2 
 - \left| \int_{\mathcal{B}} \frac{\mathrm{d}k}{2\pi}e^{ikR} u_k^\mathrm{i} v_k^\mathrm{i} \right|^2,
\end{align}

\noindent
where $\mathcal{B}$ refers to the first Brillouin zone, \textit{ie.} $\mathcal{B} = [-\pi, \pi]$.

%% file: text/appendix_14.tex
\setstretch{1.0} 

\chapter{Résumé en français}

Dans cette thèse, nous nous sommes intéressés à la propagation hors équilibre de corrélations dans des modèles quantiques sur réseau isolés
de leur environnement. Afin d'être mis hors équilibre, ces derniers sont initialement préparés dans l'état fondamental de leur Hamiltonien 
pour une certaine valeur du paramètre d'interaction et évoluent en temps avec le même Hamiltonien mais ayant une valeur différente du 
paramètre d'interaction. Ce protocole de mise hors équilibre d'un système quantique est communément appelé 'quench global et soudain'. 
L'objectif principal a été de comprendre comment l'information se propage dans un système quantique corrélé et d'expliquer notamment des
résultats \textit{a priori} contradictoires dans la littérature. Pour ce faire, nous avons employé une approche combinant à la fois des 
études analytiques mais aussi numériques. \\

En ce qui concerne l'approche analytique, celle-ci repose sur une approximation de type 'champ-moyen' ainsi que sur une théorie de
quasiparticules de type 'Bogolyubov'. Elle a notamment permis de dévoiler une expression générique des fonctions de corrélation pour 
des systèmes quantiques de particules ou de spins interagissant à courte ou longue portée sur un réseau hypercubique. Cette forme générique
consiste en une superposition cohérente d'ondes planes, représentant des paires de quasiparticles libres, se propageant dans l'espace et le
temps et pondérées par une fonction d'amplitude dépendant du quasimoment et de l'observable considérée. En utilisant la méthode de phase 
stationnaire, nous avons montré que la région causale de ces corrélations présente une structure double universelle. Cette dernière est 
composée non seulement d'une borne, appelée 'borne de corrélation' mais également d'extrema locaux dans son voisinage définissant la partie
externe et interne des corrélations résolues en temps et distance respectivement. \\
Dans le cas d'interactions de courte portée, nous avons prouvé que les deux structures se propagent ballistiquement avec des vitesses généralement
différentes et reliées à la vitesse de groupe et de phase du spectre d'excitation du Hamiltonien après quench. Plus précisément, la borne de
corrélation est caractérisée par une vitesse égale à deux fois la vitesse de groupe maximale tandis que les extrema se propagent avec une 
vitesse correspondant à deux fois la vitesse de phase au quasimoment où la vitesse de groupe est maximale. \\
Pour des interactions de longue portée de type loi de puissance, les lois d'échelle peuvent être sensiblement modifiées à cause de la possible
divergence de la vitesse de groupe. Pour ce cas spécifique correspondant au régime dit quasi-local, une double structure aux lois d'échelle algébriques
a été présentée. Bien que la borne des corrélations se propage toujours moins rapidement que ballistiquement, les extrema locaux se propagent 
plus rapidement que ballistiquement et ballistiquement pour des systèmes quantiques non gappés et gappés respectivement. Cependant pour le régime
local caractérisé par une valeur maximale bien définie de la vitesse de groupe et impliquant une décroissance rapide des interactions de longue
portée, nous avons retrouvé un comportement similaire au cas d'interactions de courte portée pour la propagation des corrélations. En effet, 
celles-ci présentent une structure double linéaire caractérisée par les mêmes vitesses de propagation qu'énoncées précédemment pour la borne de 
corrélation et les extrema locaux. \\  

Concernant l'approche numérique, des simulations ont été effectuées en utilisant des techniques de réseaux de tenseurs. Ces techniques, basées 
sur une analyse de l'intrication au sein des systèmes quantiques, fournissent des données quasi-exactes dans le cas de réseaux unidimensionnels.
Par conséquent, ces approches numériques permettent d'étudier la propagation de corrélations au-delà de l'approximation de champ moyen et ainsi
attester de la validité de nos prédictions théoriques. Par ailleurs, elles offrent la possibilité d'étudier la propagation hors équilibre de 
corrélations dans des régimes où il n'existe pas (à ce jour) de solutions théoriques. Pour conclure, ces simulations ont été effectuées pour différents
systèmes quantiques unidimensionnels et sur réseau, à savoir le modèle de Bose-Hubbard dans la phase superfluide et isolante de Mott ainsi 
que les modèles de spins XY et d'Ising transverse.